%% file: paper.tex
  \documentclass[secnum,seceqn,secthm]{myelsart}
  \usepackage{url}

\input gafd
\usepackage{amssymb}
\usepackage{graphicx}
\usepackage{fancybox}
\usepackage{psboxit}
\usepackage{bm}
\graphicspath{{./fig/}{./png/}}

\newcommand{\cal}{\mathcal}

\makeatletter
 \@addtoreset{figure}{section}
 \def\thefigure{\arabic{section}.\arabic{figure}}
 \@addtoreset{table}{section}
 \def\thetable{\arabic{section}.\arabic{table}}
\makeatother

\begin{document}

\begin{frontmatter}

\title{Astrophysical magnetic fields and nonlinear dynamo theory}

\author[ad1]{Axel Brandenburg} and
\author[ad2]{Kandaswamy Subramanian}

\address[ad1]{NORDITA, Blegdamsvej 17, DK-2100 Copenhagen \O, Denmark. \\ 
email: brandenb@nordita.dk}

\address[ad2]{Inter University Centre for Astronomy and Astrophysics,
      Post Bag 4, Pune University Campus, Ganeshkhind, Pune 411 007, India. \\ 
      email: kandu@iucaa.ernet.in}

\begin{abstract}
The current understanding of astrophysical magnetic fields
is reviewed, focusing on their generation and 
maintenance by turbulence.
In the astrophysical context this generation is usually explained by
a self-excited dynamo, 
which involves flows that can amplify a
weak `seed' magnetic field exponentially fast.
Particular emphasis is placed on the nonlinear saturation of the dynamo.
Analytic and numerical results are discussed both for small scale
dynamos, which are completely isotropic, and for large scale dynamos,
where some form of parity breaking is crucial.
Central to the discussion of large scale dynamos is the so-called
alpha effect which explains the generation of a mean field if the
turbulence lacks mirror symmetry, i.e.\ if the flow has kinetic helicity.
Large scale dynamos produce small scale helical fields as a waste
product that quench the large scale dynamo and hence the alpha effect.
With this in mind, the microscopic theory of the alpha effect
is revisited in full detail and recent results for the loss of helical
magnetic fields are reviewed.
\end{abstract}

\end{frontmatter}
\endNoHyper

\tableofcontents
\newpage

\section{Introduction}

Magnetic fields are ubiquitous in the universe.
Our most immediate encounter with magnetic fields is the
Earth's field. This field is not only useful for navigation, but it
also protects us from hazardous cosmic ray particles.
Magnetic fields play an important role in various branches of astrophysics.
They are particularly important for angular momentum transport, without which
the sun and similar stars would not spin as slowly as they do today
\cite{RW92}. Magnetic fields are responsible for the loops and arcades
seen in X-ray images of the sun and in heating the coronae of stars
with outer convection zones \cite{Par94}. They 
play a crucial role in driving turbulence in accretion discs
providing the stresses needed for accretion.
Large scale fields in these discs are also thought to be involved 
in driving jets. A field permeating a rotating black hole
probably provides one of the most efficient ways
of extracting energy to power the jets from active galactic nuclei.
Magnetic fields with micro-gauss strength and coherence scales of 
order several kilo parsecs are also observed in nearby galaxies and perhaps
even in galaxies which have just formed. The magnetic field
strength in galactic spiral arms can be up to 30 microgauss (e.g.\ in M51).
Fields of order several micro-gauss and larger,
with even larger coherence scales, are seen in clusters of galaxies.
To understand the origin of magnetic fields in all these astrophysical
systems is a problem of great importance.

The universe may not have begun magnetized. There are
various processes such as battery effects, which can lead
to a weak magnetic field, from zero initial fields. Most of
these batteries lead to field strengths much weaker than the observed
field, as will be discussed further in \Sec{battery}. 
So some way of amplifying the field is required.
This is probably accomplished by the
conversion of kinetic energy into magnetic energy, a process generally
referred to as a dynamo; see Ref.~\cite{Kra93} for a historic account.
Some basic principles of dynamos are well understood from linear theory,
but virtually all astrophysical dynamos are in a regime where the field
is dynamically important, and kinematic theory is invalid.
In recent years our understanding of nonlinear properties of dynamos has
advanced rapidly.
This is partly due to new high resolution numerical simulations which
have also triggered further developments in analytic approaches.
An example is the resistively slow saturation phase of dynamos with
helicity that was first seen in numerical simulations \cite{B01},
which then led to the development of a dynamical quenching model
\cite{FB02,BB02,Sub02,BF02b}; see \Sec{SDynamicalQuenching}.
The dynamical quenching model was actually developed much earlier
\cite{KR82}, but it was mostly applied in order to explain chaotic
behavior of the solar cycle \cite{Ruz81,SS91,Covas_etal97}.
Another example is the so-called small scale dynamo whose theory goes back
to the early work of Kazantsev \cite{Kaz68}; see \Sec{SSD}.
Again, only in recent years, with the advent of fast computers allowing
high Reynolds number simulations of hydromagnetic turbulence, the
community became convinced of the reality of the small scale dynamos.
This in turn has triggered further advances in the theoretical
understanding this problem, especially the nonlinear stages.
Also quite recent is the realization that the small scale dynamo is much
harder to excite when, for fixed resistivity, the viscosity is decreased
(i.e.\ the magnetic Prandtl number is less than unity) so that the
magnetic field is driven by a rough velocity field (\Sec{LowPrM}).

Although there have been a number of excellent reviews about dynamo
theory and comparisons with observations of astrophysical magnetic fields
\cite{kron94,beck96,vallee97,ZH97,kuls99,Widr02,Tobias02,Ossen03,Yoshi04},
there have been many crucial developments just over the past few
years involving primarily magnetic helicity. It has now become
clear that nonlinearity in large scale dynamos
is crucially determined by the magnetic
helicity evolution equation. At the same time, magnetic helicity has
also become highly topical in observational solar physics, as is evidenced
by a number of recent specialized meetings on exactly this topic
\cite{helicity_meetings}.
Magnetic helicity
emerges therefore as a new tool in both observational as well as in
theoretical studies of astrophysical magnetohydrodynamics (MHD).
This review discusses the details
of why this is so, and how magnetic helicity can be used to constrain
dynamo theory and to explain the behavior seen in recent simulations of
dynamos in the nonlinear, high magnetic Reynolds number regime.

We also review some basic properties and techniques pertinent 
to mean field (large scale) dynamos (\Sec{Revisit}), so that
newcomers to the field can gain deeper insight and are able
to put new developments into perspective.
In particular, we discuss a simplistic form of the so-called
tau approximation that allows the calculation of mean field
turbulent transport coefficients in situations where the magnetic 
fluctuations strongly exceed the magnitude of the mean field.
This is when the quasilinear theory (also known as first order smoothing
or second order correlation approximation) breaks down.
We then lead to the currently intriguing question of what saturates the
dynamo and why so much can be learned by rather simple considerations
in terms of magnetic helicity.

In turbulent fluids, the generation of large scale magnetic fields
is generically accompanied by the more rapid growth of small scale fields.
The growing Lorentz force due to these fields can back-react on the
turbulence to modify the mean field dynamo coefficients.
A related topic of great current interest is the non helical
small scale dynamo, and especially its nonlinear saturation. This could
also be relevant for explaining the origin of cluster magnetic fields.
These topics are therefore reviewed in the light of recent advances
using both analytic tools as well as high resolution simulations
(\Sec{SmallScale}).

There are obviously many topics that have been left out,
because they touch upon nonlinear dynamo theory only remotely.
Both hydrodynamic and magnetohydrodynamic turbulence are only
discussed in their applications, but there are many fundamental
aspects that are interesting in their own right;
see the text books by Frisch
\cite{FrischBook} and Biskamp \cite{BiskampBook} and
the work by Goldreich and Sridhar \cite{GS95};
for a recent review see Ref.~\cite{verma}.
Another broad research area that has been left out completely
is magnetic reconnection and low beta plasmas.
Again, we can here only refer to the text book by
Priest and Forbes \cite{PriestForbesBook}.
More close connections exist with hydrodynamic mean field theory relevant
for explaining differential rotation in stars \cite{RudigerBook}.
Even many of the applications of dynamo theory are outlined only
rather broadly, but again, we can refer to a recent text book by
R\"udiger and Hollerbach \cite{RudigerHollerbachBook} where many
of these aspects are addressed.

We begin in the next section with some observational facts
that may have a chance in finding
an explanation in terms of dynamo theory within the not too distant future.
We then summarize some useful facts of basic MHD, and also
discuss briefly battery effects to produce seed magnetic fields.
Some general properties of dynamos are discussed in Section~4.
These two sections are relatively general and can be consulted
independently of the remainder.
We then turn to small scale dynamos in Section~5.
Again, this section may well be read separately and does not
contain material that is essential for the remaining sections.
The main theme of large scale dynamos is extensively covered in Sections 6--10.
Finally, in Section~11 we discuss some applications of these ideas to various 
astrophysical systems.
Some final reflections on outstanding issues are given in Section~12.

\section{Magnetic field observations}

In this section we discuss properties of magnetic fields observed
in various astrophysical settings.
We focus specifically on aspects that are believed to be important
for nonlinear dynamo theory and its connection with magnetic helicity.
We begin with a discussion of the solar magnetic field, which
consists of small scale and large scale components.
The typical length scale associated with the large scale field is
the width of the toroidal flux belts with the same polarity
which is around $30^\circ$ in latitude, corresponding to about
$300\Mm$ ($1\Mm=1000\km$).
The pressure scale height at the bottom of the convection zone
is about $50\Mm$, and all scales shorter than that may be
associated with the small scale field.

The theory of the large scale component has been most puzzling,
while the small scale field could always be explained by
turbulence and convection shredding and concentrating the field
into isolated flux bundles.
The simultaneous involvement of a so-called small scale dynamo may provide
another source for the small scale field, which needs to be addressed.
We begin by outlining the observational evidence for large scale fields in
the sun and in stars, and discuss then the evidence for magnetic fields
in accretion discs and galaxies, as well as galaxy clusters.

\subsection{Solar magnetic fields}

\begin{figure}[t!]\begin{center}
\includegraphics[width=\textwidth]{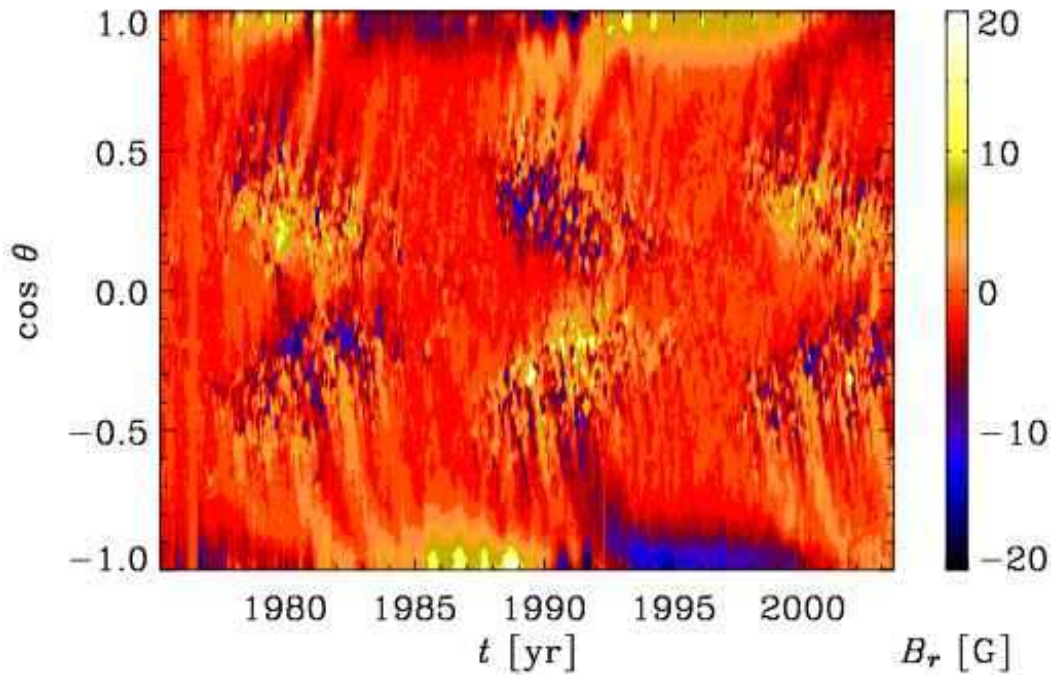}
\end{center}\caption[]{
Longitudinally averaged radial component of the observed solar
magnetic field as a function of cos(colatitude) and time.
Dark (blue) shades denote negative values and
light (yellow) shades denote positive values.
Note the sign changes both in time and across the equator
(courtesy of R.~Knaack).
}\label{stenflo88_fig2}\end{figure}

\begin{figure}[t!]\begin{center}
\includegraphics[width=.99\textwidth]{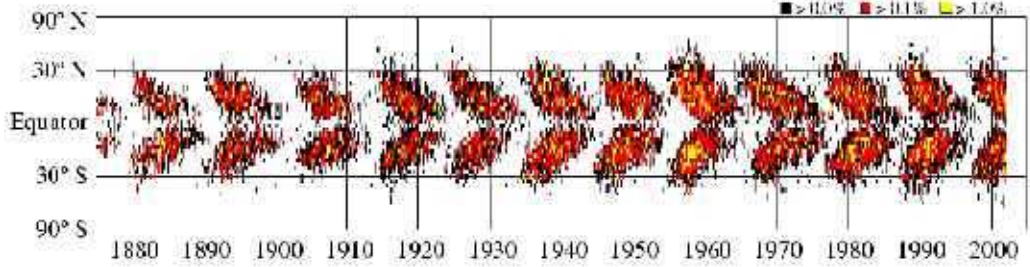}
\end{center}\caption[]{
Solar butterfly diagram showing the sunspot number in a space-time
diagram. Note the migration of sunspot activity from mid-latitudes
toward the equator (courtesy of D.~N.~Hathaway).
}\label{butterflyhathaway}\end{figure}

The sun has a magnetic field that manifests itself in sunspots
through Zeeman splitting of spectral lines \cite{Hale08}.
It has long been known that the sunspot number varies cyclically
with a period between 7 and 17 years.
The longitudinally averaged component of the radial magnetic field
of the sun \cite{Sten88,Sten94} shows a markedly regular spatio-temporal pattern
where the radial magnetic field alternates in time over the 11 year cycle
and also changes sign across the equator (\Fig{stenflo88_fig2}).
One can also see indications of a migration of the field
from mid latitudes toward the equator and the poles.
This migration is also well seen in a sunspot diagram,
which is also called a butterfly diagram,
because the pattern formed by the positions of sunspots in time and
latitude looks like a sequence of butterflies lined up along the
equator (\Fig{butterflyhathaway}).

At the solar surface the azimuthally averaged radial field is only
a few gauss ($1\G=10^{-4}\,\mbox{Tesla}$).
This is rather weak compared with the peak magnetic field in sunspots
of about $2\kG$.
In the bulk of the convection zone, because of differential rotation,
the magnetic field is believed to
point mostly in the azimuthal direction, and it is probably much
larger near the bottom of the convection zone due to an effect
known as downward pumping (\Sec{SimulationsTransport}).

\subsubsection{Estimates of the field strength in the deeper convection zone}
\label{EstimatesFieldStrength}

In the bulk of the solar convection zone the thermal energy transport is reasonably
well described by mixing length theory \cite{KW90}.
This theory yields a rough estimate for the turbulent rms velocity
which is around $u_{\rm rms}=20\m\s^{-1}$
near the bottom of the solar convection zone.
With a density of about $\rho=0.2\g\cm^{-3}$ this corresponds to an equipartition
field strength of about $3\kG$.
(The equipartition field strength is here defined as
$B_{\rm eq}=\sqrt{\mu_0\rho}\,u_{\rm rms}$, where $\mu_0$ is the
magnetic permeability.)

A similar estimate is obtained by considering the total
(unsigned) magnetic flux that emerges
at the surface during one cycle.
This argument is dubious, because one has to make an assumption about how
many times the flux tubes in the sun have emerged at the solar surface.
Nevertheless, the notion of magnetic flux (and especially unsigned flux)
is rather popular in solar physics, because this quantity is readily
accessible from solar magnetograms.
The total unsigned magnetic flux is roughly estimated to be $10^{24}\Mx$.
Distributed over a meridional cross-section of about $500\Mm$ in the
latitudinal direction
and about $50\Mm$ in radius (i.e.\ the lower quarter of the convection zone)
yields a mean field of about $4\kG$, which is in fair agreement with the
equipartition estimate above.
This type of argumentation has first been proposed in an early paper
by Galloway \& Weiss \cite{GW81}.

Another type of estimate concerns {\it not} the mean field but rather the peak
magnetic field in the strong flux tubes.
Such tubes are believed to be `stored' either just below or at the bottom
of the convection zone.
By storage one means that the field survives reasonably undisturbed for a
good fraction of the solar cycle and evolves mostly under the amplifying
action of differential rotation.
Once such a flux tube becomes buoyant in one section of the tube it rises, expands
and becomes tilted relative to the azimuthal direction owing to the
Coriolis force.
Calculations based on the thin flux tube approximation \cite{Spruit82} predict field
strengths of about $100\kG$ that are needed in order to produce the observed
tilt angle of bipolar sunspots near the surface \cite{DSC93}.

The systematic variation of the global field of the sun is important
to understand both for practical reasons, e.g.\ for space weather forecasts,
and for theoretical reasons because the solar field is a prime example
of what we call large scale dynamo action.
The 11 year cycle of the sun is commonly
explained in terms of $\alpha\Omega$ dynamo theory 
(\Secs{solutions_trends}{solar_dynamo}),
but this theory faces a number of problems that will be discussed later.
Much of the resolution of these problems focuses around magnetic helicity.
This has become a very active research field in its own right.
Here we discuss the observational evidence.

\subsubsection{Magnetic helicity of the solar field}
\label{MagneticHelicitySolarField}

\begin{figure}[t!]\begin{center}
\includegraphics[width=.99\textwidth]{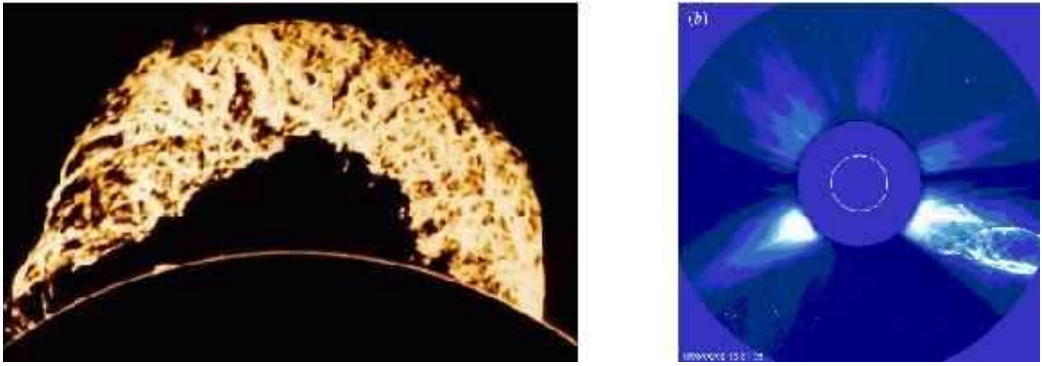}
\end{center}\caption[]{
The famous ``Grand daddy'' prominence of 4 June 1946 (left)
and a big coronal mass eruption of 2 June 1998
from the LASCO coronograph on board the SOHO satellite (right).
Note the complexity of the ejected structures, being suggestive of
helical nature.
Courtesy of the SOHO consortium. SOHO is a project of
international cooperation between ESA and NASA.
}\label{lasco46}\end{figure}

Magnetic helicity studies have become an important observational tool
to quantify the complexity of the sun's magnetic field.
Examples of complex magnetic structures being ejected from the solar
surface are shown in \Fig{lasco46}.
For a series of reviews covering the period until 1999 see Ref.~\cite{BCP99}.
The significance of magnetic helicity for understanding the nonlinear
dynamo has only recently been appreciated.
Here we briefly review some of the relevant observational findings.

The only information about the magnetic helicity of the
sun available to date is from surface magnetic
fields, and these data are necessarily incomplete.
Nevertheless, some systematic trends can be identified.

Vector magnetograms of active regions
show negative (positive) current helicity in the northern (southern)
hemisphere \cite{See90,Pev95,Bao99,Pev00}.
From local measurements one can only obtain the current
helicity density, so nothing can be concluded about magnetic helicity,
which is a volume integral.
As we shall show later (\Sec{energy_spectra}),
under the assumption of isotropy, the spectra of
magnetic and current helicity are however simply related by a wavenumber
squared factor.
This implies that the signs of current and magnetic helicities
agree if they are determined in a sufficiently narrow range of length scales.
We return to this issue in \Sec{Sfinal}.

\begin{figure}[t!]\begin{center}
\includegraphics[width=.99\textwidth]{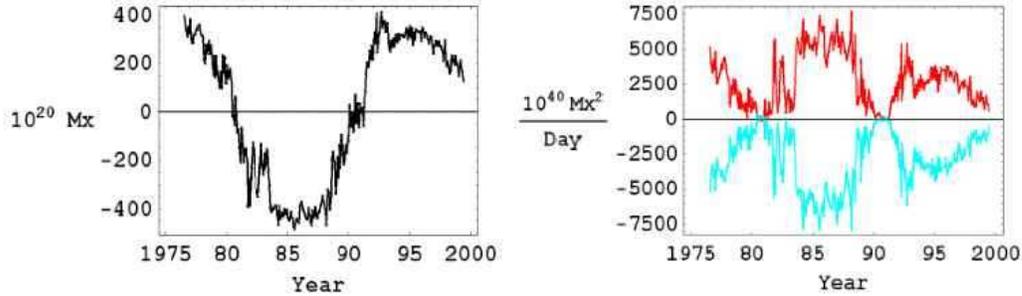}
\end{center}\caption[]{
Net magnetic flux through the solar surface at the northern hemisphere
(left hand panel) and magnetic helicity flux for northern and
southern hemispheres (right hand panel, lower and upper curves,
respectively).
Adapted from Berger and Ruzmaikin \cite{BR00}.
}\label{berger+ruz}\end{figure}

Berger and Ruzmaikin \cite{BR00} have estimated the flux of
magnetic helicity from the solar surface using magnetograms.
They discussed the $\alpha$ effect
and differential rotation as the main agents facilitating the loss of
magnetic helicity.  Their results indicate that the flux of magnetic
helicity due to differential rotation and the observed radial magnetic field
component is negative (positive) in the northern (southern) hemisphere,
and of the order of about $10^{46}\Mx^2$ integrated over the 11 year cycle;
see \Fig{berger+ruz}.

\begin{figure}[t!]\begin{center}
\includegraphics[width=.9\textwidth]{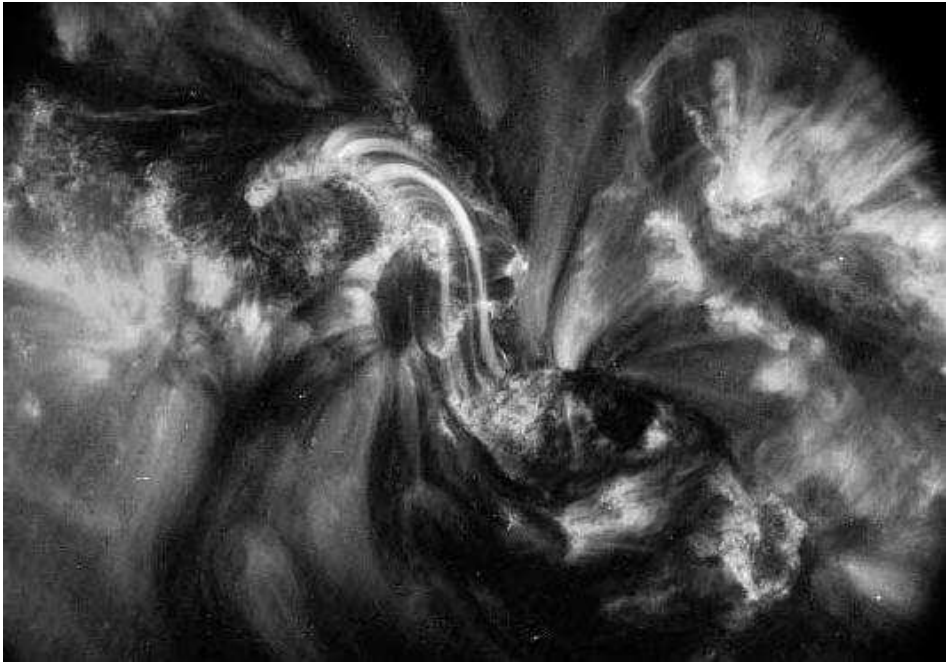}
\end{center}\caption[]{
{\sf X}-ray image at $195\,$\AA\, showing an
{\sf N}-shaped sigmoid (right-handed writhe)
of the active region NOAA AR 8668 at the northern hemisphere
(1999 August 21 at 18:51 UT).
Adapted from Gibson et al.\ \cite{Gib02}.
}\label{gibson}\end{figure}

Chae \cite{Chae00} estimated the magnetic
helicity flux based on counting the crossings of pairs of flux tubes.
Combined with the assumption that two nearly aligned flux tubes are
nearly parallel, rather than anti-parallel, his results again
suggest that the magnetic helicity is negative (positive) in the
northern (southern) hemisphere.
The same sign distribution was also found by DeVore \cite{DeVore00}
who considered magnetic helicity generation by differential rotation.
He finds that the magnetic helicity flux integrated over an 11 year
cycle is about $10^{46}\Mx^2$ both from active regions and from coronal
mass ejections.
Thus, the sign agrees with that of the current helicity obtained using
vector magnetograms.
More recently, D\'emoulin et al.\ \cite{Demoulin_etal02} showed that
oppositely signed twist and writhe from shear
are able to largely cancel, producing
a small total magnetic helicity.
This idea of a bi-helical field is supported further by studies of sigmoids
\cite{Gib02}:
an example is \Fig{gibson}, which shows a TRACE image of an
{\sf N}-shaped sigmoid (right-handed writhe) with left-handed twisted filaments
of the active region NOAA AR 8668, which is typical of the northern hemisphere.
This observation is quite central to our new understanding of nonlinear
dynamo theory \cite{BF00b,BB03} and will be addressed in more detail below
(\Sec{PhenomenologicalModel}).

\subsubsection{Active longitudes}
\label{ActiveLongitudes}

\begin{figure}[t!]\begin{center}
\includegraphics[width=.99\textwidth]{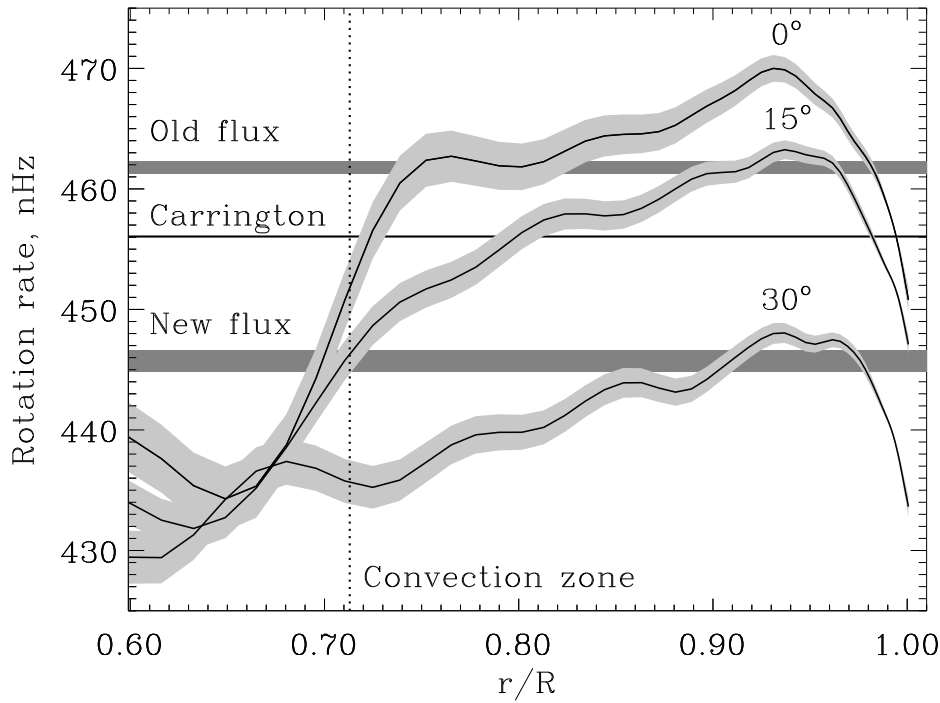}
\end{center}\caption[]{
Radial profiles of the internal solar rotation rate, as inferred from
helioseismology (sidereal, i.e.\ in a fixed frame).
The rotation rate of active zones at the beginning of the cycle
(at $\approx30^\circ$ latitude) and near the end (at $\approx4^\circ$)
is indicated by horizontal bars, which intersect the profiles of
rotation rate at $r/R_\odot\approx0.97$.
For orientation, the conventionally defined Carrington rotation period of
27.3~days (synodic value, corresponding to $424\nHz$) has been translated
to the sidereal value of $456\nHz$.
Courtesy of Benevolenskaya et al.\ \cite{Benevol99}.
}\label{bene99}\end{figure}

An important piece of information about the sun concerns the so-called
active longitudes.
These are longitudes where magnetic activity re-occurs
over long durations, exceeding even the length of the solar cycle
\cite{Vit69,Bog82,Bai87,Bai88}.
On shorter time scales of about half a year, the angular velocity of
active longitudes depends on the phase during the solar cycle, and hence
on the latitude of their occurrence.
At the beginning of the cycle, when new flux appears at high latitudes
($\pm30^\circ$ latitude), the rotation rate of these active longitudes is about
$446\nHz$.
At this latitude the rotation rate of $446\nHz$ agrees with
the value inferred from helioseismology at the fractional radius
$r/R_\odot\approx0.95$; see \Fig{bene99}.

If this magnetic activity were to come from the bottom of the convection
zone at $r/R_\odot\approx0.7$, where the rotation rate is around $435\nHz$,
it would be by $11\nHz$ too slow (\Fig{bene99}).
After half a year, the corresponding regions at $r/R_\odot\approx0.7$ and 0.95
would have drifted apart by $62^\circ$.
Thus, if the active longitudes were to be anchored at $r/R_\odot\approx0.7$, they
could not be connected with matter at this latitude; instead they
would need to be mapped to a lower latitude of about $15^\circ$,
where the rotation rate at $r/R_\odot\approx0.7$ agrees with the value of
$446\nHz$ found for the active longitudes at $30^\circ$ latitude.
Alternatively, they may simply be anchored at a shallower
depth corresponding to $r/R_\odot\approx0.95$, where the rotation rate of
these active longitudes agrees with the helioseismologically inferred value.
Similar considerations apply also to the rotation rate of old flux that
occurs at about $\pm4^\circ$ latitude.
However, here the anchoring depth is ambiguous and could be either
$r/R_\odot\approx0.97$ or in the range 0.75...0.80.
The rather unconventional suggestion of a shallow anchoring depth \cite{B05}
will be addressed further at the end of \Sec{StatusSolarDynamo}.

\subsection{Magnetic fields of late type stars}
\label{LateTypeStars}

\begin{figure}[t!]\begin{center}
\includegraphics[width=.99\textwidth]{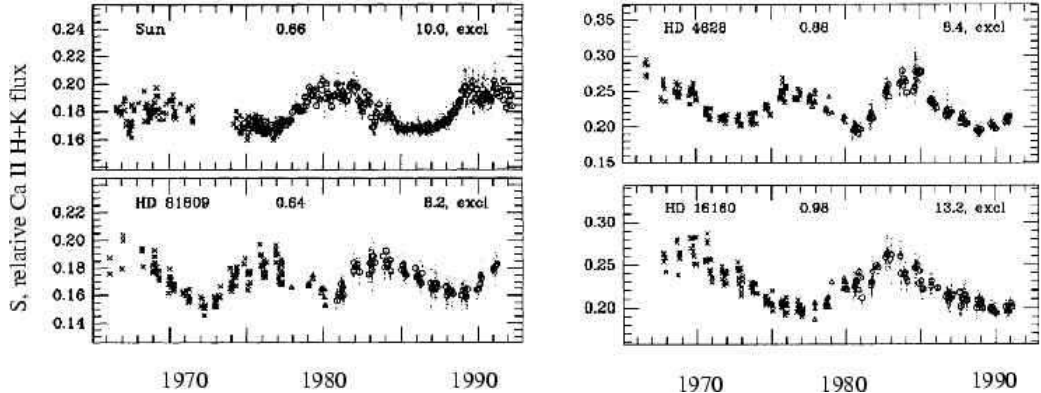}
\end{center}\caption[]{
Time traces of the relative Calcium H and K line emission, $S$,
for 4 stars (including the sun) with oscillatory activity behavior
between the years 1966 and 1992
(adapted from Baliunas et al.~\cite{Bal95})
}\label{baliunas}\end{figure}

Looking at other stars is important for appreciating that the solar dynamo
is not unique and just one particular example of a dynamo that happened
to be a cyclic one.
In fact, we now know that all stars with outer convection zones
(usually referred to as `late-type stars')
have magnetic fields whose strength tends to increase with their
angular velocity.
Some very young stars (e.g.\ T~Tauri stars) have {\it average} field strengths
of about $2\kG$ \cite{JKVK99}.
These stars are fully convective and their field varies in a more erratic fashion.
Cyclic variations are known to exist only for stars with colors $B-V$
in the range 0.57 and 1.37, i.e.\ for spectral types between G0 and K7
\cite{Bal95}.
Some examples of the time traces are shown in \Fig{baliunas}.
The sun's color is 0.66, being close to the upper (bluer) end of the mass range
where stars show cyclic activity.
For the stars in this mass range
there exists an empirical relation between three important
parameters.
One is the inverse Rossby number,
$\mbox{Ro}^{-1}\equiv2\Omega\tau_{\rm turnover}$,
where $\tau_{\rm turnover}\approx\ell/u_{\rm rms}$ is the turnover time
of the convection, estimated in terms of the mixing length, $\ell$, and
the rms velocity of the convection, $u_{\rm rms}$.
The second parameter is the ratio of cycle to rotation frequency,
$\omega_{\rm cyc}/\Omega$, where $\omega_{\rm cyc}=2\pi/P_{\rm cyc}$
and $P_{\rm cyc}$ is the cycle period ($\approx11$ years for the sun,
but ranging from 7 to 21 years for other stars).
The third parameter is the ratio of the mean chromospheric Calcium H
and K line emission to the bolometric flux, $\bra{R_{\rm HK}'}$,
which can be regarded as a proxy of the normalized magnetic field strength,
with $\bra{R_{\rm HK}'}\propto(|\bra{\BB}|/B_{\rm eq})^\kappa$ and
$\kappa\approx0.47$; see Ref.~\cite{BunteSaar93}.
These three parameters are related to each other by approximate power laws,
\EQ
\omega_{\rm cyc}/\Omega\approx c_1\mbox{Ro}^{-\sigma},\quad
\omega_{\rm cyc}/\Omega\approx c_2\bra{R_{\rm HK}'}^\nu\quad
\bra{R_{\rm HK}'}\approx c_3\mbox{Ro}^{-\mu},
\label{cyclelaws}
\EN
where $c_1=c_2 c_3^\nu$ and $\sigma=\mu\nu$.
It turns out that the slopes $\sigma$ and $\nu$ are positive for
active ({\sf A}) and inactive ({\sf I}) stars and
that both groups of stars fall on distinct branches with
$\sigma_{\sf A}\approx0.46$ and $\nu_{\sf A}\approx0.85$ for active
stars and $\sigma_{\sf I}\approx0.48$ and $\nu_{\sf I}\approx0.72$
for inactive stars \cite{BST98}.
Since $\sigma$ and $\nu$ are obtained from separate fits, there is
of course no guarantee that the relation $\sigma=\mu\nu$ will be
obeyed by the data obtained from separate fits.

\begin{figure}[t!]\begin{center}
\includegraphics[width=.99\textwidth]{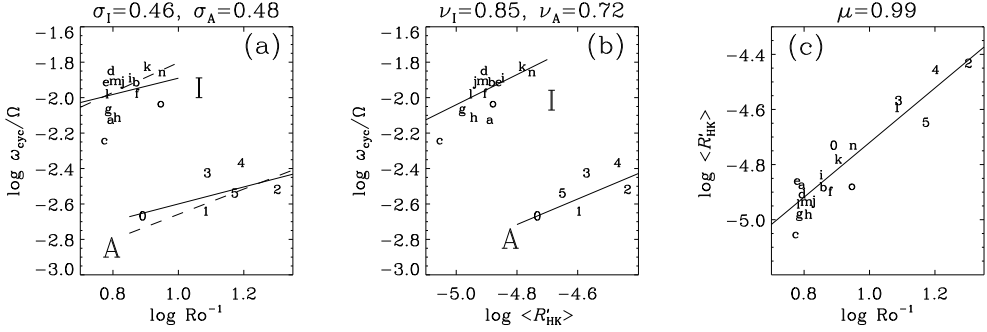}
\end{center}\caption[]{
Mutual correlations between the three quantities $\omega_{\rm cyc}/\Omega$,
$\mbox{Ro}^{-1}$, and $\bra{R_{\rm HK}'}$.
Note the two distinct branches, separate by a factor of about 6 in
the ratio $\omega_{\rm cyc}/\Omega$, with positive slope in the
first two panels.
The numbers and letters in the plots are abbreviations for
specific active and inactive stars, respectively
(adapted from Ref.~\cite{BST98}, where also a key with
the abbreviations of all stars is given).
The values of $\sigma$, $\nu$, and $\mu$, given in the titles of the three
plots, are obtained from three independent plots and hence do not obey the
relation $\sigma=\mu\nu$.
}\label{BST98}\end{figure}

In \Fig{BST98} we present scatter plots showing the mutual correlations
between each of the three quantities for all cyclic stars whose
parameters have been detected with quality parameters that were
labeled \cite{Bal95} as `good' and `excellent'.
Plots similar to the third panel of \Fig{BST98} have also been
produced for other activity proxies \cite{Vilhu84}.
This work shows that there is a relation between activity proxy and
inverse Rossby number not only for stars with
magnetic activity cycles, but for all late
type stars with outer convection zones -- even when the stars are members
of binaries \cite{SaarBran99}.

The fact that the cycle frequency depends in a systematic fashion
on either $\mbox{Ro}^{-1}$ or on $\bra{R_{\rm HK}'}$ suggests that
for these stars the dynamo has a rather stable dependence on the
input parameters.
What is not well understood, however, is the slope $\sigma\approx0.5$
in the relation $\omega_{\rm cyc}/\Omega\sim\mbox{Ro}^{-\sigma}$, and
the fact that there are two distinct branches.
We note that there is also evidence for a third branch for even more active
(`superactive') stars, but there the exponent $\sigma$ is negative
\cite{SaarBran99}.
Standard dynamo theory rather predicts that $\sigma$ is always negative
\cite{Tobias98}.
We return to a possible interpretation of the exponent $\sigma$ and the
origin of the different branches in \Sec{StellarCycles}.

\subsection{Magnetic fields in accretion discs}
\label{AccretionDiscs}

Gaseous discs spinning around some central object are frequently found in
various astrophysical settings, for example around young stars,
stellar mass compact objects (white dwarfs, neutron stars, or black holes),
or in supermassive ($10^{7} -10^{9}M_\odot$) black holes that have been
found or inferred to exist in virtually all galaxies.

Explicit evidence for magnetic fields in discs is sparse:
magnetization of meteorites
that were formed in the disc around the young sun \cite{StepinskiLevy88} or
proxies of magnetic activity such as $H_\beta$ line emission from discs in
binary stars \cite{HorneSaar91}. A direct search for 
Zeeman-induced splitting of the maser lines in the accretion
disc of the Seyfert II galaxy NGC 4258 has resulted in
upper limits of $<50\mG$ for the toroidal component of the $B$ field 
at a distance of about 0.2 pc from the central black hole \cite{maryam03}.
Faraday rotation measure (RM) maps of the central parsecs 
of quasars and radio galaxies hosting relativistic jets \cite{saikia88}
also reveal that the medium on parsec scales surrounding AGNs 
could be significantly magnetized \cite{zwaala_taylor03}.

There are two strong theoretical reasons, however, why accretion discs
should be magnetized.
First, discs are often formed in an already magnetized environment.
This is particularly clear for protostellar discs whose axes of rotation
are often aligned with the direction of the ambient field \cite{MD04}.
Second, discs with weak ambient fields are unstable to the magnetorotational
instability \cite{BH91,BH98} which, coupled with the dynamo instability, can
leads to equipartition field strengths.
In the case of protostellar discs, however, it is possible that the
magnetorotational instability only worked in its early stages.
At later stages, the parts of the disc near the midplane and
at $\sim1\AU$ distance from the central state may have become
too cold and almost neutral, so these parts of the disc may then
no longer be magnetized \cite{gammie96}.

Many accretion discs around black holes are quite luminous.
For example, the luminosity of active galactic nuclei can be as large
as 100 times the luminosity of ordinary galaxies.
Here, magnetic fields provide the perhaps only source of an instability
that can drive the turbulence and hence facilitate the conversion of
potential energy into thermal energy or radiation.
In discs around active galactic nuclei the magnetic field may either
be dragged in from large radii or it may be regenerated locally by
dynamo action.

The latter possibility is particularly plausible in the case of discs
around stellar mass black holes.
Simulations have been carried out to understand this process in detail;
see \Sec{AccretionDiscDynamos}.
Magnetic fields may also be crucial for driving outflows from discs.
In many cases these outflows may be collimated by the ambient magnetic
field to form the observed narrow jets \cite{PP92}.

\subsection{Galactic magnetic fields}
\label{GalB}

Galaxies and clusters of galaxies are currently the only 
astrophysical bodies where a large scale
magnetic field can be seen inside the body itself.
In the case of stars one only sees surface manifestations of the field.
Here we describe the structure and magnitude of galactic fields.

\subsubsection{Synchrotron emission from galaxies}

Magnetic fields in galaxies are mainly probed using radio observations
of their synchrotron emission. Excellent accounts of the
current observational status can be found in the various
reviews by Beck \cite{beck00,beck01,beck02,beck03} and references
therein. We summarize here those aspects which are relevant to 
our discussion of galactic dynamos. Some earlier reviews of
the observations and historical perspectives can be found in
Refs~\cite{beck96,vallee97,ZH97,RSS88}. A map of the total synchrotron 
intensity allows one to estimate the total interstellar
magnetic field in the plane of the sky (averaged over the volume
sampled by the telescope beam). The synchrotron emissivity also
depends on the number density of relativistic electrons, and so
some assumption has to be made about its density.
One generally assumes that the energy densities of the field and particles
are in equipartition.
(Specifically, equipartition is assumed to hold between magnetic fields 
and relativistic protons so that the proton/electron ratio enters as 
another assumption, with 100 taken as a standard value.)
In our Galaxy the accuracy of the equipartition
assumption can be tested, because we have independent measurements
of the local cosmic-ray electron energy density from direct
measurements and about the cosmic-ray proton distribution
from $\gamma$-ray data. The combination of these with the
strength of the radio continuum synchrotron emission
gives a local strength of the total magnetic field of
$6 \pm 1\muG$ \cite{strong00}, which is almost
the same value as that derived from energy equipartition 
\cite{beck01}.

The mean equipartition strength of the
total magnetic field for a sample of 74 spiral galaxies
is $\bra{B_{\rm tot}} = 9\muG$ \cite{beck00,niklas95}. The total
field strength ranges from $\bra{B_{\rm tot}} \sim 4\muG$,
in radio faint galaxies like M31 and M33 to $\bra{B_{\rm tot}} \sim 15\muG$
in grand design spiral galaxies like M51, M83 and NGC~6946 
\cite{beck03}. The strength of the total
field in the inner spiral arms of M51 is about $30\muG$.

Synchrotron radiation is intrinsically highly linearly polarized,
by $70$--$75\%$ in a completely regular magnetic field \cite{Pacho70}.
The observable polarization is however
reduced due to a number of reasons. First the magnetic field usually has 
a tangled component which varies across the telescope beam (geometrical
depolarization); second due to Faraday depolarization in 
the intervening medium and third because some part of the radio 
emission arises due to thermal continuum emission, rather than 
synchrotron emission. A map of the polarized intensity and polarization
angle then gives the strength and structure of the ordered
field, say $\meanB$ in the plane of the sky. Note that polarization
can also be produced by any random field, which is compressed
or stretched in one dimension (i.e.\ an anisotropic field which
incoherently reverses its direction frequently) \cite{laing,sokoloff98}. 
So, to make out
if the field does really have large scale order one
needs also a map of Faraday rotation measures (RMs),
as this will show large scale coherence only for ordered 
fields. Such a map also probes the strength and direction
of the average magnetic field along the line of sight.

The large scale regular field in spiral
galaxies (observed with a resolution of a few 100~pc) is
ordered over several kpc. The strength of this regular field
is typically $1...5\muG$, and up to $\sim 13\muG$ in the interarm
region of NGC~6946, which has an exceptionally strong 
large scale field \cite{BeckH96}. In our Galaxy 
the large scale field inferred from the polarization observations
is about $4\muG$, giving a ratio of regular
to total field of about $\bra{B_0}/\bra{B_{\rm tot}} \sim 0.6 - 0.7$
\cite{elly71,brow76,heiles96}.
However the value inferred from pulsar RM data is
$\bra{B_0}\approx 1.4 \pm 0.2\muG$ \cite{rand94,HQ94,ID98}, which is less than
the above estimate. This may be understood if there is 
anticorrelation between the electron density $n_e$ and the total
field $B$ \cite{beck_anvar03}.

In the context of dynamo theory it is of great interest 
to know the ratio of the regular to the random
component of the magnetic field in galaxies. This is
not easy to determine, especially because of the systematic
biases that can arise in the magnetic field estimates \cite{beck_anvar03}.
Nevertheless, current estimates suggest that
the ratio of regular to random fields is typically $1$ in interarm 
regions and $0.5$ or less in spiral arms (R.\ Beck, private
communication \cite{beck83,buc_beck91}).

\subsubsection{Global structure of galactic fields}
\label{GalB_global}

\begin{figure}[t!]\begin{center}
\includegraphics[width=\textwidth]{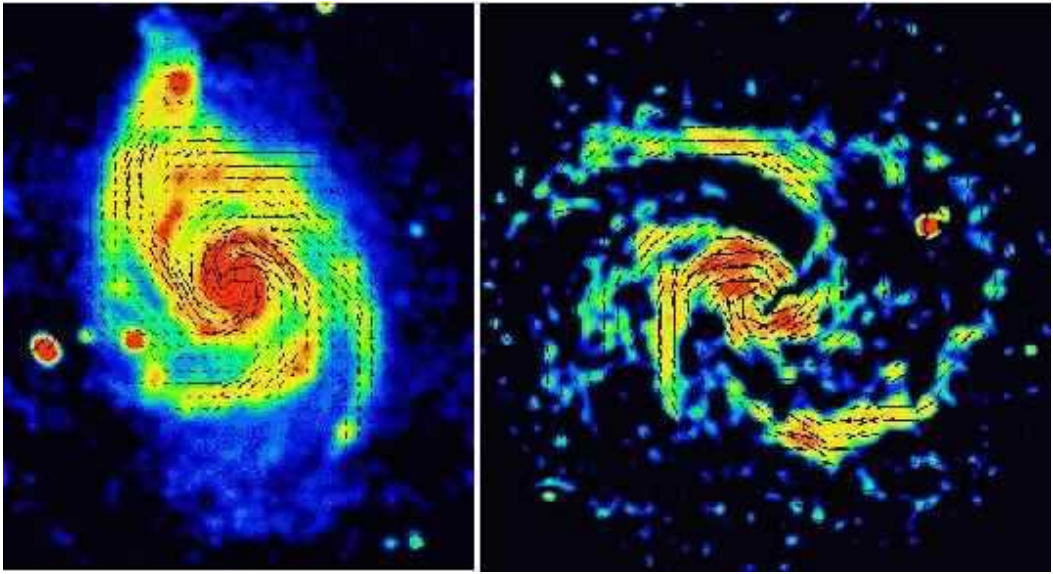}
\end{center}\caption[]{
Left: M51 in $6\cm$, total intensity with magnetic field vectors.
Right: NGC~6946 in $6\cm$, polarized intensity with magnetic field vectors.
The physical extent of the images is
approximately $28\times34\kpc^2$ for M51 (distance $9.6\Mpc$)
and $22\times22\kpc^2$ for NGC~6946 (distance $7\Mpc$).
(VLA and Effelsberg. Courtesy R. Beck.)
}\label{beck}\end{figure}

The global structure of the mean (or regular) magnetic field 
and that of the total field (mean $+$ random) are also of interest. 
The random field is almost always strongest 
within the spiral arms and thus follows the distribution of
cool gas and dust.
The regular field is generally weak within spiral arms, except
for rare cases like M51 with strong density waves.
Thus the total field is also strongest within the spiral arms
where the random field dominates.
The strongest total and regular
fields in M51 are located at the positions of the prominent dust lanes
on the inner edges of the optical spiral arms \cite{nein92,NHor96},
as expected if it were due to compression by density waves.
However, the regular field also extends far into the interarm regions. 
The regular field in M31 is nearly aligned with the spiral 
arms forming the bright `ring' of emission seen in this galaxy \cite{elly03}. 
The $\BB$ vectors of the regular
field in several other galaxies (M81, M83, NGC~1566) also
follow the optical spiral, though they are generally {\it offset} from
the optical arms. A particularly spectacular case is that of the
galaxy NGC~6946 \cite{BeckH96,frick00}; 
here the polarized emission (tracing the regular field)
is located in dominant magnetic spiral arms.
These magnetic spiral arms are interlaced and anti-correlated 
with the optical spiral structure. They  
have widths of about 500--1000~pc and
regular fields of $\sim 13\muG$. The field in these arms
is also ordered, as inferred from RM observations \cite{beck01}.
In \Fig{beck} we show $6\cm$ radio observations of the galaxies M51
and NGC~6946, with superposed magnetic field vectors. One can
clearly see that the magnetic field is ordered over large scales.

As we remarked earlier, RM observations are absolutely
crucial to distinguish between coherent and incoherent fields.
Coherence of RM on a large scale is indeed seen in a number of 
galaxies (for example, M31 \cite{elly03}, 
NGC~6946 \cite{beck01}, NGC~2997 \cite{HanB99}). 
The galaxy M31 seems to have a $20\kpc$ sized torus of emission,
with the regular field nearly aligned
with the spiral arms forming an emission `ring', with an average
pitch angle of about $-15^\circ$ \cite{elly03,beck82,han98}.
Such a field can probably be produced only by a large scale dynamo.

The structure of the regular field is described
in dynamo models by modes of different azimuthal and vertical symmetry;
see \Sec{ExcitationConditionsDisc}.
Again, Faraday rotation measure (RM) observations are crucial
for this purpose \cite{Sofue86}. The current data indicate a singly periodic
azimuthal variation of RMs, suggesting a largely axisymmetric
(ASS, $m=0$ symmetry) mean field structure in M31 \cite{beck82} and IC 342
\cite{KMB89}. There is an indication of a bisymmetric
spiral mode (BSS, $m=1$ symmetry) in M81 \cite{KMB89b}.
The field in M51 seems to be best described by a 
combination of ASS and BSS \cite{elli97}. 
The magnetic arms in NGC~6946 may be the result of a superposition
of ASS and quadrisymmetric ($m=2$) modes \cite{beck00}.
Indeed, in most galaxies the data cannot be
described by only a single mode, but require a
superposition of several modes which still cannot be
resolved by the existing observations.
It has also been noted \cite{Krause_beck98} 
that in 4 out of 5 galaxies, the radial component of the spiral field 
could be such that the field points {\it inward}.
This is remarkable in that the induction equation and the related
nonlinearities do not distinguish between solutions $\BB$ and $-\BB$.
An exception is the Hall effect where the direction matters \cite{Hol+Rue02},
but this idea has not yet been applied to the field orientation in galaxies.

The vertical symmetry of the field is much more difficult
to determine. The local field in our Galaxy is oriented
mainly parallel to the plane (cf.\ \cite{kron94,RSS88,HQ94}).
This agrees well with the results from several other
external edge-on galaxies \cite{dumke95}, where the observed 
magnetic fields are generally aligned along the discs of the galaxies. 
Further, in our Galaxy the RMs of pulsars
and extragalactic radio sources have the same sign above
and below the galactic midplane, for galactic longitudes
between $90^\circ$ and $270^\circ$, while toward the
galactic center there is a claim that the RMs change sign 
cf.\ \cite{hanman97,hanman99}. 
This may point toward a vertically symmetric field in the Galaxy away from
its central regions. Note that the determination of magnetic field
structure in the Galaxy from Faraday rotation can be complicated
by local perturbations. Taking into account such complications,
and carrying out an analysis of the Faraday rotation measures of extragalactic 
sources using wavelet transforms, one finds evidence that
the horizontal components of the regular magnetic field have
even parity throughout the Galaxy, that is the horizontal components are
similarly directed on both sides of the disc \cite{frick_etal01}.
Note that a vertically symmetric field could arise
from a quadrupolar meridional structure of the field,
while vertically antisymmetric fields could arise
from a dipolar structure. A vertically symmetric field 
also seems to be indicated from RM studies of the galaxy M31 \cite{han98}.

The discovery of isolated non-thermal filaments
throughout the inner few hundred parsecs of the galaxy
\cite{yusef-zadeh89,morris98,larosa00} with orientations
largely perpendicular to the galactic plane were 
interpreted as evidence for several milligauss level, space filling
vertical fields in the central $200\pc$ of our galaxy \cite{morris98}.
However a recent $20\cm$ survey which has found numerous linear filaments
finds them to have a wide range of orientations, which
could complicate this simple picture \cite{yousef-zadeh04,larosa04}.
The observational situation needs to be clarified.
If indeed the presence of a dipolar field in the galactic
center regions is confirmed this would provide an important
challenge for dynamo theory.

Although most edge-on galaxies have fields aligned along the disc,
several galaxies (NGC~4631, NGC~4666 and M82) have also radio halos 
with a dominant vertical field component \cite{beck00}.
Magnetic spurs in these halos are connected to star forming regions
in the disc. The field is probably dragged out by a strong, 
inhomogeneous galactic wind \cite{BDMSST93,EGRW95}.

Ordered magnetic fields with strengths similar to those in grand design
spirals have also been detected in flocculent galaxies
(M33 \cite{BuB91}, NGC~3521 and 5055 \cite{knapik00}, 
NGC~4414 \cite{soida02}), and even in irregular galaxies
(NGC~4449 \cite{chyzy00}). The mean degree of
polarization is also similar between grand design and flocculent
galaxies \cite{knapik00}. Also, a grand design spiral pattern is observed
in all the above flocculent galaxies, implying that gaseous spiral arms
are not an essential feature to obtain ordered fields. 

There is little direct evidence
on the nature of magnetic fields in elliptical galaxies, although
magnetic fields may well be ubiquitous in the hot ionized gas seen in
these galaxies \cite{moss_shukurov96}.
This may be due to the paucity of relativistic
electrons in these galaxies, which are needed to illuminate
the magnetic fields by generating synchrotron emission. Faraday rotation
of the polarized emission from background objects has been observed 
in a few cases. Particularly intriguing is the case of the gravitationally 
lensed twin quasar 0957+561, where the two images have a differential 
Faraday rotation of $\sim 100$ rad m$^{-2}$ \cite{GRB85}.
One of the images passes through the central region of a
possibly elliptical galaxy acting as the lens, and so this may indicate 
that the gas in this elliptical galaxy has ordered magnetic fields. 
It is important to search for more direct evidence for 
magnetic fields in elliptical galaxies.

\subsection{Magnetic fields in clusters of galaxies}
\label{Cluster_fields}

The most recent area of study of astrophysical magnetic fields
is perhaps the magnetic fields of clusters of galaxies. Galaxy clusters
are the largest bound systems in the universe, having
masses of $\sim 10^{14}$--$10^{15} M_\odot$ and typical sizes
of several $\Mpc$. Observations of clusters in X-rays
reveal that they generally have an atmosphere
of hot gas with temperatures $T \sim 10^7$ to $10^8$K, extending
over $\Mpc$ scales. The baryonic mass of clusters is in fact dominated
by this hot gas component. It has become clear in the
last decade or so that magnetic fields are also ubiquitous in clusters.
Succinct reviews of the observational data on cluster magnetic
fields can be found in Ref.~\cite{car_tay02,gov_fer04}. Here we gather
some important facts that are relevant in trying
to understand the origin of these fields.

Evidence for magnetic fields in clusters again comes
from mainly radio observations. Several clusters
display relatively smooth low surface brightness
radio halos, attributed to synchrotron emission from the
cluster as a whole, rather than discrete radio sources.
The first such halo to be discovered was that
associated with the Coma cluster (called Coma~C) \cite{wilson70}.
Only recently have many more been found in systematic
searches \cite{H82,GTF99,GF00,kemp_sar01}. These radio halos
have typically sizes of $\sim 1$ Mpc, steep spectral indices,
low surface brightness, low polarizations ($< 5\%$ ), and
are centered close  to the center of the X-ray emission.
Total magnetic fields in cluster radio halos, estimated using 
minimum energy arguments \cite{miley80} 
range from $0.1$ to $1\muG$ \cite{feretti99},
the value for Coma being $\sim 0.4\muG$ \cite{gio93}.
[The equipartition field will depend on the assumed
proton to electron energy ratio; for a ratio of 100,
like in the local ISM, 
the equipartition field will be larger by a factor
$\sim 100^{2/7}\approx3.7$ (R.~Beck, private communication)].

Cluster magnetic fields can also be probed using 
Faraday rotation studies of both cluster radio galaxies and also 
background radio sources seen through the cluster.
High resolution RM studies have been performed in several
radio galaxies in both
cooling flow clusters and non-cooling flow clusters
(although the issue of the existence of cooling flows
has become questionable).
If one assumes a uniform field in the
cluster, then minimum magnetic fields of $5$ to $10\muG$
are inferred in cooling flow clusters, whereas
it could be a factor of 2 lower in non-cooling flow clusters
\cite{car_tay02}. However, the observed RM is patchy
indicating magnetic field coherence scales of about
$5$ to $10\kpc$ in these clusters. If we use such
coherence lengths, then the estimated
fields become larger. For example
in the cooling flow cluster 3C295 the estimated magnetic field strength
is $\sim 12\muG$ in the cluster core \cite{allen01}, 
whereas in the non-cooling flow
cluster 3C129 the estimated field is about $6\muG$ \cite{TGAF01}.
In Hydra A, there is an intriguing trend for all the RMs to the 
north of the nucleus to be positive and to the south to be 
negative \cite{TP93}. Naively this would
indicate quite a large scale field ($100\kpc$) with strength
$\sim 7\muG$ \cite{TP93}, but it is unclear which fraction is due 
to a cocoon surrounding the radio source (cf.\ Ref.~\cite{bicknell90}).
The more tangled fields in the same cluster were inferred to
have strengths of $\sim 30\muG$ and coherence lengths $\sim 4\kpc$ \cite{TP93}.
More recently, a novel technique to analyze Faraday rotation maps
has been developed, assuming that the magnetic fields are 
statistically isotropic.
This technique has been applied to several galaxy clusters
\cite{ensslin_vogt03,vogt_ensslin03}.
This analysis yields an estimate of $3 \mu \G$ in Abell 2634, $6 \mu \G$
in Abell 400 and $12 \mu \G$ in Hydra A as conservative estimates of
the field strengths, and field correlation lengths of 
$\sim 4.9 \kpc$, $3.6 \kpc$ and $0.9 \kpc$, respectively, for these
3 clusters. (For Hydra A, a recent re-analysis of the data using an
improved RM map and revised cluster parameters, has led
to revised values of the central field of the cluster of
$7 \mu$G and correlation length of $3$ kpc, as well
as a tentative determination of a Kolmogorov type magnetic
power spectrum \cite{vogt_ensslin05}.)

\begin{figure}[t!]\begin{center}
\includegraphics[width=.65\textwidth]{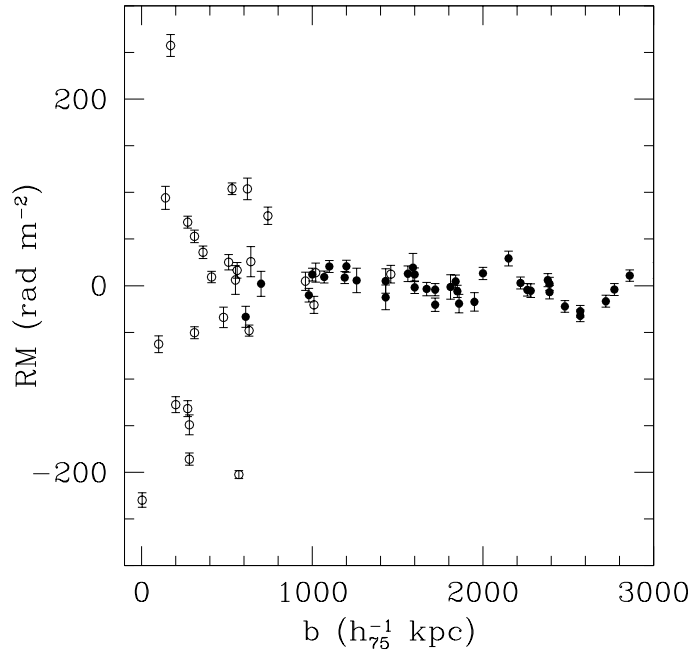}
\end{center}\caption[]{
Galaxy-corrected rotation measure plotted as a
function of source impact parameter in kiloparsecs for the sample of
16 Abell clusters. The open points represent the {\it cluster} sources
viewed through the thermal cluster gas while the closed points are the
{\it control} sources at impact parameters beyond the cluster
gas. Note the clear increase in the width of the RM distribution
toward smaller impact parameter.
Adapted from Clarke et al.\ \cite{clarke01}.
}\label{clarke+kronberg}\end{figure}

There is always some doubt whether the RMs in cluster radio
sources are produced due to Faraday rotation intrinsic
to the radio source, rather than due to the intervening 
intracluster medium. While this is unlikely in most cases \cite{car_tay02},
perhaps more convincing evidence is the fact
that studies of RMs of background radio sources seen through
clusters, also indicate several $\muG$ cluster magnetic
fields. A very interesting statistical study in this context is a recent
VLA survey \cite{clarke01}, where the RMs in and behind a
sample of 16 Abell clusters were determined. The RMs were plotted
as a function of distance from the cluster center
and compared with a control sample of RMs from field sources; 
see \Fig{clarke+kronberg}.
This study revealed a significant excess RM for sources within
about $0.5 \Mpc$ of the cluster center. Using a simple
model, where the intracluster medium consists of cells
of uniform size and field strength, but random field
orientations, Clarke et al.\ \cite{clarke01} estimate cluster magnetic fields
of $\sim 5\,(l/10\kpc)^{-1/2}\muG$, where $l$ is
the coherence length of the field.

Cluster magnetic fields can also be probed by 
comparing the inverse Compton X-ray emission and the
synchrotron emission from the same region. Note that
this ratio depends on the ratio of the background radiation energy density
(which in many cases would be dominated
by the Cosmic microwave background) 
to the magnetic field energy density. The
main difficulty is in separating out the thermal X-ray emission.
This separation can also be attempted using
spatially resolved X-ray data. Indeed, an X-ray
excess (compared to that expected from a thermal atmosphere)
was seen at the location of a diffuse radio relic source in
Abell 85 \cite{bagchi98}. This was used in Ref.~\cite{bagchi98} to derive
a magnetic field of $1.0\pm0.1\,\muG$ for this source.

Overall it appears that there is considerable evidence
that galaxy clusters are magnetized with fields ranging
from a few $\muG$ to several tens of $\muG$
in some cluster centers, and with coherence scales of order $10\kpc$.
These fields, if not maintained by some mechanism, will
evolve as decaying MHD turbulence, and perhaps
decay on the appropriate Alfv\'en time scale,
which is $\sim 10^8$ yr, much less than the age of the cluster.
We will have more to say on the possibility of dynamos
in clusters in later sections toward the end of the review.

\section{The equations of magnetohydrodynamics}

In stars and galaxies, and indeed in many other astrophysical settings,
the gas is partially or fully ionized and can carry electric currents
that, in turn, produce magnetic fields. 
The associated Lorentz force exerted on the ionized gas (also
called plasma) can in general no longer be neglected in the 
momentum equation for the gas. Magneto-hydrodynamics (MHD)
is the study of the interaction of the magnetic field and the plasma
treated as a fluid. In MHD we combine Maxwell's equations of
electrodynamics with the fluid equations, including also the
Lorentz forces due to electromagnetic fields.
We first discuss Maxwell's
equations that characterize the evolution of the magnetic field.

\subsection{Maxwell's equations}

In gaussian cgs units, Maxwell's equations can be written in the form
\EQ
{1\over c}{\partial\BB\over\partial t}
=-\nab\times\EE,\quad
\nab\cdot\BB=0,
\EN
\EQ
{1\over c}{\partial\EE\over\partial t}
=\nab\times\BB-{4\pi\over c}\JJ,\quad
\nab\cdot\EE=4\pi\rho_{\rm e},
\EN
where $\BB$ is the magnetic flux density (usually referred to
as simply the magnetic field), $\EE$ is the electric field,
$\JJ$ is the current density, $c$ is the speed of light,
and $\rho_{\rm e}$ is the charge density.

Although in astrophysics one uses mostly cgs units,
in much of the work on dynamos the MHD equations are
written in `SI' units, i.e.\ with magnetic permeability
$\mu_0$ and without factors like $4\pi/c$.
(Nevertheless, the magnetic field is still quoted often in
gauss [G] and cgs units are used for density, lengths, etc.)
Maxwell's equations in SI units are then written as
\EQ
{\partial\BB\over\partial t}
=-\nab\times\EE,\quad
\nab\cdot\BB=0,
\label{Ampere}
\EN
\EQ
{1\over c^2}{\partial\EE\over\partial t}
=\nab\times\BB-\mu_0\JJ,\quad
\nab\cdot\EE=\rho_{\rm e}/\epsilon_0,
\label{Faraday}
\EN
where $\epsilon_0=1/(\mu_0 c^2)$ is the permittivity of free space.

To ensure that $\nab\cdot\BB=0$ is satisfied at all times it is
often convenient to define $\BB=\nab\times\AAA$ and
to replace \Eq{Ampere} by the `uncurled' equation
for the magnetic vector potential, $\AAA$,
\EQ
{\partial\AAA\over\partial t}
=-\EE-\nab\phi,
\label{AmpereAAA}
\EN
where $\phi$ is the scalar potential.
Note that magnetic and electric fields are invariant under the 
gauge transformation
\EQ
\AAA' = \AAA+\nab\Lambda,
\label{gauge_trans}
\EN
\EQ
\phi' = \phi-{\partial\Lambda\over\partial t}.
\label{gauge_trans_phi}
\EN
For numerical purposes it is often convenient to choose the gauge
$\Lambda=\int\phi\,\dd t$, which implies that $\phi'=0$.
Thus, instead of \Eq{AmpereAAA} one now solves the equation
$\partial\AAA'/\partial t=-\EE$.
There are a few other gauge choices that are numerically convenient
(see, e.g., \Sec{BasicEquations}).

\subsection{Resistive MHD and the induction equation}

Using the standard Ohm's law in a fixed frame of reference,
\EQ
\JJ=\sigma\left(\EE+\UU\times\BB\right),
\EN
where $\sigma$ is the electric conductivity,
and introducing the magnetic diffusivity
$\eta=(\mu_0\sigma)^{-1}$, or $\eta=c^2/(4\pi\sigma)$
in cgs units, we can eliminate $\JJ$ from \Eq{Faraday}, so we have
\EQ
\left({1\over\eta}+{1\over c^2}{\partial\over\partial t}\right)\EE
=-\left({1\over\eta}\UU-\nab\right)\times\BB.
\label{FaradayOhm}
\EN
This formulation shows that the time derivative term (also called the Faraday
displacement current) can be neglected if the relevant time scale
over which the electric field varies, exceeds the Faraday 
time $\tau_{\rm Faraday}=\eta/c^2$.
Below we shall discuss that for ordinary Spitzer resistivity,
$\eta$ is proportional to $T^{-3/2}$ and varies between $10$ and
$10^{10}\cm^2\s^{-1}$ for temperatures between $T=10^8$ and $T=10^2\K$.
Thus, the displacement current can be neglected when the variation time scales
are longer than $10^{-20}\s$ (for $T\approx10^8\K$) and
longer than $10^{-11}\s$ (for $T\approx10^2\K$).
For the applications discussed in this review,
this condition is always met, even for neutron stars where the time
scales of variation can be of the order of milliseconds, 
but the temperatures are very high as well.
We can therefore safely neglect the displacement current and
eliminate $\EE$,
so \Eq{Faraday} can be replaced by Ampere's law
$\JJ=\nab\times\BB/\mu_0$.

It is often convenient to consider $\JJ$ simply as a short hand for
$\nab\times\BB$.
This can be accomplished by adopting units where $\mu_0=1$.
We shall follow here this convection and shall therefore simply write
\EQ
\JJ=\nab\times\BB.
\label{Current}
\EN
Occasionally we also state the full expressions for clarity.

Substituting Ohm's law into the Faraday's law of induction, and using 
Ampere's law to eliminate $\JJ$, one can write a single
evolution equation for $\BB$, which is called the induction equation:
\\

\SHADOWBOX{
\EQ
{\partial\BB\over\partial t}
=\nab\times\left(\UU\times\BB-\eta\JJ\right).
\label{Induction1}
\EN
}\\

We now describe a simple physical picture for the conductivity
in a plasma. The force due to an electric field $\EE$ accelerates
electrons relative to the ions; but they cannot move freely
due to friction with the ionic fluid, caused
by electron--ion collisions. They acquire a `terminal'
relative velocity ${\bf V}$ with respect to the ions,
obtained by balancing the Lorentz force with friction. This
velocity can also be estimated as follows. Assume that electrons 
move freely for about an electron--ion collision time $\tau_{ei}$,
after which their velocity becomes again randomized.
Electrons of charge $e$ and mass $m_e$ in free motion
during the time $\tau_{ei}$ acquire from the action of
an electric field $\EE$ an ordered speed ${\bf V} \sim \tau_{ei} e \EE/m_e$.
This corresponds to a current density $\JJ \sim e n_e {\bf V} \sim 
(n_e e^2 \tau_{ei}/m) \EE$ and hence leads to 
$\sigma \sim n_e e^2 \tau_{ei}/m_e$. 

The electron--ion collision time scale
(which determines $\sigma$) can also be estimated as follows.
For a strong collision between an electron and an ion one
needs an impact parameter $b$ which satisfies the condition
$Ze^2/b > m_e v^2$. This gives a cross section
for strong scattering of $\sigma_t \sim \pi b^2 $. Since the
Coulomb force is a long range force, the larger number
of random weak scatterings add up to give an extra `Coulomb 
logarithm' correction to make 
$\sigma_t \sim \pi (Ze^2/mv^2)^2 \ln\Lambda$,
where $\ln\Lambda$ is in the range between 5 and 20.
The corresponding mean free time between collisions is 
\EQ
\tau_{ei} \sim {1 \over n_i \sigma_t v } \sim { (k_{\rm B}T)^{3/2} 
m_e^{1/2} \over  \pi Z e^4 n_e \ln\Lambda },
\EN
where we have used the fact that
$m_e v^2 \sim k_{\rm B}T$ and $Zn_i = n_e$. Hence we obtain the estimate
\EQ
\sigma \sim {(k_{\rm B}T)^{3/2} \over  m_e^{1/2}\pi Z e^2 \ln\Lambda },
\EN
where most importantly the dependence on the electron density
has canceled out. A more exact calculation can be found,
for example, in Landau and Lifshitz \cite{LLkinetic} (Vol 10; Eq.~44.11)
and gives an extra factor of $4 (2/\pi)^{1/2}$ multiplying the 
above result. The above argument has ignored collisions between 
electrons themselves, and treated the plasma as a `lorentzian plasma'. 
The inclusion of the effect of electron--electron collisions further reduces
the conductivity by a factor of about $0.582$ for $Z=1$ to $1$ for
$Z \to \infty$; see the book by Spitzer \cite{spitzer56},
and Table~5.1 and Eqs.~(5)--(37) therein,
and leads to a diffusivity, in cgs units, of
$\eta = c^2/(4\pi \sigma)$ given by
\EQ
\eta=10^4
\left({T\over10^6\K}\right)^{-3/2}
\left({\ln\Lambda\over20}\right) {\rm cm}^2 {\rm s}^{-1}.
\label{SpitzerFormula}
\EN
As noted above, the resistivity is independent of density,
and is also inversely proportional to the temperature
(larger temperatures implying larger mean free time between
collisions, larger conductivity and hence smaller resistivity).

The corresponding expression for the kinematic viscosity $\nu$ is quite 
different. Simple kinetic theory arguments give $\nu \sim v_{\rm t} l_i$,
where $l_i$ is the mean free path of the particles which dominate
the momentum transport and $v_{\rm t}$ is their random velocity.
For a fully ionized gas the ions dominate the momentum
transport, and their mean free path $l_i \sim (n_i \sigma_i)^{-1}$,
with the cross-section $\sigma_i$, is determined
again by the ion--ion `Coulomb' interaction.
From a reasoning very similar to the above for electron--ion
collisions, we have $\sigma_i \sim \pi (Z^2e^2/k_{\rm B}T)^2 \ln\Lambda$,
where we have used $m_i v_{\rm t}^2 \sim k_{\rm B}T$. Substituting for
$v_{\rm t}$ and $l_i$, this then gives
\EQ
\nu \sim { (k_{\rm B}T)^{5/2} \over n_i m_i^{1/2}\pi Z^4 e^4 \ln\Lambda }.
\EN
More accurate evaluation using the Landau collision integral
gives a factor $0.4$ for a hydrogen plasma,
instead of $1/\pi$ in the above expression (see the end of Section~43
in Vol.~10 of Landau and Lifshitz \cite{LLkinetic}). This gives numerically
\EQ
\nu = 6.5 \times 10^{22} 
\left({T\over10^6\K}\right)^{5/2}
\left(n_i \over\cm^{-3}\right)^{-1}
\left({\ln\Lambda\over20}\right)^{-1}\cm^2\s^{-1},
\EN
so the magnetic Prandtl number is
\EQ
P_{\rm m}\equiv{\nu\over\eta}= 1.1 \times10^{-4}
\left({T\over10^6\K}\right)^4
\left({\rho\over0.1\g\cm^{-3}}\right)^{-1}
\left({\ln\Lambda\over20}\right)^{-2}.
\label{Pm}
\EN
\begin{table}[t!]\caption{
Summary of some important parameters in various astrophysical settings.
The values given should be understood as rough indications only.
In particular, the applicability of \Eq{Pm} is questionable in some cases
and has therefore not been used for protostellar discs (see text).
We have assumed
$\ln\Lambda=20$ in computing $R_{\rm m}$ and $P_{\rm m}$.
CZ means convection zone, CV discs and similar refer to cataclysmic
variables and discs around other other compact objects such as black
holes and neutron stars.
AGNs are active galactic nuclei.
Numbers in parenthesis indicate significant uncertainty due to other effects.
}\vspace{12pt}\centerline{\begin{tabular}{lcccccccccc}
& $T$ [K]
& $\rho\;[\g\cm^{-3}]$
& $P_{\rm m}$ 
& $u_{\rm rms}\;[\cm\s^{-1}]$
& $L\;[\cm]$
& $R_{\rm m}$ \\
\hline
Solar CZ (upper part) &  $10^4$  & $10^{-6}$ & $10^{-7}$ & $10^6$ &$10^{8} $& $10^{6}$ \\
Solar CZ (lower part) &  $10^6$  & $10^{-1}$ & $10^{-4}$ & $10^4$ &$10^{10}$& $10^{9}$ \\
Protostellar discs    &  $10^3$  & $10^{-10}$& $10^{-8}$ & $10^5$ &$10^{12}$&  $10$  \\
CV discs and similar  &  $10^4$  & $10^{-7}$ & $10^{-6}$ & $10^5$ & $10^7$ &   $10^4$  \\
AGN discs             &  $10^7$  & $10^{-5}$ & $10^{4}$  & $10^5$ & $10^9$  &$10^{11}$ \\
Galaxy                &  $10^4$  &$10^{-24}$ & ($10^{11}$) & $10^6$ &$10^{20}$&($10^{18}$) \\
Galaxy clusters       &  $10^8$  & $10^{-26}$&($10^{29}$)& $10^8$ &$10^{23}$&($10^{29}$)\\
\label{TAstrophysBodies}\end{tabular}}\end{table}

Thus, in the sun and other stars ($T\sim10^6\K$, $\rho\sim0.1\g\cm^{-3}$)
the magnetic Prandtl number is much less than unity.
Applied to the galaxy, using $T=10^4\K$ and
$\rho=10^{-24}\g\cm^{-3}$, $\ln\Lambda \sim 10$,
this formula gives $P_{\rm m}= 4 \times 10^{11}$.
The reason $P_{\rm m}$ is so large in galaxies
is mostly because of the very long mean free path
caused by the low density \cite{KA92}.
For galaxy clusters, the temperature of the gas is even larger
and the density smaller, making the medium much more viscous
and having even larger $P_{\rm m}$.

In protostellar discs, on the other hand, the gas is mostly neutral
with low temperatures. In this case, the electrical conductivity
is given by $\sigma = n_e e^2 \tau_{en}/m_e$, where $\tau_{en}$
is the rate of collisions between electrons and neutral particles.
The associated resistivity is $\eta=234 x_e^{-1}T^{1/2}\cm^2\s^{-1}$,
where $x_e=n_e/n_n$ is the ionization fraction and $n_n$ is the number density
of neutral particles \cite{balb_terq01}.
The ionization fraction at the ionization--recombination
equilibrium is approximately given by
$x_e =(\zeta/\beta n_n)^{1/2}$, where $\zeta$ is the ionization rate
and $\beta = 3 \times 10^{-6} T^{-1/2} \cm^3\s^{-1}$ is the
dissociative recombination rate \cite{fromangetal02,sano_stone02}. 
For a density of $\rho=10^{-10}\g\cm^{-3}$,
and a mean molecular weight $2.33m_p$ \cite{balb_terq01},
we have $n_n = 2.6 \times 10^{13} \cm^{-3}$.
Adopting for $\zeta$ the
cosmic ray ionizing rate $\zeta \sim 10^{-17}\s^{-1}$,
which is not drastically attenuated by the dense gas in the disk,
and a disc temperature $T=10^3\K$, we estimate $x_e \sim 2 \times 10^{-12}$,
and hence $\eta \sim 4 \times 10^{15} \cm^2 \s^{-1}$.

In \Tab{TAstrophysBodies} we summarize typical values of temperature
and density in different astrophysical settings and calculate the
corresponding values of $P_{\rm m}$.
Here we also give rough estimates of typical rms velocities,
$u_{\rm rms}$, and eddy scales, $L$, which allow us to calculate the
magnetic Reynolds number as
\EQ
R_{\rm m}=u_{\rm rms}/(\eta k_{\rm f}),
\label{Rmdefinition}
\EN
where $k_{\rm f}=2\pi/L$.
This number characterizes the relative importance of magnetic induction
relative to magnetic diffusion.
A similar number is the fluid Reynolds number,
$\mbox{Re}=R_{\rm m}/P_{\rm m}$, which characterizes the relative
importance of inertial forces to viscous forces.
(We emphasize that in the above table, Reynolds numbers are defined
based on the inverse wavenumber; our values may therefore be $2\pi$ times
smaller that those by other authors.
The present definition is a natural one in simulations where
one forces power at a particular wavenumber around $k_{\rm f}$.)

\subsection{Stretching, flux freezing and diffusion}
\label{SStretching}

The $\UU\times\BB$ term in \Eq{Induction1} is usually referred to as
the induction term. To clarify its role we rewrite its curl as
\EQ
\nab\times\left(\UU\times\BB\right)
=-\underbrace{\UU\cdot\nab\BB}_{\mbox{advection}}
+\underbrace{\BB\cdot\nab\UU}_{\mbox{stretching}}
-\underbrace{\BB\nab\cdot\UU}_{\mbox{compression}},
\label{stretching_term}
\EN
where we have used the fact that $\nab\cdot\BB=0$.
As a simple example, we consider the effect of a linear shear flow,
$\UU=(0,Sx,0)$ on the initial field $\BB=(B_0,0,0)$.
The solution is $\BB=(1,St,0)B_0$, i.e.\ the field component
in the direction of the flow grows linearly in time.

\begin{figure}[t!]\begin{center}
\includegraphics[width=.5\textwidth]{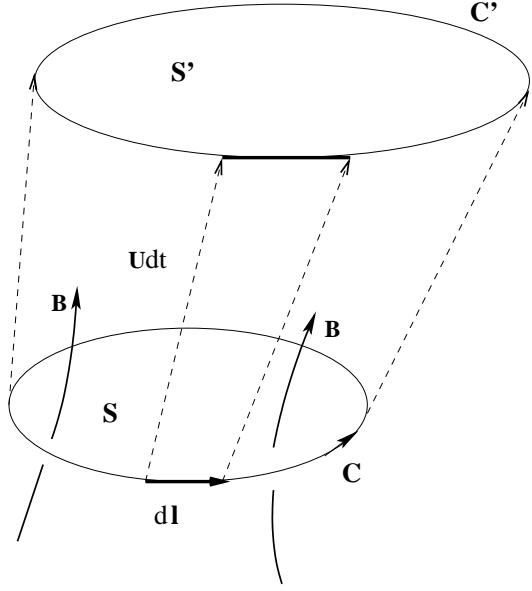}
\end{center}\caption[]{
The surface $S$ enclosed by the curve $C$ is carried by fluid motion
to the surface $S'$ after a time $\dd t$. The flux through this
surface $\Phi$ is frozen into the fluid for a perfectly conducting fluid.
}\label{flux_freeze}
\end{figure}

The net induction term more generally implies 
that the magnetic flux through a surface moving with the
fluid remains constant in the high-conductivity limit.
Consider a surface $S$, bounded by a curve $C$, 
moving with the fluid, as shown in \Fig{flux_freeze}.
Suppose we define the magnetic flux through this surface, 
$\Phi = \int_S \BB\cdot\dd\SSS$. Then
after a time $\dd t$ the change in flux is given by
\EQ
\Delta\Phi = \int_{S'} \BB(t+\dd t)\cdot\dd\SSS - 
\int_S \BB(t)\cdot\dd\SSS.
\label{Delphi}
\EN
Applying $\int\nab\cdot\BB\,\dd V=0$ at time $t+\dd t$, to
the `tube'-like volume swept up by the
moving surface $S$, shown in \Fig{flux_freeze},
we also have
\EQ
\int_{S'} \BB(t+\dd t)\cdot\dd\SSS =  \int_S \BB(t+\dd t)\cdot\dd\SSS
- \oint_C \BB(t+\dd t)\cdot (\dd\lv \times\UU \dd t) ,
\EN
where $C$ is the curve bounding the surface $S$, 
and $\dd\lv$ is the line element along $C$.
(In the last term, to linear order in $\dd t$, it does not matter
whether we take the integral over the curve $C$ or $C'$.)
Using the above condition in \Eq{Delphi}, we obtain
\EQ
\Delta\Phi = \int_S [\BB(t+\dd t) - \BB(t)]\cdot\dd\SSS
- \oint_C \BB(t+\dd t)\cdot (\dd\lv \times\UU) \dd t.
\EN
Taking the limit of $\dd t\to 0$, and noting that
$\BB\cdot(\dd\lv\times\UU)=(\UU\times\BB)\cdot\dd\lv$, we have
\EQ
{d\Phi \over \dd t} = \int_S {\partial\BB\over\partial t}\cdot\dd\SSS
- \oint_C (\UU\times\BB)\cdot\dd\lv
= -\int_S (\nabla \times \eta \JJ)\cdot\dd\SSS.
\EN
In the second equality we have used 
$\oint_C(\UU\times\BB)\cdot\dd\lv
=\int_S\nab\times(\UU\times\BB)\cdot\dd\SSS$
together with the induction equation (\ref{Induction1}).
One can see that,
when $\eta \to 0$, $\dd\Phi/\dd t \to 0$ and so $\Phi$ is constant.

Now suppose we consider a small segment of a thin flux
tube of length $l$ and cross-section $A$, in a highly
conducting fluid. Then, as the fluid moves about, conservation
of flux implies $BA$ is constant, and conservation of mass
implies $\rho A l$ is constant, where $\rho$ is the local
density. So $B \propto \rho l$. For a nearly incompressible
fluid, or a flow with small changes in $\rho$, one
will obtain $B \propto l$. Any shearing motion which increases
$l$ will also amplify $B$; an increase in $l$ leading to
a decrease in $A$ (because of incompressibility) and hence
an increase in $B$ (due to flux freezing). This effect,
also obtained in our discussion of stretching above, 
will play a crucial role in all scenarios involving
dynamo generation of magnetic fields.

The concept of flux freezing can also be derived from the 
elegant Cauchy solution of the induction equation with zero
diffusion. This solution is of use in several
contexts and so we describe it briefly below.
In the case $\eta=0$, the $\nab \times (\UU \times \BB)$
term in \Eq{Induction1} can be expanded to give
\EQ
{\DD\BB\over\DD t} = \BB\cdot\nab\UU -\BB(\nab\cdot\UU),
\EN
where $\DD/\DD t = \partial /\partial t + \UU\cdot\nab$ is
the lagrangian derivative. If we eliminate the $\nab\cdot\UU$ term
using the continuity equation for the fluid,
\EQ
{\partial\rho\over\partial t}=-\nab\cdot(\rho\UU),
\label{continuity}
\EN
where $\rho$ is the fluid density, then we can write
\EQ
{\DD\over\DD t}\left({\BB \over \rho}\right) = {\BB\over \rho}\cdot\nab\UU . 
\EN
Suppose we describe the evolution of a fluid element by
giving its trajectory as $\xx(\xx_0,t)$, where $\xx_0$ is its
location at an initial time $t_0$.
Consider further the evolution of two infinitesimally
separated fluid elements, $A$ and $B$, which, at an initial time $t=t_0$, are
located at $\xx_0$ and $\xx_0 +\delta\xx_0$, respectively. The subsequent
location of these fluid elements will be, say,
$\xx_A = \xx(\xx_0,t)$ and $\xx_B= \xx(\xx_0 +\delta\xx_0, t)$ and
their separation is $\xx_B-\xx_A = \delta\xx(\xx_0,t)$. 
Since the velocity of the fluid particles will be 
$\UU(\xx_A)$ and $\UU(\xx_A) +
\delta\xx\cdot\nab \UU$, after a time $\delta t$, the
separation of the two fluid particles will change by $
\delta t\,\delta\xx\cdot\nab \UU$. The separation vector therefore
evolves as
\EQ
{\DD \delta\xx \over\DD t} = \delta\xx\cdot\nab\UU,
\EN
which is an evolution equation identical to that satisfied by
$\BB/\rho$.
So, if initially, at time $t=t_0$, the fluid particles were on
a given magnetic field line with $(\BB/\rho)(\xx_0,t_0) = c_0 \delta\xx (t_0)
= c_0 \delta \xx_0$, where $c_0$ is a constant, then for all times 
we will have $\BB/\rho = c_0 \delta\xx$.
In other words, `if two infinitesimally close fluid particles are 
on the same line of force at any time, then they will always be on the same
line of force, and the value of $\BB/\rho$ will be proportional
to the distance between the particles' (Section~65 in Ref.~\cite{landau_mhd}).
Further, since $\delta x_i(\xx_0,t) = {\sf G}_{ij}\delta x_{0j}$, where
${\sf G}_{ij} = \partial x_i /\partial x_{0j}$, we can also write
\EQ
B_i(\xx,t)= \rho c_0 \delta x_i = {{\sf G}_{ij}(\xx_0,t)\over\det\GGGG}
\,B_{0j}(\xx_0),
\EN
where we have used the fact that 
$\rho(\xx,t)/\rho(\xx_0,t_0) = (\det\GGGG)^{-1}$.
We will use this Cauchy solution in \App{LagrangianDiffusion}.

\subsection{Magnetic helicity}
\label{MagneticHelicity}

Magnetic helicity plays an important role in dynamo theory.
We therefore give here a brief account of its properties.
Magnetic helicity is the volume integral
\EQ
H=\int_V\AAA\cdot\BB\;\dd V
\label{Hgaugedep}
\EN
over a closed or periodic volume $V$.
By a closed volume we mean one in which the magnetic field lines
are fully contained, so the field has no component normal to the
boundary, i.e.\ $\BB\cdot\nn=0$. The volume $V$ could also
be an unbounded volume with the fields falling off sufficiently
rapidly at spatial infinity.
In these particular cases, $H$ is invariant under the gauge
transformation \eq{gauge_trans}, because
\EQ
H'=\int_V\AAA'\cdot\BB'\;\dd V
=H+\int_V\nab\Lambda\cdot\BB\;\dd V
=H+\oint_{\partial V}\Lambda\BB\cdot\nnn\dd S
=H,
\label{periodic_gaugeinv}
\EN
where $\nnn$ is the normal pointing out of the closed surface $\partial V$.
Here we have made use of $\nab\cdot\BB=0$.

\begin{figure}[t!]\begin{center}
\includegraphics[width=.5\textwidth]{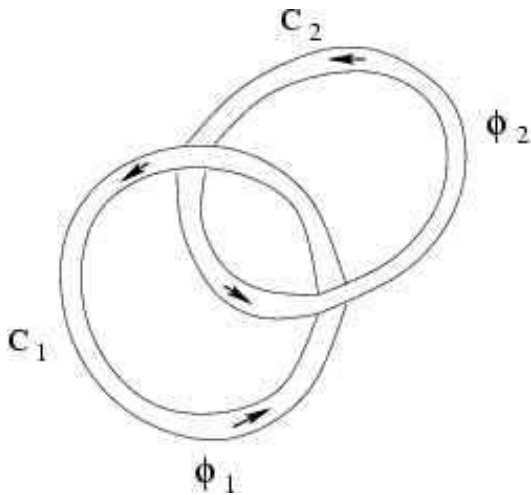}
\end{center}\caption[]{
Two flux tubes with fluxes $\Phi_1$ and $\Phi_2$ are
linked in such a way that they have a helicity $H=+2\Phi_1\Phi_2$.
Interchanging the direction of the field in one of the two
rings changes the sign of $H$.
}\label{hel_ring}
\end{figure}

Magnetic helicity has a simple topological interpretation
in terms of the linkage and twist of isolated (non-overlapping) flux tubes.
For example consider the magnetic helicity for an interlocked, but untwisted,
pair of thin flux tubes as shown in \Fig{hel_ring}, with
$\Phi_1$ and $\Phi_2$ being the fluxes in the tubes around $C_1$ and
$C_2$ respectively. For this configuration
of flux tubes, $\BB\,\dd^3x$ can be replaced by 
$\Phi_1 \dd \lv$ on $C_1$ and $\Phi_2 \dd \lv$ on $C_2$.
The net helicity is then given by the sum
\EQ
H = \Phi_1 \oint_{C_1} \AAA\cdot\dd\lv + \Phi_2 \oint_{C_2} \AAA\cdot\dd\lv,
= 2\Phi_1\Phi_2
\EN
where we have used Stokes theorem to transform
\EQ
\oint_{C_1} \AAA\cdot\dd\lv
=\!\!\int_{S(C_1)}\!\!\BB\cdot\dd\SSS\equiv\Phi_2,
\quad\quad
\oint_{C_2} \AAA\cdot\dd\lv
=\!\!\int_{S(C_2)}\!\!\BB\cdot\dd\SSS\equiv\Phi_1.
\EN
For a general pair of non-overlapping thin flux tubes, the
helicity is given by $H=\pm2\Phi_1\Phi_2$; 
the sign of $H$ depending on the relative orientation of the two tubes
\cite{Mof78}.

The evolution equation for $H$ can be derived from Faraday's law and
its uncurled version for $\AAA$, \Eq{AmpereAAA}, so we have
\EQA
{\partial \over \partial t}(\AAA\cdot\BB) 
&=& (-\EE +\nab\phi)\cdot\BB + \AAA\cdot(-\nab \times \EE)
\nonumber \\ 
&=& -2\EE\cdot\BB + \nab\cdot(\phi\BB + \AAA \times \EE).
\ENA
Integrating this over the volume $V$, 
the magnetic helicity satisfies the evolution equation
\EQ
{\dd H\over\dd t}=-2\int_V \EE\cdot\BB \dd V 
+ \oint_{\partial V}(\AAA \times \EE + \phi\BB)
\cdot\nnn\dd S = -2\eta C,
\label{magn_hel_evol}
\EN
where $C=\int_V\JJ\cdot\BB\;\dd V$ is the current helicity.
Here we have used Ohm's law, $\EE = - \UU \times \BB + \eta \JJ$,  
in the volume integral and we have assumed that
the surface integral vanishes
for closed domains. If the $\mu_0$ factor were included, this 
equation would read $\dd H/\dd t=-2\eta\mu_0 C$.

In the non-resistive case, $\eta=0$, the magnetic helicity
is conserved, i.e.\ $\dd H/\dd t=0$.
However, this does not guarantee
conservation of $H$ in the limit $\eta\rightarrow0$,
because the current helicity, $\int\JJ\cdot\BB\,\dd V$, may in principle still
become large. For example, the Ohmic dissipation rate of magnetic energy
$Q_{\rm Joule}\equiv \int \eta {\bf J}^2 \dd V$
can be finite and balance magnetic energy input by motions,
even when $\eta \rightarrow0$. This is
because small enough scales develop in the field (current sheets)
where the current density increases with decreasing $\eta$ as
$\propto\eta^{-1/2}$ as $\eta\rightarrow0$,
whilst the rms magnetic field strength, $B_{\rm rms}$, remains
essentially independent of $\eta$. Even in this case, however,
the rate of magnetic helicity dissipation {\it decreases} with $\eta$,
with an upper bound to the dissipation rate 
$\propto\eta^{+1/2}\rightarrow0$, as $\eta\rightarrow0$.
Thus, under many astrophysical conditions where $R_{\rm m}$ is
large ($\eta$ small), the magnetic helicity $H$, is almost independent
of time, even when the magnetic energy is dissipated at finite
rates.\footnote{Peculiar counter examples can however be constructed
\cite{blackman04}.
As an example, take a nonhelical large scale field together with a
small scale helical field.
Obviously, the small scale component will decay faster, and so the
magnetic helicity can decay faster than magnetic energy.
However, in the generic case where magnetic helicity is
distributed over all scales, the magnetic energy will always
decay faster than the magnetic helicity.}
This robust conservation of magnetic helicity
is an important constraint on the nonlinear evolution of
dynamos and will play a crucial role below  
in determining how large scale turbulent dynamos saturate.
Indeed, it is also at the heart of Taylor relaxation in
laboratory plasmas, where an initially unstable plasma relaxes
to a stable `force-free' state, dissipating energy, while
nearly conserving magnetic helicity \cite{taylor74}.

We also note the very important fact that the fluid velocity
completely drops out from the helicity evolution equation
\eq{magn_hel_evol}, since $(\UU \times \BB)\cdot\BB = 0$.
Therefore, any change in the nature of the fluid velocity,
for example due to turbulence 
(turbulent diffusion), the Hall effect,
or ambipolar drift (see below), does not affect magnetic helicity
conservation. 
We will discuss in more detail the
concept of turbulent diffusion in a later section,
and its role in dissipating the mean magnetic field.
However, such turbulent magnetic diffusion does {\it not} dissipate
the net magnetic helicity. This property is crucial for understanding 
why, in spite of the destructive properties of turbulence, 
large scale spatio-temporal coherence can emerge if there is helicity 
in the system.

For open volumes, or volumes with boundaries through which
$\BB\cdot\nnn \ne 0$, the magnetic helicity $H$, as defined by 
\Eq{Hgaugedep}, is no longer gauge-invariant.
One can define a gauge-invariant relative magnetic helicity
\cite{berger_field84,berger84,finn_anton85}
\EQ
H_{\rm rel}=\int_V(\AAA+\AAA^{\rm ref})\cdot(\BB-\BB^{\rm ref})\;\dd V,
\label{Hrel}
\EN
where $\BB^{\rm ref}=\nab\times\AAA^{\rm ref}$ is a reference magnetic field
that is taken to be the potential field solution
(where $\BB^{\rm ref}=\nab\varphi$ is the gradient of a potential,
so there is no current), with the boundary condition
\EQ
\nnn\cdot\BB^{\rm ref}=\nnn\cdot\BB,
\EN
i.e.\ the two fields have the same normal components.
The quantity $H_{\rm rel}$ is gauge-invariant,
because in \Eq{periodic_gaugeinv}
the term $\nnn\cdot\BB$ is replaced by $\nnn\cdot(\BB-\BB^{\rm ref})$,
which vanishes on the boundaries.

The evolution equation of the relative magnetic helicity is 
simplified by adopting a specific gauge for $\AAA^{\rm ref}$, with
\EQ
\nab\cdot\AAA^{\rm ref} = 0 , \quad 
\AAA^{\rm ref}\cdot\nnn \vert_{\partial V} = 0.
\label{aref_def}
\EN
We point out, however, that this restriction can in principle be relaxed.
When the gauge \eq{aref_def} is used for the reference field,
the relative magnetic helicity satisfies the evolution equation
\EQ
{\dd H_{\rm rel}\over\dd t}=-2\eta C
-2\oint_{\partial V}(\EE\times\AAA^{\rm ref})\cdot\dd\SSS,
\label{RelHelEvolve}
\EN
where $\dd\SSS=\nnn\,\dd S$ is the surface element.
The surface integral covers the full {\it closed} surface
around the volume $V$.
In the case of the sun the magnetic helicity fluxes from the
northern and southern hemispheres are expected to be about
equally big and of opposite sign, so they would cancel approximately
to zero.
One is therefore usually interested in the magnetic helicity flux out
of the northern or southern hemispheres, but this means that it is
necessary to include the contribution of the equator to the surface
integral in \Eq{RelHelEvolve}.
This contribution can easily be calculated for data from numerical
simulations, but in the case of the sun the contribution from the
equatorial surface is not observed. 

We should point out that it is also possible to define
magnetic helicity as linkages of flux analogous to the
Gauss linking formula for linkages of curves. We have recently
used this approach to formulate the concept of a gauge invariant
magnetic helicity density in the case of random fields, whose
correlation length is much smaller than the system size \cite{KS_AB05}.

We have emphasized earlier in this section that no net magnetic helicity
can be produced by any kind of velocity.
However, this is not true of the magnetic helicity flux which is affected
by the velocity via the electric field.
This can be important if there is differential rotation or shear which
can lead to a separation of magnetic helicity in space.
A somewhat related mechanism is the alpha effect (\Sec{LargeScale}) which
can lead to a separation of magnetic helicity in wavenumber space.
Both processes are important in the sun or the galaxy.

\subsection{The momentum equation}

Finally we come to the momentum equation, which is just the
ordinary Navier-Stokes equation in fluid dynamics supplemented
by the Lorentz force, $\JJ\times\BB$, i.e.\
\EQ
\rho{\DD\UU\over\DD t}=-\nab p+\JJ\times\BB+\ff+\FF_{\rm visc},
\label{DUDt}
\EN
where $\UU$ is the ordinary bulk velocity of the gas, $\rho$ is the density,
$p$ is the pressure, $\FF_{\rm visc}$ is the viscous force, and $\ff$
subsumes all other body forces acting on the gas, including gravity and,
in a rotating system also the Coriolis and centrifugal forces.
(We use an upper case $\UU$, because later on we shall use a lower case
$\uu$ for the fluctuating component of the velocity.)
\EEq{DUDt} has to be supplemented by the continuity equation,
\EQ
{\partial\rho\over\partial t}=-\nab\cdot(\rho\UU),
\label{continuity2}
\EN
an equation of state, $p=p(\rho,e)$, an energy equation for the internal
energy $e$, and an evolution equation for the magnetic field.

An important quantity is the adiabatic sound speed, $c_{\rm s}$, defined
as $c_{\rm s}^2=(\partial p/\partial\rho)_s$, evaluated at constant
entropy $s$.
For a perfect gas with constant ratio $\gamma$ of specific heats
($\gamma=5/3$ for a monatomic gas) we have $c_{\rm s}^2=\gamma p/\rho$.
When the flow speed is much smaller than the sound speed, i.e.\
when the average Mach number $\mbox{Ma}=\bra{\UU^2/c_{\rm s}^2}^{1/2}$ is
much smaller than unity and if, in addition, 
the density is approximately uniform,
i.e.\ $\rho\approx\rho_0={\rm const}$,
the assumption of incompressibility can be made.
In that case, \Eq{continuity} can be {\it replaced} by $\nab\cdot\UU=0$,
and the momentum equation then simplifies to
\EQ
{\DD\UU\over\DD t}=-{1\over\rho_0}\nab p+{\JJ\times\BB\over\rho_0}
+\ff+\nu\nabla^2\UU,
\label{DUDt_incomp}
\EN
where $\nu$ is the kinematic viscosity and $\ff$ is now an external body
force per unit mass.
The ratio $P_{\rm m}=\nu/\eta$ is the magnetic Prandtl number; see \Eq{Pm}.

The assumption of incompressibility is a
great simplification that is useful for many analytic considerations,
but for numerical solutions this restriction is often not necessary.
As long as the Mach number is small, say below 0.3, the weakly
compressible case is believed to be equivalent to the incompressible
case \cite{DHYB03}.

\subsection{Kinetic helicity evolution}

We introduce the vorticity $\WWW=\nab\times\UU$, and define the kinetic helicity
as $F=\int\WWW\cdot\UU\;\dd V$.
Using \Eq{DUDt_incomp}, and ignoring the magnetic field, $F$ obeys the
evolution equation
\EQ
{\dd F\over\dd t}=2\int\WWW\cdot\ff\;\dd V-2\nu\int\WWW\cdot\QQQ\;\dd V,
\EN
where $\QQQ=\nab\times\WWW$ is the curl of the vorticity.
Note that in the absence of forcing, $\ff=0$, and without
viscosity, $\nu=0$, the kinetic helicity is conserved, i.e.\
\EQ
{\dd F\over\dd t}=0\quad\mbox{(if $\nu=0$ and $\ff=0$)}.
\EN
On the other hand, in the {\it limit} $\nu\to0$
(which is different from the case $\nu=0$) the rate of kinetic helicity
production will not converge to zero with decreasing values of $\nu$.
This is a major difference to magnetic helicity conservation, where the
rate of helicity production converges to zero at low resistivity.
Ignoring compressibility effects, i.e.\ $\rho=\mbox{const}$,
this follows by assuming that both kinetic energy, $\half\int\UU^2\dd V$,
and the rate of kinetic energy dissipation,
$\epsilon=\nu\int\WWW^2\dd V$, are independent of $\nu$.
Therefore, both the magnitude of the vorticity, $|\WWW|$,
and the typical wavenumber $k_\nu$ associated with $|\WWW|\approx k_\nu|\UU|$
scale like $k_\nu\sim\nu^{-1/2}$.
Thus, $|\QQQ|\sim k_\nu|\WWW|\sim\nu^{-1}$,
so $\nu|\WWW\cdot\QQQ|\sim\nu^{-1/2}$, and hence
\EQ
\left|{\dd F\over\dd t}\right|
\sim\nu^{-1/2}\to\infty\quad\mbox{(for $\nu\to0$)}.
\EN
For comparison (as we pointed out earlier),
in the magnetic case, the current density also
diverges like $|\JJ|\sim\eta^{-1/2}$, but the rate of magnetic
helicity production is only proportional to $\JJ\cdot\BB$, and
$\eta\JJ\cdot\BB\sim\eta^{+1/2}\to0$, so
\EQ
\left|{\dd H\over\dd t}\right|
\sim\eta^{+1/2}\to0\quad\mbox{(for $\eta\to0$)}.
\EN
It is worth emphasizing again that it is 
for this reason that the magnetic helicity is such an important
quantity in magnetohydrodynamics.

\subsection{Energy and helicity spectra}
\label{energy_spectra}

Magnetic energy and helicity spectra are usually
calculated as
\EQ
M_k = \frac{1}{2}\int_{\mbox{$k$-shell}}
      \BB_{\kk}^*\cdot\BB_{\kk}\,k^2\,\dd\Omega_k,
\EN
\EQ
H_k=\frac{1}{2}\int_{\mbox{$k$-shell}}(\AAA_{\kk}^*\cdot\BB_{\kk}+
\AAA_{\kk}\cdot\BB_{\kk}^*)\,k^2\,\dd\Omega_{\kk},
\label{helspectrum_fft}
\EN
where $\dd\Omega_{\kk}$ is the solid angle element in Fourier space,
$\BB_{\kk}=\ii\kk\times\AAA_{\kk}$ is the
Fourier transform of the magnetic field,
and $\AAA_{\kk}$ is the Fourier transform of the vectors potential.
These spectra are normalized such that
\EQ
\int_0^\infty H_k\,\dd k=\bra{\AAA\cdot\BB}V\equiv H,
\EN
\EQ
\int_0^\infty M_k\,\dd k=\bra{\half\BB^2}V\equiv M,
\EN
where $H$ and $M$ are magnetic helicity and magnetic energy, respectively,
and angular brackets denote volume averages.

There is a conceptual advantage \cite{B01}
in working with the real space Fourier filtered
magnetic vector potential and magnetic field, $\AAA_k$ and $\BB_k$,
where $\BB_k=\nab\times\AAA_k$,
and the subscript $k$ (which is now a scalar!)
indicates Fourier filtering to keep only those
wavevectors $\kk$ that lie in the shell
\EQ
k-\delta k/2\le|\kk|<k+\delta k/2\quad(\mbox{$k$-shell}).
\EN
Magnetic energy and helicity spectra can then be written as
\EQ
M_k=\half\bra{\BB_k^2}V/\delta k,
\EN
\EQ
H_k=\bra{\AAA_k\cdot\BB_k}V/\delta k,
\label{helspectrum}
\EN
where angular brackets denote averages over all space.
We recall that, for a periodic domain, $H$ is gauge invariant. Since
its spectrum can be written as an integral over all space, see
\Eq{helspectrum}, $H_k$ is -- like $H$ -- also gauge invariant.

It is convenient to decompose the Fourier transformed magnetic vector
potential, $\AAA_{\kk}$, into a longitudinal component, $\hh^\parallel$,
and eigenfunctions $\hh^{\pm}$ of the curl operator.
Especially in the context of spherical domains these eigenfunctions are
also called Chandrasekhar--Kendall functions \cite{ChandrasekharKendall57},
while in cartesian domains they are usually referred to as Beltrami waves.
This decomposition has been used in studies of turbulence \cite{Waleffe93},
in magnetohydrodynamics \cite{ChristenssonHindmarshBrandenburg01},
and in dynamo theory \cite{BDS02}.
Using this decomposition we can write the Fourier transformed
magnetic vector potential as
\EQ
\AAA_{\kk}=a_{\kk}^+\hh_{\kk}^++a_{\kk}^-\hh_{\kk}^-
+a_{\kk}^\parallel\hh_{\kk}^\parallel,
\EN
with
\EQ
  \ii\kk\times\hh_{\kk}^{\pm} = \pm k \hh_{\kk}^{\pm},
  \quad\quad k = |\kk|,
\EN
and
\EQ
\bra{{\hh_{\kk}^+}^*\cdot\hh_{\kk}^+}
=\bra{{\hh_{\kk}^-}^*\cdot\hh_{\kk}^-}
=\bra{{\hh_{\kk}^{\parallel}}^*\cdot{\hh_{\kk}^{\parallel}}}=1,
\EN
where asterisks denote the complex conjugate,
and angular brackets denote, as usual, volume averages.
The longitudinal part $a_{\kk}^\parallel\hh_{\kk}^\parallel$ is
parallel to $\kk$ and vanishes after taking
the curl to calculate the magnetic field.
In the Coulomb gauge, $\nabla\cdot\AAA=0$, the longitudinal component vanishes
altogether.

The (complex) coefficients 
$a_{\kk}^\pm(t)$ depend on $\kk$ and $t$, while the eigenfunctions
$\hh_{\kk}^\pm$, which form an orthonormal set, depend only on $\kk$
and are given by
\EQ
\hh_{\kk}^\pm=\frac{1}{\sqrt{2}}{\kk\times(\kk\times\ee)\mp\ii
k(\kk\times\ee)\over k^2\sqrt{1-(\kk\cdot\ee)^2/k^2}},
\label{hhkkpm-defn}
\EN
where $\ee$ is an arbitrary unit vector that is not parallel to $\kk$.
With these preparations
we can write the magnetic helicity and energy spectra in the form
\EQ
H_k = k(|a^+|^2-|a^-|^2)V,
\label{H-apm}
\EN
\EQ
M_k = \half k^2(|a^+|^2+|a^-|^2)V ,
\label{M-apm}
\EN
where $V$ is the volume of integration.
(Here again the factor $\mu_0^{-1}$ is ignored in the definition
of the magnetic energy.)
From \Eqs{H-apm}{M-apm} one sees immediately that \cite{Mof78,BDS02}
\EQ
\half k|H_k|\leq M_k,
\label{realizability}
\EN
which is also known as the {\it realizability condition}.
A fully helical field has therefore $M_k=\pm\half k H_k$.

For further reference we now define power spectra of those components
of the field that are either right or left handed, i.e.\
\EQ
H_k^\pm=\pm k|a_\pm|^2V,\quad
M_k^\pm=\half k^2|a_\pm|^2V.
\EN
Thus, we have $H_k=H_k^++H_k^-$ and $M_k=M_k^++M_k^-$. Note that $H_k^\pm$
and $M_k^\pm$ can be calculated without explicit decomposition into
right and left handed field components using
\EQ
H_k^\pm=\half(H_k\pm2k^{-1}M_k),\quad
M_k^\pm=\half(M_k\pm\half kH_k).
\EN
This method is significantly simpler than invoking explicitly the
decomposition in terms of $a_{\kk}^\pm\hh_{\kk}^\pm$.

In \Sec{NonlinearBehavior} plots of $M_k^\pm$ will be shown and discussed
in connection with turbulence simulations.
Here the turbulence is driven with a helical forcing function
proportional to $\hh_{\kk}^+$; see \Eq{hhkkpm-defn}.

\subsection{Departures from the one--fluid approximation}

In many astrophysical settings the typical length scales are so large
that the usual estimates for the turbulent diffusion of magnetic fields,
by far exceed the ordinary Spitzer resistivity.
Nevertheless, the net magnetic helicity evolution, as we discussed
above, is sensitive to the microscopic resistivity and 
independent of any turbulent contributions.
It is therefore important to discuss in detail the foundations of
Spitzer resistivity and to consider more general cases such as the
two--fluid and even three--fluid models.
In some cases these generalizations lead to important
effects of their own, for example the battery effect.

\subsubsection{Two--fluid approximation}

The simplest generalization of the one--fluid model is
to consider the electrons and ions as separate fluids which
are interacting with each other through collisions.
This two--fluid model is also essential for deriving
the general form of Ohm's law and for describing battery effects,
that generate fields ab-initio from zero initial field.
We therefore briefly consider it below.

For simplicity assume that the ions have one charge,
and in fact they are just protons. That is the plasma is
purely ionized hydrogen. It is straightforward to generalize
these considerations to several species of ions.
The corresponding set of fluid equations, incorporating
the non-ideal properties of the fluids and the anisotropy
induced by the presence of a magnetic field, is worked
out and summarized by Braginsky \cite{brag}. For our purpose
it suffices to follow the simple treatment of 
Spitzer \cite{spitzer56}, where we take the stress tensor to be
just isotropic pressure, leaving out non-ideal terms,
and also adopt a simple form for the collision
term between electrons and protons. 
The equations of motion for the electron and proton fluids may then 
be written as
\EQ
{D_e \uu_e\over D t}
= -{\nab p_e\over n_e m_e} 
-{e \over m_e} \left ( \EE + \uu_e \times \BB \right)  
- \nab \phi_g- {(\uu_e - \uu_p) \over \tau_{ei}},
\label{elec}
\EN
\EQ
{D_i \uu_i\over D t}
= -{\nab p_i\over n_i m_i}  +
 {e \over m_i} \left ( \EE + \uu_i \times \BB \right) 
- \nab \phi_g + {m_en_e\over m_i n_i} {(\uu_e -
\uu_i) \over \tau_{ei}}.
\label{prot}
\EN
Here $D_j \uu_j /D t =\partial\uu_j/\partial t+\uu_j\cdot\nab\uu_j$ and
we have included the forces due to the pressure gradient, gravity,
electromagnetic fields and electron--proton collisions.
Further, $m_j,n_j,u_j,p_j$ are respectively the mass, number density, 
velocity, and the partial pressure of electrons ($j=e$) and 
protons ($j=i$), $\phi_g$ is the gravitational potential, 
and $\tau_{ei}$ is the $e$--$i$ collision time scale.
One can also write down a similar
equation for the neutral component $n$. Adding the $e$, $i$ and $n$ 
equations we can recover the standard MHD Euler equation.

More interesting in the present context
is the difference between the electron and proton fluid equations. 
Using the approximation $m_e/m_i\ll1$, this gives the generalized Ohms law;
see the book by Spitzer \cite{spitzer56}, and Eqs.~(2)--(12) therein,
\EQ
\EE + \uu_i \times \BB  = 
-{\nab p_e \over e n_e} +
{\JJ\over\sigma} + {1\over e n_e} \JJ\times\BB +
{m_e \over e^2} {\partial\over\partial t}
\left({\JJ\over n_e}\right),
\label{genohm}
\EN
where $ \JJ=(en_i\uu_i - en_e\uu_e)$ is the current density
and 
\EQ
\sigma = {n_e e^2 \tau_{ei} \over m_e}
\label{sig}
\EN
is the electrical conductivity.
[If $n_e \ne n_i$, additional terms arise on the RHS of \eq{genohm} 
with $\JJ$ in \eq{genohm} replaced by
$ -e \uu_i (n_e - n_i)$. These terms can usually be neglected since
$(n_e - n_i)/n_e\ll1$. Also negligible are the effects
of nonlinear terms $\propto u_j^2$.]

The first term on the RHS of Eq. (\ref{genohm}), representing the 
effects of the electron pressure gradient, is the `Biermann battery' term.
It provides the source term 
for the thermally generated electromagnetic fields \cite{Bier50,mestelrox}. 
If $\nab p_e / e n_e$ can be written as the gradient 
of some scalar function, then only an electrostatic field is induced 
by the pressure gradient. On the other hand, if this term has
a curl then a magnetic field can grow.
The next two terms on the RHS of Eq. (\ref{genohm}) are the usual 
Ohmic term $\JJ/\sigma$ and the Hall electric field 
$\JJ\times\BB/(n_e e)$, which arises due to a non-vanishing
Lorentz force. Its ratio to the Ohmic term is of order
$\omega_e \tau_{ei}$, where $\omega_e = eB/m_e$ is the electron
gyrofrequency. The last term on the RHS is
the inertial term, which can be neglected if the macroscopic
time scales are large compared to the plasma oscillation periods.

Note that the extra component of the electric field 
introduced by the Hall term is 
perpendicular to $\BB$, and so it does not
alter $\EE\cdot\BB$ on the RHS of the helicity
conservation equation \eq{magn_hel_evol}.
Therefore the Hall electric field
does not alter the volume dissipation/generation of helicity.
The battery term however can in principle contribute
to helicity dissipation/generation, but this contribution
is generally expected to be small.
To see this, rewrite this contribution to helicity generation, say
$(\dd H/\dd t)_{\rm Batt}$, using $p_e = n_e k_{\rm B} T_e$, as
\EQ
\left({\dd H\over\dd t}\right)_{\rm Batt}
= 2\int {\nab p_e\over e n_e}\cdot\BB \ \dd V
= -2\int {\ln n_e \over e} \ \BB\cdot\nab(k_{\rm B}T_e) \ \dd V,
\label{batthel}
\EN
where $k_{\rm B}$ is the Boltzmann constant, and
the integration is assumed to extend over a closed or periodic domain,
so there are no surface terms.\footnote{Note that $n_e$
in the above equation can be divided
by an arbitrary constant density, say $n_0$ to make the argument of the 
log term dimensionless since, on integrating by parts, 
$\int \ln(n_0)\BB\cdot\nab(k_{\rm B}T_e) \ \dd V =0$.}
We see from \Eq{batthel} that generation/dissipation of helicity can
occur only if there are temperature gradients parallel to
the magnetic field \cite{JiPrager96,Ji99,JiPrager02}.
Such parallel gradients are in general very small
due to fast electron flow along field lines.
We will see below that the battery effect can provide a small 
but finite seed field; this can also be accompanied by the generation of
a small but finite magnetic helicity.

From the generalized Ohm's law one can formally solve for the
current components parallel and perpendicular to $\BB$
(cf.\ the book by Mestel \cite{mestelbook}). 
Defining an `equivalent electric field'
\EQ
\EE' = {\JJ\over\sigma} + {\JJ\times\BB \over e n_e},
\EN
one can rewrite the generalized Ohms law as \cite{mestelbook}
\EQ
\JJ = \sigma \EE'_{\parallel} +\sigma_1 \EE'_{\perp} + \sigma_2 {\BB \times \EE'
\over B},
\EN
where
\EQ
\sigma_1 = {\sigma \over 1 + (\omega_e\tau_{ei})^2}, \quad
\sigma_2 = {(\omega_e\tau_{ei})\sigma \over 1 + (\omega_e\tau_{ei})^2}.
\EN
The conductivity becomes
increasingly anisotropic as $\omega_e\tau_{ei}$ increases.
Assuming numerical values appropriate to the
galactic interstellar medium, say, we have 
\EQ
\omega_e\tau_{ei} \approx 4 \times 10^{5} \left({B \over 1 \muG}\right) 
\left({T \over 10^4 K}\right)^{3/2} \left({n_e \over 1 \cm^{-3}}\right)^{-1}
\left({\ln\Lambda\over20}\right)^{-1}.
\label{omegae_tauei}
\EN
The Hall effect and the anisotropy in conductivity are
therefore important in the galactic interstellar medium
and in the cluster gas with high temperatures $T \sim 10^8 \K$ 
and low densities $n_e \sim 10^{-2} \cm^{-3}$.
Of course, in absolute terms, neither the resistivity nor the
Hall field are important in these systems,
compared to the inductive electric field or turbulent diffusion.
For the solar convection zone
with $n_e \sim 10^{18} - 10^{23} \cm^{-3}$, $\omega_e\tau_{ei} \ll 1$,
even for fairly strong magnetic fields. On the other hand, 
in neutron stars, the presence of strong magnetic fields 
$B \sim 10^{13} \G$, could make the Hall term important, especially
in their outer regions, where there are also strong density gradients.
The Hall effect in neutron stars can lead to magnetic fields undergoing 
a turbulent cascade \cite{gold_reis92}.
It can also lead to a nonlinear steepening 
of field gradients \cite{vain_chitre_olinto00} for purely toroidal fields, 
and hence to enhanced magnetic field dissipation. However, even a small
poloidal field can slow down this decay considerably \cite{holler+rue04}.
In protostellar discs, the ratio of the Hall term to microscopic diffusion
is $\sim \omega_e\tau_{en} \sim (8 \times 10^{17}/n_n)^{1/2}
(v_{\rm A}/c_{\rm s})$, where $v_{\rm A}$ and $c_{\rm s}$ are 
the Alfv\'en and sound speeds respectively
\cite{balb_terq01,sano_stone02,wardle99}. The Hall effect
proves to be important in deciding the nature of the magnetorotational
instability in these discs.

A strong magnetic field also suppresses
other transport phenomena like the viscosity and thermal
conduction perpendicular to the field. These effects
are again likely to be important in rarefied and hot
plasmas such as in galaxy clusters.

\subsubsection{The effect of ambipolar drift}
\label{ambipolar}

In a partially ionized medium the magnetic field evolution is
governed by the induction equation (\ref{Induction1}), but
with $\UU$ replaced by the velocity of the ionic component of the 
fluid, $\uu_i$. The ions experience the Lorentz force due
to the magnetic field. This will cause them to drift with respect to
the neutral component of the fluid. If the ion-neutral
collisions are sufficiently frequent, one can assume that the Lorentz force on
the ions is balanced by their friction with the neutrals. Under this
approximation, the Euler equation for the ions reduces to
\EQ
\rho_i \nu_{in} (\uu_i -\uu_n )
= \JJ \times \BB\quad\mbox{(strong coupling approximation)},
\label{ambi}
\EN
where $\rho_i$ is the mass density of ions, $\nu_{in}$ the ion-neutral
collision frequency and $\uu_n $ the velocity of the neutral particles.
For gas with nearly primordial composition and
temperature $\sim 10^4 \K$, one gets the estimate \cite{KSmn98}
of $\rho_i \nu_{in} = n_{i} \rho_n \bra{\sigma v}_{\rm eff}$,
with $\bra{\sigma v}_{\rm eff} \sim 4 \times 10^{-9} \cm^{3} \s^{-1}$,
in cgs units. Here, $n_i$ is the number density of ions and $\rho_n$ the mass
density of neutrals.

In a weakly ionized gas, the bulk velocity is
dominated by the neutrals, and \eq{ambi} substituted into the
induction equation \eq{Induction1} then leads to a
modified induction equation,
\EQ
{\partial\BB\over\partial t}
=\nab\times\left[(\UU + a\JJ \times \BB)\times\BB-\eta\JJ\right],
\label{Induction2}
\EN
where $a = (\rho_i \nu_{in})^{-1}$. The modification is therefore
an addition of an extra drift velocity, proportional to the Lorentz 
force. One usually refers to this drift velocity as ambipolar drift
(and sometimes as ambipolar diffusion) in the astrophysical community
(cf.\ Refs~\cite{mestelbook,MestelSpitzer56,Draine86} 
for a more detailed discussion).

We note that the extra component of the electric field 
introduced by the ambipolar drift is 
perpendicular to $\BB$ and so, just like the Hall term, does not
alter $\EE\cdot\BB$ on the RHS of the helicity
conservation equation \eq{magn_hel_evol}; so ambipolar drift -- like
the Hall effect -- does not alter the volume
dissipation/generation of helicity.
(In fact, even in the presence of neutrals, the magnetic field
is still directly governed by only the electron fluid velocity,
which does not alter the volume dissipation/generation of helicity.)
Due to this feature, ambipolar drift 
provides a very useful toy model for the study
of the nonlinear evolution of the mean field.
Below, in \Secs{SaturationSmallScale}{ClosureModel}, we will study a
closure model that exploits this feature.

Ambipolar drift can also be important in the magnetic field
evolution in protostars, and also in the neutral component of the 
galactic gas. In the classical (non-turbulent) picture of star formation, 
ambipolar diffusion regulates a slow infall of the gas, which
was originally magnetically supported \cite{mestelbook}; see also Chapter~11.
In the galactic context, ambipolar diffusion can lead to
the development of sharp fronts near nulls of the magnetic field.
This, in turn, can affect the rate of destruction/reconnection of the field
\cite{ax_z94,ax_z95,ax_z97}.

\subsection{The Biermann battery}
\label{battery}

Note that $\BB=0$ is a perfectly valid solution of the induction
equation (\ref{Induction1}), so no magnetic field would be generated if
one were to start with zero magnetic field.
The universe probably did not start with
an initial magnetic field. One therefore needs
some way of violating the induction equation to produce a cosmic
battery effect, and to drive currents from a state with initially
no current. There are a number of such battery mechanisms
which have been suggested 
\cite{Widr02,Bier50,mestelrox,Rees87,Sub94,kulsrud97,GR01,MG03,SS04}.
Almost all of them lead to only weak fields, much weaker
than the observed fields. Therefore, dynamo action due to a
velocity field acting to exponentiate
small seed fields efficiently,
is needed to explain observed field strengths.
We briefly comment on one cosmic battery, the Biermann battery.

The basic problem any battery has to address is how to
produce finite currents from zero currents?
Most astrophysical mechanisms use the fact that positively and
negatively charged particles in a charge-neutral universe, do not
have identical properties. For example if one considered a gas
of ionized hydrogen, then the electrons have a much smaller
mass compared to protons. This means that for a given pressure gradient
of the gas the electrons tend to be accelerated much more than the ions.
This leads in general to an electric field, which couples back
positive and negative charges. This is exactly the thermally
generated field we found in deriving the generalized Ohm's law.

Taking the curl of Eq.~(\ref{genohm}),
using Maxwell's equations (Faraday's and Ampere's law),
and writing $p_e = n_e k_{\rm B}T$, where $k_{\rm B}$
is the Boltzmann constant, we obtain
\EQ
{\partial \BB \over \partial t} =
\nab\times (\UU \times\BB) 
- \nab\times \eta\JJ
-{c k_{\rm B} \over e} {\nab n_e\over n_e}\times\nab T.
\label{modb}
\EN
Here we have taken the velocity of the ionic component
to be also nearly the bulk velocity in a completely ionized fluid,
so we put $\uu_i = \UU$. We have neglected the Hall effect and 
inertial effects as they are generally very small for the fields
one generates. 

We see that over and above the usual
flux freezing and diffusion terms we have a {\it source term}
for the magnetic field evolution, even if the initial
field were zero. This source term is non-zero if and
only if the density and
temperature gradients, $\nab n_e$ and $\nab T$, are not
parallel to each other.
The resulting battery effect, known as the Biermann battery,
was first proposed as a mechanism for the
thermal generation of stellar magnetic fields \cite{Bier50,mestelrox}.

In the cosmological context, the Biermann battery can also
lead to the thermal generation
of seed fields in cosmic ionization fronts \cite{Sub94}.
These ionization fronts are
produced when the first ultraviolet photon sources,
like quasars, turn on to ionize the intergalactic medium (IGM).
The temperature gradient in a cosmic ionization
front is normal to the front.  However, a component to the
density gradient can arise in a different direction, if the ionization
front is sweeping across arbitrarily laid down density fluctuations.
Such density fluctuations, associated with protogalaxies/clusters,
in general have no correlation to the source of the
ionizing photons. Therefore, their density gradients are not
parallel to the temperature gradient associated with the ionization
front. The resulting thermally generated
electric field has a curl, and magnetic fields
on galactic scales can grow. After compression during
galaxy formation, they turn out to have a strength 
$B \sim 3 \times 10^{-20}\G$ \cite{Sub94}. A similar effect was
considered earlier in the context of generating fields
in the interstellar medium in Ref.~\cite{Laza92}.
(This mechanism also has analogues in some laboratory
experiments, when laser generated plasmas interact with their
surroundings \cite{Stamper71,Stamper75}. Indeed, our estimate for the
generated field is very similar to the estimate in Ref.~\cite{Stamper71}.)
This field by itself falls far short of the observed microgauss strength
fields in galaxies, but it can provide a seed field,
coherent on galactic scales, for a dynamo.
Indeed the whole of the IGM is seeded with magnetic fields of small 
strength but coherent on megaparsec scales.

This scenario has in fact been confirmed in detailed numerical
simulations of IGM reionization \cite{gnedin00},
where it was found that the breakout of ionization 
fronts from protogalaxies and their propagation through the 
high-density neutral filaments that are 
part of the cosmic web, and that both generate magnetic fields. 
The field strengths increase further due to gas compression 
occurring as cosmic structures form. The magnetic field at a redshift 
$z \sim 5$ closely traces the gas density, and is highly ordered on 
megaparsec scales.
Gnedin et al.\ \cite{gnedin00} found a mean mass-weighted field strength of 
$B \sim 10^{-19}\G$ in their simulation box. 

The Biermann battery has also been shown to generate both 
vorticity and magnetic fields in oblique cosmological shocks which arise 
during cosmological structure formation 
\cite{kulsrud97,davis_widrow00}. In fact, Kulsrud et al.\ \cite{kulsrud97} 
point out that the well-known analogy between the
induction equation and the vorticity equation (without Lorentz force)
extends even to the case where a battery term is present.
Suppose we assume that the gas is pure hydrogen,
has a constant (in space) ionization fraction $\chi$,
and has the same temperature for electrons, protons and hydrogen,
it follows that $p_e = \chi p/(1+\chi)$ and $n_e = \chi \rho/m_p$.
Defining $\oo_B = e\BB/m_p$, 
the induction equation with the thermal battery term 
can then be written as
\EQ
{\partial \oo_B \over \partial t} =
\nab\times\left(\UU\times\oo_B - \eta\nab\times\oo_B\right) + 
 {\nab p \times\nab \rho \over \rho^2}{1 \over 1 + \chi}.
\label{modomb}
\EN
The last term, without the extra factor of $-(1 + \chi)^{-1}$,
corresponds to the baroclinic term in the
equation for the vorticity $\oo = \nab \times \UU$,
\EQ
{\partial \oo \over \partial t} =
\nab\times\left(\UU\times\oo-\nu\nab\times\oo\right)
- {\nab p \times\nab \rho \over \rho^2}.
\label{modombu}
\EN
So, provided viscosity and magnetic diffusivity were
negligible, both $\oo_B (1 +\chi)$ and $-\oo$
satisfy the same equation.
Furthermore, if they were both
zero initially then, for subsequent times,
we have $e\BB/m_p = - \oo/(1 +\chi)$.
Numerically, a value of $\omega \sim 10^{-15} \s^{-1}$ corresponds
to a magnetic field of about $\sim 10^{-19} \G$.

We briefly comment on the extensive work trying to generate
magnetic fields in the early universe; for example 
in a phase transition or during inflation (see for example
the reviews \cite{Widr02,Rees87,GR01,MG03} and
references therein).
The main difficulty with generating such primordial fields
in an early universe phase transition is the very small
correlation length of the generated field, which is typically
limited to a fraction of the Hubble radius at the epoch of generation.
So, even if a significant fraction of the energy density of the
universe went into magnetic fields, the field averaged over galactic
dimensions turns out to be extremely small, typically smaller than the
astrophysically generated seed fields discussed above.
One exception is if helicity is also generated, in which
can an inverse cascade can lead to an increase in the scale
of the field \cite{BEO96,field_caroll}. 

Generation of primordial fields during inflation can lead to the required 
large correlation lengths. However, one needs to break the conformal
invariance of the electromagnetic action. A number of ways
of breaking conformal invariance and generating
magnetic fields have been explored 
\cite{Widr02,GR01,MG03,turner_widrow88,ratra92}.
But the amplitude of the generated primordial field is
exponentially sensitive to the parameters. Primordial fields
generated in the early universe can also influence structure formation
in the universe if they are bigger than about a nanogauss 
\cite{kimetal95,ksjdb98a,jko98}.
Such fields can be constrained using CMB anisotropy observations
\cite{ksjdb98b,trsks01,mack02,ksjdb02,ksjdbtrs,lewis04,ks05}.
In this review we shall not treat
these issues in any detail but refer the interested reader to
the papers referred to above.

\section{Dynamos and the flow of energy}

The dynamo mechanism provides a means of converting kinetic energy
into magnetic energy.
We shall focus on the astrophysically relevant case of a turbulent
dynamo, as opposed to a laminar one.
Laminar dynamos are easier to understand -- and we shall discuss
some simple examples -- while turbulent dynamos have to be tackled
via direct numerical simulations or by stochastic methods.
Both will be discussed below.

\subsection{Energetics}
\label{Energetics}

Important insight can be gained by considering the magnetic energy equation.
By taking the dot product of \Eq{Induction1} with $\BB/(2\mu_0)$ and
integrating over the volume $V$, we obtain
\EQ
{\dd\over\dd t}\int_V{\BB^2\over2\mu_0}\,\dd V
=-\int_V\UU\cdot(\JJ\times\BB)\,\dd V
-\int_V{\JJ^2\over\sigma}\,\dd V
-\oint_{\partial V}{\EE\times\BB\over\mu_0}\,\dd\SSS.
\label{MagnEnergy}
\EN
This equation shows that the magnetic energy can be increased by doing
work against the Lorentz force, provided this term exceeds resistive
losses (second term) or losses through the surface (Poynting flux, last term).
Likewise, by taking the dot product of \Eq{DUDt} with $\rho\UU$
and integrating, one arrives at the kinetic energy equation
\EQA
{\dd\over\dd t}\int_V\half\rho\UU^2\,\dd V
=+\int_V p\nab\cdot\UU\,\dd V
+\int_V\UU\cdot(\JJ\times\BB)\,\dd V
\nonumber \\
+\int_V \rho\UU\cdot\grav\,\dd V
-\int_V 2 \nu\rho\SSSS^2\,\dd V,
\label{KinEnergy}
\ENA
where ${\sf S}_{ij}=\half(u_{i,j}+u_{j,i})-\onethird\delta_{ij}u_{k,k}$
is the traceless rate of strain tensor, and commas denote derivatives.
In deriving \Eq{KinEnergy} we
have assumed stress-free boundary conditions, so there are no
surface terms and no kinetic energy is lost through the boundaries.
Equations \eq{MagnEnergy} and \eq{KinEnergy} show that the generation
of magnetic energy goes at the expense of kinetic energy, without loss
of net energy.

\begin{figure}[t!]\begin{center}
\includegraphics[width=.99\textwidth]{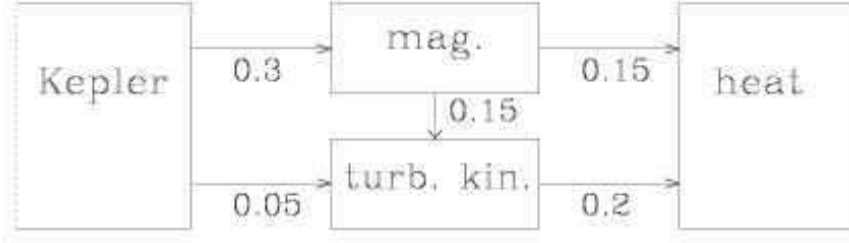}
\end{center}\caption[]{
Energy budget in a local accretion disc simulation where the turbulence
is maintained by the Balbus-Hawley instability.
The numbers on the arrows indicate the approximate energy conversion rates in
units of $\Omega E_{\rm M}$, where $\Omega$ is the angular velocity
and $E_{\rm M}$ is the steady state value of the magnetic energy.
(Adapted from Ref.~\cite{BNST95}.)
}\label{energy_disc}\end{figure}

In many astrophysical settings one can distinguish four different energy
reservoirs that are involved in the dynamo process: magnetic, kinetic,
thermal, and potential energy.
In accretion discs the magnetic energy comes ultimately from potential
energy which is first converted into kinetic energy.
This is only possible by getting rid of angular momentum via Reynolds
and/or Maxwell stresses.
Half of the potential energy goes into orbital kinetic energy and the
other half goes into turbulent kinetic energy which is then dissipated
into heat and radiation.
This requires turbulence to produce small enough length scales so that
enough kinetic energy can indeed be dissipated on a dynamical time scale.
This turbulence is most likely driven by the Balbus-Hawley (or
magneto-rotational) instability; see Ref.~\cite{BH98} for a review.
In \Fig{energy_disc} we show a typical energy diagram from a local simulation
of the Balbus-Hawley instability.
Here, the magnetic field necessary for the instability is maintained by
a dynamo process.
Most of the turbulent energy is dissipated by Joule heating \cite{BNST95}.
The magnetic energy typically exceeds the kinetic energy by a factor
of about 3 or more, but is below the thermal energy by a factor of
about 10--20; see Ref.~\cite{BNST96}.

\begin{figure}[t!]\begin{center}
\includegraphics[width=.99\textwidth]{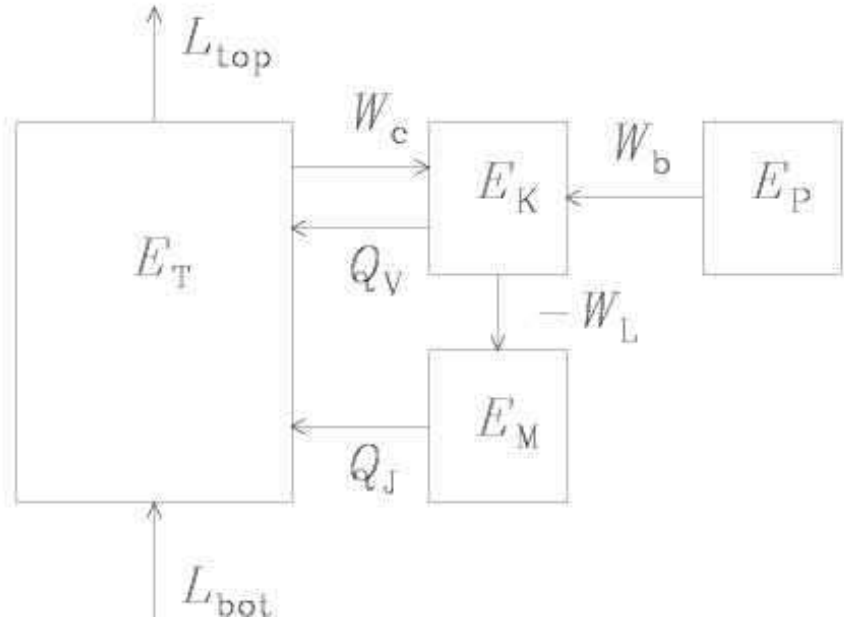}
\end{center}\caption[]{
Energy budget in a local convection simulation.
The dynamo is convectively driven by the luminosity entering from below,
giving rise to convection via work done by adiabatic compression,
$W_{\rm c}=\int p\nab\cdot\uu\,\dd V$, and through work done against
the Lorentz force, $W_{\rm L}=\int\uu\cdot(\JJ\times\BB)\,\dd V$.
Energy is being fed back from magnetic and kinetic energy to
thermal energy via Joule and viscous heating, $Q_{\rm J}$
and $Q_{\rm v}$. Some of the kinetic energy is constantly
being exchanged with potential energy $E_{\rm P}$ via
$W_{\rm b}=\int\rho\grav\cdot\uu\,\dd V$
(adapted from Ref.~\cite{BJNRST96}).
}\label{energy_convection}\end{figure}

In the case of solar convection the energy for the dynamo comes ultimately
from the nuclear reactions in the center of the star.
These acts as a source of thermal energy which gets converted into
kinetic energy via the convection instability.
The corresponding energy diagram for this case is shown in
\Fig{energy_convection}.
Potential energy does not contribute directly: it only contributes
through rearranging the mean density stratification \cite{BJNRST96}.

\subsection{Kinematic dynamos}

The onset of dynamo action can be studied in the linear
approximation, i.e.\ the velocity field is assumed to
be given (kinematic problem).
There is in general a critical value of the magnetic Reynolds number
above which the magnetic field grows exponentially.
A lot of work has been devoted to the question of whether the growth
rate can remain finite in the limit $R_{\rm m}\to\infty$ (the so-called
fast dynamo problem); see Refs~\cite{SC90,RS92,CG95} for reviews.
Fast dynamos are physically meaningful only until
nonlinear effects begin to modify the flow to limit
further growth of the field.

In the following we consider two simple examples of a dynamo.
Both are slow dynamos, i.e.\ magnetic diffusion is crucial for
the operation of the dynamo.
We also discuss the stretch-twist-fold dynamo as a qualitative
example of what is possibly a fast dynamo.

\subsubsection{The Herzenberg dynamo}

In the wake of Cowling's antidynamo theorem \cite{Cow34} the
Herzenberg dynamo \cite{Herz58} played an important role as an
early example of a dynamo where the existence of excited solutions
could be proven rigorously.
The Herzenberg dynamo does not attempt to model an astrophysical dynamo.
Instead, it was complementary to some of the less mathematical and more
phenomenological models at the time, such as Parker's migratory dynamo
\cite{Par55} as well as the observational model of Babcock \cite{Babcock},
and the semi-observational model of Leighton \cite{Leighton69},
all of which were specifically designed to describe the solar cycle.

\begin{figure}[t!]\begin{center}
\includegraphics[width=.99\textwidth]{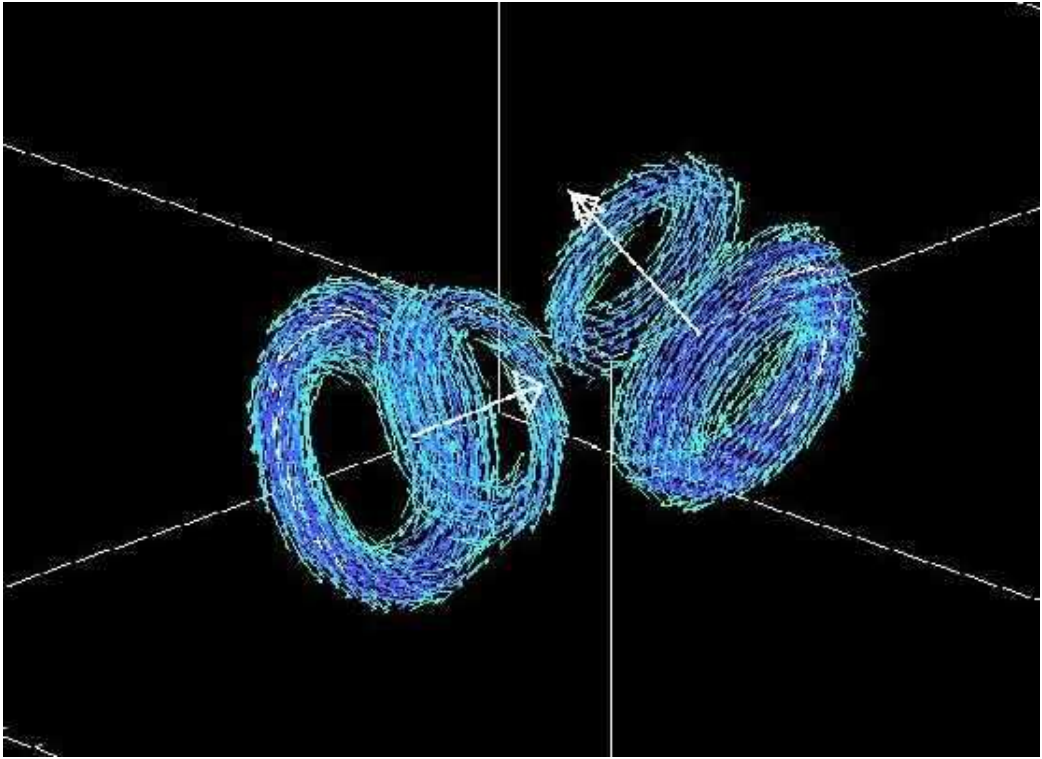}
\end{center}\caption[]{
Three-dimensional visualization of the
magnetic field geometry of the Herzenberg dynamo.
$\BB$-vectors are shown when their length exceeds
about 25\% of the maximum value
(adapted from Ref.~\cite{BMS98}).
}\label{herz_stationary}\end{figure}

The Herzenberg dynamo
is based on the mutual interaction of the magnetic fields
produced by two spinning spheres in a conducting medium.
In its simplest variant, the axes of the two spheres lie in two parallel planes
and have an angle $\varphi$ to each other; see \Fig{herz_stationary}, which shows
the field vectors from a numerical simulation of the Herzenberg
dynamo \cite{BMS98}.

Dynamo action is possible unless the angle $\varphi$ is exactly
$0^\circ$, $90^\circ$, or $180^\circ$.
For $90^\circ<\varphi<180^\circ$ nonoscillatory dynamo action
is possible.
In the limit where the radius of the spheres, $a$, is small compared
with their separation $d$, one can expand the field locally in terms
of multipoles to lowest order.
Defining a magnetic Reynolds number as $R_{\rm m}=\omega d^2/\eta$,
where $\omega$ is the spin frequency of each of the spheres,
the critical magnetic Reynolds number for dynamo
action, $R_{\rm crit}$, is found to be \cite{Mof78}
\EQ
R_{\rm crit}^{-2}=-{1\over4800}\left({a\over d}\right)^6
\sin\!^2\varphi\cos\varphi\quad\mbox{(for $90^\circ<\varphi<180^\circ$}),
\label{Herzenberg_formula}
\EN
which shows that the smallest value of $R_{\rm crit}$ is reached
for $\varphi\approx125^\circ$.
Critical magnetic Reynolds numbers are several hundreds.
However, because $R_{\rm m}$ depends quadratically on $d$, $R_{\rm crit}$
would be only around ten if we were to redefine the magnetic Reynolds
number based on some typical wavenumber; see \Eq{Rmdefinition}.
The dynamo works on the principle that each sphere winds up its
ambient field, creates thereby a strong toroidal field around itself.
Because there is an angle between the two spheres
the toroidal field of one sphere acts as a poloidal field for
the other sphere. For the toroidal field of each sphere
to propagate to the other sphere, a non-zero diffusion is
necessary, hence making this dynamo a slow dynamo.

Already back in the sixties,
the idea of the Herzenberg dynamo has been verified experimentally
\cite{LowesWilkinson1963,LowesWilkinson1968} using two conducting
cylinders embedded in a solid block of the same material.
The cylinders were in electric contact with the block through a thin
lubricating film of mercury.

The asymptotic theory of Herzenberg \cite{Herz58} assumed that
$a/d\ll1$; for excellent reviews of the Herzenberg dynamo see
Refs~\cite{Mof78,Roberts67}.
Using numerical simulations \cite{BMS98}, it has been shown that
\Eq{Herzenberg_formula} remains reasonably accurate even when
$a/d\approx1$.
These simulations also show that in the range $0^\circ<\varphi<90^\circ$
dynamo action is still possible, but the solutions are no longer steady
but oscillatory; see Ref.~\cite{BMS98} for an asymptotic treatment.
In the early papers, only steady solutions were sought, which
is the reason why no solutions were originally found for
$0^\circ<\varphi<90^\circ$.

\subsubsection{The Roberts flow dynamo}
\label{RobertsFlowDynamo}

\begin{figure}[t!]\begin{center}
\includegraphics[width=.65\textwidth]{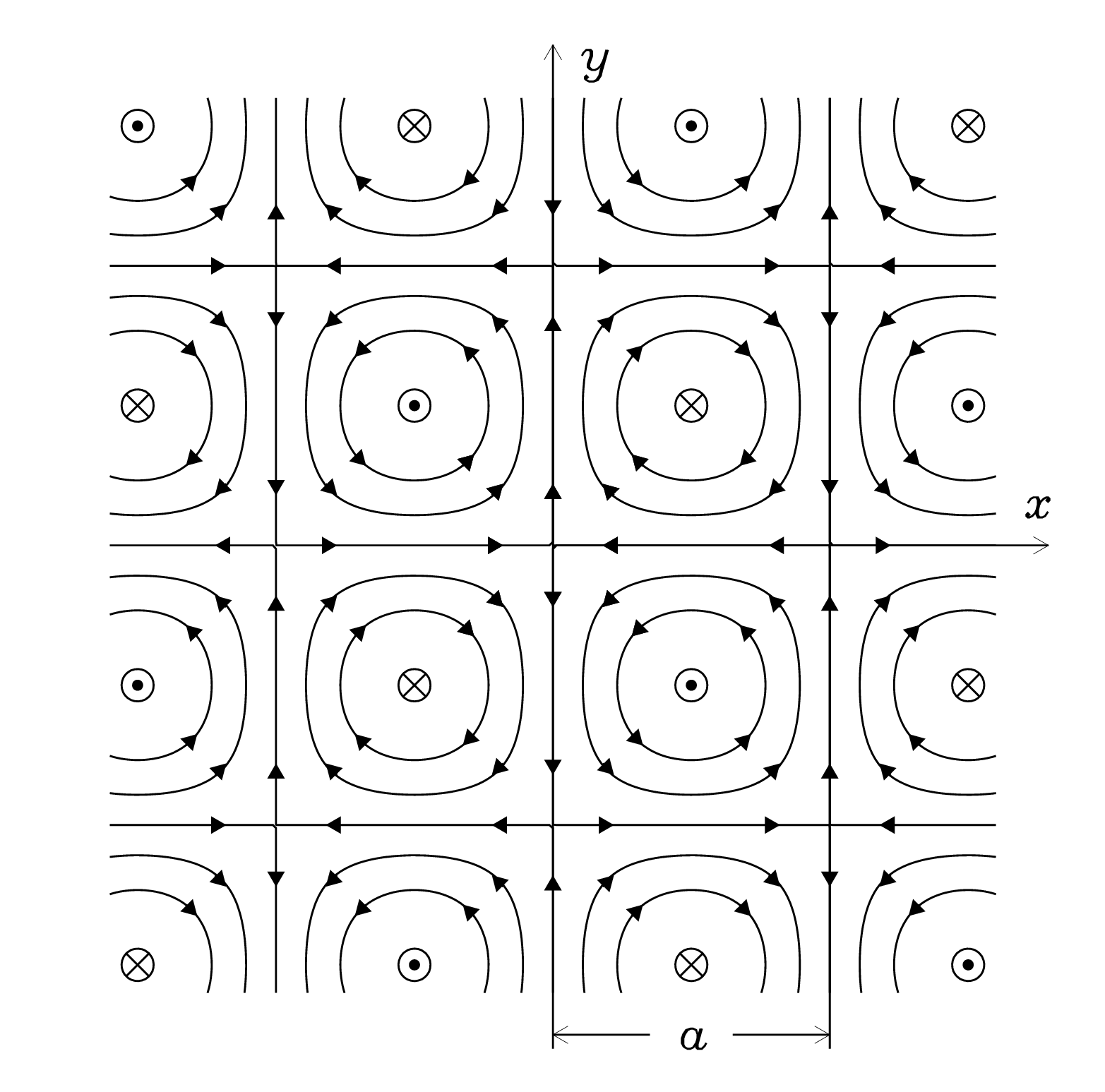}
\end{center}\caption[]{
Roberts flow pattern with periodicity $2a$, corresponding to \Eq{UUrobflow}
(adapted from Ref.~\cite{RB03}).
}\label{robflow}\end{figure}

In the early years of dynamo theory most examples were constructed and
motivated based on what seems physically possible and plausible.
An important element of astrophysical dynamos is that the flow is bounded
in space and that the magnetic field extends to infinity.
Later, and especially in recent years, these restrictions were relaxed
in may approaches.
One of the first examples is the G.\ O. Roberts dynamo
\cite{Roberts70,Roberts72}.
The flow depends on only two coordinates, $\UU=\UU(x,y)$, 
and can be written in the form
\EQ
\UU(x,y)=k_{\rm f}^{-1}\nab\times(\varphi\zz)
+k_{\rm f}^{-2}\nab\times\nab\times(\varphi\zz),
\label{UUrobflow}
\EN
with the stream function $\varphi=\sqrt{2}\,U_0\cos k_x x\cos k_y y$,
where $k_x=k_y=\pi/a$; see \Fig{robflow}.
This flow is fully helical with $\WWW=k_{\rm f}\UU$, were
$k_{\rm f}^2=k_x^2+k_y^2$ and $\WWW=\nab\times\UU$.
The flow is normalized such that $\bra{\UU^2}=U_0^2$.
While the flow is only two-dimensional (in the sense
that $\UU$ is a function only of $x$ and $y$), 
the magnetic field must be three-dimensional
for all growing solutions (dynamo effect).
The field must therefore also depend on $z$.

\begin{figure}[t!]\begin{center}
\includegraphics[width=.75\textwidth]{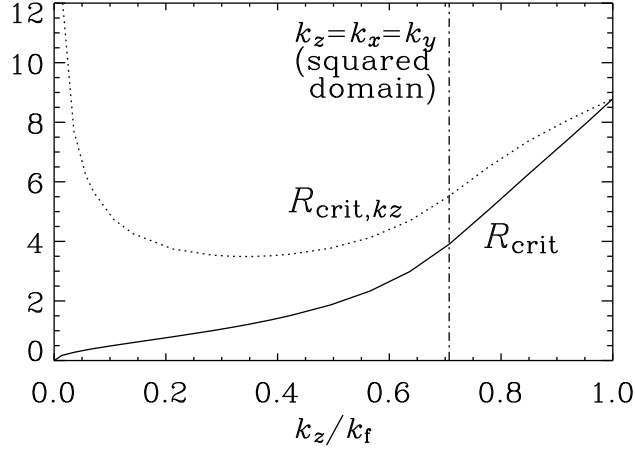}
\end{center}\caption[]{
Critical magnetic Reynolds number,
$R_{\rm crit}=U_0/(\eta k_{\rm f})_{\rm crit}$,
for the Roberts flow as a function of $k_z/k_{\rm f}$,
where $k_{\rm f}^2=k_x^2+k_y^2$.
The critical magnetic Reynolds number based on $k_z$,
$R_{{\rm crit},kz}=U_0/(\eta k_z)_{\rm crit}$, has a minimum
at $k_z\approx0.34k_{\rm f}\approx0.48k_x$ (dotted line).
The case of a squared domain with $k_x=k_y=k_z$, i.e.\
$k_z/k_{\rm f}=1/\sqrt{2}$, is indicated by the vertical
dash-dotted line.
}\label{proberts}\end{figure}

The governing equations are homogeneous with coefficients that are
independent of $z$ and $t$.
The solutions of the kinetic problem can therefore be written in the form
\EQ
\BB(x,y,z,t)={\rm Re}\left[\hat{\BB}_{k_z}(x,y)
\exp(\ii k_z z+\lambda t)\right],
\EN
where $\hat{\BB}_{k_z}$ is the eigenfunction, which is obtained by solving the
eigenvalue problem
\EQ
\lambda\hat{\AAA}_{k_z}=\UU\times\hat{\BB}_{k_z}
+\eta\left(\nabla^2-k_z^2\right)\hat{\AAA}_{k_z},
\label{eigenvalueproblem}
\EN
where
$\hat{\BB}_{k_z}=\nab\times\hat{\AAA}_{k_z}+\ii\kk_z\times\hat{\AAA}_{k_z}$
is expressed in terms of $\hat{\AAA}_{k_z}$, which is
a mixed representation of the vector potential; in real space the
vector potential would be 
${\rm Re}\left[\hat{\AAA}_{k_z}(x,y)\exp(\ii k_z z+\lambda t)\right]$.
In \Fig{proberts} we present critical values of the magnetic Reynolds
number as a function of $k_z$, obtained by solving \Eq{eigenvalueproblem}
numerically as described in Refs~\cite{RB03,PlunianRadler02}.
For $k_z=k_x=k_y\approx0.71k_{\rm f}$, the marginal state ($\lambda=0$) is
reached when $R_{\rm crit}\equiv U_0/(\eta k_{\rm f})_{\rm crit}\approx3.90$.
The larger the domain in the $z$-direction, the lower is the critical
magnetic Reynolds number.
However, the critical magnetic Reynolds number based on $k_z$,
$R_{{\rm crit},kz}=U_0/(\eta k_z)_{\rm crit}$, has a minimum
at $k_z\approx0.34k_{\rm f}\approx0.48k_x$ with $R_{{\rm crit},kz}\approx3.49$;
cf.\ \Fig{proberts}.

The horizontally averaged eigenfunction is
$\overline{\hat{\BB}}_{k_z}=(\ii,1,0)$, corresponding to a Beltrami wave
(see \Sec{energy_spectra}),
which has maximum magnetic helicity with a sign that
is opposite to that of the flow.
In the present case, the kinetic helicity of the flow is positive,
so the magnetic and current helicities of the mean field are negative.

The significance of this solution is two-fold.
On the one hand, this dynamo is the prototype of any fully
helical dynamo capable of generating a large scale field ($k_z\ll k_{\rm f}$).
On the other hand, it is a simple model of the
Karlsruhe dynamo experiment where a similar flow of liquid
sodium is generated by an arrangement of pipes with internal
`spin generators' making the flow helical.
It is also an example of a flow where the generation of the magnetic
field can be described in terms of mean field electrodynamics.

Unlike the original Roberts flow dynamo, the flow in the Karlsruhe dynamo
experiment is bounded and embedded in free space.
Within the dynamo domain, the mean field,
$\overline{\hat{\BB}}_{k_z}=(\ii,1,0)$, has only $(x,y)$-components.
The field lines must close outside the dynamo domain, giving therefore
rise to a dipole lying in the $(x,y)$-plane.
Similar fields have long been predicted for rapidly rotating
stars \cite{Rue80}.
This will be discussed in more detail in \Sec{Rapid_star+planet}.

\subsection{Fast dynamos: the stretch-twist-fold picture}

\begin{figure}[t!]\begin{center}
\includegraphics[width=.9\textwidth]{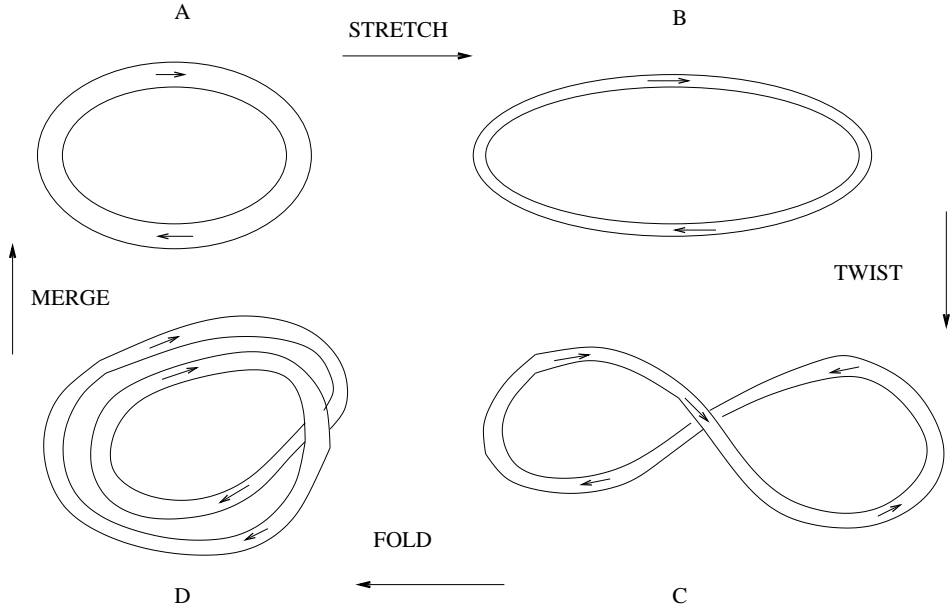}
\end{center}\caption[]{
A schematic illustration of the
stretch-twist-fold-merge dynamo.
}\label{stf_f}\end{figure}

An elegant heuristic dynamo model illustrating the
possibility of fast dynamos is what is often referred
to as the Zeldovich `stretch-twist-fold' (STF) dynamo
(see \Fig{stf_f}). This is now discussed in many books 
\cite{CG95,ZRS83} and we briefly outline it here, as it 
illustrates nicely several features of more realistic dynamos.

The dynamo algorithm starts with first stretching
a closed flux rope to twice its length preserving its volume,
as in an incompressible flow (A$\to$B in \Fig{stf_f}).
The rope's cross-section then
decreases by factor two, and because of flux freezing the
magnetic field doubles. In the next step, the rope is twisted into
a figure eight (B$\to$C in \Fig{stf_f}) and then folded 
(C$\to$D in \Fig{stf_f}) so that now there are two loops, 
whose fields now point in the same
direction and together occupy a similar volume as the original flux loop.
The flux through this volume has now doubled. The last important
step consists of merging the two loops into one (D$\to$A in \Fig{stf_f}),
through small diffusive effects.
This is important in order that the new arrangement cannot easily
undo itself and the whole process becomes irreversible.
The newly merged loops now become topologically the same as the 
original single loop, but now with the field strength scaled 
up by factor 2.

Repeating the algorithm $n$ times, leads to the
field in the flux loop growing by factor $2^n$, or at a
growth rate $\sim T^{-1}\ln2$ where $T$ is the
time for the STF steps. This makes the dynamo
potentially a fast dynamo, whose growth rate does
not decrease with decreasing resistivity.
Also note that the flux through a fixed `Eulerian surface'
grows exponentially, although the flux through any lagrangian
surface is nearly frozen; as it should be for small diffusivities.

The STF picture illustrates several other features: first we see that
shear is needed to amplify the field at step A$\to$B.
However, without the twist part
of the cycle, the field in the folded loop would cancel
rather than add coherently. To twist the loop
the motions need to leave the plane and go into the third dimension;
this also means that field components perpendicular to the loop
are generated, albeit being strong only temporarily during the
twist part of the cycle. The  source for the magnetic energy is 
the kinetic energy involved in the STF motions. 

Most discussions of the STF dynamo assume implicitly
that the last step of merging the twisted loops
can be done at any time, and that the dynamo growth rate
is not limited by this last step. This may well be true
when the fields in the flux rope are not strong enough
to affect the motions, that is, in the kinematic regime. 
However as the field becomes stronger, and if the merging 
process is slow, the Lorentz forces due to the small scale 
kinks and twists will gain in importance compared with
the external forces associated with the driving of the loop as a whole.
This may then limit the efficiency of the dynamo.

Indeed, the growing complexity of the field,
in the repeated application of the STF process, without the
merging, can be characterized particularly by the evolution
of the magnetic helicity spectrum. This is
discussed for example in Ref.~\cite{gilbert}, 
where it is pointed out that the repeated application
of the STF cycle (with the same sense for the twist part of the cycle),
under flux freezing, leads to both a large scale 
{\it writhe helicity} associated with the repeated crossings of the flux tube,
and oppositely signed {\it twist helicity} at much smaller scales.
If one does not destroy this small scale structure by diffusion,
then the Lorentz forces associated with these 
structures will interfere with the STF motions.
A somewhat similar situation holds
for the mean field turbulent dynamos, where we will find
oppositely signed (almost equal strength)
helicities being generated by the motions on different scales. 
In that case we also find the dynamo to be eventually resistively limited, 
when there is strict helicity conservation.

In this context one more feature deserves mentioning:
if in the STF cycle one twists
clockwise and folds, or twists counter-clockwise and folds
one will still increase the field in the flux rope coherently.
However, one would introduce opposite sense of writhe in these
two cases, and so opposite internal twists.
So, although the twist part of the cycle is important for the mechanism 
discussed here, the sense of twist can be random and does not require 
net helicity. This is analogous to a case
when there is really only a small scale dynamo,
but one that requires finite kinetic helicity density locally.
We should point out, however, that
numerical simulations \cite{HCK96} have shown that dynamos work
and are potentially independent of magnetic Reynolds number even if the
flow has zero kinetic helicity density everywhere.

If the twisted loops can be made to merge efficiently, the
saturation of the STF dynamo would probably proceed differently.
For example, the field in the loop may become too strong to be stretched
and twisted, due to magnetic curvature forces.
Another interesting way of saturation is that the incompressibility
assumed for the motions may break down; as one stretches the flux
loop the field pressure resists the decrease in the loop
cross-section, and so the fluid density in the loop tends to decrease
as one attempts to make the loop longer.
(Note that it is $B/\rho$ which has to increase during stretching.)
The STF picture has inspired considerable work on various
mathematical features of fast dynamos and some of this work
can be found in the book by Childress and Gilbert \cite{CG95}
which in fact has STF in its title!

\subsection{Fast ABC-flow dynamos}

ABC flows are solenoidal and fully helical with a velocity field given by
\EQ
\UU=\pmatrix{
C\sin kz+B\cos ky\cr
A\sin kx+C\cos kz\cr
B\sin ky+A\cos kx}.
\label{ABCflow}
\EN
When $A$, $B$, and $C$ are all different from zero, the flow is
no longer integrable and has chaotic streamlines.
There is numerical evidence that such flows act as fast dynamos
\cite{GallowayFrisch86}.
The magnetic field has very small net magnetic helicity
\cite{BDS02,gilbert}.
This is a general property of any dynamo in the kinematic regime and
follows from magnetic helicity conservation, as will be discussed later
(\Sec{LinearBehavior} and \Fig{Fpspec_pm_satkin}).
Even in a nonlinear formulation of the ABC flow dynamo problem,
where the flow is driven by a forcing function similar to \Eq{ABCflow}
the net magnetic helicity remains unimportant \cite{GS91,GSG91}.
This is however not surprising, because the development of net
magnetic helicity requires sufficient scale separation, i.e.\
the wavenumber of the flow must be large compared with
the smallest wavenumber in the box ($k=k_1\equiv2\pi/L$,
where $L$ is the size of the box).
If this is not the case, helical MHD turbulence behaves similarly
to nonhelical turbulence \cite{Haugen04}.
A significant scale separation also weakens the symmetries
associated with the flow and the field, and leads to a larger
kinematic growth rate, more compatible with the turnover time scale
\cite{ArchontisDorchNordlund03a}. These authors also find that  
the cigar-like magnetic field structures which develop in canonical $A=B=C$ 
flows with stagnation points, are replaced by more ribbon like structures 
in flows without such stagnation points. 

Most of the recent work on nonlinear ABC flow dynamos has focused
on the case with small scale separation and, in particular,
on the initial growth and possible saturation mechanisms
\cite{ArchontisDorchNordlund03}. In the kinematic regime, these authors
find a near balance between Lorentz work and Joule dissipation.
The balance originates primarily from small volumes where the strong
magnetic flux structures are concentrated. The net growth of the 
magnetic energy comes about through stretching and folding
of relatively weak field which occupies most of the volume.
The mechanism for saturation could involve achieving a local
pressure balance in these strong field regions \cite{Nordlund03priv}.

\section{Small scale turbulent dynamos}
\label{SmallScale}

Dynamos are often divided into small scale and large scale dynamos.
Large scale dynamos are those responsible for the solar cycle, for example.
They show large scale spatial coherence and, in the case of the sun,
they also show long-term temporal order in the sense of the 11 year cycle,
i.e.\ much longer than the time scale of the turbulent motions.
Small scale dynamos produce magnetic fields that are
correlated on scales of the order of or smaller than
the energy carrying scale of the turbulence. 
In the literature such dynamos are
sometimes also referred to as `fluctuation dynamos'.
Nonhelical turbulent flows can act as small scale dynamos,
while flows with significant amounts of kinetic helicity act as
large scale dynamos.
Inhomogeneous and anisotropic flows (e.g.\ shear flows) are potential candidates
for producing large scale dynamo action.
Of course, there may not be a
clear boundary between small and large scale dynamos
and indeed the two may interact \Sec{UnifiedTreatment}.

Small scale dynamos are potentially important for several reasons.
First, they typically have larger growth rates than large scale dynamos.
The question now arises as
to what effect this rapidly generated magnetic `noise' has on the
large scale dynamo action. Further, there could be physical settings where
large scale dynamos do not work, like in clusters of galaxies
or in elliptical galaxies where
rotation effects are negligible and hence any turbulent flows lack
helicity and persistent shear.
In such systems, turbulence may still lead to small scale
dynamo action and generate magnetic fields. Whether such fields
are coherent enough to lead to the observed cluster rotation
measures, for example, is an important question to settle.

\subsection{General considerations}
\label{GeneralConsiderations}

In a turbulent flow fluid particles random walk away
from each other with time. A magnetic field line
frozen into the fluid (assuming large $R_{\rm m}$)
will then also lengthen by this random
stretching. This leads to an increase in $B/\rho$ and for
flows with $\rho\approx\const$, the magnetic field will be amplified.
The lengthening of the field line in a given direction
also leads to a decrease in its scale in the directions
perpendicular to the stretching. As the field
strength increases, the scale of individual field structures decreases
and the Ohmic dissipation increases, until
it roughly balances the growth due to random stretching.
What happens after this?
This question of the long term behavior of the magnetic
field in a turbulent flow was first raised by Batchelor \cite{Bat50}.
He argued on the basis of the analogy of the induction equation
to the equation for vorticity, that the field will grow
exponentially if the magnetic Prandtl number is larger than unity.
This argument is dubious and the possibility of dynamo
action in turbulent flows that lack helicity was first elucidated 
in a decisive manner by Kazantsev \cite{Kaz68} for a special kind of flow. 
Numerical simulations of turbulent flows also show invariably
that dynamo action can occur for forced turbulence.
We first discuss the Kazantsev dynamo model and then
present some results from simulations of
turbulent flows.

\subsection{Kazantsev theory}
\label{SSD}

Kazantsev considered a velocity field, $\vv$, which is
an isotropic, homogeneous, gaussian random field with
zero mean and also, more importantly, which is $\delta$-correlated in time.
[We use the symbol $\vv$ instead of $\UU$ to emphasize that
$\vv$ is not the solution of the momentum equation \eq{DUDt_incomp}.]
The two point spatial correlation function of the
velocity field can be written as
$\bra{v_i({\xx},t)v_j({\yy},s)} = T_{ij}(r)\delta (t-s)$, where 
\begin{equation}
T_{ij}(r) = 
\left(\delta_{ij} -{r_i r_j \over r^2}\right)T_{\rm N}(r) +
{r_i r_j \over r^2}T_{\rm L}(r).
\end{equation}
Here $\bra{\cdot}$ denotes averaging over an ensemble of the stochastic velocity
field $\vv$, $r = \vert \xx - \yy\vert$, $r_i = x_i -y_i$ and
we have written the correlation function 
in the form appropriate for a statistically
isotropic and homogeneous tensor (cf.\ Section~34 of Ref.~\cite{landau}).
Note that homogeneity implies that the two point correlation function
depends only on $\xx-\yy$. Together with isotropy this also implies
that the correlation tensor can only contain terms proportional to
$\delta_{ij}$, $r_ir_j$ and $\epsilon_{ijk}$ and the functions multiplying
these tensors depend only on $r$.  
$T_{\rm L}(r)$ and $T_{\rm N}(r)$ are the
longitudinal and transverse correlation functions for the velocity
field. (The helical part of the velocity correlations
is assumed to be zero in this section; see however \Sec{UnifiedTreatment}.)
If $\vv$ is assumed to be divergence free, then
$T_{\rm N}$ is related to $T_{\rm L}$ via
\begin{equation}
T_{\rm N} = {1 \over 2r} {\partial \over \partial r} \left[r^2 T_{\rm L}(r)\right],
\label{tntl}
\end{equation}
with
\EQ
T_{\rm L}(0) = \onethird\int_0^t\bra{\vv(t) \cdot\vv(t')}\;\dd t' .
\EN
We will see in the next section that $T_{\rm L}(0)$ is actually
the turbulent diffusion coefficient for the mean field.

\subsubsection{Kazantsev equation in configuration space}
\label{KazantsevConfiguration}

The stochastic induction equation can now be
converted into equations for the various moments of the
magnetic field. Assume that there is no mean field
or first moment, and that the magnetic correlation
has the same symmetries as the flow; i.e.\ it is isotropic
and homogeneous. Then its equal time, two point correlation is given by
$\left<B_i({\xx},t) B_j({\yy},t) \right> = M_{ij}(r,t)$, where
\begin{equation}
M_{ij} =
\left(\delta_{ij} -{r_i r_j \over r^2}\right) M_{\rm N}(r,t)
+{r_i r_j  \over r^2} M_{\rm L}(r,t),
\label{mcor2}
\end{equation}
and $M_{\rm L}(r,t)$ and $M_{\rm N}(r,t)$ are, respectively, the longitudinal
and transversal correlation functions of the magnetic field.
Since $\nab\cdot\BB=0$,
\EQ
M_{\rm N} = {1\over2r}{\partial\over\partial r}\left(r^2 M_{\rm L}\right).
\EN
Kazantsev derived an equation for $M_{\rm L}(r,t)$ by 
deriving the equation for the $k$-space magnetic spectrum
using diagram techniques, and transforming the
resulting integro-differential equation in $k$-space into a differential
equation in $r$-space. This $k$-space equation was
also derived by Kraichnan and Nagarajan \cite{kraich_nag67}.
Subsequently, Molchanov et al.\ \cite{Mol85,Mol88}
derived this equation directly in $r$-space by the Wiener
path integral method.
We present a simple derivation of the Kazantsev equation 
in the more general
case of helical turbulence in \App{kazantsev},
following the method outlined in Subramanian \cite{subamb}.
For a nonhelical random flow we have
\begin{equation}
{\partial M_{\rm L} \over \partial t} = {2\over r^4}{\partial \over \partial r}
\left[r^4 \eta_{\rm T}(r) {\partial M_{\rm L} \over \partial r}\right] + G M_{\rm L},
\label{mleq}
\end{equation}
where $\eta_{\rm T}(r)=\eta+\eta_{\rm t}(r)$ is the sum of
the microscopic diffusivity, $\eta$, and an effective
scale-dependent turbulent magnetic diffusivity
$\eta_{\rm t}(r) = T_{\rm L}(0) - T_{\rm L}(r)$.
The term $G = -2(T_{\rm L}^{\prime\prime} + 4 T_{\rm L}^{\prime}/r)$,
where primes denote $r$ derivatives,
describes the rapid generation of magnetic 
fluctuations by velocity shear
and the potential existence of a small scale dynamo (SSD) 
{\it independent} of any
large scale field; see the book by Zeldovich et al.\ \cite{ZRS83}
and references therein.

One can look for eigenmode solutions to \Eq{mleq}
of the form $\Psi(r)\exp(2\Gamma t)  =r^2\sqrt{\eta_{\rm T}} M_{\rm L}$.
This transforms Eq. (\ref{mleq}) for $M_{\rm L}(r,t)$,
into a time independent, Schr\"odinger-type equation,
but with a variable (and positive) mass,
\begin{equation}
-\Gamma \Psi  = -\eta_{\rm T}{d^2 \Psi \over d r^2} + U_0(r)\Psi .
\label{sievol}
\end{equation}
The `potential' is
\EQ
U_0(r) = T_{\rm L}^{\prime\prime} + {2\over r}\,T_{\rm L}^{\prime}
+ {1\over2}\,\eta_{\rm T}^{\prime\prime}
-{(\eta_{\rm T}^{\prime})^2\over4\eta_{\rm T}} + {2\over r^2}\,\eta_{\rm T}
\label{U0eqn}
\EN
for a divergence free velocity field.
The boundary condition is $\Psi \to 0$ for $r \to 0, \infty$.
Note that $U_0 \to 2\eta/r^2$ as $r \to 0$,
and since $T_{\rm L}(r) \to 0$ as $r \to \infty$, 
it follows that
$U_0 \to 2[\eta + T_{\rm L}(0)]/r^2$ as $ r\to \infty$.
The possibility of growing modes with $\Gamma > 0$
is obtained, if one can have a potential well with
$U_0$ sufficiently negative in some range of $r$.
This allows for the existence of {\it bound states} with `energy'
$E = -\Gamma < 0$.
The solutions to the Kazantsev equation for various forms
of $T_{\rm L}(r)$ have been studied quite extensively by several authors
\cite{Kaz68,ZRS83,subamb,Nov83,klee86,ssd,KSmn98,RS81,SBK02}.

Suppose we have random motions correlated on a single scale $L$,
with a velocity scale $V$. Define the magnetic Reynolds
number $R_{\rm m} = VL/\eta$. (Here we define $R_{\rm m}$ using
the correlation scale $L$ instead of $L/2\pi$ since this appears most
natural, and is also commonly used, in real space treatments of the SSD.)
Such a random flow may arise if
the fluid is highly viscous, and it will also be relevant
for the viscous cut-off scale eddies in
Kolmogorov turbulence. Then one finds 
that there is a critical $R_{\rm m} = R_{\rm crit}$,
so that for $R_{\rm m} > R_{\rm crit}$, the potential $U_0$ allows for the
existence of bound states.
For $R_{\rm m} = R_{\rm crit}$, one has $\Gamma = 0$,
and this marginal stationary state is the `zero' energy
eigenstate in the potential $U_0$.  
The value of $R_{\rm crit}$ one gets 
ranges between $30 - 60$, depending on the assumed form of
$T_{\rm L}(r)$ \cite{subamb,Nov83,klee86}.  
(This $R_{\rm crit}$ corresponds to a value $30/2\pi - 60/2\pi$ 
if we were to use the corresponding wavenumber $2\pi/L$ to
define $R_{\rm m}$.)
For $R_{\rm m} > R_{\rm crit}$,
$\Gamma > 0$ modes of the SSD can be excited, and
the fluctuating field that is correlated on a
scale $L$ grows exponentially on the
corresponding `eddy' turnover time scale.
For example, suppose one adopts $T_{\rm L}(r) = (VL/3) (1 - r^2/L^2)$ for 
$r < L$, and zero otherwise, as appropriate for
a single scale flow (or the flow below the viscous
cut-off). Then, a WKBJ analysis \cite{subamb} gives the growth rate
for the fastest growing mode as 
$\Gamma = {5\over4}V/L-{\cal O}((\ln R_{\rm m})^{-2})$.

To examine the spatial structure for various
eigenmodes of the small scale dynamo,
it is more instructive to consider the function 
\EQ
w(r,t) = \bra{\BB({\xx},t) \cdot \BB({\yy},t)}
= {1\over r^2} {\partial\over\partial r}\left(r^3 M_{\rm L}\right),
\quad\mbox{$r=|\xx-\yy|$},
\label{wdef}
\EN
which measures the ensemble average of the 
dot product of the fluctuating field at two locations, 
with $w(0,t) = \bra{\BB^2}$.
We have $\int_0^{\infty} w(r) r^2 dr = \int_0^{\infty} 
(r^3 M_{\rm L})'  = 0$, since $M_{\rm L}$ is regular at the origin and vanishes 
faster than $r^{-3}$ as $r \to \infty$. Therefore
the curve $r^2w(r)$ should have zero area
under it. Since $w(0,t) = \bra{\BB^2}$, $w$ is positive near the origin.
Therefore, $\BB$ points in the same direction for small
separation. As one goes to larger values of $r$, there must
be values of $r$, say $r \sim d$, where $w(r)$ becomes negative.
For such values of $r$, the field at the origin and at a separation
$d$ are, on the average, pointing in opposite directions.
This can be interpreted as indicating that the field lines,
on the average, are curved on the scale $d$.

For the growing modes of the small scale
dynamo, one finds \cite{subamb,Nov83,klee86,ssd,KSmn98} 
that $w(r)$ is strongly peaked
within a region $r =r_d \approx L R_{\rm m}^{-1/2}$ about the origin,
for all the modes, and for the fastest growing mode,
changes sign across $r \sim L$ and rapidly decays
with increasing $r/L$. The scale $r_d$ is in fact the
diffusive scale determined by the balance of rate of the Ohmic
decay, $\eta /r_d^2$, and the growth rate $V/L$ 
due to random shearing. (Note for the single scale flow 
one has linear shear at small scales, and hence a scale-independent
shearing rate.)
Detailed asymptotic solutions
for $w(r)$ have been given in Ref.~\cite{klee86} and a WKBJ
treatment can also be found in Ref.~\cite{subamb}.
A physical interpretation of this correlation function
\cite{ZRS83,klee86,ssd} is that
the small scale field in the kinematic regime
is concentrated in structures with thickness 
$r_d$ and curved on a scale up to $\sim L$.
How far such a picture holds in the nonlinear regime is
still matter of investigation (see below).

The small scale dynamo in Kolmogorov turbulence
can be modeled \cite{subamb,KSmn98} by adopting  
$T_{\rm L}(r) = \onethird VL[1 - (r/L)^{4/3}]$
in the inertial range $l_d < r < L$,
a form suggested in Ref.~\cite{vain82}.
Here $L$ is the outer scale and $l_d \approx L \mbox{Re}^{-3/4}$ 
is the viscous cut-off scale of the turbulence, where $\mbox{Re} = VL/ \nu$ 
is the fluid Reynolds number. For Kolmogorov turbulence, the eddy
velocity at any scale $l$, is $v_l \propto l^{1/3}$,
in the inertial range. So the scale dependent diffusion
coefficient scales as $v_l l \propto l^{4/3}$.
This scaling, also referred to as Richardson's law,
is the motivation for the above form $T_{\rm L}(r)$.
Also, in order to ensure that $T_{L}'(0) =0$,
$T_{L}$ is continued from
its value at $r=l_d$ to zero,
and was taken to be zero for $r > L$.
(The exact form of the continuation has little effect on the conclusions.)
In the inertial range the potential then has the scale invariant form,
\begin{equation}
U = {v_l \over 3 l}\left[-{8 \over 9}\left({r \over l}\right)^{-2/3}
-{(4/9)(r/l)^{2/3}
\over 3/R_{\rm m}(l) + (r/l)^{4/3}}
+{6 \over R_{\rm m}(l)}\left({L^2 \over r^2}\right) \right],
\label{potinl}
\end{equation}
where $R_{\rm m}(l) = v_l l/\eta = R_{\rm m} (l/L)^{3/4}$
is the magnetic Reynolds number associated with a scale $l$.
Note that the potential $U$ (not to be confused with the velocity $\UU$)
has the same form at any scale $l$, with $R_{\rm m}(l)$
appropriate to that scale.
This suggests that conclusions about the excitation conditions
can be applied separately at different scales, $l$, provided we use
the corresponding velocity scale $v_l$ and Reynolds number $R_{\rm m}(l)$
appropriate to the scale $l$. For example, the condition
for excitation of small scale dynamo modes which are concentrated
at a scale $l$, is also $R_{\rm m}(l) = R_{\rm crit} \gg 1$. 

Note that $R_{\rm m}(l)$ decreases as one goes to smaller scales, 
and so the small scale dynamo will be first excited
when the magnetic Reynolds number at the outer scale satisfies
$R_{\rm m}(L) > R_{\rm crit}$.
For Kolmogorov turbulence described by the above
$T_{\rm L}(r)$ it was estimated \cite{subamb} using a WKBJ analysis
that $R_{\rm crit} \sim 60$. Also, the marginal mode 
which has zero growth rate, in this case, has 
$w(r)$ peaked within $r \sim L/R_{\rm m}^{3/4}$. This 
different scaling can be understood as arising due
to the fact that the shearing rate now is $(V/L)(r/L)^{-2/3}$
at any scale $r$ in the inertial range.
For the marginal mode this is balanced
by dissipation at $r=r_d$ which occurs at a rate $\eta/r_d^2$. 
 
In Kolmogorov turbulence, the cut-off scale eddies 
have $R_{\rm m}(l_d) = P_{\rm m}$,
i.e.\ the cutoff-scale magnetic Reynolds number is equal to
the magnetic Prandtl number. So, if $P_{\rm m} > R_{\rm crit}$
these eddies are themselves capable of small scale dynamo action.
The fastest growing modes then have a growth rate
$\sim v_d/l_d$, which is equal to the eddy turnover time associated
with the cut-off scale eddies, a time scale
much smaller than the turnover time of outer scale eddies.
Also, for the fastest growing mode $w(r)$ is peaked about 
a radius corresponding to the diffusive scales associated 
with these eddies, and changes sign at $r \sim l_d$. 

We have taken the scale-dependent turbulent diffusion coefficient
$T_{\rm L}(0) - T_{\rm L}(r) \propto r^n$ with $n=4/3$ to model
Kolmogorov turbulence. The index $n$ measures how `rough'
the velocity field is, with $n=2$ corresponding to a smooth
velocity field. It turns out that for growing modes one requires
$n > 1$ at least \cite{Kaz68}. Also, in a small $P_{\rm m}$ flow
the closer $n$ is to unity, 
the larger could be the critical $R_{\rm m}$ needed to excite
the small scale dynamo \cite{bold_catt03}. This may be of relevance
for understanding the results of small $P_{\rm m}$ dynamo simulations
that are described below (\Sec{LowPrM}).

\subsubsection{Kazantsev dynamo in Fourier space}
\label{KazantsevFourier}

It is instructive to study the Kazantsev problem 
in Fourier space. In $k$ space the differential equation
\eq{mleq} becomes an integro-differential equation 
for the magnetic spectrum $M(k,t)$,
which is in general difficult to solve. However, if
the magnetic spectrum is peaked on
scales much smaller than the flow, as is the case for small
fluid Reynolds number (large $P_{\rm m}$) flows, one can provide
an approximate treatment for the large $k$ regime,
$k \gg k_{\rm f}$. Here $k_{\rm f}$ is the forcing scale in case of a
single scale random flow, or the viscous scale
in case of Kolmogorov turbulence (assuming that eddies
at the cut-off scale can also induce dynamo action).
In this large $k \gg k_{\rm f}$ limit the Kazantsev equation
becomes \cite{KA92}
\EQ
{\partial M \over \partial t}
={\gamma \over 5} \left(k^2 {\partial^2 M \over \partial k^2}
- 2k{\partial M \over \partial k} + 6M \right) - 2\eta k^2 M,
\label{mkeq}
\EN
where $\gamma = -(1/6)[\nabla^2 T_{ii}(r)]_{r = 0}$,
is a measure of the rate of shearing by the flow.
In terms of $T_{\rm L}$, we have $\gamma = 7T_{\rm L}''(0) + 8T'_{\rm L}(0)/r$.
The evolution of the magnetic spectrum was analyzed
in some detail in Ref.~\cite{KA92}, and is summarized nicely
in Ref.~\cite{Schek02}. 

Suppose the initial magnetic spectrum 
is peaked at some $k=k' \ll k_{\rm f} R_{\rm m}^{1/2}$, i.e.\ at 
a wavenumber much smaller than the resistive
wavenumber, then the amplitude of each Fourier mode grows exponentially
in time at the rate ${3\over4}\gamma$. Meanwhile, the peak of
excitation moves to larger $k$, with 
$k_{\rm peak} = k'\exp({3\over5}\gamma t)$, leaving behind a power
spectrum $M(k) \propto k^{3/2}$. These features can of course
be qualitatively understood as due to the effects of 
random stretching. Once the peak reaches the resistive scale,
one has to solve again an eigenvalue problem to determine
the subsequent evolution of $M(k,t)$. Substituting
$M(k,t) = \exp^{\lambda\gamma t} \Phi(k/k_\eta)$,
where $k_\eta = (\gamma/10 \eta)^{1/2}$,
into \eq{mkeq} and demanding that $\Phi \to 0$ as $k \to \infty$,
one gets the solution \cite{Schek02}
\EQ
\Phi(k/k_\eta) = \const\times k^{3/2} K_{\nu(\lambda)}(k/k_\eta),
\quad \nu(\lambda) = \sqrt{5(\lambda -\threequarter)}.
\label{Macdonald}
\EN
Here $K_\nu$ is the Macdonald function and the eigenvalue $\lambda$
must be determined from the boundary condition at small $k$.
This is a bit more tricky in the $k$-space analysis
since the equations were simplified by taking the large $k$ limit;
fortunately the results seem independent of the exact
form of the boundary condition at small $k$ in the large 
$R_{\rm m}$ limit. In Ref.~\cite{Schek02} a zero flux boundary condition
is imposed at some $k=k_* \ll k_\eta$ and it is shown
that this fixes $\lambda \approx 3/4$ and $\nu \approx 0$.
In case we adopt $T_{\rm L}(r) = \onethird VL (1 - r^2/L^2)$, for $r < L$
and zero otherwise, one gets $\gamma = {5\over3}V/L$ and
so the growth rate is $\Gamma = {3\over4}\gamma = {5\over4}V/L$,
which agrees with the WKBJ analysis of the Kazantsev
equation obtained in \Sec{KazantsevConfiguration}.
We also get $1/k_\eta = \sqrt{6} L R_{\rm m}^{-1/2}$ which 
is of the same order as expected for the diffusive scale $r_d$ 
in the real-space treatment.

\begin{figure}[t!]\begin{center}
\includegraphics[width=.85\textwidth]{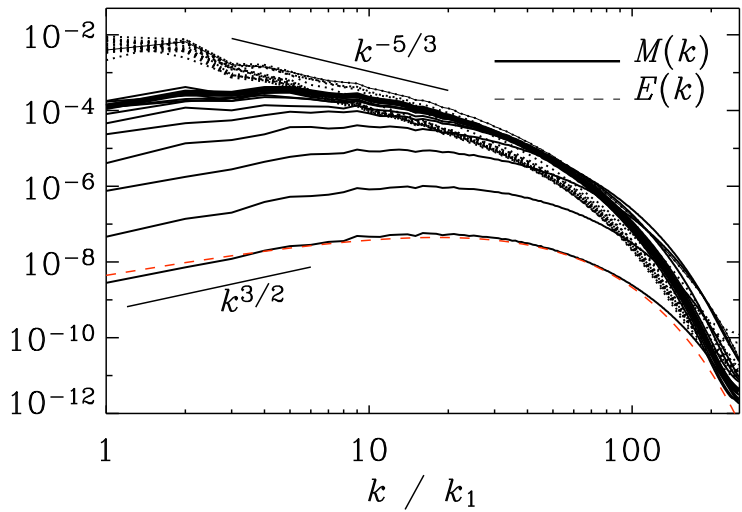}
\end{center}\caption[]{
Magnetic and kinetic energy spectra from a nonhelical turbulence
simulation with $P_{\rm m}=1$.
The kinetic energy is indicated as a dashed line (except for the first
time displayed where it is shown as a thin solid line).
At early times the magnetic energy spectrum follows the $k^{3/2}$
Kazantsev law [the dashed line gives the fit to \Eq{Macdonald}],
while the kinetic energy shows a short $k^{-5/3}$ range.
The Reynolds number is $u_{\rm rms}/(\nu k_{\rm f})\approx600$
and $512^3$ meshpoints were used \cite{Haugen04}.
The time difference between the spectra is about
$14\,(k_{\rm f}u_{\rm rms})^{-1}$.
}\label{pspec_nohel512d2}\end{figure}

\begin{figure}[t!]\begin{center}
\includegraphics[width=.85\textwidth]{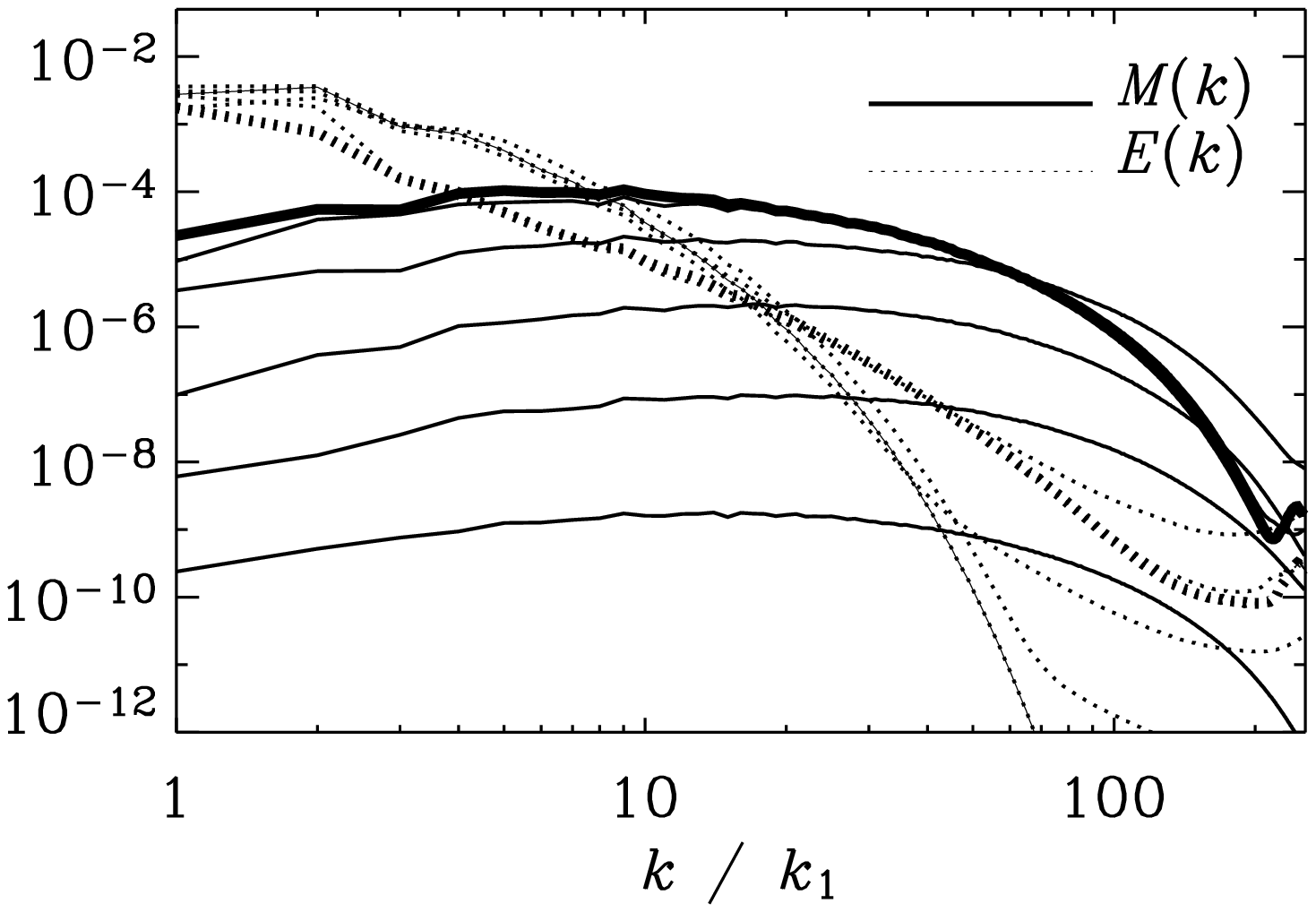}
\end{center}\caption[]{
Magnetic and kinetic energy spectra from a nonhelical turbulence
simulation with $P_{\rm m}=50$.
The kinetic energy is indicated as a dotted line (except for the first
time displayed where it is shown as a thin solid line).
The magnetic spectrum for the last time is shown as a thick line.
The Reynolds number is $u_{\rm rms}/(\nu k_{\rm f})\approx80$, and
the magnetic Reynolds number is $u_{\rm rms}/(\eta k_{\rm f})\approx4000$
and $512^3$ meshpoints were used.
Like in \Fig{pspec_nohel512d2}, the time difference between the spectra
is about $14\,(k_{\rm f}u_{\rm rms})^{-1}$.
At the end of the run the field is still not completely saturated.
}\label{pspec_largePm}\end{figure}

We show in \Figs{pspec_nohel512d2}{pspec_largePm} the time evolution
of the magnetic spectrum from simulations which are
described in detail later. One important difference between simulations
and the picture described above arises due to the presence of
power (however small) in the initial spectrum,
at the resistive scale. This leads to the magnetic
spectrum extending to the resistive scale, 
and locking onto an eigenfunction right from
the early stages of evolution, as can be seen
in \Fig{pspec_nohel512d2}. We also show the
corresponding time evolution for a high $P_{\rm m}$
simulation, which represents more closely the SSD for
when the kinetic spectrum is peaked on a single scale;
see \Fig{pspec_largePm}

\subsubsection{Further results on the Kazantsev dynamo}

A number of other interesting results have been found
for the Kazantsev problem, particularly for the case
of a single scale flow. In this case one can approximate the
velocity to be a linear random shear flow. In the regime when
the field has not yet developed small enough scales,
resistivity is unimportant and one can assume the field to be frozen 
into the flow.
The result of such passive random advection is that the
magnetic field strength develops a log-normal probability distribution function
(PDF) \cite{BoldS00,Schek_PRE02}. This log-normal form for the
PDF of $B$ can be understood from the induction
equation without the diffusion term $\dot{B}_i = B_ju_{i,j}$
where the time derivative is a lagrangian derivative. Since the
RHS is linear in $B_i$ and is multiplied by a constant random
matrix $u_{i,j}$, the ``gaussianity'' of log$B$ is expected
from the central limit theorem. This implies that the magnetic
field becomes highly intermittent.
To what extent this PDF is altered by resistivity is examined in
Ref.~\cite{CFKV99}. They find that even when resistivity is included,
the field is still intermittent and may be thought of
as being concentrated into narrow strips. 

This seems to be also borne out by a study of
the behavior of higher order $k$-space correlators \cite{naz03}.
By examining their late time evolution,
these authors find that the log-normality
of the magnetic field PDF persists in the dissipative regime.
An interpretation of their $k$-space scalings, suggests 
that the magnetic structures in physical space could look like
`ribbons' with the field directed along these ribbons.
Recent numerical work \cite{Schek03b}, however, finds that even
in the kinematic regime, the PDF of $|\BB|$ changes
character, perhaps due to having a finite box in the simulation.

Motivated by the need to understand the eventual nonlinear 
saturation of the small scale dynamo, work has been done 
by looking at the statistical properties of the Lorentz
force, in particular the component $\BB\cdot\nab\BB$
\cite{Schek02,Schek_PRE02,maron_cowley01}.
The idea is that in an incompressible flow, the effects of the
Lorentz force will be dominated by the magnetic curvature rather
than the magnetic pressure gradient.
(Note that although the magnetic pressure forces may be
much larger, they are largely balanced by thermal pressure gradients
in the incompressible limit.)
In Ref.~\cite{Schek_PRE02} it was shown that
though both magnetic energy and mean-square curvature of field lines
grow exponentially, the field strength and the curvature are anti-correlated.
Thus, regions of strong field are nearly straight,
while sharply curved fields are relatively weak.
Such a result was in fact found earlier in simulations of
dynamo generation of magnetic fields due to convection \cite{BJNRST96,BPS95}.
In these simulations the magnetic field was found to be
intermittent with relatively stronger field concentrated in 
structures which are relatively straight.
Furthermore, the saturation of the dynamo was traced back to the
component of $\BB\cdot\nab\BB$ that is perpendicular to the direction
of the field; its field-aligned component (i.e.\ the tension force)
was found to be unimportant for saturation \cite{NBJRRST92}.

The anti-correlation between the field strength and the curvature
is interpreted by \cite{Schek02,Schek_PRE02}
as showing that the field lies in folds,
which are curved on the flow scale and 
with rapid reversals on the diffusive scale.
To what extent the small scale dynamo generated field
is indeed structured in this way is still a matter 
of debate (see below). Also, one should keep in mind that 
all the above semi-analytic results on the structure
of the field pertain to the kinematic regime.
The recent simulations of Haugen et al.\ \cite{Haugen04}, discussed 
in \Sec{Simulations_SSdynamo}, show occasional field reversals,
but such events occupy only a small part of the volume.

\subsection{Saturation of the small scale dynamo}
\label{SaturationSmallScale}

How does the small scale dynamo saturate?
The answer to this question will be crucial to
understanding both the effect of the small scale
dynamo generated fields on the large scale dynamo,
as well of the relevance of these fields to explain
say cluster fields. From the discussion of the
kinematic regime, we saw that 
the field would become intermittent
and also concentrated into structures whose
thickness, in at least one dimension, is the resistive scale.
However nonlinear effects may intervene to alter this
kinematic result. There have been a number
of attempts to model the nonlinear saturation of the
small scale dynamo, none of which are entirely compelling.
We discuss these below.
 
Some effects of the growing magnetic field can arise just due to 
plasma phenomena like ambipolar drift in a partially ionized
gas \cite{KSmn98}, anisotropic viscosity \cite{Maly_Kul02}
or collisionless damping \cite{KCGS97}.
For example, collisionless processes can become important on
scales smaller than the ion mean free path
and could prevent the magnetic field from being concentrated
below such a scale \cite{KCGS97}. 
Ambipolar diffusion is particularly relevant in the galactic context,
where there could be a significant neutral component to the
interstellar medium. It changes the effective diffusivity by adding
an ambipolar diffusion component \cite{subamb,KSmn98}; see \Sec{ambipolar}
and \App{kazantsev}. The effective 
magnetic Reynolds number of the interstellar
medium, taking ambipolar diffusion into account,
decreases from about $3 \times 10^{19}$ to $R_{\rm AD} \sim 10^6$ 
\cite{KSmn98}, as the field grows to microgauss strengths.
This will then lead to an important increase of the effective diffusive
scale which, due to nonlinear ambipolar drift,
will become of order $L/R_{\rm AD}^{1/2} \sim 10^{-3} L$.
However, since $R_{\rm AD} \gg R_{\rm crit}$, 
the small scale dynamo does not saturate due to this mechanism.

\subsubsection{Saturation via artificial nonlinear drifts}
\label{nonlinear_drift}

The above discussion of ambipolar drift motivates a
nonlinear model problem which may give hints as to how the small scale
dynamo can in principle saturate even in a fully ionized 
gas \cite{sub99}. Suppose one assumes that, as the magnetic field grows
and the Lorentz force pushes on the fluid, the fluid
instantaneously responds by developing an additive `drift' component 
to the velocity, where the drift is proportional to the Lorentz force. 
The model velocity in the induction equation
is then $\uu = \vv + \vv_{\rm N}$, i.e.\ the
sum of an externally prescribed stochastic
field $\vv$, and a drift component 
$\vv_{\rm N} = a\JJ\times\BB$.
For ambipolar drift, $a$ would be related
to the properties of the partially ionized gas. 
Suppose we adopt instead $a = \tau/\rho$, where
$\tau$ is some response (or correlation) time, treated as a phenomenological
free parameter and $\rho$ is the fluid density.
This gives a model problem, where the nonlinear
effects of the Lorentz force are taken into account as
simple modification of the velocity field.
(Note that in reality it is the acceleration and not the velocity
which is proportional to the Lorentz force. Nevertheless
the above model provides a useful toy problem for examining
the nonlinear saturation effects, especially if the correlation
time $\tau$ is small compared to the Alfv\'en time.)
Adding a velocity contribution proportional
to the Lorentz force of course also leads to a problem of closure, since
in the equation for $M_{\rm L}$, the nonlinear drift term brings
in a fourth order magnetic correlation. In Ref.~\cite{subamb,sub99}
a gaussian closure is assumed (cf.\ \App{kazantsev}).

The backreaction in the form of a nonlinear drift then 
simply replaces $\eta$ by an
effective time-dependent $\eta_{\rm D} = \eta + 2aM_{\rm L}(0,t)$ in
the $\eta_{\rm T}(r)$ term of Eq.~(\ref{mleq}) 
\cite{sub99} (cf.\ \App{kazantsev}).
We thus obtain
\EQ
\frac{\partial M_{\rm L}}{\partial t} = 
\frac{2}{r^4}\frac{\partial}{\partial r}
\left[r^4 \eta_{\rm T}(r)\frac{\partial M_{\rm L}}{\partial r}\right] + 
G M_{\rm L} + K ,
\label{mleqna}
\EN
\EQ
K(r,t) = 4aM_{\rm L}(0,t) \frac{1}{r^4}\frac{\partial}{\partial r}
\left(r^4 \frac{\partial M_{\rm L}}{\partial r}\right).
\EN
Define an effective magnetic Reynolds number
(just like $R_{\rm AD}$), for
fluid motions on scale $L$, by $R_{\rm D}(t) = V L/ \eta_{\rm D}(t)$.
Then, as the energy density in the fluctuating field,
say $E_{\rm M}(t) = \threehalf M_{\rm L}(0,t)$,
increases, $ R_{\rm D}$ decreases.
In the final saturated state,
with $\partial M_{\rm L}/\partial t =0$
(obtained, say, at time $t_s$),
$M_{\rm L}$, and hence the effective $\eta_{\rm D}$ in \Eq{mleqna}
become independent of time.
Solving for this stationary state then becomes
identical to solving for the marginal (stationary) mode of the
kinematic problem, except that $R_{\rm m}$ is replaced by $R_{\rm D}(t_s)$.
The final saturated state is then the marginal eigenmode which one obtains
when $E_{\rm M}$ has grown (and $R_{\rm D}$ decreased) such that
$R_{\rm D}(t_s) = V L/[\eta + 2aM_{\rm L}(0,t_s)] = R_{\rm crit}\sim30$--$60$
(depending on the nature of the velocity field).
Also, for this saturated state we predict that $w(r)$ is 
peaked within a region $r \approx L (R_{\rm crit})^{-1/2}$
about the origin, changes sign across $r \sim L$, and then rapidly 
decays for larger $r/L$.
Further, from the above constraint it follows that
$M_{\rm L}(0,t_s) = vL /(2a R_{\rm crit})$,
where we have assumed $\eta\ll 2aM_{\rm L}(0,t_s)$. 
So, the magnetic energy at saturation is
$E_{\rm M}(t_s) = {3\over2}M_{\rm L}(0,t_s) = {3\over2}
(\rho v^2/2)\,(L/v\tau) R_c^{-1}$.
Of course, since $\tau$ is an unknown model parameter,
one cannot unambiguously predict $E_{\rm M}$.
If we were to adopt $\tau \sim L/v$, that is the eddy turnover time,
then $E_{\rm M}$ at saturation is a small fraction,
$\sim R_{\rm crit}^{-1}$ ($\ll1$), of the equipartition energy density.

Note that the mechanism for saturation is quite subtle. It is not
that the fluid velocity has been decreased by the Lorentz force,
as can be explicitly seen by looking at the kinetic energy spectra
obtained in direct simulations incorporating such a nonlinear drift 
(see \Sec{ClosureModel}). Rather, the nonlinear drift due to the
Lorentz force introduces an extra `diffusion' between the field and fluid,
effectively a growing `ambipolar' diffusion (or 
growing impedance), which leads to the dynamo saturation. 
This model for saturation may be quite simplistic, but gives
a hint of one possible property of the saturated state:
it suggests that the final saturated state of the small scale dynamo
could be (i) universal in that it does not depend on the
microscopic parameters far away from the resistive/viscous scales,
and (ii) could have properties similar to the marginal eigenmode of 
the corresponding kinematic small scale dynamo problem. 
It is of interest to check whether such a situation is indeed obtained
in simulations.

The nonlinear drift velocity assumed above is not incompressible.
Indeed ambipolar drift velocity in a partially ionized medium, 
need not have a vanishing divergence. It is the extra
diffusion that causes the dynamo to saturate. One may then
wonder what happens if we retained incompressibility of
the motions induced by the Lorentz force? Is there still
increased nonlinear diffusion?
This has been examined in Ref.~\cite{sub03} by adopting 
$\uu = \vv + \vv_{\rm N}$, with 
an incompressible $\vv_{\rm N} = a (\BB\cdot\nab\BB - \nab p)$.
Here, $p$ includes the magnetic pressure, but can be
projected out in the usual way using 
$\nab\cdot\vv_{\rm N} = 0$.
Such a model of the effects of nonlinearity is 
very similar to the quasilinear treatments
mean field dynamo saturation (\App{qnmod}).
For this form of nonlinearity, 
one gets an integro-differential equation for
the evolution of $M_{\rm L}$, which in general is not analytically
tractable. One can however make analytic headway
in two limits $r = \vert {\xx-\yy} \vert \gg l$, and $r \ll l$,
where $l(t)$ is the length scale over which $M_{\rm L}(r,t)$ is
peaked. For example, during the kinematic evolution,
$M_{\rm L}(r,t)$ is strongly peaked within a radius
$l = r_d \sim L/R_{\rm m}^{1/2}$, where $L$ is the correlation
length associated with the motions.
One gets \cite{sub03} 
\begin{eqnarray}
K(r,t)=\left\{
\begin{array}{ll}
{\displaystyle{
2aM_{\rm L}(0,t) \frac{1}{r^4}\frac{\partial}{\partial r}
\left(r^4 \frac{\partial M_{\rm L}}{\partial r}\right)
+ 8 \int_0^\infty \frac{du}{ u} (M_{\rm L}^\prime)^2}}, \quad r \ll l, \\
{\displaystyle{
-\frac{2\eta_{\rm HD}}{ r^4}\frac{\partial }{\partial r}
\left\{r^4 \frac{\partial}{\partial r} \left[
\frac{1}{r^4}\frac{\partial }{\partial r}
\left(r^4 \frac{\partial M_{\rm L}}{\partial r}\right)\right]\right\},\quad
r \gg l.}}
\end{array}
\right.
\label{zero}
\end{eqnarray}
Thus, the nonlinear backreaction term $K$ in this model problem
is like a nonlinear diffusion for small $r \ll l$
(yet partially compensated by a constant), transiting to nonlinear
hyperdiffusion for $r \gg l$ \cite{sub03}. (The hyperdiffusion
coefficient $\eta_{\rm HD}$ itself depends on $M_{\rm L}$ and
its explicit form is given in \Eq{hd}.)
In both regimes
the damping coefficients are proportional to $E_{\rm M}(t)$.
So, as the small scale field grows and $E_{\rm M}$ increases, 
the damping increases, leading to a saturated state.
Note that both diffusion and hyperdiffusion would lead
to an increase in the effective resistivity, just
as in the case of ambipolar drift.
Evidently, this property is
obtained even if one demands $\vv_{\rm N}$ to be incompressible.

A somewhat different model of the nonlinear backreaction
can be motivated, if the fluid is highly viscous with 
$\mbox{Re}\ll1 \ll R_{\rm m}$ \cite{kim99,kim00}. 
In this case $\vv_{\rm N}$ is assumed to satisfy
the equation $\nu\nab^2\vv_{\rm N} + \JJ \times \BB - \nab p = 0$,
and $\nab\cdot\vv_{\rm N} = 0$. Again the equation for $M_{\rm L}$ becomes
an integro-differential equation. The extra nonlinear
term is again simplified near $r=0$. One gets
$K(0,t) = -M_{\rm L}^2/3 - \int_0^\infty r (M_{\rm L}'(r))^2$ \cite{kim99}. 
This can be interpreted as a nonlinear reduction of the $G$ term 
governing the stretching property of the flow. 
Saturation of the small scale dynamo will then
result in a model where Lorentz forces reduce the 
random stretching, rather than increased diffusion
as in the models of Refs~\cite{sub99,sub03}.

\subsubsection{Modifying the Kazantsev spectral equation}
\label{KazantsevResistive}

Another approach to the nonlinear small scale dynamo
with a large Prandtl number $P_{\rm m}$ has been explored
in Refs~ \cite{Schek02,Schek03c}. The idea is to modify the coefficients 
of the Kazantsev equation in a phenomenologically motivated manner and then
examine its consequences.  The motivation arises from the expectation that,
as the magnetic field grows, it suppresses the dynamo action of eddies
that have energies smaller than the field. Only eddies that have
energies larger than the field are able to still amplify the
field. Also, in the kinematic regime of a small scale dynamo
with large $P_{\rm m}$, the magnetic spectrum is peaked
at large $k= k_\eta$ (the resistive wavenumber). 
With these features in mind, Schekochihin et al.\ \cite{Schek02} 
modify the $\gamma$ in the Kazantsev $k$-space equation \Eq{mkeq}
by taking it to be proportional
to the turnover rate of the smallest `unsuppressed eddy'
and study the resulting evolution, assuming the fluid
has a large $P_{\rm m}$. Specifically, they adopt
\EQ
\gamma(t) = c_1\left[\int_0^{k_s(t)} k^2 E(k) \dd k \right]^{1/2},
\quad c_2\int_{k_s(t)}^\infty  E(k) \dd k = E_{\rm M}(t), 
\EN
where $E(k)$ is the kinetic energy spectrum, $c_1$, $c_2$ are constants,
and $k_s(t)$ is the wavenumber at which the magnetic
energy equals the kinetic energy of all the suppressed eddies.

In this model, after the magnetic energy has grown exponentially
to the energy associated with the viscous scale eddies,
its growth slows down, and becomes linear in time
with $E_{\rm M}(t) = \epsilon t$, where $\epsilon = v^3/L$ is the
rate of energy transfer in Kolmogorov turbulence.
This phase proceeds until the energy reaches that of the outer scale eddies.
However, the peak in the magnetic spectrum evolves
to values of $k$ smaller than the initial $k=k_\eta$
only on the resistive time scale, which is very long
for realistic astrophysical systems. This implies
that the saturated small scale fields could have energies comparable
to the energy of the motions, but the field will be still largely incoherent. 
More recently, another model has been explored
in which saturation is achieved as a result of the
velocity statistics becoming anisotropic with respect
to the local direction of the growing field \cite{Schek03c}.

Clearly more work needs to be done to understand the saturation 
of small scale dynamos. A general feature however
in all the above models is that the small scale dynamo saturates
because of a `renormalization' of the coefficients governing
its evolution: increased nonlinear diffusion \cite{sub99},
increased diffusion plus additional hyperdiffusion \cite{sub03},
or reduced stretching \cite{Schek02,kim99,kim00,Schek03c}. 
The nature of the saturated fields is still not very clear from these
simplified models and one needs guidance also from simulations, to
which we now turn.

\subsection{Simulations}
\label{Simulations_SSdynamo}

Since the pioneering work of Meneguzzi, Frisch, and Pouquet \cite{MFP81},
small scale dynamo action has frequently been seen in direct simulations
of turbulence.
In the early years of dynamo theory,
the Kazantsev result was not yet well known and it came
somewhat as a surprise to many that kinetic helicity was not
necessary for dynamo action.
Until then, much of the work on dynamos had focused on the
$\alpha$ effect.
Indeed, the helically and non-helically forced
simulations of Meneguzzi et al.\ \cite{MFP81} in cartesian
geometry were rather seen as being
complementary to the global simulations of Gilman and Miller \cite{GM81}
in the same year.
In the global dynamo simulations the flow was driven by thermal
convection in a spherical shell in the presence of rotation, so there was
helicity and therefore also an $\alpha$ effect.
The fact that dynamos do not require kinetic helicity was perhaps
regarded as a curiosity of merely academic interest,
because turbulence in stars and galaxies
is expected to be helical.

Only more recently the topic of nonhelical MHD turbulence
has been followed up more systematically.
One reason is that in many turbulence simulations kinematic helicity
is often found to be rather weak, even if there is stratification
and rotation that should produce helicity
\cite{BJNRST96,NBJRRST92,MP89}.
Another reason is that with the advent of large enough simulations
nonhelical dynamo action has become a topic of practical reality
for any electrically conducting flow.
Local and unstratified simulations of accretion disc turbulence also
have shown strong dynamo action \cite{HGB96}.
The fact that even completely unstratified nonrotating convection
can display dynamo action has been used to speculate
that much of the observed small scale magnetic field seen at the
solar surface might be the result of a local dynamo acting only
in the surface layers of the sun \cite{Cat99}.
The other possible source of small scale magnetic fields in the
sun could be the shredding of large scale fields by the turbulence
\cite{Schuessler83,Schuessler84}.

When the magnetic field is weak enough, but the magnetic Reynolds
number larger than a certain critical value, the magnetic energy grows
exponentially.
The growth rate of the magnetic field scales with the inverse turnover
time of the eddies at the dissipative cutoff wavenumber $k_{\rm d}$, i.e.\
$\lambda\sim u_{k_{\rm d}}k_{\rm d}\sim k_{\rm d}^{2/3}\sim R_{\rm m}^{1/2}$,
where we have used $k_{\rm d}\sim R_{\rm m}^{3/4}$
(and assumed $R_{\rm m}=R_{\rm e}$).
Quantitatively, it has been found \cite{Haugen04} that, in the
range $200<R_{\rm m}<1000$,
\EQ
\lambda/(u_{\rm rms}k_{\rm f})\approx3\times10^{-3}R_{\rm m}^{1/2}
\approx0.018\times\left(R_{\rm m}/R_{\rm crit}\right)^{1/2}.
\EN
The critical magnetic Reynolds number for dynamo action is around 35
when $P_{\rm m}=1$; see Ref.~\cite{Haugen04}.
This is around 30 times larger than the critical value for helical
dynamos which is only around 1.2 \cite{B01}.
It is also larger by about a factor $3$ compared to 
$R_{\rm crit}$ obtained in the Kazantsev model, which
assumes a delta-correlated velocity field, showing that
realistic flows are less efficient compared to the 
Kazantsev model (see also below).\footnote{We recall that in this
section and throughout most of this review we have defined
$R_{\rm m}$ with respect to the wavenumber of the energy-carrying eddies.
If, instead, $R_{\rm m}$ is defined with respect to the {\it scale}
$\ell_{\rm f}=2\pi/k_{\rm f}$, as was done in Table~1 of Ref.~\cite{B01},
one has to divide his values (between 7 and 9) by $2\pi$, giving
therefore values between 1.1 and 1.4.
As another example we consider the Ponomarenko or screw dynamo with
helical motion in an infinite cylinder of radius $R$.
Here, the magnetic Reynolds number is usually defined with respect
to $R$ and the critical value is around 17.7 \cite{GaiFrei76}.}
At early times the magnetic energy spectrum follows the $k^{3/2}$
Kazantsev law, but then it reaches saturation. The kinetic energy spectrum
is decreased somewhat, such that the magnetic energy exceeds the
kinetic energy at wavenumbers $k/k_1>5$; see \Fig{pspec_nohel512d2}.

Kazantsev's theory is of course not strictly applicable,
because it assumes a delta correlated flow. Also, the $k^{3/2}$ spectrum
in the kinematic regime is obtained only for scales smaller than that
of the flow, whereas in the simulations the
velocity is not concentrated solely at the largest scale.
This may be the reason why, according to Kazantsev's theory, the growth
rate is always overestimated: in the simulations 
(Run~D2 of Ref.~\cite{Haugen04} with $R_{\rm m}=600$) the actual growth
rate of the magnetic field is only $\sim0.07\,u_{\rm rms} k_{\rm f}$.
Nevertheless, the estimate for the Kazantsev cutoff wavenumber $k_\eta$,
which is approximately where the kinematic spectrum peaks,
is still fairly accurate (see \Sec{KazantsevFourier}).

Meanwhile, simulations of nonhelical dynamos have been carried out at
a resolution of $1024^3$ meshpoints \cite{Haugen03b}.
Such simulations are nowadays done on large parallel machines using the
Message Passing Interface (MPI) for the communication between processors.
Often, spectral methods are used to calculate derivatives and to
solve a Poisson-type equation for the pressure.
Alternatively, high order finite difference schemes can be used, which
are more easily parallelized, because only data of a small number of
neighboring meshpoints need to be communicated to other processors.
In such cases it is advantageous to solve the compressible equations,
whose solutions approximate the incompressible ones when the rms velocity
is small compared with the sound speed.
One such code, that is documented and publicly available, is
the {\sc Pencil Code} \cite{PencilCode}.
Many of the simulations presented in this review have been done
using this code.
Details regarding the numerical method can also be found in Ref.~\cite{Bra03}.
Quantitative comparisons with spectral codes are presented in
Ref.~\cite{Scheko05}.

\begin{figure}[t!]\begin{center}
\includegraphics[width=.85\textwidth]{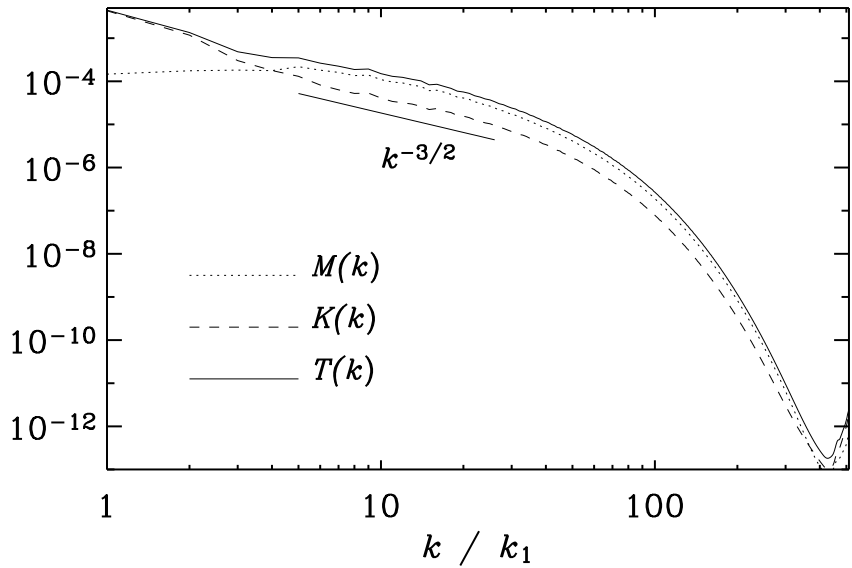}
\end{center}\caption[]{
Magnetic, kinetic and total energy spectra.
$1024^3$ meshpoints.
The Reynolds number is $u_{\rm rms}/(\nu k_{\rm f})\approx960$
(from Ref.~\cite{Haugen03b}).
}\label{power1024a}\end{figure}

In the simulations of Haugen et al.\ \cite{Haugen03b} the velocity field
is forced randomly at wavenumbers between 1 and 2, where $k=1$ is the
smallest wavenumber in a box of size $(2\pi)^3$.
These simulations begin to show indications of a small inertial range
beyond the wavenumber $k\approx8$; see \Fig{power1024a}.
The magnetic energy in the saturated state is also peaked
at about this wavenumber. Note that the semi-analytic closure models
which lead to a renormalization of the diffusion coefficient 
\cite{subamb,sub99}, 
suggest a peak of the saturated spectrum at a wavenumber 
$ k_{\rm p} \sim k_{\rm f}R_{\rm crit}^{1/2}$.
For $k_{\rm f}\approx 1.5$ and $R_{\rm crit} \sim 35$, 
this predicts $k_{\rm p}\sim 8$, which indeed seems to match the
value obtained from the simulation.

The fact that the magnetic energy spectrum peaks at $k\approx8$
(or less) implies that in the present
simulations there is not much dynamical range available before
dissipation sets in.
Furthermore, just before the dissipative subrange, hydrodynamic turbulence
exhibits a `bottleneck effect', i.e.\ a shallower
spectrum (or excess power) at large $k$.
It has been argued \cite{Falk94} that this is because of nonlocal wavevector
interactions corresponding to elongated triangles with one short wavevector
(corresponding to a long scale in the inertial range) and two long ones
(corresponding to short scales in the dissipative subrange).
These nonlocal interactions, which couple the inertial range with the
resistive subrange, limit the amount of energy that the inertial range
modes can dispose of.

While the bottleneck effect is not very pronounced in laboratory wind tunnel
turbulence \cite{SheJackson93,PKW02}, it has now become a very marked effect
in fully three-dimensional spectra available from high resolution
simulations \cite{PorterWoodwardPouquet98,Gotoh02,Kan03}.
The reason for the discrepancy has been identified as being
a mathematical consequence of the transformation between
three-dimensional and one-dimensional spectra \cite{DHYB03}.
This may also explain the shallower $k^{-3/2}$ (instead of $k^{-5/3}$)
spectra seen in \Fig{power1024a} for a narrow wavenumber interval.
Another possibility is that the true inertial range has not yet been
seen, and that asymptotic spectral equipartition may occur at still
larger wavenumbers and larger Reynolds numbers \cite{HB05}.

\begin{figure}[t!]\begin{center}
\includegraphics[width=\textwidth]{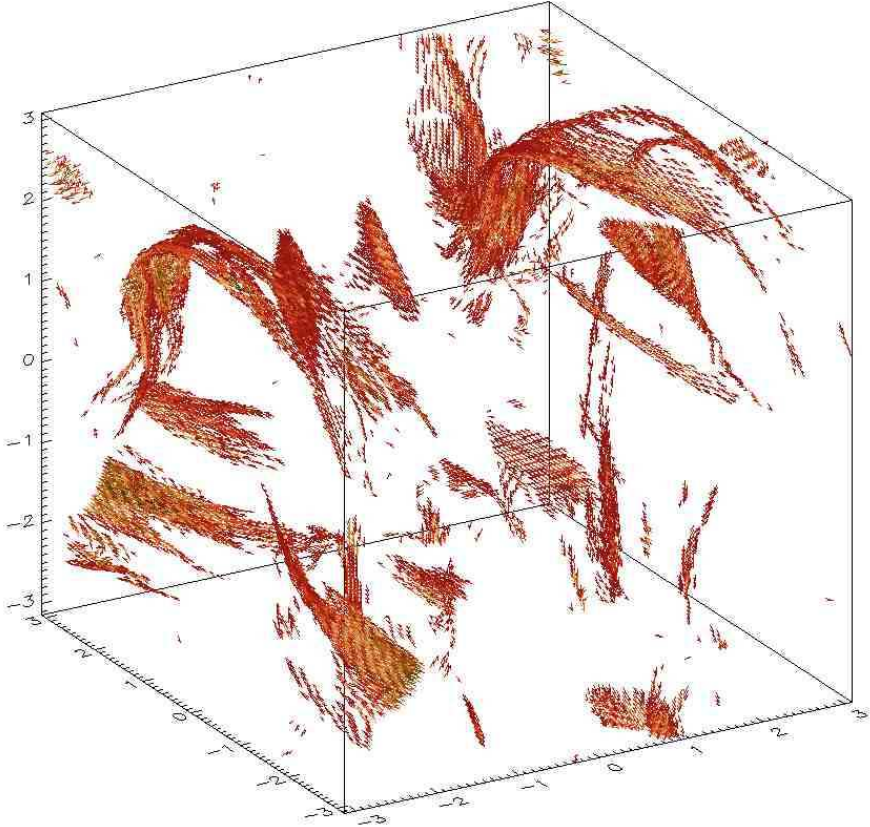}
\end{center}\caption[]{
Magnetic field vectors shown at those locations where $|\BB|>4B_{\rm rms}$.
Note the long but thin arcade-like structures extending over almost
the full domain.
The structures are sheet-like with a thickness comparable to the
resistive scale
(from Ref.~\cite{Haugen04}).
}\label{B_vec_3d}\end{figure}

\begin{figure}[t!]\begin{center}
\includegraphics[width=\textwidth]{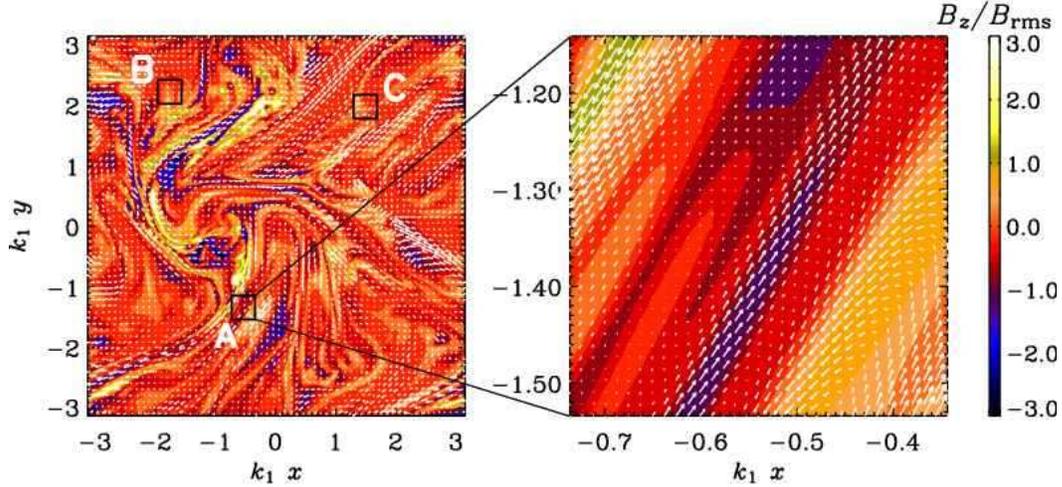}
\end{center}\caption[]{
Snapshot of the magnetic field for $P_{\rm m}=50$,
shown in a cross-section through the middle of the
computational domain at a time when the field is in a saturated state.
The data correspond to the spectra shown in \Fig{pspec_largePm}.
The field component perpendicular to the
plane of the figure is shown color coded (or in shades of gray) with black
corresponding to field pointing into the plane,
and white to field pointing out of the plane. The field in the plane of
the figure is shown with vectors whose length is proportional to the field
strength.
The right hand side shows an enlargement of the sub-domain {\sf A}
marked on the left hand side.
Note the folded structures in sub-domain {\sf A}.
}\label{VAR95slice}\end{figure}

\begin{figure}[t!]\begin{center}
\includegraphics[width=\textwidth]{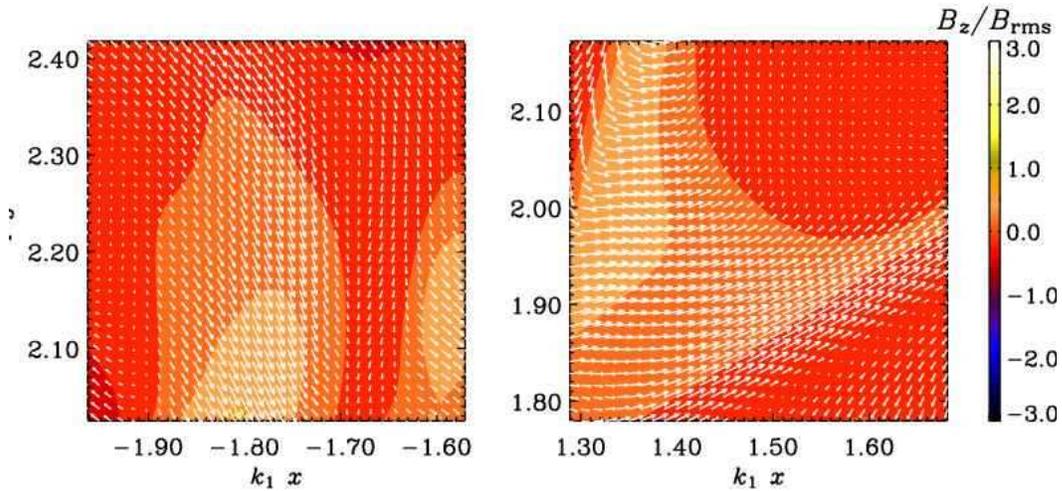}
\end{center}\caption[]{
Magnetic field in sub-domains {\sf B} and {\sf C} that were indicated
on the left hand side of \Fig{VAR95slice}.
Note the lack of folded structures.
}\label{VAR95slice2}\end{figure}

\begin{figure}[t!]\begin{center}
\includegraphics[width=.49\textwidth]{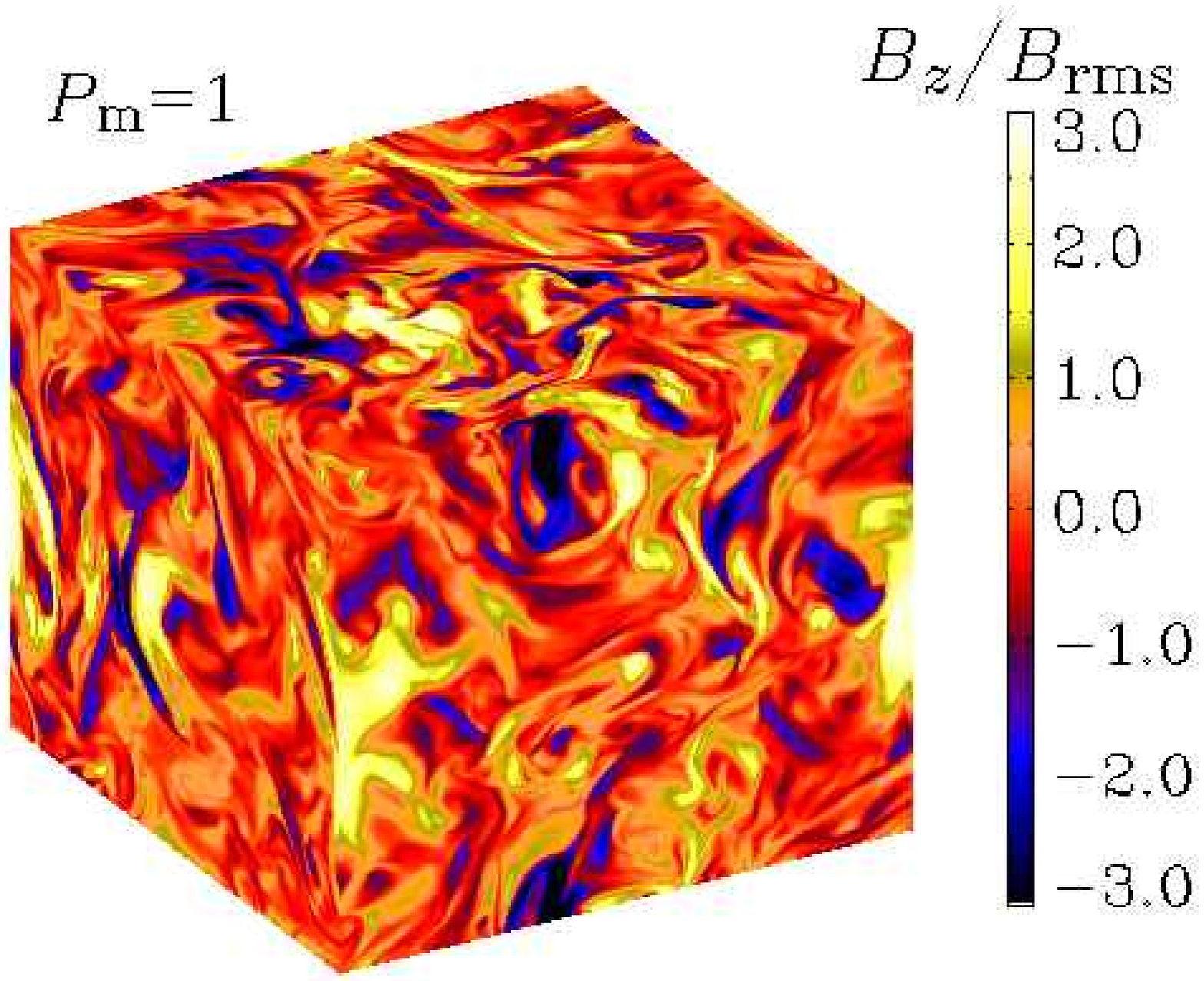}
\includegraphics[width=.49\textwidth]{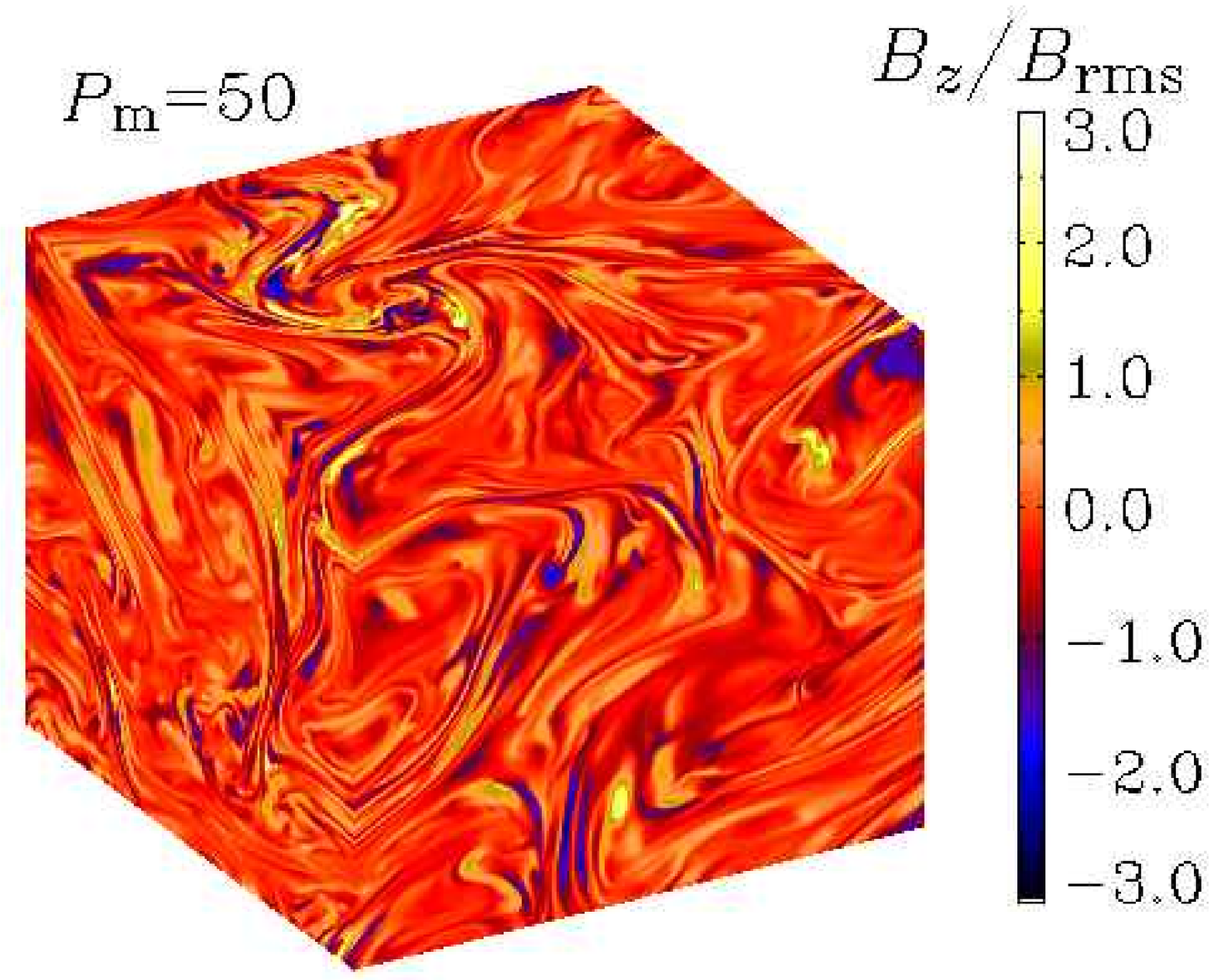}
\end{center}\caption[]{
Comparison of $B_z$ (in shades of gray shown on the periphery
of the box) for $P_{\rm m}=1$ (left) and $P_{\rm m}=50$ (right).
The data shown here correspond to the spectra shown in
\Figs{pspec_nohel512d2}{pspec_largePm}, respectively.
Note that the magnetic field for $P_{\rm m}=50$ is more intermittent
in space and less space filling than for $P_{\rm m}=1$.
}\label{pm1and50}\end{figure}

In \Fig{B_vec_3d} we show a visualization of the magnetic field vectors
at those points where the magnetic field exceeds a certain field strength.
The structures displayed represent a broad range of sizes even within
one and the same structure:
the thickness of the structures is often comparable to the resistive
scale, their width is a bit larger (probably within the inertial range)
and their length is comparable to the box size (subinertial range).
Although large scales are involved in this simulations, we must distinguish
them from the type of large scale magnetic fields seen in simulations with
kinetic helicity that will be discussed in more detail in
\Sec{SimulationsHelicalDynamos} and that are invoked to explain
the solar cycle. 
Further, although the thickness of these structures
is comparable to the resistive scale, we should keep in
mind that these represent the rare structures with 
$|\BB|>4B_{\rm rms}$, and are not volume filling.

At large magnetic Prandtl numbers the field shows folded structures
that were discussed in detail by Schekochihin et al.\ \cite{Schek04}.
An example of an arbitrarily chosen cross-section of a simulation with
$P_{\rm m}=50$ is shown in \Fig{VAR95slice},
together with enlargements of different parts of the domain.
The section {\sf A} shows a region where the field is clearly folded;
with the fairly straight field lines displaying 
rapid reversals transverse to its general direction.
On the other hand there are also many other regions
like {\sf B} and {\sf C} in the box, where the field is equally strong
but is not in resistive scale folds. This illustrates that
whereas there are folded structures, they need not be volume filling.

Folded structures are less prominent when $P_{\rm m}=1$.
A comparison of the typical field structure for $P_{\rm m}=1$ and 50 is
shown in \Fig{pm1and50}, where we show color/gray scale representations
of $B_z$ at an arbitrarily chosen moment during the saturated phase.
The magnetic Reynolds number is $u_{\rm rms}/(\nu k_{\rm f})\approx600$
and 4000 in the left and right hand panels, respectively, and the
resolution is $512^3$ meshpoints in both cases.

\begin{figure}[t!]\begin{center}
\includegraphics[width=.85\textwidth]{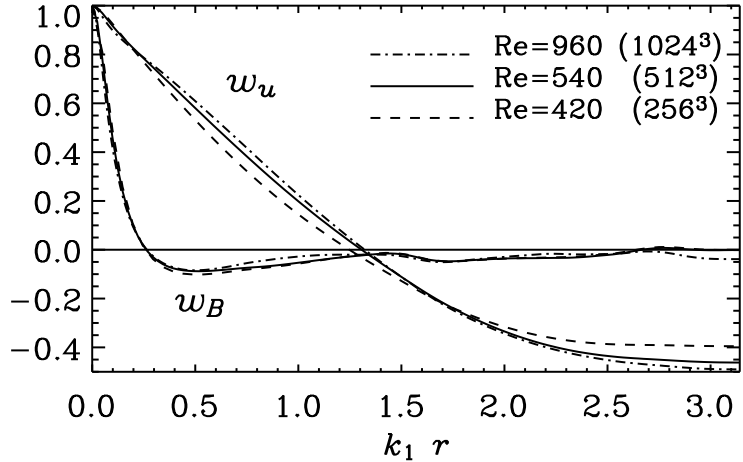}
\end{center}\caption[]{
Autocorrelation functions of magnetic field and velocity.
Note that the autocorrelation functions are nearly independent
of resolution and Reynolds number.
The velocity correlation length is $\sim3$ while
the magnetic correlation length is $\sim0.5$
(from Ref.~\cite{Haugen04}).
}\label{calc_correl}\end{figure}

For comparison with the analytic theory we plot in \Fig{calc_correl}
the correlation function \eq{wdef} of the magnetic field at saturation
(and similarly for the velocity) for runs with $P_{\rm m}=1$ \cite{Haugen04}.
Similar autocorrelation functions have also been seen in
simulations of convective dynamos \cite{BJNRST96}.
It turns out that the velocity correlation length is $\sim 3$
(50\% of the box size) while
the magnetic field correlation length is $\sim 0.5$
(8\% of the box size); see \Fig{calc_correl}.
(We recall that the box size is $2\pi$.)
Clearly, the magnetic correlation length is much shorter than
the velocity correlation length, but it is practically independent of
Re ($=R_{\rm m}$) and certainly much longer than the resistive scale,
$\sim 2\pi/k_{\rm d}\approx0.04$ (0.7\% if the box size).
The fact that $w(r)$ in the saturated state is independent
of the microscopic $R_{\rm m}$ agrees with the corresponding
prediction of the closure model involving artificial nonlinear 
drifts \cite{subamb,sub99}, discussed in \Sec{nonlinear_drift}.

\begin{figure}[t!]\begin{center}
\includegraphics[width=.9\textwidth]{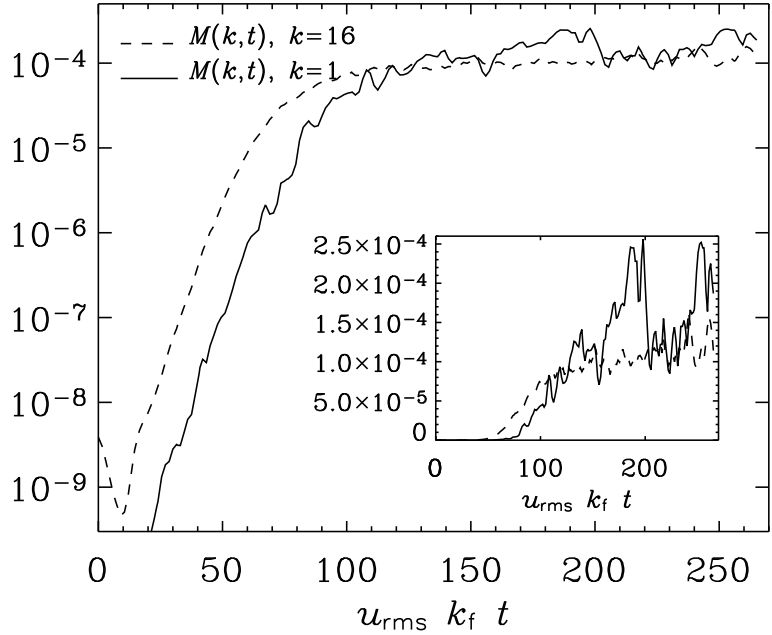}
\end{center}\caption[]{
Saturation behavior of the spectral magnetic energy at wavenumbers
$k=1$ (solid line) and $k=16$ (dashed line).
The average forcing wavenumber is $k_{\rm f}=1.5$ and
the resolution is $512^3$ meshpoints.
Note the slow saturation behavior for $k=1$
(from Ref.~\cite{Haugen04}).
}\label{ppower_time_nohel512d2}\end{figure}

\begin{figure}[t!]\begin{center}
\includegraphics[width=.8\textwidth]{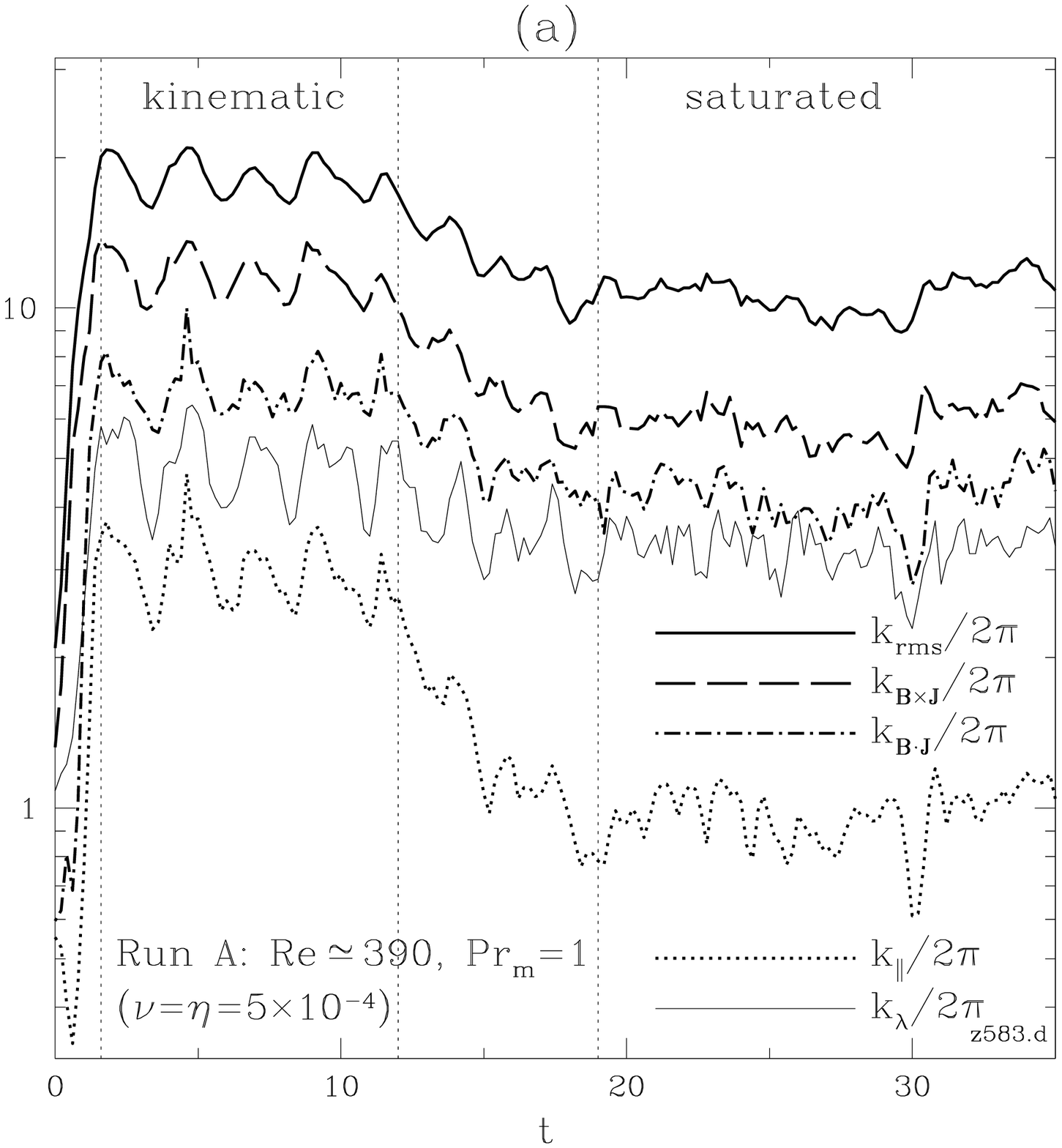}
\end{center}\caption[]{
Evolution of characteristic wavenumbers.
Note that the characteristic parallel wavenumber of the field,
$k_\parallel$, becomes comparable with the scale of the box, while the
so-called rms wavenumber, $k_{\rm rms}$, drops by almost a factor of
two as saturation sets in.
Here, $k_\parallel^2=\bra{|\BB\cdot\nab\BB|^2}/\bra{\BB^4}$ and
$k_{\rm rms}^2=\bra{|\nab\BB|^2}/\bra{\BB^2}$, where angular brackets
denote volume averages.
(The wavenumbers $k_{\BB\times\JJ}$ and $k_{\BB\cdot\JJ}$ are defined
similarly to $k_\parallel$, but with $\BB\times\JJ$ and $\BB\cdot\JJ$,
respectively, and follow a trend similar to that of $k_{\rm rms}$.)
The rms wavenumber of the flow, $k_\lambda$, is proportional to the
inverse Taylor microscale and decreases only slightly during saturation.
Courtesy A.\ A.\ Schekochihin \cite{Schek04}.
}\label{kt_z583}\end{figure}

In contrast to large scale dynamos with helicity, which are now
generally believed to have a resistively limited saturation phase
in closed or periodic domains,
the situation is less clear for nonhelical small scale dynamos; see
\Sec{KazantsevResistive} and Refs~\cite{Schek02,Schek04}.
The simulation results shown in \Fig{ppower_time_nohel512d2} seem
compatible with
a slow saturation behavior for some intermediate time span
($80<u_{\rm rms}k_{\rm f}t<200$), but not at later times
($u_{\rm rms}k_{\rm f}t>200$).

As the small scale dynamo saturates, various magnetic length scales in the
simulation increase quite sharply--some of them almost by a factor of two.
For example, in convective dynamo simulations of a layer of depth $d$
the magnetic Taylor microscale, $\sqrt{5\bra{\BB^2}/\bra{\JJ^2}}$,
increased from $0.04d$ to $0.07d$ during saturation (see Fig.~4b of
Ref.~\cite{BJNRST96}).
This increase of the characteristic length scale is in qualitative
agreement with the analytic theory of the nonlinear saturation of the
small scale dynamo (\Sec{SaturationSmallScale}).
In particular, during the kinematic stage
these length scales remain unchanged.
This can be seen from \Fig{kt_z583}, where we show the evolution of various
wavenumbers in a simulation of Schekochihin et al.\ \cite{Schek04}.
The approximate constancy of the characteristic wavenumbers
during the kinematic stage illustrates
that the simple-minded picture of the evolution of single structures
(\Sec{GeneralConsiderations}) is different from the collective
effect for an ensemble of many structures that are constantly
newly generated and disappearing.
We also note that in the saturated state the characteristic wavenumbers
are approximately unchanged, suggesting that in the nonlinear regime
there is no slow saturation phase (unlike the helical case that will be
discussed later).
During saturation, the drop of various characteristic wavenumbers is
comparable to the increase of the Taylor microscale seen in the
convective dynamo simulations \cite{BJNRST96}.

\subsection{Comments on the Batchelor mechanism and $P_{\rm m}$ dependence}
\label{LowPrM}

In an early attempt to understand the possibility of small scale dynamo
action, Batchelor \cite{Bat50} appealed to the formal analogy between
the induction equation and the vorticity equation,
\EQA
{\DD\WWW\over\DD t}&=&\WWW\cdot\nab\UU+\nu\nabla^2\WWW,\\
{\DD\BB\over\DD t}&=&\BB\cdot\nab\UU+\eta\nabla^2\BB.
\ENA
These equations imply the following evolution equations for enstrophy and
magnetic energy,
\EQA
\half{\dd\bra{\WWW^2}\over\dd t}&=&W_iW_j s_{ij}-\nu\bra{(\nab\times\WWW)^2},\\
\half{\dd\bra{\BB^2}\over\dd t}&=&B_iB_j s_{ij}-\eta\bra{(\nab\times\BB)^2},
\ENA
where $s_{ij}=\half(u_{i,j}+u_{j,i})$ is the rate of strain tensor.
Assuming that the rate of enstrophy dissipation,
$\nu\bra{(\nab\times\WWW)^2}$, is approximately balanced by the rate of
enstrophy production, $W_iW_j s_{ij}$, and that this rate is similar to
the rate of magnetic energy production, Batchelor argued that
magnetic energy would grow in time provided $\eta<\nu$, i.e.\
\EQ
P_{\rm m}\equiv\nu/\eta>1.
\EN
By now there have been several numerical investigations of
small scale dynamos that
operate in a regime where the magnetic Prandtl number is less than unity.
However, it is still not clear whether there exists a critical value of
$P_{\rm m}$ below which no dynamo action is possible \cite{Scheko_critPm},
whether the critical magnetic Reynolds number becomes independent of
$P_{\rm m}$ for small enough values \cite{bold_catt03,RogaKlee97}, or
whether, as $P_{\rm m}$ decreases, there continues to be
a rise of the critical magnetic Reynolds number \cite{Haugen04}.

\begin{figure}[t!]\begin{center}
\includegraphics[width=.8\textwidth]{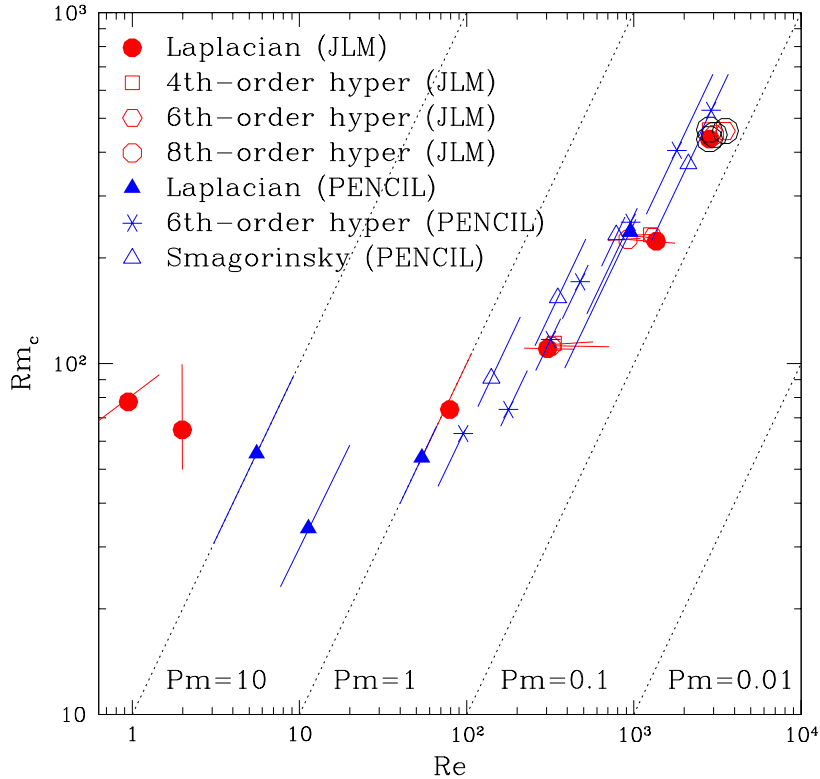}
\end{center}\caption[]{
Dependence of $R_{\rm crit}$ on $\mbox{Re}$.
``JLM'' refers to simulations done with the incompressible spectral
code written by J.~L.~Maron: runs with Laplacian viscosity, 4th-, 6th-,
and 8th-order hyperviscosity (resolutions $64^3$ to $256^3$). In this
set of simulations, hyperviscous runs were done at the same values of
$\eta$ as the Laplacian runs, so the difference between the results
for these runs is nearly imperceptible.  ``PENCIL'' refers to weakly
compressible simulations done with the {\sc Pencil Code}: runs with
Laplacian viscosity, 6th-order hyperviscosity, and Smagorinsky large-eddy
viscosity (resolutions $64^3$ to $512^3$).
Courtesy A.\ A.\ Schekochihin \cite{Scheko05}.
}\label{prmcrit}\end{figure}

Comparing with Batchelor's argument,
the main reason why his argument may not apply is that, again,
the $\WWW$ and $\BB$ fields are in general not identical, and they do in
general show quite different statistics \cite{BNST95,BJNRST96}.
(Also $\WWW$ obeys a nonlinear equation, while $\BB$ obeys a
linear equation for a given $\UU$.)
Establishing the asymptotic dependence of $R_{\rm crit}$
on $P_{\rm m}$ is important because, even though the computing power
will increase, it will still not be possible to simulate realistic
values of $P_{\rm m}$ in the foreseeable future.
Schekochihin et al.\ \cite{Scheko05} have compared the results from
two independent codes and show that there is as yet no evidence for
an asymptotic independence of $R_{\rm crit}$ on $P_{\rm m}$; see
\Fig{prmcrit}.

We note that \Fig{prmcrit} compares not only the results of two different
codes, but at the same time the results of an incompressible calculation
with one of the weakly compressible equations (Mach number about 0.1).
No significant difference is seen between the two simulations.
However, when the flow becomes transonic (Mach number about 1), the
critical magnetic Reynolds number increases by about a factor of 2;
see Ref.~\cite{HauBraMee04} for results with $P_{\rm m}=1$ and 5.

Finally, we mention one property where the $\WWW$ and $\BB$ fields do
seem to show some similarity.
For a $k^{-5/3}$ spectrum of kinetic energy the enstrophy spectrum is
proportional to $k^{1/3}$, so one may expect a similar spectrum for the
magnetic energy in the wavenumber range where feedback from the Lorentz
force can be neglected.
Such results have indeed been reported in the context of convection
\cite{BJNRST96} and forced turbulence \cite{Haugen04} during the
kinematic stage.
For forced turbulence, a $k^{1/3}$ spectrum has only been seen in the
range $k_1<k<k_{\rm p}$, where $k_{\rm p}\approx k_{\rm f}R_{\rm crit}^{1/2}$
is the wavenumber where the magnetic energy spectrum peaks; see
\Fig{power1024a} and \Sec{Simulations_SSdynamo}.
We should emphasize, however, that it is not clear that this result is
really a consequence of the (imperfect) analogy between $\WWW$ and $\BB$.

\subsection{Small and large scale dynamos: a unified treatment}
\label{UnifiedTreatment}

We have so far discussed the case of small scale dynamos where
the generated field has correlation lengths of order
or smaller than the forcing scale of the flow. In the next section
and thereafter we will discuss large scale or mean field dynamos.
The large and small scale dynamo problems are usually treated separately.
However this separation is often artificial; there is
no abrupt transition from the field correlated
on scales smaller than $L$ and that correlated on
larger scales. If we consider both large and small scale
fields to be random fields, it turns out that the equations for the
magnetic correlation functions, which involve now {\it both}
longitudinal and helical parts, are already sufficiently general to
incorporate both small and large scale dynamos.
They provide us with a paradigm to study
the dynamics in a unified fashion, which could be
particularly useful for studying the inverse
cascade of magnetic fields to scales larger than $L$.
We elaborate below.

We add a helical piece
to the two point correlation of the velocity field for the
Kazantsev-Kraichnan flow, so we have
\EQ
T_{ij}(r) =
\left(\delta_{ij}-{r_i r_j \over r^2}\right)\,T_{N}(r)
+{r_i r_j \over r^2}\,T_{L}(r)
+\epsilon_{ijk} r_k\,F(r),
\label{tijhel}
\EN
where $F(r)$ represents
the helical part of the velocity correlations.
At $r=0$, we have
\EQ
-2F(0) = -\onethird\int_0^t\bra{\vv(t)
\cdot\nab\times\vv(t')}\;\dd t' ,
\EN
indicating that $F(r)$ is related to the kinetic helicity of the flow.

Consider a system of size $S \gg L$,
for which the mean field averaged
over any scale is zero.
Of course, the concept of a large scale field still
makes sense, as the correlations between field components
separated at scales $r \gg L$, can in principle be non-zero.

Since the flow can be helical, we need to allow the magnetic field
to also have helical correlations. So, the equal-time, 
two point correlation of the magnetic field, $M_{ij}(r,t)$, is now given by
\begin{equation}
M_{ij} = \left(\delta_{ij}-{r_i r_j \over r^2}\right)\,M_{\rm N}
+{r_i r_j\over r^2}\,M_{\rm L}
+\epsilon_{ijk} r_k\,C,
\label{mcor}
\end{equation}
where $C(r,t)$ represents
the contribution from current helicity to the two-point correlation.

The Kazantsev equation can now be generalized to describe the
evolution of both $M_{\rm L}$ and $C$ 
\cite{subamb,sub99,Vain_K86,Ax_Ks00} (see 
\App{kazantsev}). We get
\EQ
{\partial M_{\rm L}\over\partial t} = {2\over r^4}{\partial \over \partial r}
\left[r^4 \eta_{\rm T} {\partial M_{\rm L} \over \partial r}\right]
+ G M_{\rm L} + 4\alpha C,
\label{mlequn}
\EN
\EQ
{\partial H \over \partial t}=-2\eta_{\rm T}C+ \alpha M_{\rm L}, \quad 
C = - \left(H'' + {4H' \over r}\right),
\label{mhequn}
\EN
where $M_{\rm L}=M_{\rm L}(r,t)$ and $C=C(r,t)$, while
$\alpha=\alpha(r)$ and $\eta_{\rm T}=\eta_{\rm T}(r)$, and
we have defined the magnetic helicity correlation function
$H(r,t)$. Note that $H(0,t) = {1\over6}\overline{\AAA\cdot\BB}$, whereas 
$C(0,t) = {1\over6}\overline{\JJ\cdot\BB}$. Also,
\EQ
\alpha(r) = -2[F(0) - F(r)] ,
\EN
and so represents the effect of the helicity in the velocity
field on the magnetic field. [We will see later in \Sec{ClosureModel} that
$\alpha(r\to\infty)$ is what is traditionally called the $\alpha$ effect].
This new term has some surprising consequences. 

Suppose $\alpha_0 = -2F(0) \ne 0$. Then
one can see from \Eqs{mlequn}{mhequn}, that
new ``regenerating'' terms arise at $r\gg L$, or for scales
much larger than the correlation scales of the flow, due to
the $\alpha$ effect. These are in the form
$\dot M_{\rm L} = .... + 4\alpha_0 C$ and $\dot H = ... + \alpha_0 M$,
which couple $M_{\rm L}$ and $C$
and lead to the growth of large scale correlations.
There is also decay of the correlations for $r \gg L$ due to
diffusion with an effective diffusion
coefficient, $\eta_{\rm T0}= \eta + T_{L}(0)$.
From dimensional analysis,
the effective growth rate is $ \Gamma_R \sim
\alpha_0/R - \eta_{\rm T0}/R^2$ for
correlations on scale $\sim R$. This is exactly
as in the large scale $\alpha^2$ dynamo to be discussed in the
next section. This also picks out a special scale
$R_0 \sim \eta_{\rm T0}/\alpha_0$ for a stationary state
(see below). Further, as the small scale dynamo, is simultaneously
leading to a growth of $M_{\rm L}$ at $r < L$,
the growth of large scale correlations can
be seeded by the tail of the small scale dynamo eigenfunction at $r > L$.
Indeed, as advertised, both
the small and large scale dynamos operate simultaneously 
when $\alpha_0 \ne 0$, and can be studied simply by solving for 
a single function $M_{\rm L}(r,t)$.

The coupled time evolution of $H$ and $M_{\rm L}$ for a non-zero
$\alpha_0$  requires numerical solution, which we discuss
later in \Sec{ClosureModel}. But interesting analytical
insight into the system can be obtained for the marginal,
quasi-stationary mode, with $\dot{M_{\rm L}} \approx 0$, $\dot{H} \approx 0$.
Note that for $\eta \ne 0$, we will find that
the above system of equations always evolves the correlation function
at the resistive time scale.
But for time scales much shorter than this, one can examine
quasi-stationary states. For $\dot{H} \approx 0$, we get
from \Eq{mhequn} $C \approx[\alpha(r)/2\eta_{\rm T}(r)] M$.
Substituting this into \Eq{mlequn} and defining once again
$\Psi =r^2\sqrt{\eta_{\rm T}}M_{\rm L}$, we get
\begin{equation}
-\eta_{\rm T}{d^2 \Psi \over dr^2} + \Psi \left [
U_0 - {\alpha^2(r) \over \eta_{\rm T}(r)} \right] = 0, 
\label{sievoln}
\end{equation}
where $U_0$ is the potential defined earlier in \Eq{U0eqn}.
We see that the problem of determining the magnetic
field correlations for the quasi-stationary mode
once again becomes the problem of determining
the zero-energy eigenstate
in a modified potential, $U = U_0 - \alpha^2/\eta_{\rm T}$.
(Note that the $U$ in this subsection is not to be
confused with velocity in other sections.)
The addition to $U_0$, due to the helical correlations,
is always negative definite. Therefore, helical correlations
tend to make bound states easier to obtain.
When $F(0) = 0$, and there is no net $\alpha$ effect,
the addition to $U_0$ vanishes at
$r \gg L$, and $U \to 2\eta_T/r^2$ at large $r$,
as before. The critical magnetic Reynolds number, 
for the stationary state,
will however be smaller than when $F(r) \equiv 0$,
because of the negative definite addition to $U_0$
(see also Ref.~\cite{kim_hughes97}).

\begin{figure}[t!]\begin{center}
\includegraphics[angle=-90,width=.8\textwidth]{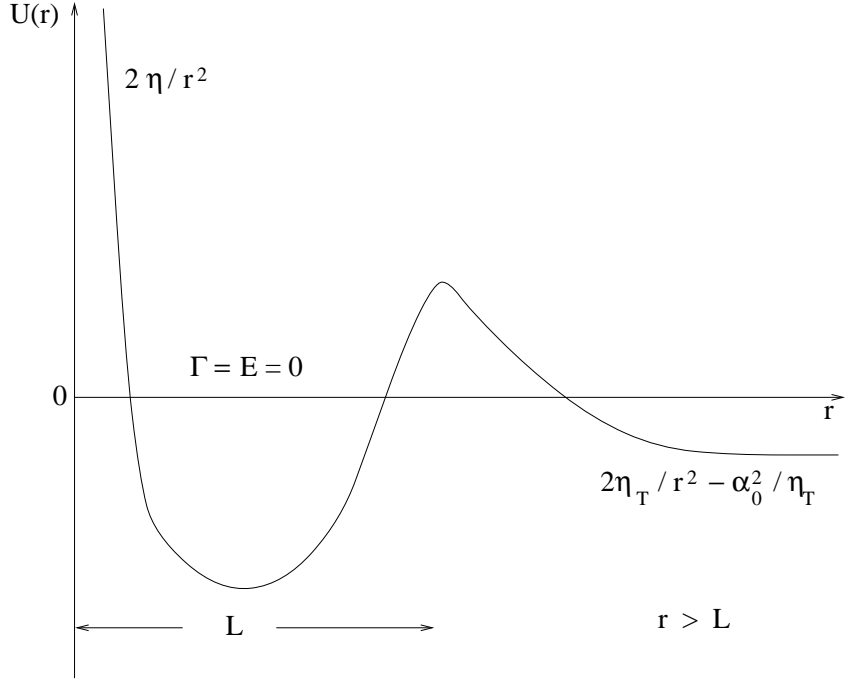}
\end{center}\caption[]{
Schematic illustration of the potential
$U(r)$ for the marginal mode in helical turbulence.
A non-zero $\alpha_0$ allows the tunneling
of the zero-energy state to produce
large scale correlations.
}\label{unified}\end{figure}

When $\alpha_0 = -2F(0) \ne 0$, a remarkable change occurs
in the potential. At $r \gg L$, where the turbulence
velocity correlations vanish, we have
$U(r) = 2\eta_{\rm T0}/r^2 - \alpha_0^2/\eta_{\rm T0}$.
So the potential $U$ tends to a negative definite constant
value of $ - \alpha_0^2/\eta_{\rm T0}$ at large $r$ (and
the effective mass changes, $1/2\eta_{\rm T} \to
1/2\eta_{\rm T0}$, which is independent of $r$.)
So there are strictly no bound states,
with zero energy/growth rate, for which the correlations
vanish at infinity. We have schematically illustrated the resulting
potential $U$ in \Fig{unified}, which is a
modification of Fig.~8.4 of
Zeldovich et al.\ \cite{ZRS83}. In fact, for a
non-zero $\alpha_0$, $U$ corresponds
to a potential which allows {\it tunneling} (of the bound state)
in the corresponding quantum mechanical problem.
It implies that the correlations are necessarily
non-zero at large $r > L$. The analytical solution
to (\ref{sievoln}) at large $r \gg L$,
is easily obtained. We have for $r \gg L$, $ M_{\rm L}(r) = {\bar M}_{\rm L}(r)
\propto r^{-3/2} J_{\pm 3/2}(\mu r)$,  
where $ \mu = \alpha_0/\eta_{\rm T0} = R_0^{-1}$. 
This corresponds to
\EQ
w(r) = {\bar w}(r)= \mu r^{-1}[C_1 \sin \mu r  + C_2 \cos \mu r ]; \quad
 r \gg L,
 \label{larb}
\EN
where $C_1,C_2$ are arbitrary constants.
Clearly for a non-zero $\alpha_0$, the correlations
in the steady state at large $r$, are like `free-particle' states,
extending to infinity!
In fact this correlation function is also the one which one
obtains if we demand the random field to be
force free with $\nab \times \BB =  \mu \BB$.
We see therefore that having helicity in the flow opens
up the possibility of generating large scale fields of scales much larger
than that of the turbulent flow. This is the feature we will elaborate 
on more in the following sections.

\subsection{Comments on anisotropy and nonlocality in MHD turbulence}
\label{GoldreichSridhar}

The usual approach to MHD turbulence is to proceed analogously to
hydrodynamic turbulence. 
Note that in hydrodynamic turbulence, one usually makes the assumption 
that the velocity field is statistically isotropic and homogeneous,
and that the nonlinear interactions in wavenumber space are local.
There is also a unique timescale associated with the
nonlinear interactions, the eddy turnover time, $\tau_k=(kv_k)^{-1}$,
where $v_k = [2kE(k)]^{1/2}$ is the turbulent velocity at wavenumber 
$k$ and $E(k)$ is the 1-D kinetic energy spectrum.
Demanding that in the `inertial range'
the energy transfer flux, $\epsilon = v_k^2/\tau_k$,
is a constant, independent of scale,
we have $v_k \propto k^{-1/3}$ and therefore the Kolmogorov
spectrum $E(k) = C \epsilon^{2/3} k^{-5/3}$ for hydrodynamic
turbulence.

There is a crucial difference in the presence of magnetic fields.
A uniform velocity field simply advects the
eddies and leaves the physical system unchanged
(Galilean invariance).
But a uniform (or large scale) magnetic field $\meanBB$, cannot be 
transformed away. Its presence supports propagation of hydromagnetic 
waves (Alfv\'en waves),
and introduces a nonlocal coupling between small and large scales.
Since these waves propagate along $\meanBB$ at the Alfv\'en speed
$V_{\rm A} = \meanB/\sqrt{4\pi\rho}$, they introduce one more time scale,
the Alfv\'en crossing time $\tau_{\rm A}(k)=(kV_{\rm A})^{-1}$, which can
play a role in determining the turbulence properties.
Furthermore, $\meanBB$ introduces a locally preferred direction,
and the turbulence can in principle be anisotropic.

The work of Iroshnikov \cite{Iroshnikov} and Kraichnan \cite{Kraichnan}
(IK) emphasized the importance of the Alfv\'en crossing time $\tau_{\rm A}(k)$.
Since the influential paper by Goldreich and Sridhar \cite{GS95} it
has become clear that anisotropy will also play a crucial
role in MHD turbulence. We briefly discuss these ideas, 
before drawing comparisons with the MHD turbulence resulting 
from dynamo action. We focus on incompressible motions, split
the magnetic field as $\BB=\meanBB +\bb$, and
write the MHD equations in a more symmetric form,
in terms of Elsasser fields $\zzz_{\pm} = \uu \pm \bb/\sqrt{4\pi\rho}$.
We have, defining $\VV_{\rm A} = \meanBB/\sqrt{4\pi\rho}$ as a vector,
\EQ
\frac{\partial \zzz_{\pm}}{\partial t} \mp \VV_{\rm A}\cdot\nab\zzz_{\pm}
= -\zzz_{\mp}\cdot\nab\zzz_{\pm} -\nab \Pi + \nu \nab^2 \zzz_{\pm}.
\label{elsas}
\EN
Here $\Pi = p/\rho + B^2/8\pi\rho$ acts to enforce
$\nab\cdot\zzz_{\pm} =0$ and we have assumed
for simplicity $\nu=\eta$.

Assuming either $\zzz_{+} =0$ or $\zzz_{-} =0$ gives exact solutions
of the ideal MHD equations. The solution with $\zzz_{-} =0$
represents the Elsasser field $\zzz_{+}$ propagating non-dispersively
in the direction of the mean field. A wave packet with $\zzz_{+} =0$,
represents $\zzz_{-}$ propagating in the direction opposite to
the mean field direction. Nonlinear interactions occur only if
there is an overlap of both type of fluctuations, $\zzz_{\pm}$.
This led Kraichnan \cite{Kraichnan} to suggest that energy
transfer in MHD turbulence results from ``collisions" between
wave packets moving in opposite directions along the mean field.
One can show from \Eq{elsas} that such collisions conserve
the individual energies $\zzz_{\pm}^2$ of the oppositely traveling
wave packets when $\nu=0$. Of course the total energy is also
conserved under ideal MHD. These two conservation
laws are equivalent to the conservation of the total energy
and the cross helicity $H_{\rm cross} = \half\int\uu\cdot\bb\,\dd^3x
={1\over8}\int (\zzz_{+}^2 - \zzz_{-}^2)\,\dd^3x$.

Suppose one assumes that equal amounts of $\zzz_+$ and $\zzz_-$ 
energies are present, and nonlinear interactions are due to oppositely 
directed wave packets. Such a collision occurs over the 
Alfv\'en crossing time of $1/(k_\parallel V_{\rm A})$, where $1/k_\parallel$
is the extent of the wave packet along $\meanBB$. The magnitude
of the nonlinear interaction term $\vert \zzz_{\mp}\cdot\nab\zzz_{\pm}\vert 
\sim k_\perp z_{k_\perp}^2$, where $z_{k_\perp}$ is the magnitude of either
Elsasser variable at wavenumber $ k_\perp$, and $1/k_\perp$ is the
extent of the wave packet transverse to $\meanBB$. (Here we have
implicitly assumed that the component of $\zzz_{\mp}$ perpendicular
to $\meanBB$ dominates over the parallel one, or that
shear Alfv\'en waves dominate pseudo Alfv\'en waves \cite{GS95,LG03}). 
Due to the nonlinear 
term, the collision will then induce a fractional change $\chi$ 
in $\zzz_{\pm}$, given by 
\EQ
\chi \sim \frac{\delta z_{k_\perp}}{z_{k_\perp}}
\sim \frac{k_\perp z_{k_\perp}}{k_\parallel V_{\rm A}}.
\EN
When $\chi \ll 1$ we are in the regime of weak turbulence.
In this case, each collision results in a small random 
change in the wave packet. Since these changes add randomly,
of order $N \sim 1/\chi^2 \gg 1$ collisions are
required for an order unity fractional change.
This implies a time scale for the cascade of energy to a smaller
scale of $\tau_{\rm cas} \sim \chi^{-2}(1/k_\parallel V_{\rm A})
\sim (k_\parallel V_{\rm A})/(k_\perp z_{k_\perp})^2$.

Let us assume,
naively following IK, that turbulence is isotropic, and 
take $z_{k_\perp} = z_k$, with wave packets
having the same parallel and perpendicular scales, i.e. 
we put $k_\parallel = k_\perp =k$ in the above $\tau_{\rm cas}$.
We demand that, far away from the injection or dissipative scales,
the energy transfer flux, $\epsilon = z_k^2/\tau_{\rm cas} = z_k^4 k/V_{\rm A}$,
is a constant, independent of scale, we get  
$z_k \sim (\epsilon V_{\rm A}/k)^{1/4}$. Noting that the 1-D energy spectrum
$E(k) \sim z_k^2/k$, we get the IK spectrum
$E(k) \sim (\epsilon V_{\rm A})^{1/2}k^{-3/2}$.
At best, the assumption of isotropy may be of use if the
large scale field is itself randomly distributed in direction.

Also implicit in the above analysis is the importance
of interactions which couple 3 waves. For weak turbulence
these contribute to energy transfer only if the 3 waves
satisfy the closure relations: $\omega_1 + \omega_2 = \omega_3$
and $\kk_1 +\kk_2 = \kk_3$. Since $\omega =V_{\rm A} \vert k_\parallel\vert$,
and the 3-mode coupling vanishes unless waves 1 and 2 propagate
in opposite directions, the closure relations imply that
either $k_{1\parallel}$ or $k_{2\parallel}$ must vanish,
and the other parallel component equals $k_{3\parallel}$.
So, for weak turbulence, 3-wave interaction do not cascade
energy along $k_\parallel$.

Now suppose we do not make the assumption of isotropy,
and keep $k_\parallel$ constant in working out
the energy flux $\epsilon = z_{k_\perp}^2/\tau_{\rm cas}$.
Then, setting $\epsilon$ to a constant in the
inertial range gives 
$z_{k_\perp} \sim (\epsilon k_\parallel V_{\rm A}/k_\perp^2)^{1/4}$.
Defining the 1-D anisotropic spectrum by $k_\perp E(k_\perp) = z_{k_\perp}^2$,
we get 
\EQ
E(k_\perp) = z_{k_\perp}^2/k_\perp \sim 
\frac{(\epsilon k_\parallel V_{\rm A})^{1/2}}{k_\perp^2}.
\EN
From the expression for $z_{k_\perp}$ we also see that
the strength of nonlinear interactions is $\chi\propto k_\perp^{1/2}$.
So, for small enough scales 
$\chi$ becomes of order unity and turbulence becomes strong.

Indeed strong MHD turbulence is the relevant one for
most astrophysical applications, not only because
of the above, but also because the stirring velocities
and fields are comparable to $\meanBB$. Recall that in weak 
Alfv\'enic turbulence, where $\chi \ll 1$, the Alfv\'en time 
$\tau_{\rm A}=(k_\parallel V_{\rm A})^{-1}$ was small compared
to the nonlinear interaction time $\tau_{\rm nl}=(k_\perp z_{k_\perp})^{-1}$.
In fact we can write $\chi = \tau_{\rm A}/\tau_{\rm nl}$.
GS argued that strong MHD turbulence exhibits what they refer to
as a {\it critical balance}, whereby these 2 timescales become
comparable, and $\chi \sim 1$. We already saw that
if $\chi$ were small then at small enough scales it
grows to order unity. On the other hand, were $\chi 
\gg 1$, for example because $k_\parallel < k_c = 
k_\perp z_{k_\perp}/V_{\rm A}$, then as the wave packets go through
each other over a distance $1/k_c < 1/k_\parallel$,
strong distortions are already introduced, thereby creating
structures with $k_\parallel = k_c$ (see the pictorial
illustration in Ref.~\cite{chandran_GS}).
As a result, $\chi$ is
driven to unity. It seems plausible therefore that something
like critical balance is a stable fixed point of the system.

Critical balance implies that the cascade time is comparable to the
other time scales, $\tau_{\rm cas} \sim \tau_{\rm A} \sim \tau_{\rm nl}$.
Assuming again a scale independent energy flux
$\epsilon = z_{k_\perp}^2/\tau_{\rm cas} = z_{k_\perp}^3k_\perp$,
we get the GS relations
\EQ
z_{k_\perp} \sim V_{\rm A} (k_\perp L)^{1/3},
\quad
k_\parallel \sim k_\perp^{2/3} L^{-1/3}.
\EN
Here we have used the estimate $\epsilon = V_{\rm A}^3/L$,
assuming that the perturbed velocities are of order $V_{\rm A}$,
and the scale of stirring is $L$. These scalings
imply that the parallel and perpendicular sizes of eddies are
correlated, with the eddies becoming highly elongated
at small scales, even if they were isotropic at
the forcing scale.
We can define the three-dimensional spectrum 
$\bar E(k_\perp,k_\parallel)$ using $\int \bar E\,\dd^3k = \sum z_{k_\perp}^2$.
Since $k_\parallel \sim k_\perp^{2/3} L^{-1/3}$, this
spectrum can be written as
$\bar E(k_\perp,k_\parallel) = A f(k_\parallel L^{1/3}/k_\perp^{2/3})$,
where $f(x)$ is a positive, symmetric function of $x$, which is of
order 1 for $x <1$ and is negligible for $\vert x\vert \gg 1$.
We can fix the dependencies in the normalization constant $A$
by changing variables in the integral over $\dd k_\parallel$ to 
$x=k_\parallel L^{1/3}/k_\perp^{2/3}$, and using the definition of $\bar E$.
We get the anisotropic Kolmogorov spectrum \cite{GS95}
\EQ
\bar E(k_\perp,k_\parallel)\sim \frac{V_{\rm A}^2}{k_\perp^{10/3} L^{1/3}}
f\left(\frac{k_\parallel L^{1/3}}{k_\perp^{2/3}}\right).
\EN
GS also derived a kinetic equation where this spectrum arises
as a stationary solution. For some succinct reviews see also 
\cite{chandran_GS,zhouetal}.

We note that in all the above discussions, the large scale
field is assumed to be given or imposed.
In this case, we see that it will introduce anisotropy at all scales, 
implying spectral nonlocality to all smaller scales.
The full extent of the possible departures from isotropy and locality
in wavenumber space, when the large scale field is generated
during dynamo action, is yet to be clarified.
It remains therefore to be seen whether in isotropically forced
turbulence (without imposed field and just the dynamo-generated
small scale field)
the field averaged over any local sub-domain really introduces
nonlocality on all smaller scales within this domain.
If so, one would expect the spectrum of the hydromagnetic turbulence
with imposed field to be similar to the deeper parts of the spectrum
in hydromagnetic turbulence without imposed field.
This does not seem to be the case, because in the latter case the
spectral magnetic energy is found to be in {\it super-equipartition}
(see \Fig{power1024a}).
This behavior is not seen in simulations with imposed field where the
spectrum of magnetic energy is rather in sub-equipartition if the field
is strong.

The effect of anisotropy due to the mean field on the turbulent
transport coefficients has been considered in a number of papers,
but no explicit connection with the Goldreich--Sridhar theory
has yet been made.
The anisotropic cascade is potentially quite important for a proper
understanding of turbulent magnetic diffusion.
Further in discussing MHD turbulence, most of the semi-analytical
works set magnetic helicity to zero. Inclusion of helicity
is important to make connection with the large scale dynamo
generated MHD turbulence.

\section{Large scale turbulent dynamos}
\label{LargeScale}

As the simulation results of \Sec{Simulations_SSdynamo} and the
unified treatment in \Sec{UnifiedTreatment} have shown, the distinction
between small and large scale dynamos is somewhat artificial.
The so-called small scale dynamo may well generate magnetic field
structures extending all the way across the computational domain;
see \Fig{B_vec_3d}.
On the other hand, the overall orientation of these structures
is still random.
This is in contrast to the spatio-temporal coherence displayed by
the sun's magnetic field, where
the overall orientation of flux tubes at a certain location
in space and time follows a regular rule and is not random.
Dynamo mechanisms that explain this will be referred to as
large scale dynamos.
Here the geometry of the domain and the presence of boundaries and
shear are important.

In this section we discuss the basic theory of large scale dynamos
in terms of the alpha effect and discuss simple models.
A discussion of large scale dynamos in terms of the inverse cascade
is presented in the next section.

\subsection{Phenomenological considerations}
\label{PhenomenologicalConsiderations}

Important insights into the operation of the solar dynamo have come from
close inspection of magnetic fields on the solar surface \cite{Babcock}.
One important ingredient is differential rotation.
At the equator the sun is rotating about 30\% faster than at the poles.
This means that any poloidal field will be sheared out and toroidal
field aligned with the direction of the shear will be generated.
Mathematically, this is described by the stretching term
in the induction equation; see \Eq{stretching_term}, i.e.\
\EQ
{\dd\BB_{\rm tor}\over\dd t}=
\BB_{\rm pol}\cdot\nab\UU_{\rm tor}+...\,.
\EN
This term describes the generation of magnetic field $\BB_{\rm tor}$
in the direction of the flow $\UU_{\rm tor}$ from a cross-stream
poloidal magnetic field $\BB_{\rm pol}$.
To an order of magnitude, the amount of toroidal field generation
from a $100\G$ poloidal field in a time interval $\Delta t=10^8\s=3\yr$ is
\EQ
\Delta\BB_{\rm tor}=
\BB_{\rm pol}\;\Delta\Omega_{\odot}\Delta t
\approx100\G\times10^{-6}\times10^8=10^4\G,
\EN
where we have used $\Omega_{\odot}=3\times10^{-6}\s^{-1}$ for the solar
angular velocity, and $\Delta\Omega_{\odot}/\Omega_{\odot}=0.3$ for the
relative latitudinal differential rotation.
So, a $10\kG$ toroidal field can be regenerated completely
from a $100\G$ poloidal field in about 3 years.
However, in the bulk and the upper parts of the solar convection zone
the poloidal fields are weaker ($3$-$10\G$), which would yield
toroidal fields on the order of $300$-$1000\G$.
This would be far too weak a field if it was to rise coherently
all the way from the bottom of the convection zone, which is still
the standard picture.
However, if the field of bipolar regions is produced locally in the upper
parts of the convection zone, as recently supposed in Ref.~\cite{B05},
a $300\G$ field might well be sufficient.
The $2\kG$ fields in sunspots could then be the result of local
compression by an ambient flow.

On the other hand, according to the standard picture (see \Sec{EstimatesFieldStrength})
the $100\kG$ field, necessary to give the right tilt of emerging flux tubes,
can perhaps be explained as the result of
stretching of more localized $\sim1\kG$ field patches that stay
coherent over a time span of about $3\yr$.

In the bulk of the solar convection zone the turnover time,
$\tau_{\rm turnover}=u_{\rm rms}/H_p$
(where $H_p$ is the pressure scale height) is only about
10 days.\footnote{The thermal flux of the sun is $4\times10^{10}\erg\cm^{-2}\s^{-1}$
A great portion of this energy flux is caused by convection and, according
to mixing length theory, $F_{\rm conv}\approx\rho u_{\rm rms}^3$.
Roughly, $\rho=0.1\g\cm^{-3}$ in the lower part of the convection
zone, so $u\approx(4\times10^{10}/0.1)^{1/2}=70\m/\s$.
The pressure scale height is ${\cal R}T/(\mu g)=50\Mm$,
so $\tau_{\rm turnover}=10\,{\rm days}$.}
In order that a toroidal field can be generated, a mean poloidal field
needs to be maintained and, in the case of the sun and other stars
with cyclic field reversals, the poloidal field itself needs to change
direction every 11 years.

\begin{figure}[t!]\begin{center}
\includegraphics[width=.99\textwidth]{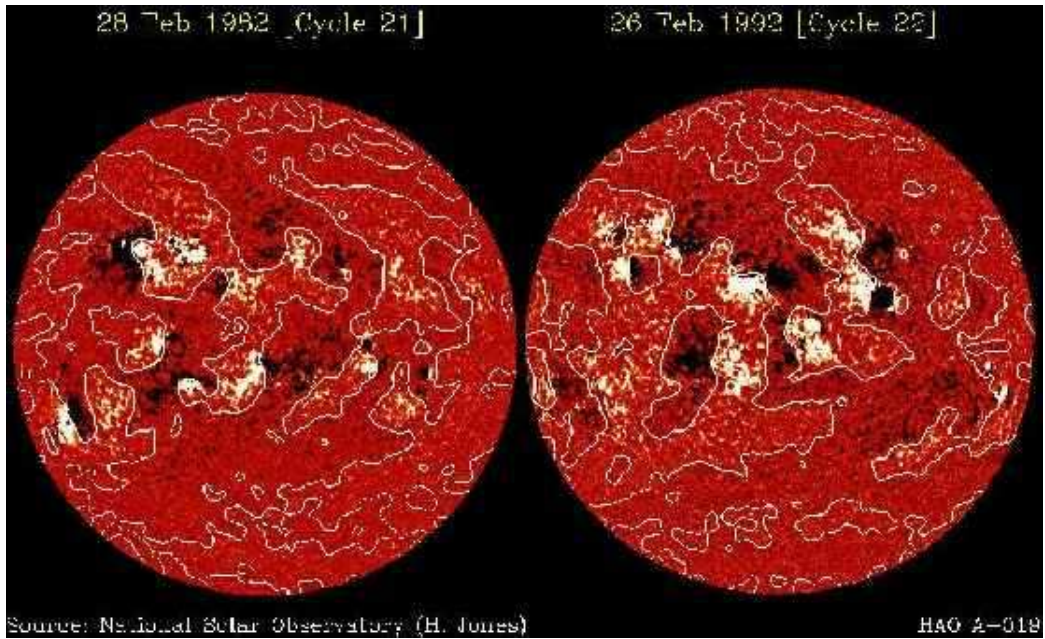}
\end{center}\caption[]{
Solar magnetogram showing bipolar regions, their
opposite orientation north and south of the equator, and
the clockwise tilt in the northern hemisphere and the
counter-clockwise tilt in the southern hemisphere.
Note that the field orientation has reversed orientation
at the next cycle (here after 10 years).
Courtesy of the High Altitude Observatory.
}\label{kittpeak}\end{figure}

Magnetic flux frequently emerges at the solar surface as
bipolar regions.
The magnetic field in sunspots is also often of bipolar nature.
It was long recognized that such bipolar regions are tilted.
This is now generally referred to as Joy's law
\cite{DSC93,Hale19}.
The sense of average tilt is clockwise in the northern hemisphere
and counter-clockwise in the southern.
This tilt is consistent with the interpretation that a toroidal
flux tube rises from deeper layers of the sun to upper layers
where the density is less, so the tube evolves in an expanding
flow field which, due to the Coriolis force, attains a clockwise swirl
in the northern hemisphere and counter-clockwise swirl in the southern
hemisphere; see \Fig{kittpeak}.

\begin{figure}[t!]\begin{center}
\includegraphics[width=.99\textwidth]{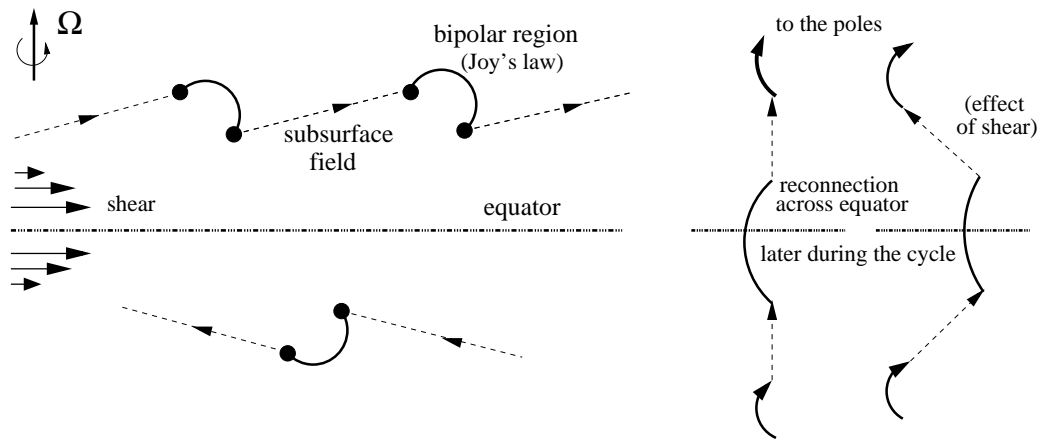}
\end{center}\caption[]{
Sketch of the Babcock-Leighton dynamo mechanism.
As bipolar regions emerge near the surface, they get tilted
in the clockwise sense in the northern hemisphere and
counter-clockwise in the southern hemisphere.
Beneath the surface this process leaves behind a poloidal
field component that points here toward the north pole on
either side of the equator.
Once the remaining subsurface field gets sheared by the surface
differential rotation, it points in the opposite direction as before,
and the whole process starts again.
}\label{babcock+leighton}\end{figure}

Observations suggest that once a tilted bipolar region has emerged at
the solar surface, the field polarities nearer to the poles drift rapidly
toward the poles, producing thereby new poloidal field
\cite{Babcock,Parker70}.
Underneath the surface, the field continues as before, but there
it is also slightly tilted, although necessarily in the opposite sense
(see \Fig{babcock+leighton}).
Because of differential rotation, the points nearest to the equator
move faster, helping so to line up similarly oriented fields \cite{Aake_priv}.
As is evident from \Fig{babcock+leighton}, a toroidal field pointing east
in the northern hemisphere and west in the southern will
develop into a global northward pointing field above the surface.

\subsection{Mean-field electrodynamics}

\begin{figure}[t!]\begin{center}
\includegraphics[width=.5\textwidth]{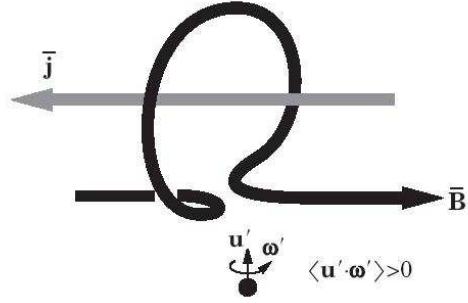}
\end{center}\caption[]{
Production of positive writhe helicity by an uprising and expanding
blob tilted in the clockwise direction by the Coriolis force in the
southern hemisphere, producing a field-aligned current $\meanJJ$ in
the opposite direction to $\meanBB$.
Courtesy of A. Yoshizawa \cite{Yoshi04}.
}\label{alpha_loop}\end{figure}

Parker \cite{Par55} first proposed the idea that the generation of a
poloidal field, arising from the systematic effects of the Coriolis
force (\Fig{alpha_loop}),
could be described by a corresponding term in the induction equation,
\EQ
{\partial\meanBB_{\rm pol}\over\partial t}=
\nab\times\left(\alpha\meanBB_{\rm tor}+...\right).
\label{ParkerPheno}
\EN
It is clear that such an equation can only be valid for averaged fields
(denoted by overbars), because for the actual fields, the
induced electromotive force (EMF) $\UU\times\BB$, would never 
have a component in the direction of $\BB$.
While being physically plausible, this approach only received general
recognition and acceptance after Roberts and Stix \cite{RS71}
translated the work
of Steenbeck, Krause, R\"adler \cite{SKR66} into English.
In those papers the theory for the $\alpha$ effect, as they
called it, was developed and put on a mathematically rigorous basis.
Furthermore, the $\alpha$ effect was also applied to spherical models
of the solar cycle (with radial and latitudinal shear) \cite{SK69a}
and the geodynamo (with uniform rotation) \cite{SK69b}.

In mean field theory one solves the Reynolds averaged equations,
using either ensemble averages, toroidal averages or, in cases
in cartesian geometry with periodic boundary conditions,
two-dimensional (e.g.\ horizontal) averages.
We thus consider the decomposition
\EQ
\UU=\meanUU+\uu,\quad
\BB=\meanBB+\bb.
\label{mean+fluct}
\EN
Here $\meanUU$ and $\meanBB$ are the mean velocity and magnetic fields,
while $\uu$ and $\bb$ are their fluctuating parts.
These averages satisfy the Reynolds rules,
\EQ
\overline{\UU_1+\UU_2}=\meanUU_1+\meanUU_2,\quad
\overline{\meanUU}=\meanUU,\quad
\overline{\meanUU\uu}=0,\quad
\overline{\meanUU_1\;\meanUU_2}=\meanUU_1\;\meanUU_2,
\EN
\EQ
\overline{\partial\UU/\partial t}=\partial\meanUU/\partial t,\quad
\overline{\partial\UU/\partial x_i}=\partial\meanUU/\partial x_i.
\EN
Some of these properties are not shared by several other averages;
for gaussian filtering $\overline{\meanUU}\neq\meanUU$, and
for spectral filtering $\overline{\meanUU\;\meanUU}\neq\meanUU\;\meanUU$,
for example.
Note that $\overline{\meanUU}=\meanUU$ implies that $\overline{\uu}=0$.

Here a comment on scale separation is in order.
The averaging procedure discussed above is valid even if there is hardly
any scale separation, i.e.\ if the averaging length of the container is
close to the eddy scale.
One concern is that an $\alpha^2$ type large scale dynamo field (i.e.\ a mean
field that is generated without shear; see below) may no longer be excited;
see \Sec{FinalFieldStrength}.
But apart from this, the absence of scale separation does not impose any
technical restrictions.
Poor scale separation does however imply that the averages
are not smooth in time and in the directions in which no averaging is
performed.
This, in turn, is reflected in the fact that turbulent transport
coefficients estimated from simulations can be very noisy.

Scale separation does become an issue, however, if one wants to use
averages for which the Reynolds rules are not obeyed.
For example, if mean field theory is to model nonaxisymmetric features
of stellar or galactic magnetic fields, one would like to define a mean
field by averaging over small volumes.
In order that the Reynolds rules remain then at least approximately
valid, scale separation must be invoked.

In the remainder we assume that the Reynolds rules do apply.
Averaging \Eq{Induction1} yields then the mean field induction equation,
\\

\SHADOWBOX{
\EQ
{\partial\meanBB\over\partial t}
=\nab\times\left(\meanUU\times\meanBB+\meanemf-\eta\meanJJ\right),
\label{Induction_meanfield}
\EN
}\\ 

\noindent
where
\EQ
\meanemf=\overline{\uu\times\bb}
\EN
is the mean EMF.
Finding an expression for the correlator $\meanemf$ in terms of the
mean fields is a standard closure problem which is at the heart of
mean field theory.
In the two-scale approach \cite{Mof78} one assumes that
$\meanemf$ can be expanded in powers of the gradients of the mean
magnetic field. This suggests the rather general expression
\EQ
\label{EiGeneral}
{\cal E}_i
=\alpha_{ij}(\ggrav,\OOO,\meanBB,...)\meanB_j
+\eta_{ijk}(\ggrav,\OOO,\meanBB,...)\partial\meanB_j/\partial x_k,
\EN
where the tensor components $\alpha_{ij}$ and $\eta_{ijk}$
are referred to as turbulent transport coefficients.
They depend on the stratification, angular velocity, and
mean magnetic field strength.
The dots indicate that the transport coefficients may also depend on correlators
involving the small scale magnetic field, for example the current
helicity of the small scale field, as will be discussed in
\Sec{SDynamicalQuenching}.
We have also kept only the lowest large scale derivative of the
mean field; higher derivative terms are expected to be smaller 
(cf.\ Section~7.2 in Moffatt \cite{Mof78}), although this may not
be justified in certain cases \cite{BS02}.

The general subject has been reviewed in many text books \cite{Mof78,KR80},
but the importance of the small scale magnetic field (or rather the small
scale current helicity) has only recently been appreciated \cite{GD,BY95},
even though the basic equations were developed much earlier \cite{KR82,PFL76}.
This aspect of the problem
will be discussed in more detail in \Sec{helicity_aquenching}.

The general form of the expression for $\meanemf$ can be determined
by rather general considerations \cite{KR80}.
For example, $\meanemf$ is a polar vector and $\meanBB$ is an axial
vector, so $\alpha_{ij}$ must be a pseudo-tensor.
The simplest pseudo-tensor of rank two that can be constructed using
the unit vectors $\ggrav$ (symbolic for radial density or turbulent
velocity gradients) and $\OOO$ (angular velocity) is
\EQ
\alpha_{ij}=\alpha_1\delta_{ij}\,\ggrav\cdot\OOO
+\alpha_2\hat g_i\hat\Omega_j+\alpha_3\hat g_j\hat\Omega_i.
\label{alphaEqn}
\EN
Note that the term $\ggrav\cdot\OOO=\cos\theta$ leads to the co-sinusoidal
dependence of $\alpha$ on latitude, $\theta$, and a change of sign at
the equator.
Additional terms that are nonlinear in $\ggrav$ or $\OOO$ enter if the
stratification is strong or if the body is rotating rapidly.
Likewise, terms involving $\meanUU$, $\meanBB$ and $\bb$ may appear if the
turbulence becomes affected by strong flows or magnetic fields.
In the following subsection we discuss various approaches to determining
the turbulent transport coefficients.

One of the most important outcomes of this theory is a quantitative
formula for the coefficient $\alpha_1$ in \Eq{alphaEqn} by 
Krause \cite{Kra67},
\EQ
\alpha_1\,\ggrav\cdot\OOO=-{\textstyle{16\over15}}
\tau_{\rm cor}^2u_{\rm rms}^2\OO\cdot\nab\ln(\rho u_{\rm rms}),
\label{KrauseEqn}
\EN
where $\tau_{\rm cor}$ is the correlation time, $u_{\rm rms}$ the root mean
square velocity of the turbulence, and $\OO$ the angular velocity vector.
The other coefficients are given by $\alpha_2=\alpha_3=-\alpha_1/4$.
Throughout most of the solar convection zone, the product $\rho u_{\rm rms}$
decreases outward.\footnote{This can be explained as follows:
in the bulk of the solar convection zone the
convective flux is approximately constant, and mixing length predicts that
it is approximately $\rho u_{\rm rms}^3$.
This in turn follows from $F_{\rm conv}\sim\rho u_{\rm rms}c_p\delta T$
and $u_{\rm rms}^2/H_p\sim g\delta T/T$ together with the expression
for the pressure scale height $H_p=(1-{1\over\gamma})c_pT/g$.
Thus, since $\rho u_{\rm rms}^3\approx\,{\rm const}$, we have
$u_{\rm rms}\sim\rho^{-1/3}$ and $\rho u_{\rm rms}^3\sim\rho^{2/3}$.}
Therefore, $\alpha>0$ throughout most of the northern hemisphere.
In the southern hemisphere we have $\alpha<0$, and
$\alpha$ varies with colatitude $\theta$ like $\cos\theta$.
However, this formula also predicts that $\alpha$ reverses sign
very near the bottom of the convection zone where $u_{\rm rms}\to0$.
This is caused by
the relatively sharp drop of $u_{\rm rms}$ \cite{Kri84}.

The basic form and sign of $\alpha$ is also borne out by simulations
of stratified convection \cite{BNPST90,OSB01}.
One aspect that was first seen in simulations is the fact that in
convection the vertical component of the $\alpha$ effect can have
the opposite sign compared with the horizontal components
\cite{BNPST90,OSB01}.
The same result has later been obtained in analytic calculations
of the $\alpha$ effect in supernova-driven interstellar turbulence
\cite{Ferriere92} and in first order smoothing approximation (FOSA)
calculations \cite{RK93}.

\subsection{Calculation of turbulent transport coefficients}
\label{CalculationTurbulentTransport}

Various techniques have been proposed for determining turbulent transport
coefficients. Even in the kinematic regime, where the changes
in the velocity field due to Lorentz forces are ignored,
these techniques have some severe uncertainties.
Nevertheless, the various techniques produce similar terms,
although the so-called minimal $\tau$ approximation (MTA) does actually
predict an extra time derivative of the electromotive force
\cite{BF02b}.
This will be discussed in full detail in \Sec{Revisit}.
We only mention here that in MTA the triple correlations are not
neglected, as they are in FOSA; see \Sec{FOSA}.
Instead, the triple correlations are approximated by quadratic terms.
This is similar in spirit to the usual $\tau$ approximation used
in the Eddy Damped
Quasi-Normal Markovian (EDQNM) closure approximation (see \Sec{edqnm}),
where the irreducible part of quartic correlations are approximated by 
a relaxation term proportional to the triple correlations.
Other approaches include direct simulations \cite{BNPST90,OSB01,CH96}
(see \Sec{SimulationsTransport}), calculations based on random waves
\cite{Mof78,Brag64,Mof70} or individual blobs \cite{Ferriere92,Stix83}
(see \Sec{RandomWavesIndividualBlobs}), or calculations based on the
assumption of delta-correlated velocity fields \cite{Kaz68,Mol88,kraichnan}
(\Sec{DeltaCorrelatedVelocityFields}).
A summary of the different approaches, their properties and limitations
is given in \Tab{Tsum}.

One of the most promising approaches is indeed MTA,
because there is now some numerical evidence from turbulence
simulations that MTA
may be valid even when the fluctuations are large
and the correlation time not very short; see
\Sec{tauApproxNonlin}, and Ref.~\cite{BraKapMoh03}.
This is of course beyond the applicability regime of FOSA.
Early references to MTA include the papers of Vainshtein and Kitchatinov
\cite{VK83} and Kleeorin and collaborators \cite{KR90,KMR96,RKR03}, but
since MTA is based on a closure hypothesis, detailed comparisons
between theory and simulations \cite{B01,BF02b,BraKapMoh03,BF03} have
been instrumental in giving this approach some credibility.

In the following we discuss the foundations of FOSA and MTA, but we
defer detailed applications and calculations of MTA to \Sec{Revisit}.

\subsubsection{First order smoothing approximation}
\label{FOSA}

The first order smoothing approximation (FOSA) or, synonymously,
the quasilinear approximation, or the second order correlation approximation
is the simplest way of calculating turbulent transport coefficients.
The approximation consists of linearizing the equations for the
fluctuating quantities and ignoring quadratic terms that would lead to
triple correlations in the expressions for the quadratic correlations.
This technique has traditionally been applied to calculating the turbulent
diffusion coefficient for a passive scalar or the turbulent viscosity
(eddy viscosity).

\begin{table}[t!]\caption{
Summary of various approaches to calculate
turbulent transport coefficients.
}\vspace{12pt}\centerline{\begin{tabular}{lcccccccccc}
Technique     &     Refs    &linear?&homogeneous?&$\bb^2/\meanBB^2$\\
\hline
EDQNM           & \cite{PFL76}            &  no   &  yes  &  large  \\
MTA             & \cite{BF02b,RKR03}      &  no   &  no   &  large? \\
simulations & \cite{BNPST90,OSB01,CH96}   &  no   &  no   &  modest/large \\
FOSA          & \cite{Mof78,KR80}         &  no   &  no   &  small  \\
stochastic  & \cite{Mol88}                &  yes  &  no   &  large  \\
random waves & \cite{Mof78,Brag64,Mof70}  &  no   &  no   &  small  \\
individual blobs &\cite{Ferriere92,Stix83}&  no   &  no   &  small  \\
\label{Tsum}\end{tabular}}\end{table}

Suppose we consider the induction equation.
The equation for the fluctuating field can be obtained by subtracting
\Eq{Induction_meanfield} from \Eq{Induction1}, so
\EQ
{\partial\bb\over\partial t}
=\nab\times\left(\meanUU\times\bb+\uu\times\meanBB+\uu\times\bb
-\meanemf-\eta\jj\right),
\label{Induction_flucts}
\EN
where $\jj=\nab\times\bb\equiv\JJ-\meanJJ$ is the fluctuating current density.
The first order smoothing approximation consists of {\it neglecting} 
the term $\uu\times\bb$ on the RHS of \Eq{Induction_flucts},
because it is nonlinear in the fluctuations.
This can only be done if
the fluctuations are small, which is a good approximation only under 
rather restrictive circumstances, for example if $R_{\rm m}$ is small.
The term $\meanemf$ is also nonlinear in the fluctuations, but it
is not a fluctuating quantity and gives therefore
no contribution, and the $\meanUU\times\bb$ is often neglected
because of simplicity (but see, e.g., Ref.~\cite{KR80}).
The neglect of the $\meanUU$ term may not be justified for systems
with strong shear (e.g.\ for accretion discs) where the inclusion
of $\meanUU$ itself could lead to a new dynamo effect,
namely the shear--current effect \cite{Roga+Klee03,Roga+Klee04,RadStep05}.
In the case of small $R_{\rm m}$, one can neglect both the 
nonlinear term $\GG\equiv\nab\times(\uu\times\bb-\overline{\uu\times\bb})$,
and the time derivative of $\bb$, resulting in a
linear equation
\EQ
\eta\nab^2\bb = -\nab\times\left(\uu\times\meanBB\right).
\EN
This can be solved for $\bb$, if $\uu$ is given. $\meanemf$
can then be computed relatively easily \cite{Mof78,KR80}.

However, in most astrophysical applications, $R_{\rm m} \gg 1$.
In such a situation, FOSA is thought to still be applicable
if the correlation time $\tau_{\rm cor}$ of the turbulence is small,
such that $\tau_{\rm cor}u_{\rm rms}k_{\rm f} \ll 1$, 
where $u_{\rm rms}$ and $k_{\rm f}$ are typical
velocity and correlation wavenumber, associated
with the random velocity field $\uu$. Under this
condition, the ratio of the nonlinear term to
the time derivative of $\bb$ is argued to be
$\sim (u_{\rm rms}k_{\rm f}b)/(b/\tau_{\rm cor}) = 
\tau_{\rm cor}u_{\rm rms}k_{\rm f} \ll 1$, and so 
$\GG$ can be neglected \cite{Mof78} (but see below). We then get
\EQ
{\partial\bb\over\partial t}=\nab\times\left(\uu\times\meanBB\right).
\label{Induction_flucts_FOSA}
\EN
To calculate $\meanemf$, we integrate $\partial\bb/\partial t$ to get
$\bb$, take the cross product with $\uu$, and
average, i.e.\
\EQ
\meanemf=\overline{\uu(t)\times{\int_0^t}\nab\times
\left[\uu(t')\times\meanBB(t')\right]\;\dd t'}.
\label{FOSA_integral}
\EN
For clarity, we have suppressed the common $\xx$ dependence
of all variables.
Using index notation, we have
\EQ
\meanemf_i(t) =\int_0^t \left[
\hat\alpha_{ip}(t,t')\meanB_p(t')+\hat\eta_{ilp}(t,t')\meanB_{p,l}(t')
\right]\,\dd t',
\EN
with $\hat\alpha_{ip}(t,t')=\epsilon_{ijk}\overline{u_j(t)u_{k,p}(t')}$
and $\hat\eta_{ilp}(t,t')=\epsilon_{ijp}\overline{u_j(t)u_l(t')}$,
where we have used $\meanB_{l,l}=0=u_{l,l}$, and commas denote partial
differentiation.
In the statistically steady state, we can assume that $\hat\alpha_{ip}$
and $\hat\eta_{ilp}$ depend only on the time difference, $t-t'$.
Assuming isotropy (again only for simplicity), these tensors must be
proportional to the isotropic tensors $\delta_{ip}$ and $\epsilon_{ilp}$,
respectively, so we have
\EQ
\meanemf(t)=\int_0^t\left[\hat\alpha(t-t')\meanBB(t')
-\hat\eta_{\rm t}(t-t')\meanJJ(t')\right]\,\dd t',
\label{alpha_eta_kernel}
\EN
where $\hat\alpha(t-t')=-\onethird\overline{\uu(t)\cdot\oo(t')}$ and
$\hat\eta_{\rm t}(t-t')=\onethird\overline{\uu(t)\cdot\uu(t')}$ are 
integral kernels, and $\oo=\nab\times\uu$
is the vorticity of the velocity fluctuation.

If we assume the integral kernels to be proportional to the delta function,
$\delta(t-t')$, or, equivalently,
if $\meanBB$ can be considered a slowly varying function of time, one arrives at
\EQ
\meanemf=\alpha\meanBB-\eta_{\rm t}\meanJJ
\label{alpha_eta}
\EN
with
\EQ
\alpha=-\onethird\int_0^t\overline{\uu(t)\cdot\oo(t')}\;\dd t'
\approx-\onethird\tau_{\rm cor}\overline{\uu\cdot\oo},
\label{FOSA_alpha}
\EN
\EQ
\eta_{\rm t}=\onethird\int_0^t\overline{\uu(t)\cdot\uu(t')}\;\dd t'
\approx\onethird\tau_{\rm cor}\overline{\uu^2},
\label{FOSA_eta}
\EN
where $\tau_{\rm cor}$ is the correlation time.
When $t$ becomes large, the main contribution to these two expressions
comes only from late times, $t'$ close to $t$, because the contributions
from early times are no longer strongly correlated with $\uu(t)$.
By using FOSA we have thus solved the problem of expressing
$\meanemf$ in terms of the mean field. The turbulent
transport coefficients $\alpha$ and $\eta_{\rm t}$ depend, respectively,  
on the helicity and the energy density of the turbulence.

One must however point out the following caveat to
the applicability of FOSA in case of large $R_{\rm m}$.
First, note that even if 
$\tau_{\rm cor}u_{\rm rms}k_{\rm f} \ll 1$, one can have 
$R_{\rm m} = (\tau_{\rm cor}u_{\rm rms}k_{\rm f})/
(\eta \tau_{\rm cor}k_{\rm f}^2) \gg 1$, because
the diffusion time $(\eta \tau_{\rm cor}k_{\rm f}^2)^{-1}$ 
can be much larger than the 
correlation time of the turbulence. As we have already discussed
in \Sec{SSD}, when $R_{\rm m} > R_{\rm crit}\sim 30$,
small scale dynamo action may take place (depending on the value
of $P_{\rm m}$) to produce exponentially growing fluctuating fields,
independent of the mean field. So the basic assumption
of FOSA of small $\bb$ relative to $\meanBB$ will be rapidly
violated and the $\uu\times\bb$ term in \Eq{Induction_flucts}
cannot be neglected.
Nevertheless, the functional form of the
expressions for the turbulent transport coefficients
obtained using FOSA seem to be not too different from
that found in simulations.
For example, it is likely that strong fluctuations produced by small
scale dynamo action do not correlate well with $\uu$ in
$\overline{\uu\times\bb}$, so they would not contribute
to $\meanemf$.
This interpretation will be developed further in \Sec{MinimalTauApprox}
on the $\tau$ approximation, which works specifically only
with those parts that do correlate.

\subsubsection{MTA -- the `minimal' $\tau$ approximation}
\label{MinimalTauApprox}

The `minimal' $\tau$ approximation is a simplified version of the
$\tau$ approximation as it has been introduced by Orszag \cite{Orszag70}
and used by Pouquet, Frisch and L\'eorat \cite{PFL76} in the context
of the Eddy Damped Quasi Normal Markovian (EDQNM) approximation.
In that case a damping term is introduced in order to express
fourth order moments in terms of third order moments.
In the $\tau$ approximation, as introduced by
Vainshtein and Kitchatinov \cite{VK83} and
Kleeorin and Rogachevskii \cite{KR90,KMR96}, one 
approximates triple moments in terms of quadratic moments
via a wavenumber-dependent relaxation time $\tau(k)$.
The `minimal' $\tau$ approximation (MTA), as it is introduced by Blackman and
Field \cite{BF02b}, is applied in real space in the two-scale approximation.
We will refer to both the above types of closures (where triple moments
are approximated in terms of quadratic moments and
a relaxation time $\tau$) as the 'minimal' $\tau$ approximation or MTA.

There are some technical similarities between FOSA and the
minimal $\tau$ approximation.
The main advantage of the $\tau$ approximation is that the fluctuations
do {\it not} need to be small and so the triple correlations are no longer
neglected.
Instead, it is assumed (and this can be and has been tested using simulations) that
the one-point triple correlations are proportional to the quadratic
correlations, and that the proportionality coefficient is an inverse
relaxation time that can in principle be scale (or wavenumber) dependent.

In this approach, one begins by considering the time derivative
of $\meanemf$ \cite{BF02b,RKR03},
\EQ
{\partial\meanemf\over\partial t}
=\overline{\uu\times\dot{\bb}}
+\overline{\dot{\uu}\times\bb},
\label{dEMFdt}
\EN
where a dot denotes a time derivative.
For $\dot{\bb}$, we substitute \Eq{Induction_flucts} and
for $\dot{\uu}$, we use the Euler equation
for the fluctuating velocity field,
\EQ
{\partial\uu\over\partial t}
= -{1\over\rho_0}\nab{p} +\ff + \FF_{\rm vis} + \HH,
\EN
where $\HH = -\uu\cdot\nab \uu + \overline{\uu\cdot\nab \uu}$
is the nonlinear term, $\ff$ is a stochastic forcing
term (with zero divergence),
and $\FF_{\rm vis}$ is the viscous force. We have also assumed
for the present that there is no mean flow ($\meanUU=0$),
and have considered the kinematic regime where
the Lorentz force is set to zero (the latter assumption will
be relaxed in \Sec{tauApproxNonlin}). All these restrictions
can in principle be lifted (see below).
For an incompressible flow, the pressure term can be eliminated
in the standard fashion in terms of the projection operator.
In practice $\ff$ correlates only weakly with $\bb$ and may therefore
be neglected, as can be the small viscous term.
The only contribution to $\overline{\dot{\uu}\times\bb}$
comes from the triple correlation
involving $\bb$ and $\HH$. The $\overline{\uu\times\dot{\bb}}$
term however has non-trivial contributions. We get
\EQ
{\partial\meanemf\over\partial t}
=\tilde{\alpha}\,\meanBB-\tilde{\eta}_{\rm t}\,\meanJJ
-{\meanemf\over\tau},
\label{demfdt}
\EN
where the last term subsumes the effects of all triple correlations, and
\EQ
\tilde\alpha=-\onethird\overline{\uu\cdot\oo}\quad\mbox{and}\quad
\tilde\eta_{\rm t}=\onethird\overline{\uu^2}
\quad\mbox{(kinematic theory)}
\label{TTau_alpbeta}
\EN
are coefficients that are closely related to the usual
$\alpha$ and $\eta_{\rm t}$ coefficients in \Eq{alpha_eta}.
We recall that in this {\it kinematic} calculation the Lorentz force
has been ignored.
Its inclusion (\Sec{tauApproxNonlin}) turns out to be extremely important:
it leads to the emergence of a small scale magnetic correction term in
the expression for $\tilde\alpha$; see \Eq{alptilde+etatilde} below.

One normally neglects the explicit time derivative of $\meanemf$
\cite{KMR96,RKR03}, and
arrives then at almost the same expression as \Eq{alpha_eta}.
The explicit time derivative can in principle be kept
\cite{BF02b}, although it becomes unimportant on time scales long
compared with $\tau$; see also Refs~\cite{BraKapMoh03,BF03} for
applications to the passive scalar problem and numerical tests.
In comparison with \Eq{alpha_eta_kernel}, we note that if one assumes
$\hat\alpha(t-t')$ and $\hat\eta_{\rm t}(t-t')$ to be proportional
to $\exp[-(t-t')/\tau]$ for $t>t'$ (and zero otherwise), one recovers
\Eq{demfdt} with the relaxation time $\tau$ playing now the role of a
correlation time.

Recently, R\"adler \& Rheinhardt \cite{RR05} have pointed out that MTA does
not reduce to FOSA even when FOSA is applicable, i.e.\ when either the
magnetic diffusivity is large or the correlation time small.
This conflict becomes particularly apparent when considering the high
conductivity limit (so the correlation time should be short for FOSA
to be valid).
In this case FOSA predicts that $\alpha$ and $\eta_{\rm t}$ depend on the
time integral over
temporal two-point correlation functions \cite{KR80}, while the MTA results
depend on the spatial two-point correlation function at a single time,
multiplied by a relaxation time.
Only in some special cases can these two quantities 
be shown to be equivalent.
On the other hand, when the viscosity is large, MTA predicts an explicit
dependence on viscosity that is not recovered under FOSA.
However, in the momentum equation there is also the forcing term whose
correlation with the magnetic field may balance that with the viscous
term \cite{RR05}.
Nevertheless, in general one cannot regard MTA as an `approximation', but
rather as a closure hypothesis that captures in a conceptually
straightforward way a number of turbulence effects, leaving the relaxation
time $\tau$ (or the Strouhal number, which is $\tau$ normalized by the
turnover time) as a free parameter.
The strongest support for MTA comes from turbulence simulations that
confirm the assumed relation between quadratic and triple correlations
and that show Strouhal numbers of the order of unity for a range
of applications \cite{BraKapMoh03,Bran+Sub05}.

We will return to the $\tau$ approximation further below
(\Sec{tauApproxNonlin}) in connection with calculating nonlinear effects
of the Lorentz force and with numerical verifications using turbulence
simulations.

\subsection{Transport coefficients from simulations}
\label{SimulationsTransport}

The main advantage of using simulations is that no approximations need to
be made other than the restriction to only moderate values of the magnetic
Reynolds number.
Most notably, this approach allows the determination of transport
coefficients in inhomogeneous systems in the presence of boundaries.
This is important in the case of the sun, where there is a relatively
sharp transition from the convection zone to the neighboring overshoot
layers.

\subsubsection{Measuring the $\alpha$ tensor}

As a preliminary step, it is useful to restrict oneself to the assumption
of an isotropic $\alpha$, ignoring also turbulent diffusion.
In that case one has $\meanemf=\alpha\meanBB$, and so one can calculate
$\alpha=\meanemf\cdot\meanBB/\meanBB^2$ as a function of
position and time.
Next, one can allow for a contribution of the form
$\meanemf=...+\ggamma\times\meanBB$, which is also called a pumping term,
because it describes the advection of mean field with the effective
velocity $\ggamma$.
It is long known that this effect expels mean magnetic field from regions
of strong turbulence \cite{Rae69,RS75}, which is also the reason why this
effect is sometimes referred to as turbulent diamagnetic effect.
The components of $\ggamma$ are related to the antisymmetric part of the
$\alpha$ tensor via $\gamma_i=-\half\epsilon_{ijk}\alpha_{jk}$.
Computationally, these components can be extracted from
$\meanemf\times\meanBB$ as
$\gamma_i={\sf M}_{ij}^{-1}(\meanemf\times\meanBB)_j$, where
the matrix ${\sf M}_{ij}=\meanB_i\meanB_j-\delta_{ij}\meanBB^2$
has to be inverted at each meshpoint.

Alternatively, one can assume that the $\alpha$ tensor is dominated by
certain components, e.g.\ its diagonal components.
For example in connection with a local cartesian model of accretion
disc turbulence the toroidal (or $y$) component
$\alpha_{yy}={\cal E}_y/\meanB_y$ has been calculated \cite{BNST95}.
We return to this in \Sec{SShearingSheet}.

In the cases discussed so far, we have to rely on the successful operation
of what corresponds to a mean field dynamo, as was indeed the case in
the accretion disc calculations.
However, another obvious method for calculating the $\alpha$ effect is to use
a simulation with
an imposed magnetic field, $\BB_0$, and to determine numerically the resulting
electromotive force, $\meanemf$.
Here it is natural to define the average as a full volume average.
For a periodic box, $\meanBB=\BB_0$.
Since such averages no longer depend on the space coordinate, there is
no mean current, i.e.\ $\meanJJ=\nab\times\meanBB=0$.

The main conclusions obtained from this approach applied to stratified
convection is that the functional
form of \Eq{KrauseEqn} is basically verified.
In particular, $\alpha$ has a positive maximum in the upper part of the
convection zone (in the northern hemisphere), changes sign with depth
\cite{BNPST90,OSB01}, varies with latitude
as expected from this equation, and is largest at high latitudes
\cite{OSB01}.
The simulations also confirm the presence of downward turbulent pumping.
Indeed, animations show that flux
tubes are regularly being entrained by strong downdrafts, then pushed
downward and amplified as the result of stretching and compression
\cite{BJNRST96,NBJRRST92,Bran94}.
The end result is a strong magnetic field at the bottom of the convection
zone where the field is expected to undergo further stretching by
differential rotation \cite{TobiasBrummellCluneToomre98}.
Recent studies have allowed a more quantitative description of
the pumping effect and the associated pump velocity
\cite{Tobias01,Brummell02}.

\begin{figure}[t!]\begin{center}
\includegraphics[width=.99\textwidth]{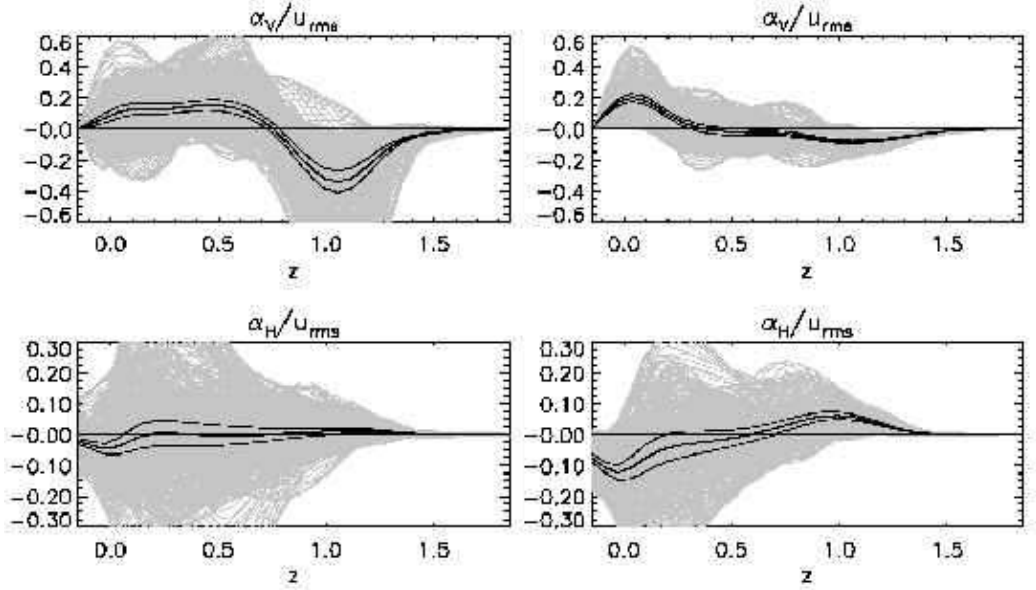}
\end{center}\caption[]{
Vertical and horizontal components of the $\alpha$ effect.
The gray lines represent different times, the thick line
is the time average, and the surrounding thin curves
indicate the error of the mean.
The simulation is carried out on the southern hemisphere.
The convection zone proper is in the range $0<z<1$, where
$z$ denotes depth.
The lower overshoot layer is in $z>1$ and the top overshoot
layer is in $z<0$.
In the upper parts of the convection zone ($0<z<0.6$) the vertical component
of $\alpha_{\rm V}\equiv\alpha_{zz}$ is positive (upper panels)
while the horizontal components,
$\alpha_{\rm H}\equiv\alpha_{xx}=\alpha_{yy}$
are negative (lower panels).
The left and right hand columns are for simulations with different
angular velocity: slower on the left (the Taylor number is $2\times10^3$)
and faster on the right (the Taylor number is $10^4$).
Courtesy M.\ Ossendrijver \cite{OSB01}.
}\label{Fossen01}\end{figure}

\begin{figure}[t!]\begin{center}
\includegraphics[width=.8\textwidth]{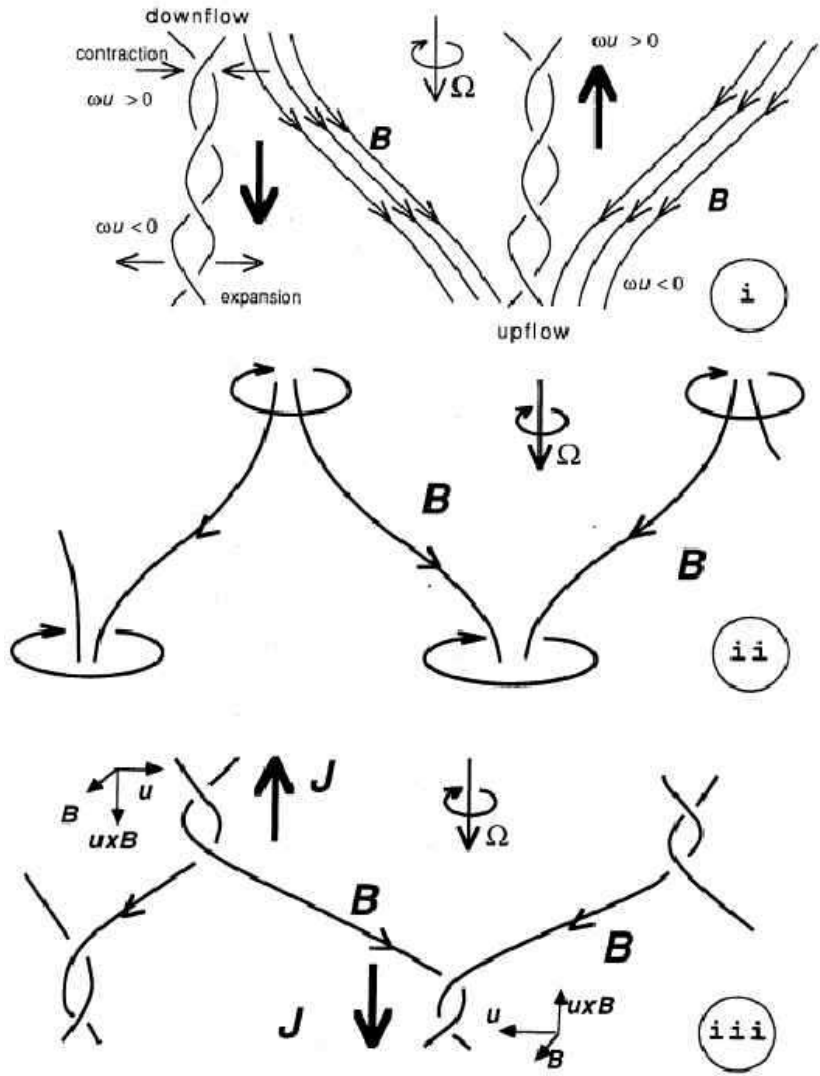}
\end{center}\caption[]{
Sketch showing the twisting of vertical magnetic field lines
by downdrafts.
The resulting electromotive force, $\emf=\uu\times\bb$, points in the
direction of the mean field, giving a positive $\alpha_{\rm V}$ in the
southern hemisphere and a negative $\alpha_{\rm V}$ in the northern
hemisphere.
Adapted from Ref.~\cite{BNPST90}.
}\label{sketchBNPST90}\end{figure}

Two surprising results emerged from the simulations.
In convection the $\alpha$ effect is extremely anisotropic
with respect to the vertical direction such that the
diagonal components of $\alpha_{ij}$ can even change sign.
While the horizontal components, $\alpha_{xx}$ and $\alpha_{yy}$, show the
expected sign and are roughly a negative multiple of the kinetic
helicity, the vertical component, $\alpha_{zz}$, shows
invariably the opposite sign \cite{BNPST90,OSB01};
see \Fig{Fossen01}.
This peculiar result can be understood by noting that vertical
field lines are wrapped around each other by individual downdrafts,
which leads to field line loops oppositely oriented than if they
were caused by an expanding updraft.

In \Fig{sketchBNPST90} we show the field line topology relevant to
the situation in the southern hemisphere.
The degree of stratification is weak, so the downdrafts at the
top look similar to the updrafts at the bottom (both are indicated
by two swirling lines).
(i) At the top and bottom boundaries the magnetic field is concentrated
in the intergranular lanes which correspond to downdrafts at the
top and updrafts at the bottom.
(ii) This leads to a clockwise swirl both at the top and at the bottom (but
counter-clockwise in the northern hemisphere); see the second panel.
(iii) This in turn causes left-handed current helicity in the upper parts and
right-handed current helicity in the lower parts, so one might expect
that $\alpha$ is negative in the upper parts and positive in the lower
parts. This is however not the case.
Instead, what really matters is $\emf=\uu\times\bb$, where $\uu$ is
dominated by converging motions (both at the top and the bottom).
This, together with $\bb$ winding in the counter-clockwise direction
around the downdraft and in the clockwise direction in the updraft, causes
$\uu\times\bb$ to point in the direction of $\meanBB$ at the top (so
$\alpha_{zz}$ is positive) and in the opposite direction at the bottom
(so $\alpha_{zz}$ is negative).
Originally, this result was only obtained for weak stratification
\cite{BNPST90}, but meanwhile it has also been confirmed for strong
stratification \cite{OSB01}.
We reiterate that a qualitatively similar result has also been obtained
in analytic calculations of the $\alpha$ effect from supernova-driven
expanding shells in the stratified galactic disc \cite{Ferriere92} and
in FOSA calculations of stratified turbulence \cite{RK93}.

\begin{figure}[t!]\begin{center}
\includegraphics[width=.99\textwidth]{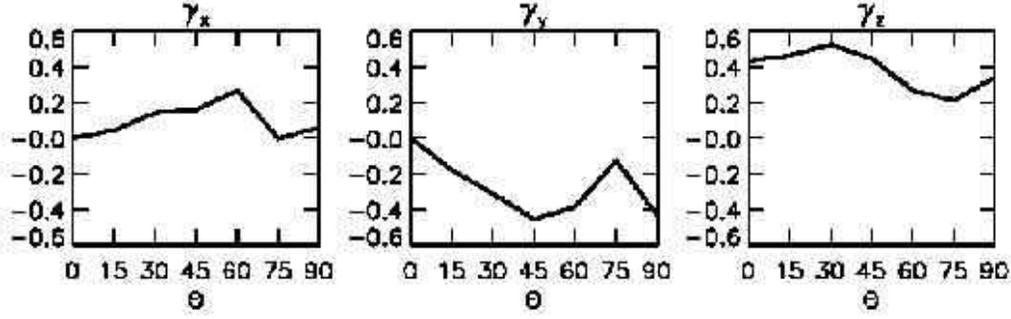}
\end{center}\caption[]{
The latitudinal dependence of the three components of the pumping velocity
($\theta=0$ corresponds to the south pole and $\theta=90$ to the equator).
The vertical pumping velocity ($\gamma_z$, right-most panel) is positive,
corresponding to downward pumping, and almost independent of latitude.
The two horizontal components of $\ggamma$ vanish at the poles.
The longitudinal pumping velocity ($\gamma_y$, middle panel) is
negative, corresponding to retrograde pumping, and the latitudinal
pumping velocity ($\gamma_x$, left panel) is positive, corresponding
to equatorward pumping.
Adapted from Ref.~\cite{OSBR02}.
}\label{Fossen02}\end{figure}

The other surprising result is that the
turbulent pumping is not necessarily restricted to the
vertical direction, but it can occur in the other two directions as well.
In general one can split the $\alpha_{ij}$ tensor into symmetric
and antisymmetric components, i.e.\
$\alpha_{ij}=\alpha_{ij}^{\rm(S)}+\alpha_{ij}^{\rm(A)}$, where the antisymmetric
components can be expressed in the form
\EQ
\alpha_{ij}^{\rm(A)}=-\epsilon_{ijk}\gamma_k,
\EN
where $\ggamma$ is the pumping velocity.
We recall that $\alpha_{ij}^{\rm(A)}\meanB_j=(\ggamma\times\meanBB)_i$,
which looks like the induction term $\meanUU\times\meanBB$,
so $\ggamma$ acts like an advection velocity.
Simulations show that $\ggamma$ has a component pointing
in the retrograde direction; see \Fig{Fossen02} and Ref.~\cite{OSBR02}.
We shall return to the theory of this term in
\Sec{HelicityInducedByRotation}; see \Eq{gamma_toroidal}.

The latitudinal component of the pumping velocity points toward
the equator and has been invoked to explain the equatorward migration
seen in the butterfly diagram \cite{Kitchatinov91}.
The equatorward pumping was found to act mostly on the toroidal
component of the mean field while the poloidal field was found to
experience a predominantly poleward pumping velocity.
This result has also been confirmed using simulations of turbulent
convection \cite{OSBR02}.

A general problem with all these calculations is that,
as the Reynolds number is increased, the fluctuations in $\meanemf$
become large and long integration times are necessary \cite{CH96}.
This is related to the problem that for large values of $R_{\rm m}$, small scale 
dynamo action becomes possible and, unless the imposed field is strong
enough, the resulting values of $\meanemf$ are only weakly linked to $\BB_0$.
Indeed, $\meanemf$ will be dominated by a rather strong noise component,
making it necessary to calculate long time averages to extract a small
average from the strongly fluctuating component.

\subsubsection{Test field procedure with finite gradients}

A general procedure for determining the full $\alpha_{ij}$ and
$\eta_{ijk}$ tensors from a simulation is to calculate the electromotive
force after applying test fields of different directions and with
different gradients \cite{Schrinner}.
There are altogether $9+27$ unknowns, if the mean field can vary in all
three directions, or $9+18$ unknowns, if it can vary only in 2 directions
(as is the case for toroidal or $y$ averages, for example).
The idea is to calculate the emf,
\EQ
\meanemf^{(q)}=\overline {\uu\times\bb^{(q)}},
\label{emftest}
\EN
for the excess magnetic fluctuations, $\bb^{(q)}$, that are due
to a given test field $\meanBB^{(q)}$.
This requires solving simultaneously a set of equations of the form
\EQ
{\partial\bb^{(q)}\over\partial t}=\nab\times\left[
\left(\meanUU+\uu\right)\times\meanBB^{(q)}
\right]+\eta\nabla^2\bb^{(q)}+\GG
\EN
for each test field $\meanBB^{(q)}$.
Here, the mean flow $\meanUU$ has been retained and
$\GG=\nab\times[\uu\times\bb^{(q)}-\overline{\uu\times\bb^{(q)}}]$
is a term nonlinear in the fluctuation.
This term would be ignored in the first order smoothing approximation,
but it can be kept in a simulation if desired.
For two-dimensional averages, for example, one has $9+18=27$ unknowns,
so one needs 9 test fields $\meanBB^{(q)}$ to calculate 9 vectors
$\meanemf^{(q)}$.
Expressing ${\cal E}_i^{(q)}$ in the form
\EQ
\meanemf^{(q)}=\alpha_{ij}\meanB^{(q)}_j+\eta_{ijk}\meanB_{j,k}^{(q)},
\EN
one arrives at a system of 27 equations for the 27 unknowns.

By choosing 3 of the test fields to be constants, one can first solve
for the 9 unknowns $\alpha_{ij}$.
The remaining coefficients in $\eta_{ijk}$ can then be obtained by
choosing test fields that vary linearly as a function of space.
This type of analysis has been applied successfully to laminar stationary
convection in a sphere exhibiting a dynamo effect \cite{Schrinner}.
In this special case no matrix inversion is necessary.
However,
in addition to problems with boundary conditions, there is the difficulty
that $\alpha_{ij}$ and $\eta_{ijk}$ may be wavenumber dependent, so it
may be better to choose only test fields with similar spatial variation
(or wavenumber).
In that case one needs to invert simple $2\times2$ matrices with coefficients
that depend on the test fields and their gradients.

Another remotely related method is to use a time-dependent magnetic
field in a successful turbulent (nonstationary) dynamo simulation.
The hope is here that the resulting mean magnetic field covers a
substantial fraction of the parameter space allowing one to calculate
meaningful moments of the form $\bra{\overline{\cal E}_i\meanB_j}$ and
$\bra{\overline{\cal E}_i\meanB_{j,k}}$.
Using a general representation of $\meanemf$ of the form \Eq{EiGeneral}
allows one to calculate the transport coefficients $\alpha_{ij}$ and
$\eta_{ijk}$ by inverting suitable correlation matrices.
This method has been applied with modest success to the problem of large
scale field generation in a local model of accretion disc turbulence
\cite{BranSok02,Kowaletal05}.

\subsubsection{Comparison of simulations with theory}

Whenever a meaningful comparison between simulations and theory
(FOSA) is possible, the agreement can be quite remarkable.
An example where this is the case is laminar convection in a
rotating spherical shell, where the velocity field from the simulation
has been inserted into the corresponding mean-field expressions
\cite{Schrinner}.

In the case of turbulent convection, where only the turbulent rms
velocity is used to scale theory to simulations, the agreement is
merely on a qualitative level.
One obvious property of simulated values of $\alpha$ is the high
degree of fluctuations \cite{CH96}.
This is because here mean fields are defined as spatial averages.
Fluctuations of the turbulent transport coefficients are normally
ignored when ensemble averages are used \cite{Hoyng87}.
As a general trend one can note that theory tends to overestimate
the magnitudes of $\alpha$, $\gamma$, and $\eta_{\rm t}$.
This can partly be explained as a consequence of catastrophic
(i.e.\ magnetic Reynolds number dependent) quenching \cite{OSB01}.
This will be discussed in detail later in this review; see
\Sec{SDynamicalQuenching}.
For now, let us note that the catastrophic quenching became particularly
obvious in simulations of isotropic homogeneous turbulence in a fully
periodic box \cite{VC92}.
Agreement between theory and simulations was only achieved when the
dynamical quenching formalism was used \cite{BB02}.
Historically, of course, neither the dynamical quenching nor the
corresponding helicity fluxes were known.
Therefore, any agreement between simulations and theory was only
a consequence of having adopted sufficiently small a magnetic Reynolds number
or, possibly, of magnetic or current helicity fluxes having been quite efficient,
so that the effect of dynamical quenching became less restrictive.

An outstanding question where much more work needs to be done
is indeed the issue of current helicity fluxes.
They will be discussed in more detail below (Sections \ref{TurbulenceAndShear},
\ref{OpenBoundaries}, \ref{MagneticHelicityLosses}, \ref{nonlinearflux}).
Their calculation is just as important as that of the other transport
coefficients, because these fluxes help alleviating the otherwise
catastrophic quenching.

\subsection{$\alpha^2$ and $\alpha\Omega$ dynamos: simple solutions}
\label{solutions_trends}

For astrophysical purposes one is usually interested in solutions
in spherical or oblate (disc-like) geometries.
However, in order to make contact with turbulence simulations in a periodic
box, solutions in simpler cartesian geometry can be useful.
Cartesian geometry is also useful for illustrative purposes.
In this subsection we review some simple cases.

\begin{figure}[t!]\begin{center}
\includegraphics[width=.99\textwidth]{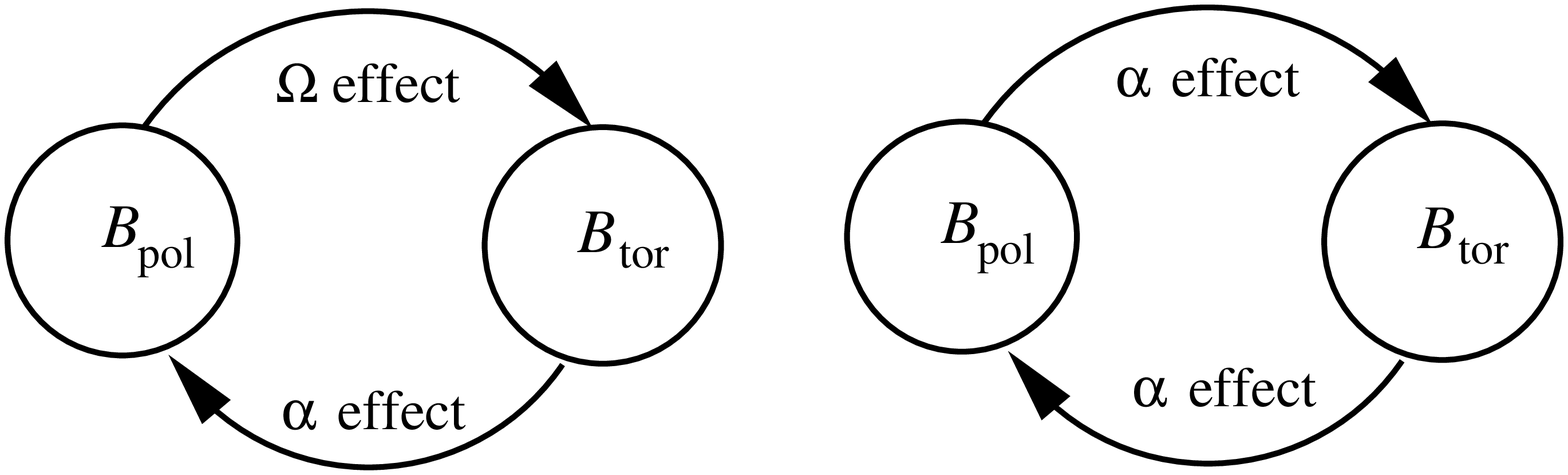}
\end{center}\caption[]{
Mutual regeneration of poloidal and toroidal fields in the case of the
$\alpha\Omega$ dynamo (left) and the $\alpha^2$ dynamo (right).
}\label{BpolBtor}\end{figure}

Mean field dynamos are traditionally divided into two groups;
$\alpha\Omega$ and $\alpha^2$ dynamos.
The $\Omega$ effect refers to the amplification of the toroidal
field by shear (i.e.\ {\it differential}\/ rotation) and its importance
for the sun was recognized very early on.
Such shear also naturally occurs in disk galaxies, since they are
differentially rotating systems \cite{Babcock}.
However, it is still necessary to regenerate the poloidal field.
In both stars and galaxies the $\alpha$ effect is the prime candidate.
This explains the name $\alpha\Omega$ dynamo;
see the left hand panel of \Fig{BpolBtor}.
However, large scale magnetic fields can also be generated by the $\alpha$
effect alone, so now also the toroidal field has to be generated by the
$\alpha$ effect, in which case one talks about an $\alpha^2$ dynamo;
see the right hand panel of \Fig{BpolBtor}.
(The term $\alpha^2\Omega$ model is discussed at the end of \Sec{Sao_model}.)

\subsubsection{$\alpha^2$ dynamo in a periodic box}
\label{Sa2_model}

\begin{figure}[t!]\begin{center}
\includegraphics[width=.8\textwidth]{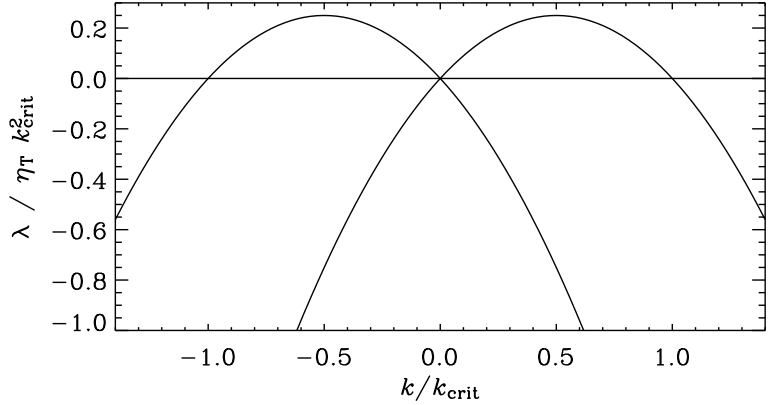}
\end{center}\caption[]{
Dispersion relation for $\alpha^2$ dynamo,
where $k_{\rm crit}=\alpha/\eta_{\rm T}$.
}\label{alpha2_disper}\end{figure}

We assume that there is no mean flow, i.e.\ $\meanUU=0$, and that
the turbulence is homogeneous, so
that $\alpha$ and $\eta_{\rm t}$ are constant.
The mean field induction equation then reads
\EQ
{\partial\meanBB \over \partial t}=
\alpha\nab\times\meanBB+\eta_{\rm T}\nabla^2\meanBB,
\quad\nab\cdot\meanBB=0,
\EN
where $\eta_{\rm T}=\eta+\eta_{\rm t}$ is the sum of microscopic
and turbulent magnetic diffusivity.
We can seek solutions of the form
\EQ
\meanBB(\xx)=
\mbox{Re}\left[\hat{\BB}(\kk)\,\exp(\ii\kk\cdot\xx+\lambda t)\right].
\EN
This leads to the eigenvalue problem
$\lambda\hat{\BB}=\alpha\ii\kk\times\hat{\BB}-\eta_{\rm T} k^2\hat{\BB}$,
which can be written in matrix form as
\EQ
\lambda\hat{\BB}=\pmatrix{
-\eta_{\rm T}k^2& -\ii\alpha k_z  &  \ii\alpha k_y \cr
 \ii\alpha k_z  &-\eta_{\rm T} k^2& -\ii\alpha k_x \cr
-\ii\alpha k_y  &  \ii\alpha k_x  &-\eta_{\rm T} k^2
}\hat{\BB}.
\EN
This leads to the dispersion relation, $\lambda=\lambda(\kk)$, given by
\EQ
(\lambda+\eta_{\rm T} k^2)
\left[(\lambda+\eta_{\rm T} k^2)^2-\alpha^2 k^2\right]=0,
\EN
with the three solutions
\EQ
\lambda_0=-\eta_{\rm T} k^2,\quad
\lambda_\pm=-\eta_{\rm T} k^2\pm|\alpha k|.
\EN
The eigenfunction corresponding to the eigenvalue
$\lambda_0=-\eta_{\rm T} k^2$ is proportional to $\kk$,
but this solution is incompatible
with solenoidality and has to be dropped.
The two remaining branches are shown in \Fig{alpha2_disper}.

Unstable solutions ($\lambda>0$) are possible for
$0 < \alpha k < \eta_{\rm T} k^2$. For $\alpha > 0$ this corresponds to
the range
\EQ
0<k<\alpha/\eta_{\rm T}\equiv k_{\rm crit}.
\EN
For $\alpha < 0$, unstable solutions are obtained for $k_{\rm crit} < k < 0$.
The maximum growth rate is at
\EQ
k_{\max}=\half k_{\rm crit}=\alpha/(2\eta_{\rm T})
\quad\quad\mbox{(maximum growth rate)}.
\EN
Such solutions are of some interest, because they have been seen
as an additional hump in the magnetic energy spectra from
fully three-dimensional turbulence simulations
(\Secs{SBeltrami}{AlphaVersusInverseCascade}).

Linear theory is only applicable as long as the magnetic field
is weak, but qualitatively one may expect that in the nonlinear
regime $\alpha$ becomes reduced (quenched), so $k_{\rm crit}$
decreases and only larger scale magnetic fields can be maintained.
This is indeed seen in numerical simulations \cite{B01}; see also
\Sec{SimulationsHelicalDynamos}.

\subsubsection{$\alpha\Omega$ dynamo in a periodic box}
\label{Sao_model}

Next we consider the case with linear shear, and assume
$\meanUU=(0,Sx,0)$, where $S=\mbox{const}$.
This model can be applied as a local model to both accretion discs ($x$
is radius, $y$ is longitude, and $z$ is the height above the midplane)
and to stars ($x$ is latitude, $y$ is longitude, and $z$ is radius).
For keplerian discs, the shear is $S=-{3\over2}\Omega$, while for
the sun (taking here only radial differential rotation into account)
$S=r\partial\Omega/\partial r\approx+0.1\Omega_\odot$ near the equator.

\begin{figure}[t!]\begin{center}
\includegraphics[width=.8\textwidth]{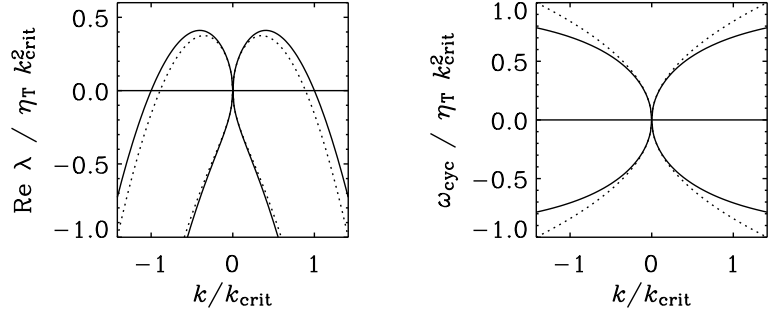}
\end{center}\caption[]{
Dispersion relation for $\alpha^2\Omega$ dynamo with
$\alpha k_{\rm crit}/S=0.35$.
The dotted line gives the result for the $\alpha\Omega$ approximation
\Eqs{lam_approx}{om_approx}.
The axes are normalized using $k_{\rm crit}$ for the full
$\alpha^2\Omega$ dynamo equations.
}\label{alpha2omega_disper}\end{figure}

For simplicity we consider axisymmetric solutions, i.e.\ $k_y=0$.
The eigenvalue problem takes then the form
\EQ
\lambda\hat{\BB}=\pmatrix{
-\eta_{\rm T}k^2& -\ii\alpha k_z  &      0         \cr
 \ii\alpha k_z+S&-\eta_{\rm T} k^2& -\ii\alpha k_x \cr
      0         &  \ii\alpha k_x  &-\eta_{\rm T} k^2
}\hat{\BB},
\label{MatrixaO}
\EN
where $\eta_{\rm T}=\eta+\eta_{\rm t}$ and $\kk^2=k_x^2+k_z^2$.
The dispersion relation is now
\EQ
(\lambda+\eta_{\rm T} k^2)\left[(\lambda+\eta_{\rm T} k^2)^2
+\ii\alpha Sk_z-\alpha^2 k^2\right]=0,
\label{disperaO}
\EN
with the solutions
\EQ
\lambda_\pm=-\eta_{\rm T} k^2\pm(\alpha^2 k^2-\ii\alpha Sk_z)^{1/2}.
\EN
Again, the eigenfunction corresponding to the eigenvalue
$\lambda_0=-\eta_{\rm T} k^2$ is not compatible
with solenoidality and has to be dropped.
The two remaining branches are shown in \Fig{alpha2omega_disper},
together with the {\it approximate} solutions
(valid for $\alpha k_z/S\ll1$)
\EQ
\mbox{Re}\lambda_\pm\approx-\eta_{\rm T} k^2\pm|\half\alpha Sk_z|^{1/2},
\label{lam_approx}
\EN
\EQ
\mbox{Im}\lambda_\pm\equiv-\omega_{\rm cyc}\approx\pm|\half\alpha Sk_z|^{1/2},
\label{om_approx}
\EN
where we have made use of the fact that $\ii^{1/2}=(1+\ii)/\sqrt{2}$.

Sometimes the term $\alpha^2\Omega$ dynamo is used to emphasize that the
$\alpha$ effect is not neglected in the generation of the toroidal field.
This approximation, which is sometimes also referred to as the $\alpha\Omega$
approximation, is generally quite good provided $\alpha k_z/S\ll1$.
However, it is important to realize that this approximation can only be
applied to axisymmetric solutions \cite{Rae86}.

\subsubsection{Eigenfunctions, wave speed, and phase relations}
\label{PhaseRelations}

We now make the $\alpha\Omega$ approximation and consider the
marginally excited solution ($\mbox{Re}\lambda=0$), which can be written as
\EQ
\mean{B_x}=B_0\sin k_z(z-ct),\quad
\mean{B_y}=\sqrt{2}\,B_0\left|{c\over\alpha}\right|
\sin[k_z(z-ct)+\varphi],
\EN
where $B_0$ is the amplitude (undetermined in linear theory),
and $c=\omega_{\rm cyc}/k_z$ is the phase speed of the dynamo wave,
which is given by
\EQ
c={\alpha S\over|2\alpha S k_z|^{1/2}}
=\pm\eta_{\rm T} k_z,
\label{PhaseSpeed}
\EN
where the upper (lower) sign applies when $\alpha S$ is positive (negative).
The sign of $c$ gives the direction of propagation of the dynamo wave;
see \Tab{Tpropagation} for a summary of the propagation directions in
different settings.

\begin{table}[t!]\caption{
Summary of propagation directions and phase relation
for $\alpha\Omega$ dynamos.
}\vspace{12pt}\centerline{\begin{tabular}{lcccccccccc}
object    &$\alpha$& $S$ &$\varphi$ & $c$ & wave propagation 
& $|\omega_{\rm cyc}|\,\Delta t$ \\
\hline
disc      & $-$ &  $-$  & $-3\pi/4$ & $+$ & away from midplane & $-3\pi/4$ \\
disc/star?& $+$ &  $-$  & $+3\pi/4$ & $-$ & equatorward        & $-3\pi/4$ \\
star?     & $-$ &  $+$  &  $-\pi/4$ & $-$ & equatorward        & $+\pi/4$  \\
star      & $+$ &  $+$  &  $+\pi/4$ & $+$ & poleward           & $+\pi/4$  \\
\label{Tpropagation}\end{tabular}}\end{table}

An important property of the $\alpha\Omega$ dynamo solutions that
can be read off from the plane wave solutions is the phase shift
of $\pm{3\over4}\pi$ (for $S<0$) and $\pm\pi/4$ (for $S>0$)
between the poloidal and toroidal fields.
It is customary \cite{Yoshimura76,Stix76} to quote instead the normalized
time delay $|\omega_{\rm cyc}|\,\Delta t=\varphi\,\sgn(c)$, by which
the toroidal field lags behind the radial field.
These values are given in the last column of \Tab{Tpropagation}.
Note that the temporal phase shift only depends on the sign of the shear
$S$ and not on $\alpha$.

\subsubsection{Excitation conditions in a sphere}
\label{ExcitationConditionsSphere}

\begin{figure}[t!]\begin{center}
\includegraphics[width=.65\textwidth]{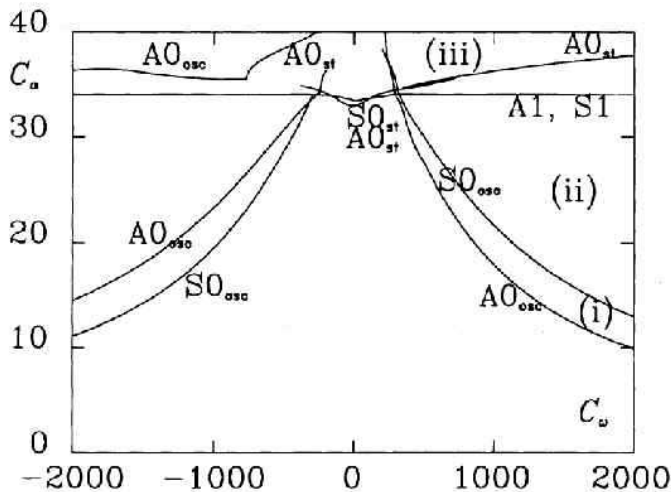}
\end{center}\caption[]{
Critical values of $C_\alpha$ versus $C_\omega$ for a dynamo
in a spherical shell. Note that near $C_\omega=200$ the nonaxisymmetric
modes {\sf S1} and {\sf A1} are more easily excited than the
axisymmetric modes {\sf S0} and {\sf A0}.
Here, $C_\omega$ (called $C_\Omega$ in the rest of this review) is defined
such that it is positive when $\partial\Omega/\partial r$ is negative,
and vice versa.
Adapted from \cite{BKMMT89}.
}\label{BKMMT89a}\end{figure}

For applications to stars it is essential to employ spherical geometry.
Over the past three decades, a number of two-dimensional and three-dimensional
models have been presented \cite{SK69a,RS72,Koe73,BKMMT89}.  The dynamo is generally characterized by two dynamo numbers,
\EQ
C_\alpha=\alpha_0 R/\eta_{\rm T},\quad
C_\Omega=\Delta\Omega R^2/\eta_{\rm T},
\EN
where $\alpha_0$ and $\Delta\Omega$ are typical values of $\alpha$
and angular velocity difference across the sphere, and $R$ is the outer
radius of the sphere.
In \Fig{BKMMT89a} we show the critical values of $C_\alpha$ (above
which dynamo action is possible) for different values of $C_\Omega$
using error function profiles,
\EQ
f_\pm(r;r_0,d)=\half\left\{1\pm{\rm erf}\left[(r-r_0)/d\right]\right\},
\EN
for $\alpha(r,\theta)=\alpha_0 f_+(r;r_\alpha,d_\alpha)\cos\theta$ and
$\Omega(r)=\Delta\Omega f_-(r;r_\Omega,d_\Omega)$, just as in the early
work of Steenbeck and Krause \cite{SK69a}, who used $r_\alpha=0.9$,
$r_\Omega=0.7$, and $d_\alpha=d_\Omega=0.075$.
On $r=R$ the field is matched to a potential field.
A detailed presentation of the induction equation in spherical harmonics
with differential rotation, meridional circulation, anisotropic $\alpha$
effect and a number of other effects is given by R\"adler \cite{Rae73}.

The solutions are classified by the symmetry properties about the equator
({\sf S} and {\sf A} for symmetric and antisymmetric fields, respectively), supplemented
by the spherical harmonic degree $m$ characterizing the number of nodes
in the azimuthal direction.
Note that for axisymmetric modes ({\sf S0} and {\sf A0}) the critical
value of $C_\alpha$ decreases if $C_\Omega$ increases, while for the
nonaxisymmetric modes (e.g.\ {\sf S1} and {\sf A1}) $C_\alpha$ is
asymptotically independent of $C_\Omega$.
This behavior for {\sf S0} and {\sf A0} is understandable because for
axisymmetric modes the excitation condition only depends on the product
of $\alpha$ effect and shear; see \Eq{lam_approx}.
For {\sf S1} and {\sf A1}, on the other hand, differential rotation
either makes the dynamo harder to excite (if $C_\Omega$ is small) or
it does not affect the dynamo at all (larger $C_\Omega$).
This is because when differential rotation winds up a nonaxisymmetric
field, anti-aligned field lines are brought close together \cite{Rae86}.
For sufficiently large values of $C_\Omega$ the field is expelled into
regions with no differential rotation ($\alpha^2$ dynamo) where the
dynamo is essentially independent of $C_\Omega$.

Generally, axisymmetric modes are easier to excite than nonaxisymmetric
modes.
There can be exceptions to this just at the junction between $\alpha^2$
and $\alpha\Omega$ dynamo behavior.
This is seen near $C_\omega=300$; see \Fig{BKMMT89a}.
Such behavior was first reported by Robert and Stix \cite{RS72}, and may
be important for understanding the occurrence of nonaxisymmetric fields
in active stars; see \Sec{StellarDynamos}.
Other potential agents facilitating nonaxisymmetric fields
include anisotropic \cite{Rae86b,Rae_etal90} and nonaxisymmetric
\cite{Moss_etal91,Big+Ruz04} forms of $\alpha$ effect and turbulent diffusivity.

Finally we note that for strong shear (large values of $|C_\Omega|$), and
$\alpha>0$ in the northern hemisphere, the most easily excited modes are
{\sf A0} (when $\partial\Omega/\partial r$ is negative, i.e.\ $C_\omega>0$
in \Fig{BKMMT89a}), and {\sf S0} (when $\partial\Omega/\partial r$
is positive).
This behavior changes the other way around, however, when the dynamo
operates in a shell whose ratio of inner to outer shell radius exceeds
a value of about 0.7 \cite{PHRoberts72,Covas_etal99}.
This is approximately the appropriate value for the sun, and it has indeed
long been recognized that the negative parity of the solar dynamo is not
always obtained from model calculations
\cite{RB95,Bonanno_etal02,DikpatiGilman01}.

\subsubsection{Excitation conditions in disc-like geometries}
\label{ExcitationConditionsDisc}

Another important class of astrophysical bodies are galactic discs as
well as discs around young stars, compact objects, or supermassive
black holes in the nuclei of active galaxies; see \Sec{AccretionDiscs}.
We consider the simplest form
of the axisymmetric mean field dynamo equations appropriate for a thin
disc \cite{RSS88},
\EQ
\dot{\meanB}_R = -(\alpha\meanB_{\phi})' + \eta_{\rm T} \meanB_R'', \quad
\dot{\meanB}_{\phi} = S\meanB_R + \eta_{\rm T} \meanB_{\phi}''.
\label{discdyn}
\EN
Here, radial derivatives have been neglected, so the problem has
become one-dimensional and can be solved separately at each radius.
Primes and dots denote $z$ and $t$ derivatives, respectively,
$\alpha=\alpha_0 f_\alpha(z)$ is a profile for $\alpha$ (asymmetric
with respect to $z=0$) with typical value $\alpha_0$,
$S = R\dd\Omega/\dd R$ is the radial shear in the disc,
and $(\meanB_R,\meanB_{\phi},\meanB_z)$ are the components
of the mean field $\meanBB$ in cylindrical coordinates.
On $z=\pm H$ one assumes vacuum boundary conditions which, in this
one-dimensional problem, reduce to $\meanB_R=\meanB_\phi=0$.
One can also impose boundary conditions on the mid-plane, $z=0$,
by selecting either symmetric (quadrupolar) fields, $B_R=\meanB_\phi'=0$,
or antisymmetric (dipolar) fields, $B_R'=\meanB_\phi=0$.
One can again define two dimensionless control parameters, 
\EQ
C_\Omega = S h^2/\eta_{\rm T}, \quad
C_\alpha = \alpha_0 h/\eta_{\rm T},
\EN
which measure the strengths of shear and 
$\alpha$ effects, respectively, where  
$h$ is a measure of the disc scale height.
(Note that $C_\Omega$ and
$C_\alpha$ used here are akin to those defined in 
\Sec{ExcitationConditionsSphere} and are identical to the
symbols $R_\Omega$ and $R_\alpha$ commonly used in galactic dynamo 
literature.) In spiral galaxies, the typical 
values are $C_\Omega \approx -10$ and $C_\alpha \approx 1$,
and so $\vert C_\Omega \vert \gg C_\alpha$.

Since $\vert C_\Omega \vert \gg C_\alpha$,
dynamo generation of axisymmetric solutions is controlled by the dynamo number 
$D = C_\Omega C_\alpha $. Exponential growth of the fields is
possible in the kinematic stage provided $\vert D \vert > D_{\rm crit}$.
Here the critical dynamo number $D_{\rm crit}\sim6$--$10$,
depending on the exact profile adopted for the $\alpha(z)$.
Here we assume $f_\alpha(z)=z/H$, where $H$ is the disc scale height.
For negative $\partial\Omega/\partial R$ and positive
values of $\alpha$ in the upper disc plane (northern `hemisphere'),
the most easily excited modes are no longer {\sf A0}, but {\sf S0}
\cite{ZRS83}.
This case is believed to be relevant to galaxies and one expects that
the most easily excited solutions for galaxies,
are modes with steady quadrupole (S0~st)
symmetry in the meridional ($Rz$) plane \cite{RSS88}.
For these modes, the growth rate in the kinematic regime can be 
approximated by,
\EQ
\gamma \approx {\eta_{\rm T} \over h^2}
\left( \sqrt{\vert D \vert} - \sqrt{D_{\rm crit}} \right).
\label{gama}
\EN
In most spiral galaxies the dynamo is supercritical
for a large range of radii, and so galactic fields
can indeed grow exponentially from small seed values.
A detailed discussion of the properties of solutions
in the galactic context is given in Ref.~\cite{RSS88}.

Further, as discussed in \Sec{ExcitationConditionsSphere}, axisymmetric
modes are easier to excite than nonaxisymmetric ones.
Although, the observed nonaxisymmetric large scale
mean field structures in some galaxies, can also be explained
by invoking nonaxisymmetric forms of $\alpha$ effect \cite{Moss_etal91},
turbulent diffusivity, or streaming motions \cite{SM93,MS91,moss96}.

\begin{figure}[t!]\begin{center}
\includegraphics[width=1.1\textwidth]{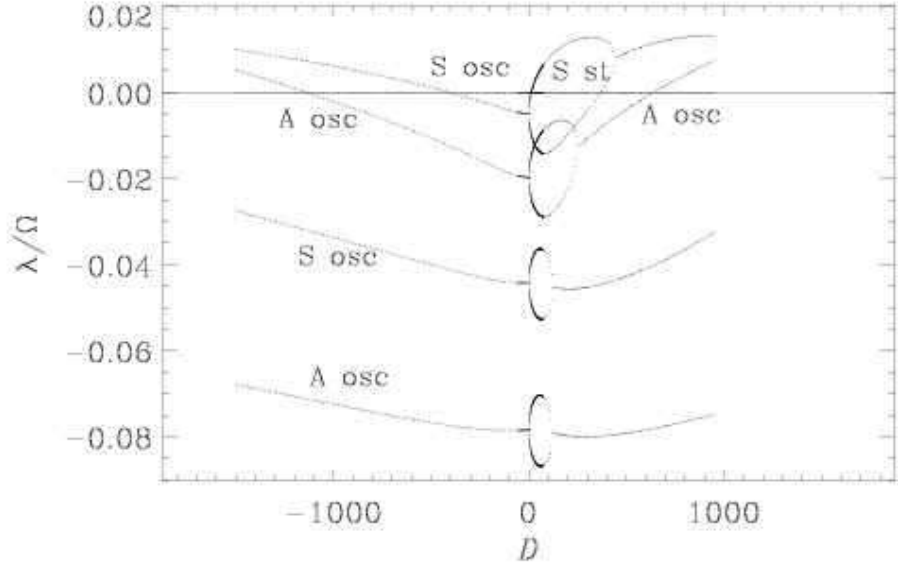}
\end{center}\caption[]{
Eigenvalues of the dynamo equations with shear in slab geometry with
radial shear.
The dynamo number, in this figure, 
is defined positive when the shear in negative and
$\alpha$ positive (opposite in sign to that in the text).
Note that for $\alpha>0$ the solution that is most easily excited is
nonoscillatory (`steady') and has even parity (referred to as {\sf
S~st}) whilst for $\alpha<0$ it is oscillatory ({\sf S~osc}).
Adapted from \cite{iceland}.
}\label{weigval_vertfield}\end{figure}

For accretion discs, $\alpha$ might be negative in the northern hemisphere
\cite{BNST95} and one therefore expects oscillatory quadrupoles (S0~osc)
\cite{iceland}.
In \Fig{weigval_vertfield} we show the growth rate of different
modes, obtained by solving \Eq{discdyn}
for both signs of the dynamo number \cite{iceland}.
In order to find all the modes, even the unstable ones, one can easily
solve \Eq{discdyn} numerically as an eigenvalue problem,
$\lambda\qq=\MMMM\qq$, where the complex growth rate $\lambda$
is the eigenvalue with the largest real part.

For illustrative purposes, we discuss the numerical technique in detail.
We introduce mesh points, $z_i=i\,\delta z$ (excluding the boundaries
at $z=0$ and $z=H$), where $\delta z=H/(N+1)$
is the mesh spacing, and $i=1,2,...,N$ denotes the position on the mesh.
The discretized eigenvector is
\EQ
\qq=(\meanB_{R1},\meanB_{\phi1},
\meanB_{R2},\meanB_{\phi2},...,
\meanB_{RN},\meanB_{\phi N})^T.
\EN
It is convenient to introduce the abbreviations
$a_i=-\alpha_i/(2\delta z)$ and $b=\eta_{\rm T}/(\delta z)^2$, so then
the second-order accurate discretized form of \Eq{discdyn} reads
\EQ
\dot{\meanB}_{R\,i} = a_{i+1}\meanB_{\phi\,i+1}-a_{i-1}\meanB_{\phi\,i-1}
 + b(\meanB_{R\,i+1}-2\meanB_{R\,i}+\meanB_{R\,i-1}),\quad
\EN
\EQ
\dot{\meanB}_{\phi\,i} = S\meanB_{R\,i}
 + b(\meanB_{\phi\,i+1}-2\meanB_{\phi\,i}+\meanB_{\phi\,i-1}),
\label{discdyn_discrete}
\EN
where $\alpha_1$, $\alpha_2$, ..., are the values of $\alpha$ at the
different mesh points.
Using symmetric boundary conditions on $z=0$, we have $B_{R\,0}=0$
and $B_{\phi\,0}'=(-3B_{\phi\,0}+4B_{\phi1}-B_{\phi2})/(2\delta z)=0$,
which is just the second order one-sided first derivative formula
\cite{Bra03}. With this we can eliminate the boundary point
$B_{\phi\,0}=\onethird(4B_{\phi1}-B_{\phi2})$,
and have, in matrix form,
\EQ
\MMMM=\pmatrix{
-2b&-\fourthird a_0& b & \onethird a_0+a_2& 0 & 0 & 0 & 0 &...\cr
 S & -\twothird b  & 0 &  \twothird b     & 0 & 0 & 0 & 0 &...\cr
 b &      a_1      &-2b&       0          & b &a_3& 0 & 0 &...\cr
 0 &       b       & S &     -2b          & 0 &  b& 0 & 0 &...\cr
 0 &       0       & b &      a_2         &-2b&  0& b &a_4&...\cr
 0 &       0       & 0 &       b          & S &-2b& 0 &  b&...\cr
...&      ...      &...&      ...         &...&...&...&...&...}.
\label{matrix}
\EN
In the lower right corner, no modification has to be applied, which
then corresponds to the vacuum boundary condition
$\meanB_R=\meanB_\phi=0$ on $z=\pm H$.
For the results shown in \Fig{weigval_vertfield} we have simply assumed
a linear profile, i.e.\ $\alpha=\alpha_0 z/H$, and $\eta_{\rm T}$ and
$S$ are constant.
The eigenvalues of the matrix $\MMMM$ can be found using standard
library routines.

\subsection{R\"adler effect and shear--current effect}
\label{ShearCurrentEffect}

An important additional contribution to the EMF is a term of the
form $\meanemf=...+\ddelta\times\meanJJ$.
This term was first derived and identified as a possible dynamo generating
term by R\"adler \cite{Rae69}, who found that in a rotating system with
non-helical turbulence $\ddelta$ is proportional to $\OO$.
Even in a non-rotating system with linear shear alone, large
scale dynamo action is possible due to the so-called shear--current effect
\cite{Roga+Klee03,Roga+Klee04,RadStep05}, where $\ddelta$ is proportional to the
vorticity $\meanWW$ of the mean flow.

\begin{figure}[t!]\begin{center}
\includegraphics[width=.7\textwidth]{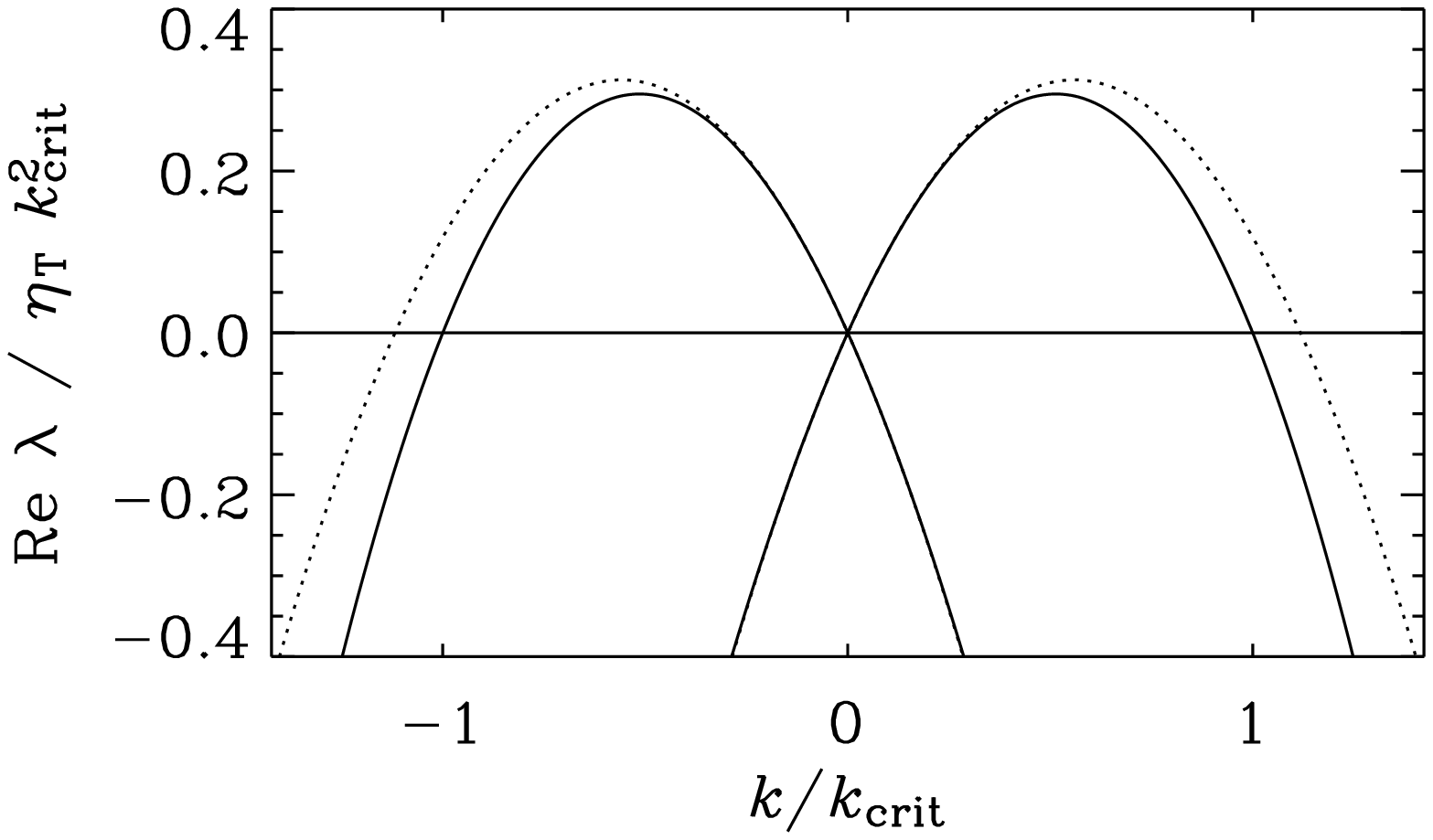}
\end{center}\caption[]{
Dispersion relation for the $\delta^2S$ dynamo with
$\delta k_{\rm crit}^2/S=0.2$.
The dotted line gives the result for the ``$\delta S$'' approximation
\Eq{dOlam_approx}.
The axes are normalized using $k_{\rm crit}$ as given by
\Eq{kcrit_delta2Omega}.
Note the similarity to the dispersion relation for the $\alpha^2$
dynamo; cf.\ \Fig{alpha2_disper}.
}\label{delta2omega_disper}\end{figure}

Numerous models with this effect have been considered in the early 1970ies
\cite{RS72,Rae86b,PHRoberts72}.
Since this term does not give a contribution to $\meanJJ\cdot\meanemf$,
it cannot provide energy to the system.
However, the presence of shear suffices to allow injection of
energy into the system to offset resistive losses and hence
to produce sustained large scale dynamo action.
The dispersion relation for such an axisymmetric ``$\delta S$'' dynamo can
easily be obtained from \Eqs{MatrixaO}{disperaO} by replacing
$\ii\alpha\to\kk\cdot\ddelta$, and its solution is
\EQ
\lambda_\pm=-\eta_{\rm T} k^2
\pm[-(\kk\cdot\ddelta)^2 k^2-(\kk\cdot\ddelta)Sk_z]^{1/2}.
\label{lam_deltaOmega}
\EN
Evidently, a necessary condition for growing solutions is that the term
in square brackets is positive, and hence that $-Sk_z/(\kk\cdot\ddelta)>k^2$.
We refer to such solutions as ``$\delta^2S$'' dynamos, so as to emphasize that
the $\ddelta$ effect enters the regeneration of both poloidal and
toroidal field.
We use here the symbol $S$ instead of $\Omega$ to emphasize that
angular velocity is unnecessary and that only shear is needed.

In the one-dimensional case with $\kk=(0,0,k_z)$ and
$\ddelta=(0,0,\delta)$, the necessary condition is $-\delta k_z^2/S>0$.
Thus, for positive shear, growing solutions are only possible for negative
values of $\delta$, and vice versa.
The critical $k_z$ may then be obtained from \Eq{lam_deltaOmega} in the form
\EQ
k_{\rm crit}=\left[-S\delta/(\delta^2+\eta_{\rm T}^2)\right]^{1/2}.
\label{kcrit_delta2Omega}
\EN
The two solution branches are shown in \Fig{delta2omega_disper},
together with the solutions of the approximate dispersion relation,
obtained by ignoring the $(\kk\cdot\ddelta)^2 k^2$ term inside the
squared brackets of \Eq{lam_deltaOmega}, i.e.\
\EQ
\lambda_\pm\approx-\eta_{\rm T} k^2\pm|-\delta Sk_z^2|^{1/2}
\quad\mbox{(for $1\gg-\delta k_z^2/S>0$)}.
\label{dOlam_approx}
\EN
Note first of all that the approximate dispersion relation is fairly good
even when $\delta k_{\rm crit}^2/S$ is not very small.
Second, the approximation is formally equivalent to the dispersion
relation for the $\alpha^2$ dynamo after replacing
$|\alpha|\to|-\delta S|^{1/2}$.

\section{Magnetic helicity conservation and inverse cascade}
\label{InverseCascade}

The mean field approach in terms of $\alpha$ effect and other
turbulent transport coefficients has been quite popular in modeling
the magnetic fields in various astrophysical bodies.
Its main shortcoming is the rather simplistic treatment of turbulence.
A very different approach has been pioneered by Frisch and coworkers
\cite{frisch} and explored quantitatively in terms of the Eddy Damped
Quasi-Normal Markovian (EDQNM) closure approximation \cite{PFL76},
which will be reviewed briefly below.
In this approach, the main mechanism producing large scale fields is
the inverse cascade of magnetic helicity toward larger scales.
The $\alpha$ effect emerges in a self-consistent
manner and, more importantly, the $\alpha$ effect is amended by a
correction term proportional to the small scale current
helicity which plays a crucial role in modern mean field theory
(\Sec{MagneticHelicityMeanFieldModels}).
We begin with an illustration of the inverse cascade mechanism
using a simple 3-mode interaction model.

\subsection{Inverse cascade in 3-mode interactions}
\label{ThreeModeInteraction}

The occurrence of an inverse cascade can be understood as the result of
two waves (wavenumbers
$\bp$ and $\qq$) interacting with each other to produce a wave of
wavenumber $\kk$.
The following argument is due to Frisch et al.\ \cite{frisch}.
Assuming that during this process magnetic energy is
conserved together with magnetic helicity, we have
\EQ
M_p + M_q = M_k ,
\label{frisch_energy}
\EN
\EQ
|H_p| + |H_q| = |H_k| ,
\label{helconspqk}
\EN
where we are assuming that only helicity of one sign is involved.
Suppose the initial field is fully helical and has the same
sign of magnetic helicity at all scales, we have
\EQ
2M_p=p|H_p|\quad\mbox{and}\quad
2M_q=q|H_q|,
\EN
and so \Eq{frisch_energy} yields
\EQ
p |H_p| + q|H_q| = 2M_k \ge k|H_k| ,
\label{helmod}
\EN
where the last inequality is just the realizability condition
\eq{realizability} applied to
the target wavenumber $\kk$ after the interaction. Using \Eq{helconspqk}
in \Eq{helmod} we have
\EQ
p|H_p|+q|H_q|\ge k(|H_p|+|H_q|).
\EN
In other words, the target wavevector $\kk$ after the interaction of
wavenumbers $\bp$ and $\qq$ satisfies
\EQ
k\le{p|H_p|+q|H_q|\over|H_p|+|H_q|}.
\label{helconspqk2}
\EN
The expression on the right hand side of \Eq{helconspqk2} is a weighted
mean of $p$ and $q$ and thus satisfies
\EQ
\min(p,q)\le{p|H_p|+q|H_q|\over|H_p|+|H_q|}\le\max(p,q),
\EN
and therefore
\EQ
k\le\max(p,q).
\EN
In the special case where $p=q$, we have $k\le p=q$, so the target
wavenumber after interaction is always less or equal to the initial
wavenumbers. In other words, wave interactions tend to transfer magnetic
energy to smaller wavenumbers, i.e.\ to larger scale. This corresponds
to an inverse cascade. The realizability condition, $\half k|H_k|\le M_k$,
was the most important ingredient in this argument.
An important assumption that we
made in the beginning was that the initial field be fully helical;
see Ref.~\cite{BDS02,MB02} for the case of fractional helicity.

\subsection{The EDQNM closure model}
\label{edqnm}

One of the earliest closure schemes applied to
the MHD dynamo problem was the eddy-damped quasi-normal Markovian 
(EDQNM) approximation. This was worked out in detail by
Pouquet, Frisch and L\'eorat (PFL) \cite{PFL76}.
EDQNM has been used frequently when dealing with fluid turbulence and
is described, for example, in Refs~\cite{Orszag70,Lesieur}.
We do not describe it in great detail.
Instead, we outline just the basic philosophy, and the crucial insights
gained from this closure. PFL assumed that 
the $\bb$ and $\uu$ fields were homogeneous, isotropic
(but helical) random fields. This is similar to
the unified treatment presented in \Sec{UnifiedTreatment}.
The large and small scale fields would again be distinguished
by whether the wavenumber $k$ is smaller or greater
than the wavenumber of the forcing. It was also
assumed that the initial $\bb$ field is statistically
invariant under sign reversal ($\bb \to - \bb$);
the MHD equations then preserve this property
and the cross helicity $\bra{\uu\cdot\bb}$
is then always zero. Now suppose the MHD equations
for the fluctuating fields are written  symbolically
as $\dot{u} = uu$, where $u$ stands for some component
of $\uu$ or $\bb$ and $\bra{u} = 0$. 
This notation is used only to illustrate the effects
of the quadratic nonlinearity; the linear dissipative
and forcing terms have been dropped since they do not
pose any specific closure problem and can be re-introduced
at the end.  Then we obtain for the second and third moments, 
again in symbolic form,
\EQ
{d \bra{uu} \over dt} = \bra{uuu} , \quad
{d \bra{uuu} \over dt} = \bra{uuuu}.
\label{moments}
\EN
The quasi-normal approximation consists of replacing the
fourth moment $\bra{uuuu}$ by its gaussian value,
that is the sum of products of second order moments.\footnote{Note that
the quasi-normal approximation 
may be more problematic when applied to MHD since the dynamo generated
$\BB$-field could initially be much more intermittent than the velocity field
(see for example the discussion of the kinematic small scale
dynamo in \Sec{SmallScale}). 
Therefore the validity of EDQNM applied to MHD is somewhat more
questionable than when applied to pure hydrodynamic turbulence.}
It turns out that such an approximation leads to
problems associated with unlimited growth of the third moment,
and the violation of the positivity constraint for
the energy spectra. This is cured by assuming that the neglected 
irreducible part of the fourth moment in \eq{moments} is in the 
form of a damping term,
which is a suitable linear relaxation operator of
the triple correlations (a procedure called eddy-damping).
One also carries out `Markovianization',
by assuming that the third moment responds to the
instantaneous product of the second moments.
The resulting third moment is substituted 
back on the RHS of the equation for second moments in 
\eq{moments}. This results in a closed equation
for the equal-time second-order moments
\EQ
{d \bra{uu} \over dt} = \theta(t) \bra{uu}\bra{uu},
\EN
where $\theta(t)$ essentially describes a relaxation
time for a given triad of interacting modes. It
is given by, 
\EQ
\theta(t) = \int_0^t \e^{-\int_\tau^t \mu(s)\,\dd s}\,\dd\tau.
\EN
Here $\mu(s)$ is the eddy damping operator and 
was written down in Ref.~\cite{PFL76} using phenomenological 
considerations. It depends on the kinetic and magnetic spectrum
and incorporates the damping effects on any mode
due to nonlinear interactions, Alfven effect and
microscopic diffusion. (In the stationary case, where $\mu=\mbox{constant}$, 
$\theta \to \mu^{-1}$.) The derived evolution equations
for the energy and helicity spectra of the random
velocity and magnetic fields, are shown to preserve the
quadratic invariants of total energy and magnetic helicity,
just as in full MHD. The complete spectral equations
under the EDQNM approximation are give in Table I
of Ref.~\cite{PFL76}, and will not be reproduced here
(see also the Appendix in Ref.~\cite{Deyoung80}).

However several crucial insights resulted from this work
about how the Lorentz force affects the large scale dynamo. 
From the EDQNM evolution equations for the kinetic and magnetic spectra,
PFL identified three important effects, all of which 
involve the coupling between widely separated scales. 
Suppose $M_k$ and $H_k$ are the magnetic
energy and helicity spectra and $E_k$, $F_k$ the
corresponding kinetic energy and helicity spectra.
And suppose we are interested in the dynamics of
the magnetic energy at wavenumber $k$ due to
velocity and magnetic fields at 
much larger scales (wavenumbers $ \le a_0 k$) and much smaller 
scales (wavenumber $\ge k/a_0$), where $a_0 \ll 1 $.
Then, in concrete terms, PFL found that the
nonlocal contributions to the evolution are
\EQ
\dot{(M_k)}_{\rm NLoc} = k\Gamma_k (E_k - M_k) + \alpha^R k^2 H_k
-2\nu^V_k k^2 M_k + \ldots,
\label{pfl_resultM}
\EN
\EQ
\dot{(H_k)}_{\rm NLoc} = (\Gamma_k/k) (F_k - k^2H_k) + \alpha^R M_k
-2\nu^V_k k^2 H_k + \ldots,
\label{pfl_resultH}
\EN
\EQ
\dot{(E_k)}_{\rm NLoc} = -k\Gamma_k (E_k - M_k) -2(\twofifth\nu^V_k + 
\nu^M_k +\nu^R_k) k^2 E_k + \ldots,
\label{pfl_resultE}
\EN
\EQ
\dot{(F_k)}_{\rm NLoc} = -k\Gamma_k (F_k - k^2H_k) -2(\twofifth\nu^V_k + 
\nu^M_k +\nu^R_k) k^2 F_k + \ldots,
\label{pfl_resultF}
\EN
where
\EQ
\Gamma_k = \fourthird k \int_0^{a_0k} \theta_{kkq} M_q \dd q,
\EN
\EQ
\alpha_k^R = -\fourthird \int_{k/a_0}^\infty \theta_{kqq} 
\left[F_k -k^2H_k \right] \dd q,
\quad \nu^V_k = \twothird \int_{k/a_0}^\infty \theta_{kqq} E_q \dd q.
\label{alp_res}
\EN
\EQ
\nu^M_k = \twothird \int_{k/a_0}^\infty \theta_{kqq} M_q \dd q, 
\quad 
\nu^R_k = {\textstyle{2\over 15}}\int_{k/a_0}^\infty 
\theta_{kqq} q{\partial \mu_q \over \partial q}( E_q -M_q) \dd q.
\EN
The first term on the RHS of \eq{pfl_resultM}, referred to by PFL as
the Alfv\'en effect, leads to equipartition of the 
kinetic and magnetic energies, at any scale
due to magnetic energy at larger scales.
This happens on the Alfv\'en crossing time of the larger scale field.
The second term, very important for what follows,
shows that the growth of large scale magnetic energy is induced
by the small scale `residual' helicity, which is the difference
between the kinetic helicity and a current helicity
contribution, due to the small scale magnetic field.
The part of $\alpha_k^R$ depending on the kinetic
helicity corresponds closely to the usual
$\alpha$ effect, once one realizes
that $\theta$ is like a correlation time.
Over and above this term PFL discovered for the first time
the current helicity contribution, the $ k^2H_k$ term in
$\alpha_k^R$. The third term gives
the turbulent diffusion of the large scale magnetic field.
Surprisingly, this term does not get affected, 
to the leading order, by nonlinear effects of the Lorentz force
(although the small scale magnetic field does affect the
diffusion of the large scale velocity field).
PFL also gave a heuristic derivation of the last two 
results, because of their potential importance.
This derivation, has since been reproduced in various forms
by several authors \cite{ZRS83,GD95,GD96} and has also been
extended to include higher order corrections \cite{sub03}.
We outline it in \App{qnmod} and
use it to discuss the nonlinear saturation of the 
large scale dynamo in \Sec{SDynamicalQuenching}.

\begin{figure}[t!]\begin{center}
\includegraphics[width=.75\textwidth]{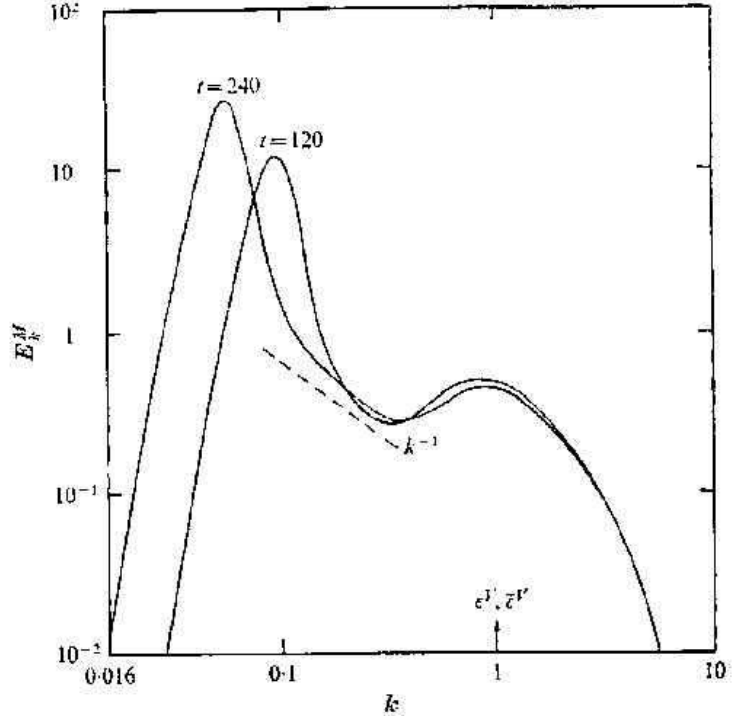}
\end{center}\caption[]{
Inverse cascade in a numerical solution of the EDQNM equations
showing the magnetic energy spectrum (here called $E_k^{\rm M}$) at two
different times.
Kinetic energy and helicity are injected at wavenumber $k=1$.
Note the peak of magnetic energy propagating toward smaller
values of $k$.
Adapted from \cite{PFL76}.
}\label{PFLfig}\end{figure}

The EDQNM equations were numerically integrated by PFL
to study the dynamo growth of magnetic energy. Most important
in the present context is the case when both kinetic energy
and helicity are continuously injected. In this case
PFL found what they described as an inverse cascade of
magnetic helicity, which led to the appearance
of magnetic energy and helicity in ever increasing scales
(limited only by the size of the system). 
\Fig{PFLfig} shows the resulting magnetic energy spectrum
at two times. One can see the build up of large scale
magnetic energy on scales much larger than the injection scale
($k=1$). PFL have also argued that this 
inverse cascade of magnetic energy resulted from a
competition between the helicity (residual $\alpha$) and
the Alfv\'en effect. We shall return to this question
in subsequent sections.
Numerical solutions to the EDQNM closure equations also reproduce
small scale dynamo action in the absence of helicity \cite{LPF81}.
However, the decisive terms describing small scale dynamo action
do not seem to appear in the nonlocal interaction terms 
extracted in \Eqss{pfl_resultM}{pfl_resultF}.

\section{Simulations of large scale dynamos}
\label{SimulationsHelicalDynamos}

For astrophysical applications, the inverse cascade approach using the PFL
model seemed too idealized compared to the $\alpha^2$ and $\alpha\Omega$
dynamo models that have been studied intensively in those years
when the PFL model was proposed.
Furthermore, there seems little scope for generalizing EDQNM to
inhomogeneous systems with rotation.
Nevertheless, the basic idea of an inverse cascade was well established
and verified by several groups \cite{B01,MFP81,PP78,KYM91,BP99}.
Only recently, however, there have been serious attempts to bridge
the gap between the PFL approach and mean field models.
In this section we review recent efforts \cite{B01} to
study helically driven hydromagnetic turbulence and to compare
with the associated $\alpha^2$ dynamo model that is applicable in
the equivalent situation, i.e.\ also with fully periodic boundary conditions.
After that, in \Sec{MagneticHelicityMeanFieldModels}, we consider in
detail the implications of the conservation of magnetic helicity to mean
field models.

\subsection{The basic equations}
\label{BasicEquations}

In order to simulate the effect of cyclonic turbulence without actually
including the physical effects that contribute to cyclonic turbulence
one can substitute the buoyancy term by an explicit body force.
The effects of stratification and rotation are therefore neglected.

A compressible isothermal gas with constant sound speed
$c_{\rm s}$, constant dynamical viscosity $\mu$, and constant magnetic
diffusivity $\eta$ is considered. To make
sure the magnetic field stays solenoidal, i.e.\ $\nab\cdot\BB=0$,
$\BB$ is expressed in terms of the magnetic vector potential $\AAA$, so the
field is written as $\BB=\nab\times\AAA$. The governing equations for
density $\rho$, velocity $\uu$, and magnetic vector potential $\AAA$,
are given by
\EQ
{\DD\ln\rho\over\DD t}=-\nab\cdot\uu,
\label{dlnrhodt}
\EN
\EQ
{\DD\uu\over\DD t}=-c_{\rm s}^2\nab\ln\rho+{\JJ\times\BB\over\rho}
+{\mu\over\rho}(\nabla^2\uu+\onethird\nab\nab\cdot\uu)+\ff,
\label{dudt}
\EN
\EQ
{\partial\AAA\over\partial t}=\uu\times\BB+\eta\nabla^2\AAA,
\label{dAdt}
\EN
where ${\rm D}/{\rm D}t=\partial/\partial t+\uu\cdot\nab$ is the advective
derivative. The current density, $\JJ=\nab\times\BB/\mu_0$, is obtained
in the form $\mu_0\JJ=-\nabla^2\AAA+\nab\nab\cdot\AAA$.
The gauge $\phi=-\eta\nab\cdot\AAA$ for the electrostatic potential
is used, and $\eta=\mbox{constant}$ is assumed, so the magnetic
diffusion term is just $\eta\nabla^2\AAA$.
Details regarding the numerical solution of these equations are analogous
to the nonhelical case and are discussed elsewhere \cite{Bra03}.
Many of the simulations presented here have been done using the
{\sc Pencil Code} \cite{PencilCode}, mentioned already in
\Sec{Simulations_SSdynamo}.

For the following it is useful to recall that each vector field can be
decomposed into a solenoidal and two vortical parts with positive and
negative helicity, respectively. These are also referred to as
Chandrasekhar-Kendall functions (cf.\ \Sec{energy_spectra}). 
It is often useful to decompose
the magnetic field into positively and negatively helical parts.
Here, we use eigenfunctions with positive eigenvalues
(i.e.\ with positive helicity) as forcing function $\ff$
of the flow. We restrict ourselves to functions selected from a finite
band of wavenumbers around the wavenumber $k_{\rm f}$, but direction
and amplitude are chosen randomly at each timestep.
Further details can be found in Ref.~\cite{B01}.
Similar work was first carried out
by Meneguzzi et al.\ \cite{MFP81}, but at the time one was barely able
to run even until saturation.
Since the nineties, a lot of work has been done on turbulent dynamos with
ABC flow-type forcing \cite{GS91,GSG91,ArchontisDorchNordlund03,BP99}.
In none of these investigations, however, the saturation behavior has
been studied and so the issue of resistively slow magnetic
helicity evolution past initial saturation remained unnoticed. It is
exactly this aspect that has now become so crucial in understanding the
saturation behavior of nonlinear dynamos. We begin by discussing first
the linear (kinematic) evolution of the magnetic field.

\subsection{Linear behavior}
\label{LinearBehavior}

Dynamo action occurs once the magnetic Reynolds number
exceeds a certain critical value, $R_{\rm crit}$.
For helical flows, $R_{\rm crit}$ is between 1 and 2.
Note that the values given in Table~1 of Ref.~\cite{B01} need to be
divided by $2\pi$ in order to conform with the definition
of the magnetic Reynolds number, $R_{\rm m}=u_{\rm rms}/(\eta k_{\rm f})$,
used throughout most of this review.
In the supercritical
case, $R_{\rm m}>R_{\rm crit}$, the field grows exponentially
with the growth rate $\lambda$, which is proportional to
the inverse turnover time,
$u_{\rm rms}k_{\rm f}$. The resistively limited saturation behavior that
will be discussed below in full detail has no obvious correspondence
in the kinematic stage when the field is too weak to affect the motions
by the Lorentz force \cite{gilbert}.
Nevertheless, there is actually a subtle effect
on the shape of the magnetic energy spectrum as $R_{\rm m}$ increases:
the magnetic energy spectrum has two bumps, each being governed
by opposite magnetic helicity.
We will explain this in more detail below, where we also show the
evolution of the two bumps (\Fig{Fpspec_pm_satkin}).
For now we just note that, while
in the weakly supercritical case the two bumps can be far apart from
each other.
However, as $R_{\rm m}$ is increased the two bumps in 
the spectra move closer together while maintaining a similar height,
decreasing thus the net magnetic helicity, as imposed by magnetic
helicity conservation.
But before we can fully 
appreciate this phenomenon, we need to discuss the effect the kinetic
helicity has on the magnetic field.

A helical velocity field tends to drive helicity in the magnetic field as well,
but in the nonresistive limit magnetic helicity conservation dictates
that $\bra{\AAA\cdot\BB}=\mbox{const}=0$ if the initial field (or at
least its helicity) was infinitesimally weak.
(Here and elsewhere, angular brackets denote volume averages.)
Thus, there must be some kind of magnetic helicity
cancelation. Under homogeneous isotropic conditions there cannot be
a spatial segregation in positive and negative helical parts.
Instead, there is a spectral segregation:
there is a bump at the forcing wavenumber and another `secondary'
bump at somewhat smaller wavenumber. The two bumps have opposite
sign of magnetic helicity such that the net magnetic helicity
is close to zero (and it would be exactly zero in the limit
$R_{\rm m}\to\infty$. At the forcing wavenumber, the sign of magnetic helicity
agrees with that of the kinetic helicity, but at smaller wavenumbers
the sign of magnetic helicity is opposite.
At small values of $R_{\rm m}$,
this secondary peak can be identified with the wavenumber where the
corresponding $\alpha^2$ dynamo has maximum growth rate; see \Sec{Sa2_model}.
Simulations seem to confirm that, as $R_{\rm m}$ increases, $k_{\rm max}$ approaches
$\half k_{\rm f}$ \cite{BDS02}, as one would expect from $\alpha^2$ dynamo theory.
Since these two peaks have opposite magnetic helicity,
their moving together in the high--$R_{\rm m}$ limit tends to
lower the net magnetic helicity, thus confirming 
earlier results \cite{gilbert,HCK96} that suggest
that the total magnetic helicity approaches zero in
the high-$R_{\rm m}$ limit.

\begin{figure}[t!]\begin{center}
\includegraphics[width=.95\textwidth]{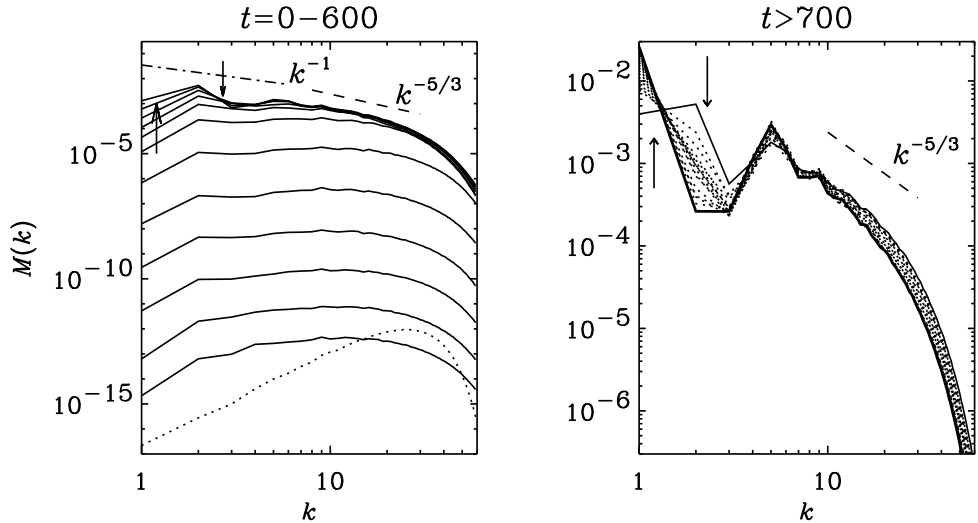}
\end{center}\caption[]{
Power spectra of magnetic energy of Run~3 of Ref.~\cite{B01}. During the initial
growth phase the field saturates at small scales first and only later at
large scales (left hand panel). Later, when also the large scale field
saturates, the field at intermediate scales ($k=2$, 3, and 4) becomes
suppressed. In the second panel, intermediate times are shown as dotted
lines, $t=700$ is shown in solid and $t=1600$ is shown as a thick solid
line. The forcing wavenumber is $k_{\rm f}=5$.
Adapted from Ref.~\cite{B03passot}.
}\label{Fpspec_growth_passot}\end{figure}

\subsection{Nonlinear behavior}
\label{NonlinearBehavior}

\begin{figure}[t!]\begin{center}
\includegraphics[width=.95\textwidth]{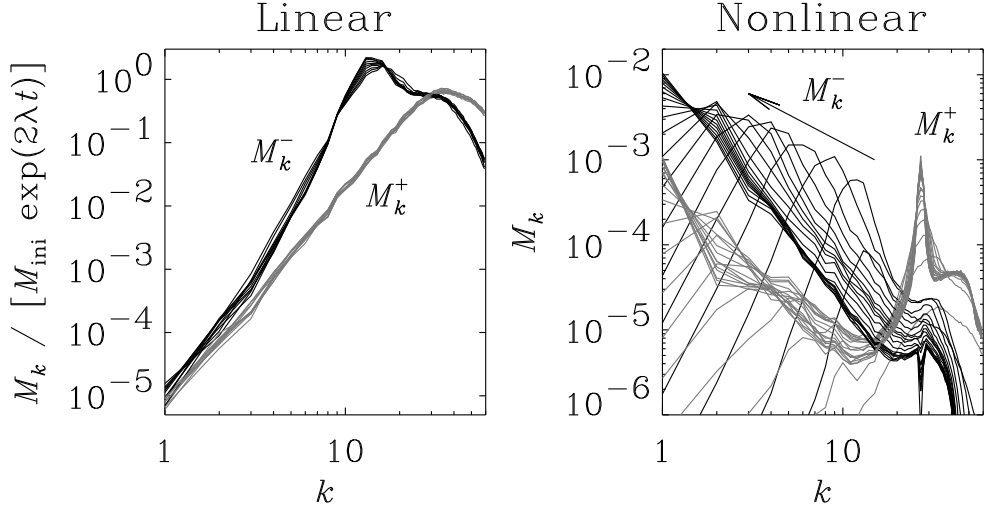}
\end{center}\caption[]{
Power spectra of magnetic energy of positively and negatively polarized
parts ($M_k^+$ and $M_k^-$)
in the linear and nonlinear regimes. The spectra in the linear
regime have been compensated by the exponential growth factor to make
them collapse on top of each other. Here the forcing wavenumber is in
the dissipative subrange, $k_{\rm f}=27$, but this allows enough scale
separation to see the inverse transfer of magnetic energy to smaller $k$.
The data are from Run~B of Ref.~\cite{BranSok02}.
}\label{Fpspec_pm_satkin}\end{figure}

Eventually, the magnetic energy stops increasing exponentially. This is
due to the nonlinear terms, in particular the Lorentz force $\JJ\times\BB$
in \Eq{dudt}, which begins to affect the velocity field. The temporal
growth of the power spectra becomes stagnant,
but the spectra saturate only partially; see
\Fig{Fpspec_growth_passot}, where we show data from a run with
forcing at wavenumber $k_{\rm f}=5$.
In the left hand panel we see that, by the time $t=600$, the power spectra
have saturated at larger wavenumbers, $k\ga3$.
However, it takes until $t\simeq1600$
for the power spectra to be saturated also at $k=1$ (right hand panel
of \Fig{Fpspec_growth_passot}). In order to see more clearly the behavior
at large scales, we show in \Fig{Fpspec_pm_satkin} data from a run with
$k_{\rm f}=27$ and compare spectra in the linear and
nonlinear regimes.
In order to clarify the different roles played by the positively and
negatively polarized components of the turbulence, we decompose the
magnetic power spectra as explained in \Sec{energy_spectra}.
The forcing has positive helicity, giving rise to a peak of
$M_k^+$ at small scales.
Magnetic helicity conservation
requires there to be energy in oppositely polarized
components, $M_k^-$.
Again, because of magnetic helicity conservation (\Sec{ThreeModeInteraction}),
the bump of $M_k^-$ can only propagate to the left, i.e.\ to larger scales.

\begin{figure}[t!]\begin{center}
\includegraphics[width=.95\textwidth]{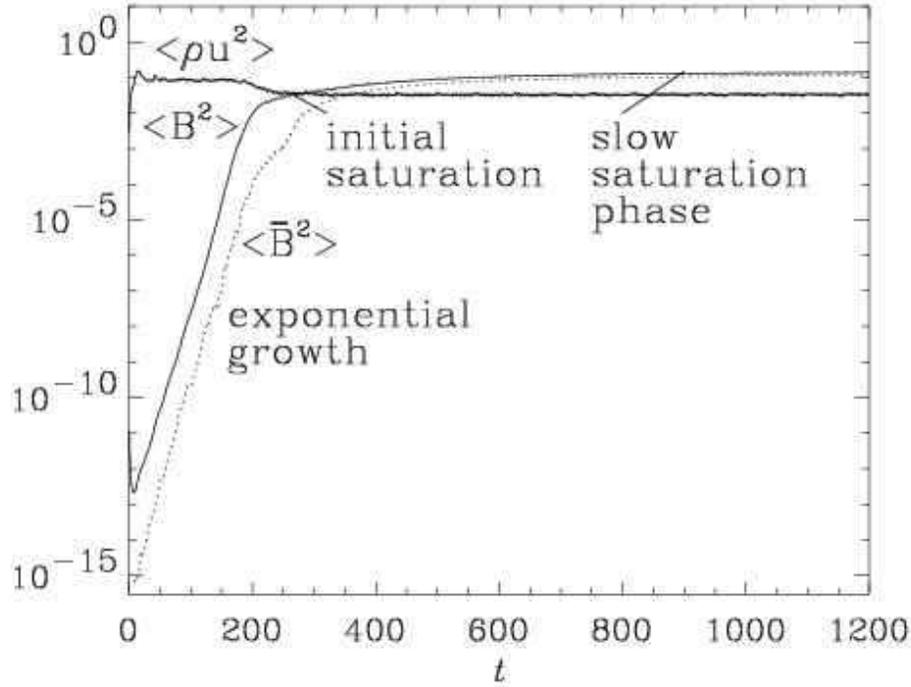}
\end{center}\caption[]{
The three stages of helical magnetic field growth: exponential growth
until initial saturation (when $\bra{\BB^2}/\mu_0=\bra{\rho\uu^2}$),
followed by a (resistively) slow saturation phase. In this plot we
have used $\mu_0=1$. The energy of the large scale magnetic field,
$\bra{\meanBB^2}$, is shown for comparison. The data are from Run~3
of Ref.~\cite{B01}.
}\label{Fpjbm_decay_nfit_run3passot}\end{figure}

In the linear regime, all spectra are just shifted
along the ordinate, so the spectra have been compensated by the
factor $M_{\rm ini}\exp(2\lambda t)$, where $\lambda$ is the growth
rate of the rms field and $M_{\rm ini}$ is
the initial magnetic energy. In the nonlinear
regime the bump on the right stays at approximately the same wavenumber
(the forcing wavenumber), while the bump on the left propagates gradually
further to the left. As it does so, and since, in addition, the amplitude of the
secondary peak increases slightly, the net magnetic helicity
inevitably increases (or rather becomes more negative in the present
case). But, because of the asymptotic magnetic helicity conservation,
this can only happen on a slow resistive time scale. This leads to
the appearance of a (resistively) slow saturation phase past the
initial saturation; see \Fig{Fpjbm_decay_nfit_run3passot}.

\subsection{Emergence of a large scale field}
\label{SBeltrami}

\begin{figure}[t!]\begin{center}
\includegraphics[width=.99\textwidth]{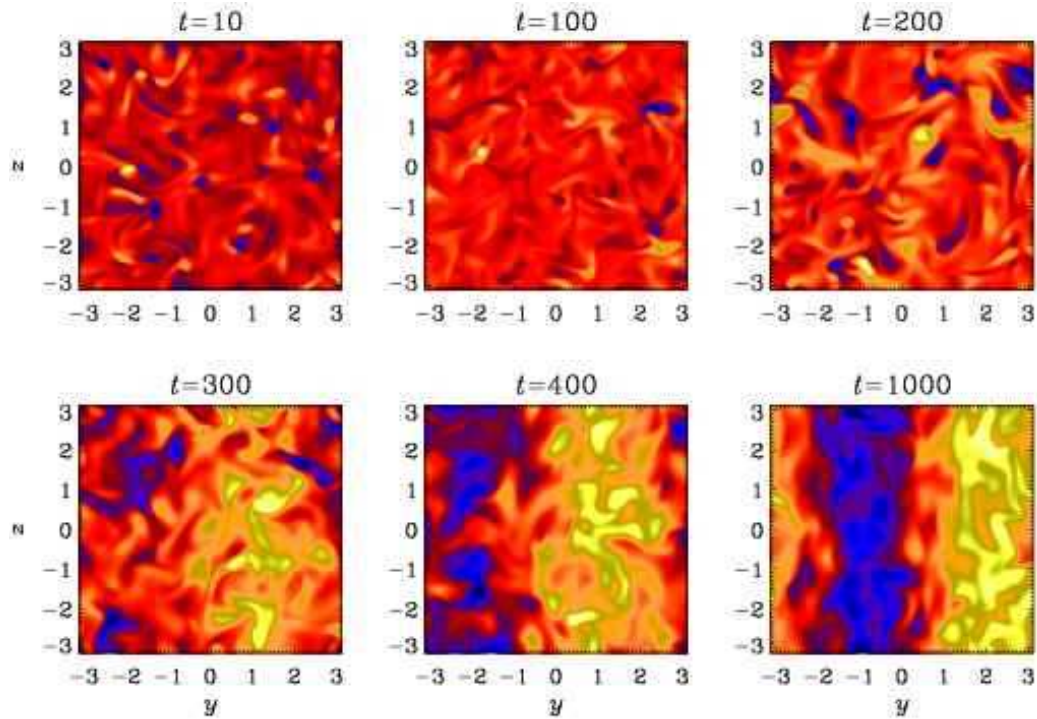}
\end{center}\caption[]{
Cross-sections of $B_x(0,y,z)$ for Run~3 of Ref.~\cite{B01}
at different times showing the gradual build-up of
the large scale magnetic field after $t=300$. Dark (light) corresponds
to negative (positive) values. Each image is scaled with respect to its
min and max values.
The final state corresponds to the second eigenfunction given in
\Eq{BeltramiEigenfunctions}, but with some smaller scale turbulence
superimposed.
}\label{pimages_hor}\end{figure}

In the simulations of Ref.~\cite{B01} the flow was forced at an intermediate
wavenumber, $k\approx k_{\rm f}=5$, while the smallest wavenumber in the
computational domain corresponds to $k=k_1=1$.
The kinetic energy spectrum peaks at $k\approx k_{\rm f}$, which is
therefore also the wavenumber of the energy carrying scale.
The turbulence is nearly fully helical with
$\bra{\oo\cdot\uu}/(k_{\rm f}\bra{\uu^2})\approx0.7...0.9$.
The initial field is random, but as time goes on it develops a large scale
component at wavenumber $k\approx k_1=1$; see \Fig{pimages_hor}.

The large scale field seen in \Fig{pimages_hor} has only one preferred
direction, which is the wavevector $\kk_1$ of $\meanBB$.
Different initial conditions can produce different directions of $\kk_1$;
see Fig.~6 of Ref.~\cite{B01}.
A suitable definition of the mean field is a two-dimensional average over
the two directions perpendicular to $\kk_1$.
(The most useful choice among the three possibilities
can only be taken a posteriori when we know the direction $\kk_1$
in which the large scale field varies.
So, in practice, one has to calculate all three possibilities and select
the right one in the end.)
The resulting large scale field is one of the following three eigenfunctions of the curl
operator whose wavevectors point along a coordinate direction, i.e.\
\EQ
\meanBB(\xx)=\pmatrix{\cos k_{\rm m}z\cr\sin k_{\rm m}z\cr 0},\quad
\pmatrix{0\cr\cos k_{\rm m}x\cr\sin k_{\rm m}x},\quad\mbox{or}\quad
\pmatrix{\sin k_{\rm m}y\cr0\cr\cos k_{\rm m}y},
\label{BeltramiEigenfunctions}
\EN
where $k_{\rm m}=k_1=1$.
These fields are force-free and are also referred to as
Beltrami fields.
The large scale field is fully helical, but with opposite
(here negative) current helicity relative to the
small scale field, i.e.\ $\meanJJ\cdot\meanBB=-k_{\rm m}\meanBB^2<0$.
This property alone allows us to estimate the saturation
amplitude of the dynamo, as will be done in the next section.

\begin{figure}[t!]\begin{center}
\includegraphics[width=.99\textwidth]{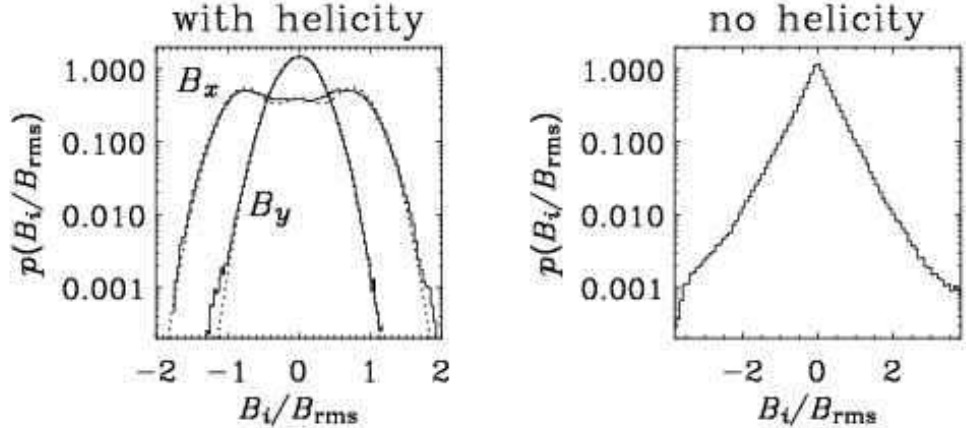}
\end{center}\caption[]{
Probability density functions of $B_x/B_{\rm rms}$ and $B_y/B_{\rm rms}$
for Run~3 of Ref.~\cite{B01} at the time of saturation (left).
For comparison, the PDFs are also shown for nonhelically forced turbulence
(right).
In the left hand plot,
the dotted lines give gaussian fits with a width of 0.27.
$B_y/B_{\rm rms}$ is fitted by a single gaussian around zero, while
$B_x/B_{\rm rms}$ is fitted by a superposition of three gaussians
(one around zero with weight 0.26, and two with weight 0.37 shifted
by $\pm0.76$ away from zero).
}\label{ppdf}\end{figure}

It is also interesting to note that the one point 
probability density functions (PDF) of
the three magnetic field components
are, to a good approximation, gaussian.
This is shown in the left hand panel of
\Fig{ppdf} where we plot one point PDFs
of $B_x$ and $B_y$ for Run~3 of Ref.~\cite{B01} at the time of saturation.
(The PDF of $B_z$ looks similar to that of $B_x$ and is not shown.)

We recall that, in this particular simulation, the $x$ and $z$ components
of the magnetic field show large scale variation in the $y$ direction.
One therefore sees two marked humps in the PDFs of $B_x$ and $B_z$, which
can be fitted by a superposition of two gaussians shifted away from zero,
together with another gaussian of lower weight around zero.
(Note that even though there is a mean $B_x$ field, the presence of the
random fluctuating component makes the notion of a PDF meaningful.
Indeed it is the presence of a mean field that distorts
the one point PDF to a double humped form, rather than a gaussian
centered at the origin).
The $y$ component of the field does not show a large scale field and
can be fitted by a single gaussian of the same width.
(In the absence of {\it large scale} dynamo action the three components of the
magnetic field would not be gaussians but stretched exponentials;
see Ref.~\cite{BJNRST96} for such results in the context of convection,
where there is only small scale dynamo action.)
In that case, the {\it modulus} of the field (i.e.\ not its individual
components) tends to have a log-normal probability density function.
For comparison we show in the right hand panel of \Fig{ppdf} the
corresponding PDFs for nonhelically forced turbulence.

\subsection{Importance of magnetic helicity}

In the following we show that for strongly helical dynamos the saturation
amplitude and the saturation time can accurately be estimated from
magnetic helicity considerations alone -- without actually solving the
induction equation explicitly.
The argument is similar in nature to the way how the descent speed of a
free-falling body can be calculated based on the consideration of
kinetic and potential energies alone, without considering the
equation of motion.
A more accurate treatment of the saturation behavior of helical dynamos
will be presented in \Secs{SDynamicalQuenching}{ComparisonSimulations}.

\subsubsection{Saturation amplitude}

Even though the current helicity of the large and small scale
fields are finite, the current helicity of the total (small scale plus
large scale) field must approach zero in the long time limit.
This is evident from the magnetic helicity equation \eq{magn_hel_evol}
which, for a closed or periodic domain, is simply
\EQ
{\dd\over\dd t}\bra{\AAA\cdot\BB}=-2\eta\bra{\JJ\cdot\BB}.
\label{helicity_eqn_closed}
\EN
Thus, in the steady state ($\dd/\dd t=0$) one has $\bra{\JJ\cdot\BB}=0$.
However, the time scale on which this can be achieved is the resistive one,
as will be discussed in the following.

Splitting the magnetic field and current density into mean and
fluctuating components, similar to \Eq{mean+fluct}, we have
\EQ
\overline{\JJ\cdot\BB}=\meanJJ\cdot\meanBB+\overline{\jj\cdot\bb},
\EN
and therefore also
$\bra{\JJ\cdot\BB}=\bra{\meanJJ\cdot\meanBB}+\bra{\jj\cdot\bb}$.
In the steady state we have
\EQ
-\bra{\meanJJ\cdot\meanBB}=\bra{\jj\cdot\bb}\quad
\mbox{(steady state)}.
\label{JBSteadyState}
\EN
Here, overbars denote suitably defined two-dimensional averages
(\Sec{SBeltrami}) and angular brackets denote volume averages.
The helical forcing tends to produce finite current helicity at small scales.
This, in turn, tends to induce finite current helicity of opposite sign
at large scales.
Depending on the degree of helicity in the forcing, the small scale field
will be more or less strongly helical.
The degree of helicity of the large scale fields depends also on other
factors such as boundary conditions and the presence of shear which
produces toroidal field quite independently of helicity.

\begin{figure}[t!]\begin{center}
\includegraphics[width=.99\textwidth]{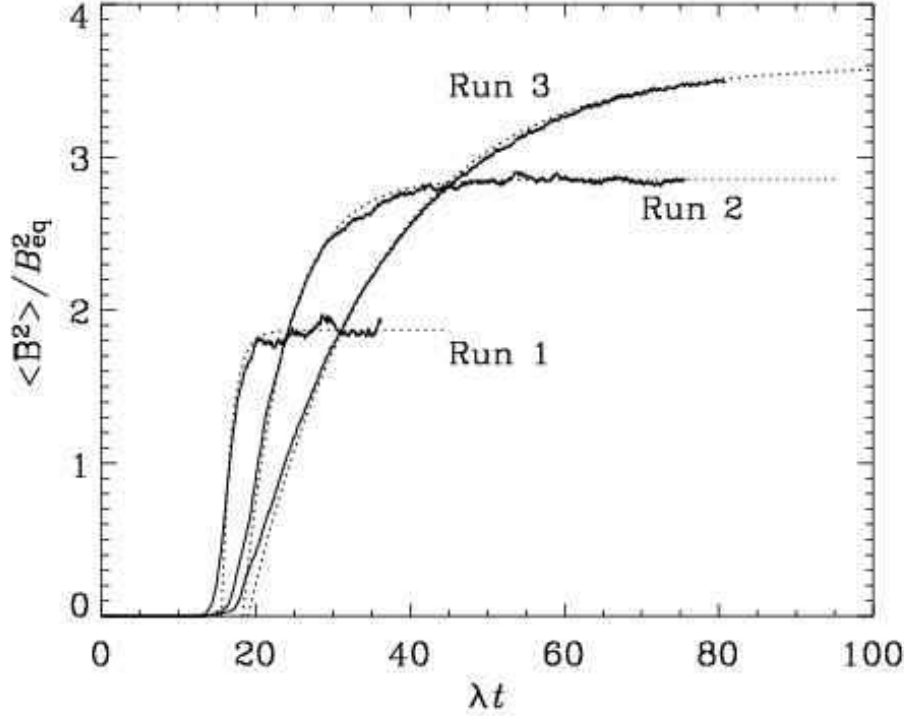}
\end{center}\caption[]{
Late saturation phase of fully helical turbulent dynamos
for three different values of the magnetic Reynolds number:
$R_{\rm m}\equiv u_{\rm rms}/\eta k_{\rm f}=2.4$, 6, and 18 for
Runs~1, 2, and 3 respectively; see Ref.~\cite{B01}.
The mean magnetic field, $\meanBB$, is normalized with
respect to the equipartition value,
$B_{\rm eq}=\sqrt{\mu_0\rho_0}u_{\rm rms}$,
and time is normalized with respect to the kinematic
growth rate, $\lambda$.
The dotted lines represent the formula \eq{helconstraint}
which tracks the simulation results rather well.
}\label{pjbm_decay_nfit}\end{figure}

The case of fractional helicities can be dealt with by introducing
efficiency factors \cite{BB02,BDS02,MB02}, but in order to explain the
basic point we just assume that
both mean and fluctuating fields are fully helical.
(This is also likely to be more fully the case when
the small scale field arises predominantly due to the tangling
of the large scale field by the helical turbulence rather than the
small scale dynamo.) So, in the fully helical case
we have $\bra{\meanJJ\cdot\meanBB}\approx\mp k_{\rm m}\bra{\meanBB^2}$
and $\bra{\jj\cdot\bb}\approx\pm k_{\rm f}\bra{\bb^2}$, where $k_{\rm f}$
is the typical wavenumber of the fluctuating field (which is close
to the wavenumber of the energy carrying scale).
This yields $k_{\rm m}\bra{\meanBB^2}=k_{\rm f}\bra{\bb^2}$, and if
the small scale field is in equipartition with the turbulent motions,
i.e.\ if $\bra{\bb^2}\approx\bra{\mu_0\rho\uu^2}\equiv B_{\rm eq}^2$, we have
\cite{B01}
\EQ
\bra{\meanBB^2}
={k_{\rm f}\over k_{\rm m}}\bra{\bb^2}
\approx{k_{\rm f}\over k_{\rm m}}B_{\rm eq}^2>B_{\rm eq}^2
\quad\mbox{(steady state)}.
\label{SteadyState}
\EN
We see that for a fully helical dynamo the large scale field at $k=k_{\rm m}$ is in
general in super-equipartition with the kinetic energy of the turbulence.
This fact is indeed confirmed by simulations which also show strong
large scale fields in super-equipartition; see \Fig{pjbm_decay_nfit}.

Obviously, the simulated values of $\bra{\meanBB^2}/B_{\rm eq}^2$ fall
somewhat short of this simplistic estimate, although the
estimate becomes more accurate in the case of larger magnetic
Reynolds number (Run~3 has
${\rm Re}=R_{\rm m}\equiv u_{\rm rms}/\eta k_{\rm f}\approx18$,
Run~2 has ${\rm Re}=R_{\rm m}\approx6$, and
Run~1 has ${\rm Re}=R_{\rm m}\approx2.4$).

\subsubsection{Saturation time}
\label{SaturationTime}

It turns out that the time dependence in \Fig{pjbm_decay_nfit} can
be well described by a fit formula that can be derived from the magnetic
helicity equation.
In the following we define the current helicities at large and small scales
as
\EQ
C_1=\bra{\meanJJ\cdot\meanBB}
\quad\mbox{and}\quad C_{\rm f}=\bra{\jj\cdot\bb},
\EN
respectively, and the magnetic helicities at large and small scales as
\EQ
H_1=\bra{\meanAA\cdot\meanBB}
\quad\mbox{and}\quad H_{\rm f}=\bra{\aaa\cdot\bb},
\EN
respectively.
Near saturation we have $|C_1|\approx|C_{\rm f}|$, but
because $C_1=k_1^2 H_1$ and $C_{\rm f}=k_{\rm f}^2H_{\rm f}$ together
with $k_1/k_{\rm f}\ll1$, we have
\EQ
|H_1|\gg|H_{\rm f}|\quad\mbox{(near saturation)}.
\label{H1ggHf}
\EN
Furthermore, as will be shown more convincingly in \Sec{Searly},
the small scale field tends to saturate on a dynamical time scale.
At late times we can therefore, to a good approximation, neglect the time
derivative of $H_{\rm f}$ relative to the time derivative of $H_1$ in
\Eq{helicity_eqn_closed}, so 
\EQ
\dot{H}_1=-2\eta k_1^2 H_1-2\eta k_{\rm f}^2 H_{\rm f}.
\label{H1eqn}
\EN
Solving \Eq{H1eqn} for $H_1(t)$, using a given value of
$H_{\rm f}=\mbox{const}$, we find
\EQ
H_1(t)=H_{\rm f}{k_{\rm f}^2\over k_1^2}\left[
1-\e^{-2\eta k_1^2(t-t_{\rm sat})}\right].
\label{helconstraint}
\EN
The time $t_{\rm sat}$ is determined by the strength of the
initial seed magnetic field, $B_{\rm ini}$; for weaker fields it
takes somewhat longer to reach saturation.
Since the growth is exponential, the dependence is only logarithmic,
so $t_{\rm sat}=\lambda^{-1}\ln(B_{\rm eq}/B_{\rm ini})$, where
$\lambda$ is the growth rate of the rms field strength of the total field.
\EEq{helconstraint} describes the evolution of $\meanBB$
quite well, as is shown in \Fig{pjbm_decay_nfit}.

Clearly, \Eq{helconstraint} is not applicable too close to $t=t_{\rm sat}$.
This is also evident from \Fig{pjbm_decay_nfit}, which shows that in
the simulations (solid lines) there is a finite mean field already at
$t=t_{\rm sat}$.
In order to describe this phase correctly, one has to
retain the time derivative of $H_{\rm f}$, as will be done in
\Secs{SDynamicalQuenching}{Sfinal} in the framework of the dynamical
quenching model.
At early times resistive effects have not yet played a role, so the
total magnetic helicity must be approximately zero, and this means that
a helical large scale field must be smaller than $B_{\rm eq}$;
see \Sec{Searly} for details.

\subsection{Alpha effect versus inverse cascade}
\label{AlphaVersusInverseCascade}

The process outlined above can be interpreted in two different ways: inverse
cascade of magnetic helicity and/or $\alpha$ effect. The two are similar in that
they tend to produce magnetic energy at scales larger than the energy-carrying
scale of the turbulence. As can be seen from
\Figs{Fpspec_growth_passot}{Fpspec_pm_satkin}, the present simulations support
the notion of {\it nonlocal} inverse transfer \cite{B01}.
However, this is not really an
inverse cascade in the usual sense, because there is no sustained flux of
energy through wavenumber space, as in the direct Kolmogorov cascade.
Instead, there is just a bump traveling in wavenumber space from a $k$
that is already smaller than $k_{\rm f}$ to even smaller values of $k$.
In that respect, the present simulations seem to differ from the
interpretation of PFL based on the
EDQNM closure approximation \cite{PFL76}.

The other interpretation is in terms of the $\alpha$ effect. We recall that
for $\alpha^2$ dynamos
there is a wavenumber $k_{\max}$ where the growth of the large scale field
is fastest; see \Sec{Sa2_model}. For reasonable estimates,
$k_{\max}$ coincides with the position of the secondary bump in the
spectrum; see Ref.~\cite{B01}, Sect.~3.5.
This can be taken as evidence in favor of
the $\alpha$ effect. In the nonlinear regime, the secondary bump travels
to the left in the spectrum (i.e.\ toward smaller $k$). In the
EDQNM picture this has to do with the equilibration of kinetic and
current helicities at progressively smaller wavenumbers, which then
leads to saturation at that wavenumber, but permits further
growth at smaller wavenumbers until equilibration occurs, and so forth.
Another interpretation is simply that, after $\alpha$ is quenched
to a smaller value, the magnetic spectrum now peaks at
$k_{\max}=\alpha/(2\eta_{\rm T})$, which is now also smaller.
At this lower wavenumber the spectrum is still not fully saturated;
the field continues to grow here until equilibration
is attained also at that scale.

\subsection{Nonlinear $\alpha$ effect}
\label{helicity_aquenching}

It is quite clear that $\alpha$ must somehow depend on the strength of
the mean field, $\meanBB$.
By considering the effect of $\meanBB$ on the correlation tensor of the
turbulence it has been possible to derive a correction to $\alpha$
of the form \cite{RS75,Mof72,Rue74}
\EQ
\alpha=\alpha_{\rm K}\left(1-\meanBB^2/B_{\rm eq}^2\right)\quad
\mbox{(for $|\meanBB|\ll B_{\rm eq}$)}.
\label{alphaquench}
\EN
For practical applications, to make sure that $|\alpha|$ decreases with
increasing field strength and to prevent $\alpha$ from changing sign,
\Eq{alphaquench} is often replaced by the fit formula \cite{IR77}
\EQ
\alpha={\alpha_{\rm K}\over1+\meanBB^2/B_{\rm eq}^2}
\quad\mbox{(conventional quenching)}.
\label{alphafitformula}
\EN
It should also be noted that fully nonlinear expressions exist; see, e.g.,
Ref.~\cite{RK93}.
However, it has long been noted that in the astrophysically relevant case,
$R_{\rm m}\gg1$, the magnitude of the fluctuating field is likely to
exceed that of the mean field, i.e.\ $\bb^2/\meanBB^2\gg1$.
Indeed, a naive application of {\it kinematic} mean field theory suggests that
\cite{KR80,Zeldovich57,KrauseRoberts73}
\EQ
\bra{\bb^2}/\bra{\meanBB^2}=R_{\rm m}\quad\mbox{(kinematic theory)}.
\label{flucts}
\EN
This result is a direct consequence of flux freezing (\Sec{SStretching})
during the compression of a uniform field of strength $\meanB$ and scale
$L$ into a sheet of `skin' thickness $d=LR_{\rm m}^{1/2}$.
It can also be derived under more general assumptions, but then only
in the two-dimensional case \cite{DurhamReview}; see \App{ZeldovichRel}.
Further in the three dimensional case, the small scale
field dynamo can also generate $\bb$ at a rate much faster than 
$\meanBB$, and unrelated to the strength of the mean field \cite{BF05}.

Nevertheless, the above argument has been used to suggest that 
the quenching formula \eq{alphafitformula} should take the small 
scale field, obeying the relation in \Eq{flucts}, into account.
Using \Eq{flucts}, this then leads to
\EQ
\alpha={\alpha_{\rm K}\over1+R_{\rm m}\meanBB^2/B_{\rm eq}^2}
\quad\mbox{(catastrophic quenching)}.
\label{VC92formula}
\EN
This formula was first suggested by Vainshtein and Cattaneo \cite{VC92}.
Only recently it has become clear that, even though \Eq{flucts} is actually no
longer valid in the nonlinear regime, \Eq{VC92formula} can indeed emerge
in a more rigorous analysis under certain circumstances \cite{BB02}.

The problem with \Eq{VC92formula} is that $\alpha$ becomes strongly
suppressed already for $\meanBB^2/B_{\rm eq}^2\ll1$.
Conversely, for the sun where
$\meanBB^2/B_{\rm eq}^2\approx1$, this means that $\alpha$ would be
negligibly small.
Equation~\eq{VC92formula} is therefore sometimes referred to as catastrophic
quenching formula.

Given the potentially catastrophic outcome of mean field theory when
applying \Eq{VC92formula} to astrophysically relevant situations,
the problem of $\alpha$ quenching has begun to attract significant
attention in the last few years.
Considerable progress has recently been made by restricting attention to
the simplest possible system that still displays an $\alpha$ effect,
but that is otherwise fully nonlinear.

\subsection{Determining alpha quenching from isotropic box simulations}

The issue of (catastrophic)
$\alpha$ quenching was preceded by the related issue of
$\eta_{\rm t}$ quenching.
Indeed, already 30 years ago concerns have been expressed \cite{Pid72}
that turbulent diffusion might not work when the magnitude of the field
is strong.
A serious argument against catastrophic $\eta_{\rm t}$ quenching came from
the measurements of decay times of sunspots where the magnetic field is
strong and yet able to decay almost on a dynamical time scale.
Estimates for the turbulent magnetic diffusivity in sunspots suggest
$\eta_{\rm t}\approx10^{11}\cm^2\s^{-1}$ \cite{KR75,PMI97,PvDG97,RK00}.

Quantitatively,
catastrophic (i.e.\ $R_{\rm m}$-dependent) $\eta_{\rm t}$ quenching was
first suggested based on two-dimensional simulations with an initially
sinusoidally modulated large scale magnetic field in the plane of the
motions \cite{CV91}.
However, these results have to be taken with caution, because constraining
the field to be in the plane of the motions is artificial
in that the interchange of field lines is then impossible.
Field lines that undergo interchanging motions
can remain nearly straight, so not much
work is involved and one would therefore not necessarily
expect catastrophic quenching if the flow were allowed to be fully
three-dimensional.
This has been confirmed using both closure models \cite{GD} and
three-dimensional simulations \cite{NGS94}.
Nevertheless, the possibility
of catastrophic quenching of $\eta_{\rm t}$ is not completely ruled out.
Simulations are not yet conclusive, as discussed below in
\Sec{ComparisonSimulations}.
However, unlike the case of catastrophic $\alpha$ quenching, which can
be explained as a consequence of magnetic helicity conservation,
there is no similar argument for an $R_{\rm m}$ dependent quenching
of $\eta_{\rm t}$.

The issue of catastrophic $\alpha$ quenching was originally motivated
by analogy with catastrophic $\eta_{\rm t}$ quenching in
two dimensions \cite{VC92}, but then backed
up by simulations with an imposed field \cite{CH96}, so $\alpha$
is calculated as the ratio of the resulting electromotive force and
the imposed magnetic field; cf.\ \Sec{SimulationsTransport}.
Another technique to measure $\alpha$
is to modify or remove a component of the mean field
in an otherwise self-consistent simulation and to describe the response of
the system in terms of $\alpha$ effect and turbulent diffusion \cite{B01}.
This technique is easily explained by looking, for example, at the
$x$-component of the $\alpha^2$ dynamo equation (\Sec{Sa2_model}),
\EQ
{\partial\meanB_x\over\partial t}=
-\alpha{\partial\meanB_y\over\partial z}
+(\eta+\eta_{\rm t}){\partial^2\meanB_x\over\partial z^2}.
\label{Bxrecovery}
\EN
If, at some point in time, the $x$ component of the mean field
is removed, i.e.\ $B_x\to B_x-\meanB_x$, then \Eq{Bxrecovery}
describes the immediate recovery of $\meanB_x$.
The recovery is described by the first term on the rhs of \Eq{Bxrecovery},
and the rate of recovery is proportional to $\alpha$.
This method allows an estimate not only of $\alpha$,
but also of $\eta_{\rm t}$ by measuring the simultaneous
temporary reduction of $\meanB_y$.
It turns out \cite{B01} that both methods give comparable results
and confirm the catastrophic quenching results \eq{VC92formula}.
This result will later be understood as a special case of the dynamical
quenching formula in the nearly steady limit for fully force-free
(force-free) fields; see \Eq{bothquenched12}.

Yet another method is to impose a nonuniform field that is a solution
of the mean field dynamo equations -- for example a Beltrami field in
the case of periodic box.
By changing the wavelength of the Beltrami field, one can determine
both $\alpha$ and $\eta_{\rm t}$ simultaneously.
This method has been used in connection with the underlying flow field
of the Karlsruhe dynamo experiment \cite{PlunianRadler02,RB03}.

\subsection{Dynamo waves in simulations with shear}
\label{SSimulationsWithShear}

The fact that dynamos exhibit cyclic behavior when there is shear
is not surprising if one recalls the type of solutions that are
possible for $\alpha\Omega$ dynamos (\Sec{Sao_model}).
On the other hand, until only a few years ago the concept of mean field
theory was only poorly tested and there was enough reason to doubt its
validity especially in the nonlinear regime.
Even in the linear regime the relevance of mean field theory
has been doubtful because the mean field can be much weaker than the
fluctuating field \cite{KA92,Hoyng87}.

\begin{figure}[t!]\begin{center}
\includegraphics[width=.9\textwidth]{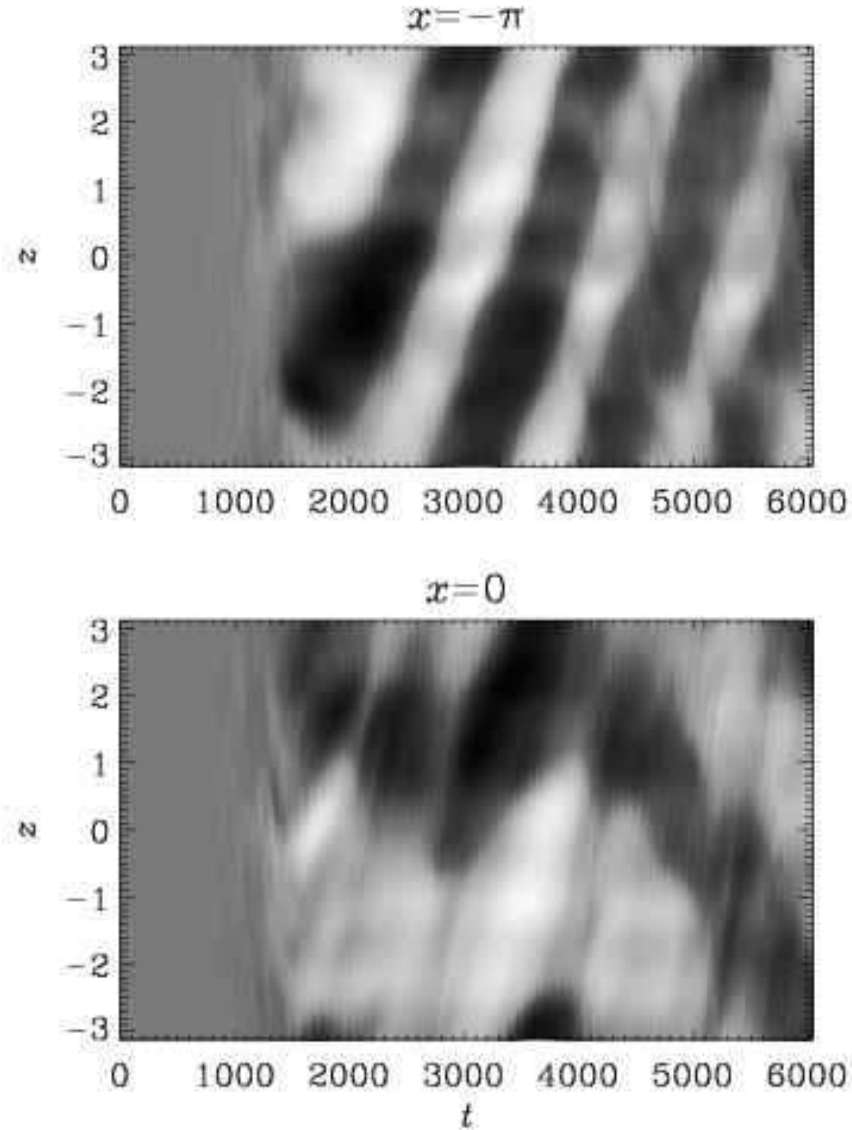}
\end{center}\caption[]{
Space-time diagram of the mean toroidal field at $x=-\pi$ (negative
local shear) and $x=0$ (positive local shear). Dark (light) shadings
refer to negative (positive) values. Note the presence of dynamo
waves traveling in the positive (negative) $z$-direction for negative
(positive) local shear
(from Ref.~\cite{BBS01}).
}\label{Fpbutter_eq}\end{figure}

Given all these reservations about the credibility of mean field theory
it was a surprise to see that cyclic dynamo action does actually
work \cite{BBS01}.
As in the simulations without shear, the emergence of a large scale
field is best seen in the nonlinear regime; see \Fig{Fpbutter_eq}.
The reason is that prior to saturation several different modes
may be excited, while in the nonlinear regime most of the modes are suppressed
by the most dominant mode.
Nonlinearity therefore has a `self-cleaning' effect \cite{B01,Ax_Ks00}.

\begin{figure}[t!]\begin{center}
\includegraphics[width=.9\textwidth]{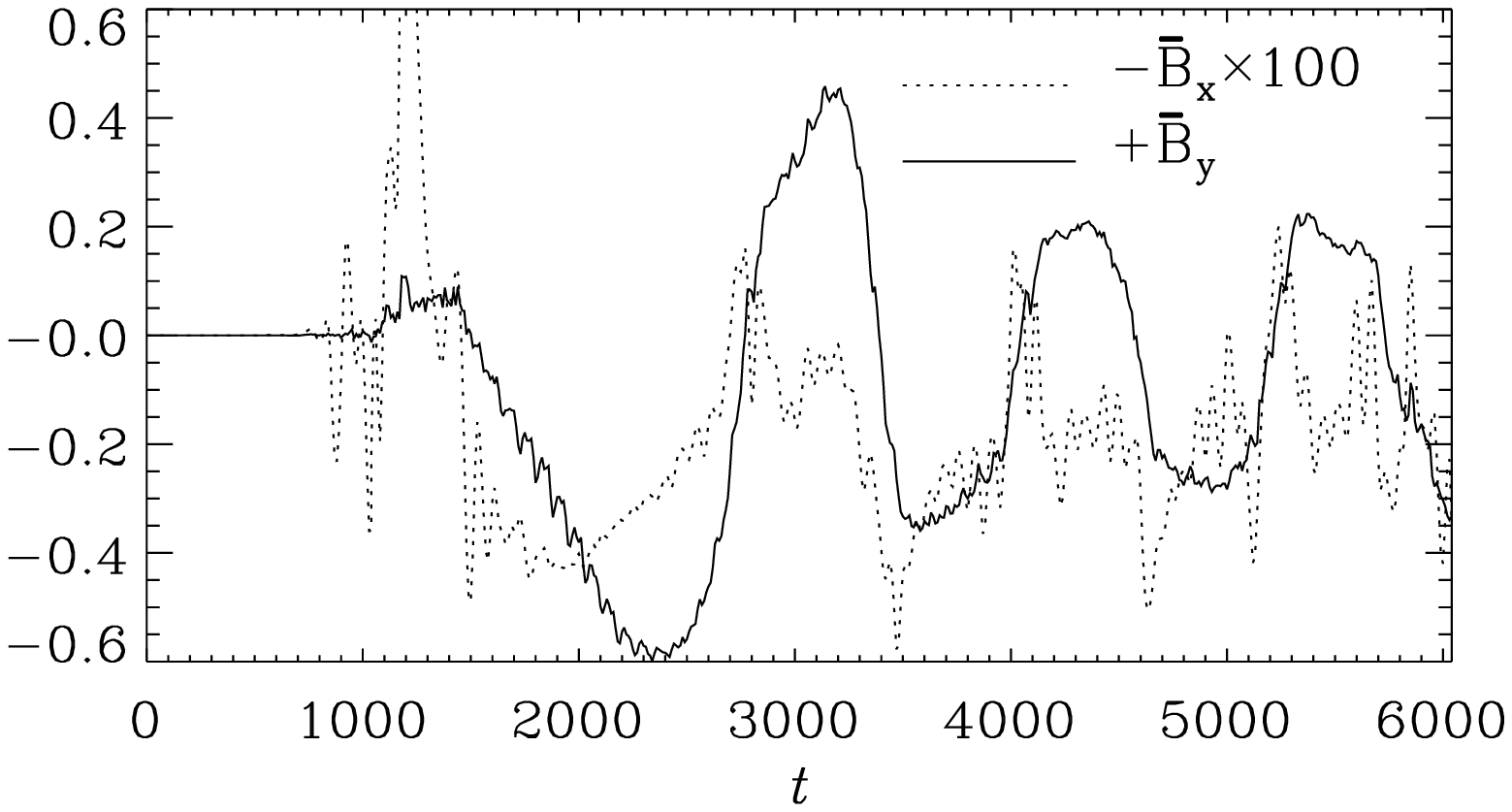}
\end{center}\caption[]{
Evolution of $\overline{B}_x$ and $\overline{B}_y$ at $x=-\pi$ and
$z=0$. Note that $\overline{B}_x$ has been scaled by a factor $-100$.
Here the overbars denote only a one-dimensional average over the
direction of shear, $y$.
(Adapted from Ref.~\cite{BBS01}.)
}\label{Fpbutter_line_eq}\end{figure}

The dynamo exhibits a certain phase relation between poloidal and
toroidal fields (see \Fig{Fpbutter_line_eq}).
This phase relation is quite similar
to what is expected from a corresponding mean field model
(see \Sec{PhaseRelations}).
Comparison with $\alpha$ quenching models in the same geometry
(see Fig.~9 of Ref.~\cite{BBS01}) produces similarly anharmonic oscillations,
as seen in \Fig{Fpbutter_line_eq}.
This strongly suggests that the simulation results can basically be described
in terms of the mean field concept.

An important question for astrophysical applications is which of the
properties of the dynamo depend on resistivity.
Certainly the late saturation behavior of the field does depend
on resistivity and satisfies the `magnetic helicity constraint'
embodied by \Eq{helconstraint}; see Fig.~8 of Ref.~\cite{BBS01}.
Subsequent simulations for different values of the magnetic Reynolds
number also seem to confirm that the cycle frequency,
$\omega_{\rm cyc}$, scales resistively and that
$\omega_{\rm cyc}/(\eta k_1^2)\approx\const={\cal O}(10)$
for $u_{\rm rms}/(\eta k_1)$ ranging from 30 to 200;
see Table~5 of Ref.~\cite{BDS02}.
On the other hand, in all cases the large scale magnetic energy exceeds
the kinetic energy by a factor that is between 20 (for smaller $R_{\rm m}$)
and 60 (for larger $R_{\rm m}$).
It is therefore clear that these models are in a very different parameter
regime than the solar dynamo where the energy of the mean field is at
most comparable to the kinetic energy of the turbulence.
Furthermore, all turbulent transport coefficients are necessarily strongly
suppressed by such strong fields even if the actual suppression is not
explicitly dependent on the magnetic Reynolds number.
A more detailed interpretation of these simulation data has been
possible by using a dynamical quenching model that will be discussed
in \Sec{ComparisonSimulations}.

\subsection{The magnetic helicity constraint in a closure model}
\label{ClosureModel}

The magnetic helicity constraint is quite general and independent of
any mean field or other model assumptions.
All that matters is that the flow possesses kinetic helicity.
The generality of the magnetic helicity constraint allows one to eliminate
models that are incompatible with magnetic helicity conservation.
In previous sections we discussed a simple unified model of
large and small scale dynamo, generalizing the Kazantsev model
to include helical velocity correlations. In 
\cite{subamb,sub99} this model was further extended to include
ambipolar diffusion as a model nonlinearity. This is a useful
toy model since the magnetic field still obeys helicity conservation
even after adding ambipolar drift. So it is interesting 
to see how in this model the $\alpha$ effect is being regulated by 
the change in magnetic helicity, and whether there are similarities 
to the dynamical quenching model under full MHD.

This issue was examined in Ref.~\cite{Ax_Ks00} by solving the
moment equations derived in \cite{subamb,sub99} numerically.
We have given in the \App{kazantsev} the derivation of the
moment equations for the magnetic correlations in the
presence of ambipolar drift. In deriving the moment equations with
ambipolar drift nonlinearity, one encounters again a closure
problem. The equations for the second moment, fortunately,
contains only a fourth order correlator of the magnetic field.
This was closed in \cite{subamb,sub99} assuming that 
the fourth moment can be written as a product of second moments.
The equations for the longitudinal correlation function $M_{\rm L}(r,t)$ 
and the correlation function for magnetic helicity density, $H(r,t)$,
then are the same as \Eqs{mlequn}{mhequn}, except
for additions to the coefficients $\eta_T(r)$ and $\alpha(r)$. We have 
\EQ
{\partial M_{\rm L} \over \partial t} = {2\over r^4}{\partial \over \partial r}
\left(r^4 \eta_{\rm N}(r) {\partial M_{\rm L} \over \partial r}\right)
+ G M_{\rm L} + 4\alpha_{\rm N}(r) C,
\label{closure1}
\EN
\EQ
{\partial H \over \partial t} =-2\eta_{\rm N} C +\alpha_{\rm N} M_{\rm L},
\label{closure2}
\EN
where
\EQ
\alpha_{\rm N}=\alpha(r) + 4aC(0,t), \quad 
\eta_{\rm N}=\eta_T(r) + 2 a M(0,t) .
\EN
Note that at large scales
\EQ
\alpha_\infty\equiv\alpha_{\rm N}(r\rightarrow\infty)=
-{\textstyle{1\over3}}\tau\bra{\oo\cdot\uu}
+{\textstyle{1\over3}}\tau_{\rm AD}\bra{\JJ\cdot\BB}/\rho_0,
\label{asuppress}
\EN
\EQ
\eta_\infty\equiv\eta_{\rm N}(r\rightarrow\infty)=
{\textstyle{1\over3}}\tau\bra{\uu^2}
+{\textstyle{1\over3}}\tau_{\rm AD}\bra{\BB^2}/\mu_0\rho_0,
\EN
where $\tau_{\rm AD}=2a\rho_0$.
Here, angular brackets denote volume averages over all space.
This makes sense because the system is homogeneous.
Expression \eq{asuppress} for $\alpha_\infty$ is very
similar to the $\alpha$ suppression formula due to the current helicity
contribution first found in the EDQNM treatment by \cite{PFL76} (see
\Sec{edqnm}). The expression for $\eta_\infty$ has the nonlinear addition
due to ambipolar diffusion.\footnote{A corresponding term from the
small scale magnetic field drops out as a consequence of the requirement
that the turbulent velocity be solenoidal. By contrast, the ambipolar
drift velocity does not obey this restriction.}
It is important to point out that the closure model, even including the above
nonlinear modifications, explicitly satisfies helicity conservation. 
This can be seen by taking $r\to 0$ in \eq{closure2}. We get
\EQ
\dot{H}(0,t) = -2\eta C(0,t), \quad 
\dd\bra{\AAA\cdot\BB}/\dd t=-2\eta\bra{\JJ\cdot\BB},
\label{helambi}
\EN  
where we have used the fact that $\bra{\AAA\cdot\BB}=6H(0,t)$, 
and $\bra{\JJ\cdot\BB}=6C(0,t)$. It is this fact that makes
it such a useful toy model for full MHD (see below). 

\begin{figure}[t!]\begin{center}
\includegraphics[width=.80\textwidth]{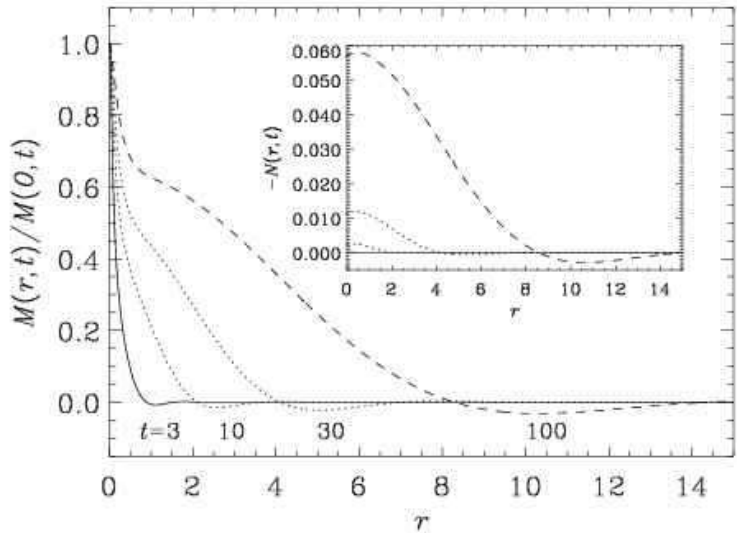}
\end{center}\caption[]{
Evolution of magnetic correlation function $M_L$ (denoted by
$M(r,t)$ in this figure) for different times, for $\eta=10^{-3}$.
The correlation function of the magnetic helicity (denoted in this figure
by $N(r,t)$), is shown in the inset.
$\eta=10^{-3}$.
}\label{Fpcors}\end{figure}

\begin{figure}[t!]\begin{center}
\includegraphics[width=.80\textwidth]{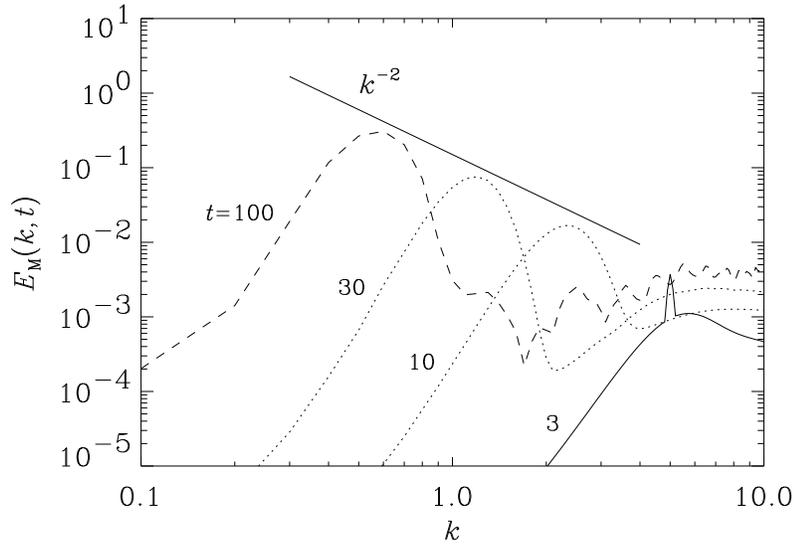}
\end{center}\caption[]{
Evolution of magnetic energy spectra.
Note the propagation of magnetic helicity and energy to progressively
larger scales. The $k^{-2}$ slope is given for orientation.
Note the similarities with \Figs{PFLfig}{Fpspec_pm_satkin}.
}\label{Fpcor}\end{figure}

Adopting functional forms of $T_{\rm L}(r)$ and $F(r)$ constructed
from the energy and helicity spectra resembling Run 3 of Ref.~\cite{B01},
\Eqs{closure1}{closure2} were solved numerically \cite{Ax_Ks00}.
In the absence of kinetic helicity, $F=0$, and without nonlinearity, $a=0$,
the standard small scale dynamo solutions are recovered.
The critical magnetic Reynolds
number based on the forcing scale is around 60 
(here we have not converted the scale into a wavenumber). 
In the presence of kinetic helicity
this critical Reynolds number decreases, confirming the general result
that kinetic helicity promotes dynamo action \cite{sub99,kim_hughes97}.
In the presence of nonlinearity
the exponential growth of the magnetic field terminates when the
magnetic energy becomes large. After that point the magnetic energy continues
however to increase nearly linearly. Unlike the case of the periodic box
the magnetic field can here extend to larger and larger
scales; see \Fig{Fpcors}. The corresponding magnetic energy spectra,
\EQ
E_{\rm M}(k,t)={1\over\pi}\int_0^L\,(kr)^3 M_{\rm L}(r,t) \,j_1(kr)\,\dd k,
\EN
are shown in \Fig{Fpcor}.

The resulting magnetic field is strongly helical and the magnetic
helicity spectra (not shown) satisfy $|H_{\rm M}(k,t)|\la(2/k)E_{\rm M}(k,t)$. 
One sees in \Fig{Fpcor} the development of a helicity wave traveling toward
smaller and smaller $k$. This is just as in the EDQNM closure model
\cite{PFL76} (see \Fig{PFLfig}) and in the simulations
of the full MHD equations \cite{B01,BS02} (see \Fig{Fpspec_pm_satkin}). 

\begin{figure}[t!]\begin{center}
\includegraphics[width=.70\textwidth]{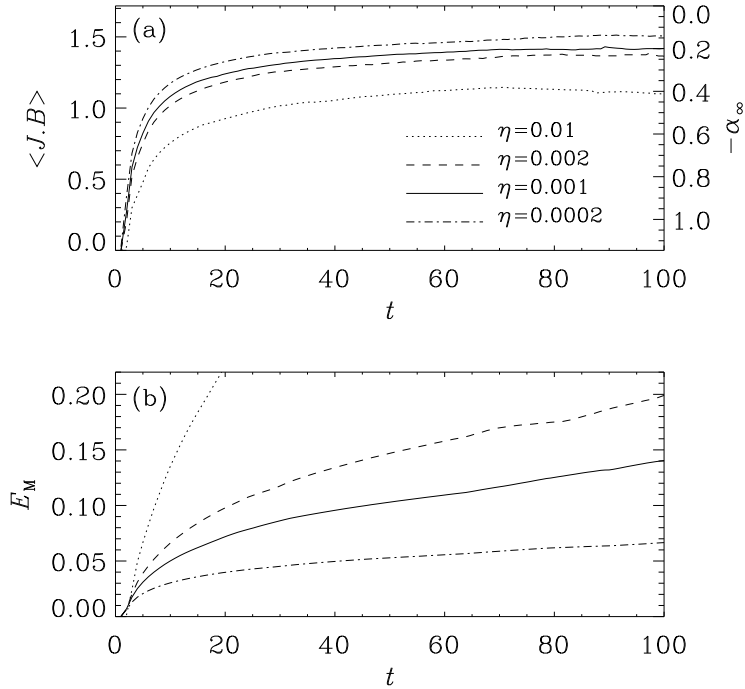}
\end{center}\caption[]{
(a) Evolution of $\bra{\JJ\cdot\BB}$ for different values of $\eta$.
The corresponding value of $\alpha_\infty$ is shown on the right hand
side of the plot. (b) The evolution of magnetic energy for the same
values of $\eta$.
}\label{Fpml}\end{figure}

In the following we address the question
of whether or not the growth of this large scale field depends on
the magnetic Reynolds number (as in \cite{B01}). To a
good approximation the wavenumber of the peak is given by
\EQ
k_{\rm peak}(t)\approx\alpha_\infty(t)/\eta_\infty(t).
\EN
This result is familiar from mean field dynamo theory
(see also Ref.~\cite{sub99}), where the marginal state
with zero growth rate, has $k = \alpha/\eta_t$, while
the fastest growing mode has $k = \alpha/2\eta_t$; see \Sec{Sa2_model}.
In our model problem, the large scale field, 
seems to go to a quasi-static state (which has no $1/2$ factor)
rather than the fastest growing mode, and
then evolve slowly (on the resistive time scale), through a 
sequence of such states. This evolution to smaller and smaller wavenumbers
is also consistent with simulations (\cite{B01}, Sect.~3.5). Note that here
$k_{\rm peak}$ decreases with time because $\alpha_\infty$ tends to a
finite limit and $\eta_\infty$ increases; see \Sec{Sa2_model}.
(This is not the case in the
box calculations where $k_{\rm peak}\ge2\pi/L$.)

As we saw from \Eq{helambi} the magnetic helicity,
$\bra{\AAA\cdot\BB}=6\,H(0,t)$, can only change if there is
microscopic magnetic diffusion and finite current helicity,
$\bra{\JJ\cdot\BB}=6\,C(0,t)$. In \Fig{Fpml} we show that, 
after some time $t=t_{\rm s}$, the current helicity
$\bra{\JJ\cdot\BB}$ reaches a finite value. This value increases somewhat
as $\eta$ is decreased. In all cases, however, 
$\tau_{\rm AD}\bra{\JJ\cdot\BB}/\rho_0$
stays below $\tau\bra{\oo\cdot\uu}$, so that $|\alpha_\infty|$
remains finite; see \eq{asuppress}. A constant $\bra{\JJ\cdot\BB}$
implies from \eq{helambi} that $\bra{\AAA\cdot\BB}$ grows 
linearly at a rate proportional to $\eta$. 
However, since the large scale field is helical, and since
most of the magnetic energy is by now (after $t=t_{\rm s}$) in the large
scales, the magnetic energy is proportional to $\bra{\BB^2}\approx k_{\rm
peak}\bra{\AAA\cdot\BB}$, and can therefore only continue to grow at a
resistively limited rate, see \Fig{Fpml}.
It is to be emphasized that this explanation is analogous to that
given in Ref.~\cite{B01} and \Sec{NonlinearBehavior} for the full MHD case;
the helicity constraint is independent of the nature of the feedback!

These results show that ambipolar diffusion (AD) provides a useful model
for nonlinearity, enabling analytic (or semi-analytic)
progress to be made in understanding nonlinear
dynamos. There are two key features that are shared both by this model and
by the full MHD equations: (i) large scale fields are the result of a
{\it nonlocal}
inverse cascade as described by the $\alpha$ effect, and (ii) after some
initial saturation phase the large scale field
continues to grow at a rate limited
by magnetic diffusion. This model also illustrates
that it is helicity conservation that is at the heart
of the nonlinear behavior of large scale dynamos; qualitatively
similar restrictions arise even for very different nonlinear
feedback provided the feedback obeys helicity conservation. 

\subsection{Nonhelical large scale turbulent dynamos with shear}
\label{TurbulenceAndShear}

Much of the discussion on large scale dynamos has focused on the
$\alpha$ effect.
In recent years attention has been drawn to the possibility of producing
large scale fields by other effects such as the shear--current or
$\meanWW\times\meanJJ$ effect \cite{Roga+Klee03,Roga+Klee04,RadStep05}.
As remarked in \Sec{ShearCurrentEffect} (where this new term was
denoted by $\ddelta$) this effect is related to
R\"adler's \cite{Rae69} $\OO\times\meanJJ$ effect in that it has
the same functional form.
At least two more possibilities have been offered for explaining large
scale fields in shearing environments without invoking kinetic helicity.
One is the incoherent $\alpha$ effect \cite{VB97} and the other one
is the Vishniac and Cho flux \cite{VC01}.
Both effects are related to the $\alpha$ effect, but there is no kinetic
helicity, so there can only be helicity fluctuations (former case) or
there can be a magnetic contribution to the $\alpha$ effect (latter case).

In simulations of realistic systems it is easily possible that a number
of effects operate simultaneously, so one cannot be sure that the
$\alpha$ effect is not also contributing.
This is different in systems where turbulence and shear are driven
by body forces such as those discussed in \Sec{SSimulationsWithShear}.

\begin{figure}[t!]\begin{center}
\includegraphics[width=.95\textwidth]{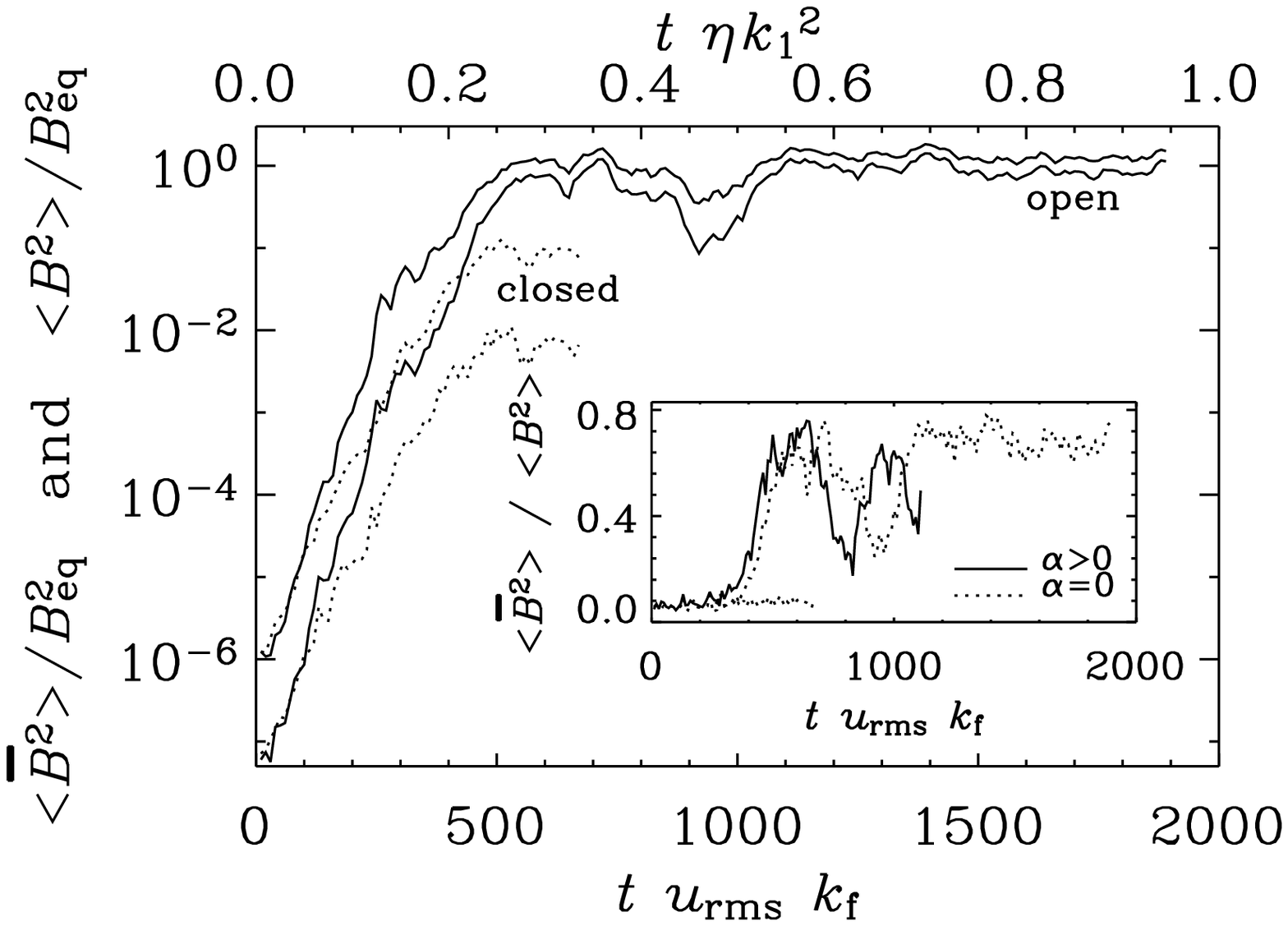}
\end{center}\caption[]{
Evolution of the energies of the total field $\bra{\BB^2}$ and of
the mean field $\bra{\meanBB^2}$, in units of $B_{\rm eq}^2$,
for runs with non-helical forcing
and open or closed boundaries; see the solid and dotted lines, respectively.
The inset shows a comparison of the ratio $\bra{\meanBB^2}/\bra{\BB^2}$
for nonhelical ($\alpha=0$) and helical ($\alpha>0$) runs.
For the nonhelical case the run with closed boundaries is also
shown (dotted line near $\bra{\meanBB^2}/\bra{\BB^2}\approx0.07$).
Note that saturation of the large scale field occurs on a
dynamical time scale; the resistive time scale is given on the
upper abscissa.
Adapted from Ref.~\cite{B05}.
}\label{pmean_comp}\end{figure}

Meanwhile, using a modified toroidal shear profile of the form
\cite{BrandenburgSandin2004}
\EQ
\meanUU\approx(0,\cos k_1x \cos k_1z,0),
\EN
it has been possible to produce strong mean fields \cite{B05,BHKS05}.
Here, mean fields are defined as toroidal averages.
In \Fig{pmean_comp} we show the evolution of the normalized magnetic
energy of the total field $\bra{\BB^2}/B_{\rm eq}^2$, and of the
mean field $\bra{\meanBB^2}/B_{\rm eq}^2$.
Both components increase first exponentially until saturation sets in.
However, the ratio between the two, $\bra{\meanBB^2}/\bra{\BB^2}$,
which is shown in the inset of \Fig{pmean_comp}, is rather small
($\approx0.07$) during the early kinematic phase, but then it
increases to values of around 0.7.
In the inset, comparison is made with the case where the turbulence
is driven with negative kinetic helicity, giving rise to a positive
$\alpha$ effect (the graph is therefore denoted by ``$\alpha>0$'').
It turns out that in both cases (with and without helicity) the
energy in the mean field is comparable

A more dramatic difference is seen in the case of closed (perfectly
conducting) outer boundaries.
Although the initial exponential increase is almost equally fast, the
field now saturates at a much lower level and it completely lacks a
mean field, i.e.\ $\bra{\meanBB^2}/\bra{\BB^2}\approx0.07$ even during
the saturated state.
This is a striking demonstration of the importance of allowing for
magnetic helicity fluxes out of the domain.
It is important that, due to the presence of shear, such fluxes can
already be driven inside the domain by the Vishniac and Cho flux,
as was demonstrated earlier \cite{BrandenburgSandin2004,KS_AB04}.
Without internal helicity transport, e.g. in the absence of shear,
open boundaries alone are not sufficient to alleviate the magnetic
helicity constraint \cite{BD01}.
We return to this issue in \Sec{OpenBoundaries}.

\section{Magnetic helicity in mean field models}
\label{MagneticHelicityMeanFieldModels}

\subsection{General remarks}

We have seen from both numerical simulations and closure models
that magnetic helicity conservation strongly constrains the
evolution of large scale fields. Due to helicity conservation,
large scale fields are able to grow eventually only on the resistive
time scale. Note that the magnetic helicity evolution does not
explicitly depend on the nonlinear backreaction due to the Lorentz
force. It merely depends on the induction equation.
It therefore provides a strong constraint on the nonlinear evolution
of the large scale field.
The effects of this constraint need to be incorporated
into any treatment of mean field dynamos.
This will be the aim of the present section,
where we solve simultaneously the mean field dynamo equation
(with the turbulent coefficients determined by the nonlinear 
backreaction) together with helicity conservation equations.  

We will see that the magnetic helicity conservation
equations for the mean and the turbulent fields, as well
as of course the mean field dynamo equation,
involve understanding the mean turbulent EMF, $\meanemf$,
in the nonlinear regime.
So we will need to model the nonlinear effects on $\meanemf$,
taking into account nonlinear backreaction effects
of the Lorentz force due to both mean and fluctuating fields.

There are two quite different forms of feedback that can arise.
One is the effect of the dynamo-generated mean field on the correlation
tensor of the turbulence, $\overline{u_i u_j}$.
The growing mean field could
cause the suppression of the $\alpha$ effect and turbulent
diffusion. Such modifications of the turbulent transport coefficients
have been calculated since the early seventies \cite{Mof72,Rue74},
adopting usually an approximation (random waves or FOSA), which linearizes the
relevant equations in the fluctuations. This approach missed an important
additional ingredient in the nonlinear backreaction due to Lorentz forces:
modifications to $\meanemf$ that involve
the fluctuating fields themselves. These arise in calculating
$\overline{\uu\times\bb}$ from terms involving the 
correlation of $\bb$ and the Lorentz force in the momentum
equation (see below).
In particular, the $\alpha$ effect gets renormalized 
in the nonlinear regime by the addition of a term
proportional to the current helicity of the fluctuating field.
This is an important effect which is crucial for the correct
description of the dynamo saturation,
but one which has been missed in much of the earlier work.
The renormalization of the $\alpha$ effect is quite general and
can occur even when the large scale dynamo effect does not involve
an $\alpha$ effect in the kinematic regime.
Examples are the $\OO\times\meanJJ$ and $\meanWW\times\meanJJ$
(or shear current) effects discussed in \Sec{ShearCurrentEffect}.

The full problem of solving the induction equation and
the momentum (Navier-Stokes) equation including the
Lorentz force simultaneously, is a formidable one.
One either takes recourse to numerical simulations or
uses rather more uncertain analytic approximations in
numerical mean field models.
The analytic treatments of the backreaction
typically involve the quasi-linear approximation or a closure
scheme to derive corrections to the mean field dynamo coefficients. 
For example the EDQNM closure suggests that
the $\alpha$ effect is renormalized by the addition
a term proportional to the current helicity of the small scale
field $\overline{\jj\cdot\bb}$, while the turbulent diffusion
is left unchanged. It would be useful to understand
this result in a simpler context. For this purpose
it can be illuminating to examine a simple heuristic treatment
of the effects of the backreaction.
However, since such a treatment is not rigorous, we defer the discussion
to \App{qnmod} and move straight to the more convincing derivation in
terms of the minimal tau approximation.

\subsection{The minimal tau approximation: nonlinear effects}
\label{tauApproxNonlin}

In this section we discuss the main aspects of the
minimal tau approximation (MTA).
Again, for the purpose of clarity, we restrict ourselves to the assumption of
isotropy \cite{BF02b,BraKapMoh03,BF03},
but the method can readily be and has been applied
to the anisotropic case \cite{RKR03,Roga+Klee00}.
A full treatment of inhomogeneous and anisotropic turbulence
is given in \Sec{Revisit}.

In order to incorporate the evolution equations for the fluctuating parts
in the expression for $\meanemf$ one can just
calculate its time derivative, rather than calculating $\meanemf$
itself \cite{BF02b}.
This way one avoids the approximate integration in
\Eqs{FOSA_alpha}{FOSA_eta}.
Thus, one calculates
\EQ
{\partial\meanemf\over\partial t}
=\overline{\dot{\uu}\times\bb}
+\overline{\uu\times\dot{\bb}},
\EN
where dots denote partial differentiation with respect to $t$.
The dominant contributions in these two terms are (using $\rho_0=\mu_0=1$)
\EQ
\overline{\dot{\uu}\times\bb}
=\onethird\overline{\jj\cdot\bb}\;\meanBB
+\overline{(\jj\times\bb)\times\bb}
-\overline{(\oo\times\uu)\times\bb}
-\overline{\nab p\times\bb}+\ldots,
\label{triplefirst}
\EN
\EQ
\overline{\uu\times\dot{\bb}}
=-\onethird\overline{\oo\cdot\uu}\;\;\meanBB
-\onethird\overline{\uu^2}\;\;\meanJJ
+\overline{\uu\times\nab\times(\uu\times\bb)}+\ldots,
\label{triplesecond}
\EN
where the first term on the right hand side of \Eq{triplefirst} and
the first two terms on the right hand side of \Eq{triplesecond} are the
usual quadratic correlation terms; all other terms are triple correlations.
(For a detailed derivation see \Sec{Revisit} below.)
Thus, we can write
\EQ
{\partial\meanemf\over\partial t}
=\tilde\alpha\meanBB-\tilde\eta_{\rm t}\meanJJ+\meanTT,
\EN
where $\meanTT$ are the triple correlation terms, and
\EQ
\tilde\alpha=-\onethird\left(\overline{\oo\cdot\uu}
-\overline{\jj\cdot\bb}\right),\quad\mbox{and}\quad
\tilde\eta_{\rm t}=\onethird\overline{\uu^2},
\label{alptilde+etatilde}
\EN
are turbulent transport coefficients that are
related to $\alpha$ and $\eta_{\rm t}$, as used in \Eq{alpha_eta}, via
$\alpha=\tau\tilde\alpha$ and $\eta_{\rm t}=\tau\tilde\eta_{\rm t}$.
Note that, at this level of approximation,
there are no free parameters in the expression for $\tilde\alpha$,
neither in front of $\overline{\oo\cdot\uu}$ nor in front of
$\overline{\jj\cdot\bb}$.
However, for strong magnetic fields, there could be quenching
functions, $g_{\rm K}(\meanBB)$ and $g_{\rm M}(\meanBB)$, in front
of both terms \cite{KMRS02}.
We should also point out that in the above derivation,
$\uu$ and $\bb$ refer to the {\it actual} small scale velocity 
and magnetic fields, and not to any perturbed field.
(This could also include any `nonhelical' SSD generated $\bb$, although we
see that this does not renormalize $\eta_{\rm t}$ and may not
contribute to the current helicity term at leading order.)
The full derivation for the more general case of slow rotation, 
weak stratification and for general magnetic and kinetic spectra 
(possibly with a $k$-dependent $\tau$), is given the \Sec{Revisit}.

As we have already emphasized earlier, the crucial step is that now, 
unlike the case of FOSA or the heuristic treatment, the triple
correlators are {\it not} neglected, but their sum is assumed to be a negative
multiple of the second order correlator, i.e.\ $\meanTT=-\meanemf/\tau$.
This assumption has been checked numerically (see below), and
a similar assumption has recently been verified
for the case of passive scalar diffusion \cite{BraKapMoh03,BF03}.
Using MTA,
one arrives then at an explicitly time-dependent equation for $\meanemf$,
\EQ
{\partial\meanemf\over\partial t}
=\tilde\alpha\meanBB-\tilde\eta_{\rm t}\meanJJ-{\meanemf\over\tau},
\EN
where the last term subsumes the effects of all triple correlations.

\begin{figure}[t!]\begin{center}
\includegraphics[width=.95\textwidth]{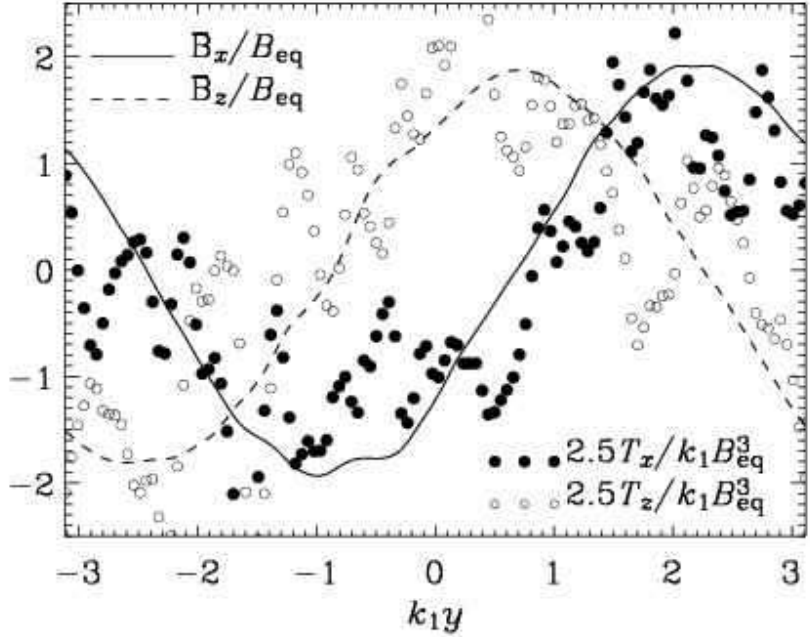}
\end{center}\caption[]{
Comparison of the spatial dependence of two components of the mean
magnetic field and the triple correlation in Run~3 of Ref.~\cite{B01}.
The magnetic field is normalized by the equipartition field strength,
$B_{\rm eq}$, and the triple correlation is normalized by
$k_1B_{\rm eq}^3$, but scaled by a factor of $2.5$ make it have a similar
amplitude as the mean field.
Note that $\meanB_x$ (solid line) correlates with $T_x$ (filled dots)
and $\meanB_z$ (dashed line) correlates with $T_z$ (open dots).
Adapted from Ref.~\cite{Bran+Sub05}.
}\label{ptriple}\end{figure}

In order to show that the assumption of a correlation between quadratic
and triple moments is actually justified we compare, using data from
Run~3 of Ref.~\cite{B01}, the spatial
dependence of the triple moments and the mean field on position.
The triple correlation is calculated as 
\EQ
\meanTT=\overline{(\jj\times\bb-\oo\times\uu-\nab p)\times\bb}
+\overline{\uu\times\nab\times(\uu\times\bb)}.
\EN
We note that in Run~3 of Ref.~\cite{B01} the mean field varied in the
$y$ direction with components pointing in the $x$ and $z$ directions
[third example in \Eq{BeltramiEigenfunctions}].
Thus, $\meanB_x(y)$ and $\meanB_z(y)$ exhibit a sinusoidal variation as
shown in \Fig{ptriple} by the solid and dashed lines, respectively.
(Note that $\meanemf$ itself has a negative correlation with $\meanBB$,
for a negative $\alpha$.)
Since $\tilde\alpha$ is negative (i.e.\ opposite to the helicity of the
forcing, which is positive), we expect a positive correlation between
$\meanTT$ and $\meanBB$.
This is indeed the case, as shown by the full and open symbols in
\Fig{ptriple}.
This demonstrates that the triple correlations are important and cannot
be neglected, as is done in FOSA.
Furthermore, since the correlation between $\meanTT$ and $\meanBB$ is
positive, and $\tilde\alpha<0$, this implies that $\tau>0$, which is
necessary for $\tau$ to be interpreted as a relaxation time.
To demonstrate that $\meanTT$ or $\meanBB$ also correlate with $\meanemf$
requires long time averaging \cite{CH96,OSBR02}, whereas here we have
only considered a single snapshot, so no time averaging was involved.

In order to determine a meaningful value of $\tau$, it is important
to get rid of the fluctuations of $\meanemf$, so time averaging is now
necessary.
The procedure is equivalent to that used in the passive scalar
case \cite{BraKapMoh03}, where a mean concentration gradient is imposed.
Here, instead of imposing a gradient of the passive scalar concentration,
one imposes a gradient in one component of the magnetic vector potential or,
what is equivalent, a uniform magnetic field $\BB_0$;
see Ref.~\cite{Bran+Sub05}.
The deviations from the imposed field are treated as fully periodic
in all three directions.

\begin{figure}[t!]\begin{center}
\includegraphics[width=.95\textwidth]{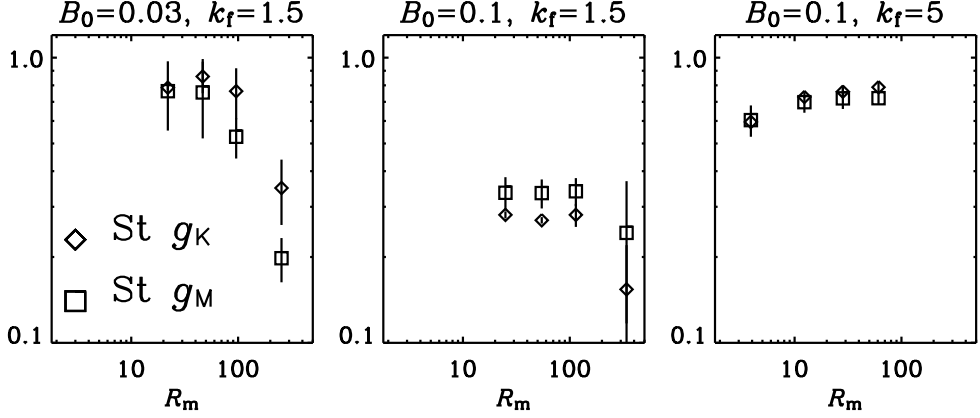}
\end{center}\caption[]{
Magnetic and kinetic Strouhal numbers as a function of $R_{\rm m}$
for different values of $B_0$ and  $k_{\rm f}$.
Here, kinetically and magnetically forced runs have been used to
calculate separately $g_{\rm K}\neq g_{\rm M}$.
Adapted from Ref.~\cite{Bran+Sub05}.
}\label{pbSt2}\end{figure}

The simulations produce average values for the three quantities,
\EQ
\alpha=\bra{\meanemf\cdot\meanBB}_t/\bra{\meanBB^2}_t,\quad
\tilde\alpha_{\rm K}=-\onethird\bra{\overline{\oo\cdot\uu}}_t,\quad
\tilde\alpha_{\rm M}=\onethird\bra{\overline{\jj\cdot\bb}}_t,
\EN
where $\bra{...}_t$ denotes combined time and volume averages.
According to MTA, these three quantities are connected to each other via
\EQ
\alpha=\tau\left(g_{\rm K}\tilde\alpha_{\rm K}
+g_{\rm M}\tilde\alpha_{\rm M}\right),
\EN
where we have allowed for the presence of additional quenching
factors, $g_{\rm K}(\meanBB)$ and $g_{\rm M}(\meanBB)$, in front of the
$\tilde\alpha_{\rm K}$ and $\tilde\alpha_{\rm M}$ factors, respectively
\cite{KMRS02}.
It turns out that for finite field strength the quenching factors,
$g_{\rm K}$ and $g_{\rm M}$, are less than unity and, more importantly,
they are slightly different from each other \cite{KMRS02}.
Using a combination of kinetically and magnetically forced turbulence
simulations, it has been possible to calculate separately these two quenching
functions multiplied by the normalized correction time,
combined with the corresponding quenching functions,
$\mbox{St}\,g_{\rm K}$ and $\mbox{St}\,g_{\rm M}$, respectively.
Here we have defined
\EQ
\mbox{St}=\tau u_{\rm rms} k_{\rm f}
\EN
as a nondimensional measure of the relaxation time \cite{landau,KR80}.
The result is shown in \Fig{pbSt2} as a function of the
magnetic Reynolds number.

\begin{figure}[t!]\begin{center}
\includegraphics[width=.85\textwidth]{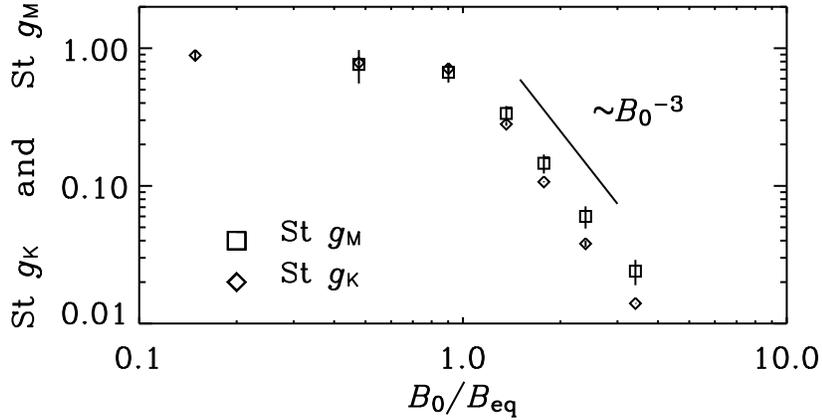}
\end{center}\caption[]{
Magnetic and kinetic Strouhal numbers as a function of $B_0/B_{\rm eq}$
for $\eta=2\times10^{-3}$ and  $k_{\rm f}=1.5$.
Kinetically and magnetically forced runs have been used to
calculate separately $g_{\rm K}\neq g_{\rm M}$.
Adapted from Ref.~\cite{Bran+Sub05}.
}\label{ptau_vs_B0}\end{figure}

We recall that in
the passive scalar case, St was found to converge to a value of about
3 in the limit of large Reynolds number and small values of $k_{\rm f}$
\cite{BraKapMoh03}.
For the present case one finds that St is approximately unity for small
field strengths, but may decrease like $B_0^{-3}$, once $B_0$ becomes
comparable with the equipartition field strength, $B_{\rm eq}$; see
\Fig{ptau_vs_B0}.

To summarize, the effect of the Lorentz forces is, to leading
order, the addition of a current helicity contribution to $\tilde\alpha$
and hence to $\alpha$.
There is an additional suppression effect which corresponds effectively
to a dependence of $\tau$ on the strength of the mean field.
This suppression seems however to be independent of or weakly dependent
on the magnetic Reynolds number
and is hence not the main limiting factor for the
growth of large scale fields when the magnetic Reynolds number is large.
(The last data point for $R_{\rm m}>100$ in the left hand panel of
\Fig{pbSt2} seems to suggest a sudden decline, but it is not clear how
reliable both this data point and the corresponding error estimate are.)

An immediate difficulty with 
having a small scale current helicity contribution to
$\alpha$ in \Eq{alptilde+etatilde} is that one
cannot directly incorporate this correction
in a mean field model, because
one only has information about the mean fields, $\meanJJ$ and $\meanBB$,
and not their fluctuations, $\jj$ and $\bb$.
The solution to this problem is to invoke the magnetic helicity equation
as an auxiliary equation to couple $\overline{\jj\cdot\bb}$ to the mean
field equations. In other words, one has to solve both the mean field dynamo
equation and the magnetic helicity equation simultaneously.
We turn to this issue in the next section.

\subsection{The dynamical quenching model}
\label{SDynamicalQuenching}

While conventional mean field theory is suitable to capture the structure
of the mean field correctly, it has become clear that simple quenching
expressions of the form \eq{alphafitformula}
or \eq{VC92formula} are unable to reproduce correctly the resistively
limited saturation phase.
Instead, they would predict saturation on a dynamical time scale,
which is not only in contradiction with simulations \cite{B01},
but it would also violate magnetic helicity conservation.
The key to modeling the late saturation behavior correctly is to ensure
that the magnetic helicity equation \Eq{helicity_eqn_closed} is satisfied
exactly at all times.

We begin with the usual mean field equation \eq{Induction_meanfield},
which has to be solved for a given form of $\meanemf$.
For $\alpha$ effect and possibly other mean field dynamos, the $\meanemf$
term produces magnetic helicity of the large scale field.
The evolution equation of magnetic helicity of the mean field $\meanBB$
can be obtained in the usual fashion from the mean field
induction equation.
We restrict ourselves here to the case of a closed domain.
In that case one obtains
\EQ
{\dd\over\dd t}\bra{\meanAA\cdot\meanBB}=
2\bra{\meanemf\cdot\meanBB}-2\eta\bra{\meanJJ\cdot\meanBB},
\label{dABdt}
\EN
which is independent of $\meanUU$. [Here, as earlier, angular
brackets denote averaging over all space, while overbars denote
suitably defined averages that could be two-dimensional
(cf.\ \Sec{SBeltrami}) or one-dimensional if there is shear
(cf.\ \Sec{SSimulationsWithShear})].
One sees that there is a source term for the mean field helicity
$\bra{\meanAA\cdot\meanBB}$, due to the presence of $\meanemf$.
The $\bra{\meanJJ\cdot\meanBB}$ term can be approximated by
$k_1^2\bra{\meanAA\cdot\meanBB}$ and corresponds to a damping term.
This merely reflects the fact that the operation of a mean field
dynamo automatically leads to the growth of linkages between the
toroidal and poloidal mean fields. Such linkages measure the helicity
associated with the mean field. One then wonders how
this mean field helicity arises? To understand this, we need to 
consider also the evolution of the small scale helicity
$\bra{\aaa\cdot\bb}$.
Since the Reynolds rules apply, we have
$\bra{\AAA\cdot\BB}=\bra{\meanAA\cdot\meanBB}+\bra{\aaa\cdot\bb}$,
so the evolution equation for $\bra{\aaa\cdot\bb}$ can be deduced by
subtracting \Eq{dABdt} from the evolution equation
\eq{helicity_eqn_closed} for the total helicity.
The fluctuating field then obeys the equation
\EQ
{\dd\over\dd t}\bra{\aaa\cdot\bb}=
-2\bra{\meanemf\cdot\meanBB}-2\eta\bra{\jj\cdot\bb},
\label{dabdt}
\EN
so the sum of \Eqs{dABdt}{dabdt} gives \Eq{helicity_eqn_closed}
without any involvement of the $\meanemf$ term.
We see therefore that the term $\bra{\meanemf\cdot\meanBB}$ merely 
transfers magnetic helicity between mean and fluctuating fields 
while conserving the total helicity!

In order to guarantee that the total helicity
evolution equation \eq{helicity_eqn_closed} is always obeyed,
\Eq{dabdt} has to be solved as an auxiliary equation along with
the mean field dynamo equation \eq{Induction_meanfield}.
From the previous two sections we have seen that $\meanemf$
is now determined by replacing the kinematic $\alpha$ effect by the
residual $\alpha$ effect, 
\EQ
\alpha=-\onethird\tau\bra{\oo\cdot\uu}+\onethird\tau\bra{\jj\cdot\bb}
\equiv\alpha_{\rm K}+\alpha_{\rm M}, 
\label{alphaTotal}
\EN
and with no immediate modifications to $\eta_{\rm t}$.
As mentioned before, subsequent modifications (i.e.\ quenching) of
$\alpha_{\rm K}$ and $\eta_{\rm t}$ occur as a direct consequence of
the decrease of turbulence intensity (which includes a decrease
of the turbulent kinetic helicity) and/or the relaxation time
(see \Fig{ptau_vs_B0}) \cite{Mof78,RS75,Rue74},
but this is not a particularly dramatic effect, because it does
not depend on $R_{\rm m}$.
More important is the $\bra{\jj\cdot\bb}$ term in $\alpha_{\rm M}$;
see \Eq{alphaTotal}.
This term can be related to $\bra{\aaa\cdot\bb}$ in \Eq{dabdt} for a given
spectrum of magnetic helicity.
For a triply-periodic domain, using isotropy, we can write
$\bra{\jj\cdot\bb} = k_{\rm f}^2 \bra{\aaa\cdot\bb}$, provided the
averages involving $\aaa$, $\bb$, and $\jj$ are governed by components with
wavenumber $k_{\rm f}$, where $k_{\rm f}$ is the approximate wavenumber of the
energy-carrying scale.
As mentioned earlier \cite{BB02}, the $k_{\rm f}$ factor may
be attenuated by a $R_{\rm m}^{1/4}$ factor to account for the fact that
for a $k^{-5/3}$ energy spectrum the $\bra{\jj\cdot\bb}$ term will pick
up contributions from smaller scales.
However, as described in \Sec{Revisit}, for such spectra,
one has to retain a $k$-dependent $\tau(k) \propto k^{-2/3}$; this
decreases for smaller scales making $\alpha_M$ (or $\alpha_K$) still 
dominated by contributions at the forcing scale, and so 
$\alpha_M$ (or $\alpha_K$) would be independent of $R_{\rm m}$ and $\mbox{Re}$.
Also, new simulations \cite{Bran+Sub05} suggest that at scales
smaller than the energy-carrying scale the field is no longer fully
helical, and hence we do not expect there to be a $R_{\rm m}^{1/4}$
factor on the effective value of $k_{\rm f}$.
The relaxation time
$\tau$ can be expressed in terms of $B_{\rm eq}^2=u_{\rm rms}^2$
using $\eta_{\rm t}=\onethird\tau u_{\rm rms}^2$.
Furthermore, $\tau$, and therefore also $\eta_{\rm t}$,  may still
depend on $|\meanBB|$ in a way as shown in \Fig{ptau_vs_B0}.
(We recall that we have used $\rho_0=\mu_0=1$ throughout.)
 
Under the assumption $\alpha_{\rm K}=\const$,
the final set of equations can be summarized in the more
compact form \cite{BB02,Sub02}
\\

\SHADOWBOX{
\EQ
{\partial\meanBB\over\partial t}=\nab\times
\left[\meanUU\times\meanBB+\alpha\meanBB
-(\eta+\eta_{\rm t})\meanJJ\right],
\label{fullset1}
\EN
\EQ
{\dd\alpha\over\dd t}=-2\eta_{\rm t} k_{\rm f}^2\left(
{\alpha\bra{\meanBB^2}-\eta_{\rm t}\bra{\meanJJ\cdot\meanBB}
\over B_{\rm eq}^2}+{\alpha-\alpha_{\rm K}\over\tilde{R}_{\rm m}}\right),
\label{fullset2}
\EN
}\\

\noindent
where $\tilde{R}_{\rm m}=\eta_{\rm t}/\eta$ is a modified definition of
the {\it microscopic} magnetic Reynolds number; cf.\ \Eq{Rmdefinition}.
Simulations of forced turbulence with a decaying large scale magnetic
field suggest
$\eta_{\rm t}\approx(0.8...0.9)\times u_{\rm rms}/k_{\rm f}$ \cite{YBR03}.
Thus,  $\tilde{R}_{\rm m}=(0.8...0.9)\times R_{\rm m}$, but for all
practical purposes the two are so close together that we assume
from now on $\tilde{R}_{\rm m}=R_{\rm m}$.
(As discussed in \Sec{MagneticHelicity}, $\eta$ cannot be replaced
by a turbulent value, so $R_{\rm m}$ is in practice really very large!)
We can now apply these equations to discuss various issues about 
the dynamical quenching of mean field dynamos.

\subsubsection{Comparison with algebraic $\alpha$ quenching}

In order to appreciate the nature of the solutions implied by
\Eqs{fullset1}{fullset2}, consider first the long-time limit
of a nonoscillatory dynamo.
In this case the explicit time
dependence in \Eq{fullset2} may be neglected
(adiabatic approximation \cite{BB02}).
Solving the resulting equation for $\alpha$ yields
\cite{KR82,GD}
\EQ
\alpha={\alpha_{\rm K}
+\eta_{\rm t} R_{\rm m}\bra{\meanJJ\cdot\meanBB}/B_{\rm eq}^2
\over1+R_{\rm m}\bra{\meanBB^2}/B_{\rm eq}^2}
\quad\mbox{(for $\dd\alpha/\dd t=0$)}.
\label{AlphaStationary}
\EN
Curiously enough, for the numerical experiments with an imposed large
scale field over the scale of the box \cite{CH96}, where $\meanBB$
is spatially uniform and therefore $\meanJJ=0$, one recovers the
`catastrophic' quenching formula \eq{VC92formula},
\EQ
\alpha={\alpha_{\rm K}\over1+R_{\rm m}\bra{\meanBB^2}/B_{\rm eq}^2}
\quad\mbox{(for $\meanJJ=0$)},
\EN
which implies that $\alpha$ becomes quenched when
$\bra{\meanBB^2}/B_{\rm eq}^2=R_{\rm m}^{-1}\approx10^{-8}$
for the sun, and for even smaller fields for galaxies.

On the other hand, if the mean field
is not imposed but maintained by dynamo action,
$\meanBB$ cannot be spatially uniform and then $\meanJJ$ is finite.
In the case of a Beltrami field \cite{B01},
$\bra{\meanJJ\cdot\meanBB}/\bra{\meanBB^2}\equiv\tilde{k}_{\rm m}$
is some effective wavenumber of the large scale field
with $\tilde{k}_{\rm m}\le k_{\rm m}$.
Since $R_{\rm m}$ enters both the numerator and the denominator,
$\alpha$ tends to $\eta_{\rm t}\tilde{k}_{\rm m}$, i.e.\
\EQ
\alpha\to\eta_{\rm t}\tilde{k}_{\rm m}\quad
\mbox{(for $\meanJJ\neq0$ and $\meanJJ\parallel\meanBB$)}.
\EN
Compared with the kinematic estimate,
$\alpha_{\rm K}\approx\eta_{\rm t}\tilde{k}_{\rm f}$,
$\alpha$ is only quenched by the scale separation ratio 
$\tilde{k}_{\rm m}/\tilde{k}_{\rm f}$.

It remains possible, however, that $\eta_{\rm t}$ is suppressed
via a quenching of $\tau$ (see \Fig{ptau_vs_B0}).
Thus, the question of how strongly $\alpha$ is quenched in the sun
or the galaxy, has been diverted to the question of how strongly $\eta_{\rm t}$
is quenched \cite{BB02}.
Note that quasi-linear treatments or MTA do not predict
$\eta_t$ quenching at the lowest order and for weak mean fields
\cite{KRP94,Roga+Klee01}.
One way to determine $\eta_{\rm t}$ and its possible quenching
is by looking at numerical solutions of cyclic dynamos
with shear ($\alpha\Omega$-type dynamos), because
in the saturated state the cycle
frequency is equal to $\eta_{\rm t}\tilde{k}_{\rm m}^2$.
The best agreement between models and simulations is achieved when
$\eta_{\rm t}$ begins to be quenched when $\bra{\meanBB^2}/B_{\rm eq}^2$
is around 0.3; see Ref.~\cite{BB02} for details.
This means that $\eta_{\rm t}$ is only quenched non-catastrophically.
This is consistent with the quenching of $\tau$; see \Figs{pbSt2}{ptau_vs_B0}.
However, more detailed work at larger magnetic Reynolds numbers needs to be done
-- preferentially in more realistic geometries that could be more readily
applied to stars and galaxies.

\subsubsection{$\alpha^2$ dynamos}

For $\alpha^2$ dynamos in a periodic box a special situation arises,
because then the solutions are degenerate in the sense that $\meanJJ$
and $\meanBB$ are parallel to each other.
Therefore, the term $\bra{\meanJJ\cdot\meanBB}\meanBB$ is the same as
$\bra{\meanBB^2}\meanJJ$, which means that in the mean EMF the term
$\alpha\meanBB$, where $\alpha$ is given by \Eq{AlphaStationary},
has a component that can be expressed as being parallel to $\meanJJ$.
In other words, the roles of turbulent diffusion (proportional to
$\meanJJ$) and $\alpha$ effect (proportional to $\meanBB$) cannot
be disentangled.
This is the {\it force-free degeneracy} of $\alpha^2$ dynamos
in a periodic box \cite{BB02}.
This degeneracy is also the reason why for $\alpha^2$ dynamos the late
saturation behavior can also be described by an algebraic
(non-dynamical, but catastrophic)
quenching formula proportional to $1/(1+R_{\rm m}\bra{\meanBB^2})$
for {\it both} $\alpha$ and $\eta_{\rm t}$, as was done in Ref.~\cite{B01}.
To see this, substitute the steady state quenching expression
for $\alpha$, from \Eq{AlphaStationary}, into the expression for $\meanemf$. 
We find
\EQA
\meanemf=\alpha\meanBB-(\eta+\eta_{\rm t})\meanJJ
={\alpha_{\rm K}
+R_{\rm m}\eta_{\rm t}\bra{\meanJJ\cdot\meanBB}/B_{\rm eq}^2
\over1+R_{\rm m}\bra{\meanBB^2}/B_{\rm eq}^2}\,\meanBB
-\eta_{\rm t}\meanJJ
\nonumber\\
={\alpha_{\rm K}\meanBB
\over 1+R_{\rm m}\bra{\meanBB^2}/B_{\rm eq}^2}
-{\eta_{\rm t}\meanJJ
\over 1+R_{\rm m}\bra{\meanBB^2}/B_{\rm eq}^2},
\label{bothquenched12}
\ENA
which shows that in the force-free case the adiabatic approximation,
together with constant (unquenched) turbulent magnetic diffusivity, becomes
equal to the pair of expressions where both $\alpha$ and $\eta_{\rm t}$
are catastrophically quenched.
This force-free degeneracy is lifted in cases with shear or when the
large scale field is
no longer fully helical (e.g.\ in a nonperiodic domain, and in particular
in the presence of open boundaries).

The dynamical quenching approach seems to be quite promising given that
it describes correctly the $\alpha^2$ dynamo found in the simulations.
There are however severe limitations that have to be overcome before
it can be used in more realistic mean field models.
Most importantly, the case of an inhomogeneous system, possibly one
with boundaries, is not solved rigorously,
although several promising approaches
have been suggested \cite{VC01,KMRS02,KMRS00,KMRS03,KR99,VLC03}.
The difficulty is to generalize \Eq{dabdt} 
to the non-homogeneous case of a mean magnetic helicity density, 
in a gauge-invariant manner. Only recently has this been attempted 
\cite{KS_AB05} by defining the magnetic helicity density of random fields 
as the density of correlated links.
An alternate possibility is to consider directly the evolution equation of
$\overline{\jj\cdot\bb}$ instead (as we do below).
The other problem with boundaries is that one still requires
a microscopic theory for the small scale losses of magnetic
helicity through the boundaries \cite{VC01,KMRS02,KMRS00}.
In any case, a more sophisticated theory should still reproduce the
homogeneous case, for which we now have now a fairly accurate
quantitative understanding.

\subsection{Saturation behavior of $\alpha^2$ and $\alpha\Omega$ dynamos}
\label{Sfinal}

In the case of homogeneous $\alpha\Omega$ dynamos, 
a major fraction of the toroidal
field can be generated by shear -- independently of helicity.
Therefore, $\bra{\meanBB^2}$ can be enhanced without producing much
$\bra{\meanJJ\cdot\meanBB}$. We discuss below the strength of the fields,
both when the final resistive saturation limit has been reached
and the case when $R_{\rm m}$ is so large that a quasi-static, 
non-resistive limit is more relevant (for example in galaxies).

\subsubsection{Final field strength}
\label{FinalFieldStrength}

In case one waits long enough, i.e.\ longer than the resistive time scale,
we noted that the final field strength is determined by the
condition $\bra{\JJ\cdot\BB} = 0$; or
\EQ
\bra{\meanJJ\cdot\meanBB} = - \bra{\jj\cdot\bb}.
\EN
In order to connect the current helicities with magnetic energies,
we proceed as follows.
First, one can quite generally relate the current and magnetic helicities
by defining characteristic wavenumbers, $k_{\rm m}$ and $k_{\rm f}$, for the
mean and fluctuating fields via
\EQ
k_{\rm m}^2=\bra{\meanJJ\cdot\meanBB}/\bra{\meanAA\cdot\meanBB},
\label{km2}
\EN
\EQ
k_{\rm f}^2=\bra{\jj\cdot\bb}/\bra{\aaa\cdot\bb}.
\label{kf2}
\EN
For a fully helical field, the same wavenumbers will also
relate the current helicity and energy in the field.
On the other hand, if the field is not fully helical,
one can introduce efficiency factors $\epsilon_{\rm m}$
and $\epsilon_{\rm f}$ that characterize the 
helicity fractions of the mean and fluctuating
fields, respectively, so we write
\EQ
\bra{\meanJJ\cdot\meanBB}/\bra{\meanBB^2} =
k_{\rm m}\epsilon_{\rm m}
\equiv \tilde{k}_{\rm m},
\label{kmtilde}
\EN
\EQ
\bra{\jj\cdot\bb}/\bra{\bb^2} = - k_{\rm f}\epsilon_{\rm f}
\equiv - \tilde{k}_{\rm f} .
\label{kftilde}
\EN
Here, $\tilde{k}_{\rm m}$ and $\tilde{k}_{\rm f}$ are 
are `effective wavenumbers' for mean and fluctuating fields, respectively.
In the final state,
$k_{\rm m}$ will be close to the smallest wavenumber in the computational
domain, $k_1$. In the absence of shear, $\epsilon_{\rm m}$ is of order
unity, but it can be less if there is shear or if the
boundary conditions do not permit fully helical
large scale fields (see below).
In the presence of shear,
$\epsilon_{\rm m}$ turns out to be inversely
proportional to the magnitude of the shear.
The value of $\epsilon_{\rm f}$ and $\tilde{k}_{\rm f}$, 
on the other hand, is determined
by small scale properties of the turbulence and is assumed known.

Both $k_{\rm m}$ and $k_{\rm f}$ are defined positive.
However, $\epsilon_{\rm m}$ can be negative which is typically the case
when $\alpha_{\rm K}<0$. The sign of $\epsilon_{\rm f}$ is defined such
that it agrees with the sign of $\epsilon_{\rm m}$, i.e.\
both change sign simultaneously and hence
$\tilde{k}_{\rm m}\tilde{k}_{\rm f}\ge0$.
In more general situations, $k_{\rm m}$ can be different from $k_1$.
Using \Eqs{kmtilde}{kftilde} together with \Eq{JBSteadyState} we have
\EQ
\tilde{k}_{\rm m}\bra{\meanBB^2}=\bra{\meanJJ\cdot\meanBB}
=-\bra{\jj\cdot\bb}=\tilde{k}_{\rm f}\bra{\bb^2},
\label{Bfin}
\EN
and so in the final saturated state,
\EQ
\bra{\meanBB^2}/\bra{\bb^2}=\tilde\kappa_{\rm f}/\tilde\kappa_{\rm m},
\label{final_state}
\EN
which generalizes \Eq{SteadyState} to the case with
fractional helicities; see also Eq.~(79) of Ref.~\cite{BDS02}.
Of course, this analysis only applies to flows with helicity. In the
nonhelical case, $\tilde{k}_{\rm m}=\tilde{k}_{\rm f}=0$,
so \Eq{Bfin} and \Eq{final_state} do not apply.

In the helical (or partially helical) case we can
determine the final field strength of the mean
and fluctuating fields.
This is possible because in periodic geometry with homogeneous $\alpha$ effect
the amplitude and energy of the dynamo wave is in general 
constant in time and does not vary with the cycle.
We now use the mean field dynamo equation
\eq{fullset1} to derive the evolution equation
for the magnetic helicity of the large scale field,
and apply it to the saturated state, so
\EQ
0=\alpha\bra{\meanBB^2}-(\eta+\eta_{\rm t})\bra{\meanJJ\cdot\meanBB}.
\label{fullset1steady}
\EN
We also use the evolution equation \eq{fullset2} for the
magnetic contribution to the $\alpha$ effect, applied again
to the saturated state,
\EQ
0={\alpha\bra{\meanBB^2}-\eta_{\rm t}\bra{\meanJJ\cdot\meanBB}
\over B_{\rm eq}^2}+{\alpha-\alpha_{\rm K}\over R_{\rm m}}.
\label{fullset2steady}
\EN
Note both equations apply also when $\meanUU\neq0$, because
the $\meanUU\times\meanBB$ term has dropped out after taking the
dot product with $\meanBB$.
We now eliminate $\alpha$ from \Eqs{fullset1steady}{fullset2steady}
and thus derive expressions for the final steady state values of 
$\bra{\meanBB^2}\equiv B^2_{\rm fin}$ in terms of $B_{\rm eq}^2$
and $\bra{\bb^2}\equiv b^2_{\rm fin}$, using \Eq{final_state}.
We get \cite{BB02},
\EQ
{B^2_{\rm fin}\over B_{\rm eq}^2}={\alpha_{\rm K}
-\eta_{\rm T}\tilde{k}_{\rm m}\over\eta_{\rm t}\tilde{k}_{\rm m}}, \quad\quad
{b^2_{\rm fin}\over B_{\rm eq}^2}={\alpha_{\rm K}
-\eta_{\rm T}\tilde{k}_{\rm m}\over\eta_{\rm t}\tilde{k}_{\rm f}}.
\label{final_field_strength}
\EN
In models where $\eta_{\rm t}$ is also quenched, both small scale and
large scale field strengths increase as $\eta_{\rm t}$ is more
strongly quenched.
(Of course, regardless of how strongly $\eta_{\rm t}$ may be quenched,
we always have from \eq{final_state}, $B^2_{\rm fin}/b^2_{\rm fin} =
\tilde\kappa_{\rm f}/\tilde\kappa_{\rm m}$ in the final state.)

One can write $B_{\rm fin}$ in an instructive way, using
the relevant dynamo control parameters for $\alpha^2$ and
$\alpha\Omega$ dynamos, defined in \Sec{PhaseRelations}.
We generalize these for the present purpose and define
\EQ
C_\alpha=\alpha_{\rm K}/(\eta_{\rm T}k_{\rm m}),\quad
C_\Omega=S /(\eta_{\rm T}k^2_{\rm m}),
\EN
where $\alpha_{\rm K}$ is the initial value of $\alpha$ due
to the kinetic helicity and $S=\Delta\Omega$ is a typical value
of the shear in an $\alpha\Omega$ dynamo.
From \Eq{fullset1steady}, the final steady state occurs
when $\alpha = \eta_{\rm T}\tilde{k}_{\rm m}\equiv\alpha_{\rm crit}$
or a critical value of the dynamo parameter $C_{\alpha, \rm crit} =
\alpha_{\rm crit}/\eta_{\rm T}k_{\rm m}$.
One can then write, using \eq{final_field_strength}, 
\EQ
B_{\rm fin}= \left({C_\alpha \over C_{\alpha, \rm crit}} -1 \right)^{1/2} 
(1 + R_{\rm m}^{-1})^{1/2} B_{\rm eq}.
\EN
For an $\alpha\Omega$ dynamo, the relevant dynamo number
is given by $D_0 = C_\alpha C_\Omega$ initially and in the
final steady state, by $D_{\rm crit} = C_{\alpha,\rm crit} C_\Omega$,
since the shear does not get affected by the nonlinear effects we consider.
In this case the final mean field strength can be written as
\EQ
B_{\rm fin} =  \left({D_0\over D_{\rm crit}} -1 \right)^{1/2}
(1 + R_{\rm m}^{-1})^{1/2} B_{\rm eq}.
\EN

Note once again that the above analysis only applies to flows with helicity. 
In the nonhelical case we have
$\alpha_{\rm K}=\tilde{k}_{\rm m}=\tilde{k}_{\rm f}=0$,
so \Eq{final_field_strength} cannot be used.
Nevertheless, without kinetic helicity one would still
expect a finite value of $\bra{\bb^2}$ because of small scale dynamo
action.
Furthermore, even in the fully helical case there can be substantial
small scale contributions.
Closer inspection of the runs of Ref.~\cite{B01} reveals,
however, that such contributions are particularly important only
in the early kinematic phase of the dynamo.

In summary, the saturation field strength depends only on the scale
separation ratio, see \Eq{final_state}, and not, for example, on the
intensity of the turbulence.

\subsubsection{Early time evolution}
\label{Searly}

\begin{figure}[t!]\begin{center}
\includegraphics[width=.8\textwidth]{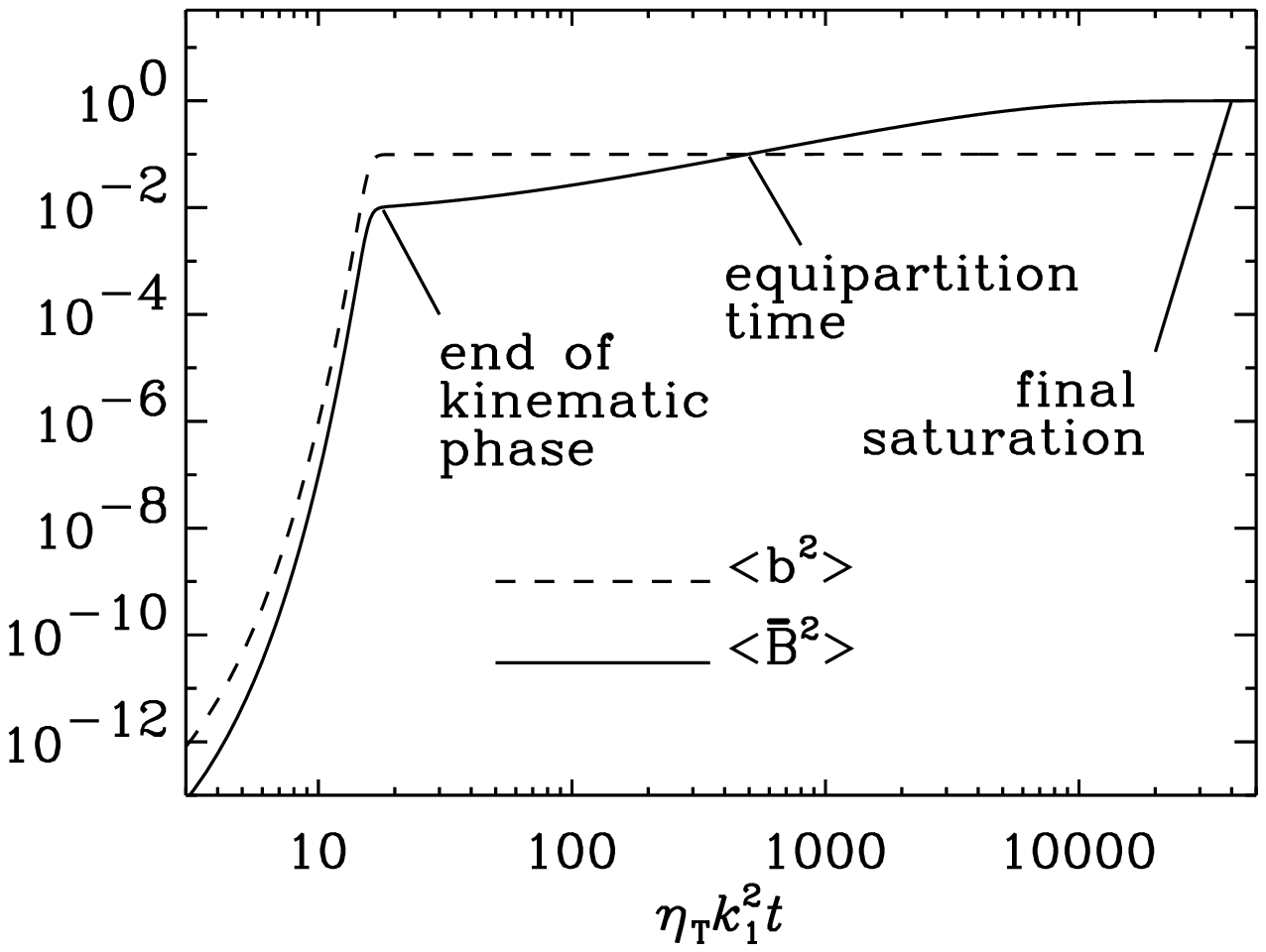}
\end{center}\caption[]{
Evolution of $\bra{\meanBB^2}$ and $\bra{\bb^2}$ (solid and dashed lines,
respectively) in a doubly-logarithmic plot for an $\alpha^2$ dynamo with
$\eta_{\rm t}=\mbox{const}$ for a case with $k_{\rm f}/k_1=10$.  Note the
abrupt initial saturation after the end of the kinematic exponential
growth phase with $\bra{\meanBB^2}/\bra{\bb^2}\sim0.1$, followed by
a slow saturation phase during which the field increases to its final
super-equipartition value with $\bra{\meanBB^2}/\bra{\bb^2}\sim10$
\cite{BD02}.
}\label{Fpcrossing}\end{figure}

During the early growth phase
the magnetic helicity varies on time scales shorter than
the resistive time, so
the last term in the dynamical $\alpha$ quenching equation
\eq{fullset2}, which is proportional to $R_{\rm m}^{-1}\to0$,
can be neglected and so $\alpha$
evolves then approximately according to
\EQ
{\dd\alpha\over\dd t}\approx-2\eta_{\rm t}k_{\rm f}^2
(\alpha-\eta_{\rm t}\tilde{k}_{\rm m})
{\bra{\meanBB^2}\over B_{\rm eq}^2}.
\label{dynquench_early}
\EN
This equation can be used to describe the end of the kinematic
time evolution when $\bra{\meanBB^2}$ grows exponentially.
(We refer to this phase as late 'kinematic' because the slow resistive
saturation has not yet set in, although of course $\alpha$ is already
becoming suppressed due to growth of the current helicity.)
We see from \eq{dynquench_early} that the 
$-2\eta_{\rm t}k_{\rm f}^2 \alpha$ term leads to a reduction of
$\alpha$. This leads to a dynamical reduction of 
$\alpha$ until it becomes comparable to
$\eta_{\rm t}\tilde{k}_{\rm m}$, shutting off any further reduction.
Therefore, the early time evolution leads to a nearly $R_{\rm m}$-independent
growth phase.
At the end of this growth phase a fairly significant large scale field
should be possible \cite{FB02,BB02,Sub02}. The basic physical 
reason is that the suppression of $\alpha$ occurs due to the growth
of small scale current helicity which, in turn, is the result of a
growth of the small scale magnetic helicity.
Since magnetic helicity is nearly
conserved for $R_{\rm m} \gg 1$, this implies a corresponding
growth of oppositely signed large scale magnetic helicity and hence the 
large scale field.
We mention, however, that numerical simulations \cite{BHD03}
have not been able to confirm a sharp transition from exponential 
to linear (resistively limited) growth,
as seen in \Fig{Fpcrossing}, which shows a numerical solution
of \Eqs{fullset1}{fullset2}; see Refs~\cite{BB02,BD02}.
The absence of a sharp cross-over from exponential to linear
growth in the turbulence simulations could be related to the fact
that several large scale modes are competing, causing an extra
delay in the selection of the final mode.

In the following discussion we restrict ourselves to the case
where $\eta$ is small or $R_{\rm m} \gg 1$. 
Magnetic helicity is then well
conserved and this conservation requires that
\EQ
\bra{\meanAA\cdot\meanBB}\approx-\bra{\aaa\cdot\bb}\quad
\mbox{(for $t\leq t_{\rm kin}$)} .
\label{helequil}
\EN
Here the time $t=t_{\rm kin}$ marks the end of the exponential
growth phase (and thus the `initial' saturation time $t_{\rm sat}$ used in
Ref.~\cite{B01}).
This time is determined by the condition that the
term in parenthesis in \Eq{dynquench_early} becomes significantly reduced,
i.e.\ $\alpha$ becomes comparable to $\eta_{\rm t}\tilde{k}_{\rm m}$.
We would like to estimate the field strengths of the
large and small scale fields by this time.
The helicity conservation constraint \eq{helequil}, 
and the definitions in \eq{km2}, \eqs{kf2}{kmtilde}, give 
\EQ
\alpha_{\rm M} = \onethird\tau\bra{\jj\cdot\bb}
= \onethird\tau k_{\rm f}^2 \bra{\aaa\cdot\bb} 
= - \onethird\tau k_{\rm f}^2 \bra{\meanAA\cdot\meanBB} 
= -\eta_{\rm t}{k_{\rm f}^2 \over k_{\rm m}^2}
{\tilde{k}_{\rm m} \bra{\BB^2} \over B_{\rm eq}^2}. 
\EN
We use this in \Eq{alphaTotal} for $\alpha$
and demand that $\alpha$, as governed by the dynamical
quenching equation \eq{fullset2}, settles down
to a quasi steady state by $t=t_{\rm kin}$.
We then get for the mean squared strength of the large scale field
at $t=t_{\rm kin}$,
\EQ
{B_{\rm kin}^2\over B_{\rm eq}^2}= {\alpha_{\rm K}
-\eta_{\rm t}\tilde{k}_{\rm m}\over\eta_{\rm t}\tilde{k}_{\rm m}}
{k_{\rm m}^2\over k_{\rm f}^2} - R_M^{-1} \approx 
{\alpha_{\rm K}
-\eta_{\rm t}\tilde{k}_{\rm m}\over\tilde{\iota}\eta_{\rm t}\tilde{k}_{\rm m}}
{k_{\rm m}^2\over k_{\rm f}^2},
\label{kinematic_field_strength}
\EN
where the effect of a finite $R_{\rm m}$ has been included through
an approximate correction factor \cite{BB02}
\EQ
\tilde{\iota}=1+R_{\rm m}^{-1}
{k_{\rm f}/\epsilon_{\rm f}\over k_{\rm m}/\epsilon_{\rm m}}.
\EN
The strength of the small scale field at $t=t_{\rm kin}$ is given by
using \eq{helequil},
\EQ
b_{\rm kin}^2= {k_{\rm f}/\epsilon_{\rm f} \over 
k_{\rm m} /\epsilon_{\rm m}} B_{\rm kin}^2
= {\alpha_{\rm K} -\eta_{\rm t}\tilde{k}_{\rm m}
\over\tilde{\iota}\eta_{\rm t}\tilde{k}_{\rm f}} B_{\rm eq}^2.
\EN
Not surprisingly, at the end of the kinematic phase the
small scale magnetic energy is almost the same as in the final state;
see \Eq{final_field_strength}. However, the large scale magnetic energy is
still by a factor $k_{\rm m}^2/k_{\rm f}^2$ smaller than in the final state
($\epsilon_{\rm m}$ may be somewhat different in the two stages).
This result was also obtained by Subramanian \cite{Sub02} using a
similar approach. 

These expressions can be further clarified by noting
that the fractional helicity of the small scale
field is likely to be similar to that of the forcing velocity field.
We can then write $\bra{\oo\cdot\uu} \approx \epsilon_{\rm f} k_{\rm f}
\bra{\uu^2}$; so $\alpha_{\rm K}\approx\eta_{\rm t}\epsilon_{\rm f}k_{\rm f}$,
and one arrives at \cite{BB02}
\EQ
{B_{\rm kin}^2\over B_{\rm eq}^2}=
{k_{\rm m}/\epsilon_{\rm m}\over\tilde{\iota}k_{\rm f}/\epsilon_{\rm f}}
\left(1-{\tilde{\kappa}_{\rm m}\over\tilde{\kappa}_{\rm f}}\right),
\label{kinematic_field_strength2}
\EN
which shows that $B_{\rm kin}$ can be comparable to and even
in excess of $B_{\rm eq}$, especially when
$\epsilon_{\rm m}$ is small (i.e.\ for strong shear).
This is of interest in connection with the question of why the
magnetic field in so many young galaxies can already have
equipartition field strengths.

As emphasized in Ref.~\cite{FB02}, even
for an $\alpha^2$ dynamo, the initial evolution to $B_{\rm kin}$
is significantly more optimistic an estimate than what could have
been expected based on lorentzian $\alpha$ quenching.
In the case of an $\alpha\Omega$ dynamo \cite{BB02,Sub02}, one
could have $\tilde{k}_{\rm m}\ll k_{\rm m}$,
and so $B_{\rm kin}$ can be correspondingly larger. In fact, for
\EQ
\epsilon_{\rm m}/\epsilon_{\rm f}\leq k_{\rm m}/k_{\rm f},
\EN
the large scale field begins to be of order $B_{\rm eq}$ and
exceeds the small scale field already during
the kinematic growth phase.

It is once again instructive to express $B_{\rm kin}$ in terms of the
dynamo control parameters, $C_\alpha$ and $C_\Omega$.
Note that for small $\eta$, or $R_{\rm m} \gg 1$, 
$\alpha_{\rm crit} =\eta_{\rm T}\tilde{k}_{\rm m} \approx 
\eta_{\rm t}\tilde{k}_{\rm m}$ and $\tilde{\iota} \approx 1$.
One can then write
\EQ
B_{\rm kin} = {k_{\rm m}\over k_{\rm f}} 
\left[{C_\alpha \over C_{\alpha, \rm crit}} -1 \right]^{1/2} B_{\rm eq}.
\EN
For an $\alpha\Omega$ dynamo with dynamo number $D_0 = C_\alpha C_\Omega$ 
and a critical value $ D_{\rm crit} = C_{\alpha, \rm crit} C_\Omega$, 
we have (assuming shear does not get affected) \cite{Sub02}
\EQ
B_{\rm kin} = {k_{\rm m}\over k_{\rm f}} 
\left[{D_0\over D_{\rm crit}} -1 \right]^{1/2} B_{\rm eq}.
\label{Bkin_shear}
\EN

We can estimate how strong the shear has to be for
the large scale field to be comparable to $B_{\rm eq}$.
Since $\alpha_{\rm K}\approx\eta_{\rm t}\epsilon_{\rm f}k_{\rm f}$, 
$C_\alpha \approx \epsilon_{\rm f} (k_{\rm f}/k_{\rm m})$,
or $k_{\rm m}/k_{\rm f} \approx \epsilon_{\rm f}/C_\alpha $.
Using $D_0 = C_\alpha C_\Omega$, we can rewrite \eq{Bkin_shear}
as
\EQ
B_{\rm kin} = \left[{C_\Omega \epsilon_{\rm f}^2 \over C_\alpha D_{\rm crit}} -
{k_{\rm m}^2\over k_{\rm f}^2} \right]^{1/2} B_{\rm eq}.
\label{Bkin_shear2}
\EN
Now $k_{\rm m}/ k_{\rm f} \ll 1$ in general.
So $B_{\rm kin}$ can become comparable to $ B_{\rm eq}$
provided the shear is strong enough that 
$C_\Omega \epsilon_{\rm f}^2 > C_\alpha D_{\rm crit}$.
For dynamo action to be possible, one also requires 
$C_\Omega C_\alpha > D_{\rm crit}$ or 
$ C_\alpha > D_{\rm crit}/C_\Omega$. Combining these two
inequalities, we see that large scale fields can become comparable to
the equipartition fields if $C_\Omega \ga D_{\rm crit}/\epsilon_{\rm f}$.
The critical dynamo number $D_{\rm crit}$ will depend on the
physical situation at hand; $D_{\rm crit} = 2$ for a one dimensional
$\alpha\Omega$ dynamo with periodic boundary conditions and
homogeneous $\alpha_{\rm K}$. 
Further, in simulations of rotating convection, $\epsilon_{\rm f}\approx0.03$
\cite{BJNRST96}; assuming that this relatively low value
of $\epsilon_{\rm f}$ is valid in more realistic simulations, we have
$C_\Omega \ga 2/\epsilon_{\rm f}\approx60$
as the condition for which the large scale field
becomes comparable to $B_{\rm eq}$.
(Note that when the large scale field
becomes comparable to $B_{\rm eq}$, quenching due to
the large scale field itself will become important, a
process we have so far ignored.)  This
condition on the shear, could be satisfied for stellar dynamos
but it is not likely to be satisfied for galactic dynamos, where typically 
$D_0/D_{\rm crit} \sim 2$ (cf.\ \Sec{galdyn}).

During the subsequent resistively limited saturation phase the energy of
the large scale field grows first linearly, i.e.\
\EQ
\bra{\meanBB^2}\approx B_{\rm kin}^2+2\eta k_{\rm m}^2(t-t_{\rm kin})
\quad\mbox{(for $t>t_{\rm kin}$)},
\label{pastkin}
\EN
and saturates later in a resistively limited fashion; see
\Eq{helconstraint}.

\subsubsection{Comparison with simulations}
\label{ComparisonSimulations}

In order to compare the dynamical quenching model with simulations of turbulent
oscillatory dynamos with shear \cite{BBS01}, it is important to
consider the same geometry and shear profile, i.e.\
sinusoidal shear, $\meanUU=(0,Sk_1^{-1}\cos k_1 x,0)$, and a mean field
$\meanBB=\meanBB(x,z,t)$. We discuss here the results of Ref.~\cite{BB02},
where simulations of Ref.~\cite{BBS01} are compared with dynamical
quenching models using either a fixed value of $\eta_{\rm t}$ or,
following earlier suggestions \cite{B01,FB02}, a turbulent diffusivity
that is quenched in the same way as $\alpha$, i.e.\
$\eta_{\rm t}\propto\alpha$.

Both simulation and model show dynamo waves traveling in the positive
$z$-direction at $x=\pm\pi$ and in the negative $z$-direction at $x=0$,
which is consistent with the three-dimensional simulations using
negative values of $\alpha_{\rm K}$, which affects the direction
of propagation of the dynamo waves.

It turns out that
in all models the values of $b^2_{\rm fin}$ are smaller than in
the simulations. This is readily explained
by the fact that the model does not take into account small scale
dynamo action resulting from the nonhelical component of the flow.
Comparing simulations with different values of $R_{\rm m}$,
the cycle frequency changes by a factor compatible
with the ratio of the two magnetic Reynolds numbers. This is not
well reproduced by a quenching expression for $\eta_{\rm t}$ that is
independent of $R_{\rm m}$. On the other hand, if
$\eta_{\rm t}$ is assumed to be proportional to
$\alpha$, then $\omega_{\rm cyc}$ becomes far
smaller than what is seen in the simulations. A possible remedy would
be to have some intermediate quenching expression for $\eta_{\rm t}$.
We should bear in mind, however, that the present model ignores
the feedback from the large scale fields. Such feedback is indeed
present in the simulations, which also show much more chaotic
behavior than the model \cite{BB02}.

In conclusion the dynamical quenching model predicts saturation
amplitudes of the large scale field that are smaller than the
equipartition field strength by a factor that is equal to the
ratio of the turbulent eddy size to the system size, and hence
independent of the magnetic Reynolds number. The field strengths
can be much larger if the dynamo involves shear and
is highly supercritical. This may be relevant to explaining the
amplitudes of fields in stars, but in galaxies the effect of shear is not
likely to be strong enough to explain large scale 
galactic fields without additional effects such as a small scale helicity 
flux (see below and \Sec{galdyn}). 
One the other hand, 
for the sun and sun-like stars, the main issue is the cycle period.
No conclusive answer can be given until there is definitive knowledge
about the quenching of the turbulent magnetic diffusivity, $\eta_{\rm t}$
and the effect of current helicity fluxes out of the domain.
The more $\eta_{\rm t}$ is quenched, the longer the cycle period can
become, unless there are significant losses of magnetic helicity
through the open boundaries that all these bodies must have.
Before discussing this in \Sec{OpenBoundaries} we first make
a few historical remarks and also address
the question of how close to being resistively limited the solar
cycle might be.
The answer is somewhat surprising: not much!

\subsubsection{Historical remarks}
\label{HistoricalRemarks}

The explicit time dependence of the $\alpha$ quenching equation
\eq{fullset2} was first proposed by Kleeorin and Ruzmaikin \cite{KR82}
in an early paper of 1982; see also the book by Zeldovich et al.\
\cite{ZRS83} and a later paper by Kleeorin et al.\ \cite{KRR95}.
For a long time the true significance of the long time scale introduced
by \Eq{fullset2} remained unclear.
It was therefore not surprising that early work focused
exclusively on explaining the chaotic nature of the solar cycle
\cite{Ruz81,SS91,Covas_etal97}.

The steady state limit \eq{AlphaStationary} of \Eq{fullset2} was
first analyzed by Gruzinov \& Diamond \cite{GD}
and Bhattacharjee \& Yuan \cite{BY95}
and so the connection with catastrophic quenching was established.
When the issue of a resistively limited saturation of $\alpha^2$ dynamos
in periodic boxes was raised \cite{B01}, the possible connection with
\Eq{fullset2} was blurred by the fact that the simulation data could
well be described by simultaneous and catastrophic quenching of $\alpha$
and $\eta_{\rm t}$.
In the paper by Field and Blackman \cite{FB02} it was first proposed
and quantitatively demonstrated that resistively slow saturation can be
explained in terms of mean field theory with a current helicity correction
to the $\alpha$ effect, as calculated by Pouquet, Frisch and L\'eorat
\cite{PFL76} already in 1976.
This result was then recast to take the form of \Eq{fullset2}, which
has to be solved simultaneously with the dynamo equation \Eq{fullset1},
and the connection with simultaneous catastrophic quenching of $\alpha$
and $\eta_{\rm t}$ was understood as a special case that applies only
to $\alpha^2$ dynamos \cite{BB02}. \EEqs{fullset1}{fullset2} were
also used to estimate the minimal mean magnetic fields which could be obtained
in galaxies, in spite of the helicity constraint \cite{Sub02},
ignoring however the effects of magnetic and current helicity fluxes.
Blackman \& Field \cite{BF00b,BF00} proposed that such helicity fluxes
could alleviate catastrophic $\alpha$ quenching.
Kleeorin and coworkers \cite{KMRS02,KMRS00,KMRS03} found that helicity
fluxes are indeed likely to lead to substantially increased saturation
field strengths, while Vishniac \& Cho \cite{VC01} suggested that
helicity fluxes could even drive a dynamo effect of its own.

The generalization of \Eq{fullset2} to spatially non-uniform systems
is nontrivial, since one has to define a gauge invariant magnetic helicity
even for systems with boundaries.
This concept is helpful in simple cases of homogeneous turbulence, but
it needs to be generalized to the more interesting inhomogeneous case,
which has only been attempted recently \cite{KS_AB05}.
This is why we have shifted the attention to considering the evolution
of current helicity instead \cite{KS_AB04}; see \Sec{nonlinearflux}.
The current helicity is gauge invariant, directly observable
and appears explicitly in the back reaction term in $\alpha$.

Initial attempts to verify the importance of helicity fluxes in open
box simulations failed \cite{BD01}, and it was only in the presence of
shear that evidence for accelerated field saturation emerged \cite{B05}.
The agreement between simulations and mean field theory with helicity
fluxes is still not fully satisfactory.
In particular, with moderately strong helicity fluxes the solutions
of the mean field model predict saturation energies that decrease
inversely proportional with magnetic Reynolds number, unless the
magnitude of the helicity flux exceeds a certain threshold \cite{BS05b}.

It should also be pointed out that the explanation of catastrophic
quenching in terms of magnetic helicity conservation is not generally
accepted; see Refs~\cite{DurhamReview,Proctor03} for recent reviews.
An alternative explanation for the catastrophic quenching phenomenon
is in terms of a suppression of chaos \cite{SuppressionOfChaos}.
Unfortunately, at the moment there is no quantitative theory that
explains the magnetic Reynolds number dependence of this suppression.

\subsection{How close to being resistively limited is the 11 year cycle?}
\label{HowClose}

In this section we present an estimate of the amount of magnetic helicity,
$H$, that is expected to be produced and destroyed during the 11 year
cycle \cite{BDS02}.
We also need to know what
fraction of the magnetic field takes part in the 11-year cycle.
Here we are only interested in the relative magnetic helicity in one hemisphere.
Following an approach similar to that of Berger \cite{berger84}, one
can bound the rate of change of magnetic helicity,
as given by \Eq{helicity_eqn_closed}, in terms of the rate
of Joule dissipation, $Q_{\rm Joule}$, and magnetic energy, $M$, i.e.\
\EQ
\left|{\dd H\over\dd t}\right|\le2\eta\bra{\JJ\cdot\BB}V
\leq2\eta\sqrt{\bra{\JJ^2}\bra{\BB^2}}V
\equiv2\sqrt{2\eta Q_{\rm Joule}M},
\EN
where $V$ is the volume, $\bra{\eta\JJ^2}V=Q_{\rm Joule}$ the Joule heat,
and $\bra{\half\BB^2}V=M$ the magnetic energy.
For an oscillatory dynamo, all three
variables, $H$, $M$, and $Q_{\rm Joule}$ vary in an oscillatory
fashion with a cycle frequency $\omega_{\rm cyc}$ of magnetic energy
(corresponding to 11 years for the sun -- not 22 years), so
we estimate
$|\dd H/\dd t|\la\omega_{\rm cyc}|H|$ and
$|\dd M/\dd t|\la Q_{\rm Joule}\la\omega_{\rm cyc} M$, and obtain
\EQ
\omega_{\rm cyc}|H|\leq2\sqrt{2\eta\omega_{\rm cyc}}\,M.
\EN
This leads to the inequality \cite{BDS02,BS02}
\EQ
|H|/(2M)\leq\ell_{\rm skin},
\label{H-M-skin}
\EN
where $\ell_{\rm skin} = \sqrt{2\eta/\omega_{\rm cyc}}$ is the skin depth,
here associated with the 11 year frequency $\omega_{\rm cyc}$.
Thus, the maximum magnetic helicity that
can be generated and dissipated during one cycle is characterized
by the length scale $|H|/(2M)$, which has to be less than
the skin depth $\ell_{\rm skin}$.

For $\eta$ we have to use the Spitzer resistivity
[see \Eq{SpitzerFormula}], so $\eta$ increases from about
$10^4\cm^2/{\rm s}$ at the base of the convection zone to about $10^7\cm^2/\s$
near the surface layers and decreases again in the solar atmosphere,
see \Eq{SpitzerFormula}.
Using $\omega_{\rm cyc}=2\pi/(11\yr)=2\times10^{-8}\s^{-1}$
for the relevant frequency at which $H$ and $M$ vary, we have
$\ell_{\rm skin}\approx10\km$ at the bottom of the convection zone
and $\ell_{\rm skin}\approx300\km$ at the top.

The value of $\ell_{\rm skin}$ should
be compared with the value $|H|/(2M)$ that can be obtained from
dynamo models \cite{BDS02}.
For a sphere (or rather a half-sphere) with open boundary conditions and volume $V$
(for example the northern hemisphere), one has to use the gauge-invariant
relative magnetic helicity of Berger and Field \cite{berger_field84}; see
\Eq{Hrel} and \Sec{MagneticHelicity}.
We assume that the dynamo saturates such that most of the magnetic
helicity is already in the large scales [cf.\ \Eq{H1ggHf}], so $H$
can be estimated from a mean field model.
We also assume that magnetic helicity is not continuously being pumped
back and forth between large and small scales.
While this remains a hypothetic possibility, it should be noted that
this has never been seen in simulations.
With these reservations in mind, we now employ
an axisymmetric mean field, which can be written as
$\meanBB=b\pphi+\nab\times(a\pphi)$, it turns
out that the relative magnetic helicity integral is simply \cite{BDS02}
\EQ
H=2\int_V ab\,\dd V\quad\mbox{(axisymmetry)}.
\label{relHm}
\EN
The results of model calculations show that \cite{BDS02},
when the ratio of poloidal field at the {\it pole} to the maximum
toroidal field inside the convection zone, $B_{\rm pole}/B_{\rm belt}$, 
is in the range consistent with observations, $B_{\rm pole}/B_{\rm belt}
=(1...3)\times10^{-4}$, then the ratio $H_{\rm N}/(2M_{\rm N}R)$ is around
$(2-5)\times10^{-4}$ for models with latitudinal shear. (Here, the subscript
`N' refers to the northern hemisphere.) This confirms
the scaling with the poloidal to toroidal field ratio \cite{BBS01},
\EQ
H_{\rm N}/(2M_{\rm N}R)={\cal O}(B_{\rm pol}/B_{\rm tor})
\ga B_{\rm pole}/B_{\rm belt}.
\EN 
Given that $R=700\Mm$ this means that 
$H_{\rm N}/(2M_{\rm N})\approx70...200\km$, which would be comparable
to the value of $\ell_{\rm skin}$ near the upper parts of the solar
convection zone.

The surprising conclusion is that the amount of mean field
helicity that needs to be generated in order to explain the large
scale solar magnetic fields is actually so small, that it may be plausible
that microscopic magnetic diffusion could still play a role in the
solar dynamo.
In other words, although open boundary effects may well be important
for understanding the time scale of the dynamo, the effect does not
need to be extremely strong.
Of course, no quantitatively accurate predictions can be made,
because the result depends on the model.

\subsection{Open boundaries}
\label{OpenBoundaries}

In the preceding sections we made the assumption that no magnetic
helicity flows through the boundaries.
This is of course unrealistic, and on the sun magnetic helicity losses
are indeed observed; see \Sec{MagneticHelicitySolarField}.
In the presence of open boundaries one has to use the relative
magnetic helicity, as defined in \Eq{RelHelEvolve}.
In this equation there emerges a surface term,
$2\oint(\EE\times\AAA^{\rm ref})\cdot\dd\SSS$,
which is due to helicity fluxes.
The importance of helicity fluxes was already demonstrated in
\Sec{TurbulenceAndShear}.
The question is how important is this effect and can it possibly alleviate
the magnetic helicity constraint by allowing the $\alpha$ effect to be
less strongly quenched \cite{BF00}.

\subsubsection{Historical remarks}

The question of how magnetic helicity losses could alleviate the
problem of catastrophic quenching and resistively limited saturation
and perhaps cycle periods remained at first unclear
when this idea was originally proposed \cite{KMRS00,BF00}.
Simulations of helically forced turbulence in an open
domain, using a vertical field boundary condition ($\BB\times\nnn=0$),
showed a shorter saturation time.
However, this was at the expense of a reduced
saturation amplitude \cite{BD01}.
This somewhat frustrating result raised concerns whether magnetic helicity
losses could even help in principle.
This was then the reason for performing the co-called vacuum cleaner
experiment where small scale magnetic fields were artificially removed
in regular intervals via Fourier filtering \cite{BDS02}.
The main insight came when it was realized that it is not so much the
boundary conditions as such that have a positive effect in alleviating
the magnetic helicity constraint, but the fact
that there is a magnetic helicity flux also {\it inside} the domain, such that
magnetic helicity can actually be transported toward the boundaries.
A leading candidate for this flux is the flux derived by 
Vishniac and Cho \cite{VC01}.
This flux will be derived and discussed further in \Sec{AnisotropicTurbulence}.

There is another complication that led to considerable confusion.
Magnetic helicity is defined as a volume integral (\Sec{MagneticHelicity}),
and, because of the gauge problem,
the concept of a magnetic helicity density does not exist.
This has actually changed and a magnetic helicity density can actually
be defined (for small scale fields) 
in terms of a density of linkages if there exists a meaningful
separation into large scale and small scale fields \cite{KS_AB05}.
Another important aspect was the realization that the quantity that we
are primarily interested in, and that directly enters the magnetic
$\alpha$ effect, is $\overline{\jj\cdot\bb}$.
In working directly with $\overline{\jj\cdot\bb}$ there is of course
no longer the comfort of dealing with a conserved quantity.
Thus, the proper evolution equation for $\overline{\jj\cdot\bb}$
can no longer be inferred from total magnetic helicity conservation
together with the evolution equation for the large scale field.
Instead, $\overline{\jj\cdot\bb}$ has to be calculated directly
using standard closure procedures \cite{KS_AB04}.
There are now strong indications that an important prerequisite for
obtaining a strong helicity flux is shear.
However, shear has only recently been included in simulations with open
boundaries; see \Sec{TurbulenceAndShear} and Ref.~\cite{B05}.
Therefore, not too much analysis has been done on this type of problem yet.

In the following we present first a phenomenological model that explains
the relative significance of large scale and small scale magnetic
helicity losses.
We then turn to speculative consequences for the observed solar field
and demonstrate that small scale losses can alleviate the quenching
problem -- at least in principle.

\subsubsection{A phenomenological model of magnetic helicity losses}
\label{PhenomenologicalModel}

It is instructive to discuss first a phenomenological description of
magnetic helicity losses.
We imagine that the loss of magnetic helicity across the boundary is
accomplished by the turbulent exchange of eddies across the boundary.
The rate of magnetic helicity loss is then proportional to some turbulent
diffusivity coefficient, $\eta_{\rm m}$ or $\eta_{\rm f}$, for the losses
from mean and fluctuating parts, respectively.
Again, we assume that the small and large scale fields are
maximally helical (or have known helicity fractions $\epsilon_{\rm m}$
and $\epsilon_{\rm f}$) and have opposite signs of magnetic helicity at small
and large scales. The details can be found in Refs~\cite{BB03,BDS02}.
The strength of this approach is that
it is quite independent of mean field theory.

We proceed analogously to \Sec{SaturationTime} where we used the
magnetic helicity equation \eq{helicity_eqn_closed} for a closed
domain to estimate the time derivative of the magnetic helicity of
the mean field, $H_1$ (or $H_{\rm m}$, which is here the same), by
neglecting the time derivative of the fluctuating field, $H_{\rm f}$.
This is a good approximation after the fluctuating field has reached
saturation, i.e.\ $t>t_{\rm sat}$.
Thus, we have
\begin{equation}
{{\rm d}H_{\rm m}\over{\rm d}t}+
\underbrace{{{\rm d}H_{\rm f}\over{\rm d}t}}_{\mbox{neglected}}=
-2\eta_{\rm m}k_{\rm m}^2H_{\rm m}
-2\eta_{\rm f}k_{\rm f}^2H_{\rm f},
\label{evolv_phenoH}
\end{equation}
where $\eta_{\rm m}=\eta_{\rm f}=\eta$ corresponds to the case
of a closed domain; see \Eq{H1eqn} in \Sec{SaturationTime}.
Assuming that surface losses can be modeled as turbulent diffusion terms,
we expect the values of $\eta_{\rm m}$ and $\eta_{\rm f}$ to be enhanced,
corresponding to losses from mean and fluctuating parts, respectively.

The phenomenological evolution equations are then written in terms of the
large and small scale magnetic energies, $M_{\rm m}$ and
$M_{\rm f}$, respectively, where we assume
$M_{\rm m}=\pm\half k_{\rm m} H_{\rm m}$ and
$M_{\rm f}=\mp\half k_{\rm f} H_{\rm f}$ for fully helical fields
(upper/lower signs apply to northern/southern hemispheres).
The phenomenological evolution equation for the energy
of the large scale magnetic field then takes the form
\begin{equation}
k_{\rm m}^{-1}{{\rm d}M_{\rm m}\over{\rm d}t}=
-2\eta_{\rm m}k_{\rm m}M_{\rm m}
+2\eta_{\rm f}k_{\rm f}M_{\rm f}.
\label{evolv_pheno}
\end{equation}
The positive sign of the term involving $M_{\rm f}$ reflects the generation
of large scale field by allowing small scale field to be removed.
After the time when the small scale magnetic field saturates, i.e.
when $t>t_{\rm sat}$, we have $M_{\rm f}\approx\mbox{constant}$,
and Eq.~(\ref{evolv_pheno}) can be solved to give
\begin{equation}
M_{\rm m}=M_{\rm f}\,{\eta_{\rm f}k_{\rm f}\over\eta_{\rm m}k_{\rm m}}
\left[1-\e^{-2\eta_{\rm m}k_{\rm m}^2(t-t_{\rm sat})}\right],
\quad\mbox{for $t>t_{\rm sat}$}.
\label{solution_pheno}
\end{equation}
This equation highlights three important aspects:
\begin{itemize}
\item{}
The time scale on which the large scale magnetic energy evolves
depends only on $\eta_{\rm m}$, not on $\eta_{\rm f}$
(time scale is shorter when $\eta_{\rm m}$ is increased).
\item{}
The saturation amplitude diminishes as $\eta_{\rm m}$ is
increased, which compensates the accelerated growth just past
$t_{\rm sat}$ \cite{BD01},
so the slope $\dd M_{\rm m}/\dd t$ is unchanged.
\item{}
The reduction of the saturation amplitude due to $\eta_{\rm m}$ can
be offset by having $\eta_{\rm m}\approx\eta_{\rm f}\approx\eta_{\rm t}$,
i.e.\ by having losses of small and large scale fields that are about
equally important.
\end{itemize}

The overall conclusions that emerge are: first, $\eta_{\rm m}>\eta$
is required if the large scale field is to evolve on a time scale other
than the resistive one and, second, $\eta_{\rm m}\approx\eta_{\rm f}$
is required
if the saturation amplitude is not to be catastrophically diminished.
These requirements are perfectly reasonable, but so far we are only
beginning to see this being also borne out by simulations
\cite{B05,BrandenburgSandin2004,BHKS05}.

An important limitation of the analysis above has been the neglect of
$\dd H_{\rm f}/\dd t$, precluding any application to early times.
Alternatively, one may directly use \Eqs{dABdt}{dabdt} and turn them
into evolution equations for $M_{\rm m}$ and $M_{\rm f}$, respectively,
\begin{equation}
k_{\rm m}^{-1}{{\rm d}M_{\rm m}\over{\rm d}t}=
2(\alpha-k_{\rm m}\eta_{\rm t}) M_{\rm m}-2\eta_{\rm m}k_{\rm m}M_{\rm m},
\label{evolv_pheno_Mm}
\end{equation}
\begin{equation}
k_{\rm f}^{-1}{{\rm d}M_{\rm f}\over{\rm d}t}=
2(\alpha-k_{\rm m}\eta_{\rm t}) M_{\rm m}-2\eta_{\rm f}k_{\rm f}M_{\rm f},
\label{evolv_pheno_Mf}
\end{equation}
where, $\alpha=\alpha_{\rm K}+\alpha_{\rm M}$ is the total $\alpha$
effect and $\alpha_{\rm M}$ the contribution from the small scale current
helicity; see \Eq{alphaTotal}.
As in \Eq{fullset2}, we can express the coefficient $\onethird\tau$
in terms of $\eta_{\rm t}/B_{\rm eq}^2$, and have in terms of energies,
$\alpha_{\rm M}=\pm \eta_{\rm t}k_{\rm f}M_{\rm f}/E_{\rm f}$ where
$E_{\rm f}=\half B_{\rm eq}^2V$ is the kinetic energy.
Equations \eqs{evolv_pheno_Mm}{evolv_pheno_Mf} can be applied at all
times, but they reduce to \Eq{evolv_pheno} in the limit of late times
when $M_{\rm f}=\mbox{const}$. 

Another caveat of the phenomenological approach is
that we have modeled the helicity fluxes with 
$\eta_{\rm m}$ and $\eta_{\rm f}$ as parameters which we can fix freely.
The physical mechanism which causes these fluxes
may not allow such a situation.
For example, if initially $H_{\rm m}=-H_{\rm f}$ so that
total helicity is zero, and the helicity fluxes involved mass
fluxes which carried both small and large scale fields
(cf.\ \Sec{bihelical} below), then the flux of small and 
large scale helicity could always be the same for all times. 
But from magnetic helicity conservation (neglecting microscopic diffusion), 
one will always have $H_{\rm m}=-H_{\rm f}$, and so
$M_{\rm m}/k_{\rm m}=M_{\rm f}/k_{\rm f}$.

\subsubsection{The vacuum cleaner experiment}
\label{VacuumCleanerExperiment}

The possible significance of open boundary conditions can be explained
as follows. We have seen that, at the end of the kinematic evolution,
the magnetic field can only change on a
resistive time scale if magnetic helicity conservation is obeyed.
The large scale field can therefore only increase if this does not involve
the generation of magnetic helicity (for example via differential rotation).
Any increase due to the $\alpha$ effect automatically implies an increase
of small scale magnetic helicity, and hence also small scale field.
However, once the small scale field has reached equipartition with the
kinetic energy, it cannot increase much further.
Any preferential loss of small scale magnetic energy would immediately
alleviate this constraint, as has been demonstrated in a numerical
experiment where magnetic field at and above the forcing wavenumber has
been removed in regular intervals via Fourier filtering \cite{BDS02}.

\begin{figure}[t!]\begin{center}
\includegraphics[width=.8\textwidth]{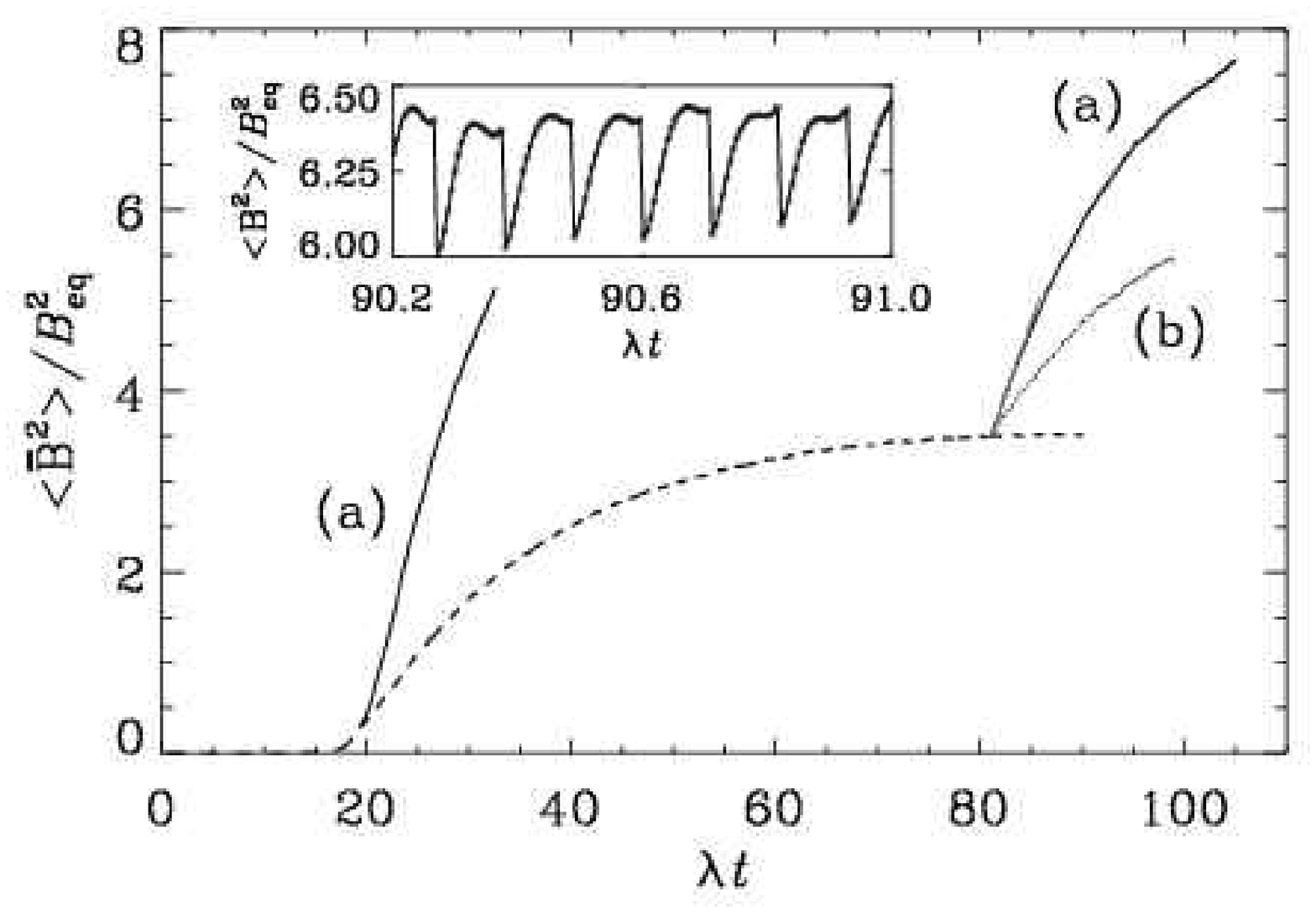}
\end{center}\caption[]{
The effect of removing small scale magnetic energy in regular time
intervals $\Delta t$ on the evolution of the large scale field (solid
lines). The dashed line gives the evolution of $\bra{\meanBB^2}$
for Run~3 of Ref.~\cite{B01}, where no such energy removal
was included. In all cases the field is shown in units of $B_{\rm
eq}^2=\rho_0\bra{\uu^2}$. The two solid lines show the evolution
of $\bra{\meanBB^2}$ after restarting the simulation from Run~3 of Ref.~\cite{B01}
at $\lambda t=20$ and $\lambda t=80$. Time is scaled with the kinematic
growth rate $\lambda$. The curves labeled (a) give the result for $\Delta
t=0.12\lambda^{-1}$ and those labeled (b) for $\Delta t=0.4\lambda^{-1}$.
The inset shows, for a short time interval, the sudden drop and subsequent
recovery of the total (small and large scale) magnetic energy in regular
time intervals.
Adapted from \cite{BDS02}.
}\label{Fpbmean}\end{figure}

In the experiment shown in \Fig{Fpbmean} the flow is forced with helical
waves at wavenumber $k=5$, giving rise to large scale dynamo action
with slow saturation at wavenumber $k=1$.
The magnetic field is then Fourier filtered in regular intervals
(between a tenth and a quarter of the $e$-folding time of the kinematic
dynamo) and the components above $k=4$ are set to zero, which is why this
is called the vacuum cleaner experiment.
Of course, the small scale magnetic energy will quickly recover and
reach again equipartition field strength, but there remains a short
time interval during which the small scale magnetic energy is in
sub-equipartition, allowing the magnetic helicity to grow
almost kinematically both at small {\it and} large scales.
The effect from each such event is small, but the effect from all events
together make up a sizeable 
enhancement to the amplitude of the generated large scale field.

\subsubsection{Speculations about boundary conditions}
\label{SpeculationsBoundaryConditions}

In order for magnetic helicity losses to have an advantageous effect
in the sun, or at
least in a more realistic simulation, it is important that the losses
of magnetic helicity occur predominantly at small scales.
It is conceivable that preferentially small scale magnetic helicity losses
may be possible in the presence of a more realistic modeling of the surface,
where magnetostatic equilibrium configurations may lose stability, leading to
the ejection of plasmoids and possibly magnetic helicity.
In this context it is important to recall that, whenever magnetic eruptions
of any form do occur, the field is always found to be strongly twisted
\cite{Demoulin_etal02,Gib02,BB03,RustKumar96,Nindos_etal03}.
It is therefore plausible that such events are an important part
of the solar dynamo.

\begin{figure}[t!]\begin{center}
\includegraphics[width=.99\textwidth]{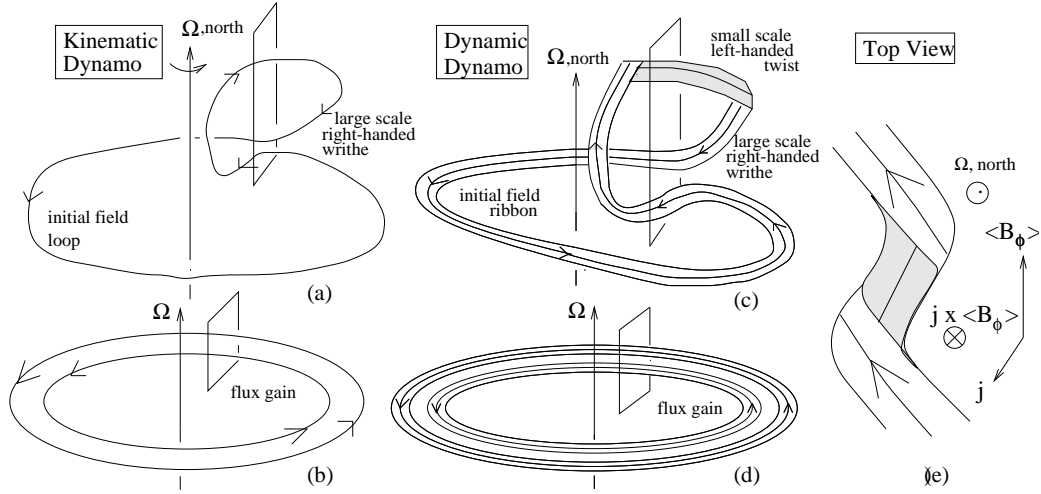}
\end{center}\caption[]{
Schematic of {\it kinematic} helical $\alpha\Omega$ dynamo in northern
hemisphere is shown in (a) and (b), whilst the {\it dynamic} helical
$\alpha\Omega$ dynamo is shown by analogy in (c) and (d).
Note that the mean field is represented as a line in (a) and (b)
and as a tube in (c) and (d).
(a) From an initial toroidal loop, the $\alpha$
effect induces a rising loop of right-handed writhe
that gives a radial field component.
(b) 
Differential rotation at the base of the loop
shears the radial component, amplifying the toroidal component,
and the ejection of the top part of loop (through coronal mass ejections)
allow for a net flux gain
through the rectangle.  
(c) Same as (a) but now with the field represented as a flux tube.
This shows how the right-handed writhe of the large scale loop
is accompanied by a left-handed twist along the tube,
thus incorporating magnetic  helicity  conservation.
(d) Same as (b) but with field represented as ribbon or tube.
(e) Top view of the combined twist and writhe
that can be compared with observed
coronal magnetic structures in active regions.
Note the {\sf N} shape of the 
right-handed large scale twist in combination with the
left-handed small scale twist along the tube.  
The backreaction force that resists the bending of the flux 
tube is seen to result from the small scale twist.
Note that diffusing the top part of the loops both allows for
net flux generation in the rectangles of (a)--(d), and alleviates 
the backreaction that could otherwise quench the dynamo.
Courtesy E.\ G.\ Blackman \cite{BB03}.
}\label{alltwist}\end{figure}

A second comment is here in order.
Looking at the coronal mass ejections depicted in \Fig{lasco46} it is
clear that fairly large length scales are involved in this phenomenon.
This makes the association with the small scale field dubious.
Indeed, the division between large and small scale fields becomes
exceedingly blurred, especially because small and large scale fields
are probably associated with one and the same flux tube structure,
as is clear from \Fig{alltwist}.

\begin{figure}[t!]\begin{center}
\includegraphics[width=.7\textwidth]{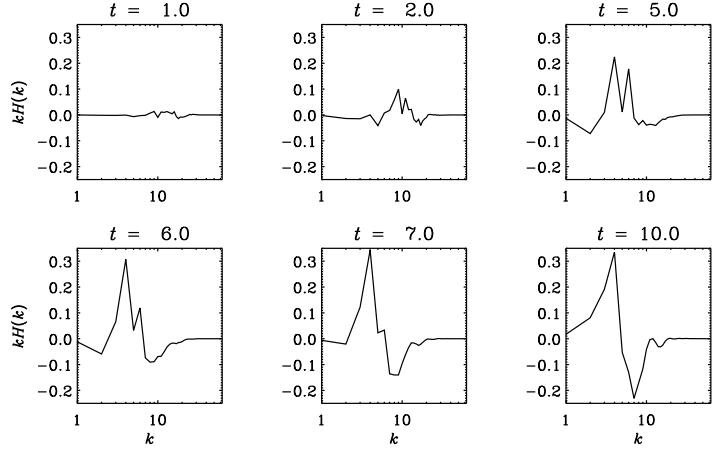}
\end{center}\caption[]{
Magnetic helicity spectra (scaled by wavenumber $k$ to give magnetic
helicity per logarithmic interval) taken over the entire computational
domain. The spectrum is dominated by a positive component at
large scales ($k=1...5$) and a negative component at small scales ($k>5$).
Adapted from Ref.~\cite{BB03}.
}\label{Fpspec}\end{figure}

\begin{figure}[t!]\begin{center}
\includegraphics[width=.7\textwidth]{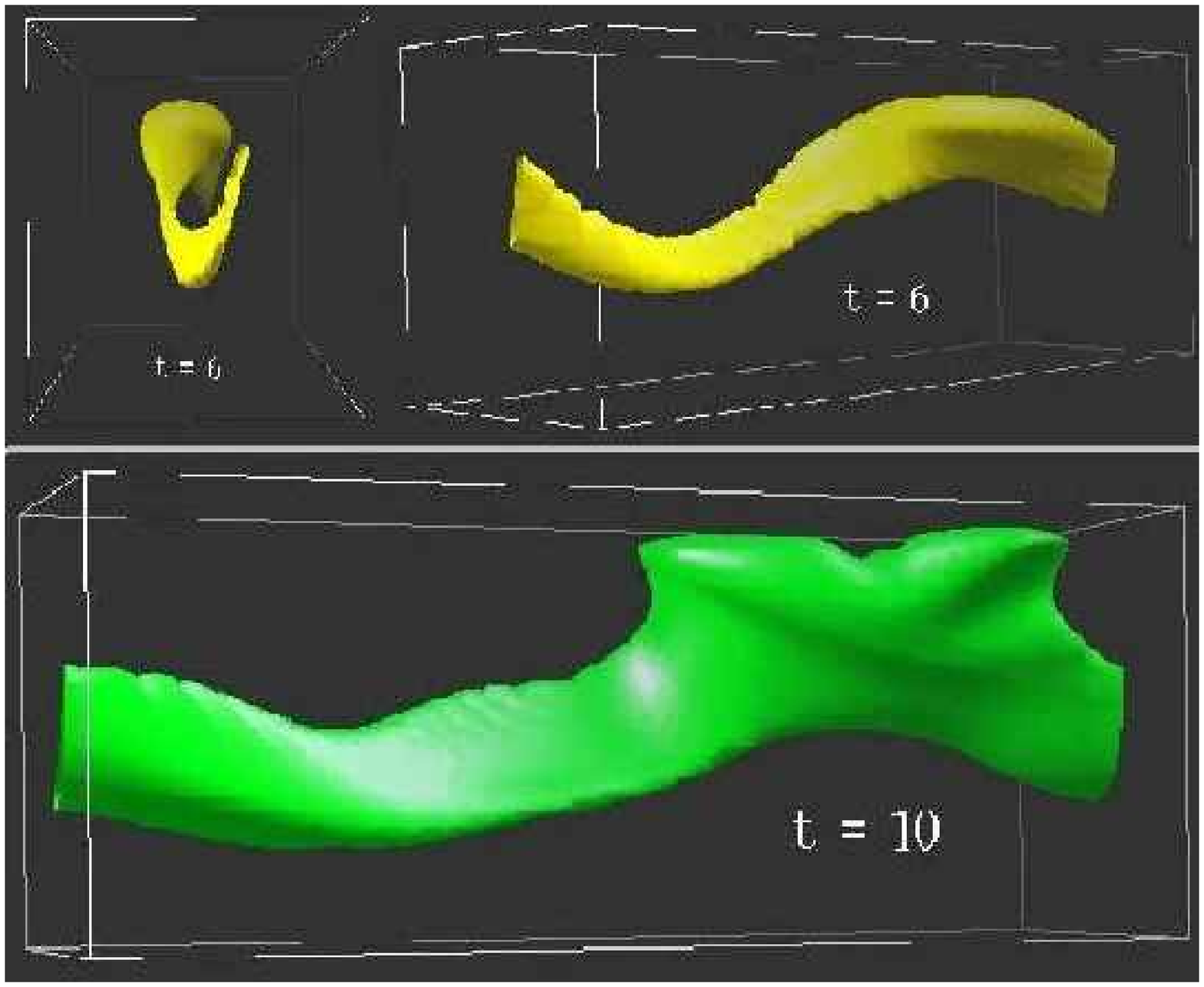}
\end{center}\caption[]{
Three-dimensional visualization of a rising flux tube in
the presence of rotation. The stratification is adiabatic such
that temperature, pressure, and density all vanish at a height
that is about 30\% above the vertical extent shown. (The actual
computational domain was actually larger in the $x$ and $z$ directions.)
Adapted from Ref.~\cite{BBS03}.
}\label{Fallviz}\end{figure}

The sketch shown in \Fig{alltwist} shows another related aspect.
Once a flux tube forms a twisted $\Omega$-shaped loop (via thermal or
magnetic buoyancy) it develops a self-inflicted internal twist in the
opposite sense.
(In the sketch, which applies to the northern hemisphere, the tilt has
positive sense, corresponding to positive writhe helicity, while the
internal twist is negative.)
These helicities can be associated with the positive $\meanJJ\cdot\meanBB$
and the negative $\overline{\jj\cdot\bb}$, that has also been confirmed
by calculating Fourier spectra from a buoyant flux tube tilting and
twisting under the influence of the Coriolis force and the vertical
density gradient \cite{BB03,BBS03}; see Fig.~\ref{Fpspec}.
Instead of visualizing the magnetic field strength, which can be strongly
affected by local stretching, we visualize the rising flux tube using
a passive scalar field that was initially concentrated along the flux
tube. This is shown in Fig.~\ref{Fallviz}.
This picture illustrates quite clearly that the $\alpha$ effect cannot
operate without (at the same time) twist helicity of the opposite sign.

\subsection{Effect of magnetic helicity losses on the $\alpha$ effect}
\label{MagneticHelicityLosses}

An important problem associated with the dynamical quenching model,
as formulated so far, is the fact that, unlike the volume average
$\bra{\meanAA\cdot\meanBB}$ over a periodic domain, the spatially
dependent magnetic helicity density $\overline{\aaa\cdot\bb}$ is
not gauge-invariant.
First, it is possible to formulate the theory directly in terms of the
small scale current helicity, $\overline{\jj\cdot\bb}$.
We should note, however, that it is possible to define
a gauge invariant magnetic helicity density
for the small scale field as a density of correlated links
and the associated flux \cite{KS_AB05}. We focus below
on the formulation in terms of the current helicity density.
Its evolution equation involves an extra divergence term 
or a current helicity flux, $\meanFF_C$.
Using the evolution equation
$\partial\bb/\partial t=-\nab\times\ee$, we can derive
an evolution equation for the small scale current helicity
in the form
\EQ
{\partial\over\partial t}\overline{\jj\cdot\bb}
=-2\,\overline{\ee\cdot\cc}-\nab\cdot\meanFF_C,
\label{jc_evolution}
\EN
where $\cc=\nab\times\jj$ is the curl of the small scale current density,
$\jj=\JJ-\meanJJ$, $\ee=\EE-\meanEE$ is the small scale electric field,
$\meanEE=\eta\meanJJ-\meanemf$ is the mean electric field, and
\EQ
\meanFF_C=\overline{2\ee\times\jj}+\overline{(\nab\times\ee)\times\bb}
\label{helflux}
\EN
is the mean current helicity flux resulting from the small scale field.

Mean-field models trying to incorporate such magnetic helicity
fluxes have already been studied by Kleeorin and coworkers
\cite{KMRS02,KMRS00,KMRS03} in the context of both galactic and
solar magnetic fields. An interesting flux of helicity
which one obtains even in nonhelical but anisotropic turbulence
due to the presence of shear
has been pointed out by Vishniac and Cho \cite{VC01}.
The presence of such a flux in non-helical shearing turbulence
has been confirmed numerically \cite{AB01}.
We will examine a general theory of such fluxes using the MTA 
in \Sec{nonlinearflux} below.

In the isotropic case,
$\overline{\ee\cdot\cc}$ can be approximated by
$k_{\rm f}^2\,\overline{\ee\cdot\bb}$.
(Here we have assumed that the effect of triple
correlation terms like $\overline{(\uu \times \bb)\cdot\nab^2\bb}$
in $\overline{\ee\cdot\cc}$ are small.)
Using FOSA or MTA, this can be replaced by
$k_{\rm f}^2(\meanemf\cdot\meanBB+\eta\overline{\jj\cdot\bb})$;
see also \Eq{fullset2}, which then
leads to the revised dynamical quenching formula
\EQ
{\partial\alpha\over\partial t}=-2\eta_{\rm t} k_{\rm f}^2\left(
{\alpha\meanBB^2-\eta_{\rm t}\meanJJ\cdot\meanBB
+\half k_{\rm f}^{-2}\nab\cdot\meanFF_C \over B_{\rm eq}^2}
+{\alpha-\alpha_{\rm K}\over R_{\rm m}}\right),
\label{fullset2flux}
\EN
where the current helicity flux,
$\meanFF_C=2\overline{\ee\times\jj}+\overline{(\nab\times\ee)\times\bb}$,
has entered as an extra term in the dynamical quenching formula
\eq{fullset2}.

Making the adiabatic approximation, i.e.\ putting the RHS of
\Eq{fullset2flux} to zero, one arrives at the algebraic quenching formula
\EQ
\alpha={\alpha_{\rm K}
+R_{\rm m}\left(\eta_{\rm t}\meanJJ\cdot\meanBB
-\half k_{\rm f}^{-2}\nab\cdot\meanFF_C\right)/B_{\rm eq}^2
\over1+R_{\rm m}\meanBB^2/B_{\rm eq}^2}
\quad\mbox{($\partial\alpha/\partial t=0$)}.
\label{AlphaStationaryFlux}
\EN
Considering again the $R_{\rm m}\to\infty$ limit, one has now
\EQ
\alpha\to
\eta_{\rm t}\tilde{k}_{\rm m}
-\half k_{\rm f}^{-2}(\nab\cdot\meanFF_C)/\meanBB^2,
\label{AlphaStationaryFlux2}
\EN
which shows that losses of negative helicity, as observed in the northern
hemisphere of the sun, would enhance a positive $\alpha$ effect
\cite{KMRS02,KMRS00,KMRS03}.
Here, $\tilde{k}_{\rm m}=\meanJJ\cdot\meanBB/\meanBB^2$ is a spatially
dependent generalization of \Eq{kmtilde} to the inhomogeneous case.
If the current helicity flux is approximated by a Fickian diffusive flux
proportional to the gradient of the small scale current helicity, i.e.\
$\meanFF_C\approx-2\eta_{\rm f}\nab\overline{\jj\cdot\bb}$, where $\eta_{\rm f}$
is an effective turbulent diffusion coefficient for the small scale current
helicity (see \Sec{PhenomenologicalModel}), the second term of
\Eq{AlphaStationaryFlux2} becomes approximately
$\eta_{\rm f}\tilde{k}_{\rm m}$, so
$\alpha$ approaches a combination of $\eta_{\rm t}\tilde{k}_{\rm m}$
and $\eta_{\rm f}\tilde{k}_{\rm m}$,
confirming again that $\alpha$ has increased.

\begin{figure}[t!]
\centering\includegraphics[width=0.95\textwidth]{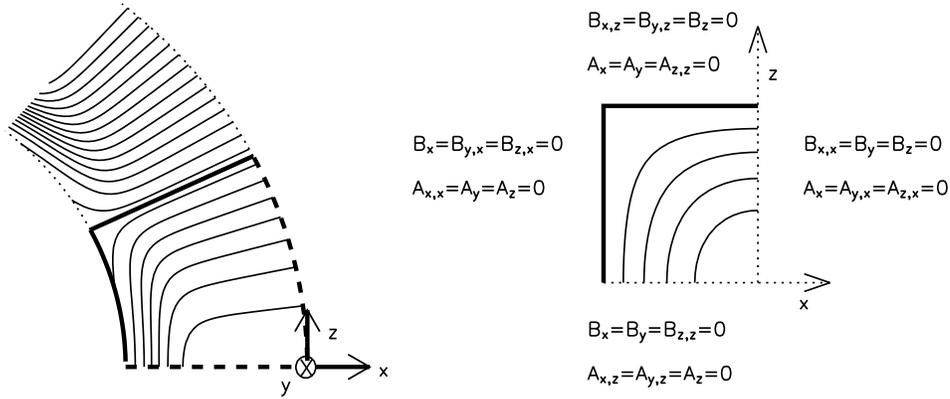}\caption{
Left: A sketch of the solar angular velocity at low latitudes with
spoke-like contours in the bulk of the convection zone merging gradually
into uniform rotation in the radiative interior.
The low latitude region, modeled in this paper, is indicated by thick
lines.
Right: Differential rotation in our cartesian model, with the equator
being at the bottom, the surface to the right, the bottom of the
convection zone to the left and mid-latitudes at the top.
Adapted from Ref.~\cite{BrandenburgSandin2004}.
}\label{sketch1}\end{figure}

In recent simulations with an imposed magnetic field
\cite{BrandenburgSandin2004}, the dependence of $\alpha$ on the magnetic
Reynolds number has been compared for both open and closed boundary
conditions using the geometry depicted on the right hand panel of
\Fig{sketch1}.
As usual, $\alpha$ was determined by measuring the turbulent
electromotive force, and hence $\alpha=\bra{\emf}\cdot\BB_0/B_0^2$.
As expected, $\alpha$ is negative when the helicity of the forcing is
positive, and $\alpha$ changes sign when the helicity of the forcing
changes sign.
The magnitudes of $\alpha$ are however different in the two cases:
$|\alpha|$ is larger when the helicity of the forcing is negative.
In the sun, this corresponds to the sign of helicity in the northern
hemisphere in the upper parts of the convection zone.
This is here the relevant case, because the differential rotation
pattern of the present model also corresponds to the northern hemisphere.

\begin{figure}[t!]
\centering\includegraphics[width=0.95\textwidth]{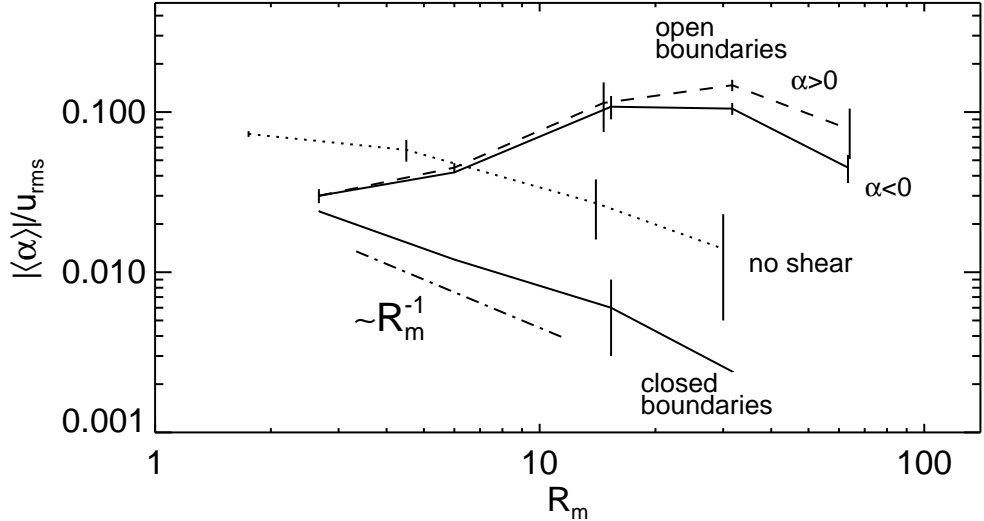}\caption{
Dependence of $|\bra{\alpha}|/u_{\rm rms}$ on $R_{\rm m}$
for open and closed boundaries.
The case with open boundaries and negative helicity is shown as a dashed line.
Note that for $R_{\rm m}\approx30$ the $\alpha$ effect
is about 30 times smaller when the boundaries are closed.
The dotted line gives the result with open boundaries but no shear.
The vertical lines indicate the range obtained by calculating
$\alpha$ using only the first and second half of the time interval.
Adapted from Ref.~\cite{BrandenburgSandin2004}.
}\label{palp_sum}\end{figure}

There is a striking difference between the cases with open and
closed boundaries which becomes particularly clear when comparing
the averaged values of $\alpha$ for different magnetic Reynolds
numbers; see \Fig{palp_sum}.
With closed boundaries $\alpha$ tends to zero like $R_{\rm m}^{-1}$,
while with open boundaries $\alpha$ shows no such decline.
There is also a clear difference between the cases with and without shear,
together with open boundaries in both cases.
In the absence of shear (dashed line in \Fig{palp_sum}) $\alpha$ declines
with increasing $R_{\rm m}$, even though for small values of $R_{\rm m}$
it is larger than with shear.

The difference between open and closed boundaries
can be interpreted in terms of a current helicity
flux through the two open boundaries of the domain.
Inspections of the actual fluxes suggests that
there is a tendency for the difference between incoming
flux at the equator (dotted line) and outgoing fluxes at outer surface
(solid line) to cancel, but the net outgoing flux is negative.
The flux for the total field is approximately four times larger than
what is accounted for by the Vishniac-Cho flux.
This might indicate that there is either another contribution to the
current helicity flux, or that the $\tau$ in the Vishniac-Cho flux
is underestimated.

\section{The microscopic theory of turbulent transport coefficients}
\label{Revisit}

In this section we describe in full detail the technique to calculate
turbulent transport coefficients in the presence of
slow rotation and weak inhomogeneity.
The technique to incorporate slow rotation and weak inhomogeneity
is the same as that introduced by Roberts and Soward \cite{RS75} long ago,
except that here we are using MTA.
The present results are basically in full agreement with a recent
calculation by R\"adler, Kleeorin and Rogachevskii \cite{RKR03},
except that here we retain the $\overline{\jj\cdot\bb}$ correction
to $\alpha$ throughout, even when this term was vanishing initially.
Indeed, in the nonlinear regime there is no
limit in which this term can be ignored,
because it is self-generated by the $\alpha$ effect.
In this section, overbars denote ensemble averages, but for all
practical purposes we may assume them to be equivalent to the
spatial averages used before.

As we have emphasized before, the ignorance of the $\overline{\jj\cdot\bb}$
feedback term in much of the mean field dynamo literature is one of the
main reasons why this theory has produced incorrect or unreliable results
and has not been in agreement with numerical simulations of hydromagnetic
turbulence \cite{B03passot}.
We are now at a point where a lot of work has to be reconsidered with
the correct feedback in place.
What needs to be done now is a careful calculation of the turbulent
transport coefficients that include, in particular, inhomogeneity and
losses through boundaries.
Here we are only able to present the initial steps in that direction.

\subsection{Transport coefficients in weakly inhomogeneous turbulence}
\label{radler_calc}

We consider the case when the correlation tensor of
fluctuating quantities ($\uu$ and $\bb$) vary slowly on
the system scale, say $\RR$. Consider the equal time,
ensemble average of the product $\overline{f(\xx_1)g(\xx_2)}$.
The common dependence of $f$ and $g$ on $t$ is assumed
and will not explicitly be stated.
We have
\EQ
\overline{f(\xx_1)g(\xx_2)}
=\int\int \overline{ \hat{f}(\kk_1) \hat{g}(\kk_2)}
\e^{\ii(\kk_1\cdot\xx_1 + \kk_2\cdot\xx_2)}
\,\dd^3k_1 \ \dd^3k_2, 
\label{twoscaledef}
\EN
where $\hat{f}$ and $\hat{g}$ are the Fourier transforms
of $f$ and $g$ respectively. (In general the Fourier transform
of any function, say $f$, will be denoted by the same symbol
with a `hat' i.e. $\hat{f}$.) 
We define the difference $\rrr = \xx_1 - \xx_2$ and the mean
$\RR = \half(\xx_1 + \xx_2)$, keeping in mind that all
two point correlations will vary rapidly with $\rrr$ but slowly
with $\RR$ \cite{RS75}. We can re-express this correlation as
\EQA
\overline{f(\xx_1)g(\xx_2)}
&=&\int\int \overline{\hat{f}(\kk + \half\KK) \hat{g}(-\kk +\half\KK)}
\ {\rm e}^{\ii(\KK\cdot\RR + \kk\cdot\rrr)}
\,\dd^3K \ \dd^3k \nonumber \\
&\equiv& \int \Phi(\hat{f},\hat{g},\kk,\RR) \ {\rm e}^{i\kk\cdot\rrr} \ 
\dd^3k, 
\ENA
where we have defined the new wavevectors
$\kk = \half(\kk_1 - \kk_2)$ and $\KK = \kk_1 + \kk_2$.
Also for later convenience we define
\EQA
\Phi(\hat{f},\hat{g},\kk,\RR)
= \int \overline{\hat{f}(\kk + \half\KK) \hat{g}(-\kk +\half\KK)}
\ {\rm e}^{i\KK\cdot\RR} \,\dd^3K .
\label{phi_def}
\ENA
In what follows we require the correlation
tensors of the $\uu$ and $\bb$ fields and the cross correlation
between these two fields in Fourier space. The velocity
and magnetic correlations, as well as the
turbulent electromotive force $\meanemf$, are given by
\vspace{-5mm}
\EQA
v_{ij}(\kk,\RR) &=& \Phi(\hat{u}_i,\hat{u}_j,\kk,\RR), \quad
\overline{\uu^2}(\RR) = \delta_{ij}\int v_{ij}(\kk,\RR) \,\dd^3k,
\label{corv_def}
\\
m_{ij}(\kk,\RR) &=& \Phi(\hat{b}_i,\hat{b}_j,\kk,\RR), \quad
\overline{\bb^2}(\RR) = \delta_{ij}\int m_{ij}(\kk,\RR) \,\dd^3k,
\label{corm_def}
\\
\chi_{jk}(\kk,\RR) &=& \Phi(\hat{u}_j,\hat{b}_k,\kk,\RR), \quad
\meanemfs_i(\RR) = \epsilon_{ijk} \int \chi_{jk}(\kk,\RR) \,\dd^3k.
\label{emf_def}
\ENA
In order to calculate these quantities we first consider the
time derivative of $\chi_{jk}$,
\EQ
{\partial\chi_{jk} \over \partial t} = 
\Phi(\hat{u}_j,\dot{\hat{b}_k},\kk,\RR) +
\Phi(\dot{\hat{u}_j},\hat{b}_k,\kk,\RR),
\label{chider}
\EN
where the dots on $\hat{b}_j$ and $\hat{u}_i$ denote a partial
time derivative.
The equation for $\bb$ is given by Eq.~(\ref{Induction_flucts}).
For the present we neglect the mean velocity term $\UU$. The Fourier
transform of this equation is then,
\EQ
\dot{\hat{b}}_k(\kk)
=\epsilon_{kpq}\epsilon_{qlm} \ii k_p \int \hat{u}_l(\kk - \kk')
\hat{\meanB}_m(\kk') \,\dd^3k'
+\hat{G}_k(\kk) -\eta k^2 \hat{b}_k(\kk),
\label{bhat}
\EN
where we have included the nonlinear term, 
$\GG = \nab\times(\uu\times\bb - \overline{\uu\times\bb})$. 
(Note that since $\overline{\uu\times\bb}$ is a mean field
quantity, it will give zero contribution to $\chi_{jk}$, when
multiplied by $\uft$ and ensemble averaged.)
Using the momentum equation for the velocity field 
$\UU$ and splitting the velocity into mean and fluctuating
parts, $\UU = \meanUU + \uu$, and neglecting for simplicity
the mean flow, $\meanUU=0$, we have, in a rotating frame,
\EQ
{\partial \uu \over\partial t} = 
-\nab p_{\rm eff} + \meanBB\cdot\nab\bb + \bb\cdot\nab\meanBB 
- 2\OO\times\uu + \HH + \ff + \nu\nab^2\uu.
\label{momentum_flucts}
\EN
Here we have assumed incompressible motions with $\nab\cdot\uu =0$, and
a constant density; we have therefore adopted units such that $\mu_0=\rho_0=1$.
We have defined the effective pressure $p_{\rm eff}$ which
contains the perturbed pressure including the magnetic
contribution, $\ff$ is the random external force, 
and the terms nonlinear in $\uu$ and $\bb$ are gathered in
$\HH = - \uu\cdot\nab\uu + \overline{\uu\cdot\nab\uu} +
\bb\cdot\nab\bb - \overline{\bb\cdot\nab\bb}$.
In the rotating frame, we also have a Coriolis force contribution
$2\OO \times \uu$.
The effective pressure term can be obtained by solving
a Poisson type equation derived by taking the divergence of
\Eq{momentum_flucts}.
In Fourier space this corresponds to multiplying the equation
by the projection operator, $P_{ij}(\kk)=\delta_{ij}-\kh_i\kh_j$,
where $\kkk=\kk/|\kk|$ is the unit vector of $\kk$, so
\EQ
{\partial\hat{u}_j\over\partial t}=P_{js}\left[
  (\widehat{\meanBB\cdot\nab\bb})_s
+ (\widehat{\bb\cdot\nab\meanBB})_s
- 2\epsilon_{slm}\Omega_l\hat{u}_m
+ \hat{f}_s + \hat{H}_s -\nu k^2 \hat{u}_s \right].
\EN
Here we have dropped the argument $\kk$ on all quantities
and, for brevity, we have introduced the notation
\EQ
(\widehat{\meanBB\cdot\nab\bb})_s=
\int \hat{\meanB}_l(\kk-\kk')(\ii k'_l)\,\hat{b}_s(\kk')\,\dd^3k',
\EN
\EQ
(\widehat{\bb\cdot\nab\meanBB})_s=
\int \hat{b}_l(\kk-\kk') (\ii k'_l)\,\hat{\meanB}_s(\kk')\,\dd^3k'
\EN
for the Fourier transforms of the magnetic tension/curvature force.
We can simplify the rotational part of the Coriolis force
by using the identity
\EQ
\epsilon_{ijk}
=\kh_i\kh_p\epsilon_{pjk}
+\kh_j\kh_p\epsilon_{ipk}
+\kh_k\kh_p\epsilon_{ijp},
\EN
noting that $P_{ij}\epsilon_{jkl}=\kh_k\kh_p\epsilon_{ipl}
+\kh_l\kh_p\epsilon_{ikp}$, and using incompressibility
$\kh_k\hat{u}_k=0$, to get
\EQ
P_{js}(\kk)\,\epsilon_{slm}\Omega_l\hat{u}_m
=\kkk\cdot\OO\,\epsilon_{jpm} \kh_p\hat{u}_m.
\EN
We are now in a position to calculate $\partial\chi_{jk}/\partial t$.
Using the induction equation for $\dot{\hat{\bb}}$ given by \eq{bhat},
the first term in \eq{chider} is given by
\EQ
\Phi(\hat{u}_j,\dot{\hat{b}_k},\kk,\RR) = 
\dot{\chi}_{jk}^{\rm K} + \overline{\uft_j\Gft_k}
-\eta (\kk+\half\ii\nab)^2\chi_{jk},
\EN
where we have substituted $\KK$ by $-\ii\nab$ and
have introduced the abbreviation
\EQA
\dot{\chi}_{jk}^{\rm K}=
\epsilon_{kpq}\epsilon_{qlm}&\int& \overline{\uft_j(\kk+\half\KK) \uft_l(-\kk
+ \half\KK - \kk')}
\,\meanBft_m(\kk') \nonumber \\
&\times& \ii(-k_p +\half K_p) \ \dd^3k'
\e^{\ii\KK\cdot\RR} \ \dd^3K.
\ENA
Further the triple correlations of the form
$\overline{\uft_j\Gft_k}$, will be either ignored (FOSA)
or replaced by the double correlation $\chi_{jk}$, divided
by a relaxation time (the $\tau$ approximation).
To evaluate the above velocity correlation in 
$\partial\chi_{jk}^K/\partial t$, we
need to bring the $\overline{\uft_j\uft_l}$ term into a form
similar to \Eq{corv_def}, so that we can replace it by $v_{jl}$.
Thus, we
define new wavevectors $\kk_1 = \kk+\half\KK$ and $\kk_2=-\kk+\half\KK + \kk'$,
and transform to new variables, ($\kk',\KK'$), where 
$\KK'= \kk_1+\kk_2 = \KK - \kk'$. We then have 
\EQA
\dot{\chi}_{jk}^{\rm K}=
\epsilon_{kpq}\epsilon_{qlm}&\int& \overline{\uft_j(\kk+ \half\kk' + 
\half\KK') \,\uft_l(-(\kk + \half\kk') + \half\KK')}
\meanBft_m(\kk') \nonumber \\ && \ii(-k_p + \half k'_p + \half K'_p)
\ \e^{\ii(\KK'+ \kk')\cdot\RR} \ \dd^3K' \ \dd^3k'.
\ENA
Using the definition of the velocity correlation function
$v_{ij}$ in \Eqs{phi_def}{corv_def} and carrying out first 
the integral over $\KK'$, and replace $\ii K'_p$ by
$\nabla_p\equiv\partial/\partial R_p$, we can write
\EQ
\dot{\chi}_{jk}^{\rm K}=
\epsilon_{kpq}\epsilon_{qlm}\int(-\ii k_p + \half\ii k'_p+\half\nabla_p)
\,v_{jl}(\kk+ \half\kk', \RR) 
\meanBft_m(\kk') \e^{\ii\kk'\cdot\RR} \dd^3k'.
\EN
Note that, since the mean field $\meanBB$ varies
only on large scales, $\meanBft(\kk')$ will be non-zero
only for small $\mod{\kk'}$. This suggests expanding $v_{jl}$
above in a Taylor series in $\kk'$, i.e.\
\EQ
v_{jl}(\kk+ \half\kk', \RR)\approx
v_{jl}(\kk,\RR)+\half k'_s{\partial\over\partial k_s} v_{jl}(\kk,\RR).
\EN
We will keep only terms that can contribute derivatives in $\RR$
no higher than the first derivative.
Integrating over $\kk'$, using
$\epsilon_{kpq}\epsilon_{qlm} = \delta_{kl}\delta_{pm} -
\delta_{km}\delta_{pl}$,
and noting that the inverse Fourier transform of
$\ii k_s'\meanBft_m(\kk')$ is
$\meanB_{m,s}\equiv\partial \meanB_m/\partial R_s$,
we get
\EQ
\dot{\chi}_{jk}^{\rm K}=
-\ii\kk\cdot\meanBB\,v_{jk} + \half\meanBB\cdot\nab v_{jk}
-v_{jl} {\partial \meanB_k \over \partial R_l} -
\half k_m{\partial v_{jk} \over \partial k_s} \meanB_{m,s}.
\EN
Now turn to the second term in \eq{chider} for 
$\partial\chi_{jk}/\partial t$. This is given by
\EQ
\Phi(\dot{\hat{u}_j},\hat{b}_k,\kk,\RR)
=\dot{\chi}_{jk}^{\rm M}
+\dot{\chi}_{jk}^\Omega
+ \overline{\Hft_j\bft_k}
-\nu (\kk-\half\ii\nab)^2\chi_{jk},
\EN
where 
\EQA
\dot{\chi}_{jk}^{\rm M} &=&
\int \ \dd^3k' \ \dd^3K \e^{\ii\KK\cdot\RR} \ 
P_{js}(\kk + \half\KK) \times \nonumber \\ 
&& [ 
\ii(k_l + \half K_l-k'_l) \ \meanBft_l(\kk') 
\ \overline{\bft_s(\kk + \half\KK - \kk') 
\bft_k(-\kk + \half\KK)} \nonumber \\
&+& (\ii k'_l) \ \meanBft_s(\kk') 
\ \overline{\bft_l(\kk + \half\KK - \kk')
\bft_k(-\kk + \half\KK)}],
\label{chi_mag}
\ENA
and 
\EQ
\dot{\chi}_{jk}^\Omega=
2\Omega_m\epsilon_{jlt}\int {p_m p_t\over\pp^2}\,
\overline{\uft_l(\kk + \half\KK)
\bft_k(-\kk + \half\KK)} \e^{\ii\KK\cdot\RR} \ \dd^3K.
\label{chi_omega}
\EN
Here we have introduced $\pp=\kk+\half\KK$ for brevity.
Once again, the triple correlations of the form
$\overline{\Hft_j\bft_k}$, will be either ignored (FOSA)
or replaced by the double correlation $\chi_{jk}$, divided
by a typical correlation time (the $\tau$ approximation).
Empirically, we know that
the term $\overline{\hat{\ff}\cdot\hat{\bb}}$ is small.
We first simplify the $\dot{\chi}_{jk}^{\Omega}$ term,
keeping only terms that are at most a first derivative
in $\RR$. For this first expand $p_m p_t/\pp^2$ as
\EQ
{(k_m + \half K_m)(k_t + \half K_t) \over (\kk + \half\KK)^2}
= {k_mk_t \over k^2} + {k_mK_t \over 2k^2} + 
{K_mk_t \over 2k^2} - {k_mk_t k_sK_s \over k^4},
\EN
keeping terms that are at most linear in $\KK$.
Then the integral over $\KK$ can be carried
out using the definition of $\chi_{ij}$.  
We get 
\EQ
\dot{\chi}_{jk}^{\Omega} =
-{\sf A}_{jl}\chi_{lk} - {\sf B}_{jlm}{\partial\chi_{lk}\over \partial R_m},
\EN
where the $\chi_{lk}$ without superscript is the full $\chi_{lk}$, and
\EQ
{\sf A}_{jl} = -2\epsilon_{jlt}{\kk\cdot\OO \over k^2}k_t,
\EN
\EQ
{\sf B}_{jlm} = \ii\epsilon_{jlm} {\kk\cdot\OO \over k^2} 
+\ii\epsilon_{jlt} {k_t\Omega_m \over k^2}
-2\ii\epsilon_{jlt} {(\kk\cdot\OO) k_t k_m \over k^4}.
\label{Bjlm}
\EN
As shown in \App{AppSimplifyChiM}, the simplification of
$\dot{\chi}_{jk}^{\rm M}$ leads to
\EQA
\dot\chi_{jk}^{\rm M} &=&
\ii\kk\cdot\meanBB\,m_{jk} + \half\meanBB\cdot\nab m_{jk}
+\meanB_{j,l} m_{lk} \nonumber \\
&-& \half \meanB_{m,s} k_m
{\partial m_{jk} \over \partial k_s}
-2{k_jk_s \over k^2} \meanB_{s,l} m_{lk}.
\label{chi_mfin}
\ENA
We can now add all the different contributions to get,
\EQA
{\partial\chi_{jk} \over \partial t} = 
{\sf I}_{jk} -&& {\sf A}_{jl}\chi_{lk} - 
{\sf B}_{jlm}{\partial\chi_{lk}\over \partial R_m }
\nonumber \\ 
&&+ \overline{\uft_j\Gft_k} + \overline{\Hft_j\bft_k}
-(\eta +\nu)k^2 \chi_{jk} -\ii(\eta - \nu)\kk\cdot\nab\chi_{jk},
\label{chi_fin}
\ENA
where
\EQA
{\sf I}_{jk} &=& 
-\ii\kk\cdot\meanBB\,(v_{jk}- m_{jk}) 
+ \half\meanBB\cdot\nab(v_{jk} + m_{jk})
+ \meanB_{j,l} m_{lk} - \meanB_{k,l} v_{jl} \nonumber \\
&-& \half \meanB_{m,s}
k_m \left[{\partial v_{jk} \over \partial k_s}
+ {\partial m_{jk} \over \partial k_s}\right]
- 2{k_jk_s \over k^2} \meanB_{s,l}m_{lk},
\label{I_jk}
\ENA
and we have only kept the terms in the microscopic diffusion
and viscosity up to first order in the large scale derivative.

The first order smoothing approximation can be
recovered by neglecting the triple correlation terms 
$\bra{\uft_j\Gft_k} + \bra{\Hft_j\bft_k}$.
Using MTA, one summarily approximates
the triple correlations as a damping term, and takes
$\bra{\uft_j\Gft_k} + \bra{\Hft_j\bft_k} \approx - \chi_{jk}/\tau(k)$,
where in general the parameter $\tau$ could be $k$-dependent.
(The microscopic diffusion terms can either be absorbed into the
definition of $\tau(k)$ or neglected for large Reynolds numbers.)
One then has
\EQ
{\partial\chi_{jk}\over\partial t}
= {\sf I}_{jk}
-{\sf A}_{jl}\chi_{lk}
- {\sf B}_{jlm}{\partial\chi_{lk}\over \partial R_m}
- {\chi_{jk}\over\tau},
\label{chi_tau_timedependent}
\EN
or
\EQ
{\partial\chi_{jk}\over\partial t}
= {\sf I}_{jk}
- {1\over\tau}{\sf D}_{jl}\chi_{lk}
- {\sf B}_{jlm}{\partial\chi_{lk}\over \partial R_m},
\label{chi_tau_timedependent2}
\EN
where ${\sf D}_{jl}=\delta_{jl} + \tau {\sf A}_{jl}$.
On time scales long compared with $\tau$,
one can neglect the time derivative of $\chi_{jk}$,
so one has \cite{BF02b}
\EQ
{\sf D}_{jl}\chi_{lk} + 
\tau {\sf B}_{jlm}{\partial\chi_{lk}\over \partial R_m }
= \tau {\sf I}_{jk}.
\label{chi_tau}
\EN
In the weakly inhomogeneous case that we are considering,
this equation can be solved iteratively. To zeroth order, 
one neglects the $\partial/\partial \RR$ terms 
on both sides of \Eq{chi_tau}.
This gives a zeroth order estimate
\EQ
\chi_{lk}^{(0)} = {\sf D}^{-1}_{lj}\,\tau {\sf I}_{jk}^{(0)},
\EN
where
\EQ
{\sf I}_{jk}^{(0)} = -\ii\kk\cdot\meanBB\,(v_{jk}- m_{jk}).
\label{I0approx}
\EN
Here the inverse matrix ${\sf D}_{jl}^{-1}$ is given by
\EQ
{\sf D}_{jl}^{-1}=\left.\left(\delta_{jl}+\Co_{\kk}\epsilon_{jlm}\kh_m
+\Co_{\kk}^2\kh_j\kh_l\right)\right/\left(1+\Co_{\kk}^2\right),
\EN
where $\Co_{\kk}\equiv2\OO\tau\cdot\kkk$ is the Coriolis number
with respect to the component of $\OO$ that is aligned with $\kk$,
and $\kkk\equiv\kk/k$ is the unit vector of $\kkk$.

To the next order, we keep first derivative terms and
substitute $\chi_{lk}^{(0)}$ in the
$\partial\chi_{lk}/ \partial R_m$ term.
This gives
\EQ
\chi_{jk} = {\sf D}_{jl}^{-1} \tau {\sf I}_{lk}
- \tau^2 {\sf D}^{-1}_{jp}  {\sf B}_{plm} 
{\partial\over\partial R_m}
\left({\sf D}^{-1}_{ls} {\sf I}_{sk}^{(0)}\right).
\label{chi_sol2}
\EN

Further, in many situations one will be concerned with
the slow rotation limit, where $\Omega\tau \ll 1$. In this
case one needs to keep only terms that are
at most linear in $\Omega\tau$. We do this below;
partial results to higher order in $\Omega \tau$ can be
found in Ref.~\cite{RKR03}.
To linear order, ${\sf D}_{jl}^{-1} = \delta_{jl} + 
2\tau\kkk\cdot\OO\epsilon_{jlm}\kh_m$. Substituting this into
\Eq{chi_sol2}, and noting that the ${\sf B}_{plm}$ term is
already linear in $\Omega$, we get
\EQ
\chi_{jk} = \tau {\sf I}_{jk} + 2\tau^2\kkk\cdot\OO\epsilon_{jlm}\kh_m
{\sf I}_{lk} - \tau^2 {\sf B}_{jlm} {\partial {\sf I}_{lk}^{(0)}\over \partial R_m }.
\label{chi_sol3}
\EN
We will work with this expression to evaluate $\meanemf$.
The first term in \eq{chi_sol3} contributes to 
$\meanemf$, even in the case when $\Omega = 0$. 
This contribution is given by
\EQA
\meanemf^{(0)}_i &=& \epsilon_{ijk}\int \tau {\sf I}_{jk} \,\dd^3k \\
&=&
\epsilon_{ijk} \int \tau 
\Big[-\ii\kk\cdot\meanBB(v_{jk}^{\rm A}- m_{jk}^{\rm A}) 
+ \half\meanBB\cdot\nab(v_{jk}^{\rm A} + m_{jk}^{\rm A}) \nonumber \\
&+& \meanB_{j,l} m_{lk} - \meanB_{k,l} v_{jl} 
- \half k_m \ \meanB_{m,s}
\left({\partial v_{jk}^{\rm A} \over \partial k_s}
+ {\partial m_{jk}^{\rm A} \over \partial k_s}\right)
- 2{k_jk_s \over k^2} \meanB_{s,l}m_{lk}\Big]\dd^3k. \nonumber
\label{emf_0}
\ENA
Note that due to the presence of the antisymmetric
tensor $\epsilon_{ijk}$ only the antisymmetric parts of 
$v_{jk}$ and $m_{jk}$ survive in some of the terms
above, and these are denoted by $v_{jk}^{\rm A}$ and $m_{jk}^{\rm A}$, 
respectively. 

To proceed we need the form of the velocity and magnetic correlation
tensors. We adopt the form relevant when these are
isotropic and weakly inhomogeneous, as discussed in detail in 
Refs~\cite{RKR03,RS75}. We take
\EQA
v_{ij} &=& \left[P_{ij}(\kk) + {\ii \over 2k^2}(k_i\nabla_j - k_j\nabla_i)
\right] E(k,\RR) \nonumber \\
&-& {1\over 2k^2} \left[\epsilon_{ijk} k_k
\left(2\ii+\!{\kk\cdot\nab\over k^2}\right) 
-(k_i\epsilon_{jlm}\!+\!k_j\epsilon_{ilm}){k_l\over k^2}\nabla_m
\right] F(k,\RR),
\label{vel_cor}
\ENA
\EQA
m_{ij} &=& \left[P_{ij}(\kk) + 
{\ii \over 2k^2}(k_i\nabla_j - k_j\nabla_i)
\right] M(k,\RR) \nonumber \\
&-& {1\over 2k^2} \left[\epsilon_{ijk} k_k
\left(2\ii+\!{\kk\cdot\nab\over k^2}\right)
-(k_i\epsilon_{jlm}\!+\!k_j\epsilon_{ilm}){k_l\over k^2}\nabla_m
\right] N(k,\RR).
\label{mag_cor}
\ENA
Here $4\pi k^2E$ and $4\pi k^2M$ are
the kinetic and magnetic energy spectra, respectively,
and $4\pi k^2F$ and $4\pi k^2N$ are the corresponding helicity spectra.
(Note that in this section we use the symbol $F(k,\RR)$ to denote
the kinetic helicity spectrum; this should not be confused with the
the coordinate space $F(r)$ defined earlier, for the helical part
of the velocity correlation function.)
They obey the relations
\EQA
&&\overline{\uu^2}\left(\RR\right) = 2 \int E(k,\RR)\,\dd^3k, \quad
\overline{\oo\cdot\uu}\left(\RR\right) = 2 \int F(k,\RR)\,\dd^3k,
\nonumber \\ 
&&\overline{\bb^2}\left(\RR\right) = 2 \int M(k,\RR)\,\dd^3k, \quad
\overline{\jj\cdot\bb}\left(\RR\right) = 2 \int N(k,\RR)\,\dd^3k.
\label{EFMN_def}
\ENA

Note that, so far, we have also made no assumptions about
the origin of the velocity and magnetic fluctuations, 
apart from their general form. The fluctuating velocity
could be driven by random forcing and also be responding
to the effects of rotation and/or the 
nonlinear effects of the Lorentz forces. 
For example, suppose one were to assume that the turbulence
were originally nonhelical, that is $E = E^{(0)}$ and 
$F$ was originally zero. The helical parts
of the velocity correlations can still be generated 
due to rotation and stratification. In this case
the helical part of the velocity correlation will no
longer be isotropic, and will reflect the anisotropy
induced by both rotation and stratification. To work out 
the corresponding modification to $v_{ij}$
one has to take into account the effect of the
Coriolis force in some approximate way. This has
been done in Ref.~\cite{RS75} by assuming that the velocity induced
by rotation is very small compared to the original
turbulent velocity and using the $\tau$ approximation
in \cite{RKR03}. Interestingly, it turns out that, 
to the lowest order in $R$ derivatives, rotation
induces a helical part to $v_{ij}$ that can be
described simply by adopting an anisotropic $F = F^{\Omega}$.
It is shown in Ref.~\cite{RKR03} that, under the $\tau$ approximation, 
this is given by
\EQ
F^{\Omega}(\kk,\RR) = -2\tau^*\left[(\kkk\cdot\OO)\,(\kkk\cdot\nab) 
- \OO\cdot\nab \right] E^{(0)}(k,\RR),
\label{rot_F}
\EN
where $\tau^*(k)$ is another correlation time, 
that could in principle be different from $\tau(k)$. 
We use this for $F$ when we discuss the effects of rotation,
although one must keep in mind that it is likely to
give only a crude and at most qualitative estimate of the
effects of rotation.\footnote{We also point out that the
velocity and magnetic correlations will become anisotropic 
when the mean field becomes strong and begins to influence
the turbulence; throughout the discussion below we do 
not explicitly take this feature into account, since the
major feed back to the $\alpha$ effect due to the
current helicity is already important for weak mean fields.
Also the anisotropy induced by rotation is already important, 
even when the mean field is weak. For a discussion of the effects of 
anisotropy induced by strong mean fields, see for example 
\cite{Roga+Klee00}.}

We substitute the velocity and magnetic correlations
given in \eq{vel_cor} and \eq{mag_cor} into \eq{emf_0}
for $\meanemf^{(0)}$. Note that the term 
proportional to $v_{jk}^{\rm A} + m_{jk}^{\rm A}$ gives zero contribution,
as the antisymmetric parts of the correlations are
odd in $k_i$ and so integrate to zero, while
doing the $k$-integration. (This continues to hold even
if $F= F^{\Omega}$.) Also the terms that have
$k$ derivatives, give zero contribution after
integrating by parts, and using $\nab\cdot\meanBB = 0$.
Further, in terms which already have $R_i$ derivatives
of $\meanBB$, one need not include the $R_i$ derivative
contributions of $v_{jk}$ and $m_{jk}$.
The remaining terms can be written as
\EQA
\meanemf^{(0)}_i &=&
\epsilon_{ijk} \int \tau 
\Big\{ \kkk\cdot\meanBB \left[
\half(\kunit_j\nabla_k - \kunit_k\nabla_j)
(E-M) - \epsilon_{jkm}\kunit_m (F-N) \right] \nonumber \\
&+& \meanB_{j,l} P_{lk}(\kk) M - \meanB_{k,l} P_{jl}(\kk) E 
- 2\kunit_j\kunit_s \meanB_{s,l} P_{lk}(\kk)M \Big\}\,\dd^3k ,
\label{emf_0a}
\ENA
where, as before, commas denote partial differentiation with respect to $R$, i.e.\
$\meanB_{j,l} \equiv \partial \meanB_j / \partial R_l$.

\subsection{Isotropic, helical, nonrotating turbulence}

In the first instance, suppose we were to assume
that the turbulence is driven by isotropic forcing
which is also helical, so that all spectral functions
depend only on $\mod{\kk}$. 
One can then do the angular parts of the $k$-integral
in \eq{emf_0a}, using a relation valid for
any ${\cal F}(k)$ of the form 
\EQ
\int \kunit_i\kunit_j {\cal F}(k) \,\dd^3k
= \onethird\delta_{ij} \int {\cal F}(k) \,\dd^3k.
\EN
We get
\EQ
\meanemf^{(0)}_i =
\epsilon_{ijk}\left[(\meanB_j\nabla_k - \meanB_k\nabla_j) 
{\tilde{E}-\tilde{M} \over 6} - \epsilon_{jkm}\meanB_m {\tilde{F}-
\tilde{N} \over 3} - \twothird \meanB_{k,j} \tilde{E} \right],
\label{emf_0b}
\EN
where we have defined
\EQA
&& \tilde{E} = \int \tau(k) E(k,\RR)\,\dd^3k, \quad 
\tilde{M} = \int \tau(k) M(k,\RR)\,\dd^3k, \nonumber \\
&& \tilde{F} = \int \tau(k) F(k,\RR)\,\dd^3k, \quad 
\tilde{N} = \int \tau(k) N(k,\RR)\,\dd^3k.
\label{tilde_E}
\ENA
For a constant $\tau(k) = \tau_0$ say, we simply have 
$\tilde{E} = \half\tau_0 \overline{\uu^2}$, 
$\tilde{M} = \half\tau_0 \overline{\bb^2}$, 
$\tilde{F} = \half\tau_0 \overline{\oo\cdot\uu}$, 
$\tilde{N} = \half\tau_0 \overline{\jj\cdot\bb}$.  
In this case we have for the turbulent EMF
\EQ
\meanemf^{(0)} = \alpha \meanBB -\eta_{\rm t}\meanJJ 
+\ggamma\times\meanBB,
\label{emf_0fin}
\EN
where
\EQ
\alpha=-\onethird \tau_0(\overline{\oo\cdot\uu}-
\overline{\jj\cdot\bb}), \quad 
 \eta_{\rm t} = \onethird \tau_0 \overline{\uu^2}, \quad
 \ggamma = - \onesixth\tau_0\nab(\overline{\uu^2}-\overline{\bb^2}),
\EN
and $\mu_0=\rho_0=1$ is still assumed. We see that the
$\alpha$ effect has the advertised current helicity
correction, but the turbulent diffusion is unaffected
by the small scale magnetic fluctuations.
There is also a turbulent diamagnetic effect
and this is affected by magnetic fluctuations,
vanishing for equipartition fields.
We should also point out that the fluctuating
velocity and magnetic fields above are the
actual fields and not that of some fiducial `original' turbulence
as for example implied in \cite{RKR03}.
Of course, we have not calculated what the values of
$\overline{\oo\cdot\uu}$, $\overline{\jj\cdot\bb}$, etc are;
one possibility is to take them from a turbulence simulations.

Note also that assuming $\tau$ independent of $k$ may be
adequate if the magnetic and kinetic spectra are dominated by a single
scale. However, it can be misleading if $F$ and $N$ are non zero
over a range of scales. In such a case it is better to adopt a
physically motivated $\tau(k)$. For example if the turbulence
is maximally helical and $F(k) \propto kE(k)$, with
$E(k) \propto k^{-11/3}$ as in Kolmogorov turbulence, then
$\tau_0 \int F(k) \,\dd k \propto \int k k^{-5/3}\dd k$ would be dominated
by the smallest scale and the kinetic and magnetic $\alpha$ effects
will be Reynolds number dependent.
However, if one takes for $\tau(k) \propto k^{-2/3}$ the eddy turnover
time in Kolmogorov turbulence, then $\int\tau(k) F(k)\,\dd^3k \propto
k^{-1/3}$ and the $\alpha$ effects (both kinetic and magnetic)
would be determined by the kinetic/current helicities at
the forcing scale of the turbulence and
independent of Reynolds number. It is important to keep
this feature in mind in interpreting expressions with constant 
$\tau_0$. (We thank Dmitry Sokoloff for this suggestion.)
We note, however, that recent simulations at higher resolution
(up to $512^3$ meshpoints) indicate that the small scale velocity
and magnetic fields are no longer fully helical beyond the forcing
wavenumber \cite{Bran+Sub05}.
Thus, the spectra of kinetic and current helicities scale
approximately like $k^{-5/3}$.

\subsection{Turbulence with helicity induced by rotation}
\label{HelicityInducedByRotation}

Now consider the case when helicity in the turbulence
is induced by rotation and stratification.
For this case, we adopt a velocity correlation function,
where $F$ is related to $E$ as in \eq{rot_F}.
For $m_{jk}$, we retain the expression given by \eq{mag_cor}.
Note that in Ref.~\cite{RKR03} the current
helicity terms have been neglected in the calculation of the effects
of rotation on $\meanemf$. There is no reason 
to make this assumption, as even if such helical magnetic
correlations were initially zero, they would be generated
dynamically, during the operation of the
large scale dynamo due to magnetic helicity
conservation. So we have to keep this term as well.
We will also only keep terms linear in $\Omega \tau$ and
linear in the $R_i$ derivative.

The first term in \eq{chi_sol3} also gives now  
an $\Omega$ dependent contribution to 
$\meanemf$, instead of just the usual isotropic
$\oo\cdot\uu$ term.
This contribution, 
can be calculated by substituting the anisotropic
$F^{\Omega}$ from \Eq{rot_F} in place of the isotropic $F$
into \Eq{emf_0a}. We have now
\EQ
\meanemf^{(0)} = \onethird \tau_0
\overline{\jj\cdot\bb} \ \meanBB -\eta_{\rm t} \meanJJ
+\ggamma\times\meanBB + \meanemf^{(\Omega 1)},
\EN
where 
\EQA
\meanemf^{(\Omega 1)}_i &=&
-\epsilon_{ijk}\epsilon_{jkm} \int \tau 
\kkk\cdot\meanBB \kunit_m F^{\Omega}\,\dd^3k
\nonumber \\
&=& \epsilon_{ijk}\epsilon_{jkm} \meanB_l \Omega_t \int \tau\tau^*
\left(\kunit_l\kunit_m\kunit_t\kunit_n\nabla_n-\kunit_l\kunit_m\nabla_t\right)
E^{(0)}(k,\RR) \,\dd^3k \nonumber \\
&=& -{\textstyle{16\over15}}\meanB_i (\OO\cdot\nab)\tilde{E}^*
+{\textstyle{4\over15}} \OO\cdot\meanBB \ \nabla_i\tilde{E}^*
+ {\textstyle{4\over15}}\Omega_i \ \meanBB\cdot\nab\tilde{E}^*,
\label{emf_om1}
\ENA
where we have used the relation valid for any ${\cal F}(k)$,
\EQ
\int\kunit_l\kunit_m\kunit_t\kunit_n {\cal F}(k)\,\dd^3k
={\textstyle{1\over15}}\left (\delta_{lm}\delta_{tn}+ \delta_{lt}\delta_{mn}
+ \delta_{ln}\delta_{mt} \right)
\int {\cal F}(k)\,\dd^3k,
\EN
and defined
\EQ
\tilde{E}^* = \int \tau(k)\tau^*(k) \ E^{(0)}(k,\RR)\,\dd^3k
\quad\left(= \half \tau_0^2 \overline{\uu^{(0)2}}\right),
\EN
the latter equality being valid for a constant $\tau=\tau^* =\tau_0$.
Note that $\meanemf^{(\Omega 1)}$ in \Eq{emf_om1} with 
$\tilde{E}^* = \half \tau_0^2 \overline{\uu^{(0)2}}$ gives
a turbulent EMF identical to the $\alpha$ effect derived by 
Krause \cite{Kra67} for the case when there is
no density stratification; see \Eq{alphaEqn} and \Eq{KrauseEqn}.
It would seem that Krause's formula has actually assumed
$\tau =\tau^*$ and also missed the two additional contributions
(the $\meanemf^{(\Omega 2)}$ and $\meanemf^{(\Omega 3)}$ terms)
to be derived below.

The second and third terms in \eq{chi_sol3} also contribute
to $\meanemf$ for non-zero $\Omega$. 
These contributions to $\meanemf$, denoted as $\meanemf^{(\Omega 2)}$ and
$\meanemf^{(\Omega 3)}$, respectively,
are calculated in \App{SecondThirdEMFOmega}.
In computing these terms we also need the following integrals
\EQA
&& \tilde{E}^{(2)} = \int \tau^2(k) E(k,\RR) \,\dd^3k, \quad
\tilde{M}^{(2)} = \int \tau^2(k) M(k,\RR) \,\dd^3k, \nonumber \\
&& \tilde{E}^{(3)} = \int \tau^2(k) k E'(k,\RR) \,\dd^3k, \quad
\tilde{M}^{(3)} = \int \tau^2(k) k M'(k,\RR) \,\dd^3k,
\label{tilde_E2}
\ENA
where primes denote derivatives with respect to $k$.
For constant $\tau(k)=\tau_0$ we have
\EQ
\tilde{E}^{(2)} = {\textstyle{1\over2}}\tau_0^2\overline{\uu^{(0)2}}, \quad
\tilde{M}^{(2)} = {\textstyle{1\over2}}\tau_0^2\overline{\bb^{2}},
\EN
together with $\tilde{E}^{(3)} = -3\tilde{E}^{(2)}$
and $\tilde{M}^{(3)} = -3\tilde{M}^{(2)}$.

The net turbulent EMF is obtained by adding all the
separate contributions,
$\meanemf_i=\meanemf^{(0)}_i+\meanemf^{(\Omega 2)}_i+
\meanemf^{(\Omega 3)}_i$, so
\EQ
\meanemf_i = \alpha_{ij} \meanB_j - \eta_{ij} \meanJ_j
+ (\ggamma\times\meanBB)_i
+ (\ddelta\times\meanJJ)_i
+ \kappa_{ijk} \meanB_{j,k},
\label{emf_rot}
\EN
where the turbulent transport coefficients, for
$k$-independent correlation times $\tau=\tau^* =\tau_0$, 
are given by,
\EQA
\alpha_{ij} = \onethird \tau_0
\delta_{ij}\overline{\jj\cdot\bb}
&-&{\textstyle{12\over15}} \tau_0^2 \Big[\delta_{ij} \OO\cdot\nab
(\overline{\uu^{(0)2}} - \onethird \overline{\bb^2})
\nonumber \\ 
&-&{\textstyle{11\over24}} (\Omega_i\nabla_j + \Omega_j\nabla_i)
(\overline{\uu^{(0)2}} + {\textstyle{3\over11}}
\overline{\bb^2})\Big],
\label{alpha_tau}
\ENA
\EQ
\eta_{ij} = \onethird \tau_0 \delta_{ij} \ \overline{\uu^{(0)2}},
\label{eta_tau}
\EN
\EQ
\ggamma = - \onesixth \tau_0 \nab \left(\overline{\uu^{(0)2}} -
\overline{\bb^2}\right)
- \onesixth \tau_0^2  \OO \times \nab
\left(\overline{\uu^{(0)2}} + 
\overline{\bb^2}\right),
\label{gamma_tau}
\EN
\EQ
\ddelta = \onesixth \OO \tau_0^2 
\left(\overline{\uu^{(0)2}} - 
\overline{\bb^2}\right),
\label{delta_tau}
\EN
\EQ
\kappa_{ijk} = {\textstyle{1\over6}}
\tau_0^2(\Omega_j \delta_{ik} + \Omega_k\delta_{ij})
\left(\overline{\uu^{(0)2}} 
+{\textstyle{7\over5}}\overline{\bb^2}\right).
\label{kappa_tau}
\EN
Note that the combination
$(\ddelta\times\meanJJ)_i+\kappa_{ijk}\meanB_{j,k}$ reduces to
\EQ
\meanemf=...
+\onethird\tau_0^2\left(\overline{\uu^{(0)2}}
+{\textstyle{1\over5}}\overline{\bb^2}\right)\nab(\OO\cdot\meanBB)
+{\textstyle{2\over5}}\tau_0^2\overline{\bb^2}\,\OO\cdot\nab\meanBB,
\label{deltakappa_tau}
\EN
so for $\overline{\bb^2}=0$ (and constant $\OO$) this combination is
proportional to $\nab(\OO\cdot\meanBB)$, and hence, if
$\uu^{(0)2}+{\textstyle{1\over5}}\overline{\bb^2}$ is constant, it
gives no contribution
under the curl -- in agreement with earlier work \cite{KRP94}.
For a comparison with Ref.~\cite{RKR03} see \App{ComparisonRKR}.

In general, $\uu^{(0)2}+{\textstyle{1\over5}}\overline{\bb^2}$ is not
constant, and so the first term in \Eq{deltakappa_tau} 
contributes to the $\alpha$ effect.
Instead of \Eq{alpha_tau} we have then
\EQA
\alpha_{ij} = \onethird \tau_0
\delta_{ij}\overline{\jj\cdot\bb}
&-&{\textstyle{12\over15}}\tau_0^2\delta_{ij} \OO\cdot\nab
(\overline{\uu^{(0)2}} - \onethird \overline{\bb^2})
\nonumber \\ 
&+&{\textstyle{1\over5}}\tau_0^2(\Omega_i\nabla_j+\Omega_j\nabla_i)
(\overline{\uu^{(0)2}}+{\textstyle{1\over3}}\overline{\bb^2})
\nonumber \\ 
&+&{\textstyle{1\over6}}\tau_0^2(\Omega_i\nabla_j-\Omega_j\nabla_i)
(\overline{\uu^{(0)2}}+{\textstyle{1\over5}}\overline{\bb^2}).
\label{alpha_tau_rev}
\ENA
The last term is antisymmetric and can therefore be included in the
second term of the expression for $\ggamma$ in \Eq{gamma_tau} which
then becomes
\EQ
\ggamma = ... - \onethird \tau_0^2  \OO \times \nab
\left(\overline{\uu^{(0)2}} + {\textstyle{3\over5}}\overline{\bb^2}\right).
\label{gamma_toroidal}
\EN
This term corresponds to a longitudinal pumping term of the
form discussed in \Sec{SimulationsTransport}; see the middle
panel of \Fig{Fossen02}.
Since $\overline{\uu^{(0)2}}$ increases outward, the longitudinal
pumping is in the retrograde direction, which is in agreement with the
simulations \cite{OSBR02}.

As mentioned earlier, even if the original
turbulence is nonhelical, one cannot assume the magnetic
part of the correlation functions to be nonhelical
(as done in Ref.~\cite{RKR03}) since, due to magnetic helicity conservation,
the magnetic helical parts can be generated
during the large scale dynamo operation.

So one still has a current
helicity contribution to the $\alpha$ effect.

In Ref.~\cite{RKR03} the $\overline{\jj\cdot\bb}$ contribution was
neglected, because the small scale field was considered to be
completely unaffected by the dynamo.
However, such a restriction is not necessary and the approach
presented above is valid even if $\overline{\bb^2}$ and
$\overline{\jj\cdot\bb}$ are affected (or even produced) by the
resulting dynamo action.
The velocity term, on the other hand, is not quite as general, which is
why we have to keep the superscript 0 in the term $\overline{\uu^{(0)2}}$.

\subsection{Nonlinear helicity fluxes using MTA}
\label{nonlinearflux}

As emphasized above, the $\overline{\jj\cdot\bb}$ term gives the
most important nonlinear contribution to the $\alpha$ effect.
In this section we present the general theory for this term.
One of our aims is also to examine possible fluxes of helicity
that arise when one allows for weak inhomogeneity in the system.
Indeed, Vishniac and Cho \cite{VC01} derived an interesting flux of helicity,
which arises even for nonhelical but anisotropic turbulence.
We derive this flux using MTA, generalizing their original derivation
to include also nonlinear effects of the Lorentz force
and helicity in the fluid turbulence \cite{KS_AB04}.
As we shall see, the Vishniac-Cho flux can also be thought of as
a generalized anisotropic turbulent diffusion. 
Further, due to nonlinear effects, other helicity
flux contributions arise which are due to the anisotropic and
antisymmetric part of the magnetic correlations.

Instead of starting with magnetic helicity, let us start with an 
equation for the evolution of the small scale current helicity,
$\overline{\jj\cdot\bb} = \epsilon_{ijk}\overline{b_i\partial_jb_k}$,
since this is explicitly gauge invariant.
As before we assume that the correlation tensor of $\bb$ varies slowly 
on the system scale $\RR$.
We then have, in terms of the Fourier components $\bft_i$,
\EQA
\overline{\jj(\xx)\cdot\bb(\xx)}
=\epsilon_{ijk}\int\int && 
\overline{\bft_i(\kk + \half\KK) 
\bft_k(-\kk +\half\KK)} \nonumber \\
&&
\ii(-k_j + \half K_j)
\ {\rm e}^{\ii\KK\cdot\RR }
\,\dd^3K \ \dd^3k.
\ENA
Here we have used the definition of correlation
functions as given by \Eq{twoscaledef}, but evaluated
at $\rrr=0$.
The evolution of $\overline{\jj\cdot\bb}$ is given by
\EQ
{\partial \over \partial t}\,\overline{\jj\cdot\bb}
=\Psi_1 + \Psi_2,
\EN
where
\EQ
\Psi_{1/2}
= \epsilon_{ijk}\int\int  
\ii(-k_j + \half K_j) 
M_{ik}^{(1/2)}(\kk,\KK)
\,{\rm e}^{\ii\KK\cdot\RR }
\,\dd^3K \ \dd^3k,
\label{psi12}
\EN
where
\EQ
M_{ik}^{(1)}=\overline{\dot{\bft}_i(\kk + \half\KK) \bft_k(-\kk +\half\KK)},
\label{psi12_M1}
\EN
\EQ
M_{ik}^{(2)}=\overline{\bft_i(\kk + \half\KK) \dot{\bft}_k(-\kk +\half\KK)}.
\label{psi12_M2}
\EN
As shown in \App{Psi1Psi2}, the final result for $\Psi_1$ and $\Psi_2$ is
\EQA
\Psi_{1/2} =&& \int \Big\{ \epsilon_{ijk}\epsilon_{ipq}\epsilon_{qlm} 
\Big[ k_pk_j \Big( \meanB_m \chi_{lk} 
+\half\ii \nabla_s\meanB_m
{\partial\chi_{lk}/\partial k_s} \Big)
\nonumber \\
&& \pm\half\ii k_p \nabla_j(\meanB_m\chi_{lk})
- \half\ii k_j \nabla_p(\meanB_m\chi_{lk})
\Big] + \meanT_{1/2}(k) \Big\}\,\dd^3k,
\label{psi12f}
\ENA
where the upper and lower signs apply to $\Psi_1$ and $\Psi_2$, respectively.
Also $\meanT_1$ and $\meanT_2$ represent the triple correlations of the small scale
$\uu$ and $\bb$ fields and the microscopic diffusion terms that one gets 
on substituting \Eq{bhat} into \Eq{psi12} respectively
(see \App{Psi1Psi2}).
Adding the $\Psi_1$ and $\Psi_2$ terms one gets 
\EQA
{\partial \over \partial t}\,\overline{\jj\cdot\bb}
= 2\epsilon_{ijk}\epsilon_{ipq}\epsilon_{qlm} \int
\Big[ k_pk_j \Big(\meanB_m \chi_{lk} +\half\ii \nabla_s\meanB_m
{\partial\chi_{lk} \over \partial k_s} \Big)
\nonumber \\
- \half \ii k_j \nabla_p(\meanB_m\chi_{lk})
\Big]\,\dd^3k +\meanT_C.
\ENA
Here we have defined $\int [\meanT_1(k) + \meanT_2(k)]\,\dd^3k = \meanT_C$.
The handling of $\meanT_C$ requires a closure approximation. But we will not
need to evaluate these terms explicitly to identify the helicity
fluxes we are interested in, i.e.\ those that couple 
$\meanemf$ and $\meanBB$.
So, we continue to write this term as $\meanT_C$.
Using $\epsilon_{ijk}\epsilon_{ipq}\epsilon_{qlm} = \epsilon_{ljk}\delta_{pm}
-\epsilon_{mjk}\delta_{pl}$ to simplify the above expression, and integrating
the $\partial\chi_{lk}/\partial k_s$ term by parts, yields
\EQA
{\partial \over \partial t}\,\overline{\jj\cdot\bb}
= && \epsilon_{ljk}\int \Big[ 2\chi_{lk} \  k_j (\kk\cdot\meanBB)
-\chi_{lk} \nabla_j(\ii \kk\cdot\meanBB)
\nonumber \\
&&
-\ii k_j \ \meanBB\cdot\nab\chi_{lk}
+ 2 \ii k_j \chi_{pk} \nabla_p\meanB_l \Big]\,\dd^3k
+\meanT_C.
\label{curhelfin}
\ENA
Note that only the antisymmetric parts of $\chi_{lk}$ contribute
in the first three terms above due to the presence of $\epsilon_{ljk}$.
We can now use \Eq{curhelfin} combined with our
results for $\chi_{lk}$ derived in the previous subsections
to calculate the current helicity evolution. We concentrate
below on nonrotating turbulence.
For such turbulence, using \Eq{chi_sol3}, we have
$\chi_{lk} = \tau {\sf I}_{lk}$, where ${\sf I}_{lk}$
is given by \Eq{I_jk}. We use this in what follows.
Let us denote the four terms in \Eq{curhelfin} by
$A_1$, $A_2$, $A_3$ and $A_4$, respectively, with
\EQ
A_1=2\epsilon_{ljk}\int \chi_{lk} k_j (\kk\cdot\meanBB) \,\dd^3k,
\quad
A_2=-\epsilon_{ljk}\int \chi_{lk} \nabla_j(\ii \kk\cdot\meanBB) \,\dd^3k,
\EN
\EQ
A_3=-\epsilon_{ljk}\int \ii k_j \meanBB\cdot\nab\chi_{lk} \,\dd^3k,
\quad
A_4=2\ii\epsilon_{ljk}\int k_j \chi_{pk} \nabla_p\meanB_l \,\dd^3k.
\EN

The first term, $A_1$, is given by 
\EQA
A_1=&&2\epsilon_{ljk} \int \tau k_j (\kk\cdot\meanBB)
\Big[-\ii\kk\cdot\meanBB(v_{lk}^{\rm A}- m_{lk}^{\rm A})
+ \half\meanBB\cdot\nab(v_{lk}^{\rm A} + m_{lk}^{\rm A})
+ \meanB_{l,s} m_{sk}
\nonumber \\
&&
- \meanB_{k,s} v_{ls}
- \half k_m \ \meanB_{m,s}
\left({\partial v_{lk}^{\rm A} \over \partial k_s}
+ {\partial m_{lk}^{\rm A} \over \partial k_s}\right)
- 2{k_lk_s \over k^2} \meanB_{s,p}m_{pk}\Big]\dd^3k.
\label{A1}
\ENA
Due to the presence of $\epsilon_{ljk}$, only
the antisymmetric parts of the tensors $v_{lk}$ and $m_{lk}$
survive, and these are denoted by $v_{lk}^{\rm A}$ and $m_{lk}^{\rm A}$
respectively. Also note that the last term above
vanishes because it involves the product $\epsilon_{ljk}k_lk_j =0$.

All the other terms of \Eq{curhelfin} already have one $R$ derivative,
and so one only needs to retain the term in $\chi_{lk} = \tau 
{\sf I}_{lk}$ which does not contain $R$ derivatives. These terms
are given by
\EQ
A_2=-\epsilon_{ljk}\int \tau \nabla_j(\ii \kk\cdot\meanBB)
\left[-\ii\kk\cdot\meanBB(v_{lk}^{\rm A}- m_{lk}^{\rm A})\right] \dd^3k,
\label{A2}
\EN
\EQ
A_3=-\nab\cdot\left(\epsilon_{ljk}\int \tau\ \ii k_j 
\left[-\ii\kk\cdot\meanBB(v_{lk}^{\rm A}- m_{lk}^{\rm A})\right] \right) \dd^3k,
\label{A3}
\EN
\EQ
A_4= 2\epsilon_{ljk}\int \tau \ii k_j  \ \nabla_p\meanB_l
\left[-\ii\kk\cdot\meanBB(v_{pk}- m_{pk})\right]\dd^3k,
\label{A4}
\EN
where we have used $\nab\cdot\meanBB =0$ to write $A_3$ as a total divergence.
We now turn to specific cases.

\subsubsection{ Isotropic, helical, nonrotating turbulence}

Let us first reconsider the simple case of 
isotropic, helical, nonrotating, and weakly inhomogeneous turbulence. 
For such turbulence we can again use \Eq{vel_cor} and \Eq{mag_cor} for
velocity and magnetic correlations. The $k$ derivative terms in $A_1$ 
in this case involve an integral over an odd number of $k_i$
and so vanishes. Only terms which involve integration
over an even number of $k_i$ survive. Also, in terms which
already involve one $R_i$ derivative, one needs to keep only
the homogeneous terms in \Eq{vel_cor} and \Eq{mag_cor}.
With these simplifications we have from \Eq{A1}
\EQA
A_1 = 2\epsilon_{ljk} \int  k^2 \kunit_j (\kkk\cdot\meanBB)\tau 
\Big[&-&\epsilon_{lkn} \kunit_n \kkk\cdot\meanBB ( F - N)
\nonumber \\
&+& \meanB_{l,s} P_{sk} M - \meanB_{k,s} P_{ls} E \Big] \ \dd^3k.
\ENA
Carrying out the angular integrals over the unit vectors $\kunit_i$ yields
\EQ
A_1 = \fourthird\meanBB^2 \int \tau k^2(F-N)\,\dd^3k 
+\twothird\meanBB\cdot\meanJJ \int \tau k^2 (M+E)\,\dd^3k.
\EN
In the case of isotropic turbulence, the second and third terms,
$A_2$ and $A_3$, are zero
because, to leading order in $R$ derivatives,
the integrands determining $A_2$ and $A_3$ have an odd number 
(3) of $\kunit_i$'s. The fourth term given by \Eq{A4} is
$A_4 = 2\epsilon_{ljk} \meanB_s \nabla_p\meanB_l 
\int \tau k^2  \kunit_j \kunit_s[E-M]\,\dd^3k$, or
\EQ
A_4 = \twothird\meanJJ\cdot\meanBB \tau\int k^2 [E-M]\,\dd^3k.
\EN
Adding all the contributions, $A_1+A_2+A_3+A_4$, we get
for the isotropic, helical, weakly inhomogeneous turbulence,
\EQ
{\partial \over \partial t}\,\overline{\jj\cdot\bb} =
\fourthird\meanBB^2 \int \tau  k^2(F-N)\,\dd^3k 
+\fourthird\meanJJ\cdot\meanBB  \int \tau  k^2 E\,\dd^3k
+\meanT_C.
\label{curisofin}
\EN
We see that there is a nonlinear correction due to the
small scale helical part of the magnetic correlation to
the term $\propto \meanBB^2$. But the nonlinear
correction to the term $\propto \meanJJ\cdot\meanBB$
has canceled out, just as for turbulent diffusion. 
Recall also that for isotropic random fields, the spectra
$H_k$ of magnetic helicity $\overline{\aaa\cdot\bb}$
and $C_k$ of current helicity $\overline{\jj\cdot\bb}$ are related
by $H_k = k^{-2} C_k$.
So the first two terms of the current
helicity evolution equation \Eq{curisofin} give exactly the
source term $-2\meanemf\cdot\meanBB$ for the magnetic helicity
evolution. Also for this isotropic case one sees that
there is no flux which explicitly depends on the mean magnetic field.

\subsubsection{Anisotropic turbulence}
\label{AnisotropicTurbulence}

Let us now consider anisotropic turbulence.
In $A_1$ term in \Eq{curhelfin}, given by \Eq{A1},
one cannot now assume the isotropic form for the
velocity and magnetic correlations.
But again, due to the presence of $\epsilon_{ljk}$, only
the antisymmetric parts of the tensors $v_{lk}$ and $m_{lk}$
survive. Also the last term in \Eq{A1} 
vanishes because it involves the product $\epsilon_{ljk}k_lk_j =0$.
One can further simplify the term involving $k$ derivatives
by integrating it by parts. Straightforward algebra, and a judicious combination
of the terms then gives
\EQA
A_1&=&2\epsilon_{ljk}\int 2\chi_{lk} \  k_j (\kk\cdot\meanBB)\,\dd^3k
=\epsilon_{ljk} \Big\{
 -2\ii \meanB_p\meanB_s \int \tau k_j k_p k_s 
(v_{lk}^{\rm A}- m_{lk}^{\rm A})\,\dd^3k
\nonumber \\
&& +2 \meanB_p \int \tau k_j k_p (  
\meanB_{l,s} m_{sk} - \meanB_{k,s} v_{ls})\,\dd^3k
+ \meanB_p \meanB_{m,j} \int \tau k_m k_p (
v_{lk}^{\rm A} +  m_{lk}^{\rm A})\,\dd^3k
\nonumber \\
&& + \nabla_s \left[\meanB_p\meanB_s \int \tau k_j k_p
(v_{lk}^{\rm A} +  m_{lk}^{\rm A}) \right] \Big\}\,\dd^3k.
\label{A1aniso}
\ENA
All the other terms $A_2$, $A_3$ and $A_4$ cannot be further simplified.
They are explicitly given by
\EQ
A_2= -\epsilon_{ljk}\meanB_p \meanB_{s,j} \int \tau k_s k_p
(v_{lk}^{\rm A}- m_{lk}^{\rm A})\,\dd^3k,
\label{A2aniso}
\EN
\EQ
A_3 = -\nabla_s\left[\epsilon_{ljk} \meanB_p \meanB_s 
\int \tau k_jk_p (v_{lk}^{\rm A}- m_{lk}^{\rm A}) \right]\,\dd^3k,
\label{A3aniso}
\EN
\EQ
A_4 =
2\epsilon_{ljk} \meanB_p \meanB_{l,s} \int \tau k_j k_p 
(v_{sk}- m_{sk})\,\dd^3k.
\label{A4aniso}
\EN
Adding all the contributions, $A_1+A_2+A_3+A_4$, we get
\EQA
{\partial \over \partial t}\,\overline{\jj\cdot\bb} &=&
2\epsilon_{jlk} \Big[
\meanB_p\meanB_s \int \tau \ii k_j k_p k_s
(v_{lk}^{\rm A}- m_{lk}^{\rm A})\,\dd^3k
\nonumber \\
&& 
+ 2 \meanB_p\meanB_{k,s} \int \tau k_j k_p v_{ls}^{\rm S}\,\dd^3k
-\meanB_p \meanB_{s,j} \int \tau k_s k_p m_{lk}^{\rm A}\,\dd^3k
\nonumber \\
&& 
-\nabla_s \Big(\meanB_p \meanB_s
\int \tau k_jk_p  m_{lk}^{\rm A} \Big) \Big]\,\dd^3k + \meanT_C.
\label{anisochel}
\ENA
Here $v_{ls}^{\rm S} = \half(v_{ls} + v_{sl})$ is the symmetric part of
the velocity correlation function.

Let us discuss the various effects contained in the above
equation for current helicity evolution.
The first term in \Eq{anisochel} represents the anisotropic
version of helicity generation due to the full nonlinear $\alpha$ effect.
In fact, for isotropic turbulence it exactly will match the
first term in \Eq{curisofin}. The second term in
\Eq{anisochel} gives the effects on helicity evolution
due to a generalized anisotropic turbulent diffusion.
This is the term which contains the Vishniac-Cho flux.
To see this, rewrite this term as
\EQA
\left.{\partial\overline{\jj\cdot\bb} \over \partial t}\right\vert_{\rm VC}&=&
4 \epsilon_{jlk} \meanB_p\meanB_{k,s} \int \tau k_j k_p v_{ls}^{\rm S}\,\dd^3k
\nonumber \\
&=& -\nab\cdot\meanFF^V 
+4\meanB_k \epsilon_{klj} \meanB_{p,s} \int \tau k_j k_p v_{ls}^{\rm S}\,\dd^3k.
\label{vischoa}
\ENA
Here the first term is the Vishniac-Cho flux,
$\meanF^{\rm VC}_s=\phi_{spk}\meanB_p\meanB_k$, where
$\phi_{spk}$ is a new turbulent transport coefficient with
\EQ
\phi_{spk}
=-4\epsilon_{jlk} \int \tau k_j k_p v_{ls}^{\rm S}\,\dd^3k
=-4\tau\overline{\omega_k \nabla_p u_s},
\label{vishflux}
\EN
the latter equality holding for a $\tau$ independent of $k$.
Obviously, only the component of $\phi_{spk}$ that is symmetric in
its second two components enters in $\meanF^{\rm VC}_s$.
The second term in \Eq{vischoa} is the effect on helicity due to `anisotropic
turbulent diffusion'. (We have not included the large scale
derivative of $v_{ls}$ to the leading order.) 
If we recall that $H_k = k^{-2} C_k$ for homogeneous
turbulence, then \Eq{vishflux} for $\meanFF^{\rm VC}$ leads
to the magnetic helicity flux of the form given 
in Eqs~(18) and (20) of Vishniac and Cho \cite{VC01}.

This split into helicity flux and anisotropic diffusion 
may seem arbitrary; some support for its usefulness comes
from the fact that, for isotropic turbulence, 
$\meanFF^V$ vanishes, while the second
term exactly matches with the corresponding
helicity generation due to turbulent diffusion, i.e.\
the $\meanJJ\cdot\meanBB$ term in \Eq{curisofin}.
Of course, we could have just retained the non-split expression in 
\Eq{vischoa}, which can then be looked at
as an effect of anisotropic turbulent
diffusion on helicity evolution. 
Also, interestingly, there is no nonlinear
correction to this term from $m_{ls}^{\rm S}$, just like there is no
nonlinear correction to turbulent diffusion in lowest order!

Finally, \Eq{anisochel} also contains terms (the last two)
involving only the antisymmetric parts of
the magnetic correlations. These terms
vanish for isotropic turbulence, but contribute
to helicity evolution for nonisotropic turbulence.
The last term gives a purely magnetic contribution
to the helicity flux, but one that depends only
on the antisymmetric part of $m_{lk}$.
Note that such magnetic correlations, even
if initially small, may spontaneously develop
due to the kinetic $\alpha$ effect or anisotropic turbulent diffusion
and may again provide a helicity flux. More work
is needed to understand this last flux term
better.

The advantage of working directly with $\overline{\jj\cdot\bb}$ evolution
is that it is the current helicity density that appears in the feedback
on the $\alpha$ effect. However a disadvantage is the appearance
of the triple correlation term $\meanT_C$ as a volume term, which cannot be
easily evaluated.
However, it has recently been possible to define,
in a gauge invariant manner, the magnetic helicity density of the small
scale random field \cite{KS_AB05}.
This can be done even in the inhomogeneous case and it is then also possible
to derive its evolution equation.
This generalizes the helicity evolution equation to the inhomogeneous case
and contains fluxes both of the Vishniac-Cho type and those 
phenomenologically invoked by Kleeorin and co-workers.  
Triple correlations also do not appear in the volume terms,
but only contribute to the flux.
This is still very much work in progress, but it is clear that
this approach may prove fruitful in the future.

In summary, MTA allows a conceptually simple and mathematically
straightforward, although technically somewhat involved analytic
treatment of the mean field transport coefficients.
The calculation of $\alpha_{ij}$, $\eta_{ij}$, $\ggamma$, $\ddelta$,
and $\kappa_{ijk}$ agrees in all important aspects with earlier
treatments \cite{RKR03}.
The by far most important new aspect is the inclusion of the
$\overline{\jj\cdot\bb}$ feedback term.
This must be coupled to a dynamical calculation of $\overline{\jj\cdot\bb}$,
which has hitherto been ignored in the vast majority of dynamo models.
The nonlinear feedback in the case of homogeneous dynamos in closed domains
is now well understood (\Sec{SDynamicalQuenching}).
However, in the case of open domains helicity fluxes need to be calculated.
Here, only partial results are available.
Clearly, more work by the various groups is required before 
we can generate a coherent picture.

\section{Discussion of dynamos in stars and galaxies}
\label{Discussion}

The applicability of the full set of mean field transport coefficients
to models of stars and galaxies is limited by various restrictions:
analytic approaches allow only weak anisotropies and suffer from uncertainties
by using approximations such as FOSA or MTA, while the results of
numerical simulations to calculate transport coefficients are difficult
to parameterize and apply only to low magnetic Reynolds numbers.
This, combined with the general lack of confidence due to a range of
different results, has resulted in a rather fragmentary usage of
a selection of various possible terms.
Thus, only partial results can be reported here.

It is clear that all the interesting effects are controlled by the degree
of anisotropy in the turbulence.
One of the important ones is rotation, without which there would be no
$\alpha$ effect and also no differential rotation or shear.
We begin with a discussion of the relative importance of rotational
effects in various astrophysical bodies and turn then to a somewhat
subjective assessment of what is the current consensus in explaining
the nature of magnetic fields in various bodies.

\subsection{General considerations}

How rapid does the rotation have to be in order that the anisotropy
effect on the turbulence becomes important?
A suitable nondimensional measure of $\Omega$ is the inverse Rossby
number, $\mbox{Ro}^{-1}=2\Omega\tau$, where $\tau$ corresponds to the
correlation time if FOSA is used, or to the relaxation time if MTA is used.
In practice, $\tau$ is often approximated
by the turnover time, $\tau_{\rm turnover}$.
In \Tab{TRossby} we give some estimates for $\mbox{Ro}^{-1}$ for various
astrophysical bodies.
For the sun, $\mbox{Ro}^{-1}$ is around 5 near the bottom of the
convection zone (but goes to zero near the surface layers).
In galaxies, but also in proto-neutron stars, $\mbox{Ro}^{-1}$ is smaller
(around unity).
Accretion discs tend to have large values of $\mbox{Ro}^{-1}$ (around 100).
This is directly a consequence of the fact that here the turbulence is
weak, as quantified by a small value of the Shakura-Sunyaev viscosity
parameter ($\alpha_{\rm SS}\approx0.01$ \cite{BNST95,HGB96,SHGB96}).
Planets also tend to have large values of $\mbox{Ro}^{-1}$, because
here the turbulence is driven by a weak convective flux, so the turnover
time is long compared with the rotation period.

It may be useful to comment on the relative meanings of `large' and `small'.
This issue may depend on the problem; one possibility is to consider
the $\alpha$ effect.
Both mean field theory \cite{RK93} and simulations \cite{OSB01} suggest
that $\alpha$ saturates when $\mbox{Ro}^{-1}\approx5$.
This indicates that values of $\mbox{Ro}^{-1}$ below 5 can be considered
as small.
In this context it is worth mentioning that the inverse Rossby
number is occasionally defined as $\tau/P_{\rm rot}$, where
$P_{\rm rot}=2\pi/\Omega$ is the rotation period, so
$\tau/P_{\rm rot}=\mbox{Ro}^{-1}/(4\pi)$.
Thus, in terms of $\tau/P_{\rm rot}$ the dividing line between large
and small would be around 0.4.

In the following we first discuss some issues connected with
understanding and modeling the solar dynamo, and then turn to
dynamos in stars and planets, and their relation to laboratory
liquid metal experiments.
Finally, we turn to dynamos in accretion discs,
galaxies and galaxy clusters.

\begin{table}[t!]\caption{
Summary of angular velocities, estimated turnover times, and
the resulting inverse Rossby number for various astrophysical bodies.
}\vspace{12pt}\centerline{\begin{tabular}{lcccccccccc}
                &     $\Omega$ [s$^{-1}$] & $\tau$  &
$\mbox{Ro}^{-1}=2\Omega\tau$ \\
\hline
Proto-neutron stars                 &  $2\times10^{3}$   &$10^{-3}\s$ &  2 \\
Discs around neutron stars          &     $10^{-2}$      &  $10^4\s$  &200 \\
Jupiter                             &  $2\times10^{-4}$  &  $10^6\s$  & 200\\
T Tauri stars                       &  $2\times10^{-5}$  &  $10^6\s$  & 40 \\
Solar convection zone (lower part)  &  $3\times10^{-6}$  &  $10^6\s$  &  6 \\
Protostellar discs                  &  $2\times10^{-7}$  &  $10^9\s$  &400 \\
Galaxy                              &     $10^{-15}$     &  $10^7\yr$ & 0.6\\
\label{TRossby}\end{tabular}}\end{table}

\subsection{The solar dynamo problem}
\label{solar_dynamo}

There are several serious shortcomings in our understanding of the sun's
magnetic field.
Although there have been many theoretical attempts,
there is as yet no solution to the solar dynamo problem.
With only a few exceptions \cite{Kuzanyan,Zhang05}, all the mean field models
presented so far have ignored the magnetic helicity issue
altogether, so it is not clear what significance such models still have.
Neglecting magnetic helicity can only be considered a reasonable
approximation if $|\overline{\jj\cdot\bb}|\ll|\overline{\oo\cdot\uu}|$, which
is probably not valid in the nonlinear regime.
We have seen earlier that an important constraint on the nonlinear dynamo
is imposed by magnetic helicity conservation.
The corresponding nonlinearity needs to be properly
incorporated into many of the solar dynamo models.

It should ultimately be possible to simulate the entire sun with its
three-dimensional turbulence, the tachocline (see \Sec{Stachocline}),
the resulting differential rotation, and the near-surface shear layer.
Several attempts have been made starting with the early work of Gilman
\cite{gilman} and Glatzmaier \cite{glatz}, and new high resolution
calculations are currently underway \cite{BrunToomre2002,Brun2004,Brunetal04}.
All these simulations have been successful in generating both small
scale and large scale magnetic fields, although they have not yet been
convincingly demonstrated cyclic behavior. Nevertheless, there is a tendency
for the toroidal field belts to propagate away from the equator, rather
than toward the equator as in the sun.
It is tempting to associate this with a positive sign of the $\alpha$ effect
(in the northern hemisphere)
that these simulations generate (even though an explicit $\alpha$ effect
is of course not invoked).

It should be emphasized that in none of the convection simulations
currently available (neither in spherical shells nor in cartesian boxes)
the magnetic Reynolds numbers are large enough to see the effect of
magnetic helicity conservation (the dynamo growth rate should be much
larger than the ohmic decay rate $\eta k_1^2$ \cite{B01},
which is hardly the case in many simulations).
We are only now at the threshold where magnetic helicity effects
begin to have a chance to show up in simulations.

\subsubsection{Magnetic helicity and cycle period}

There is the worry that in large scale simulations
magnetic helicity conservation could either
prevent cyclic behavior or it might significantly prolong the cycle
period \cite{BB03,BDS02}.
On the other hand,
if in the sun the importance of magnetic helicity conservation is
only marginally important (e.g., if magnetic helicity fluxes dominate
over resistive losses), one could imagine a prolongation of the cycle
period by a factor of about ten, which would be needed to improve
the results of conventional models, which invariably produce too short
cycle periods if magnetic helicity conservation is ignored \cite{Koe73}.
The anticipated prolongation of the cycle period might therefore
be regarded as a step in the right direction.
Conventional approaches to produce the right cycle period is to
`adjust' the ill-known parameters $\alpha_0$ and $\eta_{\rm t}$
\cite{Choudhuri90}.

The strength of the magnetic and current helicity fluxes, and hence the
degree of magnetic helicity conservation, is possibly self-regulating
via losses through the outer surface, for example such that the magnetic
helicity losses are just as strong as to affect the time scales only
slightly.
Again, this is at present quite speculative, and there is no simulation that
is able to show this.
There are only partial results \cite{BrandenburgSandin2004} suggesting
that $\alpha$ might not be quenched catastrophically if there is shear
combined with open boundaries.
It seems plausible that a significant portion of the magnetic helicity
losses from the sun occurs through coronal mass ejection (CMEs)
\cite{Chae00}.
So far, however, no turbulence simulation has realistically been able to
allow for such magnetic helicity losses, nor have simulations
to our knowledge been able
to produce phenomena that can even remotely be associated with CMEs.
It seems therefore important to study the large scale dynamo
problem in more realistic settings where CMEs are possible.

\subsubsection{Does the sun eject bi-helical fields?}
\label{bihelical}

On theoretical grounds, if most of the helicity of the solar magnetic
field is produced by the $\alpha$ effect, one would expect a certain fraction
of the solar magnetic field to be bi-helical \cite{BF00b,BB03,YB03},
in that the field that is generated by the $\alpha$ effect
has positive and negative magnetic helicity at different
scales, but hardly any net magnetic helicity.
On the other hand, if most of the sun's helicity is caused by
differential rotation, the field might equally well not be bi-helical.
(We recall that differential rotation causes segregation of magnetic
helicity in physical space, i.e.\ between north and south, while the
$\alpha$ effect causes a segregation of helicity in wavenumber space;
see \Sec{MagneticHelicity}.)
So far, however, the solar magnetic field
has not explicitly been seen to be bi-helical.
Indirectly, however, a bi-helical nature of the solar magnetic
field is indicated by the fact that bipolar regions are tilted
according to Joy's law \cite{DSC93,Hale19} (see also
\Sec{PhenomenologicalConsiderations}), suggesting the presence
of positive magnetic helicity in addition to the negative magnetic
helicity indicated by the magnetic twist found in active regions.

\begin{figure}[t!]\begin{center}
\includegraphics[width=.9\textwidth]{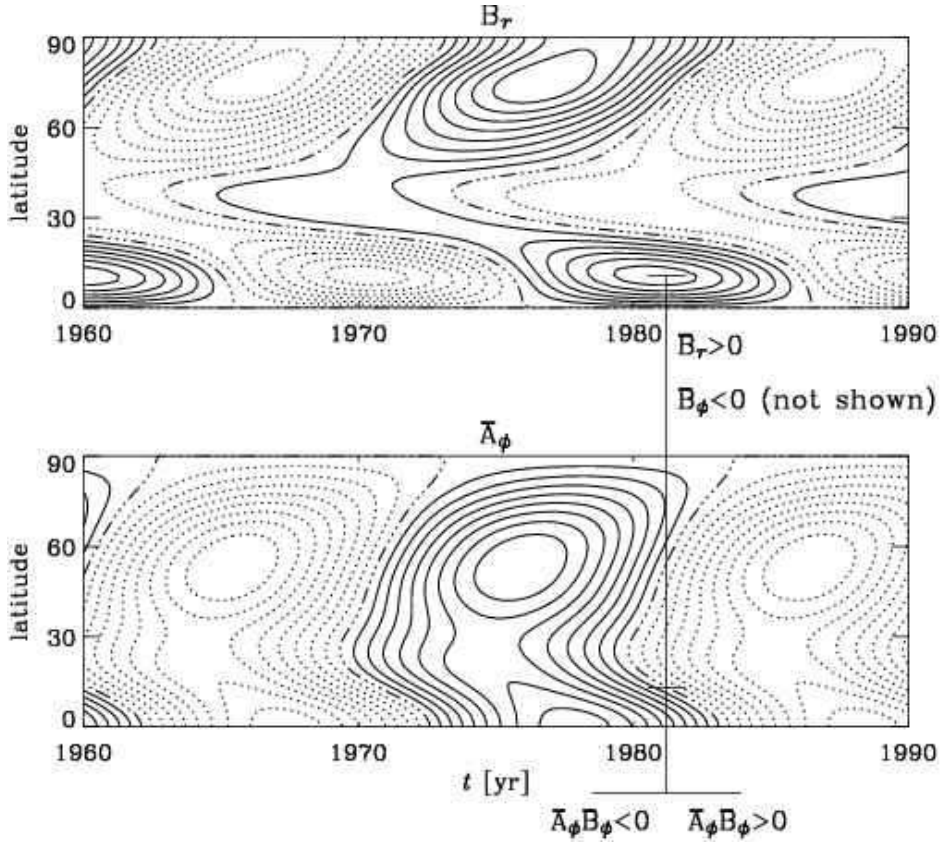}
\end{center}\caption[]{Mean dipole symmetry radial field, $\meanB_r$,
reconstructed from the coefficients of Stenflo \cite{Sten88,Sten94} (upper panel).
The corresponding toroidal component of the mean vector potential,
$\meanA_\phi$, derived from $\meanB_r$ (lower panel).
Solid contours denote positive values, dotted contours negative values.
The solar cycle maximum of 1982 is highlighted,
as is the latitude of 10$^\circ$ where $\meanB_r$ was then strongest.
The signs of various quantities at or around this epoch are
also shown (see text for more details).
Adapted from Ref.~\cite{BBS03}.
}\label{stenflo}\end{figure}

What is missing is a quantitative assessment of the relative magnitudes
of large and small scale magnetic helicities.
To get an idea about the sign and possibly the magnitude of the
relative magnetic helicity of the longitudinally averaged field
(denoted here by overbars), one
should ideally calculate $2\int\meanA_\phi\meanB_\phi\,\dd V$,
which is the gauge-invariant magnetic helicity of Berger and Field
for axisymmetric fields in a sphere (see \Sec{HowClose}).
As a first step in that direction, one can determine $\meanA_\phi$
at the solar surface from the observed radial component of the magnetic
field, $\meanB_r$, using
\EQ 
\meanB_r={1\over r\sin\theta}\,{\partial\over\partial\theta}\,
\left(\sin\theta\meanA_\phi\right),\quad r=R_\odot.
\EN
This equation can easily be solved for $\meanA_\phi$ using the modal
decomposition in spherical harmonics \cite{BBS03}; the result is shown
in \Fig{stenflo}.
Empirically, we know that $\meanB_r$ is approximately in antiphase
with $\meanB_\phi$.
Comparison of the two panels of \Fig{stenflo} suggests that
$\meanA_\phi\meanB_\phi$ was negative just before solar maximum
($t<1982\yr$) and positive just after ($t>1982\yr$).
Thus, the present attempt to assess the sign of the magnetic helicity
of the large scale field does not support nor exclude the possibility
that the large scale field has positive magnetic helicity, as suggested
by the tilt of bipolar regions.

\subsubsection{Migration of activity belts and butterfly diagram}
\label{MigrationActivityBelts}

There are still a number of other rather long standing problems.
Most important perhaps is the sense of migration of the
magnetic activity belts on either side of the equator.
This migration is traditionally believed to be associated with the
phase velocity of the dynamo wave; see \Sec{Sao_model}.
However, this would be in contradiction with $\alpha>0$ (as expected
in the northern hemisphere) and $\partial\Omega/\partial r>0$
(known from helioseismology \cite{Schou92}).

One of the currently favored models is the flux transport model
where meridional circulation is assumed to be oriented such
that, at the bottom of the convection zone, meridional circulation would advect
the dynamo wave equatorward \cite{Durney95,Choudhuri95,DC99,KRS01,%
Bonanno_etal02,DikpatiGilman01,NandyChoudhuri02}.
This proposal came originally quite as a surprise, because meridional
circulation was normally always found to make dynamos nonoscillatory
well before the anticipated advection effect could take place
\cite{KR80,RS72,Rae86b}.

Whether or not the flux transport proposal is viable cannot be
decided at present, because this model (and essentially all
other solar dynamo models available so far) lack consistency
with respect to the conservation of total (small and large
scale) magnetic helicity.
In addition, there are several other problems.
With realistic profiles of differential rotation it is difficult
to produce satisfactory butterfly diagrams \cite{Bonanno_etal02}.
It is also difficult to produce fields of dipolar rather ran quadrupolar
parity \cite{DikpatiGilman01}, although this remains debatable
\cite{Choudhuri04}.

Quite a different possibility is to invoke the near-surface shear layer
in the outer 35 megameters of the sun \cite{Howe_etal00,Thompson_etal03}.
Here the angular velocity gradient is negative, giving rise to equatorward
migration.
For this scenario to work, one has to rely on turbulent downward pumping
to prevent the magnetic field from buoyantly escaping through the surface.
However, simulations of stratified convection have clearly demonstrated
the dominance of pumping over magnetic buoyancy
\cite{BJNRST96,NBJRRST92,TobiasBrummellCluneToomre98,Tobias01,Brummell02};
see also \Sec{SimulationsTransport}.
There is obviously a lot more to be discussed in connection with this
scenario, but the situation is still premature and it
would go beyond the scope of this review.
We refer instead to a recent paper \cite{B05} devoted specifically
to this discussion.

\subsubsection{The phase relation between poloidal and toroidal fields}

Even if the problem of the migration direction of the toroidal
activity belts was solved (for example by meridional circulation;
see \Sec{MigrationActivityBelts}), there remains the problem of the
phase relation between poloidal and toroidal field, provided the
field was mainly generated at the bottom of the convection zone.
Observations suggest that $B_r$ and $B_\phi$ are in antiphase.
For example, when $B_\phi<0$ (as seen from the orientation of
bipolar regions; see \Fig{kittpeak} for Cycle~21 during the year 1982) we have
$B_r>0$ (as seen from synoptic magnetograms, see \Fig{stenflo}
at $t=1982\yr$), and vice versa \cite{Yoshimura76,Stix76}.
The basic problem is actually connected with the sense of the radial
differential rotation, i.e.\ the fact that $\partial\Omega/\partial r>0$
in low solar latitudes.
This always turns a positive $B_r$ into a positive $B_\phi$, and vice
versa, regardless of the sign of the $\alpha$ effect.
No convincing solution to this problem has yet been offered,
although it has been suggested \cite{SchlichenmaierStix1995}
that the problem might disappear in more realistic settings.
Of course, in the near-surface shear layer, $\partial\Omega/\partial r$,
so the problem with the phase relation would be solved in this scenario
where the solar dynamo works in or is shaped by the near-surface shear
layer \cite{B05}.

\subsubsection{The flux storage problem}

Since the early eighties it was realized that flux tubes with a strength
similar to or in excess of that in sunspots would float up to the surface in a
time short ($\sim50\,\mbox{days}$) compared with the dynamo time scale
($\sim10\yr$).
Therefore, magnetic buoyancy might act as an efficient sink term of mean
toroidal field.
This led to the suggestion \cite{SW80} that the dynamo may operate in
the lower part of the convection zone or just below it where magnetic
flux tubes could stay in equilibrium.
Model calculations \cite{DG86,DG88} have shown however that dynamos
in thin layers tend to produce too many toroidal flux belts in each
hemisphere \cite{MossTuominenBrandenburg90}, and that for an overshoot
layer dynamo the layer must therefore not be less than about $30\Mm$
deep \cite{RB95}, which is much more than the helioseismologically
implied thickness of the solar overshoot layer.

As mentioned above, it is also possible that the dynamo might still
work in the convection zone proper, but that turbulent pumping
brings the field continuously to the bottom of the convection zone
\cite{Bran94,TobiasBrummellCluneToomre98,Tobias01,RB95}, from
where strong toroidal flux belts can rise to the surface and form
bipolar regions (\Fig{kittpeak}).
The end result might be similar in the sense that in both cases
bipolar regions would be caused by flux tubes anchored mostly
in the overshoot layer.
It should also be emphasized that magnetic buoyancy might
not only act as a sink in a destructive sense, but it can
also contribute to driving an $\alpha$ effect
\cite{Schmitt87,FerrizMas_etal94,BrandenburgSchmitt98,Thelen00}.
A conclusive picture cannot be drawn until we have a better understanding
about things like magnetic and current helicity fluxes that now appear
quite vital for allowing the solar dynamo to work on the observed 22
year time scale.
Finally, one should bear in mind that it is also possible that the solar
field in sunspots and in active regions does not even come from very deep,
and that it originates primarily from the near-surface shear layer.
In that case the actual sunspot formation might be the result of convective
collapse of magnetic fibrils \cite{Zwaan78,SpruitZweibel79},
possibly facilitated by negative turbulent magnetic pressure effects
\cite{KMR96} or by an instability \cite{KitchatinovMazur00}
causing the vertical flux to concentrate into a tube.

\subsubsection{Significance of the tachocline and differential rotation}
\label{Stachocline}

The tachocline is the layer where the latitudinal differential
rotation turns into rigid rotation \cite{SpiegelZahn92}.
This layer is now known to be quite sharp and to coincide with
the bottom of the convection zone \cite{Howe_etal00}.
The reason the latitudinal differential rotation does not propagate
with time deeper into the radiative interior is probably connected with the
presence of a weak primordial magnetic field \cite{RK96,RK97,GM98}.

The tachocline is likely to be the place where the vertical shear gradient
plays an important role in amplifying the toroidal magnetic field.
This is perhaps not so much because the shear gradient is strongest at and
just below the tachocline, but
because the turbulent magnetic diffusivity is decreased, and because
the magnetic field is pumped into this layer from above.

In the convection zone proper the differential rotation is in rough
approximation spoke-like, i.e.\ nearly independent of spherical radius.
Simulations, on the other hand, shear a strong tendency to produce
angular velocity contours that are constant on cylinders \cite{Mie00}.
This is generally associated with the Taylor-Proudman theorem,
and is a well recognized difficulty in understanding the solar differential
rotation \cite{BMRT90,KR95,BMT92}.
It is also fairly well understood that solar-like departures from
cylindrical contours could be achieved by the baroclinic term \cite{Dur91},
$\nab s\times\nab T$, where $s$ and $T$ are specific entropy and
temperature, respectively.
In the convection zone, where the radial entropy gradient is small,
a finite baroclinic term is mostly due to the latitudinal entropy
gradient, so that
\EQ
\varpi{\partial\Omega^2\over\partial z}\approx\pphi\cdot(\nab s\times\nab T)
\approx-{1\over r}{\partial s\over\partial \theta}{\partial T\over\partial r}<0,
\EN
where $\varpi=r\sin\theta$ is the cylindrical radius,
$z=r\cos\theta$ is the distance from the equatorial plane,
and $\pphi$ is the unit vector in the azimuthal direction.
Negative values of $\partial\Omega^2/\partial z$, in turn,
require that the pole is slightly warmer than the equator
(so weak that it cannot at present be observed.
Achieving this in a simulation may require particular care in the
treatment of the outer boundary conditions.

In order to understand the computational difficulties in dealing
with the overshoot layer, we note that
in and below the tachocline the total relaxation time
is governed more strongly by the thermal time scale.
Thus, if the system is slightly thrown out of balance, it will take a thermal
time scale to reestablish a new equilibrium.
Furthermore, as long as a new equilibrium state has not yet been
reached, fairly strong amplitudes may develop, making the effective
relaxation time even longer.
While this will not be a problem for the sun, it may be a problem
for simulations.
In practice, this means that one has to be more careful setting up initial
conditions and, perhaps most importantly, one should deliberately chose
parameters whereby the thermal and acoustic time scales are not more
disparate than what can be handled in a simulation.
Nevertheless, keeping at least some representation of the tachocline,
rather than neglecting it altogether, may be crucial.
Without any representation of a tachocline it may not be possible to
argue conclusively in favor of either the tachocline scenario or
the near-surface shear layer scenario.

\subsubsection{Luminosity variations}

Cyclic variations of the magnetic field produce cyclic variations of
the luminosity through variations of the superadiabatic gradient \cite{SW80}.
Although this basic picture has been confirmed in mean field dynamo
models, the relative variations obtained are only $\delta L/L\sim10^{-6}$
\cite{BMT92}.
Larger values of a few times $10^{-3}$ can be obtained by making
the upper boundary ``partially reflecting''; see a recent paper by
Pipin~\cite{Pipin04}.
This emphasizes again that a realistic representation of the top boundary
condition can be very important.
This model also predicts variations of the hydrostatic balance and
hence of the star's quadrupole moment.
This is important for stellar dynamos that are members of a binary system,
because such variations can provide a means of determining the stars'
cycle period by measuring variations of the orbital period \cite{Lanza98}.
In \Sec{StellarDynamos} we discuss stellar dynamos in more detail.

\subsubsection{Status of different solar dynamo model scenarios}
\label{StatusSolarDynamo}

Before turning attention to stellar dynamos we briefly summarize four
different dynamo scenarios that are currently being discussed in the
context of the sun.
None of the models appear to be completely satisfactory, but the topic
is advancing rapidly, as indicated at the end of this section.
\begin{itemize}
\item
Distributed dynamo. An $\alpha$ effect is present throughout
the entire convection zone, as described by \Eq{KrauseEqn} and calculated
using a solar mixing length model \cite{Kri84}.
At the bottom of the convection zone the sign of $\alpha$ reverses
because of the sharp positive gradient of the turbulent rms velocity.
Solutions have been calculated taking also into account R\"adler's
$\OO\times\meanJJ$ effect and other expressions.
The model produces realistic butterfly diagrams including a solar
branch \cite{BT88}.
Unfortunately, the published models use angular velocity profiles that
are no longer compatible with modern helioseismological inversions.
\item
Overshoot dynamo. Here, $\alpha$ is only present in
the overshoot layer, and it is artificially suppressed at high latitudes.
Its sign in the northern hemisphere is negative because of the
sharp positive gradient of the turbulent rms velocity.
The resulting butterfly diagram looks reasonably acceptable, but
for the model to be successful, the thickness of the overshoot
layer cannot be less than $30\Mm$ \cite{RB95}, while helioseismology is
now favoring values as small as $7\Mm$ \cite{Basu97}.
\item
Interface dynamo. The sudden drop of the turbulent
magnetic diffusivity below the overshoot layer is important.
In the northern hemisphere $\alpha$ must be assumed to be negative
and finite {\it above} the convection zone.
The original model by Parker \cite{Parker93} worked with only radial
differential rotation.
When latitudinal differential rotation is included, only nonoscillatory
solutions are found \cite{Markiel+Thomas99}.
\item
Flux transport dynamo. Meridional circulation leads to
equatorward migration of magnetic activity during the course of the cycle.
The sign of $\alpha$ is positive in the northern hemisphere, and $\alpha$
is concentrated toward the upper layers of the convection zones.
The magnetic diffusivity in the bulk of the convection zone is small.
The resulting butterfly diagram is quite realistic \cite{DC99}, although the
the parity issue remains to be clarified \cite{DikpatiGilman01,Choudhuri04}.
\end{itemize}
The most popular model is currently the flux transport dynamo
scenario \cite{Bonanno_etal02,DikpatiGilman01,NandyChoudhuri02}.
However, there are a number of reasons why it might still be
worthwhile pursuing the distributed dynamo scenario.
Most important is the fact that magnetic tracers have an angular velocity
that is close to the maximum angular velocity in the sun which is,
according to helioseismology, only $35\Mm$ beneath the surface
\cite{Howe_etal00,Thompson_etal03}.
This suggests that magnetic tracers such as sunspots might not be
anchored very deeply.
There are a number of other issues that may be more easily resolved
within the framework of a distributed dynamo: instead of requiring a
field strength of $100\kG$ (typical of all dynamos where the toroidal
field emerges from the overshoot layer), only about $300\G$ may be
required if the field is generated locally within the convection zone.
Also, in the upper $35\Mm$ beneath the surface, the radial angular
velocity gradient is negative, suggesting that a locally produced
dynamo wave would migrate equatorward without invoking meridional
circulation.
These and a number of other arguments have been collected and
discussed in Ref.~\cite{B05}.
Finally, it should be recalled that dynamos in fully convective
spheres without any overshoot layer also produce magnetic activity.
This will be discussed in more detail in the next section.

\subsection{Stellar dynamos}
\label{StellarDynamos}

Looking at stars other than the sun allows us to test the dependence
of the dynamo on radius, on the thickness of the convection zone and, in
particular, on the angular velocity of the star.
In this section we summarize a few such dependencies and discuss whether they
can be reproduced by dynamo models.

\subsubsection{Fully convective stars}

Toward the less massive stars along the main sequence the thickness of
the convection zone increases relative to the stellar radius (although
in absolute units the thickness remains around $200\Mm$) until the
star becomes fully convective.
Such stars would lack a lower overshoot layer which was often thought
to be an important prerequisite of solar-like dynamos.
On the other hand, turbulent pumping would in any case tend to concentrate
the field toward the center of the star, and since the gravitational
acceleration vanishes at the center, the magnetic field can probably
still be stored for some time (if that was an issue).

\begin{figure}[t!]\begin{center}
\includegraphics[width=.75\textwidth]{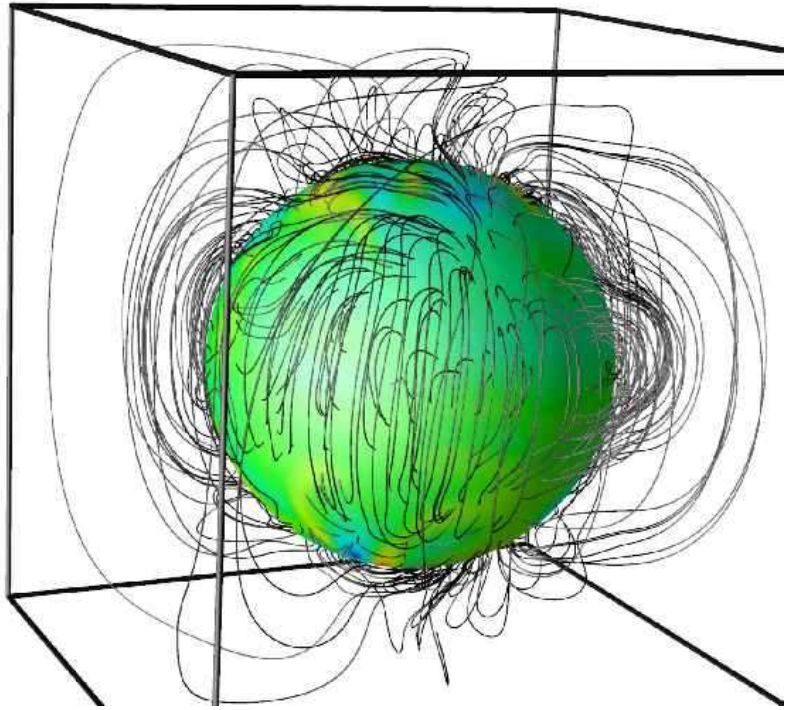}
\end{center}\caption[]{
Magnetic field lines outside a model of a fully convective M-dwarf.
Color coded is the radial field component on the surface of the star.
Courtesy of W.~Dobler \cite{Dobler05}.
}\label{half-sphere}\end{figure}

\FFig{half-sphere} shows a visualization of magnetic field lines of a
self-consistent turbulence simulation of a fully convective spherical dynamo.
In this simulation the dynamo is weakly supercritical.
Following similar approaches by other groups
\cite{Porter_etal00,Freytag_etal02,WoodwardPorterJacobs03,Dorch04,Dobler05},
the star is embedded
in a sphere, which avoids computational difficulties associated with
coordinate singularities in explicit finite difference methods using
spherical coordinates.
The star's radius is 27\% of the solar radius and the mass is
21\% of the mass of the sun.

Another class of fully convective stars are the T Tauri stars, i.e.\ stars
that have not yet settled on the main sequence.
These stars are generally known to spin very rapidly, so magnetic braking
via magnetic field lines anchored in the surrounding protoplanetary
accretion disc is usually invoked to explain the much slower rotation
rate of evolved stars such as the sun \cite{mestelbook}.
It is also possible that young stars are mainly braked by a stellar
wind if the star--disc becomes inefficient \cite{MattPudriz04,vRB04}.

\subsubsection{Models of stellar cycles}
\label{StellarCycles}

Stars exhibiting cyclic behavior can cover a broad parameter
range that allows us to test whether the dependence of cycle properties
on stellar parameters agrees with what is expected from dynamo models.
For orientation one can consider the marginally excited solution from
linear theory and it is indeed common to look at plane wave solutions
\cite{DurneyRobinson82,RobinsonDurney82,Montesinos01}.

Several results have emerged from this type of analysis.
Comparing models with different model nonlinearities ($\alpha$ quenching,
feedback on the differential rotation, and magnetic buoyancy) it turns
out \cite{NWV84}
that only models with magnetic buoyancy as the dominant nonlinearity
are able to produce a {\it positive} exponent $\sigma$ in the relation
$\omega_{\rm cyc}/\Omega=c_1\mbox{Ro}^{-\sigma}$; see \Eq{cyclelaws}.
We emphasize that it is not sufficient that $\omega_{\rm cyc}$ increases
with $\Omega$.
For example, an increase proportional to $\Omega^{0.5}$, which is found
for classical dynamo models \cite{Tobias98}, is {\it insufficient}.

Meanwhile, there has been some uncertainty regarding the correct
cycle frequency dependence on $\mbox{Ro}^{-1}$.
We recall that in \Fig{BST98} there are two separate branches,
both of which have a positive slope $\sigma$.
On the other hand, a negative slope has been found
when plotting $\omega_{\rm cyc}/\Omega$
versus the dimensional form of $\Omega$ (rather than
the nondimensional $\mbox{Ro}^{-1}=2\Omega\tau_{\rm turnover}$)
\cite{Baliunas_etal96}.
One may argue that the nondimensional form is to be preferred, because
it is more general and it also yields a smaller scatter \cite{SaarBran99}.

While ordinarily quenched dynamos tend to produce negative exponents
$\sigma$ \cite{Tobias98}, `anti-quenched' dynamos can produce the observed
exponents when the dynamo alpha and the turbulent dissipation rate,
$\tau^{-1}=\eta_{\rm t} k_z^2$, {\it increase} with field strength, e.g.\ like
\EQ
\alpha=\alpha_0|B/B_{\rm eq}|^n,\quad
\tau^{-1}=\tau_0^{-1}|B/B_{\rm eq}|^m,
\quad m,n>0.
\label{alphatau1}
\EN
The motivation behind anti-quenching lies in the realization that
the motions driving turbulent transport coefficients can be caused
by magnetic instabilities.
Examples of such magnetic instabilities include the magnetic buoyancy
instability \cite{FerrizMas_etal94,BrandenburgSchmitt98,Thelen00},
the magneto-rotational instability \cite{BH98,BNST95}, and
global instabilities of the tachocline differential rotation
\cite{DikpatiGilman01,GilmanFox97}.

Assuming that the dynamo operates in its fundamental mode
(even in the nonlinear regime) one can apply the relations
\eq{lam_approx} and \eq{om_approx} for a fixed value $k_z$,
where $k_zL=1$, where $L$ is the system size.
This leads to two algebraic relations \cite{BST98}
\EQ
\tau_0^{-1}|B/B_{\rm eq}|^m
=|\half\alpha_0\Omega'|^{1/2}\,|B/B_{\rm eq}|^{n/2},
\EN
\EQ
\omega_{\rm cyc}=\tau_0^{-1}|B/B_{\rm eq}|^m.
\EN
Dividing both equations by $\Omega$ and making use of the relation
$\bra{R_{\rm HK}'}\propto|B/B_{\rm eq}|^\kappa$
(usually $\kappa\approx0.5$; see \Sec{LateTypeStars}), we have
\EQ
2\mbox{Ro}\,\bra{R_{\rm HK}'}^{m/\kappa}
=\left|{\alpha_0\Omega'\over2\Omega^2}\right|^{1/2}
\bra{R_{\rm HK}'}^{n/2\kappa},
\EN
\EQ
\left|{\omega_{\rm cyc}\over\Omega}\right|
=2\mbox{Ro}\,\bra{R_{\rm HK}'}^{m/\kappa}.
\EN
For slow rotation $\alpha$ is proportional to $\Omega$,
while for rapid rotation it is independent of $\Omega$ \cite{RK93}.
Likewise, for $\Omega'$ the possibilities range from being
independent of $\Omega$ \cite{DonatiCameron97} to being
proportional to $\Omega^{0.7}$ \cite{Donahue_etal96}.
To account for these different possibilities, we
make the more general assumption
\EQ
\left|{\alpha_0\Omega'\over\Omega^2}\right|
\sim\mbox{Ro}^{-q},
\EN
where $q$ can be anywhere between $-0.3$ and $-2$, depending on the
assumed $\alpha(\Omega)$ and $\Omega'(\Omega)$ dependencies.
We thus have
\EQ
\mbox{Ro}\,\bra{R_{\rm HK}'}^{m/\kappa}
\sim\mbox{Ro}^{-q/2}\bra{R_{\rm HK}'}^{n/2\kappa},
\EN
\EQ
\left|{\omega_{\rm cyc}\over\Omega}\right|\sim
\mbox{Ro}\,\bra{R_{\rm HK}'}^{m/\kappa}.
\EN
Using the definitions \eq{cyclelaws}, i.e.\
$\bra{R'_{\rm HK}}\sim\mbox{Ro}^{-\mu}$ and
$|\omega_{\rm cyc}/\Omega|\sim\mbox{Ro}^{-\sigma}$,
together with $\sigma=\mu\nu$ (see \Sec{LateTypeStars}), we have
\EQ
\mbox{Ro}^{1-m\mu/\kappa}
\sim\mbox{Ro}^{-q/2-n\mu/2\kappa},
\EN
\EQ
\mbox{Ro}^{-\mu\nu}\sim
\mbox{Ro}^{1-m\mu/\kappa},
\EN
so $\mu\nu=m\mu/\kappa-1=q/2+n\mu/2\kappa$,
which yields explicit results for the dynamo exponents $m$ and $n$,
\EQ
m=\kappa(\nu+1/\mu),
\EN
\EQ
n=\kappa(2\nu-q/\mu).
\EN
Note that $m$ is independent of the rather uncertain value of $q$.
In \Tab{TBST98} we summarize the results obtained from the subset
of stars that show cycles \cite{Bal95}.
The different branches, which are separated by a factor of about 6,
might be associated with the occurrence of
different magnetic instabilities in different parameter regimes,
but no definitive proposal can be made at this point.
In this table we also give the results for an expanded sample
\cite{SaarBran99} where, in addition to the chromospheric emission,
photometric and other cycle data have been used.

\begin{table}[t!]\caption{
Summary of observable parameters characterizing stellar cycle properties
($\sigma$, $\nu$, and $\mu$) and the corresponding model parameters
$m$ and $n$, introduced in \Eq{alphatau1}; see Ref.~\cite{BST98}.
We recall that the values of $\sigma$ and $\nu$ are obtained from
separate fits and thus do not obey the relation $\sigma=\mu\nu$.
Since the scatter in the plots giving $\sigma$ is larger than in those
giving $\nu$, we discard the former in the calculation of $m$ and $n$.
In the first two rows, only stars from the original Wilson sample
\cite{Wilson78} with grades good and excellent are considered
\cite{BST98}.
This sample does not include superactive stars.
In the last three rows, stars from the expanded sample \cite{SaarBran99}
were considered.
For most of them, no calcium data are available, and therefore
only cycle periods are considered, so $\nu$ is not being determined.
}\vspace{12pt}\centerline{\begin{tabular}{lccccccccccc}
Stars  &  sample & $\sigma$  &  $\nu$  &  $\mu$  &  $m$  &  $n$  \\
\hline
inactive & grade: & (0.46) & 0.85  &  0.99 & 0.9 & 0.9--1.7 \\
  active & excellent+good & (0.48) & 0.72  &  0.99 & 0.8 & 0.8--1.6 \\
\hline
inactive & expanded &$\approx0.5$& & 1 & 0.75 & 0.5--1.5 \\
  active & sample   &$\approx0.5$& & 1 & 0.75 & 0.5--1.5 \\
superactive&Ref.~\cite{SaarBran99}&$-0.43$ & & & 0.28 & $-0.43$--0.57\\
\label{TBST98}\end{tabular}}\end{table}

Although it is well recognized that single mode approximations are
not sufficient to solve the dynamo equations as stated
[see, e.g., \Eq{fullset1}], the one-mode equations may still turn out to
be closer to physical reality, because there is now evidence
from data of accretion disc simulations that the
spectral sensitivity of the turbulent transport coefficients is highest
at small wavenumbers \cite{BranSok02}.
In other words, the multiplication $\alpha\meanBB$ in \Eq{ParkerPheno}
should be replaced by a convolution $\alpha\circ\meanBB$, where $\alpha$
would now be an integral kernel.
In wavenumber space, this would correspond to a multiplication with a
$k$-dependent $\hat\alpha$ such that $\hat\alpha(k)$ is largest for
small values of $k$.
One may hope that future simulations will shed more light on this
possibility.

\subsubsection{Rapidly rotating stars or planets}
\label{Rapid_star+planet}

An important outcome of mean field calculations in the presence
of rapid rotation (not captured by the analysis in \Sec{Revisit})
is the prediction that the $\alpha$ effect becomes
highly anisotropic and takes asymptotically the form \cite{Rue78}
\EQ
\alpha_{ij}=\alpha_0\left(\delta_{ij}-\Omega_i\Omega_j/\Omega^2\right),
\label{alpha_rapidrot}
\EN
which implies that, if the angular velocity points in the $\zz$ direction,
for example, i.e.\ $\OO=\Omega\zz$, the vertical component of $\alpha$
vanishes, i.e.\ $\alpha_{zz}\to0$ for $\mbox{Ro}^{-1}\equiv2\Omega\tau\gg1$.
For calculations of axisymmetric and nonaxisymmetric models using
large but finite values of $\mbox{Ro}^{-1}$ see Refs~\cite{RB95,MB95}.

In \Eq{alpha_rapidrot} we stated the asymptotic form of the $\alpha$
tensor in the limit of rapid rotation.
Such a highly anisotropic $\alpha$ tensor is known to lead to strong
nonaxisymmetric magnetic field configurations \cite{Rue80,RueElst94}.
Such an $\alpha$ effect has been applied to modeling the magnetic
fields of the outer giant planets that are known to have very large values
of $\mbox{Ro}^{-1}$.
Simulations using parameters relevant to the outer giant planets of our
solar system show that these bodies may have a magnetic field that
corresponds to a dipole lying in the equatorial plane
\cite{MB95,RuzmaikinStarchenko91}.

The biggest enemy of nonaxisymmetric fields is always differential
rotation, because the associated wind-up of the magnetic field brings
oppositely oriented field lines close together, which in turn leads to
enhanced turbulent decay \cite{Rae86}.
This is quite different to the case of an axisymmetric field, where the
wind-up brings equally oriented field lines together, which leads to
magnetic field enhancements; i.e.\ the $\Omega$ effect.

As the value of $\mbox{Ro}^{-1}$ increases, the $\alpha$ tensor becomes
not only highly anisotropic, but the magnitudes of all
components decreases.
This is a common phenomenon known as `rotational quenching' that affects
virtually all turbulent transport coefficients.
An important turbulent transport coefficient
that we will not say much about here is the so-called
$\Lambda$ effect (modeling the toroidal Reynolds stress),
is responsible for driving differential rotation
in stars including the sun.
Very rapidly rotating stars are therefore expected to have very
little {\it relative} differential rotation, which is indeed
observed \cite{Cameron}.
This means that the $\alpha\Omega$ dynamo will stop working and one
would therefore expect an anisotropic $\alpha^2$ dynamo mechanism to operate in
rapidly rotating bodies with outer convection zones.
This means that such stars should generate a predominantly
nonaxisymmetric field.
There are indeed numerous observational indications for this
\cite{Jetsu_etal94,Berdyugina_etal02}.
It should be noted, however, that already the large scale flow that is
generated by the magnetic field and the Reynolds stress (i.e.\ the
$\Lambda$ effect) tend to make the field nonaxisymmetric \cite{Barker_etal95}.
In addition, in rapidly rotating stars all motions tend to be mostly
in cylindrical surfaces (Taylor-Proudman theorem).
This effect is believed to cause starspots to emerge
mostly at high latitudes in rapidly rotating stars
\cite{HatzesVogt02,SchusslerSolanki92}.
Furthermore, the convective motions tend to be column-like \cite{Busse70}.
This led to the proposal of the Karlsruhe dynamo experiment where
the flow is similarly column-like.

\begin{figure}[t!]\begin{center}
\includegraphics[width=.6\textwidth]{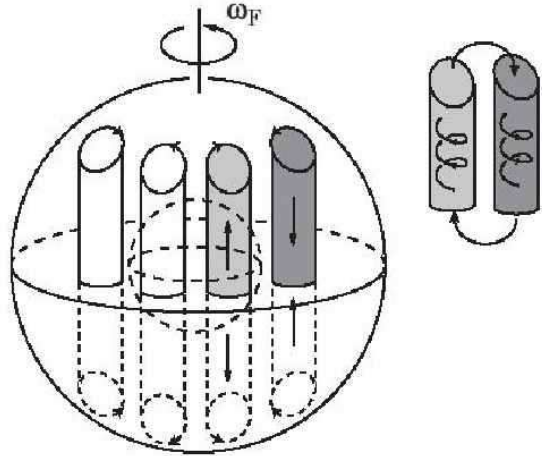}
\end{center}\caption[]{
Convection columns in a rapidly rotating spherical shell.
Courtesy of A. Yoshizawa \cite{Yoshi04}.
}\label{busse_columns}\end{figure}

\subsubsection{Connection with the Karlsruhe dynamo experiment}

\begin{figure}[t!]\begin{center}
\includegraphics[width=.75\textwidth]{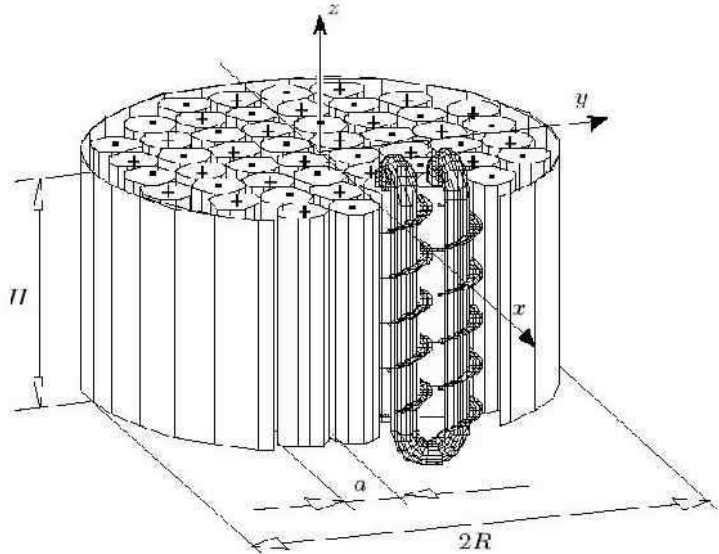}
\end{center}\caption[]{
The dynamo module of the Karlsruhe dynamo experiment.
The signs + and -- indicate that the fluid moves in the positive or negative
$z$--direction, respectively, in a given spin generator.
Courtesy R.\ Stieglitz and U.\ M\"uller \cite{stieglitzetal96}.
}\label{module}\end{figure}

In the Karlsruhe dynamo experiment, liquid
sodium is pumped upward and downward in alternating channels
\cite{stieglitzetal96}.
Each of these channels consist of an inner and an outer pipe, and the walls
of the outer one are arranged such that the flow follows a swirling
pattern; see \Fig{module}.
The swirl in all columns is such that the associated kinetic
helicity has the same sign everywhere.
In fact, locally such a flow is strongly reminiscent of the Roberts
flow (\Sec{RobertsFlowDynamo})
that is known to generate a Beltrami field in the plane
perpendicular to the direction of the pipes, i.e.\
$(\cos kz,\sin kz,0)$ if $z$ is the direction of the pipes
and the phase shift in $z$ has been ignored; see \Sec{SBeltrami}.
On a global scale such a field corresponds to a
dipole lying in the $xy$ plane, and perpendicular to the
$\zzz$ direction; see \Fig{module}.
Applied to the earth, it would therefore correspond to a
nonaxisymmetric field and would not really describe the
magnetic field of the earth, although it might be suitable for explaining
the magnetic fields of the other giant planets Uranus and Neptune \cite{MB95}
that are indeed highly nonaxisymmetric \cite{NessRadler90}.

\subsubsection{Chaos and intermittency}

The search for more complicated temporal and spatio-temporal patterns
has always been popular in mean field dynamo research.
In spatially resolved models (as opposed to one-mode approximations
\cite{JonesWeissCattaneo85}) it was for a long time difficult to find
spatially well resolved
solutions that showed chaotic or even just quasiperiodic behavior.
In fact, it was thought to be essential to invoke an additional
explicit time dependence of the form of dynamical quenching
\cite{Ruz81,SS91,Covas_etal97}.
More recently it turned out that quasiperiodic
and chaotic solutions to the mean field dynamo equations
are possible when the dynamo number is raised sufficiently
\cite{BKMMT89,TorkelssonBrandenburg94}.
Intermittent behavior has also been found when the feedback on the
mean flow is taken into account \cite{Tobias97,MossBrooke00}.
For moderately large dynamo numbers, such solutions require however
small turbulent magnetic Prandtl numbers, which is an assumption that
is not confirmed by simulations \cite{YBR03}.
Nevertheless, intermittent solutions of this or some other kind
are thought to be relevant in connection with
understanding the intermittency of the solar cycle
and the occurrence of grand minima \cite{Spiegel94,Cowas_etal01}.

The question is of course how much one can trust such detailed predictions
of mean field theory when we are still struggling to confirm the validity
of mean field theory in much simpler systems.
One may hope that in the not too distant future it will be possible
to compare mean field models with simulations in somewhat more extreme
parameter regimes where quasi-periodic and chaotic behaviors are expected
to occur.

\subsubsection{Dynamos in proto-neutron stars}

The $\sim10^{13}\G$ magnetic field of neutron stars is traditionally
thought to be the result of compressing the magnetic field of its
progenitor star.
The difficulty with this explanation is that in the early phase of a
neutron star (proto-neutron star) the neutrino luminosity was so immense
and the neutrino opacity high enough that the young neutron star must
have been convectively unstable.
Although this phase lasts for only $\sim20$ seconds, this still corresponds
to some $10^4$ turnover times, because gravity is so strong that the
turnover time is only of the order of milliseconds.
This would be long enough to destroy the primordial magnetic field
and to regenerate it again by dynamo action
\cite{ThompsonDuncan93,Bonanno03}.

If the field was initially generated by an $\alpha$ effect,
and if the associated small scale magnetic helicity has been
able to dissipate or escape through the outer boundaries, the field
must have attained some degree of magnetic helicity.
Once the turbulence has died out, such a helical field decays much
more slowly than nonhelical fields \cite{BiskampMuller99}, and it would
have attained the maximum possible scale available in a sphere.
Such a field may well be that of an aligned or inclined dipole, as observed.
There are also some interesting parallels between dynamo action in
decaying turbulence in neutron stars and that in the planned time-dependent
dynamo experiment in Perm based on decaying turbulence in liquid sodium
\cite{Dobler_etal03}.
In the Perm experiment, a rapidly spinning torus with liquid sodium will
suddenly be braked, which leads to swirling turbulence due to suitable
diverters on the wall of the torus.

\subsection{Accretion disc dynamos}
\label{AccretionDiscDynamos}

When dynamo theory was applied to accretion discs, it seemed at first
just like one more example in the larger family of astrophysical bodies
that host dynamos \cite{pud81a,pud81b,step_levy90,mang_sub94}.
Later, with the discovery of the magneto-rotational instability
\cite{BH91,BH98}, it became clear that magnetic fields are crucial for
maintaining turbulence in accretion discs \cite{HGB95,MT95}, and that this
can constitute a self-excited process whereby the magnetic field necessary
for the magneto-rotational instability is regenerated by dynamo action.

Simulations in a local (pseudo-periodic shearing box) geometry have shown that
this self-excited system can act both as a small scale dynamo if there is
no vertical density stratification \cite{HGB96,SHGB96} and as a large
scale dynamo if there is stratification \cite{BNST95,ZR00}.

From a turbulence point of view it is interesting to note that the
flow is highly anisotropic with respect to the toroidal direction.
This is true even down to the smallest scale in the simulations.
From a mean field dynamo point of view this system (with stratification
included) is interesting because it is an example of a simulation of a
dynamo where the turbulence occurs naturally and is not driven by an
artificial body force.
This simulation is also an example where an $\alpha$ effect could be
determined \cite{BD97}.

\subsubsection{Dynamo waves in shearing sheet simulations}
\label{SShearingSheet}

\begin{figure}[t!]\begin{center}
\includegraphics[width=.99\textwidth]{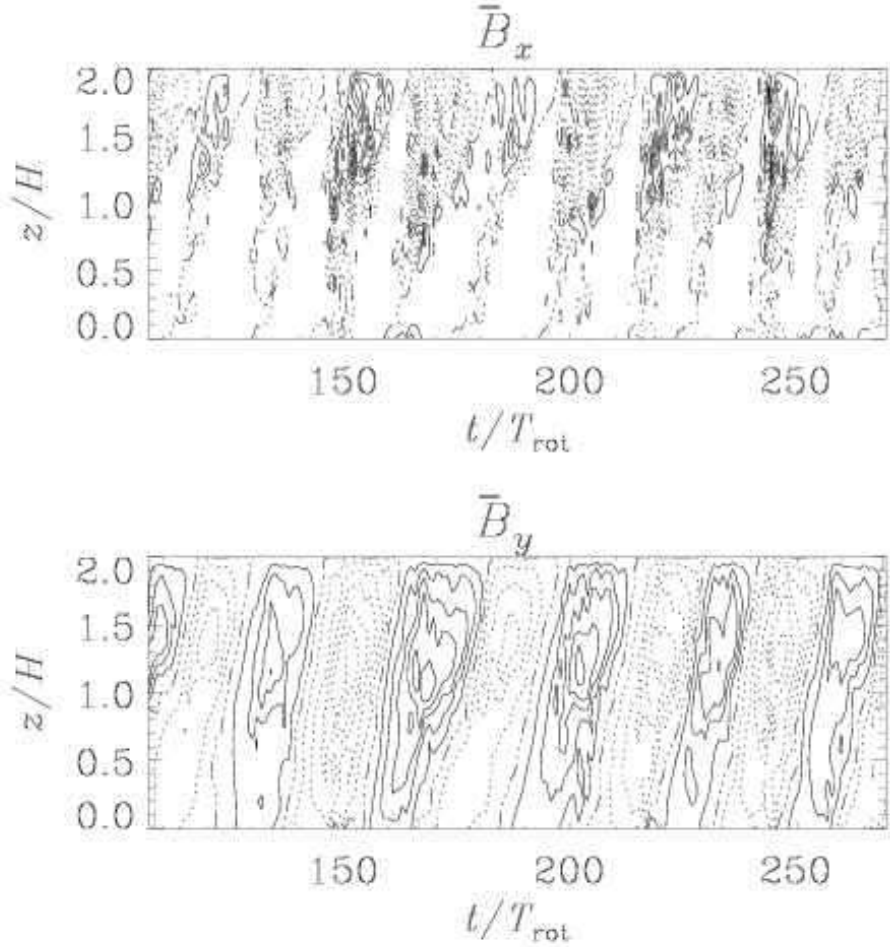}
\end{center}\caption[]{
Horizontally averaged radial and toroidal magnetic fields,
$\meanB_x$ and $\meanB_y$ respectively, as a function of time
and height, as obtained from the fully three-dimensional simulation
of a local model of accretion disc turbulence.
Time is given in units of rotational periods, $T_{\rm rot}$.
(No smoothing in $z$ or $t$ is applied.)
Dotted contours denote negative values.
Adapted from Ref.~\cite{BranSok02}.
}\label{Fbutter}\end{figure}

When the vertical field boundary condition is used, i.e.\ $B_x=B_y=0$
on $z=z_{\rm top}$ and $z_{\rm bot}$,
the horizontal flux through the box is no longer conserved, and hence the
horizontal components of the horizontally averaged field may be different
from zero, i.e.\ $\meanB_x\neq0\neq\meanB_y$, even though they may be
zero initially.
This is exactly what happened in the shearing box dynamo simulation when
these boundary conditions where used.
In \Fig{Fbutter} we show the space-time diagram (or
butterfly diagram in solar physics) of the mean magnetic field of such
a simulation \cite{BranSok02,BD97}.
In this particular simulation the symmetry of the magnetic
field has been restricted to even parity about $z=0$, so the computation has been
carried out in the upper disc plane, $0<z<L_z$, where $L_z=2H$ is the
vertical extent of the box, $H$ is the gaussian scale height of the
hydrostatic equilibrium density, and $z=0$ corresponds to the
equatorial plane.

\begin{figure}[t!]\begin{center}
\includegraphics[width=.99\textwidth]{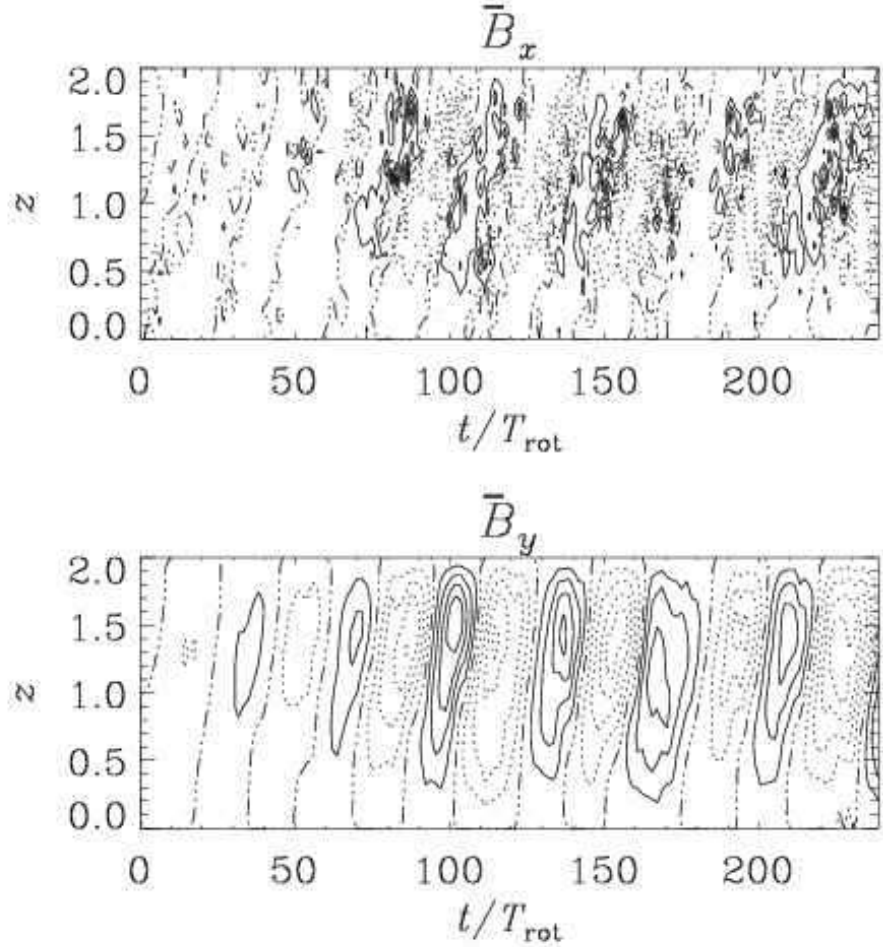}
\end{center}\caption[]{
Resulting $\meanB_x$ and $\meanB_y$ from a
mean field calculation with negative $\alpha$ effect
($\alpha_0=-0.001\Omega H$), together with a 10 times stronger
noisy components ($\alpha_N=0.01\Omega H$), and a turbulent
magnetic diffusivity ($\eta_0=0.005\Omega H^2$), producing
a cycle period of about 30 orbits.
Adapted from Ref.~\cite{BranSok02}.
}\label{Fpp_alpn=0.01}\end{figure}

It is interesting to note that the spatio-temporal behavior obtained
from the three-dimensional simulations resembles in many ways what
has been obtained using mean field models; see \Sec{Sao_model}.
Note, however, that in \Fig{Fbutter} dynamo waves propagate in the
positive $z$ direction, i.e.\ $c>0$ in \Eq{PhaseSpeed}.
This requires $\alpha<0$, which
is indeed consistent with the sign of $\alpha$ obtained earlier
by means of correlating ${\cal E}_y$ with $\mean{B_y}$ \cite{BNST95,BD97},
but it is opposite to what is expected in the northern hemisphere
(upper disc plane) from cyclonic events.

Mean-field model calculations confirm that the space-time diagram
obtained from the simulations (\Fig{Fbutter}) can be produced
with a negative $\alpha$ effect of the magnitude found by
correlating ${\cal E}_y$ with $\meanB_y$ and a turbulent magnetic
diffusivity comparable to the turbulent kinematic viscosity
obtained by measuring the total (Reynolds and Maxwell) stress
\cite{BNST95}.
The directly determined $\alpha$ effect is however so noisy, that it is
hard to imagine that it can produce a mean field that is as systematic
as that in the simulation (\Fig{Fbutter}).
However, model calculations show also that, even when the noise level of
$\alpha$ exceeds the average value by a factor of 10, the resulting mean
field is still sufficiently coherent (\Fig{Fpp_alpn=0.01}) and in fact
similar to the field obtained from the simulations \cite{BranSok02}.

A negative sign of the $\alpha$ effect in accretion discs may arise
because of two important circumstances.
First, the vertical velocity fluctuation is governed by magnetic buoyancy
and second, shear is important.
Following a simple argument of Ref.~\cite{iceland}, the $\alpha$ effect
is dominated by the contribution from the momentum equation and the
toroidal magnetic field, $B_y$,
\EQ
{\partial{\cal E}_y\over\partial t}
\sim\overline{\dot{u}_z b_x}
\sim\left({\overline{b_y b_x}\over p_0}g\right)\meanB_y
\equiv\tilde\alpha_{yy}\meanB_y,
\EN
where the vertical acceleration is assumed to be mostly due to
magnetic buoyancy, i.e.\ $\dot{u}_z\approx-(\delta\rho/\rho_0)\,g$
and $-\delta\rho/\rho_0=\delta\BB^2/(2p_0)\approx\meanB_y b_y/p_0$,
where we have linearized with respect to the fluctuations.
In accretion discs the shear is negative, i.e.\
$\partial\meanU_y/\partial x<0$, and therefore $\overline{b_y b_x}<0$.
If this effect does indeed dominate, the $\alpha$ effect will be negative,
i.e.\ $\alpha_{yy}=\tau\tilde\alpha_{yy}<0$.
Subsequent work based on FOSA confirms the possibility of a negative
value of $\alpha_{yy}$ for sufficiently strong shear \cite{RP00},
as is indeed present in accretion discs.
However, because of the competing effect to drive a positive
$\alpha$ effect from thermal buoyancy, the sign of $\alpha_{yy}$
changes to the conventional sign for weak shear.

In summary, local disc simulations suggest the possibility of a
doubly-positive feedback from both the magneto-rotational instability
and the dynamo instability, giving rise to small scale magnetic fields
(if there is no stratification) and large scale fields (if there is
vertical stratification).
The magnetic energy exceeds the kinetic energy, but is below the thermal
energy; see \Sec{Energetics}.
The latter gives information about the dimensionless value of the
Shakura-Sunyaev viscosity parameter which is typically
around 0.01 \cite{BNST95,HGB96,SHGB96}.
Large scale field generation is compatible with that from a negative
$\alpha$ which, in turn, produces oscillatory mean fields that are
symmetric about the mid-plane.
This is consistent with mean field theory, which also predicts that,
with the same vertical field boundary conditions, the field should be
oscillatory and symmetric about the midplane, i.e.\ quadrupole-like.

When the boundary conditions in the accretion disc simulations are changed
to perfectly conducting boundaries, the situation changes entirely;
the field becomes non-oscillatory and antisymmetric about the midplane,
i.e.\ dipole-like \cite{iceland}.
However, this drastic change is quite compatible with mean-field theory
which also predicts that with perfectly conducting boundaries a negative
$\alpha$ in the upper disc plane should give rise to a non-oscillatory
field that is antisymmetric about the midplane.
Having thus established this type of correspondence between simulations
and mean field theory, one is tempted to apply mean field theory with
a negative $\alpha$ in the upper disc plane to a global geometry.
It turns out that with a conducting halo, the most easily excited
solution is also antisymmetric about the midplane (i.e.\ dipolar)
and non-oscillatory.
Such models have been discussed in connection with outflows from dynamo
active accretion discs, as will be discussed further in the next section.

\subsubsection{Outflows from dynamo active discs}

\begin{figure}[t!]\begin{center}
\includegraphics[width=.99\textwidth]{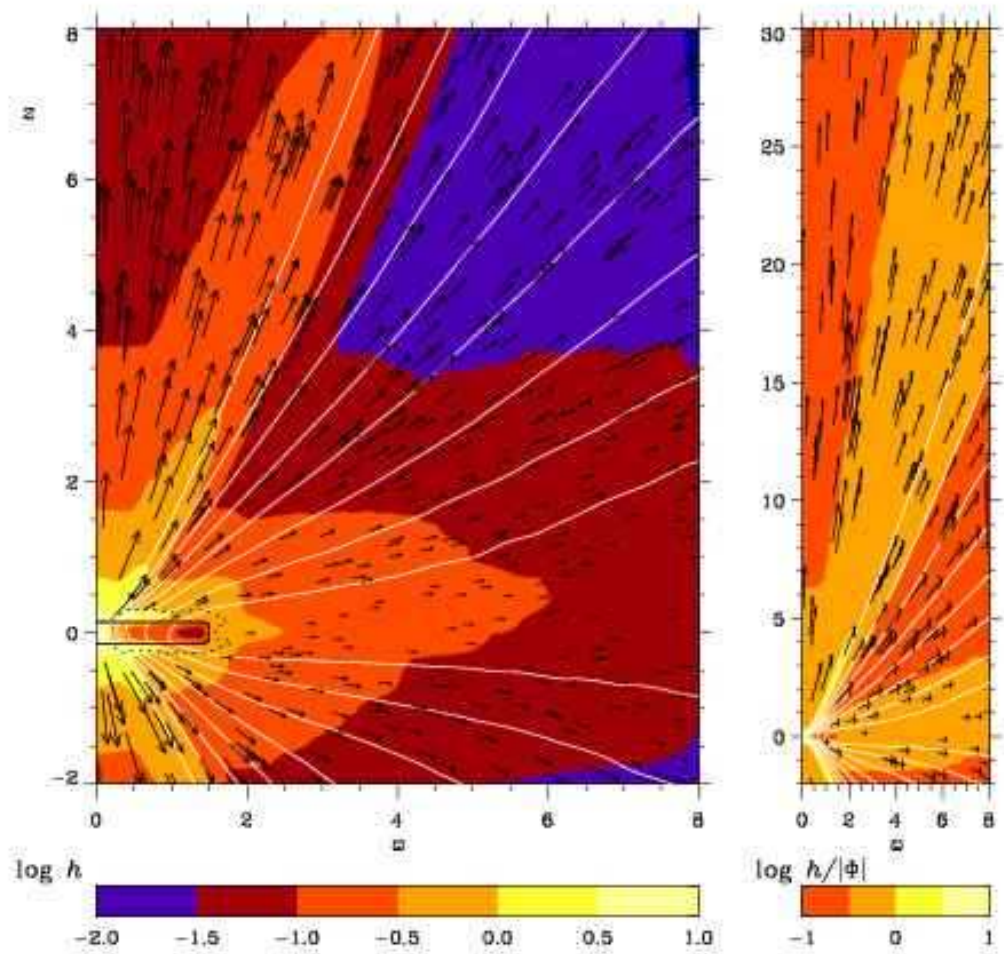}
\end{center}\caption[]{
Outflow from a dynamo active accretion disc driven by a combination
of pressure driving and magneto-centrifugal acceleration.
The disc represents a protostellar disc, whose tenuous outer
regions will are heated by the magnetic field.
This pressure contributes to hydrostatic support, which is
changed only slightly by the finite outflow velocities.
The extent of the domain is $[0,8]\times[-2,30]$ in nondimensional
units (corresponding to about $[0,0.8]\AU\times[-0.2,3]\AU$ in
dimensional units).
Left panel: velocity vectors, poloidal magnetic field lines
and gray scale representation of $h$
in the inner part of the domain.
Right panel: velocity vectors, poloidal magnetic field lines and
normalized specific enthalpy $h/|\Phi|$ in the full domain.
Adapted from Ref.~\cite{Rekowski_etal03}.
}\label{FRRRRun3c}\end{figure}

It is now commonly believed that all accretion discs
have undergone a phase with strong outflows.
One mechanism for driving such outflows is magneto-centrifugal
acceleration \cite{BlandfordPayne82}.
The magnetic field necessary might be the field that is dragged
in from the embedding environment when the disc forms, but
it might also be dynamo generated \cite{Campbell99}.
In \Fig{FRRRRun3c} we present a numerical solution of such
a model \cite{Rekowski_etal03}, where the dynamo is a mean field
$\alpha\Omega$ dynamo with negative $\alpha$ effect \cite{BNST95}.
In the presence of open boundaries considered here, the
most easily excited solution has dipolar symmetry
about the equatorial plane \cite{Bardou_etal01}.

Three-dimensional simulations of dynamo action in accretion disc
tori have confirmed that the magneto-rotational instability
can sustain a magnetic field also in such a global model
\cite{Armitage98,Hawley00,Hawley01,DeVilliersHawley03}.
These simulations are now also beginning to show the formation of outflows
that are clearly associated with large scale fields generated in the disc.
The presence of large scale fields, and in particular its vertical
component, is responsible for increased values of the  dimensionless
value of the Shakura-Sunyaev viscosity parameter which is now typically
0.1 \cite{DeVilliersHawley03}.

\subsection{Galactic dynamos}
\label{galdyn}

We saw in \Sec{GalB} that spiral galaxies have large scale 
magnetic fields of the order of a few $10^{-6}\G$, coherent on scales of 
several kpc, and also highly correlated (or anti-correlated)
with the optical spiral arms. Does the mean field turbulent dynamo 
provide a viable model for understanding the origin such fields?
We discuss below a number of observed properties 
of galactic fields which favor a dynamo origin of disc galaxy 
magnetic fields. However the strength of the field itself may
not be easy to explain, in view of the helicity constraint.

\subsubsection{Preliminary considerations}

To begin with, all the ingredients needed for large scale dynamo
action are present in disc galaxies. There is shear due to
differential rotation and so any radial component of the magnetic 
field will be efficiently wound up and amplified to produce a 
toroidal component. A typical rotation rate is
$\Omega \sim 25 \km \s^{-1} \kpc^{-1}$ (at $r\approx10\kpc$), and for a flat
rotation curve ($\meanU_\phi=\Omega R=\mbox{constant}$) the shear rate,
defined as $S=R\Omega'\equiv R\dd\Omega/\dd R$, would be the same,
i.e.\ $S=-\Omega$. 
This corresponds to a rotation time of $\sim 2\pi/\Omega
\sim 2.5 \times 10^8 \yr$. So, in a Hubble time ($\sim 10^{10} \yr$)
there have been about $40-50$ rotations. If only
shear were involved in the generation of the galactic field, 
a fairly strong primordial seed field of $\sim 10^{-7}$ G would be
required just after the galaxy formed. (We will also discuss
other problems with such a primordial hypothesis below.)

A mechanism to exponentiate the large scale
field, like for example the $\alpha \Omega$ dynamo,
is therefore desirable. An $\alpha$ effect could indeed
be present in a disc galaxy. First, the interstellar medium in
disc galaxies is turbulent, mainly due to the effect of supernovae randomly
going off in different regions. In a rotating, stratified
medium like a disc galaxy, such 
turbulence becomes helical (\Sec{HelicityInducedByRotation}).
Therefore, we potentially have an $\alpha$
effect and so we expect $\alpha\Omega$ mean field dynamo action. 
Typical parameters for this turbulence
are, a velocity scale $v \sim 10 \km \s^{-1}$, an outer scale
$l \sim 100 \pc$ giving an eddy turnover time $\tau \sim l/v
\sim 10^{7}$ yr.\footnote{Numerical simulations of such a turbulent 
multiphase interstellar
medium regulated by supernovae explosions, and including rotation,
show \cite{korpi_etal99} the multiphase gas
in a state of developed turbulence, with the warm and hot phases having
rms random velocities of $10$ and $40 \km \s^{-1}$ respectively,
and with turbulent cell size of about $60\pc$ for the warm phase.}
This yields an estimate of
the turbulent diffusion coefficient,
$\eta_{\rm t} \sim \onethird vl\sim 0.3\km\s^{-1}\kpc$
or $\eta_{\rm t} \sim 10^{26} \cm^2 \s^{-1}$.
The inverse Rossby (or Coriolis)
number is $\sim2\Omega\tau \sim 0.6$ (see \Tab{TRossby}, where
we have taken $\Omega = 30 \km\s^{-1}\kpc^{-1}$) and so
we can use \Eq{alpha_tau} to estimate 
$\alpha$ due to rotation and stratification.
This gives $\alpha \sim \tau^2\Omega (v^2/h) \sim 0.75\km \s^{-1}$, where
$h \sim 400 \pc$ is the vertical scale height of the disc. 
So the turbulence is only weakly helical; nevertheless
the degree of helicity ($\sim 5 - 10\%$) is sufficient for inducing
large scale dynamo action, because the $\Omega$ effect 
is strong enough to make the $\alpha\Omega$
dynamo supercritical.

We gave a brief discussion of the simplest form
of the mean field dynamo equations appropriate for a thin
galactic disc in \Sec{ExcitationConditionsDisc}, following 
Ref.~\cite{RSS88}; here we just gather some important facts. 
We recall that two dimensionless control parameters, 
\EQ
C_\Omega = S h^2/\eta_{\rm t}, \quad
C_\alpha = \alpha h/\eta_{\rm t},
\EN
measure the strengths of shear and 
$\alpha$ effects, respectively. 
(Note that these are identical to the $R_\Omega$ and
$R_\alpha$ commonly used in galactic dynamo literature.) 
For the above galactic parameters, the typical 
values are $C_\Omega \approx -10$ and $C_\alpha \approx 1$,
and so $\vert C_\Omega \vert \gg C_\alpha$.
Dynamo generation is controlled by the dynamo number 
$D = C_\Omega C_\alpha $, whose initial `kinematic' value
we denote by $D_0$. Exponential growth of the field is
possible in the kinematic stage, provided 
$\vert D_0 \vert > D_{\rm crit}$, where
the critical dynamo number $D_{\rm crit} \sim 6...10$,
depending on the exact profile adopted for $\alpha(z)$.
Modes of quadrupole symmetry in the meridional $(Rz)$ plane
are the easiest to excite in the case of a thin disc with
$D < 0$. Further the growth rate in the kinematic regime
given by \eq{gama} has a numerical value,
\EQ
\gamma \approx {\eta_{\rm t} \over h^2}
\left( \sqrt{\vert D \vert} - \sqrt{D_{\rm crit}} \right)
\approx (1...10)\,{\rm Gyr}^{-1}
\label{gam}
\EN
The dynamo is generally supercritical, and the value of $\gamma$
evaluated locally at different radial position is positive
for a large range of radii, and so galactic fields
can be indeed grow exponentially from small seed values.
Furthermore, in a Hubble time scale of $10^{10} \yr$, one
could exponentially grow the field by a factor of about 
${\rm e}^{30}\approx10^{13}$, provided the growth rate determined
in the kinematic regime were applicable. In this case, even a small
field of order $10^{-19}$ G would in general suffice 
as the initial seed magnetic field; with galaxies
at higher redshift requiring a larger seed magnetic field.
We shall discuss below how the nonlinear restrictions,
due to helicity conservation, may alter this picture.
Before this, we first gather some of the
observational indications which favor a dynamo origin for
the galactic field, nicely reviewed by 
Shukurov \cite{shukurov00,shukurov03}.

\subsubsection{ Observational evidence for dynamo action}

\noindent{\it The magnetic pitch angle}

The simplest piece of evidence for dynamo action is just the form
of the magnetic field lines as projected onto the disc.
It is seen that the regular magnetic field in spiral galaxies is 
in the form of a spiral with pitch angles
in the range $p= -(10^\circ\ldots30^\circ)$, where a negative $p$
indicates a trailing spiral. Note that for dynamo action one needs to 
have non-zero $\meanB_R$ and $\meanB_\phi$, or a non-zero pitch angle
for the magnetic field lines projected onto the disc.
A purely azimuthal field, with $\meanB_R=0$, will decay due to turbulent
diffusion, and the same is true for a purely radial field.
A simple estimate of the pitch angle can be obtained from
the `no-$z$' approximation \cite{SM93} to the dynamo
equations, whereby one replaces $z$ derivatives by just a division 
by $h$. Substituting also $\meanB_R,\meanB_\phi\propto\exp{\gamma t}$,
this gives 
\EQ
\left(\gamma + {\eta_{\rm t} \over h^2} \right)\meanB_R = 
-{\alpha \meanB_\phi \over h}, \quad  
\left(\gamma + {\eta_{\rm t} \over h^2} \right)\meanB_\phi = S\meanB_R. 
\label{disc_mfdyn}
\EN
One derives a rough estimate for
the ratio \cite{shukurov00}
\EQ
\tan p = {\meanB_R \over \meanB_\phi} \approx - 
\left({\alpha \over -Sh}\right)^{1/2}
= -\left({C_\alpha \over \vert C_\Omega \vert } \right)^{1/2}
\approx -{l \over h} \left({\Omega \over|S|} \right)^{1/2}.
\EN
For $l/h = 1/4$ and a flat rotation curve with $\Omega = -S$,
one then gets $p \sim -14^\circ$.
This is in the middle of the 
range of observed pitch angles in spiral galaxies. More detailed
treatments of galactic dynamos \cite{RSS88} confirm
this simple estimate. The above estimate is based
on kinematic theory, and comparison with observations
should preferably done with models including nonlinear effects.
However, similar pitch angles are also obtained when one considers
some simple nonlinear dynamo models \cite{SchultzElstnerRudiger94};
see also the case of models for M31 in Ref.~\cite{moss_anvar98}.
(The no-$z$ approximation also gives an estimate for 
$\gamma$ as in \Eq{gam} but with the crude estimate $D_{\rm crit} = 1$.)

An alternative hypothesis to the dynamo is that galactic magnetic
fields are simply strong primordial fields that are wound up 
by differential rotation (see Refs~\cite{kulsrud99,howard_kulsrud97} 
and references therein). 
In this case, detailed analysis \cite{howard_kulsrud97} shows
that differential rotation leads a field which rapidly
reverses in radius (on scales $\sim 100 \pc$). More
importantly the field is also highly wound up
with pitch angles of order $p < 1^\circ$, clearly
much smaller than the observed values. In Ref.~\cite{howard_kulsrud97},
it is argued that streaming motions in spiral arms would 
nevertheless lead to a field-aligned along the spiral; but away from
the arms the field will be nearly toroidal. Clearly,
in the case of NGC6946 it is the field in between the arms
which has a spiral form with moderately large pitch angles
with an average $p \sim 35^\circ$ for several of the magnetic arms
\cite{frick00}. Also in the case of M31, although the field 
itself occupies a ring-like region, the field lines are still 
in the form of spirals with pitch angles $\sim 10^\circ\ldots20^\circ$ 
\cite{fletcher_etal00}.  

Another possibility is that the 
primordial field is not tightly wound up because
turbulent diffusion compensates for the winding due to
shear \cite{nord_rog00}. In this case there would be a balance 
between shear and turbulent diffusion, and from the toroidal part of
\Eq{disc_mfdyn} one can estimate $\meanB_R/\meanB_\phi \sim 1/C_\omega$;
which implies $p \sim 6^\circ$ for $C_\Omega \sim 10$.
This is larger than the pitch angle determined neglecting
turbulent diffusion, but may still not be large enough
to account for the observed range of $p$ in galaxies.
Also one would need a constant source of $\meanB_R$, 
which could perhaps be due to accretion, since otherwise turbulent diffusion
(without an $\alpha$ effect) will cause $\meanB_R$ to decay. 
The galactic dynamo on the other hand provides
a natural explanation for the observed pitch angles of
the regular magnetic field. 

\noindent{\it Even vertical symmetry of the field in the Milky Way}

Another argument in favor of the galactic dynamo theory
is the observed symmetry properties of the regular magnetic field
about the galactic equator. Dynamo theory predicts that the toroidal
and radial field components should be symmetric about the equator. 
The even parity mode has a larger scale of variation in the 
vertical direction and is therefore subject to a weaker turbulent diffusion
than the odd parity mode \cite{RSS88}. A wavelet analysis of the 
Faraday rotation measures of extragalactic sources indeed indicates that
the horizontal components of the regular magnetic field
have even parity, that is they are similarly directed on both
sides of the disc \cite{frick_etal01}. 

Such an even parity field can be produced from a primordial field
if it is originally almost parallel to the disc plane.
However, such a field would still suffer the excessive winding
mentioned above. On the other hand, it is equally likely
that a primordial field has a large component parallel
to the rotation axis (entering through $z < 0$
and exiting through $z > 0$).
In this case one would have a dipole structure for the
wound up primordial field, which is not supported
by the analysis of \cite{frick_etal01}. (As mentioned
earlier, the determination of the magnetic field
structure from Faraday rotation can be complicated
by local perturbations.) Further, as discussed in 
\Sec{GalB} there was some evidence from the discovery
of linear non thermal filaments perpendicular to
the galactic plane, of a dipolar field in the central few hundred 
parsecs of our galaxy, which could be explained
on the basis of a primordial field \cite{chandran00}.
However the latest surveys for such
linear filaments \cite{yousef-zadeh04}, no longer strongly
support such a simple picture, and the observational
situation needs to be clarified.

\noindent{\it The azimuthal structure of disc galaxy fields}

The kinematic growth rates predicted by an 
axisymmetric galactic dynamo are also the largest for 
purely axisymmetric field structures. 
Of course the spiral structure induces nonaxisymmetry, and this can
enhance the growth rates of nonaxisymmetric dynamo
modes \cite{Moss_etal91,SM93,MS91,moss96}.
However this enhancement is still not such that 
one expects a widespread dominance of
nonaxisymmetric magnetic structures in disc galaxies.
Early interpretations of Faraday rotation in spiral
galaxies seemed to indicate a prevalence of
bisymmetric structures \cite{Sofue86}, which
would be difficult to explain in the framework of dynamo theory; 
but this has since not been confirmed. 
Indeed as discussed also in \Sec{GalB_global}, many galaxies
have mostly distorted axisymmetric magnetic structures,
wherein the axisymmetric mode is mixed in
with weaker higher $m$ modes. Only M81, among the nearby spirals,
remains a case for a predominantly bisymmetric magnetic structure.
The fact that it is physically interacting with a companion may
have some relevance for the origin of its bisymmetric fields 
\cite{moss96b}.

It would not be difficult to produce a predominantly bisymmetric ($m=1$)
structure by the winding up of a primordial field; more
difficult to explain would be the dominance of 
nearly axisymmetric structures.
Such configuration would require an initial primordial
field to be systematically asymmetric, with its maximum 
displaced from the disc center.
Any primordial field would in general lead to a combination
of $m=0$ and $m=1$ modes, with $m=1$ being in general more dominant.

The observed nonaxisymmetric spiral magnetic structures are also aligned
or anti-aligned with the optical spirals. This is intriguing
because of the following: A given nonaxisymmetric mode basically 
rotates at nearly the local rotation frequency, where the mode is concentrated. 
On the other hand, the optical spiral pattern rotates  
at a `pattern' frequency very different in general
compared to the local rotation frequency. It has been shown 
\cite{SM93,MS91,moss96,bykov_etal97} that 
such structures can be maintained for a
range of radii of about a few kpc around the corotation 
radius of the spiral pattern. Near the corotation radius
a frozen-in field will rotate with the spiral pattern; and
around this radius a moderate amount of the turbulent diffusion
allows modes to still rotate with the pattern frequency, instead
of with the local rotation frequency.
More intriguing are the `magnetic arms' seen in NGC6946, and 
the anti-alignment between
these magnetic arms and optical spiral arms. Explaining this feature
in a dynamo model requires an understanding of how spiral
structure influences the different dynamo control parameters 
\cite{shukurov98,moss98,Rhode_etal99}. 

Other possible evidence for dynamo action include \cite{shukurov00}
radial reversals of the magnetic field in the Milky Way,
the complicated magnetic structures that are
observed in M51, and the magnetic ring seen in M31. 
Shukurov \cite{shukurov00} has argued that
all these features can be understood in dynamo models
of these systems; whereas it is not clear how they may
be explained in models involving primordial fields.

\noindent{\it Strength of the regular magnetic field}

The regular magnetic field in disc galaxies is close
to energy equipartition with the interstellar turbulence. 
This feature directly indicates that the field is somehow
coupled to the turbulent motions; although the exact
way by which the regular field achieves equipartition
with the turbulence is still to be understood (see below).
In the case of wound up primordial fields, it is plausible that if the magnetic 
pressure exceeded the fluid pressure, then
it would rise from the disc due to buoyancy. This could be
an indirect way by which the field strength is limited to
be in rough equipartition. More difficult is to actually generate
a primordial field of the requisite strength.
In models of the generation of primordial
magnetic fields based on processes occurring in the early
universe (cf.\ Refs~\cite{turner_widrow88,ratra92,grasso_ruben01}
and references therein), the strength of the generated field
is highly uncertain, and very sensitive to the assumed parameters.

\subsubsection{Potential difficulties for galactic dynamos}

\noindent{\it Magnetic helicity constraint}

The major potential difficulty for the galactic dynamo is still the
restriction imposed by magnetic helicity conservation. We have
discussed these issues in detail in \Sec{MagneticHelicityMeanFieldModels}.
The magnetic Reynolds number in galaxies
is large enough that one would expect the total magnetic
helicity to be well conserved. The fact that shear plays
a major role in the galactic dynamo implies that the galactic
field is not maximally helical, and one can somewhat
ease the restrictions imposed by helicity conservation.
Nevertheless, if there is negligible {\it net} small scale helicity flux 
out of the galaxy,
the mean galactic field is limited to \cite{Sub02}
$\meanB\approx(k_{\rm m}/k_{\rm f}) B_{\rm eq}[(D_0/D_{\rm crit})-1]^{1/2}$
for moderately supercritical dynamo numbers;
see also \Sec{MagneticHelicityMeanFieldModels} and \Eq{Bkin_shear}.
If we adopt $D_0/D_{\rm crit} \sim 2$, $k_{\rm m}/k_{\rm f}\sim l/h$, with
a turbulence scale $l \sim 100 \pc$, and $h \sim 400 - 500 \pc$,
then the mean field strength would be $1/4$ to $1/5$ of
equipartition at saturation. 
The above estimate is a prediction of the dynamical quenching model
which has been verified mainly using periodic box simulations.
Nevertheless, even in open domains where helicity fluxes are possible,
\Eq{Bkin_shear} seems to describe the field strength near the end of
the kinematic phase reasonably well \cite{BS05b}.
This estimate also assumes the absence of $R_{\rm m}$ dependent
suppression of the Strouhal number, which does seem to be supported
by direct simulations \cite{Bran+Sub05}.

Of course a preferential
loss of small scale helicity could increase the saturation field strength
further and this has indeed been invoked by
Kleeorin and coworkers \cite{KMRS02,KMRS00,KMRS03}
and by Vishniac and Cho \cite{VC01}.
As suggested in \Sec{TurbulenceAndShear}, simulations are now beginning
to show that, in the presence of shear and open boundaries, a significant
small scale magnetic helicity flux does indeed emerge.
As demonstrated in a simulation of forced turbulence with driven
shear, strong large scale fields are generated on a dynamical time scale
when such helicity flux is possible.
However, if the helicity flux is below a certain threshold, the initially
large field strength is followed by very long term oscillations of low
magnetic field strength.
Nevertheless, these simulations are still quite idealized in that
stratification is ignored and the forcing does not represent galactic
conditions.
These deficiencies can hopefully be removed in the near future.
The theoretical underpinning
of the phenomenologically imposed helicity fluxes
\cite{KMRS02,KMRS00,KMRS03} is only beginning to be clarified \cite{KS_AB05}.
This would be another area of further investigation.

\noindent{\it Small scale magnetic noise}

As noted in \Sec{SmallScale}, the small scale dynamo
generates fluctuating magnetic fields in the kinematic stage, at a rate
faster than the large scale dynamo.
Can this rapidly generated magnetic noise suppress 
large scale galactic dynamo action?
The analytical estimate of the nonlinear $\alpha$ effect
using MTA presented in \Eq{alpha_tau} indicates that the
presence of a equipartition strength fluctuating magnetic field
does reduce the $\alpha$ effect, but only by a factor of $\sim 2/3$
for the diagonal components, and even enhances the off-diagonal
components. Also, in both MTA and quasilinear treatments (FOSA) describing the
effect of fluctuating fields, $\eta_{\rm t}$ is not renormalized at all.
Furthermore, the nonlinear behavior seen in
direct simulations of large scale dynamos \cite{B01,BDS02,BBS01}
can be understood using helicity conservation arguments alone,
without further reduction of the turbulent transport coefficients
due to magnetic noise; although similar simulations at even higher
values of the magnetic Reynolds number are desirable.
Overall, there is no strong evidence that magnetic noise
due to small scale dynamo action, catastrophically suppresses
$\alpha$ and $\eta_{\rm t}$; but this issue also needs to be 
further investigated.
Especially important to understand is whether helicity
conservation fully represents all the effects of flux freezing
or there are further effects which suppress the lagrangian
chaos and hence quench the turbulent coefficients.
For example, it would be useful to compute the turbulent
diffusion of a large scale magnetic field in the presence
of strong small scale MHD turbulence, like that 
described by Goldreich and Sridhar \cite{GS95}.

\noindent{\it Magnetic fields in young galaxies?} 

The galactic dynamo exponentiates seed mean magnetic fields
over a time scale of about $(1...10) \ {\rm Gyr}^{-1}$, depending
on the dynamo parameters. For a young galaxy which is say
formed at redshift $z=5$ and is observed at a redshift of $z=2$, 
its age is $T \sim 1.7 \ {\rm Gyr}$
in the currently popular flat Lambda dominated cosmology
(with cosmological constant contributing a density 
$\Omega_\Lambda=0.7$, matter density 
$\Omega_m = 0.3$ in units of the critical density and
a Hubble constant $H_0 = 70 \km \s^{-1} \Mpc^{-1}$).
For a mean field growth rate of say $\gamma \sim 3 \ {\rm Gyr}^{-1}$,
this corresponds to a growth by a factor $\e^{\gamma T} \sim 75$;
and so even with a seed of $\sim 10^{-9}$ G, we would have 
a mean field of only $ \sim 0.075 \muG$. So, if one sees evidence for
strong microgauss strength mean (or large scale) fields in galaxies at
high redshifts, the galactic dynamo would have difficulties
in accounting for them. If the seed fields were much weaker,
say $\sim 10^{-18}\G$, like those produced in batteries, one
would have difficulty in accounting for microgauss
strength mean fields even for moderate redshift objects,
with $z \sim 0.5$.

At present there is some tentative evidence for magnetic fields
from Faraday rotation studies of 
high redshift quasars and radio galaxies. 
A Faraday rotation map of the extended and polarized jet of 
the quasar PKS 1229-021 at $z=1.038$, has revealed
fields ordered on kpc scales and with strengths of a few microgauss
at $z=0.395$, corresponding to the position of an intervening absorption system
\cite{KPZ92}. Evidence for similarly ordered fields also
comes from rotation measure map of the quasar 3C191, which has 
an associated absorption system at $z=1.945$ similar to its emission redshift
\cite{KPZ90}. A number of high redshift radio galaxies
at $z > 2$ \cite{AK98} also show very high Faraday rotation 
of $\sim 1000\rad\m^{-2}$, which could
be indicative of strong ordered magnetic fields in their
host galaxy (although such fields could also arise from the magnetization
of a sheath around the radio lobe). There is a general
difficulty that a given RM could arise in several intervening
systems including the source and our Galaxy. This introduces
difficulties in determining magnetic fields associated
with high redshift systems \cite{PWK93}. 
A promising approach would be to use radio gravitational lenses
which show differences in Faraday rotation between their different images
\cite{GRB85,Patnaik93}. The lines of sight to the different images
follow very similar paths in the source and our galaxy, but
have large transverse separations in an intervening
object, and so could probe intervening magnetic fields better.

We should remember that the small scale dynamo could itself produce
strong magnetic fields on time scales of $\sim 10^7 \yr$. However
this would be correlated at most on the scales of the turbulence
and smaller; so it will not be able to explain fields
ordered on kpc scales.
The fields generated by a small scale dynamo could nevertheless
provide a strong seed magnetic field for the large scale dynamo
\cite{BeckPoezdShukurovSokoloff94}. Such fields can also lead to
significant Faraday rotation, but if the large scale dynamo were 
not operating, the resulting $RMs$ would not be correlated on scales 
larger than the forcing scale of the turbulence.

\subsection{Cluster magnetic fields}

We saw in \Sec{Cluster_fields} that clusters of galaxies appear to be
magnetized, with fields ranging from several $\mu$G to tens of $\mu$G
in some cluster centers, and with coherence scales 
$l \sim 10\kpc$. How are such fields generated and maintained?
Note that a tangled magnetic field, left to evolve without any forcing, 
would generate motions with a velocity of order the Alfv\'en velocity 
$v_{\rm A} = B/(4\pi \rho)^{1/2}$, and result in decaying
MHD turbulence with a characteristic decay time $\tau_{\rm decay} \sim l/v_{\rm A}$
\cite{MacLow_etal98,BiskampMueller99}.
For a $\mu$G field, in a cluster medium with densities 
$\sim 10^{-3} \cm^{-3}$ is $ v_{\rm A} \sim 70 \km \s^{-1}$ and
$\tau_{\rm decay} \sim 1.4 \times 10^8 \yr$. Although the energy
density in MHD turbulence decays as a power law,
and the decay rates can be somewhat slower 
if the field is partly helical \cite{ChristenssonHindmarshBrandenburg01},
this time scale is still much smaller than the
typical age of a cluster, which is thought to be
several billion years old. 
So one would still require some mechanism to generate  
strong enough tangled cluster magnetic fields initially, and 
preferably maintain them for cluster lifetimes. 

First, there are a number of sources of seed magnetic fields
in galaxy clusters. It is well known that the intracluster medium (ICM)
has metals which must have been generated in stars in galaxies
and subsequently ejected into the galactic interstellar medium (ISM) 
and then into the ICM. Since the ISM is likely to be magnetized with 
fields of order a few $\mu$G, this would lead to a seed
field in the ICM. The exact manner in which the ISM 
from a galaxy gets mixed into the cluster gas is uncertain;
possibly involving tidal and ram pressure stripping of the galactic gas, 
together with galactic outflows \cite{bran00}. One can roughly
estimate the seed field resulting from stripping the galactic gas, 
by using flux conservation;
that is $B_{\rm seed} \sim (\rho_{\rm ICM}/\rho_{\rm ISM})^{2/3} B_{\rm gal}$.
For $ B_{\rm gal} \sim 3 \muG$, and $\rho_{\rm ICM}/\rho_{\rm ISM}
\sim 10^{-2}...10^{-3}$, one gets $B_{\rm seed} \sim 0.1...0.03\muG$.
More difficult to estimate is the coherence scale of the seed field.
One may get even larger seed fields if cluster galaxies
have substantial magnetized outflows:
if $\sim 10^3$ galaxies have mass outflow with
$\dot{M} \sim 0.1M_\odot \yr^{-1}$ lasting for $1\Gyr$,
with a Poynting flux about 10\% of the material flux, 
and the field gets mixed into the cluster gas over a Mpc sized region,
$B_{\rm seed} \sim 0.3\muG$ would result \cite{bran00}.

Another source of seed fields is likely to be also the outflows
from earlier generation of active galaxies (radio galaxies and quasars)
\cite{rees94,MEns00,FL01,colgate_li02}.
Such outflows may leave behind magnetized bubbles of
plasma in some fraction of the intergalactic medium
(typically $\sim 10\%$ \cite{FL01}), which when 
incorporated into the ICM would seed the general
cluster gas with magnetic fields. If one assumes the cluster
gas is about $10^3$ times denser than the IGM, and blindly 
uses the enhancement of the bubble field due to
compressions during cluster formation, one can get
fields as large as $0.1 - 1 \mu$G in the ICM \cite{FL01}. 
However this is to ignore the issue of how the
field in the magnetized bubble, especially if it is predominantly
relativistic plasma from a radio galaxy, mixes with the nonmagnetized
and predominantly thermal gas during cluster formation, 
and the resulting effects on both
the field strength and coherence scales (see Ref.~\cite{ensslin02} for
the related problem of getting cosmic rays out of radio cocoons). 
It is likely that, while AGNs and galaxies provide a potentially
strong seed magnetic field, there would still be a need for their 
subsequent amplification and maintenance against turbulent decay.

In most astrophysical systems, like disc galaxies, stars and planets,
rotation plays a very crucial role in this respect; 
both in providing strong shear
and in making random flows helical, and hence leading to
large scale dynamo action. Clusters on the other hand,
are expected to have fairly weak rotation--if at all, 
so one has to take recourse to some other mechanism
for understanding cluster magnetism.
One possibility is small scale turbulent dynamo action; indeed  
most early work on the generation of cluster magnetic fields 
explored this possibility \cite{jaffe80,roland81,RSS89,RG91,deyoung92}. 
The turbulence was thought to be provided by wakes of galaxies 
moving through the intracluster medium. However galaxy wakes are 
probably inefficient. First, the gas in the galaxy would get significantly 
stripped on the first passage through the cluster, and stop
behaving as a `hard' sphere in producing a turbulent wake.
Furthermore, the wakes generated by the Bondi-Hoyle 
gravitational accretion are probably 
not pervasive enough, because the accretion radius $\sim GM/c_{\rm s}^2
\sim 0.5 \kpc\,(M/10^{11}M_\odot)(c_{\rm s}/10^3\km\s^{-1})^{-2}$ is much
smaller than galaxy radii of order $10\kpc$. 
Further, calculations using the EDQNM equations \cite{deyoung92}
gave pessimistic estimates for the generated
fields when the turbulence is induced by galactic wakes.
A different source for cluster turbulence and magnetic fields
is probably required.

Turbulence in clusters can also arise in mergers
between clusters \cite{tribble93,Roettiger_etal99a,NB99,RS01}.
In hierarchical structure formation theories, 
clusters of galaxies are thought to be assembled relatively recently.
They form at the intersection of filaments in the large scale
structure, involving major mergers between comparable mass objects and
also the accretion of smaller mass clumps. In this process, 
it is likely that clusters develop significant random
flows, if not turbulence \cite{kulsrud97,NB99,RS01}.
These would originate not only due to vorticity generation
in oblique accretion shocks and instabilities during the cluster 
formation, but also in the wakes generated during the merger with 
smaller mass subclumps. In a simulation of cluster formation,
it was found that the intracluster
medium becomes `turbulent' during cluster formation \cite{NB99},
and this turbulence
persists even after about $5$ Gyr after the last major merger.
At this time peculiar velocities are of order $400 \km \s^{-1}$ within
$1/2$ the Virial radius $r_v$; see Fig.~1 in \cite{SNB03}.
A visual inspection of the flow field
reveals `eddies'  with a range of sizes of $50 - 500 \kpc$.
Similar peculiar gas motions have also been found
in other numerical simulations of cluster formation \cite{RTM03} 
and cluster mergers \cite{RS01}. In the merger simulation \cite{RS01} 
ram pressure effects during the merger are though to displace the gas 
in the cluster core from the potential center, causing
it to become unstable. The resulting convective plumes produce
large scale turbulent motions with eddy sizes up to several hundred kiloparsecs.
Again, even after about a Hubble time
($\sim 11$ Gyr after the first core interaction),
these motions persists as subsonic turbulence, with velocities of
order $10...20\%$ of the sound speed for equal mass mergers and twice as
large for mergers with a mass ratio of $1:3$.
(In all these cases, since there is limited spatial resolution,
it may be better to call the flows random flows rather than turbulence.)

Observational evidence of intracluster turbulence is scarce. From analysis of
pressure fluctuations as revealed in X-ray observations it has been
argued \cite{Schueker04} that the integral turbulent scale in the 
Coma cluster is close to 100\,kpc, and they assume a turbulent speed 
of $250$ km s$^{-1}$ at that scale.  It may be possible, in the future,
to detect cluster turbulence via the distortions they induce in the CMB, 
as well as via Doppler broadening and shifting of metal lines in the 
X-ray spectrum \cite{SNB03}.

A quantitative assessment of the importance of such random shear
flows and perhaps turbulence, for the generation of cluster magnetic 
fields has only begun recently, by doing direct simulations 
\cite{Roettiger_etal99b,dolag99}
and also using semi-analytic estimates combined
with simulations of the small scale dynamo \cite{Sub_Shuk_Haugen05}. 
(It was also emphasized by \cite{kulsrud97} in the context of
generating protogalactic fields, that structure formation
can lead not only to seed fields but also significant
vortical motions which amplify the seed.) 
Simulations of cluster mergers \cite{Roettiger_etal99b}
showed two distinct stages of evolution of the field.
In the first stage, the field becomes quite filamentary as a result of
stretching and compression caused by shocks and bulk flows
during infall.
Subsequently, as bulk flows are replaced by more turbulent motions,
the magnetic energy increases
by an average factor $\sim 3$ to localized enhancements
of $\sim 20$.
It is argued \cite{Roettiger_etal99b} that this increase is likely
to be a lower limit, as one cannot resolve the formation of eddies
on scales smaller than half the cluster core radius.
Magnetic field evolution in a smooth particle hydrodynamics
simulation of cluster collapse has also been reported
\cite{dolag99,dolag02}; they found that compression 
and shear in random flows during cluster formation
can increase the field strengths by factors of order $10^3$.

Both these groups use ideal MHD and it would be useful
to relax this assumption.  It will be important to do further 
simulations which have the resolution to follow also the
development of turbulence and the nonlinear cascade to
small scales. This is especially important given the current lack of 
consensus (cf.\ \Sec{SmallScale}) about how the small scale dynamo saturates, 
especially in the high Prandtl number systems like the ICM.
As discussed in connection with nonlinear small scale dynamos
(\Sec{Simulations_SSdynamo}), it is likely that, once the magnetic field
has reached equipartition field strength, the power spectrum should
decrease with increasing wavenumber.
If this is not seen, the simulation may still not be sufficiently well
resolved or it may not have run for long enough, or both.
It is encouraging that in simulations of the small scale
dynamo, one gets an RM probability distribution 
peaked at zero but with a significant width, 
$\sigma_{\rm RM} \sim 100-200$ rad m$^{-2}$ as observed, 
when scaled to cluster parameters \cite{Sub_Shuk_Haugen05}.
Overall, it appears plausible that cluster magnetism is
the result of compression, random shear and turbulent
amplification of seed fields from galaxies and AGNs;
but work on understanding the efficiency and details of all 
these processes is still in its infancy.

\section{Where do we stand and what next?}

Significant progress has been made in clarifying and understanding
mean field dynamo theory in the nonlinear regime.
Only a few years ago it was a completely open question whether or not
$\alpha$ is really catastrophically quenched, for example.
What is worse, simulations were not yet available that show
whether or not mean field dynamos can work at all when the
magnetic Reynolds number is large.
A lot has changed since then and we have now begun to develop and
confirm numerically the nonlinear mean field formalism for the case
where the medium is statistically homogeneous.

One of the currently most pressing problems is to develop and test numerically
a dynamical quenching theory that is valid also in the inhomogeneous case,
where the strength of the $\alpha$ effect varies in space and changes sign,
and to the case with open boundaries allowing helicity flux to escape.
This should be studied in a geometry that is close enough to the real
case of either the solar convection zone or to galactic discs.
Concerning solar-like conditions, some progress has been made using
forced turbulence in cartesian geometry with a shear profile that
resembles that of the sun at low latitudes.
However, driving the turbulence by convection and solving the equations
in spherical geometry has so far only led to mixed success in that the
simulations are dominated by small scale fields \cite{Brun2004,Brunetal04}.
In this connection it may help to lower the magnetic Prandtl number to
suppress small scale dynamo action, keeping however the magnetic
Reynolds number large enough to allow for large scale dynamo action.

Regarding fully periodic box simulations,
there is still a case for testing more thoroughly the
dependence of turbulent transport coefficients on the magnetic Reynolds
number even in cases with closed or periodic boundaries.
An issue that is not fully resolved is whether the turbulent magnetic
diffusion is quenched in a way that depends on the magnetic Reynolds
number.
One way of clarifying this issue is by considering oscillatory large
scale dynamos with shear \cite{BBS01}, which give the possibility of
measuring directly the turbulent magnetic diffusivity via the
cycle frequency \cite{BB02}.

As far as the solar dynamo is concerned, many problems remain to be solved.
The first and perhaps most severe one is simply the lack of a
reliable theory: the mean field theory as it stands at the moment
and as it has been used even in recent years does not reproduce certain
behavior that is known from simulations \cite{B01,Mininni05}.
Again, we have here in mind issues related to magnetic helicity
conservation, which need to be dealt with using dynamical quenching
theory.
Even from a more practical point of view, if one is able to argue
that in the sun
the magnetic helicity issue can be solved in such a way that conventional
theory remains applicable, there would still be many hurdles, as outlined above.
A popular model is the flux transport model
with a suitable meridional circulation profile.
Here the preferred modes have usually quadrupolar symmetry about
the equator rather than dipolar symmetry
\cite{Bonanno_etal02,DikpatiGilman01,NandyChoudhuri02}.
However, there are a number of other problems that have revived
the idea that the solar dynamo may be a distributed one \cite{B05}.

For the galactic dynamo, one still needs to identify 
mechanisms by which fluxes
of small scale helicity can preferentially leave the
system, so as to build up the regular field
to the observed values.
However, unlike the case of the sun where we have direct evidence for
losses of helical magnetic flux through the surface, galaxies lack
such direct evidence -- even though they allow direct inspection
all they way to the midplane.
Regarding the saturation strength of the large scale field,
closed box simulations suggest that, even in the kinematic
phase, and conserving helicity, a significant
regular field can probably be built up by the mean field dynamo.
However, for systems with boundaries the field
undergoes subsequent variations on the resistive time scale
when, for long periods, the field can be extremely weak.
Such systems still seem to require strong
helicity fluxes, for example mediated by shear, for being efficient
large scale dynamos \cite{BS05b}.
For galaxies at least, the mean field dynamo theory seems to 
reasonably explain the structure of the observed fields.
In the case of clusters of galaxies, the origin of turbulence
which may be needed for dynamo action is not yet settled.
The structure of the field will also depend on an improved
understanding of how the small scale dynamo saturates.

Another important question is what role does the small scale dynamo
play relative to the large scale dynamo.
In galaxies, where the magnetic Prandtl number is large, it has been
argued that the magnetic field is dominated by small scale fields.
This is an issue that clearly requires further clarification.
Especially whether the growth of the small scale field
subtly changes the lagrangian chaos properties of the turbulence.
In solar and stellar dynamos, on the other hand, the small scale
dynamo may not work at all any more, or the critical dynamo number
may be much larger than for unit magnetic Prandtl number.
Whatever the answer, it is likely that the magnetic Prandtl number
dependence of the critical magnetic Reynolds number
can soon be settled using dedicated high-resolution
simulations.

\begin{ack}
We thank Rainer Beck, Eric Blackman, Leonid Kitchatinov, David Moss,
Karl-Heinz R\"adler, G\"unther R\"udiger, and Anvar Shukurov 
for discussions, detailed comments
and corrections to earlier versions of this review.
We acknowledge the hospitality of IUCAA and Nordita during our mutual
visits during which much of this work has flourished.
We thank the Danish Center for Scientific Computing for granting time
on the Horseshoe cluster in Odense.
\end{ack}

\appendix
\newpage
\section*{\Large Appendix}

\section{Evolution of the correlation tensor of magnetic fluctuations}
\label{kazantsev}

In this section we present a derivation 
of the Kazantsev equation in configuration space
for the more general case of helical turbulence,
incorporating also ambipolar drift as a model nonlinearity \cite{subamb}.
These equations are used to discuss the small scale
dynamo in \Sec{KazantsevConfiguration}.
They also play an important role in
\Secs{UnifiedTreatment}{ClosureModel} where kinetic helicity
drives the generation of large scale fields which, in turn,
produce small scale helical fields that act such as to
saturate the dynamo.

The derivation of the governing equations involves straightforward
but rather tedious algebra and follows the discussion in \cite{subamb}. 
We therefore only outline the steps and the
approximations below leaving out most of the algebraic details.
To make the appendix fairly self contained we repeat some of the
equations which are also given in the main text. 
We start with the induction equation for
the magnetic field, including a nonlinear ambipolar diffusion term,
written as
\EQ
{\partial B_i\over \partial t} = R^x_{ipq} U_p B_q + \eta \nab^2 B_i,
\label{basic_new}
\EN
where we have defined for later convenience, the operator
\EQ
R^x_{ipq} = \epsilon_{ilm}\epsilon_{mpq}\left({\partial \over \partial x_l}\right).
\label{op}
\EN
Here $\UU = \meanUU + \vv + \vv_{\rm N}$, where
$\meanUU$ is the mean velocity field $\vv$ is the stochastic
velocity which may be helical, and which is $\delta$- correlated in time
and $\vv_{\rm N} = a(\JJ\times\BB)\times\BB$ is the
nonlinear ambipolar diffusion component, which is used
as a model nonlinearity.

Recall that we assume $\vv$ to be an isotropic, homogeneous,
gaussian random velocity field with zero mean,
and $\delta$-correlated in time. That is 
$\bra{v_i(\xx,t)v_j(\yy,s)} = T_{ij}(r) \delta (t-s)$, with 
$T_{ij}$ as defined in \eq{tijhel},
\EQ
T_{ij}(r) =
\left(\delta_{ij}-{r_i r_j \over r^2}\right)\,T_{N}(r)
+{r_i r_j \over r^2}\,T_{L}(r)
+\epsilon_{ijk} r_k\,F(r),
\EN
where $T_L$, $T_N$ and $F$ are the longitudinal, transverse
and helical parts of the correlations respectively.
The induction equation becomes
a stochastic equation. We split the magnetic field
into mean field $\meanBB = \bra{\BB}$ and a 
fluctuating field $\bb = \BB - \meanBB$. The equation for the mean
field, for the Kazantsev model velocity field is derived in 
\App{OtherApproaches}. Here we concentrate on the evolution
of the fluctuating field. We assume $\bb$ also to be 
a homogeneous, isotropic, random field, with an
equal time two point correlation $\bra{b_i(\xx,t)b_i(\yy,t)} =
M_{ij}(r,t)$, where
${\bf r}=  \xx -\yy$, $r = \vert {\bf r} \vert$ and
\EQ
M_{ij} = M_{\rm N} \left(\delta_{ij}-\frac{r_i r_j}{r^2}\right) +
M_{\rm L}\,\frac{r_i r_j}{r^2} + C \epsilon_{ijk} r_k.
\EN
Here $M_{\rm L}(r,t)$ and $M_{\rm N}(r,t)$ are the longitudinal
and transverse correlation functions for the magnetic field
while $C(r,t)$ represents the (current) helical part of the correlations.
The evolution of $M_{ij}(r,t)$ can be got from
\EQ
(\partial M_{ij}/\partial t) =
{\partial \over\partial t} (\bra{b_i({\xx},t)b_j({\yy},t)})
= {\partial \bra{B_iB_j} \over \partial t} - 
{\partial (\meanB_i \meanB_j) \over \partial t}.
\EN
The second term in the square brackets is easy to evaluate using the
equation for the mean field 
(see below in \Sec{DeltaCorrelatedVelocityFields}). The first term is
\EQ
{\partial \over \partial t}( B_i({\xx},t) B_j({\yy},t)) = 
B_i({\xx},t){\partial  B_j({\yy},t) \over \partial t} + 
{\partial  B_i({\xx},t) \over \partial t} B_j({\yy},t). 
\label{qeq}
\EN
Suppose we define the two-point product 
$B_i({\xx},t) B_j({\yy},t) = {\cal B}_{ij}(\xx,\yy,t)$
for notational convenience.
Substitute \eq{basic_new} into \eq{qeq} and 
and let the initial value of the two-point product be
$B_i(\xx,0)B_j(\yy,0) = {\cal B}_{ij}^0$. 
Then, at an infinitesimal time 
$\delta t$ later, this product is given by the formal 
integral solution:
\begin{equation}
{\cal B}_{ij} = {\cal B}_{ij}^0
+ \int_0^{\delta t} \dd t' 
[R^x_{ipq} U^x_p {\cal B}_{qj} + R^y_{jpq} U^y_p {\cal B}_{iq}]
+\delta t [\eta \nab_x^2{\cal B}_{ij} + \nab_y^2{\cal B}_{ij}].
\label{formal_bb}
\end{equation}
For clarity, the $\xx$ and $\yy$ and $t'$ dependencies of the fields has
been suppressed (except in $\nab^2$) in the integrand.
We write down an iterative solution to this equation
to various orders in $\delta t$.
To zeroth order, one ignores the integral and puts
${\cal B}_{ij}(\xx,\yy,t') = {\cal B}_{ij}^0$
To the next order
one substitutes ${\cal B}_{ij}^0$ for ${\cal B}_{ij}$  in \Eq{formal_bb},
to get a first order iteration ${\cal B}_{ij}^{(1)}$, and then 
${\cal B}_{ij}^{(1)}$ for ${\cal B}_{ij}$, to get ${\cal B}_{ij}^{(2)}$. 
The resulting equation
is then averaged to get the $\bra{{\cal B}_{ij}}$ and
the corresponding contribution from 
$\meanB_i({\xx},\delta t)\meanB_j({\yy},\delta t)$
subtracted to get the equation for 
$M_{ij}(\delta t) = \bra{b_i({\xx},\delta t)b_j({\yy},\delta t)}$.
The presence of the $\delta$-correlated $\uu$ implies
that one has to go up to second order iteration
to get $M_{ij}(\delta t)$ correct to linear order
in $\delta t$. Then dividing by $\delta t$ and taking
the limit of $\delta t \to 0$, we get
\EQA
{\partial M_{ij} \over \partial t}
&=& \bbra{\int \ R^y_{jpq}\left[v_p({\yy},t) \ R^x_{ilm}(v_l({\xx},s) 
[M_{mq}+ \meanB_m({\xx}) \meanB_q({\yy})])\right ] ds } \nonumber\\
&+&
\bbra{\int \ R^x_{ipq}\left[ v_p({\xx},t) \ R^y_{jlm}(v_l({\yy},s) 
[M_{qm} +  \meanB_q({\xx}) \meanB_m({\yy})])\right] ds }  
\nonumber\\ &+&
\bbra{\int \ R^y_{jpq}\left[ v_p({\yy},t) \ R^y_{qlm}(v_l({\yy},s) 
M_{im})\right] ds }  \nonumber\\ &+&
\bbra{\int \ R^x_{ipq}\left[v_p({\xx},t) \ R^x_{qlm}(v_l({\xx},s) 
M_{mj})\right]  ds } \nonumber\\ 
&+& \eta [ \nabla_y^2M_{ij} + \nabla_x^2 M_{ij}] + 
R^y_{jpq}\left(\meanU_p({\yy}) M_{iq} \right) +
R^x_{ipq}\left(\meanU_p({\xx}) M_{qj} \right)
\nonumber\\ 
&+& 
R^y_{jpq}\left(\bbra{v_{Np}({\yy}) b_i({\xx})B_q({\yy})} \right) +
R^x_{ipq}\left(\bbra{v_{Np}({\xx}) B_q({\xx}) b_j({\yy})} \right). 
\label{meq}
\ENA

The first two terms on the RHS of  Eq.\ (\ref{meq})
represent the effect of velocity correlations on
the magnetic fluctuations, $M_{ij}$, and the mean field, $\meanBB$.
The next two terms give the `turbulent transport'
of the magnetic fluctuations by the turbulent velocity, the 5th and
6th terms the 'microscopic diffusion'. The 7th and 8th terms
the transport of the magnetic fluctuations by the mean velocity.
The last two nonlinear terms give the effects of the backreaction
due to ambipolar drift on the magnetic fluctuations.

For the discussions of the small scale dynamo in \Sec{SmallScale},
the unified treatment of the large scale dynamo in \Sec{UnifiedTreatment}
or the toy closure model in \Sec{ClosureModel} we do not
keep the mean field terms. (The coupling to the mean field
can be important in discussions of helicity flux.)
This also means that we can
continue to treat the statistical properties
of the magnetic fluctuations as being homogeneous and
isotropic, and use $M_{ij}({\xx}, {\yy}, t) = M_{ij}(r,t)$.

All the terms in the above equation, can be further simplified by using the
properties of the magnetic and velocity correlation functions.
In order to obtain equations for $M_{\rm L}$ and $C$, we
multiply Eq.\ (\ref{meq}) by $r^ir^j/r^2$ and $\epsilon_{ijk}r^k$
and use the identities
\begin{equation}
M_{\rm L}(r,t)= M_{ij}(r^ir^j / r^2), \quad
C(r,t) =\half M_{ij} \epsilon_{ijk}r^k/r^2.
\end{equation}
We consider some steps in simplifying the first two terms.
The first term in Eq.~(\ref{meq} ) is given by
\EQ
\bbra{\int \ R^y_{jpq}\left(v_p({\yy},t) R^x_{ilm}(v_l({\xx},s) 
M_{mq})\right ) \dd s} 
= -\epsilon_{itu}\epsilon_{ulm}\epsilon_{jrs}\epsilon_{spq}
\left [ T_{lp} M_{mq}\right]_{,rt}.
\label{appone}
\EN
For examining the evolution of $M_{\rm L}$ one needs to multiply the
above equation by $r^i r^j/r^2$. We can simplify the resulting equation
by using the identity
\EQ
r^i r^j 
{\partial^2 A \over \partial r^r \partial r^t }
= {\partial^2 (A r^i r^j )\over \partial r^r \partial r^t }
-\delta_{jt}r^i {\partial A \over \partial r^r }
- \delta_{ir}r^j {\partial A \over \partial r^t }
-\delta_{jt} \delta_{ir} A,
\EN
where $A = T^{lp} M_{mq}$. Then using
$\epsilon_{itu}\epsilon_{ulm} 
= \delta_{il}\delta_{tm} - \delta_{im}\delta_{tl}$, and the
definition of $T_{L}, T_{N} $ and $F$, straightforward algebra
gives the contribution of the first term to
$(\partial M_{\rm L} / \partial t)$
\EQ
\dot{M_{\rm L}}\vert_{1{\rm st}} = 
-{1\over r^4}{\partial \over \partial r}
\left(r^4 T_{LL}{\partial M_{\rm L} \over \partial r}\right)
+ {G \over 2} M_{\rm L} + 4 F C.
\EN
The second term of \eq{meq} gives an identical contribution.

To derive the evolution of $H$ due to these terms
multiply \eq{appone} by $\epsilon_{ijk}r^k$ .
Using the fact that the turbulent velocity and
small scale field have vanishing divergence, we
have $M_{ij,j} = 0$ and $T_{ij,j}=0$. This allows one to
simplify the contribution from the first term to 
$(\partial C / \partial t)$
\EQ
\dot{C}\vert_{1{\rm st}} 
 =-{\epsilon_{ijk} r^k \over 2 r^2} \left( T_{ij,tr}M_{tr} +
 T_{tr}M_{ij,tr} - T_{ir,t}M_{tj,r} - T_{tj,r}M_{ir,t} \right).
 \label{beg}
\EN
The first two terms on the RHS of Eq. (\ref{beg})
can be further simplified by noting that $\epsilon_{ijk}T_{ij} = 2Fr^k$ and
$\epsilon_{ijk}M_{ij} = 2Cr^k$ . We have then
\EQA
-{ \epsilon_{ijk} r^k \over 2 r^2} \left( T_{ij,tr}M_{tr} +
T_{tr}M_{ij,tr}\right) = -\Big( T_{L}C'' &+& T_{L}'
C'+ {4T_{L}C' \over r} 
+ M_{L}F'' \nonumber \\
&+&M_{L}' F'+ {4M_{L}F' \over r} \Big).
\label{ontw}
\ENA
Here prime denotes a derivative with respect to $r$.
To evaluate the contribution of the last two terms on the 
RHS of \eq{beg}, it is convenient to split up the tensors 
$M_{ij}$ and $T_{ij}$ into symmetric and antisymmetric parts 
(under the interchange of $(ij)$ ). We put a superscript 
$S$ on the symmetric part and $A$ on the antisymmetric part. 
Then we can write after some algebra
\EQA
&&{\epsilon_{ijk} r^k \over 2 r^2} \left(T_{ir,t}M_{tj,r} + 
T_{tj,r}M_{ir,t} \right] = {\epsilon_{ijk} r^k \over  r^2} 
\left[T_{ir,t}^{\rm S} M_{tj,r}^{\rm A} + T_{ir,t}^{\rm A} M_{tj,r}^{\rm S} \right)
\nonumber\\ && =
-\left( CT_{L}'' + F M_{\rm L}'' + T_{L}' C' + M_{L}' F'
+{4CT_{L}' \over r} + {4FM_{L}' \over r} \right).
\label{thfo}
\ENA
Adding the contributions from \eq{ontw} and \eq{thfo} gives
\EQ
\dot{C}\vert_{1st} 
=-{1\over r^4}{\partial \over \partial r}
\left[r^4 {\partial \over \partial r}(T_{LL}C + FM_{\rm L})\right]
\label{hfir}
\EN
The second term of \eq{meq} gives an identical contribution.
The third and fourth terms add to give a contribution
\EQ
(3{\rm rd}+4{\rm th}) 
= 4F(0)\epsilon_{jqm}(\partial M_{im}/\partial r^q) + 2T_{L}(0) 
\nabla^2 M_{ij}
\EN
to the RHS of \eq{meq}, hence justifying their being called  
`turbulent transport' of $M_{ij}$ (compare this
to the microscopic diffusion term $2\eta \nabla^2 M_{ij}$).

The last two nonlinear terms give the effects of the backreaction
due to ambipolar drift on the magnetic fluctuations.
They involve $4-$th order correlations of $\bb$.
In evaluating this term, we
make the gaussian closure approximation that the fourth order
moment of the fluctuating field can be written as a
product of 2nd moments. In this case the
nonlinear terms add to give a contribution
\EQ
\mbox{nonlinear terms} = -8aC(0,t)\epsilon_{jqm}
(\partial M_{im}/\partial r^q) + 4aM_{\rm L}(0,t) 
\nabla^2 M_{ij}
\EN
to the RHS of \eq{meq}. The
gaussian assumption of the magnetic correlations results in
the nonlinearity of this term appearing as a nonlinearity in the
coefficient, rather than the correlation function itself.
Gathering together all the terms,
we get for the coupled evolution equations for $M_{\rm L}$ and $C$ :
\EQ
{\partial M_{\rm L} \over \partial t} = {2\over r^4}{\partial \over \partial r}
\left(r^4 \eta_{\rm N} {\partial M_{\rm L} \over \partial r}\right) + 
G M_{\rm L} + 4\alpha_{\rm N} C,
\label{mleqf}
\EN
\EQ
{\partial C\over \partial t} = 
{1\over r^4}{\partial \over \partial r}
\left[r^4  {\partial \over \partial r}(2\eta_{\rm N} C 
- \alpha_{\rm N} M_{\rm L})\right].
\label{mheq}
\EN
We can also write \eq{mheq} in terms of the magnetic helicity 
correlation function $H(r,t)$, which is related to the
current helicity correlation $C(r,t)$ by
\EQ
C= - {1\over r^4}{\partial \over \partial r}
\left(r^4  {\partial H \over \partial r} \right),
\EN
\EQ
{\partial H\over \partial t}
= - 2\eta_{\rm N} C + \alpha_{\rm N} M_{\rm L}.
\label{heleq}
\EN
Here we have also defined
\EQA
&&\eta_{\rm N} = \eta + T_{LL}(0) - T_{LL}(r) + 2aM_{\rm L}(0,t),\nonumber\\ 
&&\alpha_{\rm N} = -[2F(0) - 2F(r)] + 4aC(0,t),\nonumber\\
&&G = -2(T_{\rm L}^{\prime\prime} + 4 T_{\rm L}^{\prime}/r).
\ENA
These equations generalize the Kazantsev equation to the helical
case and also include a toy nonlinearity in the form of
ambipolar diffusion. If we take the limit $r\to 0$ in \Eq{heleq},
we get $\dot{H}(0,t) = - 2\eta C(0,t)$, which is exactly the
equation for conservation (evolution) of magnetic helicity.
So our nonlinear closure model incorporates also the important
constraint provided by helicity conservation.
This set of equations is used
to discuss the nonhelical small scale dynamo (\Sec{SmallScale}),
a unified treatment of small and large scale dynamos
(\Sec{UnifiedTreatment}) and also the helicity constraint in
this toy model (\Sec{ClosureModel}).

\section{Other approaches to calculating $\alpha$ and $\eta_{\rm t}$}
\label{OtherApproaches}

In this section we present alternative approaches to calculating
turbulent transport coefficients such as $\alpha$ and $\eta_{\rm t}$.
These methods are more specialized compared to the standard approaches
presented in \Secs{CalculationTurbulentTransport}{SimulationsTransport},
but some of them are more rigorous, allowing additional insight into
the viability and fragility of $\alpha$ effect and turbulent diffusion.

\subsection{The lagrangian approach to the weak diffusion limit}
\label{LagrangianDiffusion}

In the limit of large $R{\rm m} \gg 1$, one may also think of
neglecting magnetic diffusion completely. The time evolution of
the magnetic field can then be solved exactly using the Cauchy solution,
\EQ
B_i(\xx,t)={{\sf G}_{ij}(\xx_0,t)\over\det\GGGG}
\,B_{0j}(\xx_0),
\EN
where ${\sf G}_{ij}=\partial x_i/\partial x_{0j}$ is the lagrangian
displacement matrix, $\BB_0(\xx_0)=\BB(\xx,0)$ is the initial condition, and
\EQ
\xx(\xx_0,t)=\xx_0+\int_0^t\uu^{\rm L}\left(\xx_0,t'\right)\dd t'
\label{lagrange}
\EN
is the position of an advected test particle whose original position
was at $\xx_0$. We have also defined the lagrangian velocity
$\uu^{\rm L}(\xx_0,t)=\UU(\xx(\xx_0,t))$.
Using the Cauchy solution, Moffatt \cite{Mof78} showed that 
\EQ
\alpha(t)=-\onethird\int_0^t\overline{\uu^{\rm L}(\xx_0,t)
\cdot\oo^{\rm L}(\xx_0,t')}\;\dd t',
\label{FOSA_alpha_lagr}
\EN
where $\oo^{\rm L}=\nab^{\rm L}\times\uu^{\rm L}$
is the vorticity of the lagrangian velocity fluctuation
with respect to $\xx_0$. So $\alpha(t) =0$ at the initial
time $t=0$, but the expectation is that for
times much longer than the correlation time of the turbulence, 
$t \gg \tau$, $\alpha(t)$ settles to a constant value.
One could then take the limit $t \to \infty$ in the
above integral.\footnote{The convergence of the integral to
a constant when $t\to \infty$ is not guaranteed because the above
integral has derivatives of the form
$\partial u_i/\partial x_{0j} = (\partial u_i/\partial x_m)
(\partial x_m/\partial x_{0j})$. Although $\partial u_i/\partial x_m$
is statistically stationary in time, $\partial x_m/\partial x_{0j}$
is in general not, since particles initially separated by
some $\delta \xx_{0}$, tend to wander further and further
apart in a random flow, and $\vert\delta\xx\vert \propto
\vert\delta \xx_0\vert t^{1/2}$. So secularly growing
terms may in principle contribute to the integral
determining $\alpha(t)$ making the limit $t \to \infty$
meaningless \cite{Mof78}.
}

Deriving an expression for the $\eta_{\rm t}$ coefficient is
more complicated. A naive expectation is that one will have
\EQ
\eta_{\rm t}(t)=\onethird\int_0^t\overline{\uu^{\rm L}(\xx_0,t)
\cdot\uu^{\rm L}(\xx_0,t')}\;\dd t',
\label{FOSA_eta_lagr}
\EN
analogous to the case of the effective turbulent
diffusivity of a scalar field.
One does get this term, but there are additional terms
for the magnetic field, derived for example in
Moffatt's book \cite{Mof78}. For these terms convergence to
finite values at large times is even more doubtful.
Simulations by Kraichnan \cite{Kraichnan76} suggested that
$\alpha(t)$ and $\eta_{\rm t}(t)$ converge to finite values of
order $u$ and $uL$ respectively, as $t\to\infty$,
for statistically isotropic velocity fields with gaussian
statistics.
However, numerical simulations \cite{Drummond81} using a frozen velocity
field suggest that in the limit
of large magnetic Reynolds numbers $\alpha$ tends to zero.

\subsection{Delta-correlated velocity fields}
\label{DeltaCorrelatedVelocityFields}

One of the few situations for which the kinematic mean field dynamo 
equations can be derived exactly is
for a random flow that is $\delta$-correlated in time,
as introduced by Kazantsev \cite{Kaz68}. Such a flow was also used by 
Kraichnan \cite{kraichnan} to discuss passive scalar evolution.
Such flows are of course artificial and not a solution of the momentum
equation, but they serve as an excellent example where the mathematics
can be treated exactly.

To discuss mean field dynamo action, we have to add
a helical piece to the correlation function of the
stochastic velocity field $\vv$ driving the flow.
So now we adopt $\UU=\meanUU + \vv$ in the induction equation
where, as before $\bra{v_i(\xx,t)v_j(\yy,s)} = T_{ij}(r) \delta (t-s)$, with
$T_{ij}$ as defined in \eq{tijhel}.
At $r=0$, we have 
\EQ
-2F(0) = -\onethird\int_0^t\bra{\vv(t)
\cdot(\nab\times\vv(t'))}\;\dd t'
\quad\left[\approx -\onethird\tau\overline{\vv\cdot(\nab\times\vv)}\right],
\EN
\EQ
T_{\rm L}(0) = \onethird\int_0^t\bra{\vv(t)
\cdot\vv(t')}\;\dd t'
\quad\left[\approx \onethird\tau\overline{\vv^2}\right],
\EN
where the last expressions in parenthesis would apply if we had assumed 
a small but finite correlation time $\tau$.

The induction equation is a stochastic equation
and we would like to convert it into an equation for
the mean magnetic field $\meanBB$
(see also Zeldovich et al.\ 1983, Chapter~8 \cite{ZRS83}).
Let the magnetic field at an initial time, say $t=0$, be $\BB(\xx,0)$.
Then, at an infinitesimal time $\delta t$ later, the field
is given by the formal integral solution:
\begin{equation}
\BB(\delta t) = \BB(0) +
\int_0^{\delta t} dt^{\prime} 
\left\{\nab\times [\UU(t)\times \BB(t^{\prime})]
- \eta \nab\times \BB(t^{\prime}) \right\}.
\label{formal}
\end{equation}
For clarity, the common $\xx$ dependence of $\UU$ and $\BB$,
has not been explicitly displayed above.
We write down an iterative solution to this equation
to various orders in $\delta t$.
To zeroth order, one ignores the integral and puts
$\BB^{(0)}({\xx},\delta t) = \BB({\xx},0)$. To the next order
one substitutes $\BB^0$ for $\BB$  in \Eq{formal},
to get a first order iteration $\BB^{(1)}$, and then $\BB^{(1)}$
for $\BB$, to get $\BB^{(2)}$. The resulting equation
is then averaged to get the $\meanBB({\xx},\delta t)$.
The presence of the $\delta$-correlated $\vv$ implies
that one has to go up to second order iteration
to get $\meanBB(\xx,\delta t)$ correct to linear order
in $\delta t$. This procedure yields
\begin{eqnarray}
\meanBB(\delta t) &=& \meanBB(0) - 
\delta t [\nab\times 
(\meanUU \times \meanBB(0))
- \eta \nab\times\meanBB(0)] \nonumber\\ &+&
\int_0^{\delta t}dt^{\prime} \int_0^{t^{\prime}} dt^{\prime\prime} 
\nab\times\left\{\overline{\vv(t^{\prime})\times
[\nab\times(\vv(t^{\prime\prime})\times 
\BB(0))]}\right\},
\label{sol}
\end{eqnarray}
where the overbar denotes averaging over an ensemble of the stochastic velocity
field $\vv$, as before. On using the fact that
$\vv$ at time $t$ is not correlated with the initial
magnetic field  $\BB({\xx},0)$
and taking the limit $\delta t \to 0$
we get after some straightforward algebra,
\begin{equation}
{\partial \meanBB\over \partial t} =
\nab\times \left[ \meanUU \times \meanBB 
+ \alpha \meanBB -(\eta  + \eta_{\rm t}) \nab
\times \meanBB\right]. 
\label{mean}
\end{equation}
The effect of the turbulent velocity is again to introduce the standard extra
terms representing the $\alpha$ effect with
$\alpha = -2F(0)$ and an extra turbulent contribution to the diffusion
$\eta_{\rm t} = T_{\rm L}(0)$. If one allows for weakly inhomogeneous
turbulence, one could also discover a turbulent diamagnetic effect due to 
the gradient of $\eta_{\rm t}$, which expels magnetic fields from
regions of strong turbulence. 

\subsubsection{Transport coefficients from random waves and individual blobs}
\label{RandomWavesIndividualBlobs}

We mention two additional approaches that have been important in understanding
the origin and nature of the $\alpha$ effect.
One is based on the superposition of random waves that are affected and modified
through the presence of rotation, stratification, and magnetic fields
\cite{Brag64,Mof70,Mof72,Schmitt87,Walder_etal80}, and the other is based on
the detailed analysis of individual convection cells \cite{Stix83} or blobs
from supernova explosions \cite{Ferriere92}.
The latter approach has to an $\alpha$ tensor of the form
\EQ
\alpha_{ij}=\pmatrix{
\alpha_R & -V_{\rm esc} & 0\cr
V_{\rm esc} & \alpha_\phi & 0 \cr
0 & 0 & \alpha_Z }
\EN
in cylindrical $(R,\phi,Z)$ coordinates.
The effect of individual supernova explosions is slightly different from the
effect of the so-called superbubble where one explosion has triggered several
others nearby.
The latter leads to structures more symmetric about the midplane causing
$\alpha_Z$ to be negligibly small.
For individual supernova bubbles, on the other hand, the sign of $\alpha_Z$
is found to be opposite to the sign of $\alpha_\phi$.
There is also a vertical pumping effect {\it away} from the midplane,
corresponding to the antisymmetric components of the $\alpha$ tensor,
\EQ
\gamma_i=-\half\epsilon_{ijk}\alpha_{jk}=\delta_{iZ} V_{\rm esc}.
\EN
The lack of downward pumping is possibly an artifact of neglecting the
return flow.
The motivation for neglecting the return flow is the somewhat unjustified
assumption that the magnetic field will have reconnected to their original
position by the time the return flows commences.

\section{Derivation of the Zeldovich relation}
\label{ZeldovichRel}

In this section we show that in two dimensions we have the relation
$\bra{\bb^2}/\bra{\meanBB^2}=R_{\rm m}$, where $\meanBB=\BB_0=(B_0,0,0)$
is an applied mean field.
This relation was used in \Sec{helicity_aquenching} in connection with
heuristic arguments about the nonlinear $\alpha$ effect.

Consider the evolution of the small scale magnetic vector potential
$\aaa=(0,0,a)$, where $\bb=\nab\times\aaa$, so
\EQ
{\partial a\over\partial t}+\uu\cdot\nab a+u_yB_0=\eta\nabla^2a.
\EN
Multiplication with $a$ and volume averaging yield
\EQ
{\partial\over\partial t}\bra{\half a^2}+\bra{\uu\cdot\nab(\half a^2)}
+\bra{au_y}B_0=-\eta\bra{\bb^2}.
\EN
Using stationarity, $\partial/\partial t=0$, incompressibility 
[which implies $\bra{\uu\cdot\nab(\half a^2)} = \bra{\nab\cdot(\uu a^2/2)}$],
and periodic or closed boundaries, so $\bra{\nab\cdot(\uu a^2/2)}=0$,
as well as a Fickian diffusion law for the flux,
\EQ
\bra{au_y}=-\eta_{\rm t}\nabla_y\meanA=-\eta_{\rm t}B_0,
\EN
where $(0,0,\meanA(y))$ is the corresponding vector potential
for $\BB_0$, we have
\EQ
\eta_{\rm t}B_0^2=\eta\bra{\bb^2}.
\EN
Finally, using $R_{\rm m}=\eta_{\rm t}/\eta$, we have
$\bra{\bb^2}/\bra{\meanBB^2}=R_{\rm m}$.

\section{A heuristic treatment for the current helicity term}
\label{qnmod}

In this section we present a heuristic treatment for the occurrence
of the $\overline{\jj\cdot\bb}$ correction to the $\alpha$ effect.
A more rigorous and perhaps more convincing derivation is given in
\Sec{tauApproxNonlin}.

\subsection{Taking nonlinearity into account}

Originally, when the magnetic field is weak, the
velocity is a sum of mean velocity $\meanUU$ and 
a turbulent velocity $\uu = \uu^{(0)}$ (which may be helical).
The velocity field $\uu^{(0)}$ can be thought of as obeying the
momentum (Navier-Stokes) equation without the Lorentz force. 
Then suppose we assume an ansatz that in the nonlinear 
regime the Lorentz force induces an additional nonlinear 
velocity component $\uu_{\rm N}$, 
that is, $\UU = \meanUU + \uu^{(0)} + \uu_{\rm N}$,
where $\nab\cdot\uu_{\rm N}=0$ and  
\EQ
\rho \dot{\uu}_{\rm N} =  \mu_0^{-1} ( \meanBB\cdot\nab\bb +
\bb\cdot\nab\meanBB ) - \nab p' + O(uu,bb,\nu u_{\rm N}),
\EN
where the dots denote partial time differentiation, and
$O(uu,bb,\nu u_{\rm N})$ indicates all the neglected terms, those
nonlinear in $\bb$ and $\uu = \uu^{(0)} + \uu_{\rm N}$,
and the viscous dissipation. Also, $p'$ is the perturbed
pressure including the magnetic contribution.
PFL assumed that the mean field, $\meanBB$, was strong enough that
the nonlinear and viscous terms could be neglected. 
Zeldovich et al.\ \cite{ZRS83} argued that when multiplied by $\bb$ and
averaged, the resulting triple correlations could be replaced
by double correlations, fairly close in spirit to the
minimal $\tau$ approximation; but they did not actually
perform a calculation akin to the calculation as done below.
Gruzinov and Diamond \cite{GD95,GD96} also neglected this term, calling
this a quasi-linear dynamo. This `quasilinear' treatment
is somewhat approximate; nevertheless it helps in
heuristically understanding the result of the EDQNM closure model,
in an analytically tractable fashion. 

Due to the addition of $\uu_{\rm N}$, the turbulent EMF now becomes
$\meanemf=\overline{\uu^{(0)} \times \bb } +
\overline{\uu_{\rm N} \times \bb } $,
where the correction to the turbulent EMF is 
\EQ
\meanemf_{\rm N}(t) =
\int^t \overline{
(\meanBB\cdot\nab\bb +
\bb\cdot\nab\meanBB - \nab p' )
\times \bb(t) }\,\dd t',
\label{heuristic_emf}
\EN
where (as usual) $\mu_0=\rho_0=1$ has been assumed.
One again assumes that the small scale magnetic
field $\bb(t)$ has a short correlation time, say 
$\tau_b$, and that $\tau_b$ is small
enough that the time integration
can be replaced by a simple multiplication by $\tau_b$.
We can calculate $\meanemf_{\rm N}$ either in coordinate space
or in $k$-space. The calculation, following
\cite{sub03}, is given in \Sec{closure_model}.
To the lowest order one gets
\EQ
\meanemf_{\rm N} = \onethird\tau_b \
\overline{\jj\cdot\bb} \ \meanBB,
\label{jbLowestOrder}
\EN
that is a correction to $\meanemf$ akin to the magnetic contribution
found by PFL \cite{PFL76}. Further, one verifies the result of PFL
that to the lowest order, the turbulent diffusion of the 
large scale field is not affected by nonlinear effects of the Lorentz force.
However, if one goes to higher order in the derivatives of $\meanBB$, one gets
additional hyperdiffusion of the mean field
and higher order additions to the $\alpha$ effect \cite{sub03}.

The above heuristic treatment is not 
very satisfactory, since it neglects 
possibly important nonlinear terms in the momentum equation.
At the same time the EDQNM treatment has the 
limitation that one assumes the random $\uu$ and $\bb$
fields to be homogeneous and isotropic. In the main text, we have
therefore focused on MTA, since it allows one to remove some of 
these limitations.

\subsection{Calculation of the quasilinear correction $\meanemf_{\rm N}$}
\label{closure_model}

We calculate $\meanemf_{\rm N}$ here in coordinate space
representation. We can eliminate the pressure term in
$\uu_{\rm N}$ using the incompressibility condition.
Defining a vector 
$\FF = a [\meanBB\cdot\nab\bb +
\bb\cdot\nab\meanBB]$, with $a = \tau/(\mu_0\rho_0)$,
one then gets
\begin{equation}
\meanemf_{\rm N} = \langle \FF\times\bb\rangle -
\langle [\nab(\nabla^{-2}\nab\cdot\FF)]
\times\bb\rangle ,
\label{varepn}
\end{equation}
where ${\nabla}^{-2}$ is the integral operator
which is the inverse of the Laplacian, written
in this way for ease of notation. We will write down
this integral explicitly below, using $-(4\pi r)^{-1}$ to be
the Greens function of the Laplacian. We see that
$\meanemf_{\rm N}$ has a local and nonlocal contributions.

To calculate these, we assume the small scale field to be
statistically isotropic and homogeneous,
with a two-point correlation function  $\bra{b_i(\xx,t)b_i(\yy,t)} =
M_{ij}(r,t)$, given by \eq{mcor}, 
\EQ
M_{ij} = M_{\rm N} \left(\delta_{ij}-\frac{r_i r_j}{r^2}\right) +
M_{\rm L}\,\frac{r_i r_j}{r^2} + C \epsilon_{ijk} r_k.
\EN
Recall that $M_{\rm L}(r,t)$ and $M_{\rm N}(r,t)$ are the longitudinal
and transverse correlation functions for the magnetic field
while $C(r,t)$ represents the (current) helical part of the correlations.
Since $\nab\cdot\bb=0$,
\EQ
M_{\rm N} = {1\over2r}{\partial\over\partial r} (r^2 M_{\rm L}).
\EN
We also will need the magnetic
helicity correlation, $H(r,t)$ which is given
by $C = -(H^{\prime\prime}
+ 4 H^{\prime}/r)$, where a prime ${\prime}$ denotes
derivative with respect to $r$.
In terms of ${\bf b}$, we have 
\EQ
M_{\rm L}(0,t) = \onethird\langle\bb^2\rangle, \quad
2C(0,t) = \onethird\langle\jj\cdot\bb\rangle,\quad
2H(0,t) = \onethird\langle \aaa\cdot\bb\rangle,
\EN
where $\bb=\nab\times\aaa$ and $\jj=\nab\times\bb$.
The local contribution to $\meanemf_{\rm N}$ is easily
evaluated,
\begin{equation}
\meanemf_{\rm N}^L \equiv \langle\FF\times\bb\rangle
= -a M_{\rm L}(0,t)\meanJJ
+ 2 a C(0,t)\meanBB,
\label{localeps}
\end{equation}
where $\meanJJ=\nab\times\meanBB$.

At this stage (before adding the nonlocal contribution)
there is an important correction to the $\alpha$ effect,
with $\alpha = \alpha_K + \alpha_{\rm M}$, where
\EQ
\alpha_{\rm M} = 2 a C(0,t) = \onethird\tau\bra{\jj\cdot\bb}.
\EN
(We again set $\mu_0=1$ and $\rho_0=1$ above and in what follows.)
There is also a nonlinear addition to the diffusion of the
mean field (the $-a M_{\rm L}(0,t)\nabla\times\meanBB$ term),
which as we see gets canceled by the nonlocal contribution!

Let us now evaluate the nonlocal contribution.
After some algebraic simplification, this is
explicitly given by the integral
\begin{eqnarray}
(\meanemf_{\rm N}^{\rm NL})_i({\xx},t) &\equiv&
- (\langle [\nab(\nabla^{-2} \nab\cdot\FF)]
\times {\bf b} \rangle )_i \nonumber\\
&=& 2 \epsilon_{ijk} \int 
\frac{\partial M_{mk}({\bf r},t)}{\partial r^l}
\frac{\partial \overline{B}_l ({\yy},t)}{\partial y^m}\,
\frac{r_j}{r^3}\,\frac{\dd^3r}{4\pi} ,
\label{nleps}
\end{eqnarray}
where ${\yy} = {\bf r} + {\xx}$.
Note that the mean field $\meanBB$ will in general
vary on scales $R$ much larger than the
correlation length $l$ of the small scale field. We can then use
the two-scale approach to simplify the integral in
Eq.\ (\ref{nleps}). Specifically, assuming that $(l/R) < 1$,
or that the variation of the mean field derivative in
Eq.\ (\ref{nleps}), over $l$ is small, we expand
$\partial \overline{B}_l ({\yy},t)/ \partial y^m$,
in powers of ${\bf r}$, about ${\bf r} = 0$,
\begin{equation}
\frac{\partial \overline{B}_l }{\partial y^m}
= \frac{\partial \overline{B}_l}{\partial x^m}
+ r \nunit_p \frac{\partial^2 \overline{B}_l}{\partial x^m \partial x^p}
+ \frac{ r^2 \nunit_p \nunit_q}{2}
\frac{\partial^3 \overline{B}_l}{\partial x^m \partial x^p \partial x^q }
+ \ldots,
\label{derB}
\end{equation}
where we have defined the unit vector $\nunit_i = r_i/r$ (we will
soon see why we have kept terms beyond the first term
in the expansion).
Simplifying the derivative
$\partial M_{mk}({\bf r},t)/ \partial r^l$
using Eq.\ (\ref{mcor}) and noting that
$\epsilon_{ijk}r_jr_k = 0$, we get
\begin{eqnarray}
r_j \epsilon_{ijk} \frac{\partial M_{mk}}{\partial r^l}
&=& r_j \epsilon_{ijk}\Big[r^{-1}(M_{\rm L} - M_{\rm N})\nunit_m\delta_{kl}
+ M_{\rm N}^{\prime} \nunit_l \delta_{mk} \nonumber\\
&+&  C \epsilon_{mkl} +rC^{\prime} \nunit_{\rm f} \nunit_l \epsilon_{mkf}\Big].
\label{derM}
\end{eqnarray}
We substitute (\ref{derB}) and (\ref{derM}) into
(\ref{nleps}), use
\EQ
\int \nunit_i\nunit_j {\dd\Omega\over4\pi} = \onethird\delta_{ij},
\quad 
\int  \nunit_i\nunit_j\nunit_k\nunit_l {\dd\Omega\over4\pi} =
{\textstyle{1\over15}}[\delta_{ij}\delta_{kl}+\delta_{ik}\delta_{jl}
+\delta_{il}\delta_{jk}],
\EN
to do the angular integrals in (\ref{nleps}), to get
\EQ
\meanemf_{\rm N}^{\rm NL}=a M_{\rm L}(0,t) \meanJJ
+ \frac{6a}{ 5} H(0,t) \nabla^2 \meanBB
+\frac{2a}{5}\!\left[\int_0^\infty M_{\rm L}(r,t) r\dd r \right]
\!\nabla^2\meanJJ.
\label{epsnlfin}
\EN

The net nonlinear contribution to the turbulent EMF is
$\meanemf_{\rm N} = \meanemf_{\rm N}^L
+ \meanemf_{\rm N}^{\rm NL}$, got by adding
Eq.~(\ref{localeps}) and Eq.\ (\ref{epsnlfin}).
We see first that the nonlinear diffusion term proportional to
$\nab\times\meanB$
has the same magnitude but opposite
signs in the local [Eq.~(\ref{localeps})] and nonlocal
[Eq.~(\ref{epsnlfin})] EMF's and so exactly cancels in the
net $\meanemf_{\rm N}$.
This is the often quoted result \cite{GD,PFL76,GD95,GD96}
that the turbulent diffusion is not renormalized by nonlinear
additions, in the quasi-linear approximation. However
this does not mean that there is no nonlinear
correction to the diffusion of the mean field.
Whenever the first term in an expansion is exactly zero
it is necessary to go to higher order terms. This is
what we have done and one finds that $\meanemf_{\rm N}$
has an additional hyperdiffusion $\meanemf_{\rm HD} =
\eta_{\rm HD}\nabla^2(\nab\times\meanBB)=\eta_{\rm HD}\nabla^2\meanJJ$, where
\begin{equation}
\eta_{\rm HD} = \frac{2a}{5} \int_0^\infty dr \ rM_{\rm L}(r,t).
\label{hd}
\end{equation}
Taking the curl of $\meanemf_{\rm N}$, the nonlinear
addition to the mean field dynamo equation then becomes,
\begin{equation}
\nab\times\meanemf_{\rm N} =
(\alpha_{\rm M} + h_{\rm M} \nabla^2) \nab\times\meanBB
-\eta_{\rm HD} \nabla^4 \meanBB.
\label{epsnfin}
\end{equation}
Here $\alpha_{\rm M} = {1\over3}a\bra{\jj\cdot\bb}$
is the standard nonlinear correction to the $\alpha$ effect \cite{GD,PFL76},
and $h_{\rm M} = {1\over5}a\bra{\aaa\cdot\bb}$ is an additional higher
order nonlinear helical correction derived here.

\section{Derivation of \Eq{chi_mfin}}
\label{AppSimplifyChiM}

Now turn to the simplification of the $\dot{\chi}_{jk}^M$
term in \Sec{radler_calc}. We again define wavevectors, 
$\kk_1 = \kk+\KK/2 -\kk'$ and $\kk_2=-\kk+\KK/2$,
and transform to new variables, ($\kk',\KK'$), where
$\KK'= \kk_1+\kk_2 = \KK - \kk'$. Note that
since all Fourier variables will be non-zero
only for small $\mod{\KK}$ and $\mod{\kk'}$, they
will also only be non-zero for small $|\KK'|$.
The equation \eq{chi_mag} becomes
\EQA
{\partial\chi_{jk}^M \over \partial t} &=&
\int  \dd^3K' \ \dd^3k' \e^{\ii(\KK'+\kk')\cdot\RR} \ 
P_{js}(\kk +\half\KK' +\half\kk') \nonumber \\
&&  
[\ii(-k_l +\half K'_l -\half k'_l) \ \meanBft_l(\kk') 
\ {\cal M}_{sk}(\kk - \half\kk';\KK')
\nonumber \\
&+& (\ii k'_l) \ \meanBft_s(\kk')
\ {\cal M}_{lk}(\kk - \half\kk';\KK')],
\label{chi_mag2}
\ENA
where, for notational convenience, we have defined
\EQ
{\cal M}_{sk}(\kk - \half\kk';\KK')
= \overline{\bft_s(\kk - \half\kk' + \half\KK') 
\bft_k(-(\kk -\half\kk') + \half\KK')}. 
\EN
(Note that the terms involving
$k'_l\meanBft_l(\kk')$ above are zero, 
since $\nab\cdot\meanBB=0$.)
We now expand $P_{js}(\kk +\half\KK' +\half\kk')$
about $\KK'=0$ and both $P_{js}$ and ${\cal M}_{lk}$ about
$\kk' =0$, keeping at most terms
linear in $\KK'$ and  $\kk'$ respectively to get
\EQA
{\partial\chi_{jk}^M \over \partial t} &=& \int
\Big\{ P_{js}(\kk) \ \ii k_l\meanBft_l(\kk')\left[
{\cal M}_{sk}(\kk;\KK') - {k'_t \over 2} {\partial {\cal M}_{sk}(\kk;\KK')
\over \partial k_t} \right]
\nonumber \\
&+& P_{js}(\kk) \ \ii(k'_l + \half K'_l) \ 
\left[ \meanBft_l(\kk') \ {\cal M}_{sk}(\kk;\KK')
+ \meanBft_s(\kk') \ {\cal M}_{lk}(\kk;\KK') \right]
\nonumber \\
&-& {\ii(k'_s + K'_s)k_j \over 2k^2} (\kk\cdot\meanBBft)
\ {\cal M}_{sk}(\kk;\KK') \Big\}\,
\dd^3K' \ \dd^3k' \e^{\ii(\KK'+\kk')\cdot\RR}.
\label{chi_mag3}
\ENA
We have also used here the fact that $k_sm_{sk} = \half\ii(\partial v_{sk}/
\partial R_s)$ above, to neglect terms of the form
$k'_jk_sm_{sk}$ and $K'_jk_sm_{sk}$, which will lead
to terms with two derivatives with respect to $\RR$.
(Note that in homogeneous turbulence with $\nab\cdot\bb=0$
one will have $k_sm_{sk} =0$.) 

Integrating over $\KK'$ and $\kk'$, and using the definition
of $m_{ij}$ we finally get
\EQA
{\partial\chi_{jk}^M \over \partial t} &=&
\ii\kk\cdot\meanBB\,m_{jk} + \half\meanBB\cdot\nab m_{jk}
+\meanB_{j,l} m_{lk} \nonumber \\
&-& \half \meanB_{m,s} k_m
{\partial m_{jk} \over \partial k_s}
-2{k_jk_s \over k^2} \meanB_{s,l} m_{lk}.
\ENA

\section{Calculation of the second and third terms in \eq{chi_sol3}}
\label{SecondThirdEMFOmega}

In the calculation of turbulent transport coefficients in the case
where helicity is produced by rotation and gradients of the turbulence
intensity in \Sec{HelicityInducedByRotation}, we had to evaluate the
contributions from the second and third terms in \eq{chi_sol3}
to $\meanemf$ for non-zero $\Omega$. 
The somewhat cumbersome derivation is presented below.

The contribution due to the second term is:
\EQA
\meanemf^{(\Omega 2)}_i &=& 2\epsilon_{ijk}\epsilon_{jlm}
\int \tau^2 \ \kkk\cdot\OO \ \kunit_m {\sf I}_{lk}\,\dd^3k \nonumber \\ 
&=& 2\epsilon_{ijk} \epsilon_{jlm} 
\int \tau^2 \ \kkk\cdot\OO \ \kunit_m 
\Big[-\ii\kk\cdot\meanBB\,(v_{lk}^{(0)}- m_{lk})
+ \half\meanBB\cdot\nab(v_{jk}^{(0)} + m_{jk}) \nonumber \\
&+& \meanB_{l,s} m_{sk} - \meanB_{k,s} v_{ls}^{(0)}
- \half k_s \meanB_{s,\alpha} 
\left({\partial v_{lk}^{(0)}/\partial k_\alpha}
+ {\partial m_{lk}/\partial k_\alpha}\right)\Big]\,\dd^3k.
\label{emf_om21}
\ENA
Note that due to the presence of the antisymmetric
tensor $\epsilon_{jlm}$  the last term of ${\sf I}_{lk}$ does not
contribute to $\meanemf^{(\Omega 2)}_i$. Further note
that to linear order in $\Omega$, the velocity
correlation $v_{lk}$ can be replaced by the nonhelical, 
isotropic, velocity correlation of the original turbulence 
$v_{lk}^{(0)}$. Substituting the general form of the magnetic and
velocity correlations, and noting that only terms which have
an even number of $\kunit_i$ survive the angular
integrations, we have
\EQA
\meanemf^{(\Omega 2)}_i 
&=& 2\epsilon_{ijk} \epsilon_{jlm} 
\int \tau^2 \ \kkk\cdot\OO \ \kunit_m
\Big\{-\kk\cdot\meanBB\half\kunit_k \nabla_l(E^{(0)} -M) 
+ \kk\cdot\meanBB \epsilon_{lkn}\kunit_n N  \nonumber \\
&+& \half\meanBB\cdot\nab[\delta_{lk} (E^{(0)} + M)]
+ \meanB_{l,s} P_{sk} M - \meanB_{k,s} \delta_{ls} E^{(0)}
\nonumber \\
&-& \half \kunit_s \meanB_{s,\alpha} 
[\kunit_\alpha \delta_{lk} k(E^{(0)} + M)' -
\delta_{l\alpha} \kunit_k (E^{(0)} + M)]\Big\}\dd^3k,
\label{emf_om22}
\ENA
where primes denote derivatives with respect to $k$.
The $N$ dependent term vanishes on doing the angular integrals,
and the other terms in \eq{emf_om22} simplify to
\EQA
\meanemf^{(\Omega 2)}_i &=&
-{\textstyle{1\over15}}\meanB_i(\OO\cdot\nab)(\tilde{E}^{(2)} - \tilde{M}^{(2)})
+{\textstyle{4\over15}}\OO\cdot\meanBB \
\nabla_i(\tilde{E}^{(2)}-\tilde{M}^{(2)})
\nonumber \\
&+& {\textstyle{3\over5}}\Omega_i \ \meanBB\cdot\nab(\tilde{E}^{(2)} 
+ {\textstyle{11\over9}}\tilde{M}^{(2)}) 
+ \Omega_l (\tilde{E}^{(2)} - \tilde{M}^{(2)})
\left[{\textstyle{1\over6}}\meanB_{[l,i]} + {\textstyle{7\over30}}  
\meanB_{(l,i)}\right]
\nonumber \\
&-& {\textstyle{2\over15}} \Omega_l \meanB_{(l,i)}
( \tilde{E}^{(3)} + \tilde{M}^{(3)}) .
\label{emf_om2f}
\ENA
Here, $\meanB_{(l,i)}\equiv\meanB_{l,i}+\meanB_{i,l}$,
$\meanB_{[l,i]}\equiv\meanB_{l,i}-\meanB_{i,l}$, and
$\tilde{E}^{(2)}$, ..., $\tilde{M}^{(3)}$ have been
defined in \Eq{tilde_E2}.

For a constant $\tau(k) = \tau_0$ say, we simply have
$\tilde{E}^{(2)} = \half\tau^2_0 \overline{\uu^{(0)2}}$,
$\tilde{M}^{(2)} = \half\tau^2_0 \overline{\bb^2}$,
while $\tilde{E}^{(3)}$ and $\tilde{M}^{(3)}$ depend
on the velocity and magnetic spectra.

Now consider the third term in \eq{chi_sol3}.
Using \Eqs{Bjlm}{I0approx} this is given by,
\EQA
\meanemf^{(\Omega 3)}_i &=&
-\epsilon_{ijk} \int \tau^2 \ B_{jlm} I^{(0)}_{lk,m} \dd^3k
\nonumber \\
&=& -\epsilon_{ijk} \int \tau^2 \left(
\epsilon_{jlm} \kkk\cdot\OO  
+\epsilon_{jlt} \kunit_t\Omega_m 
-2\epsilon_{jlt} \kkk\cdot\OO\,\kunit_t \kunit_m\right)
\nonumber \\ 
&\times& {\partial \over \partial R_m}\left[
\kkk\cdot\meanBB(v_{lk}^{(0)}- m_{lk})\right] \dd^3k.
\label{emf_om3}
\ENA
Using \Eqs{vel_cor}{mag_cor},
only those terms in $v_{lk}^{(0)}$ and $m_{lk}$ 
which contain an even number of $\kunit_i$
survive after doing the angular integrals over
$\kk$ in \eq{emf_om3}. Further, those terms which
contain a spatial derivative do not contribute,
since \eq{emf_om3} already contains a spatial derivative,
and we are keeping only terms up to first order in the $\RR$
derivatives. Also, in those terms that have $\epsilon_{jlt}$,
the part of the velocity and magnetic correlations
proportional to $\kunit_l$ give zero contribution.
Using these properties, and doing the resulting angular integrals, 
we get
\EQA
\meanemf^{(\Omega 3)}_i &=&
-{\textstyle{7\over15}}\OO\cdot\nab[\meanB_i(\tilde{E}^{(2)}-\tilde{M}^{(2)})]
-{\textstyle{2\over 15}}\Omega_l \nabla_i 
[\meanB_l(\tilde{E}^{(2)} - \tilde{M}^{(2)})]
\nonumber \\ 
&+& {\textstyle{1\over5}}\Omega_i \nabla_l [
\meanB_l(\tilde{E}^{(2)} - \tilde{M}^{(2)})].
\label{emf_om3f}
\ENA

\section{Comparison with the paper by R\"adler, Kleeorin, and Rogachevskii}
\label{ComparisonRKR}

The results presented at the end of \Sec{HelicityInducedByRotation}
are modified if one were to assume a power law form for the
spectrum of velocity and magnetic 
correlations in the range $ k_0 < k < k_d$
(as done in Ref.~\cite{RKR03}),
\EQ
4\pi k^2 E^{(0)} = (q-1) 
{\overline{\uu^{(0)2}} \over 2 k_0}\left({k \over k_0}\right)^{\!-q},
\quad
4\pi k^2 M = (q-1) 
{\overline{\bb^2} \over 2 k_0}\left({k \over k_0}\right)^{\!-q},
\label{spectra}
\EN
and also take $\tau(k) = \tau^*(k) = 2\tau_0 (k/k_0)^{1-q}$,
as in Ref.~\cite{RKR03}, then
\EQ
\tilde{E}^{(2)} = {\textstyle{4\over3}}\tau_0^2{\overline{\uu^{(0)2}}
\over 2}, \quad
\tilde{M}^{(2)} = {\textstyle{4\over3}}\tau_0^2{\overline{\bb^{2}} \over 2},
\EN
and hence all the $\tau_0^2$ terms
in \Eqss{alpha_tau}{kappa_tau},
would have to be multiplied by $4/3$. In this case one
also has $\tilde{E}^{(3)} = -(q+2)\tilde{E}^{(2)}$
and $\tilde{M}^{(3)} = -(q+2)\tilde{M}^{(2)}$,
so one has for the last term in \Eq{kappa_tau}
\EQ
{\textstyle{4\over30}} \left[\tilde{E}^{(3)} + \tilde{M}^{(3)}\right]
= -{\textstyle{8\over45}} (q+2) \tau_0^2 \left({\overline{\uu^{(0)2}} +
\overline{\bb^2} \over 2}\right).
\label{kappa_der}
\EN
This leads to
\EQA
\alpha_{ij} = \onethird \tau_{\rm eff}
\overline{\jj\cdot\bb}
&-&{\textstyle{16\over15}} \tau_0^2 \Big[\delta_{ij} \OO\cdot\nab
(\overline{\uu^{(0)2}} - \onethird \overline{\bb^2})
\nonumber \\ 
&-&{\textstyle{11\over24}} (\Omega_i\nabla_j + \Omega_j\nabla_i)
(\overline{\uu^{(0)2}} + {\textstyle{3\over11}}
\overline{\bb^2})\Big],
\label{alpha_tauk}
\ENA
\EQ
\eta_{ij} = \onethird \tau_0 \delta_{ij} \ \overline{\uu^{(0)2}},
\label{eta_tauk}
\EN
\EQ
\ggamma = - \onesixth \tau_0 \nab \left(\overline{\uu^{(0)2}} -
\overline{\bb^2}\right)
- {\textstyle{2\over9}} \tau_0^2  \OO \times \nab
\left(\overline{\uu^{(0)2}} + 
\overline{\bb^2}\right),
\label{gamma_tauk}
\EN
\EQ
\ddelta = {\textstyle{2\over9}} \OO \tau_0^2 
\left(\overline{\uu^{(0)2}} - 
\overline{\bb^2}\right),
\label{delta_tauk}
\EN
\EQ
\kappa_{ijk} = {\textstyle{2\over9}}
\tau_0^2(\Omega_j \delta_{ik} + \Omega_k\delta_{ij})\left\{
[\overline{\uu^{(0)2}}+{\textstyle{7\over 5}}\overline{\bb^2}]
+{\textstyle{2\over5}}(q-1)[\overline{\uu^{(0)2}}+\overline{\bb^2}]\right\}.
\label{kappa_tauk}
\EN
Here $\tau_{\rm eff} = 2\tau_0 (2-q)/(2q-3) (k_0/k_d)^{2-q}$
for the spectral dependences which have been assumed, and
$\tau_{\rm eff} = 2\tau_0 (k_0/k_d)^{1/3}$ for a Kolmogorov spectrum.
Our results for the rotation dependent terms in
$\alpha$, $\ggamma$, $\ddelta$ (and $\kappa$ when $q=1$), 
all agree exactly with the result of \cite{RKR03}; however we get
a different coefficient ($2/5$ instead of $-4/5$) in front of the term
proportional to $(q-1)$ in the expression for $\kappa_{ijk}$.

\section{Calculation of $\Psi_1$ and $\Psi_2$}
\label{Psi1Psi2}

In order to evaluate $\Psi_1$ and $\Psi_2$ in the calculation of the
nonlinear helicity fluxes of \Sec{nonlinearflux}
we use the induction equation \Eq{bhat} for the fluctuating field.
We then get for $\Psi_1$
\EQA
\Psi_1 =\int && \Big\{
\epsilon_{ijk}\epsilon_{ipq}\epsilon_{qlm}\int \int\int
\overline{\uft_l(\kk + \half\KK -\kk')
\bft_k(-\kk +\half\KK)} \meanBft_m(\kk')
\nonumber \\
&& \ii(k_p + \half K_p) 
\ii(-k_j + \half K_j) \ {\rm e}^{\ii\KK\cdot\RR } 
\,\dd^3K \ \dd^3k' + T_1(k) \Big\}\,\dd^3k.
\ENA
Here, $T_1(k)$ subsumes the triple correlations of the small scale
$\uu$ and $\bb$ fields and the microscopic
diffusion terms that one gets on substituting \Eq{bhat} into 
\Eq{psi12}. 
We transform from the variables $(\kk',\KK)$ to
a new set $(\kk',\KK')$ where $\KK' = \KK - \kk'$, use the
definition of the velocity-magnetic field correlation
$\chi_{lk}(\kk,\RR)$, and carry out the integral over $\KK'$
to write
\EQA
\Psi_1 = \int && \Big\{ 
\epsilon_{ijk}\epsilon_{ipq}\epsilon_{qlm} \int
\dd^3k' \ {\rm e}^{\ii\kk'\cdot\RR } 
\ \meanBft_m(\kk')
\ (\ii k_p + \half\ii {k'}_p + \half\nabla_p)
\nonumber \\
&&
 (-\ii k_j + \half\ii {k'}_j + \half \nabla_j ) 
\chi_{lk}(\kk - \half\kk', \RR)
+ T_1(k)\Big\}\,\dd^3k.
\ENA
Once again, since $\meanBB$ varies only on large scales,
$\meanBft(k')$ only contributes at small $k'$. One can then make
a small $k'$ expansion of $\chi_{lk}$, and do the $k'$ integral,
retaining only terms which are linear in the large scale
derivatives, to get
\EQA
\Psi_1 = \int && \Big\{ \epsilon_{ijk}\epsilon_{ipq}\epsilon_{qlm} 
\Big[ k_pk_j ( \meanB_m \chi_{lk} +\half\ii \nabla_s\meanB_m
{\partial\chi_{lk} \over \partial k_s} )
+\half\ii (k_p \nabla_j\meanB_m 
\nonumber \\
&& 
- k_j \nabla_p\meanB_m)\chi_{lk}
+ \half\ii ( k_p \nabla_j\chi_{lk} - k_j\nabla_p\chi_{lk})\meanB_m
\Big] + T_1(k) \Big\}\,\dd^3k.
\label{psi1f}
\ENA
This agrees with the result given in \Eq{psi12f} for the upper sign.

We now turn to $\Psi_2$. Again, using the induction
equation \eq{bhat} for the fluctuating field, we get for
$\Psi_2$
\EQA
\Psi_2 = \int && \Big\{
\epsilon_{ijk}\epsilon_{kpq}\epsilon_{qlm}\int \int\int
\overline{\uft_l(-\kk + \half\KK -\kk')
\bft_i(\kk +\half\KK)} \meanBft_m(\kk')
\nonumber \\ 
&&
\ii(-k_p + \half K_p) 
\ii(-k_j + \half K_j) \ {\rm e}^{\ii\KK\cdot\RR } 
\,\dd^3K \ \dd^3k' + T_2(k) \Big\}\,\dd^3k,
\ENA
where $T_2(k)$ represents the triple correlations and the microscopic
diffusion terms that one gets on substituting \Eq{bhat} into 
\Eq{psi12}. We transform from the variables $(\kk',\KK)$ to
a new set $(\kk',\KK')$ where $\KK' = \KK - \kk'$. It is also convenient to 
change from the variables $(\kk,\kk')$ to
$(-\kk,-\kk')$. Under this change we have 
\EQA
\Psi_2 =\int \Big\{ 
\epsilon_{ijk}\epsilon_{kpq}\epsilon_{qlm}&& \int\int
\overline{\uft_l(\kk + \half\kk' + \half\KK')
\bft_i(-\kk -\half\kk' +\half\KK')}
\nonumber \\
&& \meanBft^*_m(\kk')
\ \ii[k_p + \half (K'_p-k'_p) ]
\ii(k_j + \half( K'_j -k'_j) ) \nonumber \\
&&
{\rm e}^{\ii(\KK'-\kk')\cdot\RR }
\,\dd^3K' \ \dd^3k' + T_2(k) \Big\}\,\dd^3k. 
\ENA
We can now carry out the integral over $\KK'$
using the definition of the
velocity-magnetic field correlation 
$\chi_{li}(\kk,\RR)$, to get
\EQA
\Psi_2 = && \int \Big\{
\epsilon_{ijk}\epsilon_{kpq}\epsilon_{qlm} \int \dd^3k'
{\rm e}^{-\ii\kk'\cdot\RR }\ \meanBft^*_m(\kk')
\ (\ii k_p - \half\ii {k'}_p + \half\nabla_p)
\nonumber \\
&&
(\ii k_j -\half \ii {k'}_j + \half \nabla_j ) 
\chi_{li}(\kk + \half\kk', \RR)
+ T_2(k) \Big\}\,\dd^3k.
\ENA
Once again, since $\meanBB$ varies only on large scales,
$\meanBft(k')$ only contributes at small $k'$. One can then make
a small $k'$ expansion of $\chi_{lk}$, and do the $k'$ integral,
retaining only terms which are linear in the large scale
derivatives, to get
\EQA
\Psi_2 =\int && \Big\{
\epsilon_{ijk}\epsilon_{kpq}\epsilon_{qlm} 
\Big[- k_pk_j ( \meanB_m \chi_{li} +\half\ii \nabla_s\meanB_m
{\partial\chi_{li} \over \partial k_s} )
+\half\ii (k_p \nabla_j\meanB_m 
\nonumber \\
&&
+ k_j \nabla_p\meanB_m)\chi_{li}
+ \half\ii ( k_p\nabla_j\chi_{li} +  k_j\nabla_p\chi_{li})\meanB_m \Big]
+ T_2(k) \Big\}\,\dd^3k.
\label{psi2f}
\ENA
We have used here
\EQ
\int \meanBft^*_m(\kk') {\rm e}^{-\ii\kk'\cdot\RR } \dd^3k' 
= \int \meanBft_m(\kk') {\rm e}^{\ii\kk'\cdot\RR } \dd^3k'
= \meanB_m(\RR),
\EN
\EQ
\int (-\ii k'_j)  \meanBft^*_m(\kk') {\rm e}^{-\ii\kk'\cdot\RR } \dd^3k'
= \int \ii k'_j \meanBft_m(\kk') {\rm e}^{\ii\kk'\cdot\RR } \dd^3k'
= \nabla_j\meanB_m(\RR).
\EN
Interchanging the indices $i$ and $k$ in $\Psi_2$
yields the result given in \Eq{psi12f} for the lower sign.


\vfill\bigskip\noindent\tiny\begin{verbatim}
$Header: /home/brandenb/CVS/tex/mhd/phys_rep/paper.tex,v 1.622 2005/06/13 12:46:47 brandenb Exp $
\end{verbatim}

\end{document}

%% file: gafd.tex
\newcommand{\EQ}{\begin{equation}}
\newcommand{\EN}{\end{equation}}
\newcommand{\EQA}{\begin{eqnarray}}
\newcommand{\ENA}{\end{eqnarray}}
\newcommand{\eq}[1]{(\ref{#1})}
\newcommand{\EEq}[1]{Equation~(\ref{#1})}
\newcommand{\Eq}[1]{Eq.~(\ref{#1})}
\newcommand{\Eqs}[2]{Eqs.~(\ref{#1}) and~(\ref{#2})}
\newcommand{\EEqs}[2]{Equations~(\ref{#1}) and~(\ref{#2})}
\newcommand{\eqs}[2]{(\ref{#1}) and~(\ref{#2})}
\newcommand{\Eqss}[2]{Eqs.~(\ref{#1})--(\ref{#2})}

\newcommand{\Sec}[1]{Section~\ref{#1}}
\newcommand{\Secs}[2]{Sections~\ref{#1} and~\ref{#2}}

\newcommand{\App}[1]{Appendix~\ref{#1}}
\newcommand{\Fig}[1]{Fig.~\ref{#1}}
\newcommand{\FFig}[1]{Figure~\ref{#1}}
\newcommand{\Tab}[1]{Table~\ref{#1}}
\newcommand{\Figs}[2]{Figs.~\ref{#1} and \ref{#2}}

\newcommand{\bra}[1]{\langle #1\rangle}
\newcommand{\bbra}[1]{\left\langle #1\right\rangle}
\newcommand{\mean}[1]{\overline #1}
\newcommand{\meanemfs}{\overline{\cal E} {}}
\newcommand{\meanemf}{\overline{\mbox{\boldmath ${\mathcal E}$}} {}}
\newcommand{\meanA}{\overline{A}}
\newcommand{\meanB}{\overline{B}}
\newcommand{\meanT}{\overline{T}}

\newcommand{\meanJ}{\overline{J}}
\newcommand{\meanU}{\overline{U}}
\newcommand{\meanF}{\overline{\cal F}}
\newcommand{\meanFF}{\overline{\mbox{\boldmath ${\cal F}$}} {}}
\newcommand{\meanAA}{\overline{\mbox{\boldmath $A$}}}
\newcommand{\meanBB}{\overline{\bm{B}}}
\newcommand{\meanTT}{\overline{\bm{T}}}
\newcommand{\meanJJ}{\overline{\mbox{\boldmath $J$}}}
\newcommand{\meanEE}{\overline{\mbox{\boldmath $E$}}}
\newcommand{\meanUU}{\overline{\mbox{\boldmath $U$}}}
\newcommand{\meanWW}{\overline{\mbox{\boldmath $W$}}}

\newcommand{\mod}[1]{\mid\!\!#1\!\!\mid}

\newcommand{\kh}{\hat{k}}

\newcommand{\kkk}{\hat{\bm{k}}}
\newcommand{\ggrav}{\hat{\bm{g}}}
\newcommand{\pphi}{\hat{\bm{\phi}}}
\newcommand{\OOO}{\hat{\bm{\Omega}}}
\newcommand{\nnn}{\hat{\mbox{\boldmath $n$}} {}}

\newcommand{\zz}{\hat{\mbox{\boldmath $z$}} {}}

\newcommand{\kunit}{\hat{\mbox{$k$}} {}}
\newcommand{\nunit}{\hat{\mbox{$n$}} {}}
\newcommand{\uft}{\hat{\mbox{$u$}} {}}
\newcommand{\bft}{\hat{\mbox{$b$}} {}}
\newcommand{\meanBft}{\hat{\overline{\mbox{$B$}}} {}}
\newcommand{\meanBBft}{\hat{\overline{\mbox{\boldmath $B$}}} {}}
\newcommand{\Gft}{\hat{\mbox{$G$}} {}}
\newcommand{\Hft}{\hat{\mbox{$H$}} {}}

\newcommand{\rrr}{\bm{r}}
\newcommand{\xx}{\bm{x}}
\newcommand{\yy}{\bm{y}}
\newcommand{\zzz}{\bm{z}}
\newcommand{\uu}{\bm{u}}
\newcommand{\vv}{\bm{v}}

\newcommand{\RR}{\bm{R}}
\newcommand{\UU}{\bm{U}}
\newcommand{\VV}{\bm{V}}

\newcommand{\bb}{\bm{b}}
\newcommand{\qq}{\bm{q}}
\newcommand{\bp}{\bm{p}}
\newcommand{\BB}{\bm{B}}
\newcommand{\HH}{\bm{H}}

\newcommand{\AAA}{\bm{A}}
\newcommand{\aaa}{\bm{a}}
\newcommand{\jj}{\bm{j}}
\newcommand{\JJ}{\bm{J}}
\newcommand{\cc}{\bm{c}}
\newcommand{\nn}{\bm{n}}
\newcommand{\ee}{\bm{e}}

\newcommand{\ff}{\bm{f}}
\newcommand{\EE}{\bm{E}}
\newcommand{\FF}{\bm{F}}

\newcommand{\KK}{\bm{K}}

\newcommand{\GG}{\bm{G}}
\newcommand{\hh}{\bm{h}}
\newcommand{\kk}{\bm{k}}
\newcommand{\pp}{\bm{p}}
\newcommand{\lv}{\bm{l}}
\newcommand{\SSS}{\bm{S}}
\newcommand{\grav}{\bm{g}}
\newcommand{\nab}{\bm{\nabla}}
\newcommand{\OO}{\bm{\Omega}}
\newcommand{\oo}{\bm{\omega}}
\newcommand{\ggamma}{\bm{\gamma}}
\newcommand{\ddelta}{\bm{\delta}}

 {}

\newcommand{\SSSS}{\mbox{\boldmath ${\sf S}$} {}}

\newcommand{\MMMM}{\mbox{\boldmath ${\sf M}$} {}}

\newcommand{\emf}{\mbox{\boldmath ${\cal E}$} {}}

\newcommand{\QQQ}{\bm{Q}}
\newcommand{\WWW}{\bm{W}}
\newcommand{\GGGG}{{\bf G} {}}
\newcommand{\ii}{\mathrm{i}}

\newcommand{\sgn}{{\rm sgn}{}}

\newcommand{\DD}{\mathrm{D}}
\newcommand{\dd}{\mathrm{d}}
\newcommand{\const}{{\rm const}  {}}

\def\la{\mathrel{\mathchoice {\vcenter{\offinterlineskip\halign{\hfil
$\displaystyle##$\hfil\cr<\cr\sim\cr}}}
{\vcenter{\offinterlineskip\halign{\hfil$\textstyle##$\hfil\cr<\cr\sim\cr}}}
{\vcenter{\offinterlineskip\halign{\hfil$\scriptstyle##$\hfil\cr<\cr\sim\cr}}}
{\vcenter{\offinterlineskip\halign{\hfil$\scriptscriptstyle##$\hfil\cr<\cr\sim\cr}}}}}
\def\ga{\mathrel{\mathchoice {\vcenter{\offinterlineskip\halign{\hfil
$\displaystyle##$\hfil\cr>\cr\sim\cr}}}
{\vcenter{\offinterlineskip\halign{\hfil$\textstyle##$\hfil\cr>\cr\sim\cr}}}
{\vcenter{\offinterlineskip\halign{\hfil$\scriptstyle##$\hfil\cr>\cr\sim\cr}}}
{\vcenter{\offinterlineskip\halign{\hfil$\scriptscriptstyle##$\hfil\cr>\cr\sim\cr}}}}}
\def\Co{\mbox{\rm Co}}

\def\half{{\textstyle{1\over2}}}
\def\threehalf{{\textstyle{3\over2}}}
\def\onethird{{\textstyle{1\over3}}}
\def\onesixth{{\textstyle{1\over6}}}
\def\twothird{{\textstyle{2\over3}}}
\def\twofifth{{\textstyle{2\over5}}}
\def\fourthird{{\textstyle{4\over3}}}

\def\threequarter{{\textstyle{3\over4}}}

\newcommand{\G}{\,{\rm G}}
\newcommand{\rad}{\,{\rm rad}}
\newcommand{\muG}{\,\mu{\rm G}}
\newcommand{\mG}{\,{\rm mG}}

\newcommand{\nHz}{\,{\rm nHz}}

\newcommand{\kG}{\,{\rm kG}}
\newcommand{\K}{\,{\rm K}}
\newcommand{\g}{\,{\rm g}}
\newcommand{\s}{\,{\rm s}}

\newcommand{\cm}{\,{\rm cm}}
\newcommand{\m}{\,{\rm m}}
\newcommand{\km}{\,{\rm km}}

\newcommand{\Mm}{\,{\rm Mm}}
\newcommand{\Mx}{\,{\rm Mx}}
\newcommand{\pc}{\,{\rm pc}}
\newcommand{\kpc}{\,{\rm kpc}}
\newcommand{\Mpc}{\,{\rm Mpc}}
\newcommand{\yr}{\,{\rm yr}}

\newcommand{\Gyr}{\,{\rm Gyr}}
\newcommand{\erg}{\,{\rm erg}}

\newcommand{\AU}{\,{\rm AU}}

\newcommand{\yan}[5]{, #5, {Astron.\ Nachr.\ }{#2} (#1) #3--#4.}

\newcommand{\yana}[5]{, #5, {Astron.\ Astrophys.\ }{#2} (#1) #3--#4.}

\newcommand{\yanas}[5]{, #5, {Astron.\ Astrophys.\ Suppl.\ }{#2} (#1) #3--#4.}

\newcommand{\yass}[5]{, #5, {Astrophys.\ Spa.\ Sci.\ }{#2} (#1) #3--#4.}
\newcommand{\yijmpd}[5]{, #5, {Int.\ Jour.\ Mod. \ Phys. \ D}{#2} (#1) #3--#4.}
\newcommand{\ysci}[5]{, #5, {Science }{#2} (#1) #3--#4.}
\newcommand{\ysph}[5]{, #5, {Solar Phys.\ }{#2} (#1) #3--#4.}
\newcommand{\ysphS}[5]{, #5 {Solar Phys.\ }{#2} (#1) #3--#4.}
\newcommand{\yjetp}[5]{, #5, {Sov.\ Phys.\ JETP }{#2} (#1) #3--#4.}

\newcommand{\ysov}[5]{, #5, {Sov.\ Astron.\ }{#2} (#1) #3--#4.}

\newcommand{\ymn}[5]{, #5, {Monthly Notices Roy.\ Astron.\ Soc.\ }{#2} (#1) #3--#4.}
\newcommand{\ymnS}[5]{, #5 {Monthly Notices Roy.\ Astron.\ Soc.\ }{#2} (#1) #3--#4.}
\newcommand{\yqjras}[5]{, #5, {Quart.\ J.\ Roy.\ Astron.\ Soc.\ }{#2} (#1) #3--#4.}
\newcommand{\ynat}[5]{, #5, {Nature }{#2} (#1) #3--#4.}
\newcommand{\ynewa}[5]{, #5, {New\ Astronomy\ }{#2} (#1) #3--#4.}
\newcommand{\sjfm}[2]{, #2, {J.\ Fluid Mech.\ } submitted (#1).}

\newcommand{\yjfm}[5]{, #5, {J.\ Fluid Mech.\ }{#2} (#1) #3--#4.}
\newcommand{\yphl}[5]{, #5, {Phys.\ Lett.\ }{#2} (#1) #3--#4.}
\newcommand{\ypr}[5]{, #5, {Phys.\ Rev.\ }{#2} (#1) #3--#4.}
\newcommand{\yprd}[5]{, #5, {Phys.\ Rev.\ D \ }{#2} (#1) #3--#4.}
\newcommand{\ypre}[5]{, #5, {Phys.\ Rev.\ E \ }{#2} (#1) #3--#4.}
\newcommand{\yprdN}[4]{, #4, {Phys.\ Rev.\ D \ }{#2} (#1) #3.}
\newcommand{\ypreN}[4]{, #4, {Phys.\ Rev.\ E \ }{#2} (#1) #3.}
\newcommand{\yprep}[5]{, #5, {Phys.\ Rep.\ }{#2} (#1) #3--#4.}

\newcommand{\yprl}[5]{, #5, {Phys.\ Rev.\ Lett.\ }{#2} (#1) #3--#4.}
\newcommand{\yprlN}[4]{, #4, {Phys.\ Rev.\ Lett.\ }{#2} (#1) #3.}

\newcommand{\yprs}[5]{, #5, {Proc.\ Roy.\ Soc.\ Lond.\ }{#2} (#1) #3--#4.}
\newcommand{\yptrs}[5]{, #5, {Phil.\ Trans.\ Roy.\ Soc.\ }{#2} (#1) #3--#4.}

\newcommand{\yjgr}[5]{, #5, {J.\ Geophys.\ Res.\ }{#2} (#1) #3--#4.}

\newcommand{\yaj}[5]{, #5, {Astronom.\ J.\ }{#2} (#1) #3--#4.}
\newcommand{\yapj}[5]{, #5, {Astrophys.\ J.\ }{#2} (#1) #3--#4.}
\newcommand{\yapjS}[5]{, #5 {Astrophys.\ J.\ }{#2} (#1) #3--#4.}
\newcommand{\yapjs}[5]{, #5, {Astrophys.\ J.\ Suppl.\ }{#2} (#1) #3--#4.}
\newcommand{\yapjl}[5]{, #5, {Astrophys.\ J.\ Lett.\ }{#2} (#1) #3--#4.}
\newcommand{\yapjlS}[5]{, #5 {Astrophys.\ J.\ Lett.\ }{#2} (#1) #3--#4.}
\newcommand{\ypp}[5]{, #5, {Phys.\ Plasmas }{#2} (#1) #3--#4.}

\newcommand{\yannr}[5]{, #5, {Ann.\ Rev.\ Astron.\ Astrophys.\ }{#2} (#1) #3--#4.}
\newcommand{\yaraa}[5]{, #5, {Ann.\ Rev.\ Astron.\ Astrophys.\ }{#2} (#1) #3--#4.}
\newcommand{\yanar}[5]{, #5, {Astron.\ Astrophys.\ Rev.\ }{#2} (#1) #3--#4.}
\newcommand{\yanf}[5]{, #5, {Ann.\ Rev.\ Fluid Dyn.\ }{#2} (#1) #3--#4.}
\newcommand{\ypf}[5]{, #5, {Phys.\ Fluids }{#2} (#1) #3--#4.}
\newcommand{\yphy}[5]{, #5, {Physica } {#2} (#1) #3--#4.}
\newcommand{\ygafd}[5]{, #5, {Geophys.\ Astrophys.\ Fluid Dynam. }{#2} (#1) #3--#4.}

\newcommand{\yjourN}[5]{, #5, {#2} {#3} (#1) #4.}

\newcommand{\yjourS}[6]{, #6 {#2} {#3} (#1) #4--#5.}
\newcommand{\yjour}[6]{, #6, {#2} {#3} (#1) #4--#5.}

\newcommand{\pproc}[5]{, #2, In {#3} (ed.\ #4), #5 (to appear) #1.}
\newcommand{\yprocN}[6]{, #3, In {#4} (ed.\ #5), p.\ #2.\ #6, #1.}
\newcommand{\yproc}[7]{, #4, In {#5} (ed.\ #6), pp.\ #2--#3.\ #7, #1.}
\newcommand{\ybook}[3]{, {#2}.\ #3, #1.}

\newcommand{\tgafd}[2]{, #2, {Geophys.\ Astrophys.\ Fluid Dynam.\ } (#1) in preparation.}
\newcommand{\pana}[2]{, #2, {Astron.\ Astrophys.\ } (#1) in press.}

\newcommand{\spf}[2]{, #2, {Phys.\ Fluids } (#1) submitted.}
\newcommand{\sapj}[2]{, #2, {Astrophys.\ J.\ } (#1) submitted.}

\newcommand{\smn}[2]{, #2, {Monthly Notices Roy.\ Astron.\ Soc.\ } (#1) submitted.}

\newcommand{\SHADOWBOX}[1]{\shadowbox{\parbox{13.4cm}{#1}}}

%% file: paper.bbl
\begin{thebibliography}{999}
\bibitem{Aake_priv}
We believe this interpretation of the Babcock-Leighton mechanism
and a sketch similar to \Fig{babcock+leighton} is originally due
to {\AA}ke Nordlund.
\end{thebibliography}
